\documentclass[12pt,twoside]{book}
\usepackage[utf8]{inputenc}

\usepackage{libertine}
\usepackage[normalem]{ulem}
\usepackage{dsfont}
\usepackage{amsfonts}
\usepackage{amsmath}
\usepackage{amssymb}
\usepackage{amsthm}
\usepackage{mathtools}
\usepackage{bm}
\usepackage[T1]{fontenc}
\usepackage{dsfont,amsthm,enumerate,amsfonts}
\usepackage[spanish,english]{babel}
\usepackage[table,xcdraw]{xcolor}
\usepackage{graphicx}
\usepackage[font=scriptsize,labelfont=bf,justification=justified,format=plain,singlelinecheck=false]{caption}
\usepackage{subcaption}
\usepackage{floatrow}
\definecolor{myColor}{HTML}{9B3035}
\definecolor{GIQOrange}{HTML}{C74B2A}

\usepackage{apptools}
\usepackage{titlesec}
\titleformat
{\chapter} % command
[display] % shape
{\bfseries\Huge} % format
{\filleft\,\scalebox{4}{\color{lightgray}\thechapter}} % label
{0.5ex} % sep
{   \rule{\textwidth}{1pt}
    \vspace{0.5 ex}
} % before-code
[
\vspace{-0.5 ex}%
\rule{\textwidth}{1pt}
] % after-code

\usepackage[bibstyle=phys,citestyle=alphabetic,minalphanames=3,sorting=ynt,backend=biber]{biblatex}
\usepackage[hidelinks]{hyperref}
\hypersetup{
     colorlinks = true,
     citecolor = magenta,
     linkcolor = magenta,
     urlcolor = magenta}

\usepackage[a4paper,width=160mm,top=50mm,bottom=50mm,bindingoffset=6mm]{geometry}
\usepackage{fancyhdr}
\usepackage{physics}
\usepackage{ytableau}
\usepackage{graphicx}
\usepackage[export]{adjustbox}
\usepackage[compat=1.1.0]{tikz-feynman}
\usepackage{mathrsfs}
\usepackage{enumerate}
\usepackage{csquotes}
\usepackage[final]{microtype}
\usepackage{epigraph}
\usepackage{lipsum}

\usepackage[
    type={CC},
    modifier={by-nc-sa},
    version={3.0},
    imagemodifier={-eu-88x31}
]{doclicense}

\usetikzlibrary{decorations.pathreplacing}
\newcommand{\rao}{-{Stealth[length=3mm, open, round]}}
\newcommand{\rac}{-{Stealth[length=3mm, round]}}
\newcommand{\lao}{{Stealth[length=3mm, open, round]}}
\newcommand{\lac}{{Stealth[length=3mm, round]}}

\newcommand{\ball}[1]{\filldraw [color=black!, fill=white!] {#1} circle[radius=3pt] ; }
\newcommand{\Wn}[3]{\begin{tikzpicture}[baseline={([yshift=-.5ex]current bounding box.center)}]  
    \foreach \n in {1,...,#1}{
        \node at ({#2+\n*360/#1}:1cm) (n\n) {};
        \draw (0,0)--(n\n);  
    }; #3
\end{tikzpicture}}
\newcommand{\inWn}[3]{
    \foreach \n in {1,...,#1}{
        \node at ({#2+\n*360/#1}:0.8cm) (n\n) {};
        \draw #3--(n\n);  
    }; }
\newcommand{\qubit}[2]{\draw[markxi={#2}]#1 to ($#1+(0.4,0)$);
\draw ($#1+(0.2,0)$) ellipse  [x radius=0.2,y radius=0.1];}

\newcommand{\qubitl}[1]{  \qubit{#1}{0} ;
       \draw ($#1+(-0.25,0)$) to ($#1+(0,0)$);}

\newcommand{\qubitr}[1]{ \qubit{#1}{1} ;
       \draw ($#1+(0.65,0)$) to ($#1+(0.4,0)$);}

\newcommand{\inangarc}[3]{ \foreach \n in {1,...,#1}{
        \node at ($#2 +({#3+\n*180/(#1+1)}:1cm)$) (n\n) {};
        \draw #2--(n\n); 
    };}

\newcommand{\Genstate}[1]{\draw[fill=white, postaction={pattern =crosshatch}] #1 circle[radius=0.2];}
\newcommand{\Genstates}[1]{\draw #1  node[scale=1.5] {$\wr$};}
\tikzset{basel/.style n args={1}{baseline={([yshift=#1ex]current bounding box.center)}
  },
  markxi/.style n args={1}{decorate,decoration={markings,
    mark= at position #1
      with
      {\filldraw [color=black!, fill=black!]  circle(2pt);}}}
      ,markx/.style n args={1}{postaction={decoration={markings,
    mark= at position #1
      with
     with {\filldraw [color=black!, fill=black!]  circle(2pt);}},
      decorate}
      },
    markxo/.style n args={2}{postaction={decoration={markings,
    mark= at position #1
      with
      {\filldraw [color=black!, fill=black!]  circle(2pt);},
    mark= at position #2 
        with 
        { \filldraw[color=black!, fill=white!] (2.5pt,2.5pt) circle(2pt);
        }},
        decorate}
        },markxoi/.style n args={2}{  decorate,decoration={markings,
    mark= at position #1
      with
      {\filldraw [color=black!, fill=black!]  circle(2pt);},
    mark= at position #2 
        with 
        { \draw (2.5pt,2.5pt) circle[radius=0.1];
        }},
        },
    marko/.style n args={1}{postaction={decoration={markings,
    mark= at position #1 
      with 
    { \filldraw[color=black!, fill=white!] (2.5pt,2.5pt) circle(2pt);
    }},
    decorate}},
    markoi/.style n args={1}{decorate,decoration={markings,
    mark= at position #1 
      with 
    { \draw (2.5pt,2.5pt) circle[radius=0.1];
    }}}}

 \newcommand{\Kite}[1]{\scalebox{#1}{\begin{tikzpicture}[basel={-.5}]
   \draw (0,0) -- (0.5,0) ; 
   \draw[markx={0}] (0.5,0) -- (1,0.5) ;
   \draw[markx={1}] (0,0) -- (-0.5,0);
   \draw (-1,0.5) -- (-0.5,0);
   \draw (0.5,0) --(0,-1);\draw[markx={1}](0,-1.5)--(0,-1); \draw(0,-1) --(-0.5,0);
   \end{tikzpicture}}}
\newcommand{\TreeF}[1]{\scalebox{#1}{\begin{tikzpicture}[basel={-.5}]
   \draw (1,0) -- (0,0) ;
   \draw[markx={1}](-0.5,0.5) -- (0,0);
   \draw (-0.5,-0.5) -- (0,0);\draw (1.5,0.5) -- (1,0);
   \draw[markx={1}] (1.5,-0.5) -- (1,0);
	\end{tikzpicture}}}
\newcommand{\Kitetree}[1]{\scalebox{#1}{\begin{tikzpicture}[basel={-.5}]
    \draw[markx={0}](0,0) to (1,0);
    \draw(1,0) to (1.5,0.5);
    \draw(1,0) to (1.5,-0.5);
    \draw[markx={0},markx={1}](1.5,0.5) to (1.5,-0.5);
    \draw(1.5,0.5) to (1.75,0.75);
    \draw(1.5,-0.5) to (1.75,-0.75);
    \draw(-0.25,0.25) to (0,0);
    \draw(-0.25,-0.25) to (0,0);
    \draw[markxi={1}](0,0) to (1,0);
    \end{tikzpicture}}}
\newcommand{\Cuad}[1]{\scalebox{#1}{ \begin{tikzpicture}[basel={-1.5}]
        \draw[markx={1}] (-0.25,0.25) to (0,0);
        \draw (0,0) to (1,0);
        \draw [markx={0}](1,0) to (1.25,0.25);
        \draw (0,0) to (0,-1);
        \draw [markx={1}](-0.25,-1.25) to (0,-1);
        \draw (0,-1) to (1,-1);
        \draw (1,0) to (1,-1);
        \draw [markx={0}](1,-1) to (1.25,-1.25);
    \end{tikzpicture}} }
    
\newcommand{\Cam}[1]{\scalebox{#1}{\begin{tikzpicture}[basel={-.5}]
        \draw(0,0) to (1,0);
        \draw[markx={0},markx={1}] (0,0) to (0,-1);
        \draw (0,-1) to (1,-1);
        \draw (1,-1) to (2,0);
        \draw (1,0) to (2,-1);
        \draw (-0.25,0.25) to (0,0);
        \draw (0,-1) to (-0.25,-1.25);
        \draw (2,0) to (2.25,0.25);
        \draw[markx={0},markx={1}]  (1,0) to (2,0);
        \draw[markx={0},markx={1}]  (1,-1) to (2,-1);
        \draw (2,-1) to (2.25,-1.25);
    \end{tikzpicture}}}
    
 \newcommand{\CuadKite}[1]{\scalebox{#1}{ \begin{tikzpicture}[basel={-.5}]
        \draw (0,0) to (-0.25,0);
        \draw[markx={0},markx={1}](0,0) to  (1,1) ;
         \draw(0,0) to  (1,-1) ;
          \draw (1,1) to  (1.5,0.5) ;
          \draw [markx={0},markx={1}](1,-1) to  (1.5,-0.5) ;
          \draw [markx={0},markx={1}](1.5,0.5) to (2,0)  ;
           \draw (1.5,-0.5) to (2,0)  ;
            \draw (1.5,0.5) to (1.5,-0.5)  ;
            \draw (0.75,1.25) to (1,1);
            \draw (0.75,-1.25) to (1,-1);
            \draw (2.25,0) to (2,0);
    \end{tikzpicture}}}
    \newcommand{\DobKite}[1]{\scalebox{#1}{
 \begin{tikzpicture}[basel={-.5}]
 \draw (-0.5,0.5) to (0,0);
  \draw (-0.5,-0.5) to (0,0);
   \draw (-0.75,0.75) to (-0.5,0.5);
   \draw[markx={0},markx={1}] (-0.5,0.5) to (-0.5,-0.5); 
   \draw (-0.75,-0.75) to (-0.5,-0.5);
    \draw[markx={0}](0,0) to (1,0);
    \draw(1,0) to (1.5,0.5);
    \draw(1,0) to (1.5,-0.5);
    \draw[markx={0},markx={1}](1.5,0.5) to (1.5,-0.5);
    \draw(1.5,0.5) to (1.75,0.75);
    \draw(1.5,-0.5) to (1.75,-0.75);
    \draw[markxi={1}](0,0) to (1,0);
    \end{tikzpicture}}}
    \newcommand{\DobCuad}[1]{\scalebox{#1}{\begin{tikzpicture}[basel={-1.5},]
        \draw[markx={1}] (-0.25,0.25) to (0,0);
        \draw (0,0) to (1,0);
        \draw (0,0) to (0,-1);
        \draw [markx={1}](-0.25,-1.25) to (0,-1);
        \draw (0,-1) to (1,-1);
        \draw (1,0) to (1,-1);
        \draw [markx={0},markx={1}](1,-1) to (2,-1);
         \draw [markx={0},markx={1}](1,0) to (2,0);
         \draw (2,0) to (2,-1);
      \draw (2,0)  to (2.25,0.25);
       \draw (2,-1)  to (2.25,-1.25);
    \end{tikzpicture}}}
\usepackage{scalerel,stackengine}
\stackMath
\newcommand\reallywidehat[1]{%
\savestack{\tmpbox}{\stretchto{%
  \scaleto{%
    \scalerel*[\widthof{\ensuremath{#1}}]{\kern-.6pt\bigwedge\kern-.6pt}%
    {\rule[-\textheight/2]{1ex}{\textheight}}%WIDTH-LIMITED BIG WEDGE
  }{\textheight}% 
}{0.5ex}}%
\stackon[1pt]{#1}{\tmpbox}%
}
\usepackage{zx-calculus}
\definecolor{colorZxZ}{RGB}{0,0,0}
\definecolor{colorZxX}{RGB}{255,255,255}
\definecolor{colorZxH}{RGB}{255,255,0}

\newtheorem{theorem}{Theorem}
\theoremstyle{remark}

\newcommand{\myTitle}{
Kronecker states: a powerful source of multipartite maximally entangled states in quantum information}
\newcommand{\myName}{Walther Gonzalez}
\newcommand{\myProfessor}{Alonso Botero}

\newcommand{\myUniversity}{Universidad de los Andes}
\newcommand{\myFaculty}{Facultad de Ciencias}

\newcommand{\myDepartment}{Departamento de Física}
\newcommand{\myLocation}{Bogotá}

\def\straight{\tikz[baseline=-0.3ex]{
\fill (0,0) circle (1pt) coordinate (A);
\fill (1.5ex,0) circle (1pt) coordinate (B);
\draw (A)--(B);}
}
\newcommand{\lamtup}{\boldsymbol{\lambda}}
\newcommand{\rtup}{\boldsymbol{r}}

\newcommand{\omegatup}{\boldsymbol{\omega}}

\newcommand{\ctup}{\boldsymbol{c}}

\DeclarePairedDelimiter{\ceil}{\lceil}{\rceil}
\begin{document}
\frontmatter
\include{bibliography_script}
\pagestyle{empty}
\begin{center}
\,
 {\Huge \color{myColor}{\textbf{\myTitle}}}
    \vspace{0.5cm}\\
    \vspace{0.5cm}
    by\\
    \textsc{\myName}
    \vspace{15mm}
\begin{figure}[h!]
\includegraphics[width=10cm]{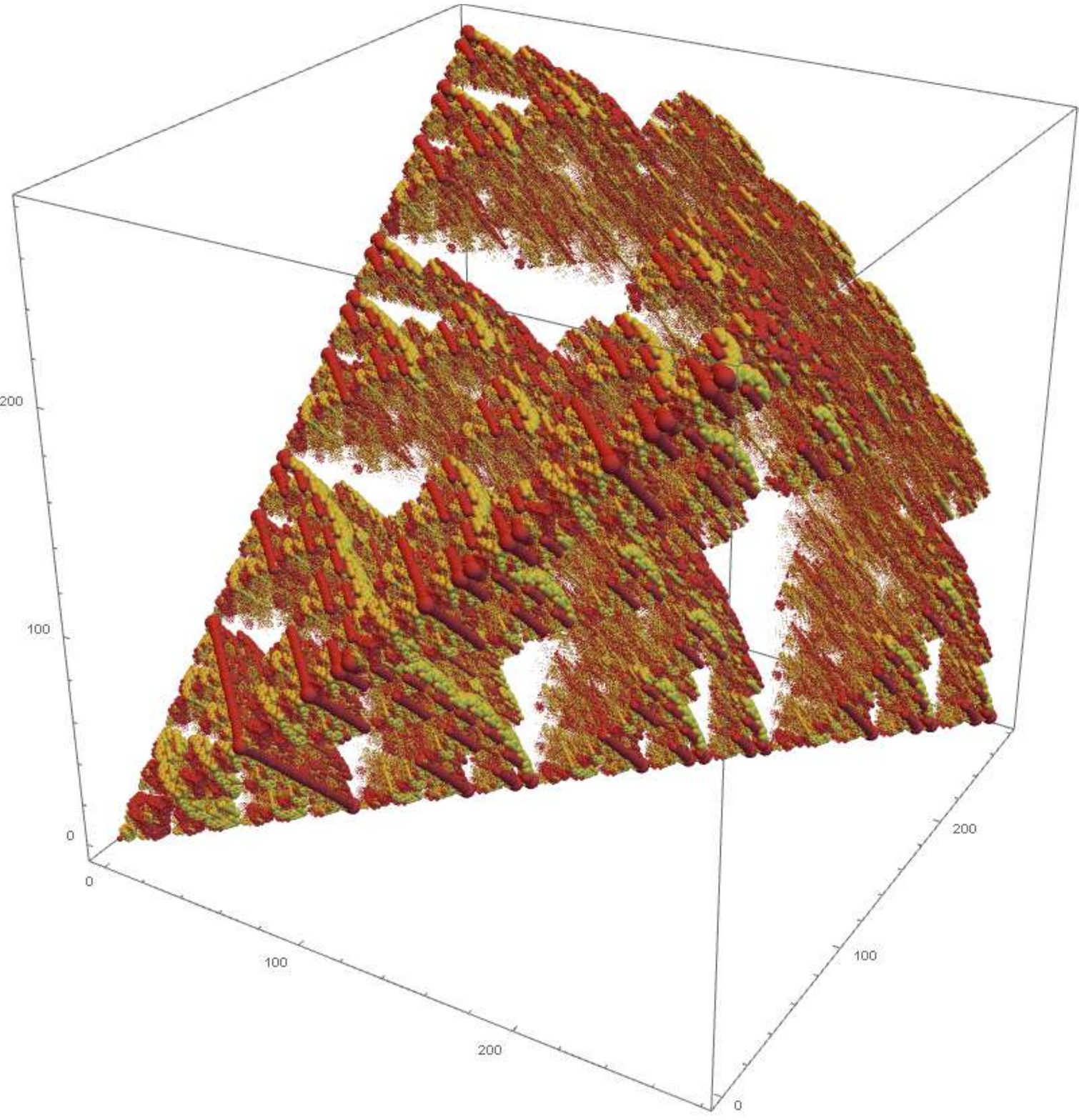}
\end{figure}
\,
\end{center}

\pagestyle{empty}
\begin{flushright}
\,
\vfill
\textit{This page is intentionally left blank}
\end{flushright}

% -------------------------------------------------- %
% This is the title page of the thesis
% -------------------------------------------------- %

\begin{titlepage}
\begin{center}
    \vspace*{1cm}
    \myUniversity\\
    \vspace{1cm}
    {\Huge \color{myColor}{\textbf{\myTitle}}}
    
    \vspace{0.25cm}
    
    \vspace{.5cm}
    by\\
    \textsc{\myName}
	\vspace{1 cm}
	
	under the supervision of \vspace{0.25cm}\\
	
	Prof. \textsc{\myProfessor} \\
	
	\vspace{1 cm}
	
	A thesis submitted in partial fulfillment of the requirements for the degree of \\
	Doctor in Physics
	\vspace{0.5cm}
	
    \myDepartment, \myFaculty\\
    	\vspace{0.5cm}
    \includegraphics[width=3cm]{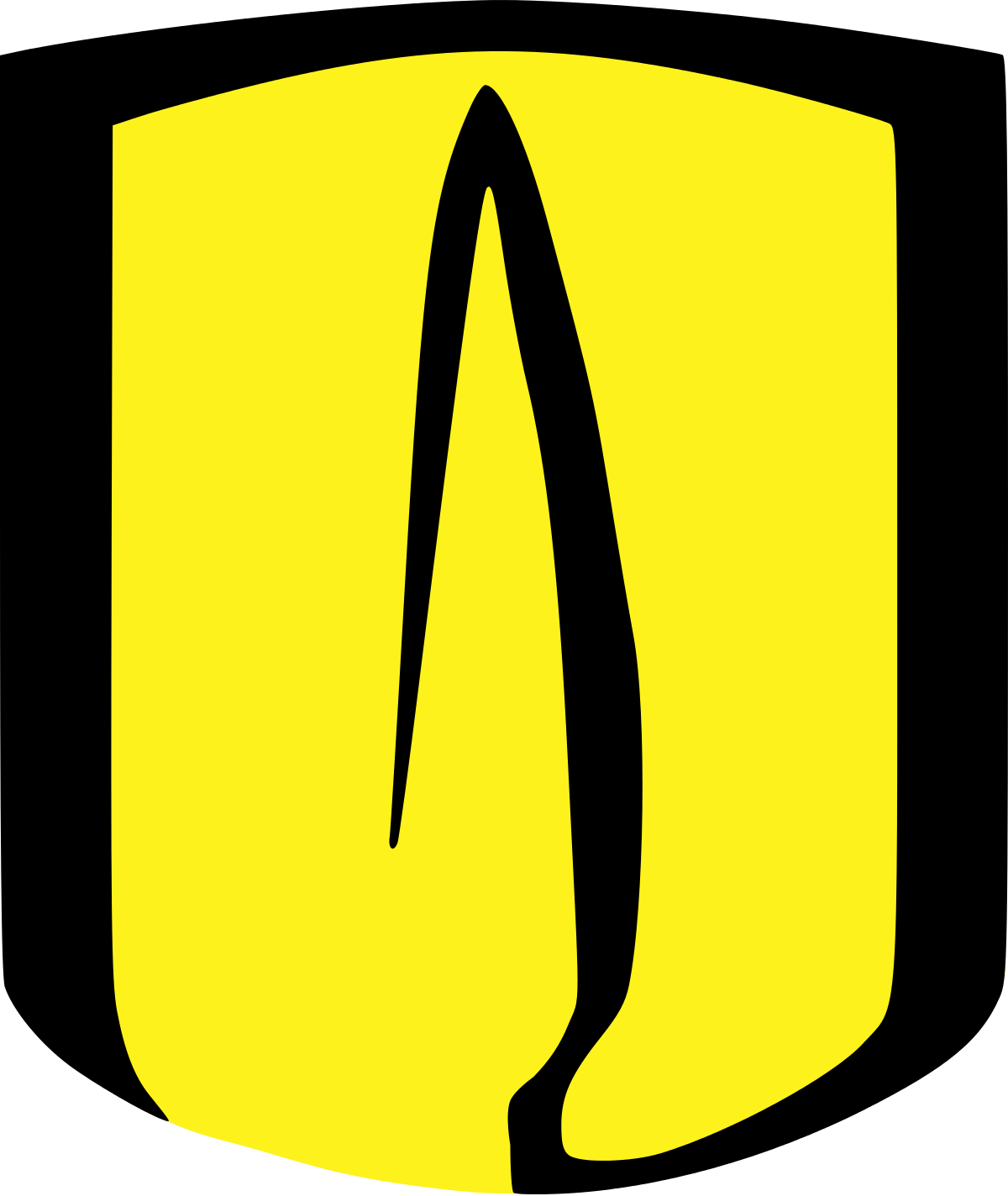}\\
    	\vspace{0.5cm}	
	\myLocation, 
\end{center}
\end{titlepage}
\pagestyle{empty}
\, 
\vfill
\begin{center}
\textit{To my son and wife, the reason of everything.}
\end{center}
\vfill
\,
\pagestyle{empty}
\begin{flushright}
\,
\vfill
\textit{This page is intentionally left blank}
\end{flushright}

\chapter*{Abstract}

In quantum information theory, maximally entangled states are essential for well-known protocols like quantum teleportation or quantum key distribution. While many of these protocols focus on bipartite entanglement, other applications, such as quantum error correction or multiparty quantum secret sharing,  are based on multipartite entanglement, precisely, on the so-called \textit{locally maximally entangled} (LME) multipartite states, where each part is maximally entangled with their complement. Such LME states appear naturally in the invariant subspaces of tensor products of irreducible representations of the symmetric group $S_n$, which we term \textit{Kronecker subspaces}, given that their dimensions are the so-called \textit{Kronecker coefficients}.
A Kronecker subspace is a vector space of LME multipartite states that we call \textit{Kronecker states}, which entangle Hilbert spaces of large dimensions. Although such states can in principle be obtained from the Clebsch-Gordan coefficients of the symmetric group, the known methods to compute these coefficients tend to be inefficient even for small values of $n$. An alternative quantum-information-based approach is inspired by entanglement concentration protocols, where Kronecker subspaces appear naturally in the \textit{isotypic decomposition} of tensor products of copies of multipartite entangled states. In this context, closed expressions have been obtained for a limited class of Kronecker states, associated with states in the so-called \textit{multiqubit W-class}. Our aim in this thesis is to extend this approach to build bases for Kronecker subspaces associated with any multiqubit system. For developing our method we first propose a graphical construction that we call ``\textit{W-state Stitching}'', where multiqubit entangled states are obtained as tensor networks built from W states. Analyzing the isotypic decomposition of copies of the graph state, an analogous set of graph Kronecker states, made from W-Kronecker states, can be obtained. In particular, the graph states of generic multiqubit states can generate any Kronecker subspace completely. Using this method, we show how to build any Kronecker subspace corresponding to systems of three and four qubits. Independently of the Kronecker state construction, the W-stitching technique has proven to be a powerful method for multiqubit entanglement classification. We hope the results of this work motivate the study of applications of Kronecker states in quantum information, and serve as a starting point for a resource theory of multipartite entanglement, with bipartite states and tripartite W states as building blocks, where the asymptotic analysis is based on Kronecker states.

\pagestyle{empty}
\begin{flushright}
\,
\vfill
\textit{This page is intentionally left blank}
\end{flushright}

\chapter*{Acknowledgments}

I would like to express my sincere gratitude to my supervisor, Prof. Alonso Botero, without whom this research would not have been possible. His unwavering patience and guidance over the years have been invaluable, and I am grateful for all that I have learned from him. Every conversation with him has been a source of new knowledge. I hope to have assimilated his vision of physics and quantum physics to some extent.\\

\noindent I would like to express my gratitude to José Mejía for generously dedicating a considerable amount of his time to explain his research in detail, which served as the foundation for the results presented in this document. Additionally, I am thankful for the valuable comments provided by Dario Egloff on the initial versions of this document. \\

\noindent I am also grateful to the Department of Physics at the University of the Andes and the Faculty of Sciences for providing me with financial support for my Ph.D. and research internship.\\

\noindent Lastly, but certainly not least, I want to express my deepest appreciation to my family, especially my wife and son, for their patience and being my primary source of inspiration throughout this process.

\pagestyle{empty}
\begin{flushright}
\,
\vfill
\textit{This page is intentionally left blank}
\end{flushright}

\tableofcontents

\mainmatter
\chapter{Introduction}
%\epigraphhead[50]{\epigraph{\textit{At the beginning there was only Chaos.}}{--- Aristophanes}}

\pagestyle{fancy}
\label{Introduction}
Entanglement is a key feature of quantum mechanics, with a broad range of applications in quantum information protocols, such as quantum teleportation \cite{Teleportation}, quantum key distribution \cite{QKD}, quantum error-correcting codes \cite{QEC}, dense coding \cite{Dense}, and many others\cite{Applications}. Despite its importance, entanglement becomes a complicated property when considering systems with more than two particles, and the richness of such systems is overwhelmed by their mathematical complexity. Finding maximally entangled states for multiple particles in Hilbert spaces of high dimensions and determining their mathematical structure has proven to be a challenge solved only for particular settings \cite{Multipartite-Entanglement}. In this research, we present a construction inspired by entanglement concentration protocols, which allows us to obtain maximally entangled states for many particles in systems of high dimensions in a systematic approach. For that, we propose a graphical construction named ``\textit{W-state Stitching}'', where $W$ states and bipartite states are used as building blocks of more general multiparticle states. Through a Schur transformation \cite{Schurtransform} on copies of the states, a vector space of maximally entangled states can be achieved, as well as corresponding algebraic expressions for it.\\

\noindent From a historical perspective, entanglement has been seen to be a fascinating and profound property of quantum systems, studied since the early stages of quantum mechanics. Entanglement refers to the non-classical correlations between two or more particles, even when physically separated. In 1935, in the paper ``Discussion of Probability Relations between Separated Systems'' \cite{Schrodinger}, Schrödinger introduced the German term ``Verschränkung'' or ``Entanglement'' in English to name this phenomenon. In Schrödinger's words, ``I would not call that \textit{one}, but rather \textit{the} characteristic trait of quantum mechanics, the one that enforces its entire departure from classical lines of thought.'' It was not until 1960 that John Bell \cite{Bell} introduced a mathematical structure and defined the ``Bell's inequality'', which constrained the strength of the correlations that could be described from local realistic theories. In subsequent experiments, physicists have found that the correlations between quantum particles violate Bell's inequality \cite{Bell}, indicating that entanglement is a real feature of nature. \\

\noindent The potential applications of entanglement in quantum information theory were not fully realized until 1991, when Artur Ekert \cite{Ekert} proposed using entangled states to enhance the security of the quantum key distribution protocol originally proposed by Bennett and Brassard \cite{QKD} seven years earlier. The following year, Bennett and Wiesner demonstrated that entanglement could also be used for quantum data compression, known as superdense coding \cite{Dense}. This allows for two bits of classical information to be transmitted using only one qubit with the help of pre-shared entanglement, leveraging the peculiarities of entanglement to achieve more efficient communication. In 1993, Bennet \cite{Teleportation} proposed a groundbreaking application of entanglement known as the teleportation protocol. The concept involves sharing an entangled state of two qubits between two parties, Alice and Bob, and using it to transmit the quantum information of a third qubit from Alice to Bob through local operations and classical communication. The remarkable feature of this protocol is that the original qubit is not physically transported but instead destroyed on Alice's side, and its state is transmitted to Bob's side through the entangled state up to applying one of four possible transformations. The discovery of quantum teleportation provided a new insight into the role of entanglement in quantum information.\\

\noindent Entanglement is the main source of many other applications in quantum information theory, such as quantum error correction codes \cite{Shor,Steane,QEC}, quantum computation speedups \cite{speedup}, quantum repeaters \cite{repeaters}, and quantum random number generators \cite{random}. While many of these applications have their roots in bipartite qubit systems, it has been demonstrated that using entangled states of higher dimensions results in stronger violations of Bell's inequalities\cite{BellQudits}, indicating stronger nonlocality. This implies that high dimensional entangled states can enhance the security and efficiency of many of the applications mentioned above \cite{QKDhigger,QuantumcryptoQutrits,QSS2}. Furthermore, the use of multipartite entangled states allows, for example, multiparty secure communication \cite{Multipartysecure}, quantum teleportation among multiple parties \cite{MPteleport}, and multiparty quantum key distribution \cite{MPQKD}, enhancing the capabilities of quantum communication networks \cite{Manybody}. Moreover, Many-body entangled states \cite{Manybody2} give rise to phenomena and topological properties not present in bipartite entangled states. Examples include topological phases of matter \cite{Phases} and long-range entanglement \cite{longrange}. Many-body entangled states are essential for exploring quantum field theory and its connection to condensed matter physics. Entanglement entropy, entanglement spectra, and entanglement Hamiltonians provide insights into the properties of quantum field theories and their critical behavior, such as conformal field theories \cite{QFT} and holography \cite{holographic}. Therefore, a better understanding of multiparticle high-dimensional entanglement is a requirement in the progress of quantum applications.\\

\noindent The question then arises: how to calculate high-dimensional multipartite states in a structured way? In this research, we propose to use the representation theory of the symmetric group $S_n$, where maximally entangled states appear in the \textit{invariant subspace} of tensor products of representations of $S_n$. We name those states as \textit{Kronecker states}, and the vector space where they belong as \textit{Kronecker subspace}. When considering a system composed of $n$ copies of some multipartite state, the Kronecker subspace appears naturally as the subspace of the total Hilbert space,  where permutations of the parts in the copied state, which are elements of $S_n$, act trivially. We present a useful framework to build general Kronecker subspaces from simpler Kronecker states that appear when analyzing the Kronecker subspace of copies of bipartite states and $W$ states \cite{sawicki2013}, for which closed expressions were already obtained. \\

\noindent Before delving into the content of this research, we need to review the mathematical structure of pure state entanglement \cite{Plenio}. For a two-particle system with $\mathcal{H}^1$ and $\mathcal{H}^2$ being the Hilbert spaces associated with each particle, if a state $\ket{\psi_{12}}$ can be written as a tensor product,
\begin{equation*}
\ket{\psi_{12}}= \ket{\psi_1}\otimes \ket{\psi_2},
\end{equation*}
where $\ket{\psi_1}\in \mathcal{H}^1$ and $\ket{\psi_2}\in \mathcal{H}^2$, then the state is said to be separable; otherwise, it is entangled. For example, the state of two qubits,
\begin{equation*}
\ket{\psi}= \frac{1}{\sqrt{2}} \left(\ket{00}+\ket{10} \right),
\end{equation*} is separable since it can be written as the tensor product of $\ket{\psi_1}=\frac{1}{\sqrt{2}} \left(\ket{0}+ \ket{1}\right)$ and $\ket{\psi_2}=\ket{0}$. On the other hand, the state
\begin{equation}
\ket{\Phi^{+}}= \frac{1}{\sqrt{2}} \left( \ket{00}+\ket{11}\right),
\label{eq:Phi+}
\end{equation}
cannot be written as a product of any two-qubit states and is, therefore, entangled. In fact, it is a maximally entangled state known as the EPR state \cite{EPR}. To understand better how this state is maximally entangled, we need to use entanglement measures, which not only identify separable and entangled states but also measure how entangled is a given quantum state. One of such entanglement measures is the \textit{entropy of entanglement}, which is defined for a pure state $\ket{\psi}$ as
\begin{equation*}
 E(\ket{\psi}):= S(\rho_1)=S(\rho_2),
 \label{eq:entanglemententropy}
\end{equation*} 
where $S(\rho)= -tr(\rho \log \rho)$ is the von-Neumann entropy, and $\rho_{i}$ is the reduced density matrix for the $i$ subsystem. By noting that individual reduced density matrices for the state in Equation \eqref{eq:Phi+} are fully mixed, one gets
\begin{equation*}
E(\ket{\Phi^{+}})=\log 2,
\end{equation*}
which is the maximum possible value for qubit systems. i.e., $\ket{\Phi^{+}}$ is a maximally entangled state. The entropy of entanglement is zero when the subsystems $1$ and $2$ are separable, and its maximum value, $\log d_i$, with $d_i$ being the dimension of the Hilbert space of the $i$-th subsystem, is achieved when the reduced density matrix is proportional to the identity. \\

\begin{figure}[ht]
    \centering
    \includegraphics[scale=0.8]{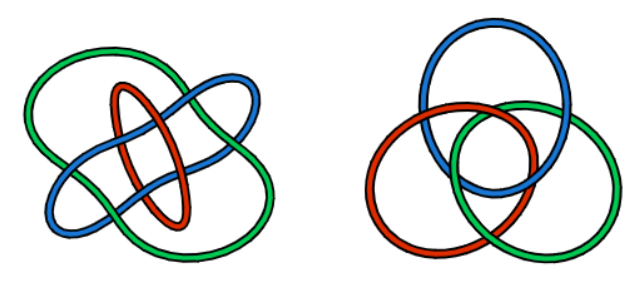}
    \caption{Graphical representation of states in \eqref{eq:WGHZ}, where each ring correspond to one of the qubits. At the left, the GHZ state is represented, and after losing one part, the remaining two are unentangled. At the right, the W state is represented, and after losing one part, the remaining two are entangled. Images taken from \cite{RWGHZ}}
    \label{fig:RWGHZ}
\end{figure}

\noindent When considering systems with more than two particles, the definition of entanglement can be straightforwardly generalized to multipartite systems. If a state with $N$ particles can be written as
\begin{equation*}
\ket{\psi}=\bigotimes_{i=1}^{N} \ket{\psi_i} \quad, \ket{\psi_i}\in \mathcal{H}^i ,\quad \forall	 i \in\{1,2\dots,N\},
\end{equation*}
with $\mathcal{H}^i$ the Hilbert space of the $i$-th particle, then the state is separable; otherwise it is entangled. The properties of entanglement entropy described before can be used to define multipartite locally maximally entangled states (LME) as those whose individual reduced density matrices are proportional to the identity:
\begin{equation*}
 \rho_i = tr_{\overline{\imath}} \ket{\psi}\bra{ \psi}	=\frac{1}{d_i} \bm{I},
\end{equation*} 
where $\overline{\imath}$ stands for the complement of part $i$ in the set of parties. One can notice that when considering systems with more than two particles, entanglement can appear in different forms. For example, when considering the three qubit case, the state
\begin{equation*}
\ket{\psi}= \frac{1}{\sqrt{2}} \left( \ket{001}+\ket{010} \right) = \frac{1}{\sqrt{2}} \ket{0} \otimes \left( \ket{01}+\ket{10} \right),
\end{equation*}
is not separable, as it is not possible to write it as a product of states in each local Hilbert space; however, it is not completely entangled as its first particle can be separated from the other two, so now we can have entanglement for some subsets of particles, and even when there is entanglement between all the parties, known as \textit{genuine entanglement}, one can find different classes of entanglement. For example, for three qubits, it is known that genuine entanglement can be separated into two classes\cite{3qubits}. When considering the states
\begin{equation}
\ket{GHZ}=\frac{1}{\sqrt{2}} \left( \ket{000} +\ket{111} \right) , \quad\ket{W}=\frac{1}{\sqrt{3}} \left( \ket{100} +\ket{010}  +\ket{001} \right) ,
\label{eq:WGHZ}
\end{equation}
they are both genuinely entangled, but their entanglement behaves very differently, for example, when one of the particles is discarded (traced out). For the $GHZ$ state, losing one of its parts makes all the entanglement lost, while for the $W$ state, after losing any of its parts, the remaining two are entangled. In figure \ref{fig:RWGHZ}, a graphical representation of these states taken from \cite{RWGHZ} is shown, where each ring represents one of the qubits. The differences between the entanglement properties of these states show different kinds of entanglement in multipartite systems. \\

\noindent A natural setup for analyzing entanglement in a multipartite system is to consider \textit{Stochastic Local Operations and Classical Communication} (SLOCC) on the states. The LOCC part of these operations can be thought of as all the possible operations that can be applied locally in a multipartite state as if all its parts are distributed in multiple laboratories and cannot interact directly. Under this setup, the different laboratories can be interconnected through a classical channel to coordinate and adjust the actions of each part of the state. In this sense, two states are said to be LOCC equivalent if they can be converted into each other through the abovementioned operations. Note how, with this construction, it is impossible to create non-local correlations between the parts of the system, meaning that the system's entanglement cannot increase. The stochastic part of SLOCC states that the conversion between states could be achieved with some non-zero probability. It can be shown that such transformation is possible between two states $\ket{\psi}$ and $\ket{\phi}$ if there exists a set of unit determinant operators  $A_i$ such that 
\begin{equation*}
A_1\otimes A_2 \otimes \dots \otimes A_N \ket{\psi}=s \ket{\phi},
\end{equation*}
with $s$ a complex scalar \cite{Multipartite-Entanglement}. The $A_i$ matrices belong to the Special Linear group of dimension $d$, $SL_{d}$, i.e., the group of unit-determinant matrices of dimension $d\times d$. In this sense, SLOCC classes are defined as the different orbits of $SL_{d_1}\otimes SL_{d_2}\otimes \dots SL_{d_N}$ in the projective Hilbert space. The projective Hilbert space is obtained by taking the ordinary Hilbert space and identifying all non-zero vectors that differ only by a scalar factor. This identification means that all vectors that are proportional to each other are considered equivalent in the projective Hilbert space.\\

\noindent By taking any state in an orbit, it is possible to transform it into a normal form by local operations of unit-determinant \cite{VerstraeteNormal}. Such normal form has the property that its local reduced density matrices are proportional to the identity, which corresponds to the definition of locally maximally entangled states. Moreover, it is shown in \cite{VerstraeteNormal} that normal forms correspond to the states of minimal norm that can be obtained under local unit-determinant operations on the initial state. When looking for normal forms, three scenarios are possible. One case is when the normal form is the zero state. Such orbits are known as the \textit{Null cone} \cite{LuqueThibon-finite}. A second option is that the normal form can only be achieved asymptotically, with an infinite sequence of steps, meaning that the normal form, also known as \textit{critical state} \cite{Bryan}, belongs to the closure of the orbit. In this case, the orbit is said to be \textit{strictly semistable} \cite{Slowik}. The last option is that the normal form is achieved in the orbit with a finite number of steps; in this case, the orbit is called \textit{polystable} \cite{Slowik}. Critical states or normal forms, are usually used as the representatives of their corresponding SLOCC class. For the three-qubit case discussed before, genuine entanglement separates into two classes known as the GHZ-SLOCC class and the W-SLOCC class, whose representatives were shown in \eqref{eq:WGHZ}. In this case, the GHZ orbit is polystable, and then the representative $\ket{GHZ}$ is chosen as the LME state in the orbit. On the other hand, the W-class belongs to the null cone, so there is no LME in the orbit. \\

\noindent In recent years, the possibility of using representation theory to find LME states has been highlighted \cite{Bryan} \cite{SlowikDesigning}. To understand how this connection is made explicitly, consider first a bipartite state $\ket{\psi} \in \mathcal{H}
^1 \otimes \mathcal{H}^2$, such that it is invariant up to a phase under the irreducible action of a given group $G$ on $\mathcal{H}^1$, 
\begin{equation*}
 X(g) \otimes I \ket{\psi} = e^{i \phi(g)} \ket{\psi} ,\quad \forall g\in G,
\end{equation*}
with $X(g)$ an irreducible representation of $G$, then note how the reduced density matrix in the first subsystem commutes with the action of the group:
\begin{equation*}
	X(g) \rho_1(\ket{\psi}) X(g) ^{\dagger} = \rho_1(\ket{\psi})  \Rightarrow [X(g),\rho_1(\ket{\psi})]=0, \quad \forall g\in G.
\end{equation*} 
We will show later in Chapter \ref{Chapter2} that if $X(g)$ is an irreducible representation, by \textit{Schur's lemma} \cite{Fulton}, $\rho_1(\ket{\psi})$ has to be proportional to the identity. This property can be easily generalized for all the parts of a multipartite entangled state; consider $\ket{\psi} \in \mathcal{H}
^1 \otimes \mathcal{H}^2 \otimes \dots \otimes \mathcal{H}^N$, such that it is invariant under the irreducible action of a given group $G$ on each of its parts
\begin{equation*}
 X^{\lambda^1}(g) \otimes X^{\lambda^2}(g)  \otimes \dots \otimes X^{\lambda^N}(g)  \ket{\psi} = e^{i \phi(g)} \ket{\psi} ,\quad \forall g\in G,
\end{equation*}
where $\lambda^i$ labels the irreducible representation acting on part $i$, then, according to Schur's lemma, each individual reduced density matrix $\rho_i(\ket{\psi})$ is proportional to the identity, showing that $\ket{\psi}$ is an LME state.\\

\noindent This setting has been used to calculate LME states and find relations between representation theory and entanglement \cite{BryanLME} \cite{SlowikDesigning}. However, the main focus has been analyzing the irreducible representations of $SU(d)$. In this research, we use the same approach by analyzing the irreducible representations of the symmetric group $S_n$. As we will see, the invariant subspace for this group is, in general, not one-dimensional, which makes it harder to compute the associated LME states. However, the richness of this structure also motivates this exploration. Having subspaces full of maximally entangled states is optimal, for example, for designing quantum codes \cite{QuantumCodes}.\\ 

 \noindent The symmetric group $S_n$ is the group of permutations of $n$ elements and its irreducible representations (henceforth \textit{irreps}) are labeled as $[\lambda]$. Our goal is to build multipartite states that are invariant under the simultaneous action of irreps of $S_n$ in each of the parts; for example, for the three-part case, we want states such that
 \begin{equation}
  D^{[\lambda^1]} (\pi) \otimes D^{[\lambda^2]} (\pi)  \otimes D^{[\lambda^3]} (\pi) \ket{\mathcal{K}_{\lambda^1\lambda^2\lambda^3}}=\ket{\mathcal{K}_{\lambda^1\lambda^2\lambda^3}} , \quad \forall \pi \in S_n, 
  \label{eq:Sninvariance}
 \end{equation}
 with $[\lambda^i]$ irreps of $S_n$ and $D^{[\lambda]}(\pi)$ the matrix representation of permutation $\pi$ in irrep$[\lambda]$. The set of states $\{\ket{\mathcal{K}_{\lambda^1\lambda^2\lambda^3}}\}$ with the property in \eqref{eq:Sninvariance} form the invariant subspace of the tensor product of irreps $([\lambda^1]\otimes [\lambda^2]\otimes [\lambda^3])^{S_n}$ i.e., the subspace where the diagonal action of $S_n$ acts trivially, whose dimension is known as the Kronecker coefficient $k_{\lambda^1\lambda^2\lambda^3}$:
 \begin{equation*}
 \dim([\lambda^1]\otimes [\lambda^2]\otimes [\lambda^3])^{S_n}=k_{\lambda^1\lambda^2 \lambda^3},
\end{equation*}
 Because of this, we name such states as \textit{Kronecker states} and the invariant subspace $([\lambda^1]\otimes [\lambda^2]\otimes [\lambda^3])^{S_n}$ as \textit{Kronecker subspace}, which will be the main object of study throughout this document.\\

\noindent Kronecker subspaces are very interesting from the quantum information point of view; they are vector spaces of dimension $k_{\lambda^1\lambda^2\lambda^3}$(which in general is greater than 1), and all vectors in them are LME states. It is worth highlighting that being a vector space, any linear combination of Kronecker states in a given set $\lambda^1\lambda^2\lambda^3$ is also a Kronecker state. Also, as these states are invariant under the irreducible actions of $S_n$, they will show up when considering setups with permutational invariance, which are very common in many quantum information protocols.\\

\noindent A relevant example where Kronecker states appear naturally is the entanglement concentration protocol proposed by Hayashi and Matsumoto \cite{Hayashi-Matsumoto}. In this protocol the goal is to take a bipartite entangled state $\ket{\psi}\in \mathcal{H}^{1}\otimes \mathcal{H}^2$, which is not maximally entangled, $E(\ket{\psi})<\log d$ (for simplicity we will take $d_1=d_2=d$), and by taking $n$ copies of it, $\ket{\psi}^{\otimes n}$, to extract a reduced quantity $n_{LME}<n$ of maximally entangled states. To achieve this, we consider the decomposition of the total Hilbert space into irreducible representations of $GL_d$ (the group of general linear transformations of dimension $d$) and $S_n$ in each part using the \textit{Schur-Weyl duality}, a powerful tool from representation theory that is used throughout this document, and that is introduced in Section \ref{Schur-Weyl}. Specifically, the $n$ copies of the Hilbert space associated with the two parts are decomposed as a diagonal action of $GL_d$ and $S_n$ irreps:
\begin{equation*}
\left((\mathcal{H}^1 \otimes \mathcal{H}^2)^{\otimes n}\right) ^{S_n}= \bigoplus_{\lambda^1 \lambda^2 \vdash n,d} \{\lambda^1\} \otimes \{\lambda^2\} \otimes ([\lambda^1] \otimes [\lambda^2] )^{S_n},
\end{equation*}
where $\{\lambda\}$ and  $[\lambda]$ respectively label irreducible representations of $GL_d$ and $(S_n)$, we also use $(\cdot)^{S_n}$ to represent the permutationally invariant subspace of  ``$\cdot$'', and $\lambda^1\lambda^2\vdash n,d$ restricts the possible irreducible representations according to the number of copies and the dimension of the original Hilbert space. Since the system is invariant under the permutation of the copies, $S_n$ irreps are restricted to the invariant subspace (or Kronecker subspace), which in this case is one-dimensional and only exists when $[\lambda^1]=[\lambda^2]=[\lambda]$, i.e., $k_{\lambda^1\lambda^2}=\delta_{\lambda^1\lambda^2}$. Then, there exists a basis transformation also known as the \textit{Schur transform} \cite{Schurtransform}, presented in Section \ref{Schurtransform}, which makes explicit the decomposition of the total Hilbert state into the diagonal form
\begin{equation*}
\left((\mathcal{H}^1 \otimes \mathcal{H}^2)^{\otimes n}\right) ^{S_n}= \bigoplus_{\lambda \vdash n,d} \{\lambda\} \otimes \{\lambda\} \otimes ([\lambda] \otimes [\lambda] )^{S_n}.
\end{equation*}
By applying the Schur transform on the copied state, this can be rewritten as:
\begin{equation*}
\ket{\psi}^{\otimes n} = \bigoplus_{\lambda \vdash n,d} \sqrt{p_{\lambda} (\psi)} \ket{\phi_{\lambda}(\psi)} \ket{\mathcal{K}_{\lambda\lambda}},
\end{equation*}
 where $p_\lambda(\psi)$ is a probability distribution that depends on the input state $\ket{\psi}$, and $\ket{\phi_{\lambda}(\psi)}$ and $\ket{\mathcal{K}_{\lambda\lambda}}$ are states in $\{\lambda\}
 \otimes\{\lambda\}
 $ and $([\lambda]\otimes[\lambda])^{S_n}$ respectively. After a projective measurement onto the irreps $\lambda$ on each part, one obtains a product state whose $GL_d$ part can be dropped to obtain $\ket{\mathcal{K}_{\lambda\lambda}}$, a maximally entangled state. We will see later in Section \ref{EntanglementHayashi} that with this protocol, the resultant Kronecker state $\ket{\mathcal{K}_{\lambda\lambda}}$ asymptotically concentrates the entanglement of the original copies $nE(\ket{\psi})$.\ \\
 
\noindent For the bipartite case it will be shown in Chapter \ref{Chapter3}, that obtaining the coefficient expansion of the Kronecker states is very simple, and corresponds to a generalization of the EPR state:
 \begin{equation*}
 \ket{\mathcal{K}_{\lambda\lambda}}= \frac{1}{\sqrt{f^{[\lambda]}}} \sum_{q} \ket	{\lambda,q}\ket{\lambda,q},
 \end{equation*}
 where $f^{[\lambda]}$ is the dimension of the $S_n$ irrep $[\lambda]$ and $q$ labels the basis elements of such irrep. However, when considering systems composed of three or more parts, obtaining closed expressions for the Kronecker states poses a significant challenge because the dimension of the Kronecker subspace is in general greater than one, which makes harder to identify the structure of the Kronecker states. One approach is to explicitly calculate the projector onto the invariant subspace $([\lambda^1]\otimes [\lambda^2] \otimes [\lambda^3])^{S_n}$. However, this method involves the computation and summation of $n!$ square matrices, each with dimensions of $f^{[\lambda^1]}\times f^{[\lambda^2]} \times f^{[\lambda^3]}$. Given that the dimensions $f$ grow exponentially with $n$, this approach quickly becomes impractical and inefficient. Another approach is to use the Clebsch Gordan Coefficients (CGC) of the symmetric group \cite{sahasrabudhe}. In Chapter \ref{Chapter3} it will be shown that when choosing real basis for irreps of $S_n$, Kronecker states can also be written as
\begin{equation*}    \ket{\mathcal{K}_{\lambda^1\lambda^2\lambda^3,i}} = \frac{1}{\sqrt{f^{[\lambda^3]}} }\sum_{q^1 q^2 q^3} C^{[\lambda^1][\lambda^2][\lambda^3],i}_{q^1q^2q^3}\ket{\lambda^1,q^1}\ket{\lambda^2,q^2}\ket{\lambda^3,q^3}.
\end{equation*}
with $ C^{[\lambda^1][\lambda^2][\lambda^3],i}_{q^1q^2q^3}$ the CGC of $S_n$, where $i$ is a label for the dimensions of the invariant subspace. In other words, finding the expansion coefficients of a Kronecker state basis is equivalent to finding a set of CGC of $S_n$. However, the algorithms available for this task are challenging to follow and typically inefficient \cite{sahasrabudhe,Doma,gao}, and computing CGCs becomes infeasible even for moderate values of $n$. For instance, we could not compute all CGC for $n=8$ and beyond on a personal computer using the available algorithms. Even for the CGC we could compute for $n=8$, the computation took several days for some sets $\lambda^1 \lambda^2 \lambda^3$.\\

\noindent Despite the difficulty exhibited by this problem, Botero and Mejia \cite{Botero} found a partial solution in a generalization of Hayashi-Matsumoto protocol to a special class of multiparticle systems. They studied the protocol using copies of states in the multipartite W-SLOCC class. By using the theory of SLOCC covariants \cite{Covariants}, they showed that the final state in the Schur transform of states in the W-SLOCC class, after projecting onto some set of local partitions $\lambda^1 \lambda^2 \lambda^3$, the resultant state is separable in the $GL_2$ and $S_n$ parts as in the bipartite case. This discovery allows for a generalization of entanglement concentration when the input states are in the W-class. Moreover, they obtained closed analytic expressions for the unique Kronecker vector for the W-class in each set of partitions $\ket{\mathcal{K}^W_{\lambda^1\lambda^2\lambda^3}}$, which is particularly important for our purposes. Nevertheless, the mathematical structure that allows closed expressions for Kronecker states also limits what can be obtained. Only one  Kronecker vector out of a space of dimension $k_{\lambda^1\lambda^2\lambda^3}$ can be obtained for each set of local partitions, and the W-SLOCC class cannot achieve some sets due to restrictions on the local spectra\cite{Polytopes}. To build completely the Kronecker subspace, the restrictions of W-class have to be overcome. \\

\noindent The main result of this work is a novel graphical method called ``\textit{W-state Stitching}'', where $W$ states of three parts ($W_3$) work as building blocks, that are stitched with bipartite states to build more complex multiqubit states, in a network-like construction. The structure of the stitching procedure can be seen graphically, and the resultant states are associated with the topology of certain graphs. By applying the Schur transform to tensor copies of these graph states, equations for general Kronecker states can be obtained from the already computed $\ket{\mathcal{K}^{W}_{\lambda^1 \lambda^2 \lambda^3}}$ for the $W$ class. Using this construction, we have built all the possible Kronecker subspaces up to $n=12$ for three parties in a personal computer. For this value, the most interesting case corresponds to a Kronecker subspace of dimension $k_{\lambda^1\lambda^2\lambda^3}=3$, where the dimension of the Hilbert space of each part is $f=275$. With four parties, we calculated all Kronecker subspaces up to $n=9$ where a value of $k_{\lambda^1\lambda^2\lambda^3\lambda^4}=39$ can be obtained.  \\

\noindent Remarkably, the W-state Stitching procedure we proposed to obtain Kronecker states becomes a very powerful tool. By identifying the symmetries of the $W_3$ states in a graphical procedure that we called ``\textit{Parameter pushing}'',  it is possible to identify which parameters in the stitched state can be used to differentiate SLOCC classes and which are irrelevant under SLOCC classification. Our exploration encompasses the three and four-qubit scenarios, offering a graphical framework for classification. This innovative method also illuminates a graphical interpretation to calculate SLOCC invariants in multiqubit settings \cite{Covariants}, which are homogeneous polynomials on the coefficients of the state that are invariant under SLOCC operations,  and as we will see in Chapter \ref{Chapter2-5}, are useful for classifying entanglement. Finding SLOCC invariants for qubit systems is not a trivial task and usually leads to complicated expressions \cite{5qubits-Thibon}. By using our graphical tools, we provide explicit graphical representations for $17$ independent SLOCC polynomial invariants in the context of five qubits, which, as far as we know, have not been obtained before. With this language, it is possible to show explicitly how these objects allow for identifying different kinds of entanglement. The elegance and efficiency of our methods simplify manual calculations of entanglement-related quantities, making them accessible to researchers new to the field. In this sense, we set the seed of a new approach to quantify and classify multipartite entanglement, where the basic units are $W_3$ states and bipartite states.\\

\noindent This document is organized as follows: in Chapter \ref{Chapter2-5}, we explore the problem of entanglement classification for qubits; starting from the two qubits case, we introduce the $LU$ and $SLOCC$ equivalence and their physical significance. We discuss $SLOCC$ invariants as tools to discern specific entanglement classes. We study the three qubits $SLOCC$ classification and show how, by calculating an invariant, it is possible to distinguish between the two genuinely entangled states $\ket{W}$ and $\ket{GHZ}$ and discuss their entanglement properties. Our exploration extends to systems of four qubits, where we discuss the parametric classification delineated in \cite{Verstraete}. This classification scheme separates the Hilbert space into nine distinct SLOCC families. We comment on the entanglement properties of some particular states and highlight the properties of the generic family known as $G_{abcd}$.  \\

\noindent Chapter \ref{Chapter2} introduces representation theory focusing on the symmetric group and the general linear group, including some useful examples. We explore the connection between angular momentum and irreducible representations of $GL_2$. In addition, we introduce Schur-Weyl duality and an algorithm implementing the Schur transform for qubit systems. We also discuss some results of Schur-Weyl duality in quantum information, such as the Keyl-Werner theorem and the entanglement concentration protocol.\\

\noindent Chapter \ref{Chapter3} introduces Kronecker states, including their definition, properties, connection with the Clebsch-Gordan coefficients of the symmetric group, and the difficulties involved in their computation. We also explore the relationship between the entanglement concentration protocol and Kronecker states. Finally, we describe the construction presented by Botero and Mejia for computing Kronecker states in the W-class.\\

\noindent In Chapter \ref{Chapter4}, we present the W-Stitching procedure to obtain multiqubit states from tripartite W states and bipartite states; we introduce the method of \textit{parameter pushing} and a set of rules obtained from the algebras of the graphical objects, that allow us to identify the relevant parameters of the construction under SLOCC orbits. We study this tool for SLOCC classification, obtaining explicit graphical constructions for any state in the cases of two, three, and four qubits.  \\

\noindent In Chapter \ref{Chapter5}, we extend the stitching procedure to W-Kronecker states and bipartite Kronecker states and, by doing so, obtain more general Kronecker states. Later, we show how applying the Schur transform in graphs allows for constructing complete Kronecker subspaces using the already known expressions for $W-$Kronecker states. We explicitly study the construction for three and four qubits, obtaining rules that allow the computation of any Kronecker state with three and four parts of length at most 2. We also explore the deep connections between multiqubit graph states and the corresponding graph in the Kronecker space. Finally, we give explicit constructions to produce the orthogonal basis of Kronecker subspaces for the three-part case up to $n=12$. We explore the limitations of some graphs for the four qubits case, showing a general graph for building any Kronecker subspace of four parts and presenting some of the most interesting results. We finish this chapter by exploring the case of five qubits and showing graphs that are good candidates to generate any Kronecker subspace of five parts.\\

\noindent  In chapter \ref{Chapter6}, we present three secondary results obtained through this investigation that are not necessary for the main objective of building Kronecker subspaces. The first result is a simple method to calculate matrices for irreducible representations of $S_n$ labeled with partitions of length two, using the Schur-Weyl duality. The second result is a graphical method, motivated by the W-state stitching to find SLOCC invariants in multiqubit settings graphically; we also show how, when using the graphs from Chapter \ref{Chapter4}, such invariants can be simplified to the point of being computed by hand in some cases. The last secondary result is obtained from a study in the multipartite W-class when applying the Schur transform on the correspondent graph states; in this case, the calculations can be made explicitly, obtaining that the set of possible spectra in this class is completely fixed by the local spectra, also implying a recurrent structure on the W-Kronecker states. \\

\noindent Finally, in Chapter  \ref{Conclusions}, we summarize the results obtained here, along with interesting open questions that leave this approach that are worth studying deeper in the future.

\chapter{Preliminaries I: Entanglement}

\label{Chapter2-5}

Entanglement is a fundamental concept in quantum physics, serving as the basis for various quantum applications. This phenomenon establishes correlations between particles, where their properties become interdependent, even when separated by vast distances, demonstrating the non-local characteristics of quantum systems. This phenomenon not only challenges classical intuition but also underpins many of the key distinctions between quantum and classical physics. \\

\noindent Entanglement becomes increasingly intricate and interesting as the number of particles within a quantum system grows. Nevertheless, formulating a comprehensive mathematical description of entanglement in such complex systems remains a formidable challenge. To harness the full potential of many-body quantum systems, it is imperative to comprehend the mathematical framework of multipartite entanglement. This thesis advances this direction by introducing a systematic methodology for computing multipartite maximally entangled states, where each component exhibits maximal entanglement with its complementary counterpart.\\

\noindent This chapter sets the stage for our exploration by establishing the basics of quantum entanglement and multipartite entanglement classification. The contents of this chapter are, therefore, not new but are based on established literature such as \cite{Plenio}\cite{Multipartite-Entanglement}\cite{Horodecki}\cite{Bennet} \cite{Nielsen}.

\section{Quantum states and entanglement}

In this document, we utilize the Hilbert space formalism to describe quantum systems. For this, we define a pure state $\ket{\psi}$ as a normalized vector in a Hilbert space $\mathcal{H}$ of complex dimension $d$, where the vector represents our complete knowledge of the quantum state. This vector can be expanded in an orthonormal basis of $\mathcal{H}$ as:
\begin{equation*}
\ket{\psi} = \sum_{i=0}^{d-1} c_i \ket{i},
\end{equation*}
where the normalization condition imposes $\sum_{i} |c_i|^2 =1$.\\

\noindent In some cases, we may not know the state of the system with certainty, but we know that with some probability $p_i$, the system is in the state $\ket{\psi_i}$. We represent this as a mixed state using the \textit{density matrix}:
\begin{equation*}
\rho= \sum_{i} p_i \ket{\psi_i}\bra{\psi_i},
\end{equation*}
which is a linear operator acting on the Hilbert space. Pure states correspond to $\rho= \ket{\psi}\bra{\psi}$, so they are rank-1 projectors on the Hilbert space. When considering the system as composed of $N$ subsystems, the Hilbert space corresponds to the tensor product of the Hilbert spaces of each subsystem
\begin{equation*}
\bm{\mathcal{H}}= \mathcal{H}^1 \otimes \mathcal{H}^2 \otimes \dots \otimes \mathcal{H}^N,
\end{equation*}
where any pure state can be written as
\begin{equation*}
\ket{\psi}=\sum_{\bm{i}} c_{\bm{i}} \ket{\bm{i}} , \quad \ket{\bm{i}}= \ket{i^1 i^2\dots i^N},
\label{eq:orthobasispsi}
\end{equation*}
each $i^j$ labels the orthonormal bases of the corresponding Hilbert space $\mathcal{H}^{j}$, and we have introduced a boldface notation that will be recurrent throughout the document to summarize sets of, or actions on, the parts of a multipartite system, in this case: $\bm{\mathcal{H}}=\otimes_{i=1}^N \mathcal{H}^{i}$.

\subsubsection{Separable and Entangled states}
With the introduced notation, we can define pure separable states as those that can be expressed as:
\begin{equation}
\ket{\psi}=\bigotimes_{i=1}^{N} \ket{\psi_i},\quad \ket{\psi_i}\in \mathcal{H}^i ,\quad \forall	 i \in\{1,2\dots,N\}.
\label{eq:Separable}
\end{equation}
For separable states, each subsystem can be described independently from the others. The state $\ket{\psi_i}$ describes completely the subsystem $i$ independently from its complement $\overline{\imath}$. \\

\noindent Now, we define pure entangled states as those pure states that cannot be written as in Equation \eqref{eq:Separable}, throughout this document we will focus on pure states. The dimension associated with separable states is given by the sum of the complex dimensions $d_i$ of each Hilbert space $\mathcal{H}^i$ minus one, by normalization; in contrast, the complex dimension of the total Hilbert space is given by
\begin{equation*}
\dim(\bm{\mathcal{H}})=\bm{d}=\prod_{i}^{N} d_i -1,
\end{equation*}
then, the Hilbert space of composed systems is predominantly composed of entangled states.\\

\noindent When considering the two-qubit case ($d_i=2$), one entangled pure state is the Bell state $\ket{\Phi^+}$ \cite{Nielsen}:
\begin{equation}
\ket{\Phi^{+}}= \frac{1}{\sqrt{2}} (\ket{00}+\ket{11}).
\label{eq:phi+}
\end{equation}
Note how, when considering the set of separable two-qubit pure states, they can be parametrized as:
\begin{equation*}
\ket{\psi^{SEP}_{12}}=c^1_0 c^2_0 \ket{00} +c^1_0 c^2_1 \ket{01} + c^1_1 c^2_0 \ket{10}+c^1_1 c^2_1 \ket{11},
\end{equation*}
where $c^i$ represents the coefficients of $\ket{\psi_i}$, then there is no possible solution for $c^1_0,c^1_1,c^2_0,c^2_1$ that can yield  $\ket{\Phi^+}$, meaning that this state is not separable; hence, it is entangled.\\

\noindent Our next objective is to quantify entanglement. To achieve this, we must establish criteria for determining when two states exhibit equal degrees of entanglement and when one state is more entangled than another. For this purpose, we will introduce the concepts of LU (Local Unitary) and LOCC (Local Operations and Classical Communication) operations.

\subsection{LU operation}
\label{LUOperations}
Let us start by defining \textit{Locally Unitary} (LU) operation as the action of unitary operators $U_i$ on each of the parts of the system. Then, we define two states as LU-equivalent if they can be obtained with certainty from each other by applying some LU operation. That is, if there exists a set of local unitary operations such that
\begin{equation*}
\ket{\phi}=U_1 \otimes U_2 \otimes \dots \otimes U_N \ket{\psi} ,
\end{equation*}
then, we say that $\ket{\psi}$  and $\ket{\phi}$  are LU-equivalent states. As entanglement describes non-local properties of the system, it is impossible to increase entanglement by local operations, so we can ensure that $\ket{\phi}$ is at least as entangled as $\ket{\psi}$. Moreover, as unitary operations are invertible, if it is possible to obtain $\ket{\phi}$ from $\ket{\psi}$ by a LU operation with certainty, then it is also possible to obtain $\ket{\psi}$ from $\ket{\phi}$ by a LU operation with certainty, so $\ket{\psi}$ is at least as entangled as $\ket{\phi}$. In conclusion, two LU-equivalent states have the same entanglement; the same can be obtained for $\sigma$ and $\rho$. LU-equivalence can also be understood by the following. When explicitly expressing the coefficient expansion of $\ket{\psi}$ in an orthonormal basis as in Equation \eqref{eq:orthobasispsi}, we observe that: 
\begin{equation*}
\bm{U} \ket{\psi}= \sum_{\bm{i}} c_{\bm{i}} \left(\bm{U} \ket{\bm{i}} \right) =\sum_{\bm{i}} c_{\bm{i}} \ket{\bm{i}'},
\end{equation*}
where $\ket{\bm{i}'}$ denotes another orthonormal basis in $\bm{\mathcal{H}}$ for expanding the state, and $\bm{U}=\bigotimes_{i=1}^N U_i$. Then, local unitary transformations can be viewed as a change in the basis of the local Hilbert spaces, so they change only the local representation of the parts of the state and not the state itself. As entanglement is a non-local correlation, it cannot change by changing the local representations. Later, we will explore how LU equivalence can classify states according to their entanglement. 
\subsection{LOCC and SLOCC operations}
Other operations that we will use are the Local Operations and Classical Communication (LOCC) \cite{LOCC},  which are defined as all possible operations that can be executed on the different subsystems with no interaction between them, as if each subsystem were located in a separate laboratory. This set of operations includes various types of measurements and allows classical communication between the laboratories, enabling the coordination of strategies for applying operations on the subsystems. Mathematically, if a state $\sigma$ is obtained with certainty from LOCC operations  on a state $\rho$, then there exists some local operators $A_i^k$ such that \cite{Plenio}:
\begin{equation*}
 \sigma=\sum_k \bm{A}^{k} \rho  {\bm{A}^{k}}^\dagger, \quad \sum_k \bm{A}^{k} {\bm{A}^{k} }^{\dagger}= \bm{I}, \quad \bm{A} ^{k}=\bigotimes_{i=1} ^N A_i ^{k} 
\end{equation*}
where the $A_i^{k}$ matrices are known as the \textit{Kraus Operators} \cite{Kraus}. This process can also be understood as the initial state $\rho$ being mapped to
\begin{equation*}
\sigma_k =\frac{\bm{A}^{k} \rho  {\bm{A}^{k}}^\dagger}{ \tr(\bm{A}^{k}\rho  {\bm{A}^{k}}^\dagger \ )} ,
\end{equation*}
with a probability given by :
\begin{equation*}
p_k= \tr(\bm{A}^{k}\rho  {\bm{A}^{k}}^\dagger \ ), \quad \sum_{k} p_k =1.
\end{equation*} 
Subsequently, the state after the LOCC can be expressed as:
\begin{equation}
\sigma=\sum_k p_k \sigma_k.
\label{eq:LOCC}
\end{equation}
In this picture, we say that after the LOCC operation, the state $\rho$ is converted to the ensemble $\sigma$ with certainty. Under this framework, all correlations that can be established between two states through LOCC operations are classical. Thus, it is impossible to increase the system's entanglement, but it is possible to reduce entanglement, for instance, by measuring the subsystems. In this sense, we know that state $\sigma$ cannot be more entangled than $\rho$, making LOCC a valuable tool for constructing a hierarchy of entangled states.\\

\noindent Additionally, it is possible to define a class of operations known as Stochastic Local Operations and Classical Communications (SLOCC), which means that LOCC can be performed with some non-zero probability. From the previous description of LOCC in Equation \eqref{eq:LOCC}, we have that with probability $p_k$, the state $\rho$ is transformed into the state $\sigma_k$. In this sense, SLOCC operations can be described as one of the possible branches within LOCC operations. Therefore, any operation of the form:
\begin{equation*}
\bm{A} \ket{\psi}, \quad \bm{A} \rho \bm{A}^{\dagger} ,
\end{equation*}
qualifies as a SLOCC operation, where $\bm{A}=\otimes_{i=1}^{N}A_i$, and $A_i$ are local operators acting on the $i$-th particle. In contrast to LOCC operations, SLOCC operations can potentially increase the entanglement of the initial state. i.e., $\sigma_k$ can be more entangled than $\rho$, although this occurs with a certain probability, $p_k$. When considering all possible scenarios, on average, the entanglement cannot increase.

\section{Entanglement measures}
Quantifying entanglement is a non-trivial task, and numerous attempts have been made to identify measures that align with physical and mathematical principles\cite{Plenio}. These measures must take the form of real positive functions that operate on states. Various desirable properties of such functions, denoted as $E(\rho)$, are outlined below:
\begin{itemize}
\item Any entanglement measure must be null for separable states:
\begin{equation*}
 E\left(\sum_{k} p_k \bm{\rho}^k \right)= E\left(
 \bigotimes_{i} \ket{\psi_i} \right)=0.
\end{equation*}
\item The entanglement measure must remain invariant under LU operations.
\begin{equation*}
E(\rho)=E(\bm{U} \rho \bm{U}^{\dagger})
\end{equation*}
\item The entanglement measure cannot, on average, increase under SLOCC operations. This property is known as ``monotonicity''. Then, if the possible outputs of an LOCC operations are defined by an assemble of density matrices $\{\sigma_k\}$, each with probability $p_k$, this property can be stated as\cite{Plenio}:
\begin{equation*}
E(\rho)\geq \sum_k p_k E(\sigma_k
),
\end{equation*}
\end{itemize}
Functions exhibiting these properties are referred to as \textit{entanglement monotones}. In the following, we will introduce some of these measures that hold relevance in the context of this document.

\subsubsection{Entropy of entanglement}
When considering a pure state composed of two subsystems, the entropy entanglement measure \cite{EntanglementConcentration} is highly relevant. This measure is defined as:
\begin{equation}
E(\rho) = S(\rho_1)=S(\rho_2) ,\quad S(\rho)=-\tr \rho (\log \rho),
\label{eq:Entropy}
\end{equation}
i.e., the Von Neumann entropy of the reduced density matrices. Such reduced density matrices correspond to $\rho_i =\tr_{\overline{\imath}} \rho$, i.e., the trace of the density matrix $\rho$ over the complementary parts of $i$. Using the notation for pure states in Equation \eqref{eq:orthobasispsi} the reduced density matrices are calculated as:
\begin{equation}
\rho_i =tr_{\overline{\imath}} \left( \rho \right) = \sum_{\overline{\imath}}  \bra{\overline{\imath}} \rho \ket{\overline{\imath}}.
\end{equation}
The entropy of entanglement defined in Equation \eqref{eq:Entropy} evaluates to zero for separable states and reaches its maximum value when the reduced density matrix is proportional to the identity:
\begin{equation*}
 S \left(\frac{1}{d} I \right) = \log d.
\end{equation*}
This measure applies to multipartite pure states to quantify entanglement between any two complementary subsystems. It is always possible to assess the entanglement of each part with its complement. As a result, one can define locally maximally entangled (LME) states \cite{BryanLME} as those whose individual reduced density matrices are proportional to the identity:
\begin{equation}
 \rho_i = tr_{\overline{\imath}} \ket{\psi}\bra{ \psi}	=\frac{1}{d_i} \bm{I}.
 \label{eq:LME}
\end{equation}
These states maximize the entropy of entanglement for each part with the full system. LME states are a remarkable class of entangled states and play a central role in many quantum protocols outlined in the introduction. Additionally, as we will see, they become special states when studying orbits under SLOCC  within the total Hilbert space.\\

\noindent For pure states, the system can be separated into two parts, $\bm{i}$, and $\overline{\bm{\imath}}$, and represented using the so-called \textit{Schmidt Decomposition} \cite{Nielsen} as:
\begin{equation*}
 \ket{\psi} = \sum_{j=1}^{d_{\bm{i}}}\sqrt{\lambda_j} \ket{j}_{\bm{i}}\ket{j}_{\overline{\bm{\imath}}}
 \label{eq:Schmidt}
\end{equation*}
where $\ket{j}_{\bm{i}}$ forms an orthonormal basis in $\bm{\mathcal{H}}^{\bm{i}}= \bigotimes_{j\in \bm{i}} \mathcal{H}^{j}$, and we assume $d_{\bm{i}}\leq d_{\overline{\bm{\imath}}}$ . The values $\lambda_j$ are known as \textit{Schmidt Coefficients}. The reduced density matrices are then given by:
\begin{equation*}
\rho_{\bm{i}}=\sum_j \lambda_j \ket{j}_{\bm{i}} \bra{j}_{\bm{i}} , \quad \rho_{\overline{\bm{\imath}}}=\sum_j \lambda_j \ket{j}_{\overline{\bm{\imath}}} \bra{j}_{\overline{\bm{\imath}}} ,
\end{equation*} 
moreover, the entanglement entropy can be calculated as:
\begin{equation}
E(\ket{\psi})=E(\rho_{\bm{i}})=E(\rho_{\overline{\bm{\imath}}})=-\sum_{j=1}^{d_{\bm{i}}} \lambda_j \log \lambda_j,
\label{entropyentanglementschmidt}
\end{equation}
this measure is usually considered the basic measure for bipartite entanglement. We will see later in subsection \ref{EntanglementHayashi} how this measure appears as the rate of entanglement concentration. 
\subsubsection{Concurrence}

For the two-qubit case, another relevant entanglement measure is the \textit{Concurrence} \cite{Concurrence}, which plays an important role in the context of the results presented here. When dealing with mixed states, the concurrence is defined as:
\begin{equation*}
C(\rho)=\max(0,\lambda_1-\lambda_2-\lambda_3-\lambda_4),
\end{equation*}
where, in this case, $\lambda_i$ are the non increasing eigenvalues of
\begin{equation*}
R=\sqrt{\sqrt{\rho}\overline{\rho}\sqrt{\rho}}, \quad \overline{\rho}= (\sigma_y\otimes \sigma_y) \rho ^{*} (\sigma_y\otimes \sigma_y)
\end{equation*}
with $\sigma_y$ representing the \textit{Pauli matrix}, and $\rho^*$ being the complex conjugate of $\rho$. Despite the apparent complexity of this definition, the concurrence can be calculated straightforwardly for pure qubit states. Consider the coefficient expansion of a bipartite pure state:
\begin{equation*}
\ket{\psi}= c_{00}\ket{00}+c_{01}\ket{01}+c_{10}\ket{10}+c_{11}\ket{11},
\end{equation*}
then, the concurrence can be expressed as the absolute value of the determinant of the density matrix:
\begin{equation}
C(\ket{\psi}\bra{\psi}) =\left|2  \left| \begin{array}{cc}
c_{00} & c_{01} \\
c_{10} & c_{11}
\end{array} \right| \right|.
\label{eq:concurrence}
\end{equation}
The concurrence exhibits behavior similar to the entropy of entanglement, being zero for separable states and maximal for locally maximally entangled (LME) two-qubit states; moreover, it is very useful to define multipartite entanglement measures, and it has the property that it is invariant under invertible determinant 1 SLOCC operations \cite{VerstraeteNormal}, that is, local operators in the Special Linear group $SL_d$. Functions with this invariance property that depend polynomially on the coefficients of the state are called \textit{SLOCC polynomial invariants}, which, as we will see in section \ref{SLOCCinvs}, are very useful functions to classify entanglement.

\subsubsection{Three-Tangle}
While the entropy of entanglement and concurrence are valuable tools for measuring entanglement in bipartite systems, they fall short when considering more complex cases, such as a three-qubit pure state. In such a scenario, it becomes essential to introduce a measure that identifies entanglement between all three parts rather than just bipartite entanglement. This is where the \textit{tangle} comes into play, defined as \cite{DistributedEntanglement}:
\begin{equation*}
\tau_{1|2}=C^2(\rho_{12}),
\end{equation*}
allowing us to quantify the residual three-partite entanglement as:
\begin{equation*}
\tau_{123}=\tau_{1|23}-\tau_{1|2}-\tau_{1|3}.
\end{equation*}
This measure is known as the \textit{three-tangle}. It has been shown that for a pure three-partite qubit state with a coefficient expansion:
\begin{equation*}
\ket{\psi}=\sum_{ijk=0}^{1} c_{ijk} \ket{ijk},
\end{equation*} 
the three-tangle can be expressed as:
\begin{equation}
\tau_{123}(\ket{\psi}) = 2\left|c_{ijk}c_{i'j'm}c_{npk'}c_{n'p'm'}\epsilon^{ii'}\epsilon^{jj'}\epsilon^{kk'}\epsilon^{nn'}\epsilon^{mm'}\epsilon^{pp'} \right|= 2HDet_3(\rho),
\label{eq:3tangle}
\end{equation}
where $\epsilon^{ii'}$ is the two-dimensional Levi-Civita tensor. The three-tangle can also be computed as the Cayley's hyperdeterminant of the three-dimensional tensor defined by $c_{ijk}$\cite{Hyperdets}. Hyperdeterminants also belong to the category of SLOCC polynomial invariants, which we will introduce in section \ref{SLOCCinvs}.\\

\noindent Generalizations of entanglement measures for four-qubit systems have been found recently \cite{entangled4}\cite{entangled42}. However, due to their complexity, they are not shown here. Nevertheless, it is noteworthy how defining such measures becomes increasingly challenging with a growing number of particles. Now, we delve into the problem of defining whether two states can or cannot be considered equivalent according to their entanglement with the tools we have introduced. 

\section{Entanglement classification}
\label{Entanglementclassification}
 When studying the problem of entanglement classification, we need to define first what criteria we will use to define two states as equivalent or as belonging to the same class. We already defined two sets of meaningful operations in this sense, and we will use them to define two classifications. These are given by LU operations and SLOCC.
 
 \subsection{LU-classification}
 As discussed in this chapter, LU operations can be interpreted as a change in the choice of orthonormal basis for each constituent Hilbert space $\mathcal{H}^i$; then, the system's entanglement cannot change under LU transformations.   Hence, we can introduce a classification known as \textit{LU-classification} to group LU-equivalent states, which, according to Section \ref{LUOperations}, are always regarded as having the same degree of entanglement. In this context, two systems of $N$ parts, i.e., N-partite states, $\ket{\psi}$ and $\ket{\phi}$, are considered equivalent if there exist unitary matrices, denoted as $U_i$, such that 
 \begin{equation*}
  \ket{\psi} =U_1\otimes U_2 \otimes \dots U_N \ket{\phi}.
 \end{equation*} 
 For bipartite systems, LU classification can be effectively carried out through Schmidt decomposition \cite{Ekert2}, which allows to write the coefficient expansion of a state in the following form:
 \begin{equation*}
 \ket{\psi} = \sum_{j=1}^{d_{\bm{i}}}\sqrt{\lambda_j} \ket{j}_{\bm{i}}\ket{j}_{\overline{\bm{\imath}}},
\end{equation*}
where $\bm{i}$ and $\overline{\bm
{\imath}}$ represent the two subsystems, and $d_{\bm{i}}$ is chosen to be the lowest dimension. Under LU operations, this Schmidt decomposition transforms as:
\begin{equation*}
U_1\otimes U_2 \ket{\psi} = \sum_{j=1}^{d_{\bm{i}}}\sqrt{\lambda_j}  U_1\ket{j}_{\bm{i}} U_2\ket{j}_{\overline{\bm{\imath}}}=\sum_{j=1}^{d_{\bm{i}}}\sqrt{\lambda_j}  \ket{j'}_{\bm{i}} \ket{j'}_{\overline{\bm{\imath}}},
\end{equation*}
where the local basis of both parts rotates while keeping the Schmidt coefficients  $\lambda_j$ invariant; as a result, two states are considered LU-equivalent if their ordered Schmidt coefficients  $\lambda_j$ are the same, allowing us to define different LU-classes for each set of  $\lambda_j$. This equivalence is consistent with the entanglement entropy measure from Equation \eqref{entropyentanglementschmidt} because two states with the same ordered Schmidt coefficients have the same entropy of entanglement. In the case of two qubits, the LU-classes are labeled by a single continuous parameter, namely $\lambda_0$. However, generalizing this classification becomes challenging, even for three qubits. It has been shown \cite{acin3qubits} that any three-qubit pure state can be written in a generalized Schmidt decomposition as:
\begin{equation*}
\ket{\psi}=\lambda_0\ket{000} +\lambda_1 e^{i \phi} \ket{100} + \lambda_2 \ket{101} +\lambda_3 \ket	{110} +\lambda_4 \ket{111},
\end{equation*}
with $\lambda_i\geq0, 0\leq \phi\leq \pi$ and $\sum_{i} \lambda_i^2=1$, requiring five parameters to label the LU-class of normalized states. While it has been demonstrated that this classification can be extended to any number of qubits \cite{LUequivalence}, it is essential to note that the number of parameters required to distinguish each class grows exponentially with the number of constituents. It is shown in \cite{Multipartite-Entanglement} that the number of parameters needed for this classification for $N$ qubits has a lower bound given by $2^{N+1}-3N-2$, being five parameters for three qubits, as stated above, and 18 parameters for four qubits. Next, we explore a coarser classification obtained from SLOCC equivalence.

\subsection{SLOCC-classification}
\label{SLOCC}

In the previous section, we introduced the SLOCC operations due to their operational significance. These operations encompass all possible transformations that can be applied locally to the parts of a state when they are not allowed to interact physically. Then, SLOCC operations cannot create entanglement between separable parties. Now, we will consider that two states $\ket{\psi}$, $\ket{\phi}$ are SLOCC equivalent  if they can be interconverted by reversible SLOCC, i.e.,
\begin{equation*}
   \ket{\psi}= \bm{A} \ket{\phi} , \quad \ket{\phi} = \bm{A}^{-1} \ket{\psi},
\end{equation*}
with this, we can restrict the set of operations $\bm{A}$ to those such that $A_i\in GL_{d}$ where $GL_{d}$ is the group of invertible matrices of dimension $d$, and for simplicity, we will assume all local Hilbert spaces to be of dimension $d$. As we want to classify the states in the projective Hilbert space, we can limit ourselves to the local actions of the Special linear group $SL_{d}$ ( the group of matrices of dimension $d\times d$ with determinant equal to one), and consider unnormalized states. From this perspective, we can ensure that two states $\ket{\psi},\ket{\phi}$ are $SLOCC$ equivalent if there exist a set of unit determinant matrices $B_i\in SL_{d}$, and a scalar $s\in \mathbb{C}$ such that:
\begin{equation*}
\bm{B} \ket{\psi} = s \ket{\phi} ,\quad \bm{B} = \bigotimes_{i=1}^{N} B_i.
\end{equation*}
\noindent It is worth noting that, by definition, SLOCC operations cannot create entanglement from unentangled states. However, due to the stochastic nature of SLOCC, there is a possibility that entanglement between the parties may increase with some non-null probability from an already entangled state, with the condition that when considering all the possible outputs with their respective probabilities, the entanglement does not increase in average. Entanglement neither can be destroyed when considering SLOCC equivalence due to the requirement of being invertible. Because of this, we cannot claim that two SLOCC equivalent states are equally entangled, but we can ensure they have the same type of entanglement. In this sense, SLOCC equivalence separates the total Hilbert space into SLOCC classes, wherein states with the same type of entanglement belong. \\
 
\noindent When studying SLOCC classification, it is useful to exploit the connection with algebraic geometry and classical invariant theory described in  \cite{Projective} \cite{Bryan}. From the algebraic geometry point of view, SLOCC classes correspond to orbits on the projective Hilbert space under the action of $SL_{d}\otimes SL_{d} \otimes \dots \otimes SL_{d}$, that can be described by using \textit{auxiliary varieties}. On the other hand, classical invariant theory permits the calculation of SLOCC \textit{invariants} and \textit{covariants}, that can be used to differentiate SLOCC classes. We will introduce here the most relevant aspects of the connection in the context of this thesis. \\

\noindent In \cite{VerstraeteNormal}, it is shown that starting from any state, we can apply local $SL_d$ operations to obtain any state with a smaller norm in the same orbit. This process can be repeated multiple times to reduce the norm as much as possible.  These states are referred to as \textit{critical states}, and have the property of being LME states \cite{VerstraeteNormal}. It is also known from \textit{Kempf-Ness} theorem \cite{Gour} that such critical states are unique in each orbit up to LU-equivalence. When this process is applied to any state, three possible cases arise.\\

\noindent The first case is when the zero state is asymptotically reached as a result of the process. These states are known as \textit{unstable orbits} \cite{LuqueThibon-finite}, and they correspond to the \textit{Nullcone} in the context of geometric invariant theory.\\

\noindent The second case is when a critical state, distinct from the null state, can be approximated arbitrarily with a finite number of steps but reached only in the infinite limit, indicating that the critical state is in the closure of the orbit. These are known as \textit{strictly semistable orbits} \cite{Slowik}. \\

\noindent The last option is when the critical state is reached within a finite sequence of steps, and they are referred to as \textit{polystable orbits}. In this scenario, critical states are representatives of their corresponding SLOCC class and are also known as \textit{Normal forms} \cite{VerstraeteNormal}. Polystable orbits are especially important in SLOCC classification because they contain the critical states, which are LME states, and also because the set of polystable orbits is of full measure in the total Hilbert space \cite{Gour}, meaning that almost all the total Hilbert space is composed by polystable orbits.\\

\noindent Now, we introduce the definitions of invariants and covariants, which are tools from classical invariant theory \cite{CIT} that can be applied to multipartite quantum systems to identify SLOCC classes.

\subsubsection{Invariants and covariants}
\label{SLOCCinvs}
SL-invariant polynomials \cite{Slowik}, from now on simply \textit{invariants}, are homogeneous polynomials in the coefficients of the state, 
\begin{equation*}
I(s\ket{\psi})=s^{k} I(\ket{\psi}),
\end{equation*}
with $s\in \mathbb{C}$ and $k$ the degree of the polynomial, such that their values do not change under SLOCC operation, this means:
\begin{equation*}
I\left(\ket{\psi}\right)= I \left(\bm{B} \ket{\psi} \right), \quad \bm{B} =\bigotimes_{i=1}^{N} B_i
\end{equation*} 
where $B_i\in SL_d$. Hence, invariants can be used to identify when two states are in different SLOCC orbits or classes, and so they are fundamental tools when performing SLOCC-classification. In general, the possible invariants for a quantum system define a ring \cite{Covariants}, for which a minimal set of generators can be obtained. The number of generators and expressions for them have been found for systems of two, three, and four qubits \cite{VerstraeteNormal} by using techniques in classical invariant theory. Another important result that connects invariants with the problem of classifying entanglement is shown in \cite{VerstraeteNormal},  where it is shown that any linearly homogeneous function of a pure state, that is invariant under SLOCC operations, is an entanglement monotone.\\

\noindent In general, it is impossible to separate the SLOCC classes of a quantum system by only using invariants. For completing a  classification scheme, it is necessary to introduce the definition of \textit{covariants} \cite{Projective}; for this, a qubit state is associated with a form:
\begin{equation*}
\ket{\psi}=\sum_{\bm{i}}\psi_{\bm{i}} \ket{\bm{i}} \Rightarrow f_\psi=\sum_{\bm{i}} \psi_{\bm{i}} \bm{x}_{\bm{i}} , \quad \bm{i} = i^1 i ^2 \dots i^N,\quad  \bm{x}_{\bm{i}} = \prod_{j=1}^{N} x^j_{i^j},
\end{equation*}
where the $x^j_{i^j}$ are binary auxiliary variables associated with the basis elements of the local Hilbert spaces. For example, any three-qubit state is mapped as:
\begin{equation*}
\ket{\psi} =\sum_{i^1i^2i^3 =0} ^1 \psi_{i^1 i ^2 i ^3 } \ket{i^1 i^2 i^3} \Rightarrow f_\psi= \sum_{i^1i^2i^3 =0} ^1 \psi_{i^1 i ^2 i ^3 } x^1_{i^1} x^2 _{i^2} x^3_{i^3}.
\end{equation*}
With this, a covariant is defined as a polynomial acting on the coefficients of the state and the auxiliary variables, $C(\ket{\psi},\bm{x}_{\bm{i}})$, such that
\begin{equation*}
C(\ket{\psi},\bm{x}_{\bm{i}})=C(\bm{B}\ket{\psi},\bm{B}^{-1}\bm{x}_{\bm{i}}),  \qquad \bm{B} \in SL_d \otimes SL_d \otimes \dots \otimes SL_d.
\end{equation*}
In the following, we will explicitly present the SLOCC classification for the cases of two, three, and four qubits, where invariants and covariants play an important role.

\subsubsection{Two-qubit SLOCC classification}
Consider a two-qubit state in its normal form, given by the Schmidt decomposition:
\begin{equation}
\ket{\psi}=\sqrt{p_0}\ket{00}+\sqrt{1-p_0}\ket{11},
\label{eq:twoqubitsSchmidt}
\end{equation} 
with $p_0\leq \frac{1}{2}$. This state can be operated on with the SLOCC corresponding to:
\begin{equation*}
  B_1\otimes B_2=  \frac{1}{\sqrt{2}} \left( \frac{1}{\sqrt{p_0}} \ket{0}\bra{0} +\frac{1}{\sqrt{1-p_0}} \ket{1}\bra{1} \right) \otimes \bm{I},
\end{equation*}
to obtain:
\begin{equation*}
  B_1\otimes B_2 \ket{\psi}=  \ket{\Phi^{+}}.
\end{equation*}
where $\ket{\Phi^{+}}$ is the state shown in Equation \eqref{eq:phi+}, and due to its entropy of entanglement:
\begin{equation*}
E(\ket{\Phi^{+}})=\log 2,
\end{equation*}
we can conclude that it is a maximally entangled state in two qubits. It is important to note that the transformation shown above can only be performed if $p_0\neq0$. Consequently, the two-qubit Hilbert space is divided into two SLOCC classes: the class of product states $(p_0=0)$, with a representative $\ket{00}$, and the class of entangled states $(p_0\neq 0)$, where the representative is chosen to be $\ket{\Phi^{+}}$.

\noindent It is important to note that this classification is coarser than the LU classification. For the case of two qubits, there are infinitely many $LU$ classes, each labeled by $p_0$, whereas there are only two SLOCC classes. The concurrence is the only invariant for two-qubit systems and can be used to distinguish between these two classes. Consider the coefficient expansion of two states:
\begin{equation*}
\ket{\psi}=\sum_{i,j}^{d} T_{i,j} \ket{ij} , \quad \ket{\phi} =\sum_{i,j}^{d} T'_{i,j} \ket{ij},
\end{equation*}
where $T$ and $T'$ are matrices $2\times2$ with the coefficient of the states. If these states are SLOCC equivalent, then there exist some unit-determinant operators $B_1$ and $B_2$ and a non-zero scalar $s$ such that
\begin{equation*}
B_1\otimes B_2 \ket{\psi}=s\ket{\phi}.
\end{equation*}
Then:
\begin{equation*}
s T'=B_1 T B_2^T.
\end{equation*}
By computing the determinant of both sides, we have:
\begin{equation}
\det(s T')= \det(B_1) \det(T) \det(B_2)=\det (T),
\label{eq:Determinant}
\end{equation}
implying that:
\begin{equation*}
s^{d} \det(T')=\det(T).
\end{equation*}
Note that $\det(T)$ is the concurrence, defined in Equation \eqref{eq:concurrence}; this implies that either both concurrences $C(\ket{\psi}\bra{\psi}), C(\ket{\phi}\bra{\phi})$ are equal to zero, or neither of them is zero. It is important to observe that this invariant separates the two classes for two qubits, where $C(\ket{\psi}\bra{\psi})=0$ for separable states and $C(\ket{\psi}\bra{\psi})\neq0$ for entangled states. The same analysis can be applied to the entropy of entanglement: it is zero for the separable class and non-zero for the entangled states. However, the entropy of entanglement is not a polynomial invariant, as it cannot be written as a polynomial in the coefficients of the state. 

\subsubsection{Three qubit SLOCC classification}
\label{Threequbitsclassification}
For classifying the SLOCC classes for three qubits, we can start by checking whether any of the parts of a given state is entangled with the other two or not by using the entropy of entanglement. Note that if any part is separable, it is not possible to entangled it with the others by SLOCC operations. By only considering this, we have the following SLOCC classes:

\begin{itemize}
\item Separable: The first case is when none of the parts is entangled with the other two, corresponding to a completely separable state. For this, all the individual entropies of entanglement are equal to zero:
\begin{equation*}
S(\rho_1) = S(\rho_2)=S(\rho_3)=0.
\end{equation*}
This class is usually labeled as $A-B-C$, and its representative can be chosen to be:
\begin{equation*}
\ket{\psi_{A-B-C}}=\ket{000}.
\end{equation*}
\item Bipartite entanglement: The second case is when only one of the qubits is not entangled with the complement. In this case, its corresponding entropy of entanglement is zero, but the other two are not. For example, when the first part is separable, one has:
\begin{equation*}
S(\rho_1)=0, \, S(\rho_2)\neq 0, \, S(\rho_3)\neq 0.
\end{equation*}
These are labeled as the $A-BC,\,B-AC$, and $C-AB$ classes. The representative of the first one can be chosen to be:
\begin{equation*}
\ket{\psi_{A-BC}}= \frac{1}{\sqrt{2}} \ket{0} \left(\ket{00}+\ket{11} \right),
\end{equation*}
and similar for $B-AC$ and $C-AB$.
\end{itemize}
We are left with all the states where none of the individual entropies of entanglement is equal to zero
\begin{equation*}
S(\rho_1)\neq0 ,S(\rho_2)\neq0, S(\rho_3) \neq 0,
\end{equation*} 
comprising all the genuinely entangled states. However, these states define two separated SLOCC orbits, which can be discriminated by the three-tangle in Equation \eqref{eq:3tangle}. 	The two orbits have as representatives \cite{3qubits} the W and GHZ states
\begin{equation*}
\ket{W} = \frac{1}{\sqrt{3}} \left(\ket{100}+\ket{010}+\ket{001} \right), \quad \ket{GHZ} =\frac{1}{\sqrt{2}} \left(\ket{000}+\ket{111} \right).
\end{equation*}
Hence, their classes are named the \textit{W-SLOCC} class and the \textit{GHZ-SLOCC} class respectively. Both classes are genuinely entangled, but when measuring the three-tangle in the representatives, one has:
\begin{equation*}
\tau_{123}(\ket{W})=0, \quad \tau_{123}(\ket{GHZ})\neq 0,
\end{equation*}
showing that they belong to different orbits. One can notice that the W-SLOCC class belongs to the null cone since applying SLOCC operations, it is possible to get arbitrarily close to the zero state \cite{VerstraeteNormal}:
\begin{equation*}
\lim_{t\rightarrow \infty} \left( \begin{array}{cc}
1/t & 0 \\
0  & t 
\end{array} \right)^{\otimes 3} \ket	{W} =0,
\end{equation*} 
while the GHZ-SLOCC class is a polystable orbit, the only full measure orbit in the total Hilbert space of three qubits \cite{Gour}, with critical state $\ket{GHZ}$ being an LME state. Any other three-qubit state with the property of being an LME state (i.e., individual reduced density matrices are proportional to the identity) has to be LU equivalent to $\ket{GHZ}$ by the Kempf-Ness theorem \cite{Gour}, that ensures that LME states are unique up to LU transformations in closed orbits. The classification of SLOCC classes of three qubits can also be performed by calculating the local entropies of entanglement and the three-tangle, as shown in Table \ref{tab:threequbitsclass}. However, this classification can be performed by using invariants and covariants, as shown in \cite{Zimmerman}\cite{Freudenthal}, where a set of covariants can replace the local entropies of entanglement, and the three-tangle corresponds to the only polynomial invariant under $SL_2\otimes SL_2 \otimes SL_2$ action.\\ 
\begin{table}[h!]
\begin{tabular}{|c|c|c|c|c|}
\hline
                   & $S(\rho_1)$    & $S(\rho_2)$    & $S(\rho_3)$    & $D$      \\ \hline
$\ket{\psi_{A-B-C}}$ & $0$      & $0$      & $0$      & $0$      \\ \hline
$\ket{\psi_{A-BC}}$  & $0$      & $\neq 0$ & $\neq 0$ & $0$      \\ \hline
$\ket{\psi_{B-AC}} $ & $\neq 0$ & $0$      & $\neq 0$ & $0$      \\ \hline
$\ket{\psi_{C-AB}}$  & $\neq 0$ & $\neq 0$ & $0$      & $0$      \\ \hline
$\ket{W_3}$          & $\neq 0$ & $\neq 0$ & $\neq 0$ & $0$      \\ \hline
$\ket{GHZ_3^2} $     & $\neq 0$ & $\neq 0$ & $\neq 0$ & $\neq 0$ \\ \hline
\end{tabular}
\caption{SLOCC classification for three qubits.}
\label{tab:threequbitsclass}
\end{table}

\noindent The GHZ and W  states exhibit entanglement in distinct ways \cite{Multipartite-Entanglement}, emphasizing the relationship between SLOCC classification and entanglement characterization. The GHZ state features entanglement among all three particles, measured by the three-tangle, while the W-state only has entanglement between all 2-1 qubit pairs. These differences influence the properties of each class. When one qubit of the GHZ state is measured on the computational basis, the remaining two qubits collapse into a separable state. However, if one qubit is measured on the basis defined by:
\begin{equation*}
    \ket{+}=\frac{1}{\sqrt{2}}\left(\ket{0}+\ket{1}\right) , \quad  \ket{-}=\frac{1}{\sqrt{2}}\left(\ket{0}+\ket{1}\right),
   % \label{eq:+-basis}
\end{equation*}
both outcomes occur with a probability of $1/2$, and the remaining two qubits are projected into the Bell states $\ket{\Phi^+}$ and $\ket{\Phi^-}$, respectively. Consequently, the GHZ state can be deterministically transformed into an EPR state. If one qubit of the GHZ state is discarded, the reduced density matrix for the other two qubits is described by the equation:
\begin{equation*}
\rho_{23}(GHZ)= \frac{1}{2} \left(\ket{0}\bra{0} +\ket{1}\bra{1} \right),
\end{equation*}
representing an unentangled bipartite mixed state. In contrast, for the W state, if one of the qubits is measured in the computational basis, there is a $2/3$ probability that the remaining two qubits are projected into the Bell state $\ket{\Psi^{+}}$, and with a probability of $1/3$ they are projected to the separable state $\ket{00}$. Therefore, obtaining an EPR pair with certainty is not feasible. However, if one qubit is discarded, the reduced density matrix becomes:
\begin{equation*}
\rho_{23}(W)= \frac{1}{3}\ket{0}\bra{0} + \frac{2}{3} \left(\ket{\Psi^+}\bra{\Psi^+} \right),
\end{equation*}
representing an entangled mixed state. In this regard, the entanglement of the W class proves to be more robust under particle loss compared to the GHZ class. The method presented in Chapter \ref{Chapter4} is a construction that uses bipartite entangled states and W states to build more general multipartite states.

\subsubsection{Four qubit classification}
When dealing with states of four qubits, the complexity of the problem increases dramatically, leading to an infinite number of SLOCC classes. Nevertheless, a study by Verstraete and others \cite{Verstraete} demonstrates how SLOCC classes can be separated into nine parametric families based on their normal forms. These normal forms had one mistake that was corrected in \cite{NormCorrect}; the corrected normal forms are detailed in Table \ref{table:fourqubitsclass}. However, it is essential to understand that these families hold different meanings than the classes discussed earlier. Within the same family, one may encounter states with different entanglement properties. For example, the family denoted as $L_{abc_2}$ is typically entangled, but when $a = b = c = 0$, the state $L_{000_2} = \ket{0110}$ becomes completely separable. \\
\begin{table}[h!]
\begin{tabular}{|c|}
\hline
      $L_{0_{3\oplus	\bar{1}}0_ {3\oplus	\bar{1}}}=\ket{0000}+ \ket{0111}$\\ \hline
      $L_{0_{7\oplus	\bar{1}}}=\ket{0000}+\ket{1011}+ \ket{1101}+\ket{1110}$\\ \hline
      $L_{0_{5\oplus	\bar{3}}}=\ket{0000}+ \ket{0101}+\ket{1000}+\ket{1110}$\\ \hline
      $L_{a_20_{3\oplus	\bar{1}}}=a\left(\ket{0000}+\ket{1111} \right)+\ket{0011}+\ket{0101}+ \ket{0110}$\\ \hline
      $L_{a_4}=a\left(\ket{0000}+ \ket{1111}+\ket{0101}+\ket{1010}\right)+i\ket{0001}-i\ket{1011}$\\ \hline
      $L_{ab_3}=a\left(\ket{0000}+ \ket{1111} \right) +\frac{a+b}{2}\left(\ket{0101}+\ket{1010}\right)+\frac{a-b}{2}\left(\ket{0110}+\ket{1001}\right)$\\ 
$      +\frac{i}{\sqrt{2}}\left( \ket{0001}+\ket{0010}-\ket{0111}-\ket{1011}\right)$ \\ \hline
      $L_{a_2b_2}=a\left(\ket{0000}+ \ket{1111}\right)+b\left( \ket{0101}+\ket{1010}\right)+\ket{0110}+\ket{0011}$\\ \hline
      $L_{abc_2}=\frac{a+b}{2}\left(\ket{0000}+ \ket{1111}\right)+\frac{a-b}{2}\left(\ket{0011}+ \ket{1100}\right) +c \left(\ket{0101}+\ket{1010} \right)+\ket{0110}$\\ \hline
      $G_{abcd}=\frac{a+d}{2}\left(\ket{0000}+ \ket{1111}\right)+\frac{a-d}{2}\left(\ket{0011}+ \ket{1100}\right)$ \\
      $+\frac{b+c}{2}\left(\ket{0101}+ \ket{1010}\right)+\frac{b-c}{2}\left(\ket{1001}+ \ket{0110}\right)$\\ \hline
\end{tabular}
\caption{Normal forms of SLOCC families for four qubits in the classification of \cite{Verstraete}.}
\label{table:fourqubitsclass}
\end{table} 
\\

\noindent Within each parametric family, the choices of parameters $a, b, c, d$ generally correspond to different classes within that family. One prominent family is the $G_{abcd}$, the only full-measure family. All the other families depend on up to three parameters, making them infinitely smaller than the $G_{abcd}$ family. The normal form of the state shown in Table \ref{table:fourqubitsclass} has the unique property that its one-particle density matrices are proportional to the identity, thus making them LME states. Interestingly, when $a = b = c = d$, the state becomes a product of two $EPR$ states, once again underscoring how the entanglement properties of each family depend on the choices of parameters.\\

\noindent Similarly to the case of three qubits, invariants, and covariants can be employed to differentiate between various SLOCC classes and families of four qubit states. In the four-qubit scenario, the invariants form a ring with four generators and 170 covariants \cite{LuqueThibon-finite}. These generators are denoted as $B, L, M, D_{xy}$, where $B$ is a polynomial invariant of degree 2, and $L$ and $M$ are polynomials of degree 4, while $D_{xy}$ has a degree of 6. However, to find these expressions, the classical invariant theory was utilized, which required the computation of a set of covariants to derive the invariants, which is beyond the scope of this document. In their work \cite{LuqueThibon-tame}, Holweck, Luque, and Thibon introduced a classification scheme based on these invariants and a subset of covariants to determine the belonging of any state to a specific family from Table \ref{table:fourqubitsclass}. We employ this classification to demonstrate that any four-qubit state can be constructed using the method presented in Chapter \ref{Chapter4}. \\

\noindent In the case of five-qubit systems, there is no known SLOCC classification, and although it is known that there are $17$ generators for the algebra of invariants \cite{5qubits-Thibon}, closed expressions to compute them for all cases have yet to be determined. In Chapter 7, we present a novel method offering a graphical interpretation of these invariants, making it possible to construct and work with them in multi-qubit systems. We also provide graphical representations for complete sets of independent invariants for systems ranging from two to five qubits.\\

\noindent There is a nice connection between the number of non-null invariants and critical states. For the cases of two qubits, the only invariant is the concurrence, and the SLOCC class, with no null concurrence, can be represented with an LME state, namely $\ket{\Phi^{+}}$. In the case of three qubits, the only invariant is the three-tangle, and the SLOCC class with non-null three-tangle is the GHZ-class, and it is the unique class that can be represented with an LME state, namely the GHZ state. In the scenario of four qubits, there are four independent polynomial invariants, and the only family where all of the invariants are truly independent is the $G_{abcd}$ family, which is the only family represented by an LME state. This observation is not a coincidence; LME states are not only important to label SLOCC closed orbits, but LME states can also be obtained from imposing invariance on states under the action of the irreducible representation of a given group. The main focus of this research is to calculate explicitly a family of LME states that arise from the invariance under the action of the symmetric group. In the next chapter, we will introduce the tools from representation theory relevant to this construction.\\

\chapter{Preliminaries II: Representation Theory}
\label{Chapter2}
The main goal of this research is to develop a systematic method to calculate a basis for Kronecker subspaces, composed of Locally Maximally Entangled (LME) states, which we name Kronecker states. A Kronecker subspace is an invariant subspace of the tensor product of irreducible representations of the symmetric group in $n$ elements, denoted as $S_n$. A strong relationship between $S_n$ and $GL_d$, the General Linear group of dimension $d$, is exploited to construct these Kronecker states. This relation is made explicit by the \textit{Schur-Weyl duality}  that we will introduce in Section \ref{Schur-Weyl}. \\

\noindent In chapter \ref{Chapter5}, we will present our findings, but before we delve into the results, we will introduce some fundamental concepts related to groups and representation theory. We will discuss the irreducible representations of both $S_n$ and $GL_d$. Subsequently, we will elucidate the Schur-Weyl Duality, explaining how it can be applied, focusing on the case with $d=2$. This approach will use the \textit{Schur transform}, presented in Section \ref{Schurtransform}, a crucial tool for the results presented in this research.\\

\noindent In this chapter, we will confine our discussion to the most pertinent aspects to the context of this study. For those interested in more comprehensive discussions into groups and representation theory, we recommend consulting resources such as \cite{Fulton}, \cite{chenbook}, and \cite{Sagan}, which provide similar approaches and extensive coverage of these topics. 
\section{Groups and Representations}
\subsection{Groups}
A group is a mathematical structure consisting of a set $G$ and a binary operation ``$\cdot$'' that takes any two elements $g_1$ and $g_2$ of $G$ to form a third element $g_3 = g_1 \cdot g_2$, such that the following axioms are satisfied:
\begin{itemize}
    \item {\bf Closure:} For any pair of elements $g_1$ and $g_2$ in the group $G$, $g_1 \cdot g_2$, must also belong to the set $G$.
    \item {\bf Associativity:} The binary operation must be associative, which means that for all elements $g_1$, $g_2$, and $g_3$ in the group $G$, it holds that $ (g_1 \cdot g_2) \cdot g_3 = g_1 \cdot (g_2 \cdot g_3)$.
    \item {\bf Identity element:} There exists an element $e \in G$, called the identity element, such that for all $g \in G$, $g \cdot e = e \cdot g = g$.
    \item {\bf Inverse element:} For each element $g$ within the group, a corresponding element $g^{-1}$ must exist in the group, referred to as the inverse element of $g$. The inverse element satisfies the condition that $g \cdot g^{-1} = g^{-1} \cdot g = e$.
\end{itemize}

\noindent We will focus on two specific groups, namely, the Symmetric Group on $n$ elements, $S_n$, and the General Linear group of dimension $d$, $GL_d$.

\subsubsection*{The Symmetric group $S_n$}

Consider the set $S_n$ defined as the set of $n!$ permutations of $n$ distinct elements, i.e., all the possible re-orderings of the elements and the binary operation given by the composition of permutations. We use the cycle notation for permutations:
\begin{equation*}
\pi=(p_1,p_2,p_3,\dots p_k),
\end{equation*}
whose action sends each element $p_i$ to the $p_{i+1}$ position, and $p_k$ is sent to the $p_1$ position. For example, when the permutation $(135)$ acts on a set of $n=5$ elements labeled $\{A,B,C,D,E\}$, it rearranges the elements as follows:
\begin{equation*}
(135) \{A,B,C,D,E\}=\{E,B,A,D,C\}.
\end{equation*}
Several essential facts about cycles within permutations are worth noting and will be valuable for later discussions:

\begin{itemize}
\item Any permutation in $S_n$ can be expressed as the product of disjoint cycles. The cycle structure, denoted by $\rho$, is a list of non-increasing lengths of these cycles. Any elements that remain unpermuted are included as cycles of length 1. For example, in $n=5$, the permutation $\pi=(135)$, must be written as $\pi=(135)(2)(4)$ and then, its cycle structure is given by $\rho[(135)(2)(4)]=[311]$.

\item Cycles with only two elements are called "transpositions." When two adjacent elements are swapped, it is called an "adjacent transposition."

\item Every element in $S_n$ can be represented as a product of adjacent transpositions. 

\item We can represent permutations as products of disjoint cycles. 
\end{itemize}

Let us now examine how the axioms of a group are satisfied by the set of permutations:

\begin{itemize}
\item {\bf Closure:} The composition of permutations in $S_n$ results in another permutation of $n$ elements. For instance, consider the composition of $(23)$ and $(135)$. It involves permuting first with $(135)$ and then permuting the result with $(23)$. The outcome is another permutation, represented as $(1235)$:
\begin{equation*}
(23)(135) \{A,B,C,D,E\}= (23) \{E,B,A,D,C\} =\{E,A,B,D,C\} = (1235) \{A,B,C,D,E\},
\end{equation*}
\item {\bf Associativity:} The composition operation is naturally associative. 
\item {\bf Identity element:} The identity element $e$ can be expressed as $e = (1)(2)\ldots(n)$, where each cycle of length one indicates that the elements remain in their original positions. Consequently, under the action of $e$, no element within the set is moved or permuted.
\item{ \bf Inverse element:}  For each permutation $\pi$, there is always an inverse element $\pi^{-1}$ that effectively reverses the permutation. For instance, if $\pi = (1235)$, its inverse is $\pi^{-1} = (1532)$:
\begin{equation*}
(1532)(1235)\{A,B,C,D,E\}= (1532) \{E,A,B,D,C\} = \{A,B,C,D,E\}.
\end{equation*}
\end{itemize}
By satisfying these axioms, it is evident that $S_n$ indeed forms a group known as the symmetric group. 

\subsubsection{The General Linear group $GL_d$}

Now, consider the set $GL_d$ composed of the invertible complex square matrices of dimension $d$. The binary operation defined for $GL_d$ is matrix multiplication. Then, for the group axioms, we have:
\begin{itemize}
\item {\bf Closure:} The product of two matrices in $GL_d$ results in another invertible matrix of dimension $d$, showing that the set is closed under matrix multiplication.

\item {\bf Associativity:} Matrix multiplication is inherently associative. 
\item {\bf Identity element:} The identity element in $GL_d$ corresponds to the identity matrix of dimension $d$, typically denoted as $I_d$. This identity matrix acts as the multiplicative identity, satisfying $A\cdot I_d = I_d \cdot A=A$ for all $A\in GL_d$.
\item  {\bf Inverse element:} For every matrix $A \in GL_d$ there exists an inverse matrix $A^{-1} \in GL_d$ such that $A\cdot A^{-1}=A^{-1} \cdot A= I_d$.
\end{itemize}
In summary, $GL_d$ indeed forms a group known as the General Linear group. \\

\noindent Now, let us introduce some essential definitions for groups, providing examples whenever possible from the context for $S_n$ and $GL_d$.

\subsubsection{Conjugacy classes}
Let $G$ be a group and $x$ any group element. We define the conjugacy class of $x$ as:
\begin{equation*}
c_x =\{ g\in G|g= h x h^{-1},\text{for some } h \in G \},
\end{equation*}
where $hxh^{-1}$ is known as the \textit{conjugation operation}. This operation defines an equivalence relation that separates the group $G$ in conjugacy classes.\\

\noindent Conjugacy operation in $GL_d$ let the set of eigenvalues invariant, then, the conjugacy classes of $GL_d$ separate the matrices according to their sets of eigenvalues $a=(a_1,a_2,\dots, a_d)$ up to permutations.\\

\noindent For $S_n$, it turns out that conjugacy operation only relabel the disjoint cycles that uniquely define the permutation, not changing their sizes \cite{Zeilinger}. This means that in each conjugacy class $c_x$, all permutations have the same cycle structure as $x$. It is natural to label each distinct conjugacy class according to the cycle structure of its elements. For that, let us introduce the definition of \textit{partition} to label such conjugacy classes. 

\subsubsection{Partitions}
\label{Partitions}
 The partitions of $n$ are the different lists $\lambda$ of non increasing positive integers $\lambda_i$ such that $\sum_{i} \lambda_i =n$. There is a correspondence between all the partitions of $n$ and the possible cycle structures of $S_n$. For example, for $S_4$, there are five partitions: 
 \begin{equation*}
\lambda^1 = (4) ,\ , \lambda^2 =(3,1) , \, \lambda^3=(2,2) , \, \lambda^4 = (2,1,1) , \, \lambda^5 =(1,1,1,1),
 \end{equation*}
  which correspond to the $5$ cycle structures and hence conjugacy classes of $S_4$. It is customary to gather repeated numbers as powers, letting the partitions in the previous example be:
  \begin{equation*}
\lambda^1 = (4) ,\, \lambda^2 =(3,1) , \, \lambda^3=(2,2) , \, \lambda^4 = (2,1^2) , \, \lambda^5 =(1^4).
 \end{equation*}
Partitions will be very important in this document because they are also used to label the irreducible representations of $GL_d$ and $S_n$, as we will show later.
\subsubsection{The Group Table}
The group table is a very useful tool to understand the structure of a finite group. It lists the result of the binary operation between the elements in the group. For instance, the elements of $S_3$ are $\{e, (12), (13), (23), (123), (132)\}$ and the group table is shown below:
\begin{equation}
    \begin{array}{|c|c|c|c|c|c|c|}
    \hline
\cdot &\cellcolor[HTML]{C0C0C0} e    & \cellcolor[HTML]{C0C0C0}(12)  & \cellcolor[HTML]{C0C0C0}(13)  & \cellcolor[HTML]{C0C0C0}(23)  &\cellcolor[HTML]{C0C0C0}(123) & \cellcolor[HTML]{C0C0C0}(132) \\\hline
\cellcolor[HTML]{C0C0C0}e     & e    & (12)  & (13)  & (23)  & (123) & (132) \\\hline
\cellcolor[HTML]{C0C0C0}(12)     & (12)  & e     & (132) & (123) & (23)  & (13)  \\\hline
\cellcolor[HTML]{C0C0C0}(13)     & (13)  & (123) & e    & (132) & (12)  & (23)  \\\hline
\cellcolor[HTML]{C0C0C0}(23)     & (23)  & (132) & (123) & e    & (13)  & (12)  \\\hline
\cellcolor[HTML]{C0C0C0}(123)    & (123) & (13)  & (23)  & (12)  & (132) & e    \\\hline
\cellcolor[HTML]{C0C0C0}(132)    & (132) & (23)  & (12)  & (13)  & e    & (123)  \\ \hline
\end{array}
\label{eq:S3table}
\end{equation}
Note that in each row and column, the group elements appear once.
\subsubsection{Subgroups}
A subset $G_s$ of $G$ is said to be a subgroup of $G$ if it satisfies all the conditions for a group and is denoted by $G \supset G_s$. If $G_s\supset G_s'$, then together they form a chain group $G \supset G_s \supset G_s'$.\\

\noindent $GL_d$ has many interesting subgroups. For example, when considering the set of square matrices of dimension $d$ with unit determinant, they satisfy the conditions for being a group, known as the \textit{special linear} group $SL_d$, so we have $GL_d \supset SL_d$. The group of unitary matrices of dimension $d$, i.e., the \textit{Unitary} group, $U_d$, is also a subgroup of $GL_d$. $U_d$ also has a subgroup, the group of unitary matrices of dimension $d$ with unit determinant, known as the \textit{special unitary} group, $SU_d$, so there is a group chain corresponding to $GL_d\supset U_d \supset SU_d$.\\

\noindent For finite groups, subgroups can be read from the group table as any combination of rows/columns where the corresponding elements appear once for each row/column. For $S_3$, the subgroups are $\{e\}, \{e,(12)\}, \{e,(13)\}, \{e,(23)\}, \{e,(123),(132)\}$. In $S_n$, the group chain $S_n \supset S_{n-1} \supset S_{n-2} \supset \cdots \supset S_2$ is very relevant, as we will see later.

\subsubsection*{Homomorphism and Isomorphism}
The mathematical structure represented by a group is an abstract concept that can be thought of as something separated from the definition chosen for the group itself \cite{chenbook}. The definitions of homomorphisms and isomorphisms help us understand this group structure feature. For this, define a map $\Phi$ that takes elements on a group $G$ and map them to elements on another group $G'$, this is written as:
\begin{equation*}
\Phi: G \rightarrow G',
\end{equation*} 
then $\Phi$ is said to be an homomorphism if for any $g_1,g_2,g_3$ such that $g_1 \cdot g_2=g_3$ it holds that:
\begin{equation*}
\Phi(g_1)\cdot \Phi(g_2) = \Phi(g_3),
\end{equation*}
which can be understood as the map preserving the multiplication rule of the original group. If there exists any homomorphism $\Phi$ between $G$ and $G'$, we can say that $G$ is homomorphic to $G'$, and denote this relation as:
\begin{equation*}
G\rightarrow G'.
\end{equation*} 
Any group has a homomorphism where all the elements are mapped to the identity, i.e.:
\begin{equation*}
\Phi(g)=e, \quad  \forall g\in G,
\end{equation*}
where the map preserves the multiplication rule. However, some homomorphisms are less trivial. For example, in the symmetric group, we have $S_3\rightarrow S_2$ when considering the map that acts as:
\begin{equation*}
\Phi((123))=\Phi((132))=\Phi(e)=e, \quad \Phi((12))=\Phi((13))= \Phi((23)),
\end{equation*}
preserving the multiplication rule for all the elements in $S_3$.\\

\noindent Another interesting homomorphic relation is $GL_d\rightarrow SL_d$, where $SL_d$ is the group of unit-determinant matrices with dimensions $d\times d$, and the map $\Phi$ acts on any element $A\in GL_d$ as:
\begin{equation*}
\Phi(A) = \frac{1}{\det(A)^{1/d}} A,
\end{equation*}
which also preserves the multiplication.\\

\noindent There is a special kind of homomorphism when there is a one-to-one correspondence between the elements of $G$ and the elements of $G'$, i.e., no two elements of $G$ are mapped to the same element in $G'$ or vice-versa. In this case, we say that $\Phi$ is an isomorphism; hence, $G$ and $G'$ are isomorphic groups. We denote this relation as:
\begin{equation*}
G\approx G'.
\end{equation*}
For finite groups, this means that the group tables of $G$ and $G'$ are equal, up to some relabeling of the elements. Two isomorphic groups represent the same abstract group.\\

\noindent For example, consider the set of symmetry transformations in the plane that leave the vertices of an equilateral triangle to be fixed, also known as the \textit{Dihedral} group, $D_3$. Its elements correspond to three reflections $\sigma_1,\sigma_2,\sigma_3$, two rotations $C_3,C_3^2$, and the identity $e$ as shown in the following figure:
\begin{equation*}
 \begin{tikzpicture}
 \draw[\lac \rac] (-0.5,2.4) to (0.5,2.4);
 \draw (0,2.6) node {$\sigma_1$};
  \draw (90:2)--(210:2)--(-30:2)--cycle;
 \draw[dashed] (0,2.5) to (0,-1.5);
 \draw[\lac \rac] (-2.8,-1.3) to (-2,-1.7);
 \draw (-2.5,-1.8) node {$\sigma_3$};
  \draw[dashed] (-2.5,-1.5) to (2.4,1.5);
   \draw[dashed] (2.5,-1.5) to (-2.4,1.5);
   \draw[\lac \rac] (2.8,-1.3) to (2,-1.7);
 \draw (2.5,-1.8) node {$\sigma_2$};
 \draw[\lac-] (0.5, 2) to[bend left=50] (2,-1);
  \draw (2.5,0.5) node {$C_3$};
  \draw[\lac-] (-0.5, 2) to[bend right=50] (-2,-1);
  \draw (-2.5,0.5) node {$C_3^2$};
 \end{tikzpicture}
\end{equation*}
 The group table of $D_3$ is:
 \begin{equation*}
    \begin{array}{|c|c|c|c|c|c|c|}
    \hline
\cdot & \cellcolor[HTML]{C0C0C0}e    &\cellcolor[HTML]{C0C0C0} \sigma_1  & \cellcolor[HTML]{C0C0C0}\sigma_2  &\cellcolor[HTML]{C0C0C0} \sigma_3  &\cellcolor[HTML]{C0C0C0} C_3 & \cellcolor[HTML]{C0C0C0}C_3^2 \\\hline
\cellcolor[HTML]{C0C0C0} e        & e     & \sigma_1  & \sigma_2  & \sigma_3  & C_3 & C_3^2 \\\hline
\cellcolor[HTML]{C0C0C0} \sigma_1     & \sigma_1  & e     & C_3^2 & C_3 & \sigma_3  & \sigma_2  \\\hline
\cellcolor[HTML]{C0C0C0} \sigma_2     & \sigma_2  & C_3 & e    & C_3^2 & \sigma_1  & \sigma_3  \\\hline
\cellcolor[HTML]{C0C0C0} \sigma_3     & \sigma_3  & C_3^2 & C_3 & e     & \sigma_2  & \sigma_1  \\\hline
\cellcolor[HTML]{C0C0C0} C_3    & C_3 & \sigma_2  & \sigma_3  & \sigma_1  & C_3^2 & e     \\\hline
\cellcolor[HTML]{C0C0C0} C_3^2    & C_3^2 & \sigma_3  & \sigma_1  & \sigma_2  & e     & C_3  \\ \hline
\end{array}.
\end{equation*}
 Note how this group table is the same as \eqref{eq:S3table} after the relabeling:
 \begin{equation*}
    e\rightarrow e, \sigma_1 \rightarrow (12), \sigma_2 \rightarrow (13), \sigma_3 \rightarrow (23) ,C_3\rightarrow (123) , C_3^2 \rightarrow (132).
 \end{equation*}
 Then, it is clear that $D_3$ and $S_3$ are isomorphic; they are the same abstract group. One theorem that highlights the importance of studying the $S_n$ group instead of any other finite group is \textit{Cayley's theorem}.

\subsubsection{Cayley's Theorem}
\label{Cayleys}
 \textit{Every finite group $G$ is isomorphic to a subgroup of the permutation group $S_{|G|}$.}\\
 
 \noindent The proof of this theorem is very simple. Label the elements of any group $G=\{g_1,g_2 , \dots , g_{|G|}\}$, then write the group table. Each row of the table will correspond to a permutation $\pi$ of $G$ because being $G$ a group, each element $g$ appears once in each row, in a position defined by the multiplication rule. Then, we can associate each row with the permutation $\pi$ of $S_{|G|}$. Therefore, every group $G$ can be embedded in $S_{|G|}$. \\
 
 \noindent Cayley's theorem makes explicit the importance of the symmetric group when studying finite groups.

\subsection{Representations}
A representation is a homomorphism of a group $G$ to a set of matrices in $GL_d$ for some $d$. The homomorphism is given by a map
\begin{equation*}
    X: G \rightarrow GL_d
\end{equation*}
such that 
\begin{itemize}
    \item $X(e)=I$, where $I$ is the identity matrix.
\item $X(g_1\cdot g_2)=X(g_1) \cdot X(g_2) \quad \forall g_1,g_2 \in G$.
\end{itemize}
All groups have a trivial representation where each element is mapped to the number 1, a representation of dimension 1. 
\begin{equation*}
    X^{triv}(e)=1, X^{triv}(g_1)\cdot X^{triv}(g_2)=(1) \cdot (1)=1=X^{triv}(g_1 \cdot g_2),
\end{equation*}
where we used a superscript to label the representation as it will be usual, and $triv$ stands for the trivial representation. For $S_n$, it is possible to build a natural representation of dimension $d=n$, where the matrix $X^{def}(\pi)$ associated with a permutation $\pi$ consists of the identity matrix after exchanging rows as $\pi$ indicates. For example, for $S_3$, we have:
\begin{equation*}
    S_3: \quad X^{def}(12)= \left( \begin{array}{ccc}
       0  &1&0  \\
        1 & 0 &0 \\
        0 & 0&1 
    \end{array}  \right)   \quad  , \quad X^{def}(123)= \left( \begin{array}{ccc}
       0  &0&1  \\
        1 & 0 &0 \\
        0 & 1&0
    \end{array}  \right).
\label{eq:Definingrepresentation}
\end{equation*}
This representation is the \textit{defining representation} of $S_n$.
Representations are a powerful tool for studying groups, as they allow us to understand the structure and properties of a group through its action on matrices. In physics, representations are used extensively in fields such as particle physics \cite{chenbook}, gravitation\cite{Repsingravitation}, and more specifically in quantum mechanics \cite{Repinquantum}, where vectors and matrices represent states and actions on states in a vector space. \\

\noindent The vector space $V$ defining the basis of the representation is known as the representation space. In the previous example, the representation space is $V=\mathbb{C}\{\ket{1},\ket{2},\ket{3}\}$ where:
\begin{equation*}
    \ket{1}= \left(\begin{array}{c}
         1 \\
         0 \\
         0
    \end{array} \right), \quad  \ket{2}= \left(\begin{array}{c}
         0 \\
         1 \\
         0
    \end{array} \right), \quad  \ket{3}= \left(\begin{array}{c}
         0 \\
         0 \\
         1
    \end{array} \right),
\end{equation*}
and $\mathbb{C}\bm{S}$ denotes the vector space generated by the elements in $\bm{S}$ over $\mathbb{C}$. For the previous example, the representation space considers all the possible linear combinations 
\begin{equation*}
 c_1 \ket{1} + c_2 \ket{2} + c_3 \ket{3},
\end{equation*}
with $c_i\in \mathbb{C}$. Another useful representation for $S_n$ is the regular representation, where the representation space is defined as $V=\mathbb{C} S_n$, i.e., the action of the group on itself. This representation has dimension $n!$, and we can define the orthonormal basis elements as:
\begin{equation}
\{\ket{e}, \, \ket{(12)}, \, \ket{(13)}, \, \ket{(23)}, \, \ket{(123)} , \, \ket{(132)} \}.
\label{eq:regbas}
\end{equation}
Then, the action of the group in these elements defines the representation matrices. For example, for $X^{reg}(12)$, we have:
\begin{equation}
	X^{reg} (12) = \left(\begin{array}{cccccc}
	0 & 1 & 0 & 0 &0 &0 \\
	1 & 0 & 0 & 0 &0 &0 \\
	0 & 0 & 0 & 0 &1 &0 \\
	0 & 0 & 0 & 0 &0 &1 \\
	0 & 0 & 1 & 0 &0 &0 \\
	0 & 0 & 0 & 1 &0 &0 \\
\end{array}	 \right).
\label{eq:regrep}
\end{equation}
For the group $GL_d$, a recurrent representation in this document is the \textit{standard representation}, where each matrix $A\in GL_d$ is represented by itself. In this case, the representation space is defined by $d$ orthonormal  vectors as  $V=\mathbb{C}\{\ket{0}, \ket{1} ,\dots, \ket{d-1}\}$. \\ 

\noindent There are many possible representations for a given group. One of the objectives of representation theory is to find the basic pieces that can be used to represent a group, known as irreducible representations. 

\subsubsection*{Reducibility}
Representations can be broken down into smaller pieces that are easier to analyze. Reducibility is an important concept in representation theory because it allows us to study complex representations by analyzing simpler ones. For that, define $W$ as a subspace of a representation space $V$, with representation $X$ such that 
\begin{equation*}
    \ket{w}\in W \rightarrow X(g) \ket{w} \in W \quad \forall g \in G
\end{equation*}
This subspace is also known as an \textit{invariant subspace}, which is a representation space on its own.\\

\noindent For example, consider the defining representation for $S_n$ with $V=\mathbb{C}\{\ket{1},\ket{2},\dots,\ket{n}\}$, and take the one dimensional supspace given by $W=\mathbb{C}\{\ket{1}+\ket{2}+\dots \ket{n}\}$. The action of any $\pi$ over vectors in $W$ will be:
\begin{equation*}
    X(\pi) c(\ket{1}+\ket{2}+\dots + \ket{n}) =c(\ket{\pi(1)}+\ket{\pi(2)}+\dots + \ket{\pi(n)})=c(\ket{1}+\ket{2}+\dots + \ket{n}) \in W
\end{equation*}
So $W$ is an invariant subspace of $V$. Note that the representation in this subspace is given by $X(\pi)=1, \forall \pi \in S_n$, in other words, the trivial representation. 

\noindent A representation space $V$ is said to be \textit{reducible} if it contains a nontrivial invariant subspace $W$ ($V$ and 0 are trivial invariant subspaces). Otherwise $V$ is said to be \textit{irreducible}.\\

\noindent For the regular representation, it is also possible to identify the trivial representation on the invariant subspace $W=\mathbb{C}\{\ket{(e)}+\ket{(12)}+\dots \} $. However, the regular representation is even more interesting because it is known that all the irreducible representations appear in it. 

\subsubsection*{Complete reducibility}
The idea is to reduce any representation as much as possible. Note how the representation space can be separated into subrepresentations, and the process can be repeated until we end up with only irreducible representations, then 
\begin{equation*}
    V=W^{1}\oplus W^{2} \oplus \dots \oplus W^{k}
    \label{eq:completereduct}
\end{equation*}
with $W^{i}$ irreducible representations. After identifying these irreducible representations, the matrix representations on $V$ can be written in a block diagonal form:
\begin{equation*}
    X(g) \cong \left(\begin{array}{ccccc}
    X^{\lambda^1}(g) &  & & &  \\
     & X^{\lambda^2}(g) & & & \\
     & & . & &\\
     & & & . & \\
     & & & & X^{\lambda^k}(g)
\end{array} \right).
\label{eq:BlockDiag}
\end{equation*}
Where $X^{\lambda^i}(g)$ is the representation matrix of $g$ in the irreducible representation $W^{\lambda^i}$, and the symbol "$\cong$" is used to make explicit that a change of basis is needed for the block-diagonalization of the matrix. We will see that for the groups we are interested in, the irreducible representations can be labeled by partitions; hence, the ${\lambda^i}$ label will appear naturally. In general, when making this reduction, some of the irreducible representations appear more than once, so we can state that a representation can be decomposed in unique irreducible representations $\lambda^i$ as:
\begin{equation}
X(g)= \bigoplus_{i} X^{\lambda^i}(g) \otimes I_{m_i},
\label{eq:Irrepdecomp}
\end{equation}
where $m_i$ is the number of times that the irreducible representation $\lambda^i$ appears in this decomposition, also known as the \textit{multiplicity}.
\subsection{Irreducibility}
The next task is introducing a tool to determine whether a representation is reducible or irreducible. We need to introduce first some definitions from representation theory:
\subsubsection*{Isomorphic representations}
 First, we define a \textit{group homomorphism} as a linear map between two representations of a group $G$ that respects the group structure. Specifically, let $(V,X)$ and $(W,Y)$ be two representations of $G$, where $V$ and $W$ are the representation spaces, and $X$ and $Y$ are the homomorphic maps from $G$ to $V$ and $W$ respectively. A linear map $T:V\to W$ is called a group homomorphism if it satisfies the following condition for all $g\in G$ and $\ket{v}\in V$:
\begin{equation*}
T (X(g) \ket{v})=Y(g) (T \ket{v}),
\end{equation*}
which implies that
\begin{equation*}
T X(g) =Y(g) T.
\label{eq:homomorphism}
\end{equation*}
We define a \textit{group isomorphism} as a bijective group homomorphism, meaning that the map $T$ is invertible. Therefore, two representations spaces $V$ and $W$ with maps $X$ and $Y$ are isomorphic when there exists a transformation $T$ such that
\begin{equation*} Y(g)=TX(g)T^{-1} \quad \forall g\in G.
\end{equation*}
In this sense, two representations are isomorphic or equivalent when they only differ by a change of basis.
With this, we can introduce Schur's Lemma, which is a very important result of representation theory and will be recurrent in the context of this document.
\subsubsection{Schur's Lemma}\label{Schur'sLemma}

\textit{Let $V$ and $W$ be two irreducible representation spaces of a group $G$. If $T:V\rightarrow W$ is a group homomorphism, then either $T$ is a group isomorphism or $T$ is the zero map.}\\

\noindent This Lemma has two important corollaries:

\begin{itemize}
\item Let $X$ and $Y$ be two irreducible representations of $G$. If $T$ is a matrix such that $TX(g)=Y(g)T$ for all $g\in G$, then either $T$ is invertible or $T$ is the zero matrix.
\item Let $X$ be an irreducible representation of $G$. Then, the only matrices $T$ that commute with $X(g)$ for all $g\in G$ are those of the form $cI$. This can be seen as follows: if $T$ commutes with $X(g)$, then $TX(g)=X(g)T$ for all $g\in G$. Thus, $(T-cI)X(g)=X(g)(T-cI)$ for any $c$. If $c$ is equal to $\lambda_T$, an eigenvalue of $T$, then $T-cI$ is not invertible, and then by the previous corollary $T-cI=0$. Therefore, $T$ must be of the form $cI$. In conclusion, for any irreducible representation of $G$, we have:
\begin{equation*}
TX(g)T^{-1}=X(g), \quad  \forall g \in G \rightarrow T =c I.
\end{equation*}
\end{itemize}
 This lemma and its corollaries give rise to one important property for irreducible representations, the \textit{great orthogonality theorem}. Consider two irreducible representations of the same group $X^{\lambda^1}, X^{\lambda^2}$ then, the great orthogonality theorem states for their matrix elements that \cite{Keppeler}:
\begin{equation}
\frac{1}{|G|}\sum_{g\in G} \overline{ X^{\lambda^1}(g)_{i,k}}X^{\lambda^2}(g)_{l,j}=\delta_{\lambda^1,\lambda^2}\frac{\delta_{i,j}\delta_{k,l}}{d}.
\label{eq:orthoirreds}
\end{equation}
Where $d=\dim(X^{\lambda^1})$. The great orthogonality theorem is the cornerstone of the \textit{character theory}, which allows us to identify when a given representation is reducible or not. Furthermore, when the representation is reducible, the character theory explicitly gives the decomposition into irreducible representations from now on \textit{irreps}.
\subsubsection{Characters}
For a representation $X^{\lambda}$ of a group $G$, the character of some element $g\in G$ is defined as the trace of the matrix representation:
\begin{equation*}
\chi^{\lambda} (g)=\tr\left(X^{\lambda} (g) \right).
\end{equation*} 
Some interesting properties of characters must be highlighted. Consider two isomorphic representations $X^{\lambda},X^{\tilde{\lambda}}$, then there exists a basis transformation $T$ such that:
\begin{equation*}
X^{\tilde{\lambda}}(g)= T X^{\lambda}(g) T^{-1} , \quad \forall g \in G,
\end{equation*}
then, the character of two isomorphic representations  are the same:
\begin{equation*}
\chi^{\tilde{\lambda}} (g) = \tr	\left( X^{\tilde{\lambda}}(g)\right) =\tr	\left(TX^{\lambda}(g)T^{-1}\right)=\tr	\left(T^{-1}TX^{\lambda}(g)\right) =\tr	\left( X^{\lambda}(g)\right)=\chi^{\lambda}(g).
\end{equation*}
Under the same procedure, it is easy to check that the character is a class function, i.e., its value is the same for all the elements in the same conjugacy class. Consider $g_1,g_2$ belonging to the same conjugacy class; then, there is some $h\in G$ such that $g_2=h g_1 h^{-1}$, then for any representation $X^{\lambda}$ we have:
\begin{equation*}
\chi^{\lambda}(g_2)= \tr( X^{\lambda}(g_2)) =\tr( X^{\lambda}(h g_1 h^{-1}))= \tr( X^{\lambda}(h)X^{\lambda}(g_1)X^{\lambda}(h^{-1}) ) =\tr( X^{\lambda}(g_1))=\chi^{\lambda}(g_1).
\end{equation*}
Class functions are very relevant in representation theory, and it is known that characters of irreps can be used to generate the set of class functions\cite{Zeilinger}. However, this discussion goes beyond the scope of this document.\\

\noindent The great orthogonality theorem has an important consequence in terms of characters. First, let us define the inner product between characters:
\begin{equation*}
\langle \chi^{\lambda^1}|\chi^{\lambda^2} \rangle =\frac{1}{|G|} \sum_{g\in G} \overline{\chi^{\lambda^1}(g)} \chi^{\lambda^2} (g).
\end{equation*}
Then, we have from equation \eqref{eq:orthoirreds} that if $X^{\lambda^1}$ and $X^{\lambda^2}$ are irreps of $G$ then:
\begin{equation}
\langle \chi^{\lambda^1}|\chi^{\lambda^2} \rangle =\frac{1}{|G|} \sum_{g\in G} \sum_{j,k} \overline{X^{\lambda^1}(g)_{jj}} X^{\lambda^2} (g)_{kk}=\frac{\delta_{\lambda^1 \lambda^2}}{d}\sum_{j,k} \delta_{jk}\delta_{jk} = \delta_{\lambda^1 \lambda^2},
\label{eq:CharOrtho}
\end{equation}
which is known as the \textit{orthogonality relation} of characters. We can show how characters can identify whether some representation is reducible. Consider a reducible representation $X^{\mu}$ of a group $G$, then there exists a basis transformation $T$ such that:
\begin{equation*}
TX^{\mu}(g) T^{-1} = \bigoplus_{i} X^{\lambda^{i}}(g) \otimes I_{m_{i,\mu}}, \quad \forall g \in G,
\end{equation*}
where $m_{i,\mu}$ is the multiplicity of $\lambda^{i}$ in $\mu$, and the sum runs over all the irreps of $G$. Then, the character of $X^{\mu}$ is:
\begin{equation*}
\begin{gathered}
\chi^{\mu}(g)=\tr(X^{\mu}(g))= \tr(TX^{\mu}(g) T^{-1} ) = \tr(\bigoplus_{i} X^{\lambda^{i}}(g) \otimes I_{m_{i,\mu}}) \\ =\sum_{i} m_{i,\mu} \tr(X^{\lambda^{i}}(g))=\sum_{i} m_{i,\mu} \chi^{\lambda^{i}}(g).
\end{gathered},
\end{equation*}
when taking the inner product, we have:
\begin{equation*}
\braket{\chi^\mu}{\chi^\mu}=\frac{1}{|G|}\sum_{g\in G} \overline{\chi^{\mu}(g)} \chi^{\mu}(g) =\frac{1}{|G|}\sum_{g\in G} \sum_{ij} m_{i,\mu}m_{j,\mu} \overline{\chi^{\lambda^i}(g)} \chi^{\lambda^{j}}(g),
\end{equation*}
by summing $g$ and using equation \eqref{eq:CharOrtho} we get:
\begin{equation*}
\braket{\chi^\mu}{\chi^\mu}=\sum_{i} m_{i,\mu}^2.
\end{equation*}
If $\mu$ is irreducible, then its decomposition into irreps can only have one term with $m_{i,\mu}=1, \mu\cong\lambda^i$; then
\begin{equation*}
\braket{\chi^\mu}{\chi^\mu}=1.
\end{equation*}
only for irreps. When this inner product is larger than one, $\mu$ is composed of more than one irrep. Characters can also be used to identify the multiplicity of a given irrep in the diagonal decomposition \cite{Zeilinger}:
\begin{equation}
\braket{\chi^\mu}{\chi^{\lambda^i}}=m_i.
\label{eq:multiplicity}
\end{equation}
With this tool, it is possible to break down any representation in its irreps, even without knowing the change of basis that block-diagonalizes the matrix. One interesting case for us is the decomposition into irreducible representations of the tensor product representation obtained by taking the tensor product of two representations. 

\subsubsection{Tensor Product representation}
Consider we take two irreducible representations of a group $G$, $\lambda^1,\lambda^2$, and for each element $g$ we build the matrix
\begin{equation*}
X^{\lambda^1\otimes \lambda^2} (g)= X^{\lambda^1}(g) \otimes X^{\lambda^2} (g),
\end{equation*}
then, the map $X^{\lambda^1\otimes \lambda^2} (g)$ is also a representation. This is because, firstly, the identity element is mapped to the identity matrix:
\begin{equation*}
X^{\lambda^1\otimes \lambda^2} (e)= X^{\lambda^1}(e) \otimes X^{\lambda^2} (e) = I_{d_1} \otimes I_{d_2}= I_{d_1\times d_2}, 
\end{equation*}
being $d_i$ the dimension of the irrep $\lambda_i$. Secondly, this map respects the multiplication rule of the group:
\begin{equation*}
\begin{gathered}
X^{\lambda^1\otimes \lambda^2} (g_1\cdot g_2)= X^{\lambda^1}(g_1 \cdot g_2) \otimes X^{\lambda^2} (g_1 \cdot g_2) =  X^{\lambda^1}(g_1 ) X^{\lambda^1}(g_2 )  \otimes X^{\lambda^2} (g_1)  X^{\lambda^2}(g_2 ) \\
=\left(X^{\lambda^1}(g_1 ) \otimes X^{\lambda^2} (g_1)  \right) \left( X^{\lambda^1}(g_2 )  X^{\lambda^2}(g_2 ) \right) =X^{\lambda^1\otimes \lambda^2} (g_1) X^{\lambda^1\otimes \lambda^2} (g_2).
\end{gathered}
\end{equation*}
This representation is known as the \textit{Tensor product representation} and will be the most relevant construction throughout this document. This kind of construction appears naturally in quantum mechanics when considering multipartite systems. We usually define a representation for each of the parts of the system, but when thinking of the system as a whole, it will be represented by the tensor product of the individual representations.\\

\noindent In general, the tensor product representation is reducible, i.e., it can be decomposed diagonally in irreps. This diagonalization can be seen as:
\begin{equation*}
\lambda^1\otimes \lambda ^2 = \bigoplus_i  \lambda^i \otimes  I_{m_{i,\lambda^1\otimes \lambda^2}},
\end{equation*}
where $m_{i,\lambda^1\otimes \lambda^2}$ is the multiplicity of irrep $\lambda^i$ in the decomposition of $\lambda^1 \otimes \lambda^2$. This multiplicity can be calculated from equation \eqref{eq:multiplicity} as:
\begin{equation*}
m_{i,\lambda^1\otimes \lambda^2}=\braket{\chi^{\lambda^1\otimes \lambda^2}}{\chi^{\lambda^{i}}}= \frac{1}{|G|} \sum_{g} \overline{\chi^{\lambda^1}(g) \chi^{\lambda^2} (g)} \chi^{\lambda^{i}}(g).
\end{equation*}
For the symmetric group $S_n$, this multiplicity receives the name of \textit{Kronecker coefficient}, and we label it as $k_{\lambda^1\lambda^2\lambda}$. For this group, we will write the diagonal decomposition as:
\begin{equation}
[\lambda^1] \otimes [\lambda^2] = \bigoplus_{\lambda\vdash n} [\lambda] \otimes I_{k_{\lambda^1\lambda^2\lambda }}, 
\label{eq:DiagSn}
\end{equation}
where irreps are labeled by the partitions of $n$ as $[\lambda]$,  as we will explain in the next section. In this group the multiplicity, i.e., the Kronecker coefficient, can be calculated as:
\begin{equation}
k_{\lambda^1\lambda^2\lambda}= \frac{1}{n!} \sum_{\pi \in S_n} \chi^{[\lambda^1]} (\pi)\chi^{[\lambda^2]} (\pi)\chi^{[\lambda]} (\pi).
\label{eq:KronCoef}
\end{equation} 
This value will appear again later when characterizing the subspace in the tensor product of irreps of $S_n$, over which $S_n$ acts trivially. i.e., the \textit{invariant subspace}. Besides the multiplicity of irreps in the tensor product decomposition, we will also be interested in the change of basis that allows block-diagonalization. We know that there exists some unitary matrix $U$ such that for all $g\in G$:
\begin{equation*}
U X^{\lambda^1\otimes \lambda^2} (g) U^{-1} = \bigoplus_{i} \bigoplus_{s_i=1}^{ m_{i,\lambda^1\otimes \lambda^2} } X^{\lambda^i} (g),
\end{equation*}
where $s_i$ label the different subspaces corresponding to the same $\lambda^i$ in the decomposition. In terms of the basis of each representation this can be written as \cite{Keppeler}:
\begin{equation}
\sum_{j=1}^{d_{\lambda^1}}\sum_{k=1}^{d_{\lambda^2}} C^{\lambda^1\lambda^2\lambda^{i},s}_{l_j^1,l_k^2,l^i} \ket{\lambda^1,l^1_{j}} \ket{\lambda^2,l^2_{k}}=\ket{\lambda^{i},l^i,s},
\label{eq:CGC}
\end{equation}
where $l^i_j$ label the basis elements of irrep $\lambda^i$, and $C^{\lambda^1\lambda^2\lambda^{i},s}_{l_j^1,l_k^2,l^i} $ are known as the \textit{Clebsch Gordan coefficients} (CGC) of the group $G$. CGC are the matrix elements of $U$, hence, the unitarity conditions of $U$:$U^{\dagger }U =I= U U^{\dagger}$, can be understood as orthogonality relations of CGC:
\begin{equation*}
\begin{gathered}
\sum_{s,\lambda^{i},l^{i}} C^{\lambda^1\lambda^2\lambda^{i},s}_{l_j^1,l_k^2,l^i} \overline{C^{\lambda^1\lambda^2\lambda^{i},s}_{l_{j'}^1,l_{k'}^2,l^i}}= \delta_{jj'}\delta_{kk'}, \\
\sum_{jk} C^{\lambda^1\lambda^2\lambda^{i},s}_{l_j^1,l_k^2,l^i} \overline{C^{\lambda^1\lambda^2\lambda^{i'},s'}_{l_{j}^1,l_{k}^2,{l'}^{i'}}}= \delta_{ii'}\delta_{ss'}\delta_{ll'}.
\end{gathered}
\end{equation*}
CGC are important mathematical objects that are generally hard to compute. Some closed results are known for $SU_2$, $SU_3$, $SL_3$, but despite some efforts \cite{sahasrabudhe} \cite{Mohammed}\Cite{Doma}, there are, to the best of our knowledge, no closed expressions for $S_n$, and the known approaches are strictly mathematical. One of the main results of this research is a physics-based method to calculate CGC for the symmetric group $S_n$ using algebraic expressions and a recursive construction.\\

\noindent Now, we will focus on defining the irreps of $S_n,$ and the irreps appearing in the tensor product of $n$ copies of the standard representation of $GL_d$, which are known as the \textit{polynomial representations} of $GL_d$\cite{Keppeler}.

\section{Irreducible representations of $GL_d$ and $S_n$}
It is a well-known result of representation theory that the different irreducible representations of $S_n$ and the polynomial representations of $GL_d$ can be labeled by partitions $\lambda$. Such partitions can be represented graphically by the \textit{Young diagrams}\cite{Sagan}, allowing us to interpret graphically the dimensions and other properties of the corresponding irreps. 

\subsection{Young diagrams}
Young diagrams are in one-to-one correspondence with partitions. The Young diagram for a partition of $n$ given by $\lambda=(\lambda_1,\lambda_2,\dots,\lambda_k)$ corresponds to a left-aligned diagram with $n$ boxes and $k$ rows, where in the $i$-th row, there are $\lambda_i$ boxes. For example, for all the partitions of $n=4$, the different Young diagrams are:
\begin{equation*}
\begin{gathered}
\lambda^1=(4) = \ydiagram{4}, \, \lambda^2=(3,1) = \ydiagram{3,1},\, \lambda^3=(2,2) = \ydiagram{2,2}, \,\\  \lambda^4=(2,1,1)= \ydiagram{2,1,1}, \, \lambda^5=(1,1,1,1) = \ydiagram{1,1,1,1}
\end{gathered}.
\end{equation*}
Each of these diagrams will label the different irreducible representations of $S_n$ and $GL_d$; the connection between irreps and partitions has been explained multiple times in the literature \cite{Balachandran} \cite{Fulton} \cite{Sagan}. However, this explanation is out of the scope of this work, and we will focus on explaining how the Young diagrams permit labeling the basis elements in each irrep of $S_n$ and $GL_d$ groups. Later, in Section \ref{Schurtransform}, these labels will acquire a physical significance that is what is relevant for us.

\subsection{Irreducible representations of $GL_d$}
\label{GLd}
Let us start by recalling that $GL_d$ is the group of invertible matrices of dimensions $d\times d$, with entries over the complex numbers. This group has a representation in itself, which is an irreducible representation known as the \textit{standard representation}. A matrix $A\in GL_d$ is represented by itself in the standard representation of dimension $d$. We will be interested in the irreducible representations that appear in the diagonal decomposition of the tensor product of $n$ copies of the standard representation, known as the \textit{polynomial representations} of $GL_d$\cite{Keppeler}. \\

\noindent One useful way of approach the polynomial representations of $GL_d$ is thinking on the standard representation of $GL_d$ as belonging to a vector space defined by $\ket{i}$, where $i$ can take values from $0$ to $d-1$. We will represent the standard representation as a Young diagram of $n=1$, i.e., a single box that represents the index in the standard representation:
\begin{equation*}
 \ydiagram{1} \Rightarrow \begin{ytableau} i \end{ytableau}.
\end{equation*}
We are interested in the irreducible representations that appear when decomposing tensor products of some irrep with one copy of the standard representation. In terms of Young diagrams, it corresponds to considering all the Young diagrams obtained by adding one box in the different rows to the initial irrep. For the case of two copies, the product basis will be represented by a two indices vector space $\ket{i_1i_2}$. In terms of Young diagrams, we can add the second box in two different ways:
\begin{equation*}
\begin{ytableau} i_1 \end{ytableau} \otimes \begin{ytableau} i_2 \end{ytableau}= \begin{ytableau} i_1 &  i_2 \end{ytableau} \oplus \begin{ytableau} i_1 \\ i_2 \end{ytableau} .
\end{equation*} 
The first case corresponds to adding one box to the first row. This irrep is labeled as $\{2\}$ where we introduced the notation $\{\lambda\}$ for the irreps of $GL_d$. The second case corresponds to adding one box to the second row, and is labeled as $\{1,1\}$. It is not possible to add boxes in the third row, then these are the only two possibilities. Each resultant Young diagram is an irreducible representation that specifies a rule for symmetrizing the indices in the product basis of the copies of the standard vector space. For this, indices in the rows are symmetrized, and indices in the columns are anti-symmetrized. For the example with two copies, the two Young diagrams permit to obtain two in general unnormalized with the symmetries of the vector of two indices $\ket{i_1i_2}$:
\begin{equation*}
\{2 \} :
\qquad \begin{ytableau}
i_1 & i_2
\end{ytableau}  \Rightarrow \ket{i_1i_2}+\ket{i_2i_1}, \qquad \{1,1\} :\begin{ytableau}
i_1\\
i_2
\end{ytableau} \Rightarrow  \ket{i_1i_2} -\ket{i_2i_1}. 
\end{equation*} 
We will focus on the irreps of $GL_2$, so the indices can only take values $0$ and $1$, then Young diagrams encode the different symmetries in the computational basis. For the symmetric irrep we can define three independent unnormalized vectors in the product basis with the symmetry given by the Young diagram:
\begin{equation}
\{2\}:
i_1i_2 = 00 \Rightarrow 2\ket{00}, \quad i_1i_2 = 01  \Rightarrow \ket{01}+\ket{10}, \quad   i_1i_2 = 11 \Rightarrow 2\ket{11}.
\end{equation}
Note that choosing the values $i_1i_2=10$ lead to the same vector as for  $i_1i_2=01$, so, the obtained states are not independent. The three states in the previous equation, when normalized, define a basis for an invariant subspace in the tensor product representation that correspond to an irreducible representation. This can be seen as follows: first, let us name the normalized vectors according to the respective irrep and to the number of ones of the values for the ordered indices $i_1i_2$, also known as the \textit{weight} $\omega$,  as $\ket{\lambda,\omega}$. For example, $i_1i_2=01 \Rightarrow \ket{\{2\},1}$. Then, we have:
\begin{equation}
\ket{00}= \ket{2,0}, \quad \frac{1}{\sqrt{2}}\left(\ket{01}+ \ket{10} \right)= \ket{2,1}, \quad \ket{11}= \ket{2,2}
\label{eq:basis2}
\end{equation}
Now, define an element $A$ in $GL_2$:
\begin{equation*}
A= \left( \begin{array}{cc}
a_{00}  & a_{01} \\
a_{10} & a_{11} \
\end{array} \right).
\end{equation*}
Next, let act the tensor product of two copies of $A$ on each of the vectors defined in Equation \eqref{eq:basis2} after normalize them. For the first one we have:
\begin{equation*}
\begin{gathered}
A\otimes A \ket{2,0} = A\ket{0} \otimes  A\ket{0} =(a_{00}\ket{0}+a_{10} \ket{1})\otimes(a_{00}\ket{0}+a_{10} \ket{1})\\
= a_{00}^2 \ket{00}+ a_{00}a_{10} \ket{01} + a_{00}a_{10} \ket{10}+  a_{10}^2 \ket{11} = a_{00}^2 \ket{2,0}+ \sqrt{2} a_{00}a_{10} \ket{2,1} +  a_{10}^2 \ket{2,2}.
\end{gathered}
\end{equation*}
This shows that the resultant state belong to the subspace defined by $\ket{2,0},\ket{2,1}$ and $\ket{2,2}$. We can do the same with the other two vectors in Equation \eqref{eq:basis2}:
\begin{equation*}
\begin{gathered}
A\otimes A \ket{2,1}= \sqrt{2}a_{00}a_{01}\ket{2,0} + (a_{00}a_{11}+ a_{01}a_{10}) \ket{2,1}+\sqrt{2}a_{10}a_{11}\ket{2,2},\\
A\otimes A \ket{2,2}= a_{01}^2 \ket{2,0}+ \sqrt{2} a_{01}a_{11} \ket{2,1} +  a_{11}^2 \ket{2,2}.
\end{gathered}
\end{equation*}
Therefore, the vectors in Equation \eqref{eq:basis2}  define an invariant subspace, which can be shown it corresponds to an irreducible representation. We can make the same analysis with the irrep corresponding to $\{1,1\}$. Note first that with the values $0$ and $1$ for the indices, only one independent normalized state can be obtained:
\begin{equation*}
\{1,1\}:  \frac{1}{\sqrt{2}} \left(\ket{01} -\ket{10} \right)  = \ket{(1,1),1}.
\end{equation*}
Note that when acting with $A\otimes A$ we have:
\begin{equation*}
A\otimes A \ket{(1,1),1} =\left(a_{00}a_{11}-a_{01}a_{10} \right)\ket{(1,1),1} = \det (A) \ket{(1,1),1}.
\end{equation*}
Thus, the one-dimensional subspace defined by $\ket{(1,1),1}$ is also an invariant subspace, corresponding to another irreducible representation. Note that with the two irreps represented by $\{2\}$ and $\{1,1\}$ we complete the four-dimensional space, defined by $GL_2\otimes GL_2$. The dimensions of the irreps can be obtained easily from the corresponding Young diagrams. For this, we fill the diagram with numbers from $0$ to $d-1$ such that they are not decreasing horizontally and are strictly increasing vertically, in this way we ensure to only consider independent basis elements. Filling young diagrams with these restrictions corresponds to \textit{semi-standard Young tableaux} (SSYT). For the two irreps in the example we have:
\begin{equation*}
 \{2\}:  \begin{ytableau}
0 &0
\end{ytableau} ,\quad \begin{ytableau}
0 &1
\end{ytableau},\quad \begin{ytableau}
1 &1
\end{ytableau} ,\quad \{1,1\} : \quad \begin{ytableau}
0\\
1
\end{ytableau}.
\end{equation*} 
In this sense, the number of SSYT in a Young diagram representing the partition $\lambda$ is equal to the dimension of the irrep $\{\lambda\}$. Let us now consider the case when decomposing three copies of the standard representation. We can do this by first adding one box to the standard representation, and then adding the other one. This process can be seen graphically as:
{\small 
\begin{equation*}
\begin{ytableau} i_1 \end{ytableau} \otimes \begin{ytableau} i_2 \end{ytableau} \otimes\begin{ytableau} i_3 \end{ytableau}  =  \left( \begin{ytableau} i_1 & i_2 \end{ytableau} \oplus \begin{ytableau} i_1 \\ i_2\end{ytableau} \right) \otimes \begin{ytableau} i_3 \end{ytableau}  = \begin{ytableau} i_1 & i_2 & i_3  \end{ytableau} \oplus \begin{ytableau} i_1 &i_2 \\ i_3 \end{ytableau} \oplus \begin{ytableau} i_1 & i_3 \\ i_2 \end{ytableau} \oplus \begin{ytableau} i_1 \\ i_2 \\ i_3 \end{ytableau}.
\end{equation*}}
\noindent First, note that for $GL_2$, the last irrep $\{1,1,1\}$ is not admissible, because there is no way to put numbers from $0$ to $1$ strictly increasing along three boxes. Then, the possible irreps for $n$ copies of $GL_2$ will be those with at most two rows. We will label those partitions as $\lambda \vdash n,2$, i.e., partitions of $n$ with at most two rows. From the previous decomposition into irreps, it must be noticed that the same irrep appears twice, $\{\lambda\}=\{2,1\}$, which means that two copies of the same irrep appear in the decomposition of the tensor product into irreps. Note that one irrep was obtained by symmetrizing first $i_1$ with $i_2$ and then anti-symmetrizing $i_1$ with $i_3$, while the other is obtained by firs anti-symmetrizing $i_1$ with $i_2$ and then symmetrizing $i_1$ with $i_3$. However, as they are represented by the Young diagram, they are isomorphic representations. We will see later, that these multiplicities are determined by the irreducible representations of $S_n$. In this case the corresponding symmetries for the different irreps are represented by the unnormalized vectors in the computational basis $\ket{i_1i_2i_3}$ as:
\begin{equation}
\begin{gathered}
\{3\}:\begin{ytableau}
i_1 & i_2 & i_3 
\end{ytableau}  \Rightarrow \ket{i_1i_2i_3}+\ket{i_2i_1i_3} +\ket{i_3i_2i_1} +\ket{i_1i_3i_2} +\ket{i_2i_3i_1} +\ket{i_3i_1i_2}  , \\
\{2,1\}_1:\begin{ytableau}
i_1 & i_2 \\
 i_3 
\end{ytableau}  \Rightarrow \ket{i_1i_2i_3} + \ket{i_2 i_1 i_3} - \ket{i_3 i_2 i_1}- \ket{i_2 i_3 i_1}, \\
\{2,1\}_2:\begin{ytableau}
i_1 & i_3 \\
 i_2 
\end{ytableau}  \Rightarrow \ket{i_1i_2i_3}+ \ket{i_3 i_2 i_1} - \ket{i_2 i_1 i_3} - \ket{i_2 i_3 i_1}. 
\end{gathered}
\label{eq:symmYoung}
\end{equation}
Note that the two multiplicities of $\{2,1\}$ labeled by subscripts are isomorphic up to a relabeling of the indices, i.e., they can be obtained from each other acting with an element of $S_n$ on the indices. The SSYT for both irreps are:
\begin{equation*}
\begin{gathered}
\{3\}: \begin{ytableau}
0 & 0 & 0 
\end{ytableau} , \,  \begin{ytableau}
0 & 0 & 1 
\end{ytableau} , \,  \begin{ytableau}
0 & 1 & 1 
\end{ytableau} ,\, \begin{ytableau}
1 & 1 & 1 
\end{ytableau} \\
\{21\}: \begin{ytableau}
0 & 0 \\
 1 
\end{ytableau} , \,  \begin{ytableau}
0 & 1 \\
1 
\end{ytableau} \\
\end{gathered}
\end{equation*}
Then, the dimension of irrep $\{3\}$ is four, while for the irrep $\{21\}$ is two, which appears twice in the decomposition of three copies of the standard representation. The basis elements of the irreps can be obtained easily and expressed in the $\ket{\lambda,\omega}$ notation for free multiplicity cases and $\ket{\lambda,\omega,s}$ for the cases with multiplicity with $s$ the multiplicity index :
{\small
\begin{equation*}
\begin{gathered}
\{3\}:\quad  \ket{000}= \ket{3,0}, \quad  \frac{1}{\sqrt{3}}\left(\ket{001}+\ket{010}+\ket{100}\right) = \ket{3,1}, \\
\frac{1}{\sqrt{3}}\left(\ket{011}+\ket{101}+\ket{110}\right) = \ket{3,2},  \quad \ket{111} = \ket{3,3}, \\
\{21\}_1: \quad  \frac{1}{\sqrt{6}}\left(2\ket{001}-\ket{100}-\ket{010}\right)= \ket{(21),1,1}, \quad \frac{1}{\sqrt{6}}\left(\ket{011}+\ket{101}-2\ket{110}\right) =  \ket{(21),2,1}, \\
\{21\}_2: \quad  \frac{1}{\sqrt{2}}\left(\ket{010}-\ket{100}\right)= \ket{(21),1,2}, \quad \frac{1}{\sqrt{2}}\left(\ket{011}-\ket{101}\right) =  \ket{(21),2,1}.
\end{gathered}
\end{equation*} }
It is not hard to check again that all the irreps are invariant subspaces; but more interestingly, we can note that the action of the tensor product on any multiplicity of irrep $\{2,1\}$ transforms as the single copy of $A$ multiplied by the determinant of $A$:
\begin{equation*}
A\otimes A \otimes A \cdot (c_1 \ket{(21),1,1} +c_2 \ket{(21),2,1} ) \cong \det(A)\cdot  \tilde{A} (c_1\ket{\tilde{0}}+c_2\ket{\tilde{1}}) ,
\end{equation*}
with $c_1$ and $c_2$ complex coefficients, and $\tilde{A}$ is the same matrix $A$ but re-interpreted as acting on the two-dimensional Hilbert space defined by $\ket{\tilde{0}}=\ket{(21),1,1}$ and $\ket{\tilde{1}}=\ket{(21),2,1}$. This is a general property of irreps with number of rows equal to the dimension $d$. Each full column correspond to one determinant, $\det(A)$, multiplied by the irrep corresponding to erasing the full columns in the Young diagram. \\

\noindent In conclusion, Young diagrams permit to separate the irreducible representations of the tensor product on copies of the standard representation. Each partition $\lambda$ label one symmetrizing rule that corresponds to an irreducible representation of $GL_d$, and the possible SSYT in the Young diagram label the basis elements of the irrep. However, for this group, it is common to relate the dimension with the character of the identity element. \\

 \noindent In general, The characters of $GL_d$ irreps denoted by $\{\lambda\}$ can be obtained as Schur polynomials on the set of eigenvalues of the matrix $A\in GL_d$ to be represented. We label such a set of eigenvalues as $a$. Schur polynomials are symmetric functions that can be computed as \cite{Schupoly}:
 \begin{equation}
 s_{\lambda}(a)= \frac{ \left|\begin{array}{cccc}
 a_1^{\lambda_1+d-1} & a_2^{\lambda_1+d-1} &\cdots & a_d^{\lambda_1+d-1} \\
 a_1^{\lambda_2+d-2} & a_2^{\lambda_2+d-2} &\cdots & a_d^{\lambda_2+d-2} \\
 \vdots & \vdots & \ddots & \vdots \\
 a_1^{\lambda_d} & a_2^{\lambda_d} &\cdots & a_d^{\lambda_d} \\
 \end{array} \right|}{\prod_{1\leq j \leq k \leq d} (a_j -a_k)}.
 \label{eq:Schurpoly}
 \end{equation}
 Then, the dimension of an irrep labeled by $\{\lambda\}$ can be calculated as:
 \begin{equation*}
 \dim(\{\lambda\})=s_{\lambda} (\bm{1}),
 \end{equation*}
 where $\bm{1}=(1,1,\dots,1)$, and the evaluation is done after expressing Equation \eqref{eq:Schurpoly} as a polynomial. It is possible to find bounds for Schur polynomials as:
 \begin{equation}
  a_1^{\lambda_1} a_2^{\lambda_2} \cdots a_d^{\lambda_d} \leq s_{\lambda}(a) \leq s_{\lambda}(\bm{1}) a_1^{\lambda_1} a_2^{\lambda_2} \cdots a_d^{\lambda_d}
  \label{eq:Schurasympt}
 \end{equation}
 which will be useful when analyzing asymptotic behaviors.\\
 
\noindent Now, consider any SSYT of two rows; then, the entries in the first $\lambda_2$ columns are fixed, and we are left with $n-2\lambda_2$ boxes that, when imposing the restrictions of SSYT, can be filled in $n-2\lambda_2+1$ ways. For example, consider $\lambda=(5,2)$, then the possible SSYT are:
\begin{equation}
\begin{ytableau}
0 & 0 & 0 & 0 & 0\\
1 & 1
\end{ytableau} , \, \begin{ytableau}
0 & 0 & 0 & 0 & 1\\
1 & 1
\end{ytableau}, \, \begin{ytableau}
0 & 0 & 0 & 1 & 1\\
1 & 1
\end{ytableau}, \, \begin{ytableau}
0 & 0 & 1 & 1 & 1\\
1 & 1
\end{ytableau},
\label{eq:SSYT52}
\end{equation} 
which corresponds to:
\begin{equation*}
 \dim(\{5,2\})= n-2\lambda_2+1= 7-4+1=4.
 \end{equation*}
This fact can also be seen from the Schur polynomial, which for partitions of two parts is reduced to \cite{Schupoly}:
\begin{equation*}
s_{(\lambda_1,\lambda_2)}(a_1,a_2)= \sum_{i=\lambda_2}^{n-\lambda_2}  a_1^i a_2 ^{n-i},
\end{equation*}
then, by replacing in the sum and evaluating $a_1=a_2=1$ we have:
\begin{equation*}
\dim( \{\lambda_1,\lambda_2\})=s_{(\lambda_1,\lambda_2)}(1,1)= n-2\lambda_2 +1.
\end{equation*}
It is important to note how, given a partition $\lambda\vdash n,2$, each SSYT is unambiguously defined by the number of $1$s in the SSYT, $\omega$. Then, the basis elements defined by the SSYT from Equation \eqref{eq:SSYT52} could also be labeled as:
\begin{equation*}
\ket{\{\lambda\},\omega}\Rightarrow \ket{\{5,2\},2}, \, \ket{\{5,2\},3}, \,\ket{\{5,2\},4}, \,\ket{\{5,2\},5}.
\end{equation*}
When having diagrams labeled by $\{\lambda\}$ with the number of rows being equal to the dimension $d$, each of the columns with $d$ rows corresponds to an anti-symmetrizing of all the different indices of the tensor representation of the matrices $A$, which is by definition the determinant of the matrix. Then, for the case of $d=2$, each column with two rows can be replaced by a factor of $\det(A)$. For example, for the matrix representation $V^{\{\lambda\}}(A)$ in the irrep $\{\lambda\}$, this is:
\begin{equation*}
V^{\{\lambda_1,\lambda_2\}} (A)= \det(A)^{\lambda_2} V^{\{n-2\lambda_2\}} (A), \qquad \text{e.g.,} \quad V^{\{5,2\}}(A)=( \det(A))^{2} V^{\{3\}}.
\end{equation*}
When the matrices to be represented are with unit-determinant, or when we are just not interested in the value of the determinant, we can focus only on the non-full columns to analyze the representation. Having defined the notation for the basis elements of $GL_2$, we still have to build the irreducible representations obtained from the tensor product representations of the standard representation. However, this process will be introduced later in Section \ref{Schurtransform}, as it is a fundamental tool in this research. Now, we will present the irreducible representations of the other group relevant to this document, the Symmetric group $S_n$.  

\subsection{Irreducible representations of $S_n$}
\label{IrrepsSn}
In a similar way as Young diagrams permit to identify irreducible representations of $GL_d$ from the tensor copies of the standard representation, they also permit to identify irreducible representations of $S_n$ from the regular representation that we introduced in Equations \eqref{eq:regbas} and \eqref{eq:regrep}. However, we will not present this approach. Instead, we will focus on how the irreducible representations of $S_n$ appear in the decomposition of the tensor product of copies of the standard representation of $GL_d$. For this, let us define the \textit{standard Young tableaux} (SYT) associated to a Young diagram of $n$ boxes as the possible ways of filling all the boxes with a unique number from $1$ to $n$, such that the numbers $1,2,\dots,n$ appear strictly increasing along rows and columns. For example, for the partition $\lambda=(422)$:
\begin{equation*}
\begin{ytableau}
3 & 1 &7 &4 \\
6 & 2 \\
8 & 5
\end{ytableau} \notin SYT  ,\quad  \begin{ytableau}
1& 3 &4 &7 \\
2 & 6 \\
5 &8
\end{ytableau} \in SYT.
\end{equation*}
It turns out that the irreducible representations of $S_n$ are labeled by the different partitions of $n$ similar to the irreps of $GL_d$; however, for $S_n$ the basis elements are labeled by the SYT, as for $GL_d$ they are labeled by the SSYT.  Then, each possible SYT in a Young diagram $\lambda$, label one of the basis elements of the irreducible representation that will be denoted by $[\lambda]$.\\
 
\noindent Note from the example in Equation \eqref{eq:symmYoung} that, two multiplicities of the irrep $\{2,1\}$ of $GL_d$ appeared in the decomposition on irreps of the tensor product of three copies of the standard representation. The two multiplicities correspond to the two different ways of achieving the Young diagram $\lambda=(2,1)$, by adding one box at the time:

\begin{equation}
\begin{ytableau}
i_1 
\end{ytableau} \Rightarrow \begin{ytableau}
i_1 &i_2
\end{ytableau} \Rightarrow \begin{ytableau}
i_1 & i_2 \\
i_3
\end{ytableau} , \qquad \begin{ytableau}
i_1 
\end{ytableau} \Rightarrow \begin{ytableau}
i_1 \\
i_2
\end{ytableau} \Rightarrow \begin{ytableau}
i_1 & i_3 \\
i_2
\end{ytableau}.
\end{equation}
The sub-indices of the indices in the computational basis follow the rules of SYT. In this sense, the multiplicites of $GL_d$ are labeled by the basis elements of irreps of $S_n$. In this case, for $S_n$ the dimension of the irrep $[2,1]$ is two, because there are two different ways of building the Young diagram.
Each path represents one different process of symmetrizing anti-symmetrizing the indices of the product basis of $GL_d$ that arrive at the same partition $\lambda$. \\

\begin{figure}
\includegraphics[scale=1.5]{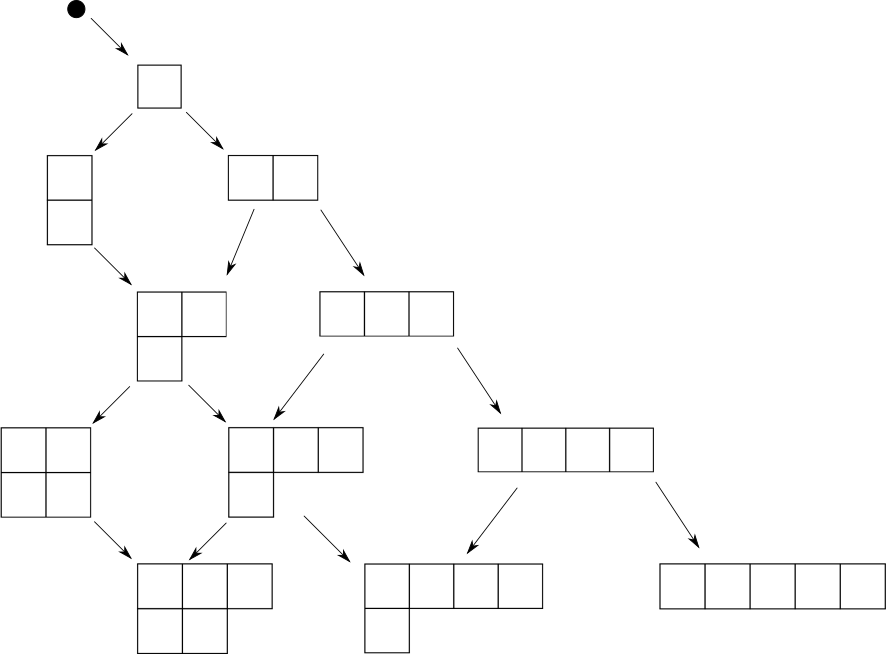}
\caption{Simplified Young Lattice up to $n=5$, where only partitions with at most two rows are considered. From the first box to any $\lambda$, there are exactly $f^{[\lambda]}$ paths, each corresponding to one SYT and hence, to one Young-Yamanouchi symbol.}
\label{fig:Younglattice}
\end{figure}

\noindent The dimension of an irrep $[\lambda]$ can be obtained by counting how many SYT can be obtained in the young diagram $\lambda$. For example, given the partition $[3,2]$, there are five possible SYT:
\begin{equation}
\begin{gathered}
q^{[3,2]}_1=
\begin{ytableau}
1 & 2 &3  \\
4 & 5 \\
\end{ytableau} , \, q^{[3,2]}_2=
\begin{ytableau}
1 & 2 &4  \\
3 & 5 \\
\end{ytableau} , \, q^{[3,2]}_3=
\begin{ytableau}
1 & 2 &5  \\
3 & 4 \\ 
\end{ytableau}, \\q^{[3,2]}_4=
\begin{ytableau}
1 & 3 &4  \\
2 & 5 \\
\end{ytableau} , \, q^{[3,2]}_5=
\begin{ytableau}
1 & 3 &5  \\
2 & 4 \\
\end{ytableau},
\end{gathered}
\label{eq:basis32}
\end{equation}
where we used $q_i^{[\lambda]}$ to label each SYT and so, each basis element. Then, we conclude that $[3,2]$ is an irrep of dimension five, i.e.,
\begin{equation*}
f^{[3,2]}=5,
\end{equation*}
where $f^{[\lambda]}$ is the dimension of irrep $[\lambda]$. \noindent Despite the utility of SYT to label the basis elements of irreps in $S_n$, there are different notations that may be useful. One recurrent notation in this document is to use a modification of the \textit{Yamanouchi symbols} \cite{Wilson}, where each $SYT$ will be represented by a list of length $n$, first denote $r^{q}(i)$ to be the row where number $i$ appears in the SYT labeled as $q$, then, the $i$-th number in the list will correspond to $r^{q}(i)-1$. For example, for the basis elements of $[3,2]$ defined in equation \eqref{eq:basis32} we have:
\begin{equation*}
\begin{gathered}
q_1^{[3,2]}=\{0,0,0,1,1\}, \, q_2^{[3,2]}=\{0,0,1,0,1\}, \, q_3^{[3,2]}=\{0,0,1,1,0\}, \\
\,q_4^{[3,2]}=\{0,1,0,0,1\} , \, q_5^{[3,2]}=\{0,1,0,1,0\}.
\end{gathered}
\end{equation*}
Another relevant point of view for the basis elements of irreps of $S_n$ is that they can also be understood as a possible path to obtain the Young diagram $\lambda$ by adding one box in each step, and the Yamanouchi symbol tells in which row the box should be added. This point of view permits us to see that the base defined by SYT is adapted to the chain group of $S_n$: $S_n\supset S_{n-1} \supset \dots \supset S_2 \supset S_1$, when taking the first $n'$ elements in any Yamanouchi symbol representing a basis element of $[\lambda]$ in $S_n$, it will correspond to a valid Yamanouchi symbol of some irrep $[\lambda']$ in $S_{n'}$. For example, the $q_4^{[3,2]}$ base can be understood as the following path of partitions to achieve $[3,2]$:
\begin{equation*}
q_4^{[3,2]}=\{0,1,0,0,1\} : \ydiagram{1} \Rightarrow \ydiagram{1,1} \Rightarrow \ydiagram{2,1} \Rightarrow \ydiagram{3,1} \Rightarrow \ydiagram{3,2}.
\end{equation*}
Thinking on the basis elements of irreps of $S_n$ as paths in the possible set of partitions is a useful point of view when relating its action to tensor products of $SL_2$ or, equivalently, when analyzing the problem of addition of spin $1/2$  particles,  as we will see later. In particular, when considering only partitions of at most two rows (on which we will focus later), it is useful to think about all possible paths in a simplified \textit{Young's lattice} \cite{Sagan}, where a structure of partitions can be obtained by starting at some partition of $n$ and connecting it with the partitions of $n+1$ that are obtained when adding one box in the first row or one box in the second row to the corresponding initial Young diagram. The simplified Young's lattice is shown in Figure \ref{fig:Younglattice} up to $n=5$.\\

\noindent With this graphical tool, each basis element of $[\lambda]$ corresponds to one of the paths that end up in partition $\lambda$. The Young Yamanouchi symbol, labeling the basis element, corresponds to the direction of the steps in the path, being $q_i=0$ ($q_i=1$) when the $i$-th step is taken to the right (left) in the lattice. With this graphical representation, the base $q_4^{[3,2]}$ can be seen as:\\

\begin{equation}
q_4^{[3,2]}=
\begin{ytableau}
 1 & 3 & 4 \\ 
 2& 5  
 \end{ytableau}
 =\{0,1,0,0,1\}=\includegraphics[scale=0.8,valign=c]{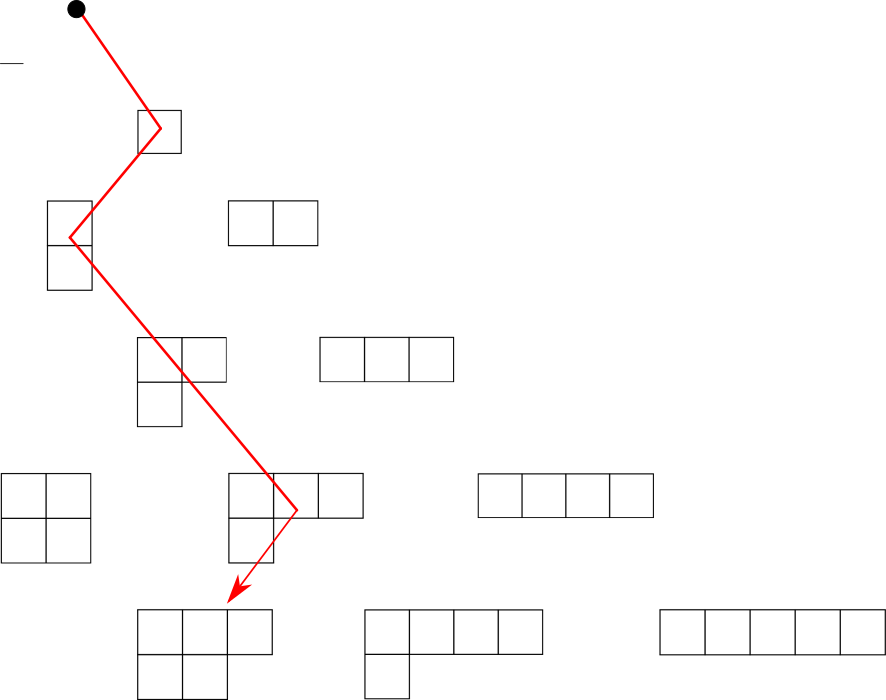}.
 \label{eq:IrrepsLabels}
\end{equation}

\noindent The three ways to label the basis elements of $S_n$, i.e., SYT, Young Yamanouchi symbols, and Young lattice's paths, will be used through this document, as each has advantages that will be exploited depending on the context. \\ 

\noindent Finding the dimension of an irreducible representation of $S_n$ is reduced to the problem of determining how many standard Young tableaux there are for a given partition $\lambda$. This process can be understood from a combinatorial point of view. First, we take the Young diagram $\lambda$ and fill each box with the number of boxes below and to the right of it, plus $1$, known as the \textit{hook lengths}. Then, the diagram will be filled with these numbers $h_i$ called hooks. The \textit{hook formula}  \cite{Zeilinger} can then be used to calculate the dimension of the irreducible representation associated with $\lambda$.
 \begin{equation*}
   f^{[\lambda]}:= \frac{n!}{\prod_i h_i}.
   \label{eq:hook}
 \end{equation*}
 For example, the young diagram $[3,2]$ filled with the hook lengths is: 
 \begin{equation*}
     \begin{ytableau}
     4 & 3& 1 \\
     2  &1 \\ 
        \end{ytableau}
     \Rightarrow f^{[3,2]}= \frac{5!}{4*3*2*1*1}=5.
 \end{equation*}
Using the hook formula \eqref{eq:hook}, we can understand how the dimension $f^{\lambda}$ of an irreducible representation grows asymptotically. If we fill each row of the Young diagram with the numbers $1,2,\cdots,\lambda_i$, then it is clear that:
\begin{equation*}
f^{[\lambda]}\leq \frac{n!}{\prod_i \lambda_i !}.
\end{equation*}
On the other hand, we can also define the numbers $v_i=\lambda_i+k+i$, where $k$ is the length of partition $\lambda$, leading to the lower bound:
\begin{equation*}
f^{[\lambda]}\geq \frac{n!}{\prod_i v_i!}.
\end{equation*}
In the asymptotic limit when $n\rightarrow\infty$ with $k$ fixed, we have $v_i \approx \lambda_i$, so:
\begin{equation}
f^{[\lambda]} \sim \frac{n!}{\prod_i \lambda_i !}.
\label{eq:asymptotichook}
\end{equation}
Where ``$\sim$'' stands for asymptotic behavior. Using Stirling's approximation, we can obtain asymptotically, 
\begin{equation}
f^{[\lambda]}\sim \exp(nH(\bar{\lambda})),
\label{eq:asymptoticsflambda}
\end{equation}
where $H(\bar{\lambda})$ is the Shannon entropy of the normalized partition $\bar{\lambda}=\lambda/n$. This shows that dimensions of irreps of $S_n$ grow exponentially with $n$.\\

\noindent With the basis elements of irreps of $S_n$ defined, it is possible to obtain a matrix representation for each element of $S_n$ in each irrep using the \textit{Young-Yamanouchi algorithm}. In this algorithm, analyzing permutational relations of the SYT, it is possible to find the matrix entries of all the adjacent transpositions of $S_n$. With these matrix representations, obtaining any other element of $S_n$ through matrix multiplication is possible, as adjacent transpositions are generators of $S_n$. I.e., any element of $S_n$ can be obtained by multiplications of adjacent transpositions. We will briefly outline the Young-Yamanouchi algorithm here. However, in Section \ref{myirreps}, we introduce a more relevant method to build the matrix representations of $S_n$.
\subsubsection{Young-Yamanouchi algorithm}
\label{Young-Yamanouchi}
For this algorithm, we start by defining the \textit{axial distance} $\rho_{q}(i)$ in a standard Young tableau defined by $q$ as the number of steps to go from the box filled with $i$ to the box filled with $i+1$, with the convention that the steps must be taken horizontally and vertically only. Steps going up or to the right are positive, and steps going down or to the left are negative. For example, consider the following SYT:
\begin{equation*}
q^{[4,2,1]}_{1}=\begin{ytableau}
1 & 3 & 4 & 7\\
2 & 6 \\
5
\end{ytableau},
\end{equation*}
then, for going from the box filled with $1$ to the box filled with $2$, we need to take a step going down, i.e., $\rho_{q^{[4,2,1]}_{1}} (1)=-1$, for going from $2$ to $3$ we can take one step up and one step to the right, i.e.,  $\rho_{q^{[4,2,1]}_{1}} (2)=2$, all axial distances in this SYT are:
\begin{equation*}
 \rho_{q^{[4,2,1]}_{1}} (1)=-1 , \, \rho_{q^{[4,2,1]}_{1}} (2)=2 , \,  \rho_{q^{[4,2,1]}_{1}} (3)=1, \,
\rho_{q^{[4,2,1]}_{1}} (4)=-4, \,
 \rho_{q^{[4,2,1]}_{1}} (5)=2, \,
  \rho_{q^{[4,2,1]}_{1}} (6)=3.
\end{equation*}
We also have to list all the SYTs that are related through adjacent transpositions. For example, if we consider another SYT of $[4,2,1]$:
\begin{equation*}
q_2^{[4,2,1]}= \begin{ytableau}
1 & 3 & 4 & 7\\
2 & 5 \\
6
\end{ytableau},
\end{equation*}
this one can be obtained from $q_1^{[4,2,1]}$ by applying the adjacent transposition of $5$ with $6$, i.e.,
\begin{equation*}
(56)q_1^{[4,2,1]} =q_2^{[4,2,1]}.
\end{equation*}
With this, the entries of the matrix representation of adjacent transpositions $(i,i+1)$ can be obtained as:
 \begin{equation*}
     D^{[\lambda]}_{qq'}(i,i+1) = \bra{q} (i,i+1) \ket{q'}=\left\{ \begin{array}{cc}
        \frac{1}{\rho_q(i)}  &  q=q' \\
        & \\
        \frac{\sqrt{\rho^2_q(i)-1}}{|\rho_q(i)|}  &  q=(i,i+1)q' \\
        &\\
        0 & otherwise
     \end{array} \right. ,
     \label{eq:Young-Yamanouchi}
 \end{equation*}
where $D^{[\lambda]}(i,i+1)$ is the matrix representation of adjacent transposition $(i,i+1)$ in the irreducible representation $\lambda$. \\
 
 \noindent Now, we will use the Young-Yamanouchi algorithm to compute the two-dimensional representation in $S_3$ corresponding to $[2,1]$. The two SYTs and their corresponding Yamanouchi symbols are
 \begin{equation*}
      q_1^{[2,1]}:=\begin{ytableau}
        1 & 2 \\
        3
      \end{ytableau} \Rightarrow \{0,0,1\} ,\quad q_2^{[2,1]}:= \begin{ytableau}
        1 & 3 \\
        2
      \end{ytableau} \Rightarrow \{0,1,0\}.
 \end{equation*}
When considering the Young Lattice, these basis are represented by the only two paths to get to the partition $[2,1]$:
\begin{equation*}
\includegraphics[scale=1.2,valign=c]{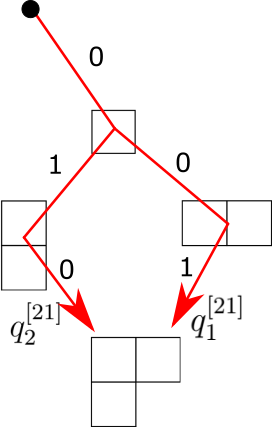},
\end{equation*}
we now labeled each step with its corresponding value in the Yamanouchi symbol, being 0 to the right and $1$ to the left. Note that the only relation by adjacent transpositions in this case is:
 \begin{equation*}
     (23)q_1 =q_2 .
 \end{equation*}
The axial distances in these basis are:
 \begin{equation*}
     \rho_{q_1^{[2,1]}}(1)=1, \quad    \rho_{q_1^{[2,1]}}(2)=-2 \quad   \rho_{q_2^{[2,1]}}(1)=-1 \quad   \rho_{q_2^{[2,1]}}(2)=2,
 \end{equation*}
 then, the matrices of adjacent transpositions using equation \eqref{eq:Young-Yamanouchi} can be obtained as:
 \begin{equation*}
     D^{[2,1]}(12)=\left( \begin{array}{cc}
        1  & 0 \\
         0 & -1
     \end{array}\right) \quad, \quad  D^{[2,1]}(23)=\left( \begin{array}{cc}
       - \frac{1}{2}  & \frac{\sqrt{3}}{2} \\
         \frac{\sqrt{3}}{2} & \frac{1}{2}
     \end{array}\right)
     \label{eq:irep2321}
 \end{equation*}
 where the entries in the matrices are ordered as ${q_1^{[2,1]},q_2^{[2,1]}}$. These matrices are built in a way that the chain structure of the symmetric group is made explicitly; note how for $D^{[2,1]}(12)$  the diagonal form corresponds to 
\begin{equation*}
D^{[2,1]}(12)=D^{[2]}(12) \oplus D^{[1^2]}(12).
\end{equation*}
With the two matrices in \eqref{eq:irep2321}, the remaining four matrices of $S_3$ can be obtained from the group table in \eqref{eq:S3table}:
\begin{equation*}
\begin{gathered}
     D^{[2,1]}(e)=D^{[2,1]}(12) \cdot D^{[2,1]}(12)=\left( \begin{array}{cc}
        1  & 0 \\
         0 & 1
     \end{array}\right) , \\
      D^{[2,1]}(123)= D^{[2,1]}(12)\cdot D^{[2,1]}(23)=\left( \begin{array}{cc}
       - \frac{1}{2}  & \frac{\sqrt{3}}{2} \\
         -\frac{\sqrt{3}}{2} & -\frac{1}{2}
     \end{array}\right), \\
     D^{[2,1]}(132)= D^{[2,1]}(23) \cdot  D^{[2,1]}(12)=\left( \begin{array}{cc}
       - \frac{1}{2}  & -\frac{\sqrt{3}}{2} \\
         \frac{\sqrt{3}}{2} & -\frac{1}{2}
     \end{array}\right), \\  D^{[2,1]}(13)=D^{[2,1]}(12) \cdot  D^{[2,1]}(132))\left( \begin{array}{cc}
       - \frac{1}{2}  & -\frac{\sqrt{3}}{2} \\
        - \frac{\sqrt{3}}{2} & \frac{1}{2}
     \end{array}\right).
 \end{gathered}
 \end{equation*}
 From these matrix representations, it can be checked that the multiplication rules for the group are satisfied, and from the characters, i.e., the traces of the matrices, it is possible to check that this is an irreducible representation with Equation \eqref{eq:orthoirreds}.\\
 
\noindent We are now in a position to present a connection between $S_n$ and $GL_d$ that highlights the importance of these groups when analyzing quantum systems, where it is made explicit that quantum systems of $n$ particles, each in a Hilbert space of dimension $d$, allow a decomposition of the total Hilbert space in irreducible representations of $S_n$ and $GL_d$.

 \section{Schur-Weyl Duality} \label{Schur-Weyl}
 Consider a system with $n$ qudits, with a total Hilbert space $\mathcal{H}=(\mathbb{C}^d)^{\otimes n}$. The product basis is then
 \begin{equation*}
     \ket{i_1,i_2,\dots i_n} , \quad i_j \in\{0,1,2, \dots ,d-1\},
 \end{equation*}
 The group $GL_d$ acts with its standard representation in each of the qudits, and the group $S_n$ acts by permuting the $n$ parts. Then, their actions in $\mathcal{H}$ can be described as:
 \begin{equation*}
 \begin{gathered}
     U(\pi)  \ket{i_1,i_2,\dots i_n} =\ket{i_{\pi^{-1}(1)},i_{\pi^{-1}(2)},\dots i_{\pi^{-1}(n)}},\\
     A^{\otimes n} \ket{i_1,i_2,\dots i_n} = A \ket{i_1}\otimes A\ket{i_2} \otimes \dots \otimes A\ket{i_n}.
 \end{gathered}
 \end{equation*}
 with $A\in GL_d$ and $U(\pi)$ is the matrix that permutes the basis elements according to $\pi$. First, note that $U(\pi)$ being a representation of $S_n$, can be decomposed in irreps of $S_n$ according to \eqref{eq:Irrepdecomp} as:
\begin{equation*}
U(\pi) =\bigoplus_{\lambda \vdash n} I_{m_{[\lambda]}} \otimes D^{[\lambda]}(\pi),
\end{equation*} 
 where $\lambda\vdash n$ restricts the representation to partitions of $n$, and $m_{[\lambda]}$ is the multiplicity of irrep $[\lambda]$ in the decomposition of $U$. Similarly for $A^{\otimes n}$, it can be decomposed as irreducible representations of $GL_d$ labeled by partitions of $n$ with at most $d$ rows, i.e.,
 \begin{equation*}
A^{\otimes n} =\bigoplus_{\lambda \vdash n,d} V^{\{\lambda \}} \otimes I_{m_{\{\lambda\}}},
\end{equation*}
with  $m_{\{\lambda\}}$ the multiplicity of irrep $\{\lambda\}$ in $A^{\otimes n}$, and $V^{\{\lambda\}}(A)$ the matrix representation of $A$ in $\{\lambda\}$.
\textit{Schur-Weyl Duality} states that, due to the action of $GL_d$ and $S_n$ commute, they can be block-diagonalized simultaneously, with both irreps labeled by the same partition $\lambda$, implying that irreps of $S_n$ and $GL_d$ decompose completely the total Hilbert space. Schur-Weyl duality can be represented as:
\begin{equation}
(\mathbb{C}^{d})^{\otimes n} \cong \bigoplus_{\lambda} \{\lambda \} \otimes [\lambda].
\label{eq:Schur-Weyl}
\end{equation}
When considering the actions of $U(\pi)$ and $A^{\otimes n}$ this can be read as:
\begin{equation*}
U(\pi)A^{\otimes n} = A^{\otimes n} U(\pi) =\bigoplus_{\lambda\vdash n,d} V^{\{\lambda\}}(A) \otimes D^{[\lambda]} (\pi),
\end{equation*}
which means that $U(\pi)$ acts independently in $S_n$ irreps and $A^{\otimes n}$ acts independently in $GL_d$ irreps. When considering each action separatedly, for $U(\pi)$ we have:
\begin{equation*}
U(\pi) =\bigoplus_{\lambda\vdash n,d} I_{s_{\lambda}(\bm{1})} \otimes D^{[\lambda]}(\pi),
\end{equation*}
meaning that the multiplicity of irrep $[\lambda]$ in $U$ decomposition, is given by the dimension of irrep $\{\lambda\}$ (in $GL_d$). Similarly, for $A^{\otimes n}$ we have:
  \begin{equation}
     A^{\otimes n} = \bigoplus_{\lambda\vdash n,d} V^{\{\lambda\}}(A) \otimes I_{f^{[\lambda]}},
     \label{Schur-WeylGL}
 \end{equation}
 where now the multiplicity of irrep $\{\lambda\}$ is given by the dimension of irrep $[\lambda]$ (in $S_n$). This is a notion that we introduced previously; the different paths for obtaining an irreducible representation $\{\lambda\}$ in $GL_d$, i.e., the different choices in the orders to symmetrizing and anti-symmetrizing the indices of the computational basis, correspond to a different multiplicity of the same irrep. On the other hand, each path to build the Young diagram $\lambda$, labels a basis element of the irrep $[\lambda]$ in $S_n$. Then, there are as many multiplicities of the irrep $\{\lambda\}$ in $GL_d$ as basis elements of irrep $[\lambda]$ in $S_n$.  We already explored this property of the decomposition into irreps of tensor products of copies of $GL_d$ in Equation \eqref{eq:symmYoung}, where the irrep $\{2,1\}$ of $GL_2$ appeared twice in the decomposition into irreps. This is exactly the dimension of irrep $[2,1]$  in $S_3$, $f^{[2,1]}=2$.\\
 
\noindent One important consequence of Schur-Weyl duality is that, there is some transformation that takes the representations $U(\pi)$ and $A^{\otimes n}$ into their diagonal block forms, i.e., their decomposition into irreps simultaneously, known as the \textit{Schur transform}.
\section{Schur Transform}
\label{Schurtransform}
Schur-Weyl duality states that copies of a Hilbert space can be decomposed simultaneously into irreps of $GL_d$ and $S_n$. The operation that makes this decomposition explicit is the \textit{Schur transform} \cite{Schurtransform}. In this section, we will show how this transformation can be made explicit when restricting $GL_d$ to $SL_2$, permitting the use of the CGC of $SL_2$ for the diagonalization. This restriction is strong but meaningful in quantum mechanics, allowing us to analyze any multi-qubit system. The procedure and language presented here are mainly based on \cite{Botero}.\\

\noindent We want to find the transformation between the product basis of $n$ qubits to the basis of irreps of $SL_2$ and $S_n$. In this setup, the basis for the initial Hilbert space, defined by $n$ copies of the standard representation $SL_2\otimes SL_2 \otimes \dots \otimes SL_2$, can be chosen to be:
\begin{equation*}
\ket{s_1,s_2,s_3,\dots,s_n}, \quad s_i \in  \{0,1 \},
\end{equation*}
i.e., each basis element will be labeled by a binary sequence of length $n$. We often refer to this as the \textit{computational basis}. \\

\noindent On the other side, we want to end up in a simultaneous basis of $SL_2$ and $S_n$, and, as we know from Schur-Weyl duality, these are simultaneously labeled by the same partition $\lambda$. Then, this basis will be labeled as:
\begin{equation*}
\ket{\lambda,\omega,q},
\end{equation*}
where $\omega$ labels the basis elements of $\{\lambda\}$ according to the weight of the representing SSYT, and $q$ labels the basis elements of $[\lambda]$, usually with the modified Yamanouchi symbol. We refer to this basis as the \textit{Schur basis}. As the length of the partitions is at most $2$ (because $d=2$), we will simplify the notation by using only the value of $\lambda_2$ to label $\lambda$ and state the $n$ value. For example, instead of using $[31]$, we will use $[1]$ in $n=4$. In summary, finding the Schur transform is the same as finding the coefficients that relate the computational basis with the Schur basis:
\begin{equation*}
\ket{s_1,s_2,s_3,\dots,s_n} = \sum_{\lambda=0}^{n/2} \sum_{\omega,q} \bm{\Gamma}_{s}^{\lambda,\omega,q}\ket{\lambda,\omega,q}.
\end{equation*}  
This transformation can be done by refining the well-known process of angular momentum addition in quantum mechanics, particularly in the case of spin $1/2$.
\subsubsection*{Angular momentum addition}
Systems of angular momentum $j$ can be studied as representations in $SL_{2j+1}$, and the process of angular momentum addition is nothing but the decomposition into irreps of tensor products of irreps of $SL_{2j+1}$. Note that the example at the beginning of subsection \ref{GLd} corresponds to the problem of adding two spins $1/2$. The symmetric representation in this example it is the subspace usually known as \textit{the triplet}, while the anti-symmetric representation is \textit{the singlet} \cite{Grifit}. In this context, $j$ values label the irreps, and $m$ values label the basis elements of irrep $j$. Angular momentum addition consists of rewriting the tensor product as a sum of possible values of total angular momentum:
\begin{equation*}
\ket{j_1,m_1}\ket{j_2,m_2} = \sum_{J} \braket{J,M} {j_1,m_1;j_2,m_2} \ket{J,M}, \quad  M=m_1+m_2,
\end{equation*}
where $\braket{J,M}{j_1,m_1;j_2,m_2}$ are the CGC of $SL_{2j+1}$. When restricting to qubit systems, a good strategy is to consider what happens to the system when adding one qubit at a time. In that particular case, the angular momentum addition can be expressed as:
\begin{equation}
\begin{gathered}
\ket{j_1,m_1}\ket{1/2,\pm 1/2} =  \braket{j_1+1/2,m_1 \pm 1/2} {j_1,m_1;1/2,\pm 1/2}\ket{j_1+1/2,m_1 \pm 1/2} \\+\braket{j_1-1/2,m_1 \pm 1/2}{j_1,m_1;1/2,\pm 1/2} \ket{j_1-1/2,m_1 \pm 1/2},
\end{gathered}
\label{eq:angmomadd}
\end{equation}
and for this case, the CGC have a nice closed expression \cite{CGCJ}:
    \begin{equation*}
       \braket{J,M}{j_1,1/2;m_1,\pm 1/2}=  \begin{array}{|c|c|c|} 
        \hline
        J     &m_2=1/2 &m_2=-1/2  \\ \hline
        j_1+1/2     & \sqrt{\frac{j_1+m_1+1}{2j_1+1}}& \sqrt{\frac{j_1-m_1+1}{2j_1+1}}   \\ \hline
        j_1-1/2 &  -\sqrt{\frac{j_1-m_1}{2j_1+1}} & \sqrt{\frac{j_1+m_1}{2j_1+1}}\\ \hline
        \end{array}.
\end{equation*}
With this, it is possible to decompose the tensor product of many qubits into a sum on a total angular momentum basis. The first thing we want to do is to translate this process to the language that we use for $SL_2$ representations. By fixing the number of parts $n$, the relation between the notations is simple:
\begin{equation*}
J=\frac{n}{2}-\lambda, \quad M=\frac{n}{2}-\omega,
\end{equation*} 
and the restriction for $M$, $-J\leq M \leq J$, is translated to $\lambda \leq \omega \leq n-\lambda$. Under this change of notation, the CGC for  adding $1/2$ spin corresponds to:

\begin{equation*}
 \braket{j,m}{j_1,1/2;m_1,\pm 1/2}= \braket{\lambda,\omega'+s_n}{\lambda',0;\omega',s_n}= \Gamma ^{\lambda,\omega,n}_{\lambda',s_n},
 \label{eq:CGCGL}
\end{equation*}
with 
\begin{equation*} 
\Gamma_{\lambda,s_n}^{\lambda',\omega,n}=
        \begin{array}{|c|c|c|} 
        \hline
        \lambda     &s_n=0 &s_n=1  \\ \hline
        \lambda'    & \sqrt{\frac{n-\lambda-\omega}{n-2\lambda}}& \sqrt{\frac{\omega-\lambda}{n-2\lambda}}   \\ \hline
       \lambda'+1 &  -\sqrt{\frac{\omega-\lambda+1}{n-2\lambda+2}} & \sqrt{\frac{n-\lambda-\omega+1}{n-2\lambda+2}}\\ \hline
        \end{array},
\end{equation*}
where $s_n$ refers to the binary value added in the last step in the process. Then, the process of adding one qubit shown in Equation \eqref{eq:angmomadd} can be written, ommiting the $\lambda$ label for the qubit added (which is always $\lambda=0$) as:
\begin{equation*}
\ket{\lambda',\omega'}\ket{s_n} = \Gamma^{\lambda',\omega'+s_n,n}_{\lambda',s_n} \ket{\lambda',\omega'+s_n} + \Gamma^{\lambda',\omega'+s_n,n}_{\lambda'+1,s_n} \ket{\lambda'+1,\omega'+s_n}.
\end{equation*}
Let us use this as an example to decompose a simple tensor product. Consider we want to find the coefficients for the decomposition of the tensor product of $\ket{\lambda=2,\omega=2}$ with $n=5$, and a single qubit, say $\ket{1}$ (or $\ket{\lambda=0,\omega=1}$). In young diagrams, we are looking for coefficients corresponding to the following decomposition:
\begin{equation*}
\ydiagram{3,2} \otimes \ydiagram{1} \cong \ydiagram{4,2} \oplus \ydiagram{3,3}.
\end{equation*}
Note how with our notation, when starting from $\lambda'$, we can only obtain $\lambda=\{\lambda',\lambda'+1\}$ when adding one qubit. By using the angular momentum addition, we have the following:
\begin{equation*}
\ket{2,2}\ket{0,1}= \Gamma^{2,3,6}_{2,0} \ket{2,3} +\Gamma^{2,3,6}_{3,0} \ket{3,3} ,
\end{equation*}
where we ommited $\lambda$ and $\omega$, but the values are always ordered as $\ket{\lambda,\omega}$. It is worth highlighting that the possible final states are the ones such that $\omega=\omega'+s_n$, i.e., the $\omega$ value is fixed. Calculating both CGC, we obtain the following:
\begin{equation*}
\ket{2,2}\ket{0,1}=\sqrt{\frac{3-2}{6-2*2}}  \ket{2,3} +\sqrt{\frac{6-2-3+1}{6-2*2+2}}\ket{3,2} = \sqrt{\frac{1}{2}} \ket{2,3} +\sqrt{\frac{1}{2}}\ket{3,3}.
\end{equation*}
This process would correspond to the diagonalization of $SL_2$ irreducible representations; however, it does not distinguish all the permutational inequivalent ways to obtain the initial state of $n-1$ particles. For the previous example, the base  $\ket{2,2}$, i.e.,
\begin{equation*}
\begin{ytableau}
   0 & 0 & 0  \\
   1 & 1
\end{ytableau} ,
\end{equation*}
can be obtained in several ways from one qubit system:
\begin{equation*}
\begin{gathered}
q_1 =
\begin{ytableau}
   0 
\end{ytableau} \Rightarrow \begin{ytableau}
   0 & 0 
\end{ytableau} \Rightarrow \begin{ytableau}
   0 & 0 & 0 
\end{ytableau} \Rightarrow \begin{ytableau}
   0 & 0 & 0  \\
   1 
\end{ytableau} \Rightarrow \begin{ytableau}
   0 & 0 & 0  \\
   1 & 1
\end{ytableau}, \\
q_2 =
\begin{ytableau}
   0 
\end{ytableau} \Rightarrow \begin{ytableau}
   0 & 0 
\end{ytableau} \Rightarrow \begin{ytableau}
   0 & 0 \\
   1 
\end{ytableau} \Rightarrow \begin{ytableau}
   0 & 0 & 0  \\
   1 
\end{ytableau} \Rightarrow \begin{ytableau}
   0 & 0 & 0  \\
   1 & 1
\end{ytableau} ,\\
q_3 =
\begin{ytableau}
   0 
\end{ytableau} \Rightarrow \begin{ytableau}
   0 & 0 
\end{ytableau} \Rightarrow \begin{ytableau}
   0 & 0 \\
   1 
\end{ytableau} \Rightarrow \begin{ytableau}
   0 & 0   \\
   1  &1 
\end{ytableau} \Rightarrow \begin{ytableau}
   0 & 0 & 0  \\
   1 & 1
\end{ytableau}, \\
q_4 =
\begin{ytableau}
   0 
\end{ytableau} \Rightarrow \begin{ytableau}
   0 \\
   1
\end{ytableau} \Rightarrow \begin{ytableau}
   0 & 0 \\ 
   1 
\end{ytableau} \Rightarrow \begin{ytableau}
   0 & 0 & 0  \\
   1 
\end{ytableau} \Rightarrow \begin{ytableau}
   0 & 0 & 0  \\
   1 & 1
\end{ytableau}, \\
q_5 =
\begin{ytableau}
   0 
\end{ytableau} \Rightarrow \begin{ytableau}
   0 \\
   1  
\end{ytableau} \Rightarrow \begin{ytableau}
   0 & 0 \\
   1
\end{ytableau} \Rightarrow \begin{ytableau}
   0 & 0  \\
   1 & 1
\end{ytableau} \Rightarrow \begin{ytableau}
   0 & 0 & 0  \\
   1 & 1
\end{ytableau},
\end{gathered}
\end{equation*}
each with different permutational symmetries, representing the different basis elements of $n=5$, $[\lambda]=[2]$. For the Schur transform, this permutational symmetry matters. The only difference with angular momentum addition is that keeping a register of the different ways to get to the final partition will be necessary, and such a register turns out to be the basis elements of $S_n$ irreps. Note that each path of partitions will also mean a path of CGC used in the construction, and the resultant transformation will also depend on the $S_n$ basis elements, which is necessary if we want to diagonalize its action simultaneously. 

\subsubsection{Schur transform on $d=2$}
First, let us redefine the CGC in terms of where the new box is being added to the initial Young Tableaux $\lambda'$ in each step of the angular momentum addition:
\begin{equation}
\Gamma_{q_n,s_n}^{\lambda,\omega,n}=
        \begin{array}{|c|c|c|} 
        \hline
        \lambda     &s_n=0 &s_n=1  \\ \hline
        q_n=0    & \sqrt{\frac{n-\lambda-\omega}{n-2\lambda}}& \sqrt{\frac{\omega-\lambda}{n-2\lambda}}   \\ \hline
       q_n=1 &  -\sqrt{\frac{\omega-\lambda+1}{n-2\lambda+2}} & \sqrt{\frac{n-\lambda-\omega+1}{n-2\lambda+2}}\\ \hline
        \end{array}
        \label{eq:Gammas}.
\end{equation}
In this sense, the coefficients of the Schur basis for $n$ elements can be obtained from the coefficients of the Schur basis for $n-1$ elements by adding one qubit. Then, one gets:
\begin{equation*}
\ket{\lambda',\omega',q'}\ket{s_n} = \Gamma^{\lambda',\omega'+s_n,n}_{0,s_n} \ket{\lambda', \omega'+s, q'0} +  \Gamma^{\lambda'+1,\omega'+s_n,n}_{1,s_n} \ket{\lambda', \omega'+s, q'1},
\end{equation*}
where $q'0$ ($q'1$) refers to the sequence obtained from the initial sequence $q'$ after concatenating a 0 (1) at the end. A very useful relation that can be obtained from here is the inner product in the Schur basis of $n$ with those of $n-1$:
\begin{equation}
\braket{\lambda,\omega,q}{\lambda',\omega',q'}\ket{s_n}= \Gamma^{\lambda,\omega,n}_{q_n,s_n} \delta_{\omega,\omega'+s_n}\delta_{q'q_n,q}\delta_{\lambda,\lambda'+q_n},
\label{eq:innerprodschur}
\end{equation}
with $q_n$ being the last value of the sequence $q$, the appearing deltas can be understood as an equivalent of \textit{selection rules} from angular momentum addition. Now, consider that we want to perform the Schur transform in a binary sequence of $n$ elements, i.e., to find the coefficients of:
\begin{equation*}
\ket{s_1,s_2,s_3,\dots,s_n} = \sum_{\lambda=0}^{n/2} \sum_{\omega,q}\bm{\Gamma}_{s}^{\lambda,q} \ket{\lambda,\omega,q}.
\end{equation*} 
Due to angular momentum addition selection rules, it is clear that $\omega$ is fixed to be $\omega=\sum_{i=1}^{n} s_i$. It is useful to define $\omega_i = \sum_{j=1}^{i} s_i$, i.e., the partial weight of the sequence at step $i$. By picking one set $\lambda,q$, we can see that the coefficient we want to find corresponds to the product of CGC defined by the sequence and the path $q$. For example, consider the sequence $\ket{010011}$, and we want to calculate the coefficient corresponding to $\lambda=2,q=\{0,1,0,0,0,1\}$. This path can be seen as:
\begin{equation*}
\begin{ytableau}
   0 
\end{ytableau} \Rightarrow \begin{ytableau}
   0 \\ 1 
\end{ytableau} \Rightarrow \begin{ytableau}
   0 &0\\
    1 
\end{ytableau}  \Rightarrow
\begin{ytableau}
   0 &0 &0 \\
    1 
\end{ytableau} \Rightarrow \begin{ytableau}
   0 &0 & 0 & 1\\
    1 
\end{ytableau} \Rightarrow 
\begin{ytableau}
   0 &0& 0& 1\\
    1  & 1
\end{ytableau},
\end{equation*}
where at $i$ step, one box filled with $s_i$ is added to the $q_i-1$ row. This path can also be seen from the Young lattice, where the Young diagrams now are filled in each step with the elements of the sequence $s$:
\begin{equation*}
\includegraphics[scale=1]{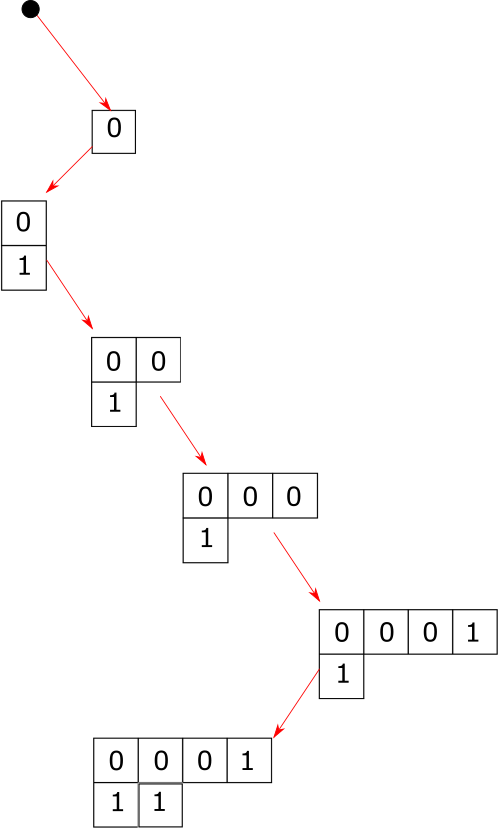}
\end{equation*}
From this point of view, some steps are not possible when the sequence is fixed. For example, step number four cannot be done to the left; otherwise, we end up in a Young Tableau:
\begin{equation*}
\begin{ytableau}
 0 & 0 \\
 1 & 0 
\end{ytableau},
\end{equation*}
which is not an SSYT in $GL_d$ as the second column is not strictly increasing. Given a possible path as the previous one, it is then translated to a sequence of CGC to be multiplied:
\begin{equation*}
\begin{gathered}
\begin{ytableau}
   0 
\end{ytableau} \xRightarrow{\Gamma^{1,1,2}_{1,1}} \begin{ytableau}
   0 \\ 1 
\end{ytableau} \xRightarrow{\Gamma^{1,1,3}_{0,0}}\begin{ytableau}
   0 &0\\
    1 
\end{ytableau}  \xRightarrow{\Gamma^{1,1,4}_{0,0}}
\begin{ytableau}
   0 &0 &0 \\
    1 
\end{ytableau}\\
  \xRightarrow{\Gamma^{1,2,5}_{0,1}} \begin{ytableau}
   0 &0 & 0 & 1\\
    1 
\end{ytableau} \xRightarrow{\Gamma^{2,3,6}_{1,1}} 
\begin{ytableau}
   0 &0& 0& 1\\
    1  & 1
\end{ytableau}
\end{gathered},
\end{equation*}
With this, we can read the coefficient of the Schur transform to be:
\begin{equation*}
\bm{\Gamma}_{010011}^{2,\{0,1,0,0,0,1\}}=\Gamma^{1,1,2}_{1,1} \cdot \Gamma^{1,1,3}_{0,0}\cdot \Gamma^{1,1,4}_{0,0} \cdot \Gamma^{1,2,5}_{0,1} \cdot \Gamma^{2,3,6}_{1,1} = \sqrt{\frac{1}{2}}\cdot 1 \cdot \sqrt{\frac{1}{2}}\cdot \sqrt{\frac{1}{3}} \cdot \sqrt{\frac{1}{2}}= \frac{1}{2 \sqrt{6}}.
\end{equation*}
The same process should be done for all the partitions $\lambda$ and basis elements $q$. In general, we will have:
\begin{equation*}
\bm{\Gamma}^{\lambda,q}_s = \prod_{i=1}^{n} \Gamma^{\lambda_i,\omega_i,i}_{q_i,s_i},
\end{equation*}
where $\lambda_i=\sum_{j=1}^{i}q_i$ is the partial partition at step $i$. Then, the Schur transform can be performed as:
\begin{equation}
\ket	{s}= \sum_{\lambda\sim \omega_s} \ket{\lambda,\omega_s} \sum_q \bm{\Gamma}^{\lambda,q}_{s} \ket{\lambda,q},
\label{eq:Schurtransformsequence}
\end{equation} 
where $\omega_s$ is the weight of the full sequence, i.e., $\omega_s=\sum_{i=1}^{n} s_i$, and $\lambda \sim \omega_s$ restricts the appearing partitions to those such that $\lambda \leq \omega_s \leq n-\lambda$, due to the natural restriction of $SL_2$ irreps. Equivalently, we have that the coefficients for transforming one basis into the other are given by:
\begin{equation}
\braket{\lambda,\omega_s,q}	{s}= \bm{\Gamma}^{\lambda,q}_{s}.
\label{eq:Schurtransfsequence}
\end{equation}
Summarizing, the Schur transform can be performed as follows: given a sequence $s$, we start by choosing $\lambda$ and $q$, which is understood as a path in the Young lattice, then, according to this path, we perform angular momentum addition (By calculating $\Gamma$) at each step in the path, and at the end, we multiply the chain of $\Gamma$'s defined by the sequence and the path. Next, we repeat the process on another $q$ until we scan all the possible $\lambda$s and $q$s. In each path step, we can check if the filled diagram is a valid SSYT; if not, its corresponding coefficient in the Schur transform is zero. We have found this perspective very useful for thinking about the Schur transform. With the Schur transform of the sequences that define the product basis, it is straightforward to apply to states on such a basis.\\

\noindent Schur transform can be performed over all the sequences defining the computational basis, and the coefficients $\bm{\Gamma}^{\lambda,q}_s$ build a unitary matrix that under conjugation simultaneously diagonalizes the actions of  $GL_2$ and $S_n$. For example, when applying the Schur transform on the computational basis of three qubits, we can obtain the following matrix:
\begin{equation*}
        \bm{\Gamma}=\left( \begin{array}{cccccccc}
        1& & & & & & & \\ & \sqrt
    {\frac{1}{3}}&\sqrt
    {\frac{1}{3}}& & \sqrt
    {\frac{1}{3}}&& & \\
    & & & \sqrt
    {\frac{1}{3}} & & \sqrt
    {\frac{1}{3}} & \sqrt
    {\frac{1}{3}}& \\ &&&&&&& 1\\ & \sqrt
    {\frac{2}{3}}&-\sqrt
    {\frac{1}{6}}& & -\sqrt
    {\frac{1}{6}} & & & \\
     & &\sqrt
    {\frac{1}{2}}  &  & -\sqrt
    {\frac{1}{2}} & & & \\
    & & & \sqrt
    {\frac{1}{6}}& & \sqrt
    {\frac{1}{6}}& -\sqrt
    {\frac{2}{3}} &\\
     & & & \sqrt
    {\frac{1}{2}} & & -\sqrt
    {\frac{1}{2}} & & \\ 
        \end{array} \right).
    \end{equation*}
 Rows are labeled by the computational basis, and columns by the Schur basis. When conjugating this matrix on three tensor products of elements $A\in GL_d$, one gets:
  \begin{equation*}
       \bm{\Gamma} ( A ^{\otimes 3})  \bm{\Gamma}^{\dagger}= \left(\begin{array}{ccc}
            V^{\{0\}}(A) &  & \\
             &  V^{\{1\}} (A)& \\
             & & V^{\{1\}} (A)
        \end{array} \right) .
    \end{equation*}
Similarly, with any permutation of $S_3$ acting on the computational basis:
    \begin{equation*}
       \bm{\Gamma} U(\pi)  \bm{\Gamma}^{\dagger}= \left(\begin{array}{cccccc}
            D^{[0]}(\pi) &  & & & &\\
             &  D^{[0]}(\pi) & & & & \\
             && D^{[0]}(\pi)& & & \\
             & & &  D^{[0]}(\pi) & &\\
             & & & &  D^{[1]}(\pi)& \\
             & & & & &  D^{[1]}(\pi)
        \end{array} \right),
\end{equation*}
where it is worth highlighting that multiplicities of irreps in $S_n$ ($GL_2$) are the dimension of irreps in $GL_2$($S_n$), which is stated by Schur-Weyl duality. This diagonalization can build $S_n$ irreducible representations without using the Young-Yamanouchi algorithm, as described in Section \ref{myirreps}.
    
\section{Schur-Weyl Duality in Quantum information}
In quantum mechanics, it is common to encounter systems that exhibit invariance under permutations; for example, we can model systems composed of identical copies of the same state, where permuting these copies does not alter the overall system. For this kind of setup, the total Hilbert space is invariant under permutations, so under Schur-Weyl duality, it must only contain the trivial representation in $S_n$. In this context, we will explore two applications in the field of quantum information that harness this symmetry to yield valuable insights and solutions.

  \subsection{Keyl-Werner Theorem}
\label{Keyl-WernerTheorem}
Quantum state tomography is a common problem in quantum information that involves reconstructing an unknown quantum state by making measurements on many identically prepared copies of the state. For a quantum system described by a density matrix $\rho$, finding the eigenvalues of $\rho$ gives important information about the system. These eigenvalues, denoted by $a=(a_1,a_2,\dots,a_d)$, with $a_i\geq a_{i+1}$, label the conjugacy class in $GL_d$ that gathers all equivalent states. One way to approach this problem is through the Keyl-Werner estimation theorem, as described in \cite{Keyl-Werner}. The theorem shows how Schur-Weyl duality can be used to obtain $a$ with high precision in a single measurement over $n$ copies of the state.\\

\noindent Consider the Schur-Weyl transform on $n$ copies of $\rho$, which, by permutational symmetry, must be of the form:
\begin{equation*}
\rho^{\otimes n}= \bigoplus_{\lambda} V^{\{\lambda\}}(\rho) \otimes I_{f^{[\lambda]}}.
\end{equation*}
as shown in Equation \eqref{Schur-WeylGL}. Note how the corresponding part to the symmetric group is just the identity due to permutation invariance. This setup aims to determine $a$ with high precision in a single measure on the $n$ copies. For this, note that a collective measurement of $\lambda$ on the $n$ copies is described by a set of orthogonal projectors $\{\mathcal{P}_\lambda\}$, and the probability of obtaining an output $\lambda$ is
\begin{equation*}
p(\lambda|\rho^{\otimes n})= \operatorname{tr} \left(\mathcal{P}_\lambda \rho ^{\otimes n} \right)= \operatorname{tr} \left(V^{\{\lambda\}}(\rho) \otimes I_{f^{[\lambda]}} \right)= \operatorname{tr} \left(V^{\{\lambda\}} (\rho)\right) \otimes \operatorname{tr} \left( I_{f^{[\lambda]}} \right)= s_\lambda(a)f^{[\lambda]},
\end{equation*}
where in the last equality, we used the Schur polynomial $s_\lambda(a)$ in Equation \eqref{eq:Schurpoly} for the character of $\{\lambda\}$ representation,  and the fact that the trace of the identity element in irrep $[\lambda]$ is the dimension of the irrep $f^{[\lambda]}$ in Equation \eqref{eq:hook}. We can then use the asymptotics in $f^{[\lambda]}$ and $s_{\lambda} (a)$ shown in Equations \eqref{eq:asymptotichook} and \eqref{eq:Schurasympt} to obtain the bounds:
\begin{equation*}
\left(\prod_i a_i ^{\lambda_i} \right)\left( \frac{n!}{\prod_i v_i} \right)\leq s_\lambda(a)f^{[\lambda]} \leq s_\lambda(\bm{1}) \left( \prod_i a_i ^{\lambda_i} \right)\left(\frac{n!}{\prod_i \lambda_i} \right),
\end{equation*}
where $\lambda_i$ and $v_i$ are defined in \eqref{eq:hook}.
Then, $s_{\lambda}(a)f^{[\lambda]}$, follows a large deviation law that can be obtained by Stirling's approximation to be:
\begin{equation}
s_\lambda(a)f^{[\lambda]} \sim e^{-nD(\bar{\lambda}||a)}
\label{eq:KeylWerner}
\end{equation}
where $D(\bar{\lambda}||a)$ is the relative entropy between the set of eigenvalues $a$ and the normalized partition $\bar{\lambda}=\lambda/n$: 
\begin{equation}
    D(\bar{\lambda}||a)= \sum_{i}  \overline{\lambda_i} \log \left(\frac{\overline{\lambda_i}}{a_i} \right),
    \label{eq:Relative}
\end{equation}
which is $0$ only when $a=\bar{\lambda}$. Then, the probability $p(\lambda|\rho^{\otimes n})$ will be asymptotically concentrated in the normalized partition $\bar{\lambda}$ that corresponds to the set of eigenvalues $a$. In this sense, the output of the $\lambda$ measurement, with high probability, will give an output $\bar{\lambda}=a$, allowing a single measure to determine $a$. \\

\subsection{Bipartite entanglement concentration}
\label{EntanglementHayashi}
In several quantum information applications, the protocols are designed to use LME states as inputs. However, preparing perfect LME states may be a challenging task. Because of this, it is useful to consider one protocol that takes many entangled states (but not necessarily LME) and concentrates the entanglement in some reduced quantity of LME states. This process is known as \textit{Entanglement Concentration}, and the first protocol was proposed by \cite{EntanglementConcentration}; however, we will present the Hayashi-Matsumoto protocol \cite{Hayashi-Matsumoto} which uses Schur-Weyl Duality and the Keyl-Werner theorem.\\

\noindent This protocol exploits the fact that when considering the tensor product of two irreps of $S_n$, the invariant subspace is one-dimensional. For a bipartite state, the total Hilbert space corresponds to $\mathcal{H}=\mathcal{H}^1\otimes \mathcal{H}^2$, when taking $n$ copies of the bipartite state, the copied total Hilbert space is:
\begin{equation*}
(\mathcal{H}^{\otimes n})^{S_n} =\left({\mathcal{H}^{1}} ^{\otimes n}  \otimes {\mathcal{H}^{2}} ^{\otimes n} \right) ^{S_n},
\end{equation*}
where $(\cdot)^{S_n}$ refers to the subspace invariant under the permutation of the copies. We can state Schur-Weyl duality for the copied Hilbert space of each part. According to Equation \eqref{eq:Schur-Weyl}, we have:
\begin{equation*}
\left({\mathcal{H}^{1}} ^{\otimes n}  \otimes {\mathcal{H}^{2}} ^{\otimes n} \right) ^{S_n} \cong  \left( \left(\bigoplus_{\lambda^1\vdash n,d} \{\lambda^1\}\otimes [\lambda^1]\right) \otimes\left(\bigoplus_{\lambda^2\vdash n,d} \{\lambda^2\}\otimes [\lambda^2]\right)\right) ^{S_n}.
\end{equation*}
We can group the corresponding parts of $GL_d$ and $S_n$, and noting that the permutation invariance only affects the $S_n$ irreps, we have: 
\begin{equation*}
\left({\mathcal{H}^{1}} ^{\otimes n}  \otimes {\mathcal{H}^{2}} ^{\otimes n} \right) ^{S_n} \cong  \bigoplus_{\lambda^1,\lambda^2 \vdash n,d} \left( \{\lambda^1\}\otimes \{\lambda^2\}\right) \otimes\left([\lambda^1] \otimes [\lambda^2]\right) ^{S_n} .
\end{equation*}
The invariant subspace $\left([\lambda^1]\otimes[\lambda^2]\right)^{S_n}$ refers to the subspace that acts trivially under $S_n$ i.e., copies of the trivial representation $[n]$ in the tensor product $[\lambda^1]\otimes[\lambda^2]$. We saw previously that the multiplicity of such copies is known as the Kronecker coefficient and can be calculated from the characters as:
\begin{equation*}
k_{\lambda^1\lambda^2 n} = \frac{1}{n!} \sum_{\pi \in S_n} \chi^{[\lambda^1]} (\pi) \chi^{[\lambda^2]} (\pi) \chi^{[n]} (\pi),
\end{equation*}
as all the characters are $1$ in the trivial representation we have:
\begin{equation*}
k_{\lambda^1\lambda^2 n} = \frac{1}{n!} \sum_{\pi \in S_n} \chi^{[\lambda^1]} (\pi) \chi^{[\lambda^2]} (\pi) = \delta_{\lambda^1 \lambda^2}.
\end{equation*}
In the last step, we used the orthogonality of the characters from equation \eqref{eq:CharOrtho}. Then, the invariant subspace only exists when $\lambda^1=\lambda^2=:\lambda$ and it is a one-dimensional subspace. Finally, the Schur-Weyl duality for this setup can be simplified to:
\begin{equation*}
\left({\mathcal{H}^{1}} ^{\otimes n}  \otimes {\mathcal{H}^{2}} ^{\otimes n} \right) ^{S_n} \cong  \bigoplus_{\lambda\vdash n,d} \left( \{\lambda\}\otimes \{\lambda\}\right) \otimes\left([\lambda] \otimes [\lambda]\right) ^{S_n} 
\end{equation*}
This simplification is only possible for bipartite systems, and it is used in Hayashi-Matsumoto protocol to obtain maximally entangled states from copies of entangled (but not maximally entangled) bipartite states in any dimension. We will outline the mathematical structure of this protocol for the two qubits case, where some of the techniques we introduced before appear and prove to be useful.\\

\noindent It was shown in equation \eqref{eq:twoqubitsSchmidt} that any entangled state of two qubits can be written in the normal form:
\begin{equation*}
    \ket{\psi}=\sqrt{p_0} \ket{00}+ \sqrt{1-p_0}\ket{11}.
\end{equation*}
It can be seen that the entanglement entropy of this state is given by:
\begin{equation*}
E(\ket{\psi})= -p_0 \log_2 p_0 - (1- p_0) \log_2 (1-p_0)=H_2(p_0),
\end{equation*}
where $H_2$ is the binary entropy function. By taking $n$ copies of the state, the coefficient expansion corresponds to:
\begin{equation}
    \ket{\psi}^{\otimes n}= \sum_{\omega=0}^{n} (p_0) ^{(n-\omega)/2} \left(1-p_0\right)^{\omega/2} \sum_{s \sim\omega}\ket{ s }\ket{s},
    \label{eq:phinexpansion}
\end{equation}
where $s$ are binary sequences of length $n$, and $s\sim\omega$ denotes all the binary sequences with $\omega$ ones. Now, let us focus on the Schur transform of the sequences. According to equation \eqref{eq:Schurtransformsequence}, we have:
\begin{equation*}
   \sum_{s \sim\omega} \ket{ s }\ket{s}=\left(\sum_{\lambda}\ket{\lambda,\omega_s}\sum_{q} \bm{\Gamma}^{\lambda,q}_{s}  \ket{\lambda,q} \right) \left(\sum_{\lambda'}\ket{\lambda',\omega_s}\sum_{q'} \bm{\Gamma}^{\lambda',q'}_{s} \ket{\lambda',q'}\right),
\end{equation*}
expanding and grouping the $GL_2$ parts and the $S_n$ parts of the two terms on the right hind side, we have:
\begin{equation}
   \sum_{s \sim\omega} \ket{ s }\ket{s}=\sum_{\lambda,\lambda'}\ket{\lambda,\omega}\ket{\lambda',\omega} \left(\sum_{q,q'}\bm{\Gamma}^{\lambda,q}_{s} \bm{\Gamma}^{\lambda',q'}_{s}   \ket{\lambda,q}\ket{\lambda',q'}\right),
   \label{eq:HMst2}
\end{equation}
Now, we know that as the sum on the left considers all the sequences with the same weight $\omega$, it is invariant under the simultaneous permutation of any of the $n$ elements in both sequences, i.e.,
\begin{equation*}
 \sum_{s \sim\omega} \ket{ s }\ket{s}=\pi \otimes \pi  \sum_{s \sim\omega} \ket{ s }\ket{s},
\end{equation*}
then, for the $S_n$ part of the Schur transform, we have:
\begin{equation}
\sum_{q,q'}\prod_{i,j=1}^{n}\bm{\Gamma}^{\lambda,q}_{s} \bm{\Gamma}^{\lambda',q'}_{s} \ket{\lambda',q'}=\sum_{q,q'}\bm{\Gamma}^{\lambda,q}_{s} \bm{\Gamma}^{\lambda',q'}_{s}  D^{\lambda} (\pi)\ket{\lambda,q} D^{\lambda'} (\pi)\ket{\lambda',q'}.
\label{eq:HMpt}
\end{equation}
It is possible to define a linear map known as \textit{partial transpose} \cite{Hayashi-Matsumoto} such that:
\begin{equation*}
t(\ket{\phi_1}\ket{\phi_2})=\ket{\phi_1}\bra{\phi_2}, \quad \ket{\phi_1}\ket{\phi_2}=t^{-1}(\ket{\phi_1}\bra{\phi_2}),
\end{equation*}
then we can rewrite RHS of Equation \eqref{eq:HMpt} as:
\begin{equation*}
 t^{-1} \left(D^{\lambda}(\pi) \left(\sum_{q,q'}\bm{\Gamma}^{\lambda,q}_{s} \bm{\Gamma}^{\lambda',q'}_{s} \ket{\lambda,q} \bra{\lambda',q'}\right)   D^{\lambda'}(\pi) ^{\dagger} \right).
\end{equation*}
Note that this is invariant under any $\pi\in S_n$ and as $\lambda$ is an irrep, we know from Schur's lemma that, first, $\lambda=\lambda'$, otherwise the corresponding summand is null, and when $\lambda=\lambda'$ the summand has to be proportional to the identity. Furthermore, it is not hard to show that the proportionality constant is $1$, then, one gets replacing in \eqref{eq:HMpt}:
\begin{equation*}
\left(\sum_{q,q'}\bm{\Gamma}^{\lambda,q}_{s} \bm{\Gamma}^{\lambda',q'}_{s}   \ket{\lambda,q}\ket{\lambda',q'}\right) = t^{-1}\left(\sum_{q} \ket{\lambda,q}\bra{\lambda,q} \right) \delta_{\lambda\lambda'} =\sum_{q} \ket{\lambda,q} \ket{\lambda,q} \delta_{\lambda\lambda'},
\end{equation*}
using this result in Equation \eqref{eq:HMst2} we finally obtain for the Schur transform:
\begin{equation*}
     \sum_{s \sim\omega} \ket{ s }\ket{s} =
\sum_{\lambda,q} \ket{\lambda,\omega,q}\ket{\lambda,\omega,q}.
\end{equation*}
From Equation \eqref{eq:phinexpansion}
we get:
\begin{equation*}
    \ket{\psi}^{\otimes n}= \sum_{\lambda} \left( \sum_{\omega=0}^{n} (p_0) ^{(n-\omega)/2} (1-p_0)^{\omega/2} \ket{\lambda,\omega}\ket{\lambda,\omega} \right) \left(\sum_{q} \ket{\lambda,q}\ket{\lambda,q}\right),
\end{equation*}
where $s_{\lambda}(p_0)$ is the Schur polynomial defined in Equation \eqref{eq:Schurpoly}. By normalizing both terms and including the restriction from $GL_2$, $\lambda\geq \omega \geq n-\lambda$, we get:
\begin{equation}
    \ket{\psi}^{\otimes n}=\sum_{\lambda} \sqrt{f^{[\lambda]}s_{\lambda}(p_0)}\left(\sum_{\omega=\lambda}^{n-\lambda} \frac{(p_0) ^{(n-\omega)/2} (1-p_0)^{\omega/2}  }{\sqrt{s_\lambda(p_0)}}\ket{\lambda,\omega}\ket{\lambda,\omega}  \right)\left(\frac{1}{\sqrt{f^{[\lambda]}}}\sum_{q} \ket{\lambda,q}\ket{\lambda,q} \right),
\label{eq:Hayashi}
\end{equation}
This transformation can be seen as
\begin{equation*}
    \ket{\psi}^{\otimes n}= \sum_{\lambda} \sqrt{f^{[\lambda]}s_{\lambda}(p_0)} \ket{\Phi_{\lambda\lambda}(p_0)} \ket{\mathcal{K}_{\lambda\lambda}},
\end{equation*}
where $\ket{\Phi_{\lambda\lambda}(p_0)}$ , $\ket{\mathcal{K}_{\lambda\lambda}}$ are normalized entangled states in $\{\lambda\}\otimes \{\lambda\}$ and $[\lambda]\otimes [\lambda]$ representations respectively, and $f^{[\lambda]}s_{\lambda}(p_0)$ corresponds to the probability of getting $\lambda$ as an outcome of a projective measurement in the irreps. The Hayashi-Matsumoto protocol consists on taking $n$ copies of any bipartite state and measuring the corresponding irrep on at least one of the parts. After this projective measurement, the total state ends up with a vector that is separable in the partition of $\{\lambda\}\otimes \{\lambda\}$ and $([\lambda]\otimes[\lambda])^{S_n}$: 
\begin{equation*}
    \ket{\psi}^{\otimes n} \xrightarrow[]{P_\lambda} \ket{\Phi_{\lambda\lambda}(p_0)} \ket{\mathcal{K}_{\lambda\lambda}}.
\end{equation*}
The part corresponding to the invariant subspace $([\lambda]\otimes[\lambda])^{S_n}$ is what we call a \textit{Kronecker state}, which is locally maximally entangled. Recalling that the individual reduced density matrices of Kronecker states are proportional to the identity, its entropy of entanglement, from Equation \eqref{eq:entanglemententropy}, is given by 
\begin{equation*}
E(\ket{\mathcal{K}_{\lambda\lambda}})=S(\frac{1}{f^{[\lambda]}} \bm{I}_f^{[\lambda]})=\log_{2} f^{\lambda}.
\end{equation*}
From Equation \eqref{eq:KeylWerner}, the probability is asymptotically concentrated in the partition $\lambda^*/n=p_0$. By Using \eqref{eq:asymptoticsflambda} we have:
\begin{equation*}
E(\ket{\mathcal{K}_{\lambda^* \lambda^*}})=log_2 f^{\lambda ^*} \sim log_2 (2^{nH(\lambda^*/n)})=n H(p_0),
\end{equation*}
showing how all the entanglement from the $n$ initial copies is concentrated asymptotically. \\

\noindent When considering a generalization of the Hayashi-Matsumoto protocol for systems with more particles, one quickly finds that this is not possible. When considering $n$ copies of a three-qubit state, for example, by applying the Schur transform in the $n$ copies of each part, the most general form is given by:
\begin{equation}\label{eq:Threequbits}
\ket{\psi}^{\otimes n}=\bigoplus_{ \lambda^1\lambda^2\lambda^3 } \sqrt{p(\lambda^1\lambda^2\lambda^3|\psi)}\left[ 
%\sqrt{p({\lambda^1\lambda^2\lambda^3}|\psi^{\otimes n})}
\sum\limits_{i=1}^{k_{\lambda^1\lambda^2\lambda^3}}\ket{\Phi_{\lambda^1\lambda^2\lambda^3,i}(\psi)}\otimes\ket{\mathcal{K}_{\lambda^1\lambda^2\lambda^3,i}} \right],
\end{equation}
where $\ket{\Phi_{\lambda^1\lambda^2\lambda^3,i}(\psi)}$ is some normalized state in $V_{\lambda^1}\otimes V_{\lambda^2} \otimes V_{\lambda^3}$, which depends on the parameters of the input state $\ket{\psi}$, $\ket{\mathcal{K}_{\lambda^1\lambda^2\lambda^3,i}}$ is a normalized Kronecker state asociated to the $i$-th multiplicity, and $p(\lambda^1\lambda^2\lambda^3|\psi)$ is the probability of projecting in a particular set $\lambda^1\lambda^2\lambda^3$. After a projective measurement in any set $\lambda^1\lambda^2\lambda^3$, the resultant state is generally not separable between the $GL_2$ and $S_n$ parts, and hence it is not possible to extract the entanglement from the Kronecker states $\ket{\mathcal{K}_{\lambda^1\lambda^2\lambda^3,i}}$. However, we will see that for states in the multipartite W-class, the generalization is possible, and makes possible not only the generalization of Hayashi-Matsumoto protocol to multiparticle systems, but also gives explicit expressions for calculating Kronecker states for more than two qubits.\\

\noindent These applications show not only the utility of considering Schur-Weyl duality in permutational invariant setups but also how Kronecker states naturally emerge as the bearers of entanglement when examining the asymptotic behavior of such systems. In the next chapter, we will formally introduce Kronecker states and outline various approaches to their explicit calculation while addressing the primary challenges associated with these methods. Furthermore, we will introduce the generalized entanglement concentration protocol that allows for the exact computation of Kronecker states within the W-SLOCC class. This protocol will be pivotal for the results we present in the final chapter.
\chapter{Kronecker states}
\label{Chapter3}
In the introduction, we emphasized the significance of multiparticle maximally entangled states in high dimensions as they enable the implementation and improvement of several quantum information protocols. Due to the increasing complexity of quantum systems, it is essential to have a better understanding of the entanglement structure of many particles. To address this issue, we propose studying the maximally entangled subspace corresponding to the permutationally invariant subspace of the tensor product of irreducible representations of $S_n$. We refer to this subspace as the Kronecker Subspace, as the Kronecker coefficient determines its dimension, and the states belonging to this subspace are called Kronecker states. The primary objectives of this research are to comprehend the mathematical structure of Kronecker states and to develop a tool for constructing the Kronecker subspace. \\

\noindent In this chapter, we formally define Kronecker states and highlight their properties, which make them special LME states, and justify why we focus on obtaining them. Later, we explore a naive approach to calculating them, showing how, in essence, calculating Kronecker states is a complex problem. We explore the existing connection between Kronecker states and Clebsch Gordan Coefficients (CGC) of the symmetric group and discuss some of the known methods to calculate the CGC. We also show that the problem of calculating Kronecker states is equivalent to the problem of finding the CGC of the symmetric group $S_n$. We finish this chapter by introducing a method to calculate a specific kind of Kronecker states that appear when generalizing the entanglement concentration protocol presented in Section \ref{EntanglementHayashi} to multipartite states in the W-class. This method ends up with an analytical expression that is easy to follow for calculating this kind of Kronecker states, which we name \textit{W-Kronecker states}. However, this approach has limitations as it only allows us to calculate a small part of the total Kronecker subspace. These limitations are overcome with the method we will introduce in Chapter \ref{Chapter5}, where we build general multiqubit states using only W-states and bipartite states. From this construction general Kronecker states can be obtained from \textit{W-Kronecker states} that can be calculated explicitly.
\section{Definition}
As we saw in the last chapter, the Schur-Weyl duality states that $n$ tensor copies of a $d-$dimensional Hilbert space  $\mathcal{H}\simeq \mathbb{C}^d$ can be decomposed into a direct sum of coordinated irreducible representations (irreps) of the General Linear group $GL_d$ and the Symmetric group $S_n$,
\begin{equation}
\mathcal{H}^{\otimes n}\cong\bigoplus_{\lambda\vdash  n,d} \{\lambda\} \otimes [\lambda],
\label{eq:isotypic}
\end{equation}
where $\{\lambda\}$ are irreps of $GL_d$, $[\lambda]$  are irreps of $S_n$, and $\lambda\vdash n,d$ restricts the irreps to partitions of $n$ with length at most $d$. Equation \eqref{eq:isotypic} is known as the \textit{isotypic decomposition} \cite{Bacon}. It is worth noting, that the dimension of $\{\lambda\}$, the number of SSYT is given by the Schur polynomial (or equivalently the Weyl formula \cite{Fulton} evaluated at $1^{\otimes n}$) and grows polynomially in $n$ , while the dimension of $[\lambda]$ is given by the \textit{hook formula} in Equation \eqref{eq:hook}, and grows exponentially with $n$ as $ \exp(n H(\bar{\lambda}))$\cite{christandl_spectra_2006}, with $H$ the Shannon entropy, and $\bar{\lambda}$ the reduced partition $\bar{\lambda}=\lambda/n$. \\

\noindent Now, consider $n$ copies of a state with $N$ parts, whose Hilbert space corresponds to $\bm{\mathcal{H}}=\otimes_{i=1}^{N} \mathcal{H}^i$, where $\mathcal{H}^i$ labels the Hilbert space associated to each particle. Then, we can apply Schur-Weyl duality locally to the copies of each particle Hilbert space:
\begin{equation*}
\bm{\mathcal{H}} ^{\otimes n}= {\mathcal{H}^{1}}^{\otimes n} \otimes  \dots \otimes {\mathcal{H}^{N}}^{\otimes n} = \left(\bigoplus_{\lambda^1} \{\lambda^1\} \otimes [\lambda ^1] \right) \otimes \dots \otimes \left(\bigoplus_{\lambda^N} \{\lambda^N\} \otimes [\lambda ^N] \right),
\end{equation*} 
we can group this multilocal isotypic decomposition as:
\begin{equation*}
\bm{\mathcal{H}}^{\otimes n}\cong \bigoplus_{\lamtup} \{\lamtup\} \otimes [\lamtup],
\end{equation*}
where we extended the boldface notation to partitions and irreps as: $\lamtup=(\lambda^{1},\dots,\lambda^{N})$, $\{\bm{\lambda}\}=\bigotimes_{i} \{\lambda^i\}$ , $[\lamtup]=\bigotimes_{i} [\lambda^{i}]$, and all $\lambda^{i}\vdash n,d$.  When considering  permutationally invariant systems of $n$ copies, i.e., systems that are invariant under the reordering of the $n$ copies, we see that they lie in the symmetric part of the total Hilbert space $ \left( \bm{\mathcal{H}}
 ^{\otimes n} \right)^{S_n}$, where $(\cdot ) ^{S_n}$ represents invariance under $S_n$. Therefore the multilocal isotypic decomposition  reads
\begin{equation*}
    \left( \bm{\mathcal{H}} ^{\otimes n}\right)^{S_n} \cong \bigoplus_{\lamtup} \{\lamtup\} \otimes [\lamtup]^{S_n},
\end{equation*}
where $[\lamtup]^{S_n}$ is the invariant subspace of $[\lamtup]$ under the diagonal action of $S_n$, i.e., a subspace where simultaneous actions of $S_n$ in its $N$ parts act trivially. We can find the dimension of such invariant subspace by calculating how many copies of the trivial representation $[n]$ appear in the decomposition of the tensor product $[\bm{\lambda}]$. We know from Equation \eqref{eq:multiplicity} that this multiplicity can be obtained from the characters as:
\begin{equation*}
\dim([\bm{\lambda}]^{S_n})= \frac{1}{|n!|} \sum_{\pi \in S_n} \chi^{[\lambda^1]}(\pi) \chi^{[\lambda^1]}(\pi)\chi^{[\lambda^2]}(\pi) \dots \chi^{[\lambda^N]}(\pi)\cdot \chi ^{[n]}(\pi),
\end{equation*}
as $\chi ^{[n]}(\pi)$ is always $1$, then this dimension corresponds to a \textit{Generalized Kronecker coefficient}, as it was defined for the case of three irreps in Equation \eqref{eq:KronCoef}, i.e.:
\begin{equation}
k_{\bm{\lambda}}=\dim([\bm{\lambda}]^{S_n})= \frac{1}{|n!|} \sum_{\pi \in S_n} \chi^{[\lambda^1]}(\pi) \chi^{[\lambda^1]}(\pi)\chi^{[\lambda^2]}(\pi) \dots \chi^{[\lambda^N]}(\pi).
\label{eq:GenKron}
\end{equation}
This value can also be understood as the multiplicity of any of the irreps $[\lambda^i]$ in the decomposition of the tensor product of the complementary irreps, $[\bm{\lambda}^{\bar{i}}]$, in the sense that we defined the Kronecker coefficient in Equation \eqref{eq:KronCoef}. As the dimension of the invariant subspace is given by the Kronecker coefficient, we name the states in the invariant subspace $[\bm{\lambda}]^{S_n}$,  \textit{Kronecker states}, 
\begin{equation*}
    \ket{\mathcal{K}_{\lamtup}} \in [\lamtup]^{S_n}.
\end{equation*}
Kronecker states are the main focus of this research. They are maximally entangled states when considering the bipartition of one part against the others (as we will prove soon), also known as Locally Maximally Entangled (LME) states. We also name the invariant subspace $[\bm{\lambda}]
^{S_n}$ as the \textit{Kronecker subspace}. It is a very special vector space. In general, the Kronecker coefficient is greater than one, meaning that Kronecker subspace can be spanned by $k_{\bm{\lambda}}>1$ orthogonal Kronecker states. Furthermore, being elements of a vector space, they have the property that any complex linear combination of $K$ Kronecker states is another Kronecker state:
\begin{equation*}
c_1\ket{\mathcal{K}_{\bm{\lambda},1}}+c_2\ket	{\mathcal{K}_{\bm{\lambda},2}}+ \dots c_K \ket{\mathcal{K}_{\bm{\lambda},K}} \in [\bm{\lambda}]^{S_n}, \quad \forall c_i \in \mathbb{C}, \ket	{\mathcal{K}_{\bm{\lambda},i}} \in [\bm{\lambda}]^{S_n}.
\end{equation*}  
Another important property is that by definition, these states are invariant under the diagonal action of $S_n$, i.e.:
\begin{equation*}
D^{[\lambda^{1}]}(\pi) \otimes\dots \otimes D^{[\lambda^{N}]}(\pi)\ket{\mathcal{K}_{\lamtup}}=\ket{\mathcal{K}_{\lamtup}} ,  \quad \forall \pi \in S_n,
\end{equation*}
with $D^{\lambda}(\pi)$ the matrix representation of the element $\pi$ in the irrep $\lambda$. This property is important on its own, as there is a set of operations (in fact, a group) that leave Kronecker states invariant; this makes Kronecker states good candidates to perform quantum information tasks. For example, \textit{Werner states} \cite{Wernerstates}, which are invariant under the simultaneous action of $U_d$ in each of the parts, have been used for quantum information applications such as quantum data hiding\cite{Hiding}, and entanglement teleportation \cite{Entanglementteleportation}, and have also been used to study fundamental quantum problems such as nonadditivity of entanglement \cite{nonadditivity}. However, exploring applications of Kronecker states goes beyond the scope of this research. \\

\noindent The invariance described before can also be used to show that Kronecker states are LME states. Note that the reduced density matrices in each part, $\rho_i(\mathcal{K}_{\lamtup})=tr_{\overline{\imath}} (\ket{\mathcal{K}_{\lamtup}}\bra{\mathcal{K}_{\lamtup}})$, i.e., the partial trace of the density matrix in all the parts but $i$, remain invariant under conjugacy with elements of $S_n$:
\begin{equation*}
 D^{[\lambda^{i}]}(\pi)  \rho_i(\mathcal{K}_{\lamtup}){D^{[\lambda^{i}]}}^{\dagger} (\pi)=\rho_i(\mathcal{K}_{\lamtup}).
\end{equation*}
As $[\lambda^{i}]$ is an irrep of $S_n$ by Schur's lemma, the reduced density matrix is proportional to the identity in the irrep $[\lambda^i]$:
\begin{equation*}
\rho_i(\mathcal{K}_{\lamtup}) =c \bm{I}_{f^{[\lambda^{i}]}},
\end{equation*}
with $f^{[\lambda^{i}]}$ the dimension of irrep $[\lambda^{i}]$. As the density matrix has trace equal to one, and the trace of the identity is the dimension of the irrep, the proportionality constant can be found by tracing in both sides to be $\frac{1}{f^{[\lambda]}}$, then we get
\begin{equation*}
\rho_i(\mathcal{K}_{\lamtup}) =\frac{1}{f^{[\lambda^{i}]}} \bm{I}_{f^{[\lambda^{i}]}}.
\label{eq:reducedkron}
\end{equation*}
As the reduced density matrix is proportional to the identity in all of the irreps, Kronecker states are maximally entangled states when considering the bipartition of one part against the other as defined in Equation \eqref{eq:LME}. LME states are an important resource in quantum information protocols such as quantum teleportation \cite{Teleportation}, quantum key distribution \cite{QKD}, or quantum error correction\cite{QEC}.\\

\noindent Despite the promising properties of Kronecker states, studying their possible applications in quantum information is very difficult without a clear understanding of their mathematical structure. The main purpose of this document is to introduce a method to calculate explicitly any multipartite Kronecker state for sets of partitions $\bm{\lambda}$ where each irrep is at most of two parts ($d=2$), which we will see soon, is a highly non-trivial task.

\section{Obtaining Kronecker states}
In this section, we will show an algorithm to calculate Kronecker states to understand how complex this process can be. We start with the simplest cases and explore some examples for more complicated ones. Let us first consider systems of two parts; for this case, the \textit{Kronecker coefficient} is given by
\begin{equation}
    k_{\lambda^1\lambda^2}= \frac{1}{n!} \sum_{\pi \in S_n} \chi ^{[\lambda^1]}(\pi) \chi ^{[\lambda^2]}(\pi) = \delta_{\lambda^{1}\lambda^{2}},
    \label{eq:KronCoef2}
\end{equation}
following from the orthonormality of characters seen in Equation \eqref{eq:CharOrtho}. Then, the Kronecker subspace only exists for $\lambda^1=\lambda^2=:\lambda$, and it is one-dimensional, i.e., there is only one Kronecker state. To find this state, we expand it in its coefficients as:
\begin{equation*}    \ket{\mathcal{K}_{\lambda\lambda}}= \sum_{i,j} K_{i,j}\ket{\lambda,q_i}\ket{\lambda,q_j}.
\end{equation*}
By definition, this state is invariant under any permutation $\pi$ acting simultaneously in each part, that is:
\begin{equation*}
\ket{\mathcal{K}_{\lambda\lambda}}= D^{[\lambda]} (\pi) \otimes D^{[\lambda]}(\pi) \ket{\mathcal{K}_{\lambda,\lambda}}= \sum_{i,j} K_{i,j}D^{[\lambda]} (\pi) \otimes D^{[\lambda]}(\pi) \ket{\lambda,q_i}\ket{\lambda,q_j}.
\end{equation*}
Summing over all permutations in both sides we get:
\begin{equation*}
n! \ket{\mathcal{K}_{\lambda\lambda}}=\sum_{i,j} K_{i,j} \sum_{\pi} \left(D^{[\lambda]} (\pi) \otimes D^{[\lambda]}(\pi)  \right) \ket{\lambda,q_i}\ket{\lambda,q_j}.
\end{equation*}
For the tensor product, we can apply Equation \eqref{eq:orthoirreds}, which also shows how this state only exists for $\lambda^1=\lambda^2$. By doing this, we get:
\begin{equation*}
n! \ket{\mathcal{K}_{\lambda,\lambda}}=n!\sum_{i,j} K_{i,j} \frac{1}{f^{[\lambda]}} \sum_{kl} \delta_{l,i}\delta_{l,j}\ket{\lambda,q_k}\ket{\lambda,q_k},
\end{equation*}
which leads to the following expression:
\begin{equation*}
\ket{\mathcal{K}_{\lambda,\lambda}}=  \frac{\sum_l K_{l,l}}{f^{[\lambda]}} \sum_{k}  \ket{\lambda,q_k}\ket{\lambda,q_k},
\end{equation*}
as $\sum_{l} K_{l,l}$ is a constant, and we are looking for a normalized state, the only Kronecker state is then:
\begin{equation}
\ket{\mathcal{K}_{\lambda,\lambda}}=  \frac{1}{\sqrt{f^{[\lambda]}}} \sum_{i}  \ket{\lambda,q_i}\ket{\lambda,q_i} .
\label{eq:2partKronecker}
\end{equation}
This expression is the canonical form of any Kronecker state of two parts. Note how we already calculated them in Equation \eqref{eq:Hayashi} for the Hayashi-Matsumoto protocol as the carriers of entanglement in the asymptotic case.\\

\noindent The problem of calculating Kronecker states becomes significantly harder for three or more parts. As there is no general expression for Kronecker states in different sets $\bm{\lambda}=\lambda^1\lambda^2\lambda^3$ of irreps, Kronecker states must be built for each possible set $\bm{\lambda}$. One possible approach is to use the \textit{projector} onto the invariant subspace to obtain a set of Kronecker states. \\

\noindent For any representation $X$ of a group $G$, it is possible to build the projector $P_X^{\lambda}$ onto the carrier space of any irreducible representation $\lambda$ as \cite{chenbook}:
\begin{equation}
P_X^{\lambda} = \frac{\dim(\lambda)}{|G|} \sum_{g \in G} \chi^{\lambda} (g) X(g).
\label{eq:Projector}
\end{equation} 
For our purposes, we want to build the projector onto the invariant subspace $[n]$ of the tensor product representation $[\lambda^1]\otimes[\lambda^2]\otimes[\lambda^3]$. Following Equation \eqref{eq:Projector}, we get:
\begin{equation}
\mathcal{P}_{[\bm{\lambda}]} ^{[n]}= \frac{1}{n!} \sum_{\pi \in S_n}  D^{[\lambda^1]}(\pi) \otimes D^{[\lambda^2]}(\pi) \otimes D^{[\lambda^3]}(\pi) ,
\label{eq:ProjSn}
\end{equation}
where we have used the fact, as pointed out before, that $\chi^{[n]}(g) =1$. With this, a possible way to construct Kronecker states is to calculate such a projector and apply it to any state in $[\bm{\lambda}]$. This process ensures that the resultant state belongs to the invariant subspace, and hence, it is a Kronecker state:
\begin{equation*}
P_{[\bm{\lambda}]} ^{[n]} \ket{\psi_{\bm{\lambda}}} \in [\bm{\lambda}]^{S_n} , \quad \forall \ket{\psi_{\bm{\lambda}}}\in [\bm{\lambda}].
\end{equation*}
Let's use this method for the simplest case with $n=3$ and $[\lambda^1]=[\lambda^2]=[\lambda^3]=[21]$. Then, we need all the representation matrices for $[21]$, which can be obtained by the Young Yamanouchi algorithm, or, from the procedure described to build matrix representations proposed in Section \ref{myirreps}. Using the shorthand notation where we label each irrep only by the second element of the partition, $[\lambda]=[\lambda_2]$, the six matrices are:
\begin{equation*}
\begin{gathered}
    D^{[1]}(e)= \left(\begin{array}{cc}
     1    & 0 \\
      0   & 1
    \end{array} \right), \quad  D^{[1]}(12)= \left(\begin{array}{cc}
     1    & 0 \\
      0   & -1
    \end{array} \right), \quad  D^{[1]}(23)= \left(\begin{array}{cc}
     -\frac{1}{2}    & \frac{\sqrt{3}}{2} \\
     \frac{\sqrt{3}}{2}   & \frac{1}{2}
    \end{array} \right) \\
      D^{[1]}(13)= \left(\begin{array}{cc}
     -\frac{1}{2}    & -\frac{\sqrt{3}}{2} \\
    - \frac{\sqrt{3}}{2}   & \frac{1}{2}
    \end{array} \right), \quad  D^{[1]}(123)= \left(\begin{array}{cc}
     -\frac{1}{2}    & \frac{\sqrt{3}}{2} \\
     \frac{\sqrt{3}}{2}   & -\frac{1}{2}
    \end{array} \right), \quad  D^{[1]}(231)= \left(\begin{array}{cc}
     -\frac{1}{2}    & -\frac{\sqrt{3}}{2} \\
    - \frac{\sqrt{3}}{2}   & \frac{1}{2}
    \end{array} \right).
    \end{gathered}
\end{equation*}
where the two basis elements are the Yamanouchi symbols $q^{[1]}_{1}=\{001\},q^{[1]}_{2}=\{010\}$. With these matrices, the projector can be constructed, yielding:
\begin{equation}    \mathcal{P}_{[1]\otimes[1]\otimes[1]}^{[3]}= \frac{1}{3!} \sum_{\pi \in S_3} D^{[1]}(\pi) \otimes  D^{[1]}(\pi) \otimes  D^{[1]}(\pi) = \left( \begin{array}{cccccccc}
    \frac{1}{4} &0 &0 & -\frac{1}{4} &0 & -\frac{1}{4} &- \frac{1}{4} &0 \\0&0&0&0&0&0&0&0\\0 &0&0&0&0&0&0&0\\ -\frac{1}{4} & 0&0 &\frac{1}{4} & 0& \frac{1}{4} & \frac{1}{4}&0   \\ 0&0&0&0&0&0&0&0 \\ -\frac{1}{4} & 0&0 &\frac{1}{4} & 0& \frac{1}{4} & \frac{1}{4}&0 \\-\frac{1}{4} & 0&0 &\frac{1}{4} & 0& \frac{1}{4} & \frac{1}{4}&0 \\0&0&0&0&0&0&0&0 \end{array}\right).
    \label{eq:Projex}
\end{equation}
 Note from Equations \eqref{eq:GenKron} and \eqref{eq:ProjSn}, that the Kronecker coefficient can also be calculated as:
\begin{equation*}
 k_{\bm{\lambda}}= \tr(\mathcal{P}_{[\bm{\lambda}]} ^{[n]});
\end{equation*} 
then, for the set $\bm{\lambda}=(1,1,1)$ in $n=3$, the Kronecker coefficient can be obtained to be $k_{(1,1,1)}=\operatorname{tr} \left(\mathcal{P}_{[1]\otimes[1]\otimes[1]}^{[0]} \right)=1$. There is only one Kronecker state in the invariant subspace $([1]\otimes[1]\otimes [1])^{S_3}$, which can be obtained by applying the projector to any state whose projection is not null. For example, one possible state corresponds to the first element of the product basis labeled as:
\begin{equation*}
\ket{\bm{\lambda},\bm{}q^{\bm{\lambda}}}=\ket{111,q^{[1]}_1 q^{[1]}_1q^{[1]}_1},
\end{equation*}
which corresponds to the first column of $\mathcal{P}_{[1]\otimes[1]\otimes[1]}^{[3]}$ in Equation \eqref{eq:Projex}
. When acting on it with the projector and normalizing the result, one gets the Kronecker state:
\begin{equation*}
    \ket{\mathcal{K}_{111}}=\frac{1}{2} \left( \ket{111,q^{[1]}_{1}q^{[1]}_{1}q^{[1]}_{1}} -\ket{111,q^{[1]}_{1}q^{[1]}_{2}q^{[1]}_{2}}-\ket{111,q^{[1]}_{2}q^{[1]}_{1}q^{[1]}_{2}}-\ket{111,q^{[1]}_{2}q^{[1]}_{2}q^{[1]}_{1}} \right).
\end{equation*}
It can be noticed that any other choice of state to project on will to $\ket{\mathcal{K}_{111}}$ up to a constant, which correspond to the same state. Some remarkable results arise from this simple example. First, note that the irrep $[21]$ is of dimension $2$, so we can consider this state a LME state of 3 qubits. A clearer visualization is obtained when dropping the $\bm{\lambda}$ label, and making the change of notation $q_1^{[1]}\rightarrow 0, q^{[1]}_2\rightarrow 1$ in the kets. The obtained Kronecker state corresponds to:
\begin{equation}
    \ket{\mathcal{K}_{111}}= \frac{1}{2} \left(\ket{000}- \ket{011}-\ket{101} -\ket{110} \right).
    \label{eq:Kron3ex}
\end{equation}
It is easy to check that the reduced density matrices in each part are proportional to the identity:
\begin{equation*}
    \rho_{1}=\rho_{2}=\rho_{3} = \frac{1}{2} \left(\ket{0}\bra{0}+\ket{1}\bra{1} \right) = \frac{\bm{I}_2}{2},
\end{equation*}
in correspondence with \eqref{eq:reducedkron}, showing again that this is a LME state. Now, recalling from Section \ref{Entanglementclassification}, we know that the Hilbert space of three qubits contains only one stable orbit corresponding to the GHZ-class, i.e., all critical states (or LME states) are LU-equivalent to the GHZ state. The equivalence can be shown in this case by performing a multi-local rotation given by
\begin{equation*}
    \ket{0'}=\frac{\ket{0}+ \ket{1}}{\sqrt{2}}, \quad \ket{1'}=\frac{\ket{0}-i \ket{1}}{\sqrt{2}}.
\end{equation*}
Then, the obtained Kronecker state can be rewritten as:
\begin{equation*}
    \ket{\mathcal{K}'_{111}}= \frac{1}{\sqrt{2}} \left(\ket{0'0'0'}+ \ket{1'1'1'}\right).
\end{equation*}
This result is expected; it is not possible to obtain an LU inequivalent state to the GHZ state within the tensor product of three Hilbert spaces of dimension two.\\

\noindent This approach for calculating Kronecker states is easy to follow and to replicate for other values of $n$ and $\bm{\lambda}$; however, the complexity of this approach relies on building of the projector $\mathcal{P}_{[\bm{\lambda}]}^{[n]}$, a squared matrix obtained by adding $n!$ matrices with dimension $\left(f^{[\lambda^1]} \times f^{[\lambda^2]} \times f^{[\lambda^3]} \right)^2$, where each irrep dimension $f^{[\lambda]}$ grows exponentially with $n$ as shown in Equation \eqref{eq:asymptoticsflambda}. This approach becomes unfeasible very fast; for example, the first case where the Kronecker coefficient for three irreps is greater than one with irreps of $d=2$ is in $n=6$ with $[\lambda^1]=[\lambda^2]=[\lambda^3 ]=[4,2]$ (or simply $[2]$ using the short notation). In this case, the dimensions of the irreps are all $f^{[2]}=9$, so we need to calculate and add $6!=120$ square matrices of dimension $729\times 729$, and then look for two states in $[2]\otimes[2]\otimes[2]$ that project onto two different Kronecker states, which can be orthogonalized to obtain a basis for $([2]\otimes[2]\otimes[2])^{S_6}$. This method becomes a major task even for small values of $n$. 

\section{Kronecker states and Clebsch-Gordan coefficients}
\label{CGC}
Obtaining closed expressions for the basis of the Kronecker subspaces is a highly non-trivial task. For the goal of obtaining meaningful Kronecker subspaces, i.e., Kronecker subspaces of many particles with high dimensions and high Kronecker coefficient, the projector approach cannot be considered. When looking for other options to calculate Kronecker states from representation theory, it can be seen that Kronecker states correspond with the Clebsch-Gordan coefficients (CGC) of the symmetric group up to a factor.\\

\noindent We are interested in calculate states in the invariant subspace of the tensor product of three irreps, $\left( [\lambda^1]\otimes [\lambda^2]\otimes [\lambda^3] \right)^{S_n}$. For analyzing this subspace we can start by coupling first the irreps $[\lambda^1]$ and $[\lambda^2]$, and decompose the tensor product into intermediate irreps $[\lambda]$ as shown in Equation \eqref{eq:DiagSn}:
\begin{equation*}
    ([\lambda^1]\otimes[\lambda^2]\otimes [\lambda^3]) ^{S_n} =\left( \left(\bigoplus_{\lambda \vdash n} [\lambda]_{[\lambda^1]\otimes[\lambda^2]} \otimes I_{k_{\lambda^1\lambda^2\lambda}} \right)\otimes [\lambda^3] \right)^{S_n} ,
\end{equation*}
where $[\lambda]_{[\lambda^1]\otimes[\lambda^2]} $ recalls that these copies of $[\lambda]$ are the ones appearing in the decomposition of $[\lambda^1]\otimes[\lambda^2]$. Then we can repeat the process by now decomposing the tensor product of $[\lambda^3]$ with each of the copies of $[\lambda]$,
\begin{equation}
 ([\lambda^1]\otimes[\lambda^2]\otimes [\lambda^3]) ^{S_n}  = \bigoplus_{\lambda \vdash n} \left( [\lambda]_{[\lambda^1]\otimes[\lambda^2]}  \otimes [\lambda^3] \right)^{S_n}  \otimes I_{k_{\lambda^1 \lambda^2 \lambda}}.
\end{equation} 
Then, from Equation $\eqref{eq:KronCoef2}$ we know that the invariant subspace from the tensor product of two irreps of $S_n$ only exists when both irreps are the same, and is multiplicity free. So, $\left( [\lambda]_{[\lambda^1]\otimes[\lambda^2]}  \otimes [\lambda^3] \right)^{S_n} =\left( [\lambda^3]_{[\lambda^1]\otimes[\lambda^2]}  \otimes [\lambda^3] \right)^{S_n} \delta_{\lambda\lambda^3}$. Substituting in the previous equation, we have:
\begin{equation*}
    ([\lambda^1]\otimes[\lambda^2]\otimes [\lambda^3]) ^{S_n} = \left( [\lambda^3]_{[\lambda^1]\otimes[\lambda^2]}  \otimes [\lambda^3] \right)^{S_n} \otimes I_{k_{\lambda^1\lambda^2\lambda^3}} .
\end{equation*}
Then, the problem can be thought of as looking for $k_{\lambda^1\lambda^2\lambda^3}$ copies of two parts Kronecker states. We also know that, for bipartite systems, Kronecker states have the canonical form shown in Equation \eqref{eq:2partKronecker}, then, this invariant subspace can be spanned by $k_{\lambda^1\lambda^2\lambda^3}$ Kronecker states, each built from a copy of $[\lambda^3]$ in the decomposition of $[\lambda^1]\otimes [\lambda^2]$, where each Kronecker state can be written as:
\begin{equation}    \ket{\mathcal{K}_{\lambda^1\lambda^2\lambda^3,i}} = \frac{1}{\sqrt{f^{[\lambda^3]}} }\sum_{q^{[\lambda^3]}}\ket{\lambda^3,q^{[\lambda^3]},i}_{\lambda^1\lambda^2} \ket{\lambda^3,q^{[\lambda^3]}},
\label{eq:414}
\end{equation}
where $q^{[\lambda^3]}$ labels the basis elements of irrep $[\lambda^3]$, and $\ket{\lambda^3,q^{[\lambda^3]},i}_{\lambda^1\lambda^2}$ refers to the basis elements of the $i$-th copy of $\lambda^3$ in the decomposition of $[\lambda^1]\otimes [\lambda^2]$. However, we want to expand the Kronecker state in the product basis $\ket{\lambda^1\lambda^2\lambda^3,q^{[\lambda^1]}q^{[\lambda^2]}q^{[\lambda^3]}}$, and for doing that, we need to relate the basis elements of $\ket{\lambda^3,q^{[\lambda^3]},i}_{\lambda^1\lambda^2}$ with the product basis where it came from, $\ket{\lambda^1\lambda^2,q^{[\lambda^1]}q^{[\lambda^2]}}$. As described previously in Equation \eqref{eq:CGC} the relation of those basis is given by the CGC, in this case, of the Symmetric group as:
\begin{equation*}
\ket{\lambda^{3},q^{[\lambda^3]},i}=\sum_{q^{[\lambda^1]}q^{[\lambda^2]}} C^{[\lambda^1][\lambda^2][\lambda^{3}],i}_{q^{[\lambda^1]},q^{[\lambda^2]},q^{[\lambda^3]}} \ket{\lambda^1,q^{[\lambda^1]}} \ket{\lambda^2,q^{[\lambda^2]}}.
\end{equation*}
By plugging this in Equation \eqref{eq:414} we get that Kronecker states can be calculated from CGC of $S_n$ as:
\begin{equation*}    \ket{\mathcal{K}_{\lambda^1\lambda^2\lambda^3,i}} = \frac{1}{\sqrt{f^{[\lambda^3]}} }\sum_{q^{[\lambda^1]}q^{[\lambda^2]}q^{[\lambda^3]}}  C^{[\lambda^1][\lambda^2][\lambda^{3}],i}_{q^{[\lambda^1]},q^{[\lambda^2]},q^{[\lambda^3]}} \ket{\lambda^1\lambda^2\lambda^3,q^{[\lambda^1]}q^{[\lambda^2]}q^{[\lambda^3]}}.
\end{equation*}
We can also consider the coefficient expansion of Kronecker states as represented by:
\begin{equation*}    \ket{\mathcal{K}_{\lambda^1\lambda^2\lambda^3,i}} = \sum_{q^{[\lambda^1]}q^{[\lambda^2]}q^{[\lambda^3]}}  K_{\lambda^1\lambda^2\lambda^{3},q^{[\lambda^1]}q^{[\lambda^2]}q^{[\lambda^3]},i} \ket{\lambda^1\lambda^2\lambda^3,q^{[\lambda^1]}q^{[\lambda^2]}q^{[\lambda^3]}}.
\end{equation*}
Then, the relation between coefficients of Kronecker states and CGC of $S_n$ is given by:
\begin{equation}
K_{\lambda^1\lambda^2\lambda^{3},q^{[\lambda^1]}q^{[\lambda^2]}q^{[\lambda^3]},i}= \frac{1}{\sqrt{f^{[\lambda^3]}}} C^{[\lambda^1][\lambda^2][\lambda^{3}],i}_{q^{[\lambda^1]},q^{[\lambda^2]},q^{[\lambda^3]}}. 
\label{eq:CGCKron}
\end{equation}
In other words, finding the coefficient expansion of Kronecker states in the product basis of the irreps is equivalent to the problem of finding the CGC of the symmetric group corresponding to the same set of irreps. This remarkable connection allows some progress when trying to calculate Kronecker states. For example, in \cite{CGCMucha}, expressions for specific sets $\bm{\lambda}$ with $k_{\bm{\lambda}}=1$ are obtained; however, there is no general expression for cases with multiplicity, i.e., $k_{\bm{\lambda}}>1$. In \cite{chenbook}, the authors introduce the \textit{Eigenfunction method} to analyze irreps of $S_n$. Through building \textit{class operators}, which correspond to the sum of all the operators in the same conjugacy class, they show that by a proper choice of linear combinations of class operators, it is possible to build a matrix $M$. Then, they show that CGC can be obtained as coefficients of eigenvectors of $M$ in the product basis of two irreps, namely $[\lambda^1]\otimes [\lambda^2]$. Compared with the projector method, this one is more efficient because the matrix to be built being of dimension $(f^{[\lambda^1]}\otimes f^{[\lambda^2]})^2$. However, the complexity again grows when looking for the eigenvectors of such matrix. Another problem of this approach is that, as the authors point out in \cite{gao}, the separation of the different copies of $[\lambda^3]$ in $[\lambda^1]\otimes[\lambda^2]$ is arbitrary and can only be partially solved by imposing symmetries in the construction, which becomes harder as the multiplicity of the copies grows. In \cite{Isoscalar}, the authors use the \textit{isoscalar factors} (ISF), which relate representations of $S_n$ with those of $S_{n-1}$, to find expressions for CGC depending on ISF. They explore this construction and obtain expressions for calculating ISF and hence CGC for certain sets $\bm{\lambda}$ with specific relations between the $\lambda^i$. They focus on cases with no multiplicity and state that the method can be used even with multiplicity; however, there is no clear interpretation of the multiplicity separation. A more recent advance in this direction is shown in \cite{Doma}, where the authors enlarge the list of sets allowed by the ISF method and introduce a list of symmetries in CGC. However, they also focus on cases with no multiplicity, and the problem of multiplicity separation is not described. All of these methods have at least one of the following problems:
\begin{itemize}
\item The method becomes very inefficient rapidly with the dimensions of the irreps.
\item The method is not general for the possible sets $\bm{\lambda}$.
\item The method cannot be used with $k_{\bm{\lambda}}>1$ or does not lead to a meaningful way to separate the multiplicities. 
\end{itemize}
One objective of this thesis is to provide an alternative algorithm that allows calculating Kronecker states, and hence Clebsch Gordan coefficients of $S_n$ for any set $\bm{\lambda}$ with each $\lambda^i$ of at most two parts, whose complexity grows slowly compared with the existing methods, and where the multiplicity problem is solved partially by the nature of the construction. In fact, for many sets $\bm{\lambda}$, we can set the first base of the invariant subspace to correspond to the only Kronecker state appearing in the multipartite W-class, which can be calculated from an analytical expression as we will see in next section.

\section{W-Class entanglement concentration and W-Class Kronecker states}
\label{WKronecker}

\noindent Bipartite Kronecker states naturally arise in the bipartite entanglement protocol proposed by Hayashi-Matsumoto \cite{Hayashi-Matsumoto} described in Section \ref{EntanglementHayashi}. As was discussed there, a multipartite generalization is not directly obtained, due to the multiplicity of the invariant subspaces, making it impossible to separate the states belonging to $GL_d$ irreps from those belonging to $S_n$ irreps. Remarkably, a generalization of the protocol is possible when the states used as input belong to the multipartite $W$-class \cite{sawicki2013}, which are generalizations of the W-class for the three-qubit case. In \cite{Botero}, Botero and Mejía consider the multipartite entanglement concentration protocol by replicating the idea of Hayashi and Matsumoto but restricting the input states to belong to the multipartite W-class. It is shown that under this restriction, the corresponding subspace of $\{\bm{\lambda}\}$ is always one-dimensional. Meaning that when taking $n$ copies of a state in the W-class, the Schur transform leads to a unique state in $\{\bm{\lambda}\}$ and some linear combination of Kronecker states in $[\bm{\lambda}]$ (which by definition is again a Kronecker state) for which the authors obtain an analytical expression. As this result is a cornerstone in the method that we propose, to build general Kronecker states, we will summarize here the results obtained in \cite{Botero} and show how the expression for the unique Kronecker state is applied in simple cases.\\

\noindent First, we need to define the $N$-partite W class as the orbit under SLOCC operations of the $\ket{W_N}$ state:
\begin{equation}
\ket{W_N} = \frac{1}{\sqrt{N}}  \sum_{i=1}^{N} \ket{\boldsymbol{1}_i }, 
\label{eq:WN}
\end{equation}
where $\boldsymbol{1}_i$  is a sequence with a ``1" at position $i$ and zeros in the remaining $(N-1)$ positions. It is known that any state in this class is, up to LU transformations,  completely specified  by the set of eigenvalues of the reduced density matrices on each part, $\rtup$ \cite{sawicki2013}, and 
is LU-equivalent to the state \cite{yu2013} :
\begin{equation}\label{eq:ecstatepsi}
\ket{\psi^W} = \sqrt{c_{\rtup}^{0}}|\boldsymbol{0} \rangle + \sum_{i=1}^{N}\sqrt{c_{\rtup}^{i}}|\boldsymbol{1}_i \rangle, 
\end{equation}
where $\boldsymbol{0}$ is a sequence of $N$ zeros, and $c_{\rtup}^{i}$ are real coefficients with $\sum_{i=0}^{N}c_{\rtup}^{i}=1$.  We show later in Equation \eqref{eq:cspec} that these coefficients can be obtained from the set of local spectra $\bm{r}$. The normal form in Equation \eqref{eq:ecstatepsi} can be used to compute a unique unnormalized state corresponding to the $GL_2$ part of the Schur transform. For this, Mejia and Botero calculate explicitly the \textit{SLOCC covariants} \cite{Covariants} of any state in the normal form and show that by fixing $n$ and $\bm{\lambda}$, the obtained covariant is unique up to a constant. By using a mapping between homogeneous polynomials and states \cite{Fulton}, a unique unnormalized state is obtained from this covariant as:
\begin{equation}
\label{eq:phigen2}
\ket{\reallywidehat{\Phi}_{\bm{\lambda}}(\psi^W)}= n!^{-(N-2)/2}
   \sum_{\omega^{0}=0}^{n} \dfrac{(c_{\rtup}^{0})^{\omega^{0}/2} } {\omega^{0}!} \sum_{\omegatup} {\ctup_{\rtup}}^{\omegatup/2} \sqrt{A_{\lamtup,\omegatup}}
 \ket{ \lamtup, \omegatup}.
\end{equation}
with a list of weights $\bm{\omega}=\omega^1\omega^2\dots \omega^N$ restricted by $\sum_{i=0}^N \omega^i =n$, and
 \begin{equation}
 \bm{c}_{\bm{r}}^{\bm{\omega}/2}= \prod_{i=1}^N ({c_{\bm{r}}^{i}}) ^{\omega_i/2}   , \quad    A_{\bm{\lambda},\bm{\omega}}=\prod_{i=1}^N A_{\lambda^i,\omega^i } \quad , \quad  A_{\lambda,\omega}=\frac{(n-\lambda-\omega)!}{(\omega-\lambda)!}.
 \label{eq:Alambda}
 \end{equation}
\noindent Given a set of marginal spectra $\bm{r}$, and a set of irreps $\bm{\lambda}$, the state  $\ket{\reallywidehat{\Phi}_{\bm{\lambda}}(\psi^W)}$ is defined uniquely. It corresponds to the unique possible state in $\{\bm{\lambda}\}$ that appears in the Schur transform of $\ket	{\psi^W}^{\otimes n}$. Recall from Equation \eqref{eq:Threequbits} that the multi-local Schur transform on multipartite qubit states generically corresponds to:
\begin{equation*}
\ket{\psi}^{\otimes n}=\bigoplus_{ \lamtup \vdash n,2} \sqrt{p(\bm{\lambda}|\psi)}\left[\sum_{i=1}^{k_{\lamtup}}\ket{\Phi_{\lamtup,i}(\psi)}\ket{\mathcal{K}_{\lamtup,i}} \right].
\end{equation*}
This expression is simplified when using W-class states
\begin{equation}
     \ket{\psi^{W}}^{\otimes n} =\bigoplus_{\bm{\lambda} \in \Lambda_n^{W}} \eta_{\bm{\lambda}} \ket{\reallywidehat{\Phi}_{\bm{\lambda}} \left(\psi^{W} \right)}\ket{\mathcal{K}^W_{\bm{\lambda}}},
     \label{eq:WSchur}
\end{equation}
where $\eta_{\bm{\lambda}}$ is a factor relating the unnormalized state $\ket{\reallywidehat{\Phi}_{\bm{\lambda}}(\psi^W)}$ and the normalized state $\ket{\Phi_{\bm{\lambda}}(\psi^W)}$ as :
 \begin{equation*}
        \ket{\Phi_{\bm{\lambda} }(\psi^W)}=\frac{ \eta_{\bm{\lambda }}}{\sqrt{p(\bm{\lambda}|\psi^W)}} \ket{\reallywidehat{\Phi}_{\bm{\lambda}}(\psi^W)},
  \end{equation*}
 and $\ket{\mathcal{K}^W_{\bm{\lambda}}}$ is a unique Kronecker state corresponding to the W-class, which we will call \textit{W-Kronecker state}.  $\Lambda_n^{W}$ is the set of all the $\lamtup$ that lie in the sector of the W in the \textit{entanglement polytope} \cite{Polytopes}, defined by:
\begin{equation} \label{eq:Wlambda}    2\lambda^{i}\leq \sum_{j=1}^N \lambda^{j} \leq n.
\end{equation}
 We will discuss this restriction in depth in the next subsection. From Equation \eqref{eq:WSchur}, it is possible to note how the Hayashi-Matsumoto algorithm can be generalized for states in the W class. When making a projective measurement on the local partitions, this measurement corresponds to a projection to a given set $\bm{\lambda}$ as:
\begin{equation}
\ket{\psi^{W}}^{\otimes n}  \xrightarrow{P^{\bm{\lambda}}}  \ket{\Phi_{\lamtup}(\psi^W)}\ket{\mathcal{K}_{\lamtup}^W},
\end{equation} 
 where now the state in $\{\bm{\lambda}\}$ is separable from the corresponding Kronecker state, which is a maximally entangled state. For more details on this process, the reader should refer to the original paper \cite{Botero}.\\

\noindent We are mainly interested in the process of obtaining an analytical expression for $\ket{\mathcal{K}^W_{\bm{\lambda}}}$ that is also shown in the original paper \cite{Botero}, but by its relevance, we will deduce it completely here. It will be useful to write the Kronecker states associated with the $W$ class in their coefficient expansion as:
\begin{equation} \ket{\mathcal{K}^W_{\bm{\lambda}}}= \sum_{\bm{q}} K^W_{\bm{\lambda
    ,q} }\ket{\bm{\lambda},\bm{q}},
    \label{eq:Kronbasis}
\end{equation}
where each basis element of irrep $\lambda^{i}$ in $S_n$ is labeled (as discussed in Section \ref{IrrepsSn}) with the corresponding (modified) Yamanouchi symbol i.e., a binary list $q^{i}=\{q_1^{i},q_2^{i},\dots ,q_n^{i}\}$ that represents one possible valid way to obtain the Young Tableaux associated to $\lambda^{i}$ by adding boxes in the first row $(q_i=0)$ or in the second row $(q_i=1)$ in each step. An example of the different labeling schemes used for the basis elements of $S_n$ is shown in Equation \eqref{eq:IrrepsLabels}. We also use here the boldface notation  for the set of basis elements $\bm{q}=(q^{1},q^{2} ,\dots, q^{N})$.\\

\noindent The Schur transform shown in Equation \eqref{eq:WSchur} can be used on any state in the W-class, and as the Kronecker state appearing in the decomposition does not depend on the specific state used as input, we can choose any. As the state with the most simple structure is the $\ket{W_N}$ state, then, from now on, we pick the state to be the W state itself, i.e., $\ket{\psi^W}=\ket{W_N}$  defined in Equation \eqref{eq:WN}. For this state, the expression in Equation 
\eqref{eq:phigen2} is simplified to:
\begin{equation}
\ket{\reallywidehat{\Phi}_{\bm{\lambda}} (W_N)}=\frac{n!^{-(N-2)/2}}{\sqrt{N^n}} \sum_{\bm{\omega}}  \sqrt{A_{\bm{\lambda},\bm{\omega}}}\ket{\bm{\lambda},\bm{\omega}}.
\label{eq:Wphi}
\end{equation}
This expression can be used to obtain a recursive relation between W-Kronecker states. We can think of the Schur transform of $n$ copies of $\ket{W_N}$ in two equivalent ways; the first is to apply Equation \eqref{eq:WSchur}
directly, obtaining:
\begin{equation}
     \ket{W_N}^{\otimes n} = \bigoplus_{\lamtup \in \Lambda_n^W}\eta_{\lamtup}\ket{\reallywidehat{\Phi}_{\bm{\lambda}}(W_N)}\ket{\mathcal{K}^W_{\lamtup}},
        \label{eq:Wcopies1}
\end{equation}
and as a recursive construction from the Schur transform on $n-1$ copies of $\ket{W_N}$,
\begin{equation}
    \ket{W_N}^{\otimes n} =\underbrace{ \left( \bigoplus_{\lamtup' \in \Lambda_{n-1}^{W_N}} \eta_{\lamtup' }\ket{\reallywidehat{\Phi}_{\bm{\lambda}'}(W_N)}\ket{\mathcal{K}^W_{\lamtup'}}\right)}_{\ket{W_N}^{\otimes (n-1)}}\otimes \ket{W_N},
      \label{eq:Wcopies2}
\end{equation}
where $\bm{\lambda}'$ refer to sets of irreps of $S_{n-1}$. We can match the expression in Equation \eqref{eq:Wcopies1} with Equation \eqref{eq:Wcopies2}, and  replace the corresponding expansions of $\ket{\reallywidehat{\Phi}(W_N)}$ states from Equation \eqref{eq:Wphi}. By fixing some set $\lamtup$ we get:
 \begin{equation*}
  \begin{gathered}
  n!^{-(N-2)/2}
     \eta_{\bm{\lambda}}   \left(\sum_{\bm{\omega}}  \sqrt{A_{\bm{\lambda},\bm{\omega}}} \right)  \ket{\mathcal{K}^W_{\bm{\lambda},\bm{q}}}=\\(n-1)!^{-(N-2)/2}\bigoplus_{\bm{\lambda}'<\bm{\lambda}} \eta_{\bm{\lambda}'} \left( \left(\sum_{\bm{\omega}'}  \sqrt{A_{\bm{\lambda}',\bm{\omega}'}}  \right) \ket{\mathcal{K}^W_{\bm{\lambda}',\bm{q}'}}\right) \otimes \left( \sum_{i=1}^{N}\ket{\boldsymbol{1}_i }\right),
  \end{gathered}
  \label{eq:Rec1}
\end{equation*}
where $\bm{\lambda}'<\bm{\lambda}$ restrict the partitions of $n-1$, appearing on the right-hand side of the equation to those $\bm{\lambda'}$ that can reach $\bm{\lambda}$ by adding one box either up or down in the Young Tableaux for each ${\lambda'}^{i}$. When expanding the W-Kronecker states in its coefficients as shown in equation \eqref{eq:Kronbasis}, the previous expression can be simplified to:
\begin{equation*}
  \begin{gathered}
  n^{-(N-2)/2}
     \eta_{\bm{\lambda}}   \sum_{\bm{\omega},\bm{q}}  \sqrt{A_{\bm{\lambda},\bm{\omega}}} K^{W}_{\bm{\lambda},\bm{q}}  \ket{\bm{\lambda}, \bm{\omega},\bm{q}}=\\\bigoplus_{\bm{\lambda}'<\bm{\lambda}} \eta_{\bm{\lambda}'} \left( \sum_{\bm{\omega}',\bm{q}'}  \sqrt{A_{\bm{\lambda}',\bm{\omega}'}}  K^{W}_{\bm{\lambda'},\bm{q}'}\ket{\bm{\lambda}',\bm{\omega}',\bm{q}'}\right) \otimes \left( \sum_{i=1}^{N}\ket{\boldsymbol{1}_i }\right),
  \end{gathered}
  \label{eq:Rec2}
\end{equation*}
that by projecting with an element of the total Schur basis, $\bra{\bm{\lambda},\bm{\omega},\bm{q}}$,  corresponds to:
\begin{equation}
  \begin{gathered}
  n^{-(N-2)/2}
     \eta_{\bm{\lambda}}    \sqrt{A_{\bm{\lambda},\bm{\omega}}} K^{W}_{\bm{\lambda},\bm{q}}=\\\bigoplus_{\bm{\lambda}'<\bm{\lambda}} \eta_{\bm{\lambda}'}  \sum_{\bm{\omega}',\bm{q}'}  \sqrt{A_{\bm{\lambda}',\bm{\omega}'}}  K^{W}_{\bm{\lambda'},\bm{q}'} \left( \sum_{i=1} ^N \bra{\bm{\lambda},\bm{\omega},\bm{q}}\ket{\bm{\lambda}',\bm{\omega}',\bm{q}'}\ket{\boldsymbol{1}_i }\right) ,
  \end{gathered}
  \label{eq:Rec3}
\end{equation}
Note that the inner product appearing in the last equation is in fact a product of the CGC used for the Schur transform, for example the term corresponding to $i=1$ is:
\begin{equation*}
\bra{\bm{\lambda},\bm{\omega},\bm{q}}\ket{\bm{\lambda}',\bm{\omega}',\bm{q}'}\ket{100\dots} =\braket{\lambda^1, \omega^1,q^1}{\lambda'^1, \omega'^1,q'^1}\ket{1}\prod_{i=2}^N \braket{\lambda^i,\omega^i, q^i}{\lambda'^i,\omega'^i, q'^i}\ket{0}.
\end{equation*}
The inner products at the right are the CGC for the Schur basis that we introduced in the previous chapter. From Equation \eqref{eq:innerprodschur}, we can notice that $\bm{\lambda}'$ $\bm{q}'$ are fixed by $\bm{\lambda}$ and $\bm{q}$, and $\bm{\omega}'$ is fixed by $\omega$ and $\bm{1_i}$. We get:
\begin{equation*}
\bra{\bm{\lambda},\bm{\omega},\bm{q}}\ket{\bm{\lambda}',\bm{\omega}',\bm{q}'}\ket{\bm{1}_i}= \Gamma^{\lambda^i,\omega^i ,n}_{q^i _n,1}\delta_{\omega^i,\omega'^{i+1}} \delta_{q^i, q'^iq^i_n} \delta_{\lambda^i, \lambda'^i+q^i_n}\prod_{j\neq i} \Gamma^{\lambda^j,\omega^j ,n}_{q^j _n,0} \delta_{\omega^j,\omega'j} \delta_{q^j, q'^jq^j_n} \delta_{\lambda^j, \lambda'^j+q^j_n}.
\end{equation*}
By plugging this expression in Equation \eqref{eq:Rec3}, the sums on $\bm{\lambda}', \bm{\omega}'$ and $\bm{q}$ collapse, obtaining:
\begin{equation}
  n^{-(N-2)/2}
     \eta_{\bm{\lambda}} K^{W}_{\bm{\lambda},\bm{q'}\bm{q_n}}= \left(  \sum_{i=1} ^N \sqrt{\frac{A_{\bm{\lambda}-\bm{q_n},\bm{\omega}-\bm{1}_i}}{A_{\bm{\lambda},\bm{\omega}}}}     \Gamma_{q_n ^i,1}^{\lambda^i,\omega^i,n} \prod_{j\neq i} \Gamma^{\lambda^j,\omega^j,n}_{q_n^j ,0} \right)\eta_{\bm{\lambda}-\bm{q_n}} K^{W}_{\bm{\lambda}-\bm{q_n},\bm{q}'}.
     \label{eq:Rec4}
\end{equation}
It is useful to define a re-scaled coefficient:
\begin{equation*}
\hat{K}^W_{\bm{\lambda},\bm{q}} = n! ^{-(N-2)/2} \eta_{\bm{\lambda}} K^W_{\bm{\lambda},\bm{q}},
\end{equation*} 
then, Equation \eqref{eq:Rec4} can be written as:
\begin{equation}
\hat{K}^W_{\bm{\lambda},\bm{q}'\bm{q_n}}= F_{\bm{\lambda},\bm{q}_n,n} \hat{K}^W_{\bm{\lambda}-\bm{q_n},\bm{q}'}
\label{eq:RecKron}
\end{equation}
where $F_{\bm{\lambda},\bm{q}_n,n} $ is a recurrence factor that can be calculated as:
\begin{equation}
F_{\bm{\lambda},\bm{q}_n,n} = \sum_{i=1} ^N \sqrt{\frac{A_{\bm{\lambda}-\bm{q_n},\bm{\omega}-\bm{1}_i}}{A_{\bm{\lambda},\bm{\omega}}}}     \Gamma_{q_n ^i,1}^{\lambda^i,\omega^i,n} \prod_{j\neq i} \Gamma^{\lambda^j,\omega^j,n}_{q_n^j ,0}.
\label{eq:Flambda2}
\end{equation}
From the definitions of $A_{\lambda,\omega}$ in Equation \eqref{eq:Alambda}, and the CGC in Equation \eqref{eq:Gammas}, one can show that for any combination of $q_n$ and $s_n$ one gets:
\begin{equation*}
\sqrt{\frac{A_{\lambda',\omega'}}{A_{\lambda,\omega}}} \Gamma_{q_n,s_n}^{\lambda,\omega,n}=  \frac{1+s_n(\omega-q_n(n-\lambda+1)-(1-q_n)\lambda-1)}{\sqrt{n-2\lambda+2q}},
\end{equation*}
Note that in the $i$-th term of the sum in Equation \eqref{eq:Flambda2}, only the $i$-th term has $s^{i}_n=1$ and all the others are $s^{j}_n=0$, then, we have:
 \begin{equation*}
 F_{\bm{\lambda},\bm{q}_n,n} = \sum_{i=1} ^N \frac{\omega^i-q^i_n(n-\lambda^i+1)-(1-q^i_n)\lambda^i}{\prod_j \sqrt{n-2\lambda^j+2q_n ^j }}.
 \end{equation*}
This expression can be simplified by noting that $\sum_{i}^{N}\omega^i=n$, then, the final expression for the recurrence factor is:
 \begin{equation}
F_{\bm{\lambda},\bm{q}_n,n}= \frac{ n-\sum_{i=1}^{N} \left[q_n^{i} ( n-\lambda ^{i}+1 )-(1-q_n^{i} ) \lambda ^{i} \right] }{\prod_i \sqrt{n -2\lambda ^{i} +2 q_n ^{i}}}.
\label{eq:Flambda}
\end{equation}
This expression, along with the recurrence relation in Equation \eqref{eq:RecKron}, allow us to obtain unnormalized W-Kronecker states in the product basis of irreps of $S_n$:
\begin{equation}
\ket{\hat{\mathcal{K}}^W_{\bm{\lambda}}}= \sum_{\bm{q}} \hat{K}^W_{\bm{\lambda},\bm{q}} \ket{\bm{\lambda},\bm{q}},
\label{eq:UnnormKron}
\end{equation}
from W-Kronecker states in the product basis of irreps of $S_{n-1}$. By noting that any Kronecker state is connected through this process to the unique W-Kronecker state in $S_1$ , i.e.,
\begin{equation*}
\ket{\mathcal{K}^W_{\bm{0}}}=\ket{\bm{0},\bm{0}}.
\end{equation*}
then any scaled coefficient $\hat{K}^W_{\bm{\lambda}\bm{q}}$ can be obtained from iterating the recurrence equation from 1 to $n$, then, by introducing the notation $\lambda^{i,j}=\sum_{k=1}^{j} q^{i}_k$  for the partial partitions at step $j$ for each basis element $q^{i}$, we have the expression:
\begin{equation} \label{eq:KroneckerW}
\hat{K}^W_{\lamtup,\bm{q}}=\prod_{j=2}^{n} F_{\bm{\lambda}^j,\bm{q}_j,j}=\prod_{j=2} ^{n} \frac{ j- \sum_{i=1}^N \left[ q_j^{i}(j+1-\lambda^{(i),j})-(1-q_j^{i})\lambda^{i,j}\right]  }{\prod_{i=1}^N \sqrt{n-2\lambda^{i,j}+2q_j^{i}}} .
\end{equation}
With this, we can compute the W-Kronecker states for any $N$ and $n$ up to normalization. By defining a set $\bm{\lambda}$ for which we want to calculate the correspondent W-Kronecker state, we can use the previous equation for calculating all the coefficients for the possible basis $\bm{q}$ and build the unnormalized state $\ket{\hat{\mathcal{K}}^W_{\bm{\lambda}}}$ according to Equation \eqref{eq:UnnormKron}, then we can normalize the obtained state to get $\ket{\mathcal{K}^W_{\bm{\lambda}}}$.\\

\noindent This process is very efficient for calculating Kronecker states, and hence Clebsch Gordan Coefficients of $S_n$, as there is no need to build any matrix representation. In general, for the three parts case, we will be building vectors of dimension $f^{[\lambda^1]}\times f^{[\lambda^2]}\times f^{[\lambda^3]}$ (without square) directly from its coefficients, which shows a clear advantage over the mentioned methods in the previous section. The only required knowledge from representation theory for using this equation is to understand the relation between Young diagrams and labels of irreps of $S_n$. However, this approach is insufficient for calculating all possible Kronecker states with irreps of at most two parts. Two restrictions must be highlighted:
\begin{itemize}
\item One of the most interesting properties of Kronecker states is that they define a vector space, and any linear combination of states in such vector space is naturally a Kronecker state. However, the expression obtained was achieved precisely because in this particular class, the appearing Kronecker state is unique. So, with this approach, it is impossible to find more than one Kronecker state in each set $\bm{\lambda}$. Using this method as the unique tool to calculate Kronecker states will not permit exploring the rich structure of Kronecker subspaces.

\item These W-Kronecker states are obtained from the  Schur transform on states belonging to the N-partite W class. As pointed out, the sets of partitions appearing in the decomposition of any state in this class are restricted by the polytope $\Lambda_n^W$ defined in Equation \eqref{eq:Wlambda}. It is easy to find sets of partitions outside of this polytope. For example, consider n=5, and $\bm{\lambda}=222$ (i.e., $[\lambda^1]=[\lambda^2]=[\lambda^3]=[32]$), where the Kronecker coefficient is $k_{\bm{\lambda}}=1$. Even though there exists a Kronecker state in the invariant subspace, it cannot be obtained from Equation \eqref{eq:KroneckerW} because the set $\bm{\lambda}$ is out of the polytope $\Lambda_n^W$, the set of lower values of the partitions add to more than $n$:
\begin{equation*}
\sum_{i=1}^3 \lambda ^i =6,
\end{equation*}
to understand why the W-class has this restriction, let us discuss the entanglement polytope \cite{EntanglementPolytope}.
\end{itemize}

\subsection*{Entanglement Polytope}
One recurrent problem in quantum mechanics that is related to the Kronecker coefficient is the \textit{One-body quantum marginal problem} \cite{marginalK}, in the pure state case. Consider that we have a set of $N$ reduced density matrices $\{\rho_i\}$, also called \textit{marginal states}, and we want to know if there exists some possible N-partite pure state $\ket{\psi}$ such that its reduced density matrices are the set $\{\rho_i\}$. This problem is very complex in general, but it has a well-known solution for the case of qubits \cite{Higuchi}. In this case, the N-partite pure state $\ket{\psi}$ only exists if the set of smallest eigenvalues (we consider the eigenvalues to be in descending order as $a=(a_1a_2$)) $\{a^i_2 \}$ of the reduced density matrices  $\{\rho_i\}$ satisfy polygonal inequalities:
\begin{equation*}
2\cdot a^{i}_2 \leq \sum_{j=1} ^{N} a^{j} _2 , 
\end{equation*}
this set of inequalities defines a polytope that is upper bounded by the planes corresponding to $a^{i}_2=1/2$ because it is the maximal value of the minimal eigenvalue. Any set of reduced density matrices with eigenvalues outside of this polytope is not compatible with any pure state. It is worth noting how any separable qubit in the state has fixed its minimal eigenvalue to zero, shedding light on how this polytope is related to the entanglement on the state \cite{EntanglementPolytope}. Moreover, in \cite{WPoly}, it was shown that the marginal spectra of states belonging to the W class correspond to a restriction of the total polytope. This restriction is given by the plane defined by:
\begin{equation*}
 \sum_{i=1}^N a_2 ^i\leq 1,
\end{equation*}
then, it is possible to use the set of marginal spectra, $a_2$ to decide if a state can or cannot belong to the W class.\\  

\noindent Remarkably, this result is strongly connected with representation theory. In \cite{Nonzero}, by using the Keyl Werner theorem, it is shown that if three density matrices are compatible with a three-partite pure state, then there exists a normalized set of irreps of $S_n$, $\bar{\bm{\lambda}}=\bar{\lambda^1}\bar{\lambda^2 }\bar{\lambda^3}$, with $\bar{\lambda^i}=\lambda^i/n$, arbitrarily close to $a_2=a_2^1a_2^2a_2^3$, such that that the Kronecker coefficient is not null $k_{\bm{\lambda}}>0$. This result implies that the polytope of admissible marginal spectra is equal to the set of normalized partitions with a Kronecker coefficient different from zero, called $\overline{\text{KRON}}$. Due to the connection with the marginal problem, we can define two interesting polytopes regarding irreps. The first one is the total polytope $\overline{\text{KRON}}$, where belong all the sets of irreps $\bm{\lambda}$ with non-null Kronecker coefficient, $k_{\bm{\lambda}}>0$, that is defined by the following: 
\begin{equation}
 \overline{\text{KRON}} : \left\{ \bm{\lambda} \,|\,2\cdot \lambda^{i} \leq \sum_{j=1} ^{N} \lambda^{j} , \quad \forall \lambda^i\right\} .
 \label{eq:KRON}
\end{equation}
And the polytope corresponding to the W class, named $\Lambda_n^W$ is defined as \cite{Botero}:
\begin{equation}
 \Lambda_n^W : \left\{ \bm{\lambda} \,|\,2\cdot \lambda^{i} \leq \sum_{j=1} ^{N} \lambda^{j} \leq N , \quad \forall \lambda^i\right\} .
 \label{eq:WLambda}
\end{equation}
	
\begin{figure}
\includegraphics[scale=0.5]{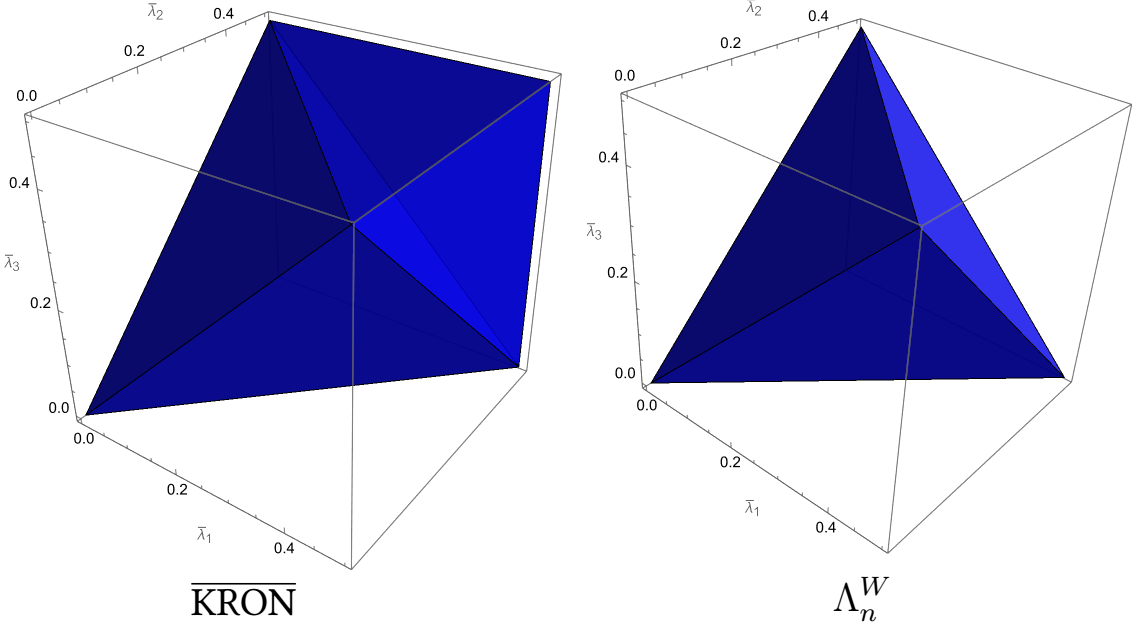}
\caption{Representation of polytopes $\overline{\text{KRON}}$ and $\Lambda_n^W $ for the three-parts case, each axis represents the second element of the normalized partition labeling the individual irreps. It can be seen that irreps with no null Kronecker coefficient can not be obtained from W-class states.}
\label{fig:Polytope}
\end{figure}

In figure \ref{fig:Polytope}, both polytopes are represented in the three-parts case; it is clear that there is a region of $\overline{\text{KRON}}$ that states in the W-class cannot achieve. In terms of the Kronecker state construction, this means that sets of irreps with a Kronecker coefficient different from zero can not be obtained from states in the W class.\\

\noindent Despite these restrictions, we can think of the approach of building W-Kronecker states as an efficient protocol to find one of the $k_{\bm{\lambda}}$ Kronecker states in the admissible sets of partitions$\bm{\lambda} \in \Lambda_n^W$. In the following, we illustrate a simple example to show how Equation \eqref{eq:KroneckerW} is used to calculate W-Kronecker states.

\subsection*{Example of W-Kronecker state calculation}

For this example, we want to build the Kronecker state corresponding to the invariant subspace in $[\bm{\lambda}]=[1]\otimes[1] \otimes [2]$ for $n=4$ (or in the long notation $[\bm{\lambda}]=[31]\otimes [31]\otimes [2^2]$). First, let us label the basis elements of the two irreps. For $[1]$, we have three basis elements:
\begin{equation*}
q^{[1]}_1 = \{0,0,0,1\} , \quad q^{[1]}_2 = \{0,0,1,0\}, \quad q^{[1]}_3= \{0,1,0,0\}.
\end{equation*} 
It is also useful to use the Young lattice representation; for this case, we have the three paths (labeled the same as their corresponding $q$):
\begin{equation*}
\includegraphics[scale=1,valign=c]{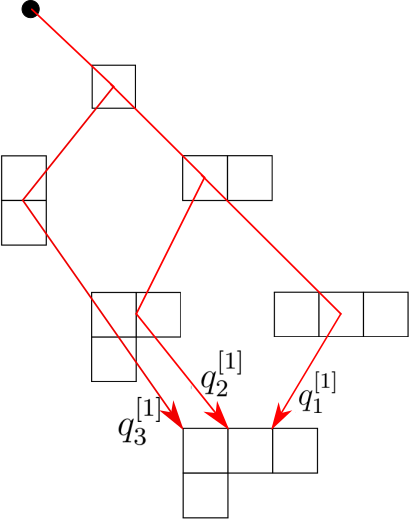}
\end{equation*}
Similarly, for the irrep $[2^2]$ we have:
\begin{equation*}
q_1^{[2]}=\{0,0,1,1\} ,\quad q_2^{[2]}=\{0,1,0,1\} ,
\end{equation*}
or equivalently within the Young lattice:
\begin{equation*}
\includegraphics[scale=1,valign=c]{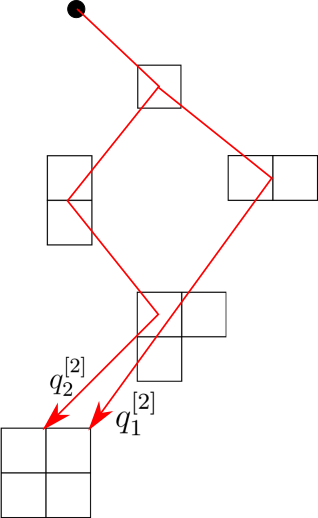}.
\end{equation*}
Now, we need to choose some set $\bm{q}$ to calculate its coefficient. Due to the dimensions of the irreps, there are $f^{[\lambda^1]}\times f^{[\lambda^2]} \times f^{[\lambda^3]}= 3\times 3 \times 2=18$ basis elements in the product basis, which we will label as $\bm{q}_i$ with $i\in \{1,\dots,18\}$. Let us start with $\bm{q}_1=q_1^{[1]}q_1^{[1]}q_1^{[2]}$. In order to find the construction of this coefficient from the recursive equation, we can find the needed factors easily by looking simultaneously at the Young lattice of the three basis elements:
\begin{equation*}
\ket{\bm{\lambda,\bm{q}^1}}=\includegraphics[scale=1,valign=c]{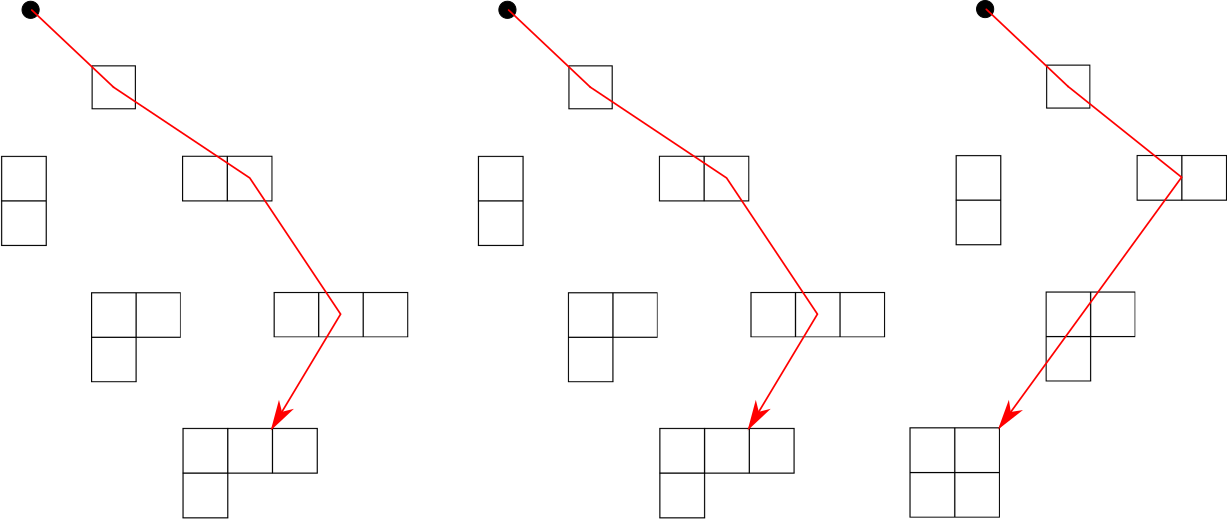}.
\end{equation*}
From this multiple lattice, it is possible to read the sequence of $F_{\bm{\lambda}^j,\bm{q}_j,j}$ to be used in the construction. For this, we look at the line corresponding to $n=j$ and read from it the partial partitions $\bm{\lambda}^j$, and the step taken to reach those from $n'=j-1$. In this case, for example, for $j=2$, the three partitions are at $\bm{\lambda}^{2}=(000)$, and all parts arrived there from taking a step to the right, so $\bm{q}_2=(000)$. For $j=3$, the set of partial partitions is $\bm{\lambda}^3 =(001)$, and the first two parts take the step to the right, while the third one take the step to the left, so $\bm{q}_3=(001)$. To finish, the last set of partitions is $\bm{\lambda}^4 =\bm{\lambda}=(112)$ and the steps were $\bm{q}_4=(111)$. Then, for this coefficient, we have:
\begin{equation*}
\hat{K}^W_{\bm{\lambda},\bm{q}^1}= F_{(000),(000),2} \cdot F_{(001),(001),3} \cdot F_{(112),(111),4} ,
\end{equation*}
which can be calculated from Equation \eqref{eq:Flambda} to be:
\begin{equation*}
\hat{K}^W_{\bm{\lambda},\bm{q}^1}= \frac{1}{2\sqrt{2}} \cdot 0 \cdot \frac{-7}{4\sqrt{2}}=0.
\end{equation*}
Note how the third \textit{multi-step} gives a null value; this can be used to discard all the basis $\bm{q}^i$ that have this step; however, in this case, the only base with such step is $\bm{q}^1$. This multi-step should also be discarded beforehand as the partial partition $\bm{\lambda}^3=(001)$ is out of $\Lambda^W_3$. This condition should be checked in each multi-step. The next basis element in a lexicographic ordering is $\bm{q}^2=q_1^{[1]}q_1^{[1]}q_2^{[2]}$ i.e.,
\begin{equation*}
\ket{\bm{\lambda,\bm{q}^2}}=\includegraphics[scale=1,valign=c]{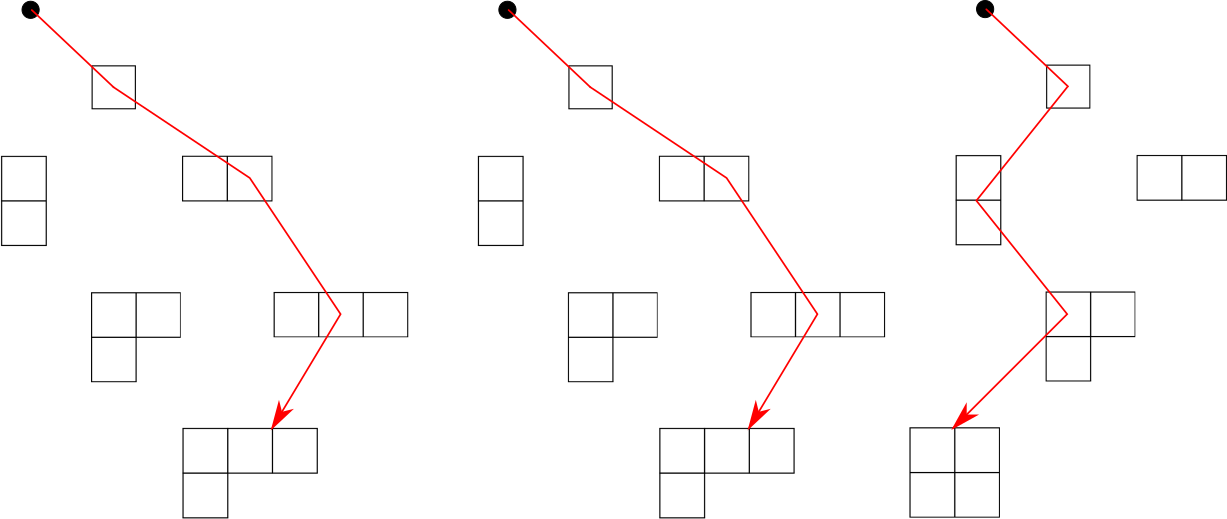}.
\end{equation*}
It can be seen that the set of partial partitions $\bm{\lambda}^2=(001)$ is out of $\Lambda^W_2$, showing this coefficient is also null, and by the same reason, some other basis elements give null coefficients, those bases are:
\begin{equation*}
\begin{gathered}
\bm{q}^2=q_1^{[1]}q_1^{[1]}q_2^{[2]},\qquad 
\bm{q}^4=q_1^{[1]}q_1^{[2]}q_2^{[2]},\\
\bm{q}^8=q_1^{[2]}q_1^{[1]}q_2^{[2]}, \qquad 
\bm{q}^{10}=q_1^{[2]}q_1^{[2]}q_2^{[2]}.
\end{gathered}
\end{equation*}
The next coefficient to calculate is for the basis element $\bm{q}^3=q_1^{[1}q_2^{[1]}q_1^{[2]}$. For this one, we have:
\begin{equation*}
\ket{\bm{\lambda,\bm{q}^3}}=\includegraphics[scale=1,valign=c]{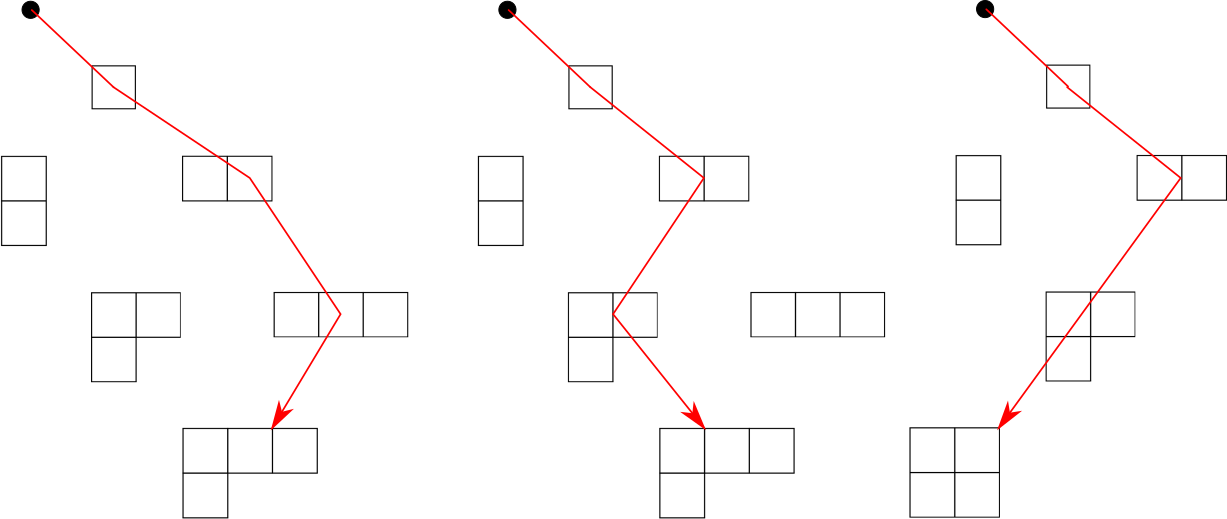}.
\end{equation*}
The coefficient is calculated as:
\begin{equation*}
\hat{K}^W_{\bm{\lambda},\bm{q}^3}=F_{(000),(000),2} \cdot F_{(011),(011),3} \cdot F_{(112),(101),4} =\frac{1}{\sqrt{2}}\cdot \frac{-1}{\sqrt{3}} \cdot -1 = \frac{1}{\sqrt{6}}.
\end{equation*}
This term is the first non-zero coefficient for the W-Kronecker state. As the two first irreps are the same, and the Equation \eqref{eq:Flambda} is symmetric, we also have:
\begin{equation*}
\hat{K}^W_{\bm{\lambda},\bm{q}^7}=\frac{1}{\sqrt{6}}
\end{equation*}
where $\bm{q}^7=q_2^{[1}q_1^{[1]}q_1^{[2]}$. For the next coefficient to calculate, $\bm{q}^5=q^{[1]}_1q^{[1]}_3q_1^{[2]}$, the second set of partial partitions is $\bm{\lambda}^2=(010)$ that is out of $\Lambda^W_2$, hence, its coefficient is zero. By the same multi-step, the coefficient of $\bm{q}^{11}=q_2^{[1]}q_3^{[1]}q_1^{[1]}$ is also zero, and by symmetry, the coefficients of $\bm{q}^{13}=q^{[1]}_3q^{[1]}_1q_1^{[2]}$ and $\bm{q}^{15}=q^{[1]}_3q^{[1]}_2q_1^{[2]}$. The last null coefficient corresponds to $\bm{q}^{18}=q^{[1]}_3q^{[1]}_3q_2^{[2]}$, whose second partial partition $\bm{\lambda}^2=(111)$ is out of $\Lambda^{W}_2$. The remaining coefficients are not zero and can be calculated with the process described, obtaining:
\begin{equation*}
\begin{gathered}
\hat{K}^W_{\bm{\lambda},\bm{q}^6}= \hat{K}^W_{\bm{\lambda},\bm{q}^{14}} =\frac{1}{\sqrt{6}},\\
\hat{K}^W_{\bm{\lambda},\bm{q}^9}=- \hat{K}^W_{\bm{\lambda},\bm{q}^{12}} =-\hat{K}^W_{\bm{\lambda},\bm{q}^{16}}=- \hat{K}^W_{\bm{\lambda},\bm{q}^{17}} = \frac{1}{2\sqrt{3}}, 
\end{gathered} 
\end{equation*}
where the basis elements correspond to:
\begin{equation*}
\begin{gathered}
\bm{q}^6= q^{[1]}_1q^{[1]}_3q_2^{[2]}, \qquad 
\bm{q}^9= q^{[1]}_2q^{[1]}_2q_1^{[2]},  \qquad 
\bm{q}^{12}= q^{[1]}_2q^{[1]}_3q_2^{[2]}, \\
\bm{q}^{14}= q^{[1]}_3q^{[1]}_1q_2^{[2]}, \qquad
\bm{q}^{16}= q^{[1]}_3q^{[1]}_2q_2^{[2]}, \qquad 
\bm{q}^{17}= q^{[1]}_3q^{[1]}_3q_1^{[2]}. 
\end{gathered}
\end{equation*}
We can put everything together to write the ``unnormalized'' Kronecker state as:
\begin{equation}
 \begin{gathered}
    \ket{\hat{\mathcal{K}}^W_{112}}= \frac{1}{\sqrt{6}}\left(\ket{112,q^{[1]}_1q^{[1]}_2q^{[2]}_1}+\ket{112,q^{[1]}_1q^{[1]}_3q_2^{[2]}}+\ket{112,q_2^{[1]}q_1^{[1]}q_1^{[2]}}+\ket{112,q_3^{[1]}q_1^{[1]}q_2^{[2]}}\right)\\+\frac{1}{2\sqrt{3}} \left(\ket{112,q_2^{[1]}q_2^{[1]}q^{[2]}_1}-\ket{112,q_2^{[1]}q_3^{[1]}q_2^{[2]}}-\ket{112,q_3^{[1]}q_2^{[1]}q_2^{[2]}}-\ket{112,q_3^{[1]}q_3^{[1]}q_1^{[2]}} \right),
 \end{gathered}
 \label{eq:WKron112}
 \end{equation}
 in this particular case, one can check that $\ket{\hat{\mathcal{K}}^W_{112}}=\ket{\mathcal{K}^W_{112}}$ i.e., the resultant state is already normalized. It is always possible to understand the obtained Kronecker states as LME states for multi-qudits, where each qudit dimension corresponds to the dimension of $f^{[\lambda^i]}$. For this example, the obtained Kronecker state corresponds to an LME state of two qutrits and one qubit. The mapped state is obtained by dropping the $\bm{\lambda}$ label and making the assignations:
  \begin{equation*}
  q^{[1]}_1 \rightarrow 0, \, q^{[1]}_2 \rightarrow 1, \, q^{[1]}_3 \rightarrow 2, \, q^{[2]}_1 \rightarrow 0, \, q^{[2]}_2 \rightarrow 1, 
  \end{equation*}
  then, the state becomes:
 \begin{equation*}
    \ket{\hat{\mathcal{K}}^W}= \frac{1}{\sqrt{6}}\left(\ket{010}+\ket{021}+\ket{100}+\ket{201}\right)+\frac{1}{2\sqrt{3}} \left(\ket{110}-\ket{121}-\ket{211}-\ket{220} \right).
 \end{equation*} 
\noindent This process can be used to find one set of CGC for each partition set $\bm{\lambda} \in \Lambda_n^W$ by using Equation \eqref{eq:CGCKron} normalizing the obtained state. For example,  from the obtained state for $n=4$, $\ket{\mathcal{K}^W_{112}}$,  which casually is already normalized, we can read the CGC by multiplying each coefficient by $\sqrt{f^{[\lambda^3]}}=\sqrt{2}$ obtaining: 
\begin{equation*}
\begin{gathered}
    C^{[1][1][2]}_{q_1^{[1]},q_2^{[1]},q_1^{[2]}} =C^{[1][1][2]}_{q_1^{[1]},q_3^{[1]},q_2^{[2]}} =C^{[1][1][2]}_{q_2^{[1]},q_1^{[1]},q_1^{[2]}} =C^{[1][1][2]}_{q_3^{[1]},q_1^{[1]},q_2^{[2]}}= \frac{1}{\sqrt{3}},\\
     C^{[1][1][2]}_{q_2^{[1]},q_2^{[1]},q_1^{[2]}} =-C^{[1][1][2]}_{q_2^{[1]},q_3^{[1]},q_2^{[2]}} =-C^{[1][1][2]}_{q_3^{[1]},q_2^{[1]},q_2^{[2]}} =-C^{[1][1][2]}_{q_3^{[1]},q_3^{[1]},q_1^{[2]}}= \frac{1}{\sqrt{6}}.
\end{gathered}
\end{equation*}
These values agree with the ones in \cite{gao}, where the authors show tables of CGC values up to $S_6$. With the method shown here, we calculated exactly (in rational form) all W-Kronecker states up to $S_{13}$, where vectors with more than six million coefficients appear.\\

\noindent To summarize this chapter, we recall that the main goal of this document is to present a procedure to calculate Kronecker states for any set $\bm{\lambda}$, where each $\lambda^i$ consists of at most two parts, which are the possible irreps when studying qubit systems. Kronecker states constitute	 the invariant subspace of the tensor product of irreps of $S_n$:
\begin{equation*}
\ket{\mathcal{K}_{\bm{\lambda}}} \in \left([\lambda^1] \otimes [\lambda^2] \otimes\dots \otimes  [\lambda ^N] \right)^{S_n}.
\end{equation*}
A vector space with a dimension given by the Kronecker coefficient $k_{\bm{\lambda}}$. This problem can be reformulated by exploiting the connection between Kronecker states and Clebsch Gordan Coefficients of the symmetric group. This connection is made explicit in terms of the coefficients of Kronecker states in Equation \eqref{eq:CGCKron}. However, finding such CGC is not a simple task. The procedure we will introduce later to calculate Kronecker states must also be understood as a tool to compute the CGC efficiently.\\ 

\noindent One efficient method for a special class of Kronecker states is obtained from the entanglement concentration protocol. When analyzing this problem in bipartite systems, we obtain an expression for all bipartite Kronecker states shown in Equation \eqref{eq:2partKronecker} by applying the Schur transform to $n$ copies of bipartite states. It is highlighted that the entanglement of the $n$ copies is concentrated in the Kronecker states. A multipartite generalization of the protocol can be obtained when using states in the multipartite W class. Despite the multiplicity of the invariant subspace, $k_{\bm{\lambda}}$, the mathematical structure of states in this class implies that, as a result of the Schur transform, the obtained state in $\{\bm{\lambda}\}$ is unique for each set $\bm{\lambda}$. This makes it possible to separate the obtained state from the corresponding Kronecker state in $[\bm{\lambda}]^{S_n}$ after a projection onto the set of partitions $\bm{\lambda}$.\\

\noindent More relevant for this work is that the Schur transform, in this case can be done by focusing on the W state itself, where a recurrence construction can be obtained from Equations \eqref{eq:RecKron},  \eqref{eq:Flambda} and \eqref{eq:KroneckerW} to calculate any Kronecker state of $N$ parts corresponding to the W class. Despite the simplicity and efficiency of this approach,
there are two problems. First,  the sets $\bm{\lambda}$ allowed for this construction are defined by $\Lambda^{W}_n$ in Equation \eqref{eq:WLambda}, making it impossible to find Kronecker states for some sets $\bm{\lambda}$. The second limitation is that this construction for sets with $k_{\bm{\lambda}}>1$ does not provide more than one Kronecker state; it calculates only one, which we call the W-Kronecker state.\\

\noindent The main goal of this thesis is to present a procedure for calculating Kronecker subspaces in any set $\bm{\lambda}$, which requires to obtain $k_{\bm{\lambda}}$ orthogonal Kronecker states. The procedure we will introduce can be used for any set $\bm{\lambda}$, and we can ensure that by using our algorithm, it is possible to build a complete basis of $k_{\bm{\lambda}}$ orthogonal Kronecker states, allowing the construction of any Kronecker subspace that appear in the Schur basis of qubit systems. The thought process leading to this result is rather simple: we know how to calculate Kronecker states of bipartite states (Equation \eqref{eq:2partKronecker}), and we also know how to calculate Kronecker states in the W class, where the W state is a simple representative (Equation \eqref{WKronecker}). If we want to calculate general Kronecker states appearing in the Schur basis of more general states, we can do it by first building multipartite general states using only bipartite states and W-states and then looking at the corresponding Schur transform to find the corresponding Kronecker states. In the next chapter, we will show the first step of this procedure in a construction that we name \textit{W-state stitching}.

\chapter{Multipartite state construction from W state stitching}
\label{Chapter4}
In this section, we present a graphical method to build general multipartite qubit states using $W_3$ states and bipartite states. In this construction, we build graph states, where vertices represent $W_3$ states and inner edges are bipartite states. In this way, multiqubit states can be obtained by projecting bipartite states, which we call \textit{stitches}, with copies of $W_3$ states, where the parts to be projected can be read from the corresponding graph, and the coefficients defining the stitches are used to modify the obtained state. Furthermore,  using a suitable representation of the stitches as elements of $SL_2\times \mathbb{Z}_2$ (matrices of dimension $2\times 2$ with determinant either $1$ or $-1$), it is possible to define a set of graphical rules, that can be used to identify the parameters of the stitches that are relevant for SLOCC classification. With this, it is possible to find the \textit{minimal graph} from which different SLOCC classes can be obtained. \\

\noindent The idea of representing quantum states in a graphical notation is not new; for example, the very well-known approach of graph states \cite{Graphstates} uses vertices to represent qubits and edges to represent the interaction between them; this construction has a broad range of applications in quantum error correction and quantum computation. Another very relevant graphical method is the  $ZW$ calculus \cite{ZW}, which we learned about during the final stage of this investigation. In this method, processes are represented by diagrams, where each part of the system is represented by a line that can interact with other elements or lines. The nature of interactions is given by the elements and connections in each path. This construction is very rich, and it has been shown that it is complete, meaning that any quantum operation in the Hilbert space formalism can be expressed in  the graphical calculus \cite{ZXW}. Our construction shares many features with the ZW-calculus. However, the focus of our construction is to identify multiqubit SLOCC classification, which has not been explored deeply in ZW-calculus \cite{ZW} \cite{SLOCCZW}. So it involves some modifications of the ZW-calculus tailored to our aims. The relation between the ZW notation and our graphical representation is presented in Appendix \ref{Appendix04}. This mapping allow us to ensure that with the W-state stitching it is possible to obtain any multiqubit state. \\

\noindent In the ZW calculus, $W$ states are special states as they are for us, and some of the rules we present here have also been obtained in this construction. However, what differs in our construction is that we exploit a useful feature of bipartite entangled states that fits nicely with the stabilizer group of $W$ states; namely,  that any possible entangled bipartite state can be parametrized up to a scale factor, in terms of a $2\times 2$ matrix of unit-determinant acting in one of the parts of the maximally entangled state $\bra{\Phi^+}$. By using the invariance properties of $\bra{\Phi^+}$, the state can be reparametrized in one of two possible ways:either as two upper triangular matrices acting on different parts of the state $\bra{\Phi^{+}}$, or, as an upper triangular matrix, acting on one of the parts of the state $\bra{\Psi^{+}}$. Using this parametrization in the graph states, such upper triangular matrices can be read as acting on the $W_3$ states. We exploit the symmetries of $W$ states under local actions of upper triangular matrices to reinterpret these actions in a way that simplifies graph states according to SLOCC equivalence. This construction leads to graph states that serve as representatives of SLOCC classes. There is also a difference in how our graphs must be read compared to those of ZW calculus. Our language emphasizes states over processes; this essential difference means that our graphs do not have a direction associated with them, allowing us to ``move'' objects through the graph. This freedom facilitates  SLOCC classification.\\

\noindent  We begin this chapter by introducing the stitching procedure and the graphical notation that will be utilized. This notation was adapted to the one used in $ZW$ calculus to avoid confusion for readers familiar with the method. Subsequently, we explore a graphical interpretation of the symmetries within the $W_3$ state, which we refer to as \textit{Parameters pushing}, which permits us to identify what parameters used in the multiqubit construction are relevant under SLOCC classification. Applying this technique allows us to simplify the various graphs that can be generated, reducing the relevant parameters under the SLOCC classification for each graph. We explore this construction for the cases of two, three, and four qubits, finding graphs representing states in all the possible SLOCC classes.\\

\section{The W-state Stitching Procedure}
	\textit{Stitching} is a method for constructing multiqubit states using graph states with two building blocks: the $W_3$ state as kets and generally entangled bipartite qubit states as bras. Given that states only differing by a complex scale factor represent the same physical object, we will work with unnormalized states to simplify the calculations. First, let us introduce the objects and the language that will be used in the construction.

\subsection{Objects and notation}
The bipartite state, or from now on, \textit{the stitch}, can be written  in general as:
\begin{equation*}
\bra{\varsigma}= \varsigma_{00}\bra{00}+\varsigma_{01}\bra{01}+\varsigma_{10}\bra{10}+\varsigma_{11}\bra{11}.
\end{equation*}
This state can be written in an equivalent form by considering a matrix action on one of the parts of a maximally entangled state $\bra{\Phi^{+}}$ as:
\begin{equation}
    \bra{\varsigma}= \bra{\Phi^{+}} I\otimes  \bm{\varsigma} , \quad  
     \bm{\varsigma} := \left( \begin{array}{cc}
          \varsigma_{00}&\varsigma_{01}  \\
         \varsigma_{10} & \varsigma_{11}
     \end{array}\right),
     \label{eq:Stitch}
\end{equation}
where $\bra{\Phi^{+}}$ is the (unnormalized) Bell state \cite{Nielsen}:
\begin{equation*}
\bra{\Phi^{+}}=  \bra{00} +\bra{11}.
\end{equation*}
Note how Equation \eqref{eq:Stitch} materializes the fact that for two qubits, any entangled state is connected to the maximally entangled state through SLOCC operations. In particular, we can restrict $\bm{\varsigma}$ to be unit-determinant i.e., $\bm{\varsigma}\in SL_2$. By doing this, there are three different ways to parametrize the matrix. The first is by fixing $\varsigma_{11}=\frac{1+\varsigma_{01}\varsigma_{10}}
{\varsigma_{00}}$:
\begin{equation}
\bm{\varsigma^1} =\left( \begin{array}{cc}
          \varsigma_{00}&\varsigma_{01}  \\
         \varsigma_{10} & \frac{1+\varsigma_{01}\varsigma_{10}}{\varsigma_{00}}
     \end{array}\right).
 \label{eq:sigmaunit}
\end{equation}
Generically, the unit-determinant matrix $\bm{\varsigma^1}$ admits an LU decomposition, that is a product of a lower triangular matrix and an upper triangular matrix of the form:
\begin{equation*}
\bm{\varsigma^1} = \left( \begin{array}{cc}
         1&0  \\
         \frac{\varsigma_{10}}{\varsigma_{00}}& 1
     \end{array}\right) \cdot \left( \begin{array}{cc}
          \varsigma_{00}&\varsigma_{01}  \\
        0 & \frac{1}{\varsigma_{00}}
     \end{array}\right)
\end{equation*}
The key idea behind parameter pushing construction will be to use symmetries of the multipartite states $\bra{\Phi^+}$ and $\ket{W_3}$, to \textit{move} actions from one of the parts of the state to another (or to others). In this case, what we consider is the invariance of $\bra{\Phi^+}$, that is:
\begin{equation}
\bra	{\Phi^{+}}=\bra	{\Phi^{+}}A \otimes \left( A^{-1}\right)^{T}\Rightarrow  \bra	{\Phi^{+}} A \otimes I =\bra	{\Phi^{+}} I \otimes A ^{T}.
\label{eq:Phisym}
\end{equation}
with $A\in GL_2$. Thus, we can move an operator acting to the right on one part to the other part by transposing it. Then, the stitch $\bra{\varsigma^1}$, can be parametrized by two upper triangular matrices acting on the parts of $\bra{\Phi^{+}}$ as:
\begin{equation}
 \bra{\varsigma^1}= \bra{\Phi^{+}} I\otimes \left( \begin{array}{cc}
         1&0  \\
         \frac{\varsigma_{10}}{\varsigma_{00}}& 1
     \end{array}\right) \cdot \left( \begin{array}{cc}
          \varsigma_{00}&\varsigma_{01}  \\
        0 & \frac{1}{\varsigma_{00}}
     \end{array}\right) =\bra{\Phi^{+}} \left( \begin{array}{cc}
         1& \frac{\varsigma_{10}}{\varsigma_{00}}  \\
        0& 1
     \end{array}\right)\otimes  \left( \begin{array}{cc}
          \varsigma_{00}&\varsigma_{01}  \\
        0 & \frac{1}{\varsigma_{00}}
     \end{array}\right).
     \label{eq:stitch2}
\end{equation}
Upper triangular matrices of unit-determinant will be recurrent objects that depend only on two parameters, so we will denote them by:
\begin{equation}
u(v,w)= \left( \begin{array}{cc}
          v&w  \\
        0 & \frac{1}{v}
     \end{array}\right).
     \label{eq:udef}
\end{equation}
We can also denote upper triangular matrices with diagonal fixed to 1 as:
\begin{equation}
\mathcal{A}(w)=u(1,w)= \left( \begin{array}{cc}
          1&w  \\
        0 & 1
     \end{array}\right).
     \label{eq:Adef}
\end{equation}
In terms of $u(v,w)$ and $\mathcal{A}(w)$, we can rewrite Equation \eqref{eq:stitch2} as:
\begin{equation}
\bra{\varsigma^1}=\bra{\Phi^{+}}\mathcal{A}\left(\frac{\varsigma_{10}}{\varsigma_{00}} \right) \otimes u(\varsigma_{00},\varsigma_{01}).
\label{eq:paramstitch}
\end{equation}
Note how this parametrization requires that $\varsigma_{11} \neq 0$ and $\varsigma_{00} \neq 0$. These two cases require a different parametrization. The first case is when the last entry of the matrix is $\varsigma_{11}=0$. In that case, the unit-determinant matrix is parametrized as:
\begin{equation*}
\bm{\varsigma}^2=\left( \begin{array}{cc}
         \varsigma_{00}&\frac{-1}{\varsigma_{10} } \\
         \varsigma_{10}& 0   \end{array}\right) ,
\end{equation*}
 after the LU decomposition on this matrix, we have the following:
\begin{equation*}
\bm{\varsigma}^2=\left( \begin{array}{cc}
        1&0  \\
         \frac{\varsigma_{10}}{\varsigma_{00}}& 1   \end{array}\right)\cdot \left( \begin{array}{cc}
         \varsigma_{00}&-\frac{1}{\varsigma_{10}}  \\
         0& \frac{1}{\varsigma_{00}}   \end{array}\right),
\end{equation*}
which, by the symmetry of $\bra{\Phi^{+}}$, can be expressed as:
\begin{equation*}
\bra{\varsigma^2}=\bra{\Phi^{+}} \mathcal{A}\left(\frac{\varsigma_{10}}{\varsigma_{00}} \right) \otimes  u (\varsigma_{00},\varsigma_{01}),
\end{equation*}
where $\varsigma_{01}=-\frac{1}{\varsigma_{10}}$;
therefore, this exception can be parametrized exactly as in Equation \eqref{eq:paramstitch}. The other exception is when the first entry of the matrix is $\varsigma_{00}=0$. In this case, the matrix is:
\begin{equation*}
\bm{\varsigma^3}= \left( \begin{array}{cc}
         0&  \frac{1}{\varsigma_{10}} \\
        \varsigma_{10}& \varsigma_{11}   \end{array}\right) ,
        \label{eq:sigmaunit2}
\end{equation*}
where now we chose the matrix to have a determinant equal to $-1$. We can do this since the sign of the determinant changes by multiplying the matrix by the imaginary number $i$, so, up to a scale factor, this matrix is equivalent to a matrix in $SL_2$. The approach for this matrix is slightly different because it does not admit an LU decomposition, but we can write it as the $\sigma_x$ matrix acting on an upper diagonal matrix as:
\begin{equation*}
\bm{\varsigma^3}= \sigma_x \cdot \left( \begin{array}{cc}
        \varsigma_{10}& \varsigma_{11} \\
        0&  \frac{1}{\varsigma_{10}}   \end{array}\right).
\end{equation*}
We can use the symmetry of $\bra{\Phi^{+}}$ to move the action of $\sigma_x$ to the other part, as $\sigma_x$ matrix is symmetric, it does not change when changing the part where is acting on the $\bra{\Phi^{+}}$ state. After using the symmetry of $\bra{\Phi^{+}}$, and using the notation of \eqref{eq:udef} we have:
\begin{equation}
\bra	{\varsigma^3} = \bra{\Phi^{+}} \sigma_x \otimes u(\varsigma_{10},\varsigma_{11}) = \bra{\Psi^{+}} I \otimes u(\varsigma_{10},\varsigma_{11}) ,
 \label{eq:paramstitch2}
\end{equation}
where in the last step we let $\sigma_x$ act on one of the parts of $\bra{\Phi^{+}}$ to get $\bra{\Psi^{+}}$. This parametrization is, in essence, different from \eqref{eq:paramstitch}. Up to this point, we know that we can parametrize any bipartite entangled state in one of two ways: The first is the result of acting with upper triangular matrices on each part of $\bra{\Phi^+}$, which we call the \textit{$\Phi$ stitch}, or, the second,   the result of acting with one upper triangular matrix on one of the parts of a $\bra{\Psi^{+}}$, which we call the \textit{$\Psi$-stitch}. Before continuing, let us define another class of upper diagonal matrix that will appear recurrently, namely diagonal matrices of unit-determinant:
\begin{equation}
\mathcal{B}(v)=u(v,0)= \left( \begin{array}{cc}
        v& 0 \\
        0&  \frac{1}{v}   \end{array}\right) .
\label{eq:Bdef}
\end{equation}
Note how any upper diagonal matrix $u(v,w)$ can be decomposed as a product of $\mathcal{A}(vw)$  and $\mathcal{B}(v)$ :
\begin{equation*}
\mathcal{A}(vw) \cdot \mathcal{B}(v) = \left( \begin{array}{cc}
        v& w \\
        0&  \frac{1}{v}   \end{array}\right) = u(v,w).
\end{equation*}
The other key elements of the construction are the states to be stitched, the $W_3$ states. The unnormalized $W_3$ state is:
\begin{equation*}
\ket{W_3} =\ket{100}+\ket{010}+\ket{001}.
\end{equation*}
The choice of using W states in our construction was firstly motivated by the fact that we know how to calculate W-Kronecker states, and that could be exploited for more general cases as we will see in the next chapter. However, it will be seen soon that W states symmetries fit nicely with the purpose of identifying SLOCC equivalences. Note that using only bipartite states, it is impossible to obtain states of more than two qubits using projections. Therefore, it is necessary to include three-qubit states in the construction. To reproduce entangled states, we cannot use states with separable qubits, so we are left with the $ GHZ$ state and the $W$ state of three parts as the simplest candidates. We will show later that picking the $W$ state for the construction is useful, as it has a \textit{stabilizer group} related with the two possible parametrizations of the stitches in Equations \eqref{eq:paramstitch} and  \eqref{eq:paramstitch2}. Also, by using the construction with the $W_3$ state, the $GHZ$-state can be obtained so that we can focus on the construction with the $W_3$ state.\\

\noindent Now, we will introduce a graphical notation for all the objects so far defined. First, we represent the $W_3$ state as a vertex with three edges as:
\begin{equation*}
    \ket{W_3}=\begin{tikzpicture}[basel={-.5}]
  \draw[markx={1}] (0,0) to (0.5,0);
  \draw (0.5,0) to (1,0.5);
  \draw (0.5,0) to (1,-0.5);
  \draw (-0.5,0) node {$A$};
   \draw (1.2,0.5) node {$B$};
    \draw (1.2,-0.5) node {$C$};  
\end{tikzpicture},
\end{equation*}
where $A,B$ and $C$ label the three parts of the state. This is the same notation for W states as for ZW calculus, except that, since our construction involves states and not processes, the direction of the edges have no meaning. This is in contrast to the ZW-calculus, where the directions are related to the direction of the flow of time in the process. In this construction, each edge corresponds to one of the parts of the represented state. W vertices are joined by bipartite states, which will be  represented by two-edge objects. As we already know, any bipartite stitch can be written in terms of $\bra{\Phi^{+}}$ and $\bra{\Psi^{+}}$, which are related by the matrix operation $\sigma_x$, we can define a graphical representation for them as edges with or without a $\bullet$:
\begin{equation*}
\bra	{\Phi^{+}}=  \begin{tikzpicture}[basel=-.55];
\Genstates{(-0.2,0)};
\draw(0,0) to (0.8,0);\Genstates{(1,0)};
\end{tikzpicture}, \quad  \bra{\Psi^{+}}=  \begin{tikzpicture}[basel=-.55]
\Genstates{(-0.2,0)};
\draw[markx=0.5](0,0) to (0.8,0);
\Genstates{(1,0)};
\end{tikzpicture},
\end{equation*}
where the $\bullet$ symbol represents the matrix $\sigma_x$, and the two extremes of each line represent one of the parts of the state. The symbol ``$\wr$'' at the extremes is used to specify that these indices are to be contracted to something else to be defined as bras. For example:
\begin{equation}
\begin{tikzpicture}[basel={-.5},every node/.style={scale=0.6}] 
  \draw (0,0) to (-0.5,0.5);
  \draw (0,0) to (-0.5,-0.5);
    \draw[markx={0},markx={1}] (0,0) -- (2,0);
  \draw (2,0) to (2.5,0.5);
  \draw (2,0) to (2.5,-0.5);
   \draw [line width=0.5pt, double distance=1.5pt, arrows = {-Latex[length=0pt 2 0]},gray] (1,-0.2) -- (1,-0.7);
  \draw (1,-1) node {$\bra{\Phi^{+}}$};  
\end{tikzpicture},
\end{equation}
where the inner line is a bra, as is contracting W states in both sides. We always will interpret external edges as indices of a ket state. When having these objects free, i.e., without being connected to something else, they correspond to:
\begin{equation*}
\ket	{\Phi^{+}}= \begin{tikzpicture}[basel=-.5]
\draw(0,0) to (0.8,0);
\end{tikzpicture}, \quad  \ket{\Psi^{+}}= \begin{tikzpicture}[basel=-.5]
\draw[markx=0.5](0,0) to (0.8,0);
\end{tikzpicture}.
\end{equation*}
For the different upper diagonal matrices, we define the graphical representations:
\begin{equation*}
    \mathcal{B}(v) =\tikz {\filldraw [color=black!, fill=white!] (0,0) circle[radius=3pt];
    \node[above=1mm] {$v$} (0.1,0.2);}, \quad  \mathcal{A}(w)= \tikz{
   \draw [\rao]
           (0,0) -- (0.2,0) ; \draw (0.2,0.3) node {$w$}
   }
   \quad  u(v,w)= \tikz{
   \draw  [\rac]           (0,0) -- (0.2,0)  ; \draw (0.1,0.3) node {$v,w$};
  }.
\end{equation*}
This notation was chosen to represent with clarity in which part each matrix is acting; we will show later that \textit{balls} ($\mathcal{B}$) can be slid along the lines so their actions can be unambiguously understood as acting on any of the parts; this is not the same with the \textit{arrows} ($\mathcal{A},u$), which in general have a well-defined direction of action. Often we refer to these actions as \textit{decorations} on the lines.\\

\noindent By using the graphical notation, the two stitches parametrized in \eqref{eq:paramstitch} and \eqref{eq:paramstitch2} can be represented. The first stitch, named the $\Phi$-stitch, is a $\bra{\Phi^{+}}$ state (represented by a line) with an upper triangular matrix of diagonal 1 acting in one part (represented by a \textit{white arrow} on one of the extremes), and an upper triangular matrix in the other part (represented by a \textit{black arrow} on the other extreme). The second stitch, named the $\Psi$-stitch, is a $\bra{\Psi^{+}}$ state (represented by a line with a $\bullet$) with an upper triangular matrix acting on one of the parts (represented by a black arrow on one of the extremes). Then, the graphical representations of the stitches are:
\begin{equation*}
   \bra{\varsigma^\Phi} =  \begin{tikzpicture} [basel={-1.5},every node/.style={scale=0.6}]
   \Genstates{(-0.2,0)};
       \draw [\lao \rac]            (0,0) -- (2,0) ;   
   \draw (0,0.3) node {$\varsigma_{10}/\varsigma_{00}$};
   \draw (2,0.3) node {$\varsigma_{00},\varsigma_{01}$};   
   \Genstates{(2.2,0)};
   \end{tikzpicture} , \quad \bra{\varsigma^\Psi} =\begin{tikzpicture}
        [basel={-1.5},every node/.style={scale=0.6}]
        
   \Genstates{(-0.2,0)};
   
   \Genstates{(2.22,0)};
        \draw [\rac]        (0,0) -- (2,0) ;  ;\draw[markxi={0.5}] (0,0)--(2,0) ;
        \draw (2,0.3) node {$\varsigma_{10},\varsigma_{11}$};         \end{tikzpicture} ,
        \label{eq:stitch2g}
\end{equation*}
where decorations on the ends of the edges mean that the matrix represented by the object is acting on the part represented by that end. It is worth recalling that by the symmetries of $\bra{\Phi^+}$, we can move actions from one part to another by transposing them; this means that symmetric matrices such as  $\bullet$ and $\tikz {\filldraw [color=black!, fill=white!] (0,0) circle[radius=3pt];}$ can be moved freely along edges, and interpreted in any of the parts of the edge unambiguously. Therefore, these objects are usually drawn in the middle of lines, but they can be moved through the lines if needed. Before delving into the properties of the objects that we will use, let us define how we use them to build multiqubit states.
    
\subsection{Multiqubit construction}

The idea behind the stitching process is to build a graph state $\ket{G}$ by taking the tensor product of $\omega$ copies $\ket{W_3}$ states and \emph{stitching} them together by projecting with a number $s$ of bipartite $\bra{\varsigma}$ states, between parts of $W_3$ states. The graph state $\ket{G}$ is:
\begin{equation*}    \ket{G}=\bigotimes_{i=1}^{s}\bra{\varsigma^{i}}\bigotimes_{i=1}^{\omega}\ket{W_3},
\label{eq:GraphConstruction}
\end{equation*}
where the contractions between $\bra{\varsigma}$ and $\ket{W_3}$ are determined by the topology of the graph state $G$. The simplest example is shown in Figure \ref{fig:Tree4}, where two $W_3$ states are stitched with a two-particle state $\bra{\varsigma^{}}$ to obtain a four-qubit state, and, from now on, when the parameters are not shown in the graph, it means that all parameters are generic and independent. Notice that only uncontracted parties are labeled. Due to the projections, actions on edges of bipartite states can also be understood as actions on the correspondent edges of the $W_3$ state in the vertex. For example, the state in Figure \ref{fig:Tree4} is:
\begin{equation*}
\begin{gathered}
\bra{\Phi^{+}}_{EF} \mathcal{A}(w_2)_E \otimes u(v,w_2)_F \ket{W_3}_{ABE} \ket{W_3}_{CDF} \\
=\bra{\Phi^{+}}_{EF} \left(\bm{I}\otimes \bm{I} \otimes  \mathcal{A}(w_2)_E \right) \ket{W_3}_{ABE} \left(\bm{I} \otimes \bm{I}\otimes  u(v,w_2)_F  \ket{W_3}_{CDF} \right),
\end{gathered}
\end{equation*}
where we label the parts as on the left side of the figure and reinterpret the actions of arrows on $\bra{\Phi^{+}}$ as the same actions on the correspondent parts of the $W_3$ states.\\

\begin{figure}[h!]
\centering
\begin{tikzpicture}[basel={-.5},every node/.style={scale=0.6}]  
  \draw[markx={1}] (-0.5,0) to (-1,0);
  \draw (-1,0) to (-1.5,0.5);
  \draw (-1,0) to (-1.5,-0.5);
  \draw (-1.7,0.7) node {$A$};
  \draw (-1.7,-0.7) node {$B$};
  \draw (-0.5,-0.2) node {$E$};
    \draw[\lao \rac ] (0,0) -- (2,0);
 \draw[markx={1}] (2.5,0) to (3,0);
  \draw (3,0) to (3.5,0.5);
  \draw (3,0) to (3.5,-0.5);
  \draw (3.5,0.7) node {$C$};
  \draw (3.5,-0.7) node {$D$};
  \draw (2.2,-0.2) node {$F$};
\end{tikzpicture} = 
\begin{tikzpicture}[basel={-.5},every node/.style={scale=0.6}]  
  \draw [markx={0}](-1,0) to (-1.5,0.5);
  \draw (-1,0) to (-1.5,-0.5);
  \draw (-1.7,0.7) node {$A$};
  \draw (-1.7,-0.7) node {$B$};
    \draw[\lao \rac ] (-1,0) -- (0,0);
  \draw[markx={0}] (0,0) to (0.5,0.5);
  \draw (0,0) to (0.5,-0.5);
  \draw (0.5,0.7) node {$C$};
  \draw (0.5,-0.7) node {$D$};
\end{tikzpicture}
\caption{Graph in tree form, stitching two $W_3$ states with a stitch $\varsigma$, to obtain a four-qubit state. The contraction occurs in the E and F parties of the $\ket{W_3}$ states, which are not labeled in the resulting $\ket{G}$ state.}
\label{fig:Tree4}
\end{figure}

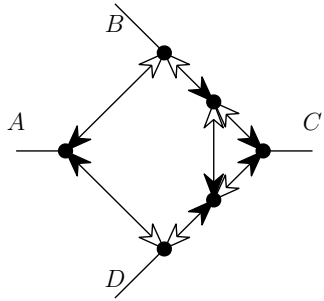
\begin{figure}[h!]
\centering
\scalebox{1.3}{
     \begin{tikzpicture}[baseline={([yshift=-.5ex]current bounding box.center)},every node/.style={scale=0.6}]
    \draw[\lao \rac](1,1) -- (0,0);
     \draw[\lao \rac](1,-1) -- (0,0);
      \draw[\lao \rac](1,1) -- (1.5,0.5);
       \draw[\lao \rac](1,-1) -- (1.5,-0.5);
    \draw[\lao \rac](1.5,-0.5) -- (2,0);
    \draw[markxi={0}](1.5,0.5) -- (2,0);
    \draw[markxi={0}](1.5,-0.5) -- (2,0);
    \draw[\lao \rac](1.5,0.5) -- (2,0);
      \draw[\lao \rac](1.5,0.5) -- (1.5,-0.5);
\draw[markx={1}](-0.5,0) -- (0,0);
\draw[markx={0}](2,0) -- (2.5,0);
\draw[markx={0}](1,1) -- (0.5,1.5);
     \draw[markx={0}](1,-1) -- (0.5,-1.5);
     \draw (-0.5,0.3) node {$A$};
      \draw (0.5,1.3) node {$B$};
       \draw (2.5,0.3) node {$C$};
        \draw (0.5,-1.3) node {$D$};
    \end{tikzpicture}}
\caption{An example of the Stitching procedure, where six $W_3$ states are stitched by seven $\varsigma$ states, leaving four free edges, corresponding to a four-qubit state.}
\label{fig:GenStitch}
\end{figure}

\noindent The state obtained in the stitching procedure depends on the structure of the graph and the parameters of the stitch. Such graphs allow inner loops and as much structure as desired. The number of qubits in the resultant state corresponds to the number of external edges in the final state, which is given by:
\begin{equation}
N=3 \omega -2s.
\label{eq:Numberofparts}
\end{equation}
Let us use the graph in Figure \ref{fig:Tree4} to build a four-qubit state and show how multiqubit states are obtained in this process. For this example, take:
\begin{equation*}
\bra{\varsigma} =  \begin{tikzpicture} [basel={-1.5},every node/.style={scale=0.6}]
   \Genstates{(-0.2,0)};   
   \Genstates{(2.2,0)};
       \draw [\lac-]            (0,0) -- (2,0) ;   
   \draw (0,0.3) node {$v,w$};
   \end{tikzpicture} = \bra{\Phi^{+}} \left( \begin{array}{cc}
        v& w \\
        0&  \frac{1}{v}   \end{array}\right) \otimes \bm{I}.
\end{equation*}
Then, the graph state to calculate is:
\begin{equation*}
\begin{tikzpicture}[basel={-.5},every node/.style={scale=0.6}]  
  \draw [markx={0}](-1,0) to (-1.5,0.5);
  \draw (-1,0) to (-1.5,-0.5);
  \draw (-1.7,0.7) node {$A$};
  \draw (-1.7,-0.7) node {$B$};
    \draw[\lac-] (-1,0) -- (0,0);
  \draw[markx={0}] (0,0) to (0.5,0.5);
  \draw (0,0) to (0.5,-0.5);
  \draw (0.5,0.7) node {$C$};
  \draw (0.5,-0.7) node {$D$};
  \draw (-0.8,0.3) node {$v,w$};
\end{tikzpicture}.
\end{equation*}
When this contraction is mapped back to the vector and matrix notation, one obtains:
\begin{equation*}
\left(\bra{00}_{EF} +\bra{11}_{EF} \right) (\bm{I}\otimes \bm{I}\otimes \left( \begin{array}{cc}
        v& w \\
        0&  \frac{1}{v}   \end{array}\right)  \left(\ket{100_E}+\ket{010_E}+\ket{001_E} \right) \left (\ket{100_F}+\ket{010_F}+\ket{001_F}\right),
\end{equation*}
where we chose to contract the last part of each $W_3$ labeled with $E$ and $F$ respectively. By doing this contraction, one gets:
\begin{equation}
\begin{tikzpicture}[basel={-.5},every node/.style={scale=0.6}]  
  \draw [markx={0}](-1,0) to (-1.5,0.5);
  \draw (-1,0) to (-1.5,-0.5);
  \draw (-1.7,0.7) node {$A$};
  \draw (-1.7,-0.7) node {$B$};
    \draw[\lac-] (-1,0) -- (0,0);
  \draw[markx={0}] (0,0) to (0.5,0.5);
  \draw (0,0) to (0.5,-0.5);
  \draw (0.5,0.7) node {$C$};
  \draw (0.5,-0.7) node {$D$};
  \draw (-0.6,0.3) node {$v,w$};
\end{tikzpicture}  = \begin{array}{c}
v \left(\ket{1010}+ \ket{1001}+ \ket{0110} +\ket{0101} \right) + \\
w \left(\ket{0010}+\ket{0001} \right) + \frac{1}{v}\ket{0000}
\end{array} .
\label{eq:exampletree}
\end{equation}
We could calculate this state by building the vector directly without the need for the graph; however, we will endow the graphical notation with a set of rules that will allow us to simplify the calculations in a way that is much more transparent than when using braket notation. Stitching W vertices together allows us to build multiqubit states with any number of parts (or external edges). For example, in Figure \ref{fig:GenStitch}, a graph where six $\ket{W_3}$ states are stitched using seven $\bra{\varsigma}$ states results in a four qubit state ($6\cdot 3-2\cdot7=4$).
We can ensure that with W stitching, it is possible to build any multiqubit state by exploiting the connection with the ZW calculus. ZW-calculus is known to be complete \cite{SLOCCZW} for qubit systems. As any construction in ZW-calculus can be mapped to our language and vice-versa, we know that our language is also complete. In Appendix \ref{Appendix04}, we explain completely the connection between languages. \\

\noindent It can be noticed from the example in Figures \ref{fig:Tree4} and \ref{fig:GenStitch} that if one wants to explore all the possible states obtained from a given graph, the process becomes complicated very quickly. For the first one, we need to check what states are obtained with three complex parameters (one from the white arrow and two from the black arrow), and for the second example, $21$ complex parameters must be considered. However, not all parameters are important. Recalling our goal of obtaining general Kronecker states, one can notice that in the Schur basis expansion of $n$ copies of a multiqubit state,
\begin{equation*}\label{eq:Nqubits}
\ket{\psi}^{\otimes n}=\bigoplus_{ \lamtup \vdash n,2} \sqrt{p(\bm{\lambda}|\psi)}\left[\sum_{i=1}^{k_{\lamtup}}\ket{\Phi_{\lamtup,i}(\psi)}\ket{\mathcal{K}_{\lamtup,i}} \right],
\end{equation*}
SLOCC operations act only in the $GL_2$ basis without changing the Kronecker subspace. Therefore, for exploring such subspace, we need to focus on identifying the SLOCC inequivalent states obtained from each graph. This process leads to many simplifications, as we shall now see.

\subsection*{SLOCC equivalent graph states}

To identify inequivalent SLOCC states, we must identify when two graph states are SLOCC equivalent. Introducing the following notation to represent a generic graph state of $N$ parts:
\begin{equation*}
\begin{tikzpicture}[basel=-.5,every node/.style={scale=0.5}]
\def\array{$1$,$2$,$$,$$,$$,$$,$$,$$}    \def\arrayx{$$,$$,$$,$$,$$,$$,$$,$N$}
    \foreach [count=\n] \x in \array{
        \node at ({90+\n*180/(8+1)}:0.6cm) (n\n) {\x};
        \draw (0,0)--(n\n); 
        };
         \foreach [count=\n] \x in \arrayx{
        \node at ($(0,0)+({-90+\n*180/(8+1)}:0.6cm)$) (n\n) {\x};
        \draw (0,0)--(n\n); 
        };
\Genstate{(0,0)};
\end{tikzpicture}.
\end{equation*}
Note that SLOCC operations are proportional to some unit-determinant matrices acting locally on the qubits of the state. Graphically, these actions can be represented as stitches on external vertices. Then, any SLOCC operation can be written up to a scale factor as stitches in external edges, and hence, two graphs that only differ by the decorations on the external edges are SLOCC equivalent. For example, given any graph of five qubits, we have the equivalence 
\begin{equation*}
\begin{tikzpicture}[basel=-.5,every node/.style={scale=0.3}]
\inWn{5}{0}{(0,0)};
\Genstate{(0,0)};
\end{tikzpicture} \cong
\begin{tikzpicture}[basel=-.5,every node/.style={scale=0.3}]
\draw (.5,-.5) to (0,0);
\Genstate{(0,0)};
\draw[\lac \rao] (1,0) to (0.2,0);
\draw[\lac \rao] (.4,.9) to (0.1,0.13);
\draw[\lac \rao] (-.9,.5) to (-0.16,0.1);
\draw[markxi={.5}](-.7,-.7) to (-0.13,-0.1);
\draw[\rac] (-.7,-.7) to (-0.13,-0.1);
\end{tikzpicture} ,
\end{equation*}
where ``$\cong$''  represents SLOCC equivalence. Thus, any invertible operation on external edges is irrelevant under SLOCC equivalence. Moreover,  we will see that many inner actions, or decorations acting on inner edges, lead to the same SLOCC orbits; for example, recall the state from the example in Equation \eqref{eq:exampletree}, and the state obtained when stitching with $\bra{\varsigma}=\bra{\Phi^{+}}$: 
\begin{equation}
 \begin{tikzpicture}[baseline={([yshift=-.5ex]current bounding box.center)},every node/.style={scale=0.6}]  
 \foreach \n in {1,...,3}{
        \node at ($(0,0)+({240+\n*360/3}:0.7cm) $)(n\n) {};
        \draw (0,0)--(n\n); }
    \draw (0,0) -- (1,0);
   \foreach \n in {1,...,3}{
        \node at ($(1,0)+({60+\n*360/3}:0.7cm) $)(n\n) {};
        \draw[markx={0},markx={1}] (0,0) to (1,0); 
        \draw (1,0)--(n\n); }
        \draw (-0.5,0.5) node {$A$};
         \draw (-0.5,-0.5) node {$B$};
 \draw (1.5,0.5) node {$C$};
  \draw (1.5,-0.5) node {$D$};
\end{tikzpicture} = \ket{0000} +\ket{W_2}\ket{W_2}.
\label{eq:exampletree2}
\end{equation}
It turns out that both states belong to the same SLOCC class, namely $L_{a00_2} (L_{abc_2} , b=c=0)$ in the classification of \cite{Verstraete}of Table \ref{table:fourqubitsclass}. In fact, the state in Equation \eqref{eq:exampletree} can be obtained from the state in Equation \eqref{eq:exampletree2} with the following SLOCC operation:
\begin{equation*}
v \cdot \begin{tikzpicture}[basel={-.5},every node/.style={scale=0.6}]  
 \foreach \n in {1,...,3}{
        \node at ($(0,0)+({240+\n*360/3}:0.7cm) $)(n\n) {};
        \draw (0,0)--(n\n); }
    \draw (0,0) -- (1,0);
   \foreach \n in {1,...,3}{
        \node at ($(1,0)+({60+\n*360/3}:0.7cm) $)(n\n) {};
         \draw[markx={0},markx={1}] (0,0) to (1,0); 
        \draw (1,0)--(n\n); }
        \draw (-0.5,0.5) node {$A$};
         \draw (-0.5,-0.5) node {$B$};
 \draw (1.5,0.5) node {$C$};
  \draw (1.5,-0.5) node {$D$};

  \ball{(-0.1,-0.2)};
  \draw[\rac] (-0.3,0.5) to (0,0);
  \draw (0.2,0.5) node {$\frac{1}{v},w$};
    \draw (0,-0.5) node {$\frac{1}{v}$};
\end{tikzpicture} =\begin{tikzpicture}[basel={-.5},every node/.style={scale=0.6}]  
 \foreach \n in {1,...,3}{
        \node at ($(0,0)+({240+\n*360/3}:0.7cm) $)(n\n) {};
        \draw (0,0)--(n\n); }
    \draw[\lac-] (0,0) -- (1,0);
   \foreach \n in {1,...,3}{
        \node at ($(1,0)+({60+\n*360/3}:0.7cm) $)(n\n) {};
        \draw (1,0)--(n\n); }
          \draw[markx={0},markx={1}] (0,0) to (1,0); 
        \draw (-0.5,0.5) node {$A$};
         \draw (-0.5,-0.5) node {$B$};
         \draw(0.2,0.3) node {$(v,w)$};
 \draw (1.5,0.5) node {$C$};
  \draw (1.5,-0.5) node {$D$};
\end{tikzpicture}.
\label{eq:exampletree3}
\end{equation*}
This simple example shows how some of the parameters of the stitches can be absorbed in SLOCC operations. In more complex cases, as, for example, the graph in Figure \ref{fig:GenStitch}, identifying the relevant information will prove to be necessary as there are $21$ complex parameters in the stitches, and in four qubits, the most general SLOCC class depends only on three complex parameters. In the next section, we present the tool that allows us to move parameters within a graph toward external edges to identify the relevant parameters in the stitching procedure.
\section{Parameter pushing}
In order to ``clean'' a graph state from information that can be absorbed in SLOCC operations, we need to exploit symmetries that allow us to rewrite operations, as we already did in Equation \eqref{eq:Phisym}, when using the symmetry of $\ket{\Phi^+}$ to parametrize the stitches. Similarly, the $W_3$ state has an interesting symmetry regarding $u$ operations that can be used to switch between inner and outer actions. \\

\noindent One of the reasons why we use W states rather than GHZ states is because the stabilizer group of W state is larger than the stabilizer group of the $GHZ$ when considering SLOCC operations. It is known that the dimension of the stabilizer set is related to the measure of the SLOCC orbits in the projective Hilbert space \cite{Bryan} \cite{Gour}. As $GHZ$ is the full measure orbit for three qubits, its stabilizer group must be of lower dimension than for the W state which is an orbit of zero measure. This relation can be easily seen by building the stabilizer of both states. For the GHZ state, the stabilizer group is the set of unit-determinant matrices $B_1,\,B_2,\,B_3$, and a complex scalar $s$ such that
\begin{equation*}
B_1\otimes B_2 \otimes B_3 \ket{GHZ} = s \ket{GHZ}.
\end{equation*}
The solution to the previous equation is achieved by the following set of discrete operations \cite{StabGHZ}:
\begin{equation*}
\sigma_x\otimes \sigma_x \otimes \sigma_x , \qquad I \otimes \sigma_z \otimes \sigma_z, \qquad \sigma_z \otimes I \otimes \sigma_z, 
\end{equation*}  
and one continuous, two-parameter operation\cite{Gour}:
\begin{equation*}
 \mathcal{B}(v_1)  \otimes \mathcal{B}(v_2) \otimes \mathcal{B}\left(\frac{1}{v_1 v_2} \right),
\end{equation*}
with $\mathcal{B}(v)$ as defined in Equation \eqref{eq:Bdef}. Then, the most general set of operations $B_1,B_2,B_3$ that leave  the $GHZ$ state invariant (up to normalization) is defined by at most two complex parameters corresponding to the two parameters in the last equation. On the other hand, for the $W$ state, the solution to
\begin{equation*}
B_1\otimes B_2 \otimes B_3 \ket{W} = s \ket{W},
\end{equation*}
is uniquely given by
\begin{equation}
u(v,w_1)\otimes u(v,w_2) \otimes u(v,-w_1-w_2) \ket{W} = v\ket{W},
\label{eq:WInvariance}
\end{equation}
with $u(v,w)$ as defined in Equation \eqref{eq:udef}. Note that this solution has three complex parameters, $v,\,w_1$, and $w_2$, which makes its stabilizer dimensionally larger than the stabilizer of the $\ket{GHZ}$ state. The extra parameter makes the orbit of the W state of measure zero	 in the projective Hilbert space of three qubits, as opposed to the orbit of the GHZ state, which is full measure. \\

\noindent In terms of the stitching method, this larger invariance of W states implies that a larger set of SLOCC operations can be rewritten in a useful form than if we used the GHZ state. In general, what we want to do is rewrite the action of an upper diagonal matrix $u$ in one of the parts by corresponding upper diagonal matrices in the other two parts; from Equation \eqref{eq:WInvariance}, this can be done as:
\begin{equation}
    u(v,w) \otimes I \otimes I \ket{W_3} = v \cdot I \otimes u(1/v ,w_1) \otimes u(1/v,w_2)  \cdot \ket{W_3}
    \label{eq:Wstabilizer},
\end{equation}
where $w_1+w_2=w$. In this construction, we will set $w_1=w,w_2=0$. This property is understood as that of \textit{Pushing} the action of an upper diagonal matrix on one of the parts to an upper diagonal matrix on one of the other parts, and a diagonal matrix in the remaining one. Graphically, the property in Equation \eqref{eq:Wstabilizer} is:
\begin{equation*}
   \begin{tikzpicture}[basel={-0.5},every node/.style={scale=0.6}]]
   \draw [\rac]            (0,0) -- (0.5,0) ;
   \draw[markx={1}] (1,0.5) -- (0.5,0);
  \draw (1,-0.5) -- (0.5,0);
   \draw (0.3,0.3) node  {$v,w$};
	 \end{tikzpicture} = v\cdot  \begin{tikzpicture}[basel={-0.5},every node/.style={scale=0.6}]]
   \draw  (0,0) -- (0.5,0) ;
  \draw [\rac](1,0.5) -- (0.5,0);
   \draw[markx={1}]  (1,-0.5) -- (0.5,0);
   \ball{(0.8,-0.3)}
   \draw (0.7,0.7)  node {$(1/v,w)$};
   \draw (0.7,-0.7) node {$1/v$};
\end{tikzpicture}.
  \label{eq:rule1}
\end{equation*}
This property is very powerful given that the stitches can be parametrized in terms of upper diagonal matrices, which are precisely the decorations that describe the symmetry of the $W_3$ state (This symmetry can also be understood in terms of the $W_N$ state, but we will focus on $W_3$). Note that $\mathcal{A}$ and $\mathcal{B}$ are just special cases of $u$, so the following properties are also true:
\begin{equation*}
     \begin{tikzpicture}[basel={-0.5},every node/.style={scale=0.6}]
   \draw[markx={1}]  (0.5,0) -- (1,0) ;
   \ball{(0.7,0)};
   \draw (1.5,0.5) -- (1,0);
   \draw (1.5,-0.5) -- (1,0);
   \draw (0.7,0.3) node {$v$};
\end{tikzpicture} = v\cdot  \begin{tikzpicture}[basel={-0.5},every node/.style={scale=0.6}]
   \draw[markx={1}]   (0.5,0) -- (1,0) ;
   \draw (1.5,0.5) -- (1,0);
   \draw (1.5,-0.5) -- (1,0);
   \ball{(1.3,0.3)};
  \ball{(1.3,-0.3)};   
   \draw (1.3,0.7) node {$1/v$};   
   \draw (1.3,-0.7) node {$1/v$};
\end{tikzpicture} , \quad  \begin{tikzpicture}[basel={-0.5},every node/.style={scale=0.6}]
   \draw [\rao]            (0.5,0) -- (1,0) ;
   \draw[markx={1}]  (1.5,0.5) -- (1,0);
   \draw (1.5,-0.5) -- (1,0);
   \draw (0.8,0.3) node {$w$};
\end{tikzpicture} = \begin{tikzpicture}[basel={-0.5},every node/.style={scale=0.6}]
   \draw [markx={1}]  (0.5,0) -- (1,0) ;
   \draw [\rao](1.5,0.5) -- (1,0);
   \draw (1.5,-0.5) -- (1,0);
    \draw (1.1,0.5) node {$w$};
\end{tikzpicture}.
   \label{eq:rule2}
\end{equation*}
Using this property, which we will call \textit{parameter pushing}, the example shown in equation \eqref{eq:exampletree3} is an obvious application of the symmetry. Parameter pushing is the main tool that we have in the construction to identify what parameters are relevant under the SLOCC classification in the W states stitching. To show how important pushing is, let us go back again to the graph state in Figure \ref{fig:Tree4}, and remember that any stitch can be parametrized in either one of the two forms in Equation \eqref{eq:stitch2g}, the $\Phi$-stitch and the $\Psi$-stitch. Then, all the states that can be obtained from the $\Phi$-stitch are:
\begin{equation}
\begin{tikzpicture}[basel={-0.5},every node/.style={scale=0.6}]
   \draw[\lac \rao] (0,0) -- (1,0) ; 
   \draw (-0.25,0.25) -- (0,0);
   \draw (-0.25,-0.25) -- (0,0);\draw (1.25,0.25) -- (1,0);
   \draw (1.25,-0.25) -- (1,0);\draw[markxi={0},markxi={1}] (0,0)--(1,0) ;
   \draw (-0.35,0.25) node{$1$};
    \draw (-0.35,-0.25) node{$2$};
     \draw (1.35,0.25) node{$3$};
      \draw (1.35,-0.25) node{$4$}; 
      \draw (0.2,0.3) node {$v,w_1$};
     \draw (0.9,0.25) node {$w_2$};      
\end{tikzpicture} = v\cdot  \begin{tikzpicture}[basel={-0.5},every node/.style={scale=0.6}]
   \draw(0,0) -- (1,0) ;
   \draw[\rac](-0.25,0.25) -- (0,0);
   \draw (-0.25,-0.25) -- (0,0);\draw[\rao] (1.25,0.25) -- (1,0);
   \draw (1.25,-0.25) -- (1,0);\draw[markxi={0},markxi={1}] (0,0)--(1,0) ;
   \draw (-0.35,0.25) node{$1$};
    \draw (-0.35,-0.25) node{$2$};
     \draw (1.35,0.25) node{$3$};
      \draw (1.35,-0.25) node{$4$}; 
      \ball{(-0.2,-0.2)};
      \draw (0,-0.4) node {$\frac{1}{v}$};
       \draw (0,0.5) node {$\frac{1}{v},w_1$};
     \draw (1,0.4) node {$w_2$};  
\end{tikzpicture} \cong \begin{tikzpicture}[basel={-0.5},every node/.style={scale=0.6}]
   \draw(0,0) -- (1,0) ;
   \draw(-0.25,0.25) -- (0,0);
   \draw (-0.25,-0.25) -- (0,0);\draw(1.25,0.25) -- (1,0);
   \draw (1.25,-0.25) -- (1,0);\draw[markxi={0},markxi={1}] (0,0)--(1,0) ;
   \draw (-0.35,0.25) node{$1$};
    \draw (-0.35,-0.25) node{$2$};
     \draw (1.35,0.25) node{$3$};
      \draw (1.35,-0.25) node{$4$};   
\end{tikzpicture} = \ket{0000} +\ket{\Psi^{+}}\ket{\Psi^{+}}.
\label{eq:tree1}
\end{equation}
In the first step, we pushed both arrows through the $W_3$ states, and in the second step, we just dropped all the outer operations, as they are SLOCC operators. Then, regardless of the three parameters of the stitch, all the states are in the same $SLOCC$ class, whose graphical representative is $\begin{tikzpicture}[basel={-0.5}, every node/.style={scale=0.6}]
   \draw(0,0) -- (1,0) ;
   \draw(-0.25,0.25) -- (0,0);
   \draw (-0.25,-0.25) -- (0,0);\draw(1.25,0.25) -- (1,0);
   \draw (1.25,-0.25) -- (1,0);\draw[markxi={0},markxi={1}] (0,0)--(1,0) ;
   \draw (-0.35,0.25) node{$1$};
    \draw (-0.35,-0.25) node{$2$};
     \draw (1.35,0.25) node{$3$};
      \draw (1.35,-0.25) node{$4$};   
\end{tikzpicture} $. We can also check to see what classes are obtained when using the $\Psi$-stitch:
\begin{equation}
\begin{tikzpicture}[basel={-0.5},every node/.style={scale=0.6}]
   \draw[\lac-] (0,0) -- (1,0) ; 
   \draw (-0.25,0.25) -- (0,0);
   \draw (-0.25,-0.25) -- (0,0);\draw (1.25,0.25) -- (1,0);
   \draw (1.25,-0.25) -- (1,0);\draw[markxi={0},markxi={.5},markxi={1}] (0,0)--(1,0) ;
   \draw (-0.35,0.25) node{$1$};
    \draw (-0.35,-0.25) node{$2$};
     \draw (1.35,0.25) node{$3$};
      \draw (1.35,-0.25) node{$4$}; 
      \draw (0.2,0.3) node {$v,w$};
\end{tikzpicture} = v\cdot  \begin{tikzpicture}[basel={-0.5},every node/.style={scale=0.6}]
   \draw(0,0) -- (1,0) ;
   \draw[\rac](-0.25,0.25) -- (0,0);
   \draw (-0.25,-0.25) -- (0,0);\draw (1.25,0.25) -- (1,0);
   \draw (1.25,-0.25) -- (1,0);\draw[markxi={0},markxi={.5},markxi={1}] (0,0)--(1,0) ;
   \draw (-0.35,0.25) node{$1$};
    \draw (-0.35,-0.25) node{$2$};
     \draw (1.35,0.25) node{$3$};
      \draw (1.35,-0.25) node{$4$}; 
      \ball{(-0.2,-0.2)};
      \draw (0,-0.4) node {$\frac{1}{v}$};
       \draw (0,0.5) node {$\frac{1}{v},w$}; 
\end{tikzpicture} \cong \begin{tikzpicture}[basel={-0.5},every node/.style={scale=0.6}]
   \draw(0,0) -- (1,0) ;
   \draw(-0.25,0.25) -- (0,0);
   \draw (-0.25,-0.25) -- (0,0);\draw (1.25,0.25) -- (1,0);
   \draw (1.25,-0.25) -- (1,0);\draw[markxi={0},markxi={.5},markxi={1}] (0,0)--(1,0) ;
   \draw (-0.35,0.25) node{$1$};
    \draw (-0.35,-0.25) node{$2$};
     \draw (1.35,0.25) node{$3$};
      \draw (1.35,-0.25) node{$4$}; 
\end{tikzpicture}.
\label{eq:tree2}
\end{equation}
A simple calculation shows that the graph state $\begin{tikzpicture}[basel={-0.5},every node/.style={scale=0.6}]
   \draw(0,0) -- (1,0) ;
   \draw(-0.25,0.25) -- (0,0);
   \draw (-0.25,-0.25) -- (0,0);\draw (1.25,0.25) -- (1,0);
   \draw (1.25,-0.25) -- (1,0);\draw[markxi={0},markxi={.5},markxi={1}] (0,0)--(1,0) ;
   \draw (-0.35,0.25) node{$1$};
    \draw (-0.35,-0.25) node{$2$};
     \draw (1.35,0.25) node{$3$};
      \draw (1.35,-0.25) node{$4$}; 
\end{tikzpicture}$ is the unnormalized four qubit W-state $\ket{W_4}$. This process shows how by using the $\Psi$ stitch on the state, there are no relevant parameters under SLOCC, but the two final states in Equations \eqref{eq:tree1} and \eqref{eq:tree2} are SLOCC inequivalent. They belong to different classes according to \cite{Verstraete}. We identified that the two ways of constructing the state, with three and two complex parameters, respectively, correspond only to two unique SLOCC inequivalent states. This identification is the essence of the approach that we will use. However, pushing by itself is not enough to analyze more complex graphs. In the next section, we will introduce a set of rules that can be obtained from the algebras of the objects defined for the construction and some that can be obtained from the construction itself that will help to make the graphical process even more useful.

\section{Stitching and Pushing rules}
\label{rules}
 W-state stitching and the procedure of parameter pushing based on the symmetries of W-states can be used together to develop a graphical tool to perform calculations on multipartite states and for calculating SLOCC invariants as we will see in Section \ref{Invariants}. Here, we present a set of rules that summarize the most important properties of this construction:
%Voy a dejar todos los labels al final de cada codigo para que sea facil ponerlos o quitarlos
\begin{enumerate}[(i)]
\item \textbf{Graphical definitions:} First, we present the graphical representations of the elementary objects of the construction:
\begin{enumerate}[1.]
\item {\bf Vertices:} Any vertex on the construction is representing a $W_3$ state:
\begin{equation*}
\ket{W_3}=\Wn{3}{0}{\draw[markxi=0](0,0) to (0.2,0);}.
\end{equation*}
\item {\bf Edges:} Any clean edge, i.e., line without decorations, represents the bipartite maximally entangled state $\ket{\Phi^{+}}$, or, if connected to something else, the $\bra{\Phi^{+}}$,  where each end of the edge represents one of the parts of the state:
\begin{equation*}
\ket{\Phi^{+}}=\begin{tikzpicture}[basel=-.5]
        \draw(0,0) to (1,0);
    \end{tikzpicture} , \qquad 
\bra{\Phi^{+}}=\begin{tikzpicture}[basel=-.5]
\Genstates{(-0.1,0)}
        \draw(0,0) to (1,0);
        \Genstates{(1.1,0)}
    \end{tikzpicture}  ,
\end{equation*}
where we use the ``$\wr$'' symbols  to represent that the element is a bra, contracting kets in both sides.
\item {\bf Operations:} There are three basic operations from which any other invertible operation can be obtained up to a scale factor:
\begin{equation*}
\sigma_x=\bullet ,\quad \quad \mathcal{B}(v)= \begin{tikzpicture}[basel=-1.5,every node/.style={scale=0.6} ]
     \draw (-0.2,0) to (0.2,0);
        \ball{(0,0)};        
        \draw(0,0.3) node {$v$};
    \end{tikzpicture}, \quad \quad  \mathcal{A}(w)=  \begin{tikzpicture}[basel=-1.3,every node/.style={scale=0.6}]
     \draw[\rao] (-0.2,0) to (0.2,0); 
        \draw(0.1,0.2) node {$w$};
    \end{tikzpicture},
\end{equation*}
where $\mathcal{B}$ and $\mathcal{A}$ matrices are defined in Equations \eqref{eq:Bdef} and \eqref{eq:Adef}. The arrows point to the part where they are acting on. 
\end{enumerate}
Although any other construction can be obtained from the ones defined above, it will be useful to introduce symbols derived from the elemental ones:
\begin{enumerate}[1.]\addtocounter{enumii}{3}
\item {\bf Qubit:} We show in the Appendix \ref{Appendix03} that qubit $\ket{1}$ can be obtained by stitching together two parts of the same $\ket{W_3}$ state with a clean $\bra{\Phi^{+}}$, which leads to a natural representation of $\ket{1}$:
\begin{equation*}
    \ket{1}= \tikz{\qubitl{(0,0)}} .
\end{equation*}
\item {\bf Composed objects:} Other recurrent objects that can be obtained from composing the previous ones are:
\begin{equation*}
\ket{\Psi^{+}}=\begin{tikzpicture}[basel=-.5]
        \draw[markx=0.5](0,0) to (1,0);
    \end{tikzpicture}, \quad  \quad u(v,w)=\begin{tikzpicture}[basel=-1.3,every node/.style={scale=0.6}]
     \draw[\rac] (-0.2,0) to (0.2,0); 
        \draw(0.1,0.2) node {$v,w$};
    \end{tikzpicture} =\begin{tikzpicture}[basel=-1.3,every node/.style={scale=0.6}]
     \draw[\rao] (-0.2,0) to (0.4,0); 
      \ball{(0,0)};
        \draw(0.3,0.2) node {$\frac{w}{v}$};
        \draw(0,0.2) node {$v$};
    \end{tikzpicture} , \quad \quad \ket{0} = \tikz{\qubitl{(0,0)};
  \draw[markx={1}](-0.5,0)to (-0.25,0)} .
\end{equation*}
The $u$ matrix is defined in Equation \eqref{eq:udef}, and we usually refer to the lines with $\bullet$ on them as \textit{$\bullet$-lines}.
\item {\bf The Z-ball:} There is a special case for balls when the argument is the imaginary number $i$. In that case, the matrix is proportional to the Pauli operator $\sigma_z$, and we define a special notation for it:
\begin{equation*}
    \mathcal{B}(i)=\begin{tikzpicture}[basel=-1.7,every node/.style={scale=0.6} ]
     \draw (-0.2,0) to (0.2,0);
        \ball{(0,0)};        
        \draw(0,0.3) node {$i$};
    \end{tikzpicture}=i \cdot \begin{tikzpicture}[basel=-1,every node/.style={scale=0.6} ]
     \draw[marko={0.3}] (-0.2,0) to (0.2,0);
    \end{tikzpicture}=i \sigma_z.
\end{equation*}

\end{enumerate}
    \item \textbf{Stitch-Arrows equivalence:} Any entangled bipartite state $\bra{\varsigma}$ with a coefficient matrix of unit-determinant as in Equation \eqref{eq:sigmaunit} can be parametrized in one of the two following forms:
    \begin{equation*}
     \bra{\varsigma^{\Phi}}= 
  \begin{tikzpicture}[basel={-1.4},every node/.style={scale=0.6} ]
  \draw[\lao \rac] (0,0) to (1,0);
  \draw (0,0.3) node{$\varsigma_{10}$};
  \draw (1,0.3) node{$\varsigma_{00},\varsigma_{01}$};
  \Genstates{(-0.2,0)};
  \Genstates{(1.2,0)};
    \end{tikzpicture} , \quad \quad \bra{\varsigma^{\Psi}}=
    \begin{tikzpicture}[basel={-2},every node/.style={scale=0.6} ]
         \draw[ \rac] (0,0) to (1,0);
         \draw[markxi={0.5}] (0,0) to (1,0);
         \draw (1,0.4) node{$\frac{1}{\varsigma_{01}},\varsigma_{11}$};
           \Genstates{(-0.2,0)};
  \Genstates{(1.2,0)};
    \end{tikzpicture} ,
\end{equation*}
named the $\Phi$-stitch and the $\Psi$-stitch respectively. It is worth recalling here that when stitches are contracted with W states, the actions of arrows are understood as acting on the contracted parts of the W states.
\item \textbf{Multiplication rules:} The algebras of the objects defined in (i) have the following multiplication rules:
\begin{enumerate}[1.]
\item Balls and white arrows have a nice structure because they define two groups under matrix multiplication: balls are the group of diagonal matrices with determinant 1, and white arrows are the group of upper triangular matrices with diagonal equal to 1. Then, when multiplying balls, we get another ball, and when multiplying white arrows, we get another white arrow. The ways that matrix multiplication combines such objects are:
\begin{equation*}
 \begin{tikzpicture}[basel={-1.5},every node/.style={scale=0.6}]
        \draw(-0.1,0) to (0.6,0); 
        \ball{(0.1,0)};
        \ball{(0.4,0)};      

        \draw (0.1,0.3) node {$v_1$};
        \draw (0.4,0.3) node {$v_2$};
\end{tikzpicture}=\begin{tikzpicture}[basel={-1.5},every node/.style={scale=0.6}]
        \draw(0,0) to (0.6,0); 
        \ball{(0.3,0)};
        
        \draw (0.3,0.3) node {$v_1v_2$};
    \end{tikzpicture} 
    , \quad \quad \begin{tikzpicture}[basel={-1.5},every node/.style={scale=0.6}]
        \draw[\rao](0.3,0) to (0.5,0);
        \draw[\rao](0,0) to (0.3,0);
        
        \draw (0.1,0.3) node {$w_1$};
        \draw (0.4,0.3) node {$w_2$};
    \end{tikzpicture} =\begin{tikzpicture}[basel={-1.5},every node/.style={scale=0.6}]
        \draw[\rao](0,0) to (0.5,0);
         \draw (0.5,0.3) node {$w_1+w_2$};
    \end{tikzpicture} .
    \end{equation*}
    \item As shown above, in (i.5), black arrows can be obtained by multiplying balls and white arrows. However, this decomposition can be done in two equivalent forms:
    \begin{equation*}    
\begin{tikzpicture}[basel={-1.5},every node/.style={scale=0.6}]
        \draw[\rac](0,0) to (0.5,0);
        
        \draw (0.4,0.3) node {$v,w$};
    \end{tikzpicture} =
    \begin{tikzpicture}[basel={-1.8},every node/.style={scale=0.6}]
        \draw[\rao](0,0) to (0.7,0);
        \ball{(0.2,0)}
        
        \draw (0.2,0.3) node {$v$};
        \draw (0.6,0.3) node {$w/v$};
\end{tikzpicture}=\begin{tikzpicture}[basel={-1.5},every node/.style={scale=0.6}]
        \draw[\rao](0,0) to (0.5,0);
        \ball{(0.6,0)}
        
        \draw (0.3,0.3) node {$vw$};
        \draw (0.7,0.3) node {$v$};
    \end{tikzpicture},
    \end{equation*}
    with these decompositions, the algebras of balls and white arrows define any operation on black arrows.
 \item The last two operations are $\sigma_x$ and $\sigma_z$, that, being Pauli matrices, are involutory matrices, so when multiplying two of each, we get:   
    \begin{equation*}
 \begin{tikzpicture}[basel={-.5},every node/.style={scale=0.6}]
        \draw[markx={0.3},markx={0.7}](0,0) to (0.5,0);
    \end{tikzpicture} =\begin{tikzpicture}[basel={-.5},every node/.style={scale=0.6}]
        \draw(0,0) to (0.5,0);
    \end{tikzpicture} , \quad \quad \begin{tikzpicture}[basel={-1},every node/.style={scale=0.6}]
        \draw[marko={0.1},marko={0.5}](0,0) to (0.5,0);
    \end{tikzpicture} =\begin{tikzpicture}[basel={-.5},every node/.style={scale=0.6}]
        \draw(0,0) to (0.5,0);
    \end{tikzpicture} .
\end{equation*}
\end{enumerate}
\item \textbf{Sliding rules:} Due to the symmetry of $\bra{\Phi^{+}}$ shown in Equation \eqref{eq:Phisym}, symmetric operators such as $\bullet$ and balls can be applied in any of the parts. Graphically, that means that those objects can be slid through empty lines:
\begin{equation*}
    \begin{tikzpicture}[basel={-1.5},every node/.style={scale=0.6}]
        \draw(0,0) to (0.5,0);    \ball{(0,0)};   
        
        \draw (-0.1,0.3) node {$v$}; 
\end{tikzpicture}=\begin{tikzpicture}[basel={-1.5},every node/.style={scale=0.6}]
        \draw(0,0) to (0.5,0); 
        \ball{(0.5,0)};  
        \draw (0.6,0.3) node {$v$};
    \end{tikzpicture}, \quad  \quad
     \begin{tikzpicture}[basel={-.5},every node/.style={scale=0.6}]
        \draw[markx={0.1}](0,0) to (0.5,0);     
\end{tikzpicture}=\begin{tikzpicture}[basel={-.5},every node/.style={scale=0.6}]
        \draw[markx={0.9}](0,0) to (0.5,0);  
    \end{tikzpicture} , \quad \quad \begin{tikzpicture}[basel={-1},every node/.style={scale=0.6}]
        \draw[marko={0.1}](0,0) to (0.5,0);
    \end{tikzpicture} =\begin{tikzpicture}[basel={-1},every node/.style={scale=0.6}]
        \draw[marko={0.5}](0,0) to (0.5,0);
    \end{tikzpicture}
 \end{equation*}
 \item {\bf Commutation rules} As balls and $\bullet$ can be slid on the lines, sometimes we will want to move balls through $\bullet$s, which corresponds to the commutation operation. Such commutation rules are:
 \begin{equation*}
 \begin{tikzpicture}[basel={-1.5}, every node/.style={scale=0.6}]
        \draw[markx={0.5}](0,0) to (0.5,0); 
        \ball{(0,0)};     

        \draw (0,0.3) node {$v$};
    \end{tikzpicture}=\begin{tikzpicture}[basel={-1.8},every node/.style={scale=0.6}]
        \draw[markx={0.5}](0,0) to (0.5,0); 
        \ball{(0.5,0)};

        \draw (0.5,0.3) node {$1/v$};
    \end{tikzpicture} , \quad \quad \begin{tikzpicture}[basel={-.8},every node/.style={scale=0.6}]
        \draw[marko={0.1},markx={0.5}](0,0) to (0.5,0);
    \end{tikzpicture} = - \, \begin{tikzpicture}[basel={-.8},every node/.style={scale=0.6}]
        \draw[markxo={0.5}{0.6}](0,0) to (0.5,0);
    \end{tikzpicture}
 \end{equation*}
 We include here the fact that on $\bullet$-lines, arrows can be switched from one part to another as:
\begin{equation*}
\begin{tikzpicture}[basel={-1.3},every node/.style={scale=0.6}]
        \draw[\lac-](0,0) to (0.5,0);
        \draw[markxi={0.8}](0,0) to (0.5,0);
        
        \draw (0.1,0.2) node {$v,w$};  
\end{tikzpicture}=\begin{tikzpicture}[basel={-1.8},every node/.style={scale=0.6}]
        \draw[\rac](0,0) to (0.5,0);
          \draw[markxi={0.2}](0,0) to (0.5,0);
          
        \draw (0.4,0.3) node {$1/v,w$};
    \end{tikzpicture}.
    \end{equation*}

\item \textbf{Pushing rules:} The symmetries of W-states allow to \textit{push} operations in one of the parts to the other two as:
\begin{enumerate}[1.]
\item A black  arrow with parameters $(v,w)$ can be pushed through W vertices, creating a black arrow with parameters $(1/v,w)$ in one edge and a ball with $1/v$ as a parameter in the other edge, and a scale factor $v$:
 \begin{equation*}
        \begin{tikzpicture}[basel={-1},every node/.style={scale=0.6}]
        \draw[\rac](0,0) to (0.3,0);
        \draw[markx={0}] (0.3,0) to (0.6,0.3);
        \draw (0.3,0) to (0.6,-0.3);
        \draw (0.1,0.3) node {$v,w$};
    \end{tikzpicture}= v \cdot \begin{tikzpicture}[basel={-0.5},every node/.style={scale=0.6}]
        \draw(0,0) to (0.3,0);
        \draw[\lac-] (0.3,0) to (0.6,0.3);
        \draw[markx={0}] (0.3,0) to (0.6,-0.3);
        \ball{(0.5,-0.2)};
        \draw (0.3,0.5) node {$(1/v,w)$};
         \draw (0.5,-0.5) node {$1/v$};
    \end{tikzpicture}
    \end{equation*}
    \item Balls are a special black arrow where the second parameter is set to $w=0$. Then, when pushing a ball with parameter $v$ through a $W$ vertex, it results in one ball with parameter $1/v$ in each opposite edge and a scale factor $v$. We apply this to generic balls and Z-balls as:
    \begin{equation*}
     \begin{tikzpicture}[basel={-1},every node/.style={scale=0.6}]
        \draw(0,0) to (0.3,0);
        \draw [markx={0}](0.3,0) to (0.6,0.3);
        \draw (0.3,0) to (0.6,-0.3);
        \ball{(0.1,0)};
        \draw (0.1,0.3) node {$v$};
    \end{tikzpicture} = v\cdot\begin{tikzpicture}[basel={-0.5},every node/.style={scale=0.6}]
        \draw(0,0) to (0.3,0);
        \draw[markx={0}] (0.3,0) to (0.6,0.3);
        \draw (0.3,0) to (0.6,-0.3);
        \ball{(0.5,0.2)};
        \ball{(0.5,-0.2)};
         \draw (0.5,0.4) node {$1/v$};
          \draw (0.5,-0.4) node {$1/v$};
    \end{tikzpicture} , \quad  \quad \begin{tikzpicture}[basel={-1},every node/.style={scale=0.6}]
        \draw[marko={0.2}](0,0) to (0.3,0);
        \draw [markx={0}](0.3,0) to (0.6,0.3);
        \draw (0.3,0) to (0.6,-0.3);
    \end{tikzpicture} = -\, \begin{tikzpicture}[basel={-0.5},every node/.style={scale=0.6}]
        \draw(0,0) to (0.3,0);
        \draw[markxo={0}{0.3}] (0.3,0) to (0.6,0.3);
        \draw[marko={0.3}] (0.3,0) to (0.6,-0.3);
        \end{tikzpicture}.
\end{equation*} 
\item   White arrows are a special black arrow where the first parameter is set to $v=1$. Then, when pushing a white arrow through a $W$ vertex results in the same arrow acting on one of the opposite edges:
\begin{equation*}
 \begin{tikzpicture}[basel={-1},every node/.style={scale=0.6}]
        \draw[\rao](0,0) to (0.3,0);
        \draw[markx={0}] (0.3,0) to (0.6,0.3);
        \draw (0.3,0) to (0.6,-0.3);
        \draw (0.1,0.3) node {$w$};
    \end{tikzpicture}=\begin{tikzpicture}[basel={-1},every node/.style={scale=0.6}]
        \draw(0,0) to (0.3,0);
        \draw[\lao-] (0.3,0) to (0.6,0.3);
        \draw[markx={0}] (0.3,0) to (0.6,-0.3);
        \draw (0.3,0.3) node {$w$};
    \end{tikzpicture}.
    \end{equation*}
\end{enumerate}
\end{enumerate}
Up to this point, we introduced the rules that arise from the definitions of the basic objects and their algebras. In the next part, we will introduce some rules obtained when analyzing the construction in simple setups. Most of these rules are obtained in the Appendix \ref{Appendix03}.
\begin{enumerate}[(i)]
\addtocounter{enumi}{6}
\item \textbf{Operations on qubits:} First, we want to summarize how the qubits, introduced in (i.4) and (i.5), interact with other objects in the construction. 
\begin{enumerate}[1.]
\item First, the interaction between qubits is very simple:
\begin{equation*}
 \begin{tikzpicture}[basel={-.5},every node/.style={scale=0.6}]
            \qubitl{(0,0)};
            \qubitr{(-1.3,0)};
            \draw[markx={0},markx={1}](-0.7,0) to (-0.2,0);
        \end{tikzpicture}=
        \begin{tikzpicture}[basel={-.5},every node/.style={scale=0.6}]
            \qubitl{(0,0)};
            \qubitr{(-0.9,0)};
        \end{tikzpicture}=1 , \quad \quad  \begin{tikzpicture}[basel={-.5},every node/.style={scale=0.6}]
            \qubitl{(-0.25,0)};
            \qubitr{(-1.3,0)};
            \draw[markx={0.5}](-0.7,0) to (-0.5,0);
        \end{tikzpicture} =0.
\end{equation*}
\item When qubits are contracted with the $W$ vertex, we obtain a $\ket{\Psi^+}$, when contracting with a $\bra{0}$, or the separable state $\ket{00}$ when contracting with $\bra{1}$:
\begin{equation*}
 \begin{tikzpicture}[basel={-0.5}]
   \inangarc{2}{(-0.5,0)}{90};
   \draw[markx={0},markx={0.8}] (-0.5,0) to (-0.25,0);
    \qubitl{(0,0)};
    \end{tikzpicture} =\begin{tikzpicture}[basel={-.5}]]
    \draw[markx={0.5}] (0,0.2) to[bend left=90]  (0,-0.2) ;
\end{tikzpicture} , \quad \quad \begin{tikzpicture}[basel={-0.5}]
   \inangarc{2}{(-0.5,0)}{90};
   \draw[markx={0}] (-0.5,0) to (-0.25,0);
    \qubitl{(0,0)};
    \end{tikzpicture} =\begin{tikzpicture}[basel={-.5}]]
   \draw[markx=0.5](-0.5,0.2) to(0,0.2);\qubitl{(0,0.2)};\draw[markx=0.5](-0.5,-0.2) to(0,-0.2);\qubitl{(0,-0.2)}
\end{tikzpicture} 
\end{equation*}
\item The actions with the set of operations can be obtained from the matrix representations to be:
\begin{equation*}
    \begin{gathered}
 \begin{tikzpicture}[basel={-1.5},every node/.style={scale=0.6}]
  \qubitl{(0,0)};
  \draw[markx={1}](-0.7,0) to (-0.25,0);
  \ball{(-0.5,0)};
  \draw (-0.5,0.3) node {$v$};
\end{tikzpicture} = v \cdot \begin{tikzpicture}[basel={-0.5}]
  \qubitl{(0,0)};
  \draw[markx={1}](-0.5,0) to (-0.25,0);
\end{tikzpicture} , \quad \quad  \begin{tikzpicture}[basel={-1.5},every node/.style={scale=0.6}]
  \qubitl{(0,0)};
  \draw(-0.7,0) to (-0.25,0);
  \ball{(-0.5,0)};
  \draw (-0.5,0.3) node {$v$};
\end{tikzpicture} = \frac{1}{v} \cdot \begin{tikzpicture}[basel={-0.5}]
  \qubitl{(0,0)};
\end{tikzpicture} ,\\
\begin{tikzpicture}[basel={-2},every node/.style={scale=0.6}]
  \qubitl{(0,0)};
  \draw[markx={1}](-0.5,0)to (-0.25,0);
  \draw[\rao](-1,0) to (-0.5,0);
  \draw (-0.8,0.3) node {$w$};
\end{tikzpicture} = \begin{tikzpicture}[basel={-0.7}]
  \draw[markx=0.5](-0.5,0) to(0,0);\qubitl{(0,0)};\end{tikzpicture} , \quad \quad \begin{tikzpicture}[basel={-2},every node/.style={scale=0.6}]
  \qubitl{(0,0)};
  \draw[\rao](-0.75,0) to (-0.25,0);
  \draw (-0.5,0.3) node {$w$};
\end{tikzpicture} = w\cdot \begin{tikzpicture}[basel={-0.7}]
  \qubitl{(0,0)};
  \draw[markx={1}](-0.5,0)to (-0.25,0);\end{tikzpicture} + \begin{tikzpicture}[basel={-0.7}]
  \qubitl{(0,0)};\end{tikzpicture},\\
  \begin{tikzpicture}[basel={-2},every node/.style={scale=0.6}]
  \qubitl{(0,0)};
  \draw[markx={1}](-0.5,0)to (-0.25,0);
  \draw[\rac](-1,0) to (-0.5,0);
  \draw (-0.8,0.3) node {$v,w$};
\end{tikzpicture} =v\cdot  \begin{tikzpicture}[basel={-0.7}]
  \draw[markx=0.5](-0.5,0) to(0,0);\qubitl{(0,0)};\end{tikzpicture} , \quad  \quad \begin{tikzpicture}[basel={-2},every node/.style={scale=0.6}]
  \qubitl{(0,0)};
  \draw[\rac](-0.75,0) to (-0.25,0);
  \draw (-0.5,0.3) node {$v,w$};
\end{tikzpicture} = w\cdot \begin{tikzpicture}[basel={-0.7}]
  \qubitl{(0,0)};
  \draw[markx={1}](-0.5,0)to (-0.25,0);\end{tikzpicture} + \frac{1}{v}\begin{tikzpicture}[basel={-0.7}]
  \qubitl{(0,0)};\end{tikzpicture}.
        \end{gathered}
    \end{equation*}
\end{enumerate}
and clearly $\bullet$ action exchange between $\ket{0}$ and $\ket{1}$.  
\item \textbf{Ball self-destruction :} When moving a ball through a loop, according to sliding rules (iv.2) and commutation rules (v.2), the argument of the ball is inverted by each $\bullet$ in the path. Then, we can start by separating the ball $v$ into two balls $\sqrt{v}$ by using multiplication rules (iii.1). After that, when pushing one of the balls through the vertices of the loop, if the number of $\bullet$'s is odd, the ball can be self-destroyed inside the loop. However, throughout this process, by the pushing rule (vi.2), one ball is pushed to each external edge of the loop and a factor appears, which must be considered. For example:
    \begin{equation*}
        \begin{tikzpicture}[basel={-2},every node/.style={scale=0.6}]
        \draw (-0.2,0.2) to (0,0);
        \draw (1,0.2) to  (0.8,0);
        \draw (-0.2,-1) to (0,-0.8);
        \draw (1,-1) to (0.8,-0.8);
   \draw[markx={0},markx={1}] (0,0) to (0.8,0);
   \draw[markx={0},markx={1}] (0.8,0) to (0.8,-0.8);
   \draw[markx={0},markx={0.5},markx={1}] (0,-0.8) to (0.8,-0.8);
   \draw[markx={0},markx={1}] (0,0) to (0,-0.8);
   \ball{(0.4,0)};
   \draw (0.4,0.3) node{$v$};
\end{tikzpicture}\stackrel{iii}{=} 
   \begin{tikzpicture}[basel={-2},every node/.style={scale=0.6}]
        \draw (-0.2,0.2) to (0,0);
        \draw (1,0.2) to  (0.8,0);
        \draw (-0.2,-1) to (0,-0.8);
        \draw (1,-1) to (0.8,-0.8);
   \draw[markx={0},markx={1}] (0,0) to (0.8,0);
   \draw[markx={0},markx={1}] (0.8,0) to (0.8,-0.8);
   \draw[markx={0},markx={0.5},markx={1}] (0,-0.8) to (0.8,-0.8);
   \draw[markx={0},markx={1}] (0,0) to (0,-0.8);
   \ball{(0.2,0)};
   \ball{(0.6,0)};
   \draw (0.2,0.3) node{$\sqrt{v}$};
    \draw (0.6,0.3) node{$\sqrt{v}$};
\end{tikzpicture} \stackrel{v}{=}\sqrt{v} \cdot 
 \begin{tikzpicture}[basel={-2},every node/.style={scale=0.6}]
        \draw (-0.2,0.2) to (0,0);
        \draw (1,0.2) to  (0.8,0);
        \draw (-0.2,-1) to (0,-0.8);
        \draw (1,-1) to (0.8,-0.8);
   \draw[markx={0},markx={1}] (0,0) to (0.8,0);
   \draw[markx={0},markx={1}] (0.8,0) to (0.8,-0.8);
   \draw[markx={0},markx={0.5},markx={1}] (0,-0.8) to (0.8,-0.8);
   \draw[markx={0},markx={1}] (0,0) to (0,-0.8);
   \ball{(0.4,0)};
   \draw (0.4,0.3) node{$\sqrt{v}$};
    \ball{(1,0.2)};       \draw (1,0.5) node{$1/\sqrt{v}$};
      \draw (1,-0.4) node{$\frac{1}{\sqrt{v}}$};
      \ball{(0.8,-0.4)};  
\end{tikzpicture} 
 \stackrel{v}{=} \dots \stackrel{v}{=} \sqrt{v} \cdot \begin{tikzpicture}[basel={-1.5},every node/.style={scale=0.4}]
        \draw (-0.4,0.4) to (0,0);
        \draw (1.2,0.4) to  (0.8,0);
        \draw (-0.4,-1.2) to (0,-0.8);
        \draw (1.2,-1.2) to (0.8,-0.8);
   \draw[markx={0},markx={1}] (0,0) to (0.8,0);
   \draw[markx={0},markx={1}] (0.8,0) to (0.8,-0.8);
   \draw[markx={0},markx={0.5},markx={1}] (0,-0.8) to (0.8,-0.8);
   \draw[markx={0},markx={1}] (0,0) to (0,-0.8);
   \ball{(1,0.2)};   
   \ball{(-0.2,0.2)};   
   \ball{(1,-1)};   
   \ball{(-0.2,-1)};
     \draw (1,0.5) node{$1/\sqrt{v}$};
      \draw (1,-1.3) node{$\sqrt{v}$};
       \draw (-0.2,-1.3) node{$\sqrt{v}$};
        \draw (-0.2,0.5) node{$1/\sqrt{v}$};
        \ball{(0.2,0)};
   \ball{(0.6,0)};
   \draw (0.2,0.3) node{$\frac{1}{\sqrt{v}}$};
    \draw (0.6,0.3) node{$\sqrt{v}$};
\end{tikzpicture}\stackrel{iii}{=}\sqrt{v}\cdot \begin{tikzpicture}[basel={-1.5},every node/.style={scale=0.6}]
        \draw (-0.4,0.4) to (0,0);
        \draw (1.2,0.4) to  (0.8,0);
        \draw (-0.4,-1.2) to (0,-0.8);
        \draw (1.2,-1.2) to (0.8,-0.8);
   \draw[markx={0},markx={1}] (0,0) to (0.8,0);
   \draw[markx={0},markx={1}] (0.8,0) to (0.8,-0.8);
   \draw[markx={0},markx={0.5},markx={1}] (0,-0.8) to (0.8,-0.8);
   \draw[markx={0},markx={1}] (0,0) to (0,-0.8);
   \ball{(1,0.2)};   
   \ball{(-0.2,0.2)};   
   \ball{(1,-1)};   
   \ball{(-0.2,-1)};
     \draw (1,0.5) node{$1/\sqrt{v}$};
      \draw (1,-1.3) node{$\sqrt{v}$};
       \draw (-0.2,-1.3) node{$\sqrt{v}$};
        \draw (-0.2,0.5) node{$1/\sqrt{v}$};
\end{tikzpicture}, 
    \end{equation*}
where we specified over the ``='' sign the rule used in each step.
 \item \textbf{$W_N$ states and $W_N$ reordering:}
 W states have a nice mathematical structure that allows the calculation of an N-partite W state as:
 \begin{equation*}
  \begin{tikzpicture} [baseline={([yshift=-.5ex]current bounding box.center)},every node/.style={scale=0.6}] \def\array{$1$,$2$,,,,,,$N_1$}    \def\arrayx{$N-N_1$,,,,,,$2$,$1$}
    \foreach [count=\n] \x in \array{
        \node at ({90+\n*180/(8+1)}:0.6cm) (n\n) {\x};
        \draw (0,0)--(n\n); 
        };
        \draw[markx={0},markx={0.5},markx={1}] (0,0)--(1,0);
         \foreach [count=\n] \x in \arrayx{
        \node at ($(1,0)+({-90+\n*180/(8+1)}:0.6cm)$) (n\n) {\x};
        \draw (1,0)--(n\n); 
        };
        \draw (0.5,0) node {$\times$};
\end{tikzpicture}  =\begin{tikzpicture} [basel={-1.5},every node/.style={scale=0.6}] \def\array{$1$,$2$,,,,,,}    \def\arrayx{,,,,,,,$N$}
    \foreach [count=\n] \x in \array{
        \node at ({90+\n*180/(8+1)}:0.6cm) (n\n) {\x};
        \draw (0,0)--(n\n); 
        };
         \foreach [count=\n] \x in \arrayx{
        \node at ($(0,0)+({-90+\n*180/(8+1)}:0.6cm)$) (n\n) {\x};
        \draw (0,0)--(n\n); 
        };
        \draw[markxi={0}](0,0) to (0,1);
        \end{tikzpicture},
\end{equation*}
 where we generalized the notation of $W_3$ to the N-partite W state as a vertex with $N$ edges. This construction can be reversed to write the $W_N$ state as the stitch of any two subsets of $N$ with the $\bullet$ line, then all of the possible constructions are equivalent, allowing one to replace any two constructions freely. The most simple case is:
 \begin{equation*}
    \begin{tikzpicture}[basel={-0.5},every node/.style={scale=0.6}]
   \draw (0,0) -- (0.5,0) ; 
   \draw (0.5,0) -- (0,0) ;
   \draw (-0.25,0.25) -- (0,0);
   \draw (-0.25,-0.25) -- (0,0);\draw (0.75,0.25) -- (0.5,0);
   \draw (0.72,-0.25) -- (0.5,0);\draw[markxi={0},markxi={0.5},markxi={1}] (0,0)--(0.5,0) ;
   \draw (-0.35,0.25) node{$1$};
    \draw (-0.35,-0.25) node{$2$};
     \draw (0.85,0.25) node{$3$};
      \draw (0.85,-0.25) node{$4$}; 
\end{tikzpicture} = \begin{tikzpicture}[basel={-0.5}]
    \draw[markx={0.5}] (-0.2,-0.2) to (0.2,0.2);
    \draw (-0.2,0.2) to (0.2,-0.2);
\end{tikzpicture} =\begin{tikzpicture}[basel={-0.5},every node/.style={scale=0.6}]
   \draw (0,0) -- (0.25,0.25) ; 
   \draw (-0.25,0.25) -- (0,0) ;
   \draw (0,0) -- (0,-0.5);
   \draw (0,-0.5) -- (0.25,-0.75);\draw (0,-0.5) -- (-0.25,-0.75);
  \draw[markxi={0},markxi={0.5},markxi={1}] (0,0)--(0,-0.5) ;
   \draw (-0.35,0.25) node{$1$};
    \draw (-0.35,-0.75) node{$2$};
     \draw (0.35,0.25) node{$3$};
      \draw (0.35,-0.75) node{$4$};
\end{tikzpicture}
\end{equation*}
Replacing the first form with the last one has proven to be very useful to simplify states. Note how the generalization of notation for $W_N$ states is consistent with the two qubits case, because $\ket{W_2}$ is represented with a $\bullet$ and two edges, that is the same representation of $\ket{\Psi^{+}}$, but $\ket{W_2}$ and $\ket{\Psi^{+}}$ are exactly the same state:
\begin{equation*}
\ket{W_2}=\ket{\Psi^{+}}= \begin{tikzpicture}[basel=-.5]
        \draw[markx=0.5](0,0) to (1,0);
    \end{tikzpicture}.
\end{equation*}
\item \textbf{One-qubit constructions:} When stitching two parts of the same W state, only two SLOCC inequivalent states of one qubit appear. These shapes appear very frequently in more complex constructions, so it is worth to state them as rules:
\begin{equation*}
      \begin{tikzpicture}[basel={-1},every node/.style={scale=0.6}]
        \draw[markx={1}] (0.25,0) to (0.5,0);
       \draw[markx={0.9}] (0.5,0)to[bend left=70] (1.5,0);
         \draw (0.5,0)to[bend right=70] (1.5,0);
            \ball{(1,0.25)};
            \draw (1,0.5) node {$v$};
    \end{tikzpicture} = \left( \frac{1+v^2}{v}\right) \cdot \begin{tikzpicture}[basel={-.6}]
      \qubitl{(0,0)} ;
      \draw[markx={1}](-0.5,0) to (-0.25,0);
   \end{tikzpicture} , \quad \quad \quad  \begin{tikzpicture}[basel={-0.8}]
        \draw [markx={1}](0.25,0) to (0.5,0);
       \draw[markxo={0.9}{0.5}] (0.5,0)to[bend left=70] (1.5,0);
         \draw (0.5,0)to[bend right=70] (1.5,0);
    \end{tikzpicture} =0.
\end{equation*}
\item \textbf{Two-qubit constructions:} When stitching two parts of two different W states, the set of SLOCC inequivalent constructions are:
\begin{enumerate}[1.]
\item When both stitches are $\Phi$-stitches, there are two inequivalent cases, one with a normal ball that leads to an entangled state, and another with the Z-ball, leading to a separable state:
\begin{equation*}
\begin{tikzpicture}[basel={-1.5},every node/.style={scale=0.6}]
            \draw[markx={1}](0,0) to (0.25,0);
            \draw(0.25,0) to[bend right=90] (0.75,0);
            \draw(0.25,0) to[bend left=90]  (0.75,0);
            \draw[markx={0}] (0.75,0) to (1,0); 
\ball{(0.5,0.15)};\draw (0.5,0.4) node{$v$};\end{tikzpicture}= \sqrt{1+v^2} \cdot \begin{tikzpicture}[basel={-2},every node/.style={scale=0.6}]
\draw (0,0) -- (0.5,0);
\ball{(0.25,0)}
\draw (0.25,0.3) node {$\frac{\sqrt{1+v^2}}{v}$};    
\end{tikzpicture}, \quad \quad \begin{tikzpicture}[basel={-0.8}]
            \draw[markx={1}](0,0) to (0.25,0);
            \draw(0.25,0) to[bend right=90] (0.75,0);
            \draw[marko={0.5}](0.25,0) to[bend left=90]  (0.75,0);
            \draw[markx={0}] (0.75,0) to (1,0); 
            \end{tikzpicture} =\begin{tikzpicture}[basel={-0.5}]
                \qubitl{(0,0)};
                \qubitr{(0.5,0)};
                \end{tikzpicture}.
\end{equation*}
\item When there is one $\Phi$-stitch and one $\Psi$-stitch, only entangled states can be obtained:
\begin{equation*}
\begin{tikzpicture}[basel={-0.5},every node/.style={scale=0.6}]
            \draw[markx={1}](0,0) to (0.25,0);
            \draw(0.25,0) to[bend right=90] (0.75,0);
            \draw[markx={0.5}](0.25,0) to[bend left=90]  (0.75,0);
            \draw[markx={0}] (0.75,0) to (1,0); 
            \end{tikzpicture}= \begin{tikzpicture}
                \draw[markx={0.5}](0,0) to (0.5,0);
            \end{tikzpicture} 
\end{equation*}
\item When both stitches are $\Psi$-stitches, we have two options, one with generic balls, leading to a separable state, and one with the Z-ball, where the state is canceled:
\end{enumerate}
\begin{equation*}
     \begin{tikzpicture}[basel={-1.5},every node/.style={scale=0.6}]
            \draw[markx={1}](0,0) to (0.25,0);
            \draw[markx={0.5}](0.25,0) to[bend right=90] (0.8,0);
            \draw[markx={0.5}](0.25,0) to[bend left=90]  (0.8,0);
            \draw[markx={0}] (0.8,0) to (1.05,0); \ball{(0.7,0.15)};
            \draw (0.7,0.35) node{$v$};
            \end{tikzpicture} = \left( \frac{1+v^2}{v}\right)\begin{tikzpicture}[basel={-0.5}]\qubitl{(0,0)};
                \qubitr{(0.5,0)};
                \draw[markx={1}] (-0.5,0) to (-0.25,0);
                 \draw[markx={0}](1.2,0) to (1.45,0);
            \end{tikzpicture}, \quad \quad 
      \begin{tikzpicture}[basel={-0.8}]
            \draw[markx={1}](0,0) to (0.25,0);
            \draw[markx={0.5}](0.25,0) to[bend right=90] (1.25,0);
            \draw[markxo={0.5}{0.6}](0.25,0) to[bend left=90]  (1.25,0);
            \draw[markx={0}] (1.25,0) to (1.5,0); 
            \end{tikzpicture} = 0.
\end{equation*}
\end{enumerate}
This set of rules is used to simplify a graph state and identify what parameters on the stitches are relevant under SLOCC classification. Most rules without parameters introduced here are also used in ZW calculus \cite{ZW}, but the parametric ones are original.

\subsection{Examples:}
\label{Examples}

We will show in two examples how the rules allow identifying relevant parameters in a graph state and also make explicit when separable parts appear in the graph. First, consider the following graph state:
\begin{equation*}
 \begin{tikzpicture}[basel={-2},every node/.style={scale=0.6}]
        \draw (-0.2,0.2) to (0,0);
        \draw[markxi={0},markxi={1}] (0,0) to (1,0);
        \draw (0,0) to (1,0);
        \draw(0,0) to (0,-1);
        \draw[markxi={0},markxi={1}] (0,-1) to (1,-1); \draw (0,-1) to (1,-1);
        \draw (1,0) to (1,-1);
        \draw (1,-1) to (1.2,-1.2);
        \draw (1,0) to (1.2,0.2);
        \draw (-0.2,-1.2) to (0,-1);
\end{tikzpicture}.
\end{equation*}
Now, we must consider the different ways of filling this graph with the stitches in rule (ii). We will show explicitly the construction where all the stitches are $\Phi$-stitches:
\begin{equation*}
   \begin{tikzpicture}[basel={-1},every node/.style={scale=0.6}]
        \draw (-0.2,0.2) to (0,0);
        \draw[markxi={0},markxi={1}] (0,0) to (1,0);
        \draw[\lac\rao] (0,0) to (1,0);
        \draw[\lac\rao] (0,0) to (0,-1);
        \draw[markxi={0},markxi={1}] (0,-1) to (1,-1); \draw[\lao\rac] (0,-1) to (1,-1);
        \draw[\lao\rac] (1,0) to (1,-1);
        \draw (1,-1) to (1.2,-1.2);
        \draw (1,0) to (1.2,0.2);
        \draw (-0.2,-1.2) to (0,-1);
\end{tikzpicture}
  \stackrel{vi}{\propto}  \begin{tikzpicture}[basel={-1},every node/.style={scale=0.6}]
        \draw[\rac] (-0.2,0.2) to (0,0);
        \draw[markxi={0},markxi={1}] (0,0) to (1,0);
        \draw(0,0) to (1,0);
        \draw(0,0) to (0,-1);
        \draw[markxi={0},markxi={1}] (0,-1) to (1,-1); \draw(0,-1) to (1,-1);
        \draw(1,0) to (1,-1);
        \draw[\lac-] (1,-1) to (1.2,-1.2);
        \draw[\lao-] (1,0) to (1.2,0.2);
        \draw[\rao] (-0.2,-1.2) to (0,-1);

        \ball{(0.5,0)};
        \ball{(0.5,-1)};
\end{tikzpicture},
\end{equation*}
where we already applied the pushing rules (vi) to \textit{push} out the black arrows in external vertices, leaving, for each corner, a ball inside, while the white arrows can be pushed out freely. As we are interested in SLOCC classification, we can drop external decorations as they are SLOCC operations; because of this, from now on, we will omit the decorations in external edges and use the $\cong$ symbol:
\begin{equation*}
\begin{tikzpicture}[basel={-1},every node/.style={scale=0.6}]
        \draw[\rac] (-0.2,0.2) to (0,0);
        \draw[markxi={0},markxi={1}] (0,0) to (1,0);
        \draw(0,0) to (1,0);
        \draw(0,0) to (0,-1);
        \draw[markxi={0},markxi={1}] (0,-1) to (1,-1); \draw(0,-1) to (1,-1);
        \draw(1,0) to (1,-1);
        \draw[\lac-] (1,-1) to (1.2,-1.2);
        \draw[\lao-] (1,0) to (1.2,0.2);
        \draw[\rao] (-0.2,-1.2) to (0,-1);

        \ball{(0.5,0)};
        \ball{(0.5,-1)};
\end{tikzpicture}\cong
     \begin{tikzpicture}[basel={-1},every node/.style={scale=0.6}]
        \draw (-0.2,0.2) to (0,0);
        \draw[markxi={0},markxi={1}] (0,0) to (1,0);
        \draw(0,0) to (1,0);
        \draw(0,0) to (0,-1);
        \draw[markxi={0},markxi={1}] (0,-1) to (1,-1); \draw(0,-1) to (1,-1);
        \draw(1,0) to (1,-1);
        \draw(1,-1) to (1.2,-1.2);
        \draw(1,0) to (1.2,0.2);
        \draw(-0.2,-1.2) to (0,-1);

        \ball{(0.5,0)};
        \ball{(0.5,-1)};
\end{tikzpicture}\stackrel{iii,iv,vi}{\cong} \begin{tikzpicture}[basel={-1},every node/.style={scale=0.6}]
        \draw (-0.2,0.2) to (0,0);
        \draw[markxi={0},markxi={1}] (0,0) to (1,0);
        \draw(0,0) to (1,0);
        \draw(0,0) to (0,-1);
        \draw[markxi={0},markxi={1}] (0,-1) to (1,-1); \draw(0,-1) to (1,-1);
        \draw(1,0) to (1,-1);
        \draw(1,-1) to (1.2,-1.2);
        \draw (1,0) to (1.2,0.2);
        \draw(-0.2,-1.2) to (0,-1);

        \ball{(0.5,0)};
\end{tikzpicture}.
\end{equation*}
In the last step, one of the balls was moved along the inner lines, using sliding rules (iv), then, pushed through vertices (vi), and multiplied with the other to combine them into just one ball, using multiplication rules (iii). With this process, we have simplified the problem of SLOCC classification from 12 original complex parameters in the four stitches to only one. It has to be highlighted that this parameter is relevant for SLOCC classification; by modifying its value, it is possible to achieve three SLOCC inequivalent families:
\begin{equation*}
\begin{gathered}
    \begin{tikzpicture}[basel={-1},every node/.style={scale=0.6}]
        \draw (-0.2,0.2) to (0,0);
        \draw[markxi={0},markxi={1}] (0,0) to (1,0);
        \draw(0,0) to (1,0);
        \draw(0,0) to (0,-1);
        \draw[markxi={0},markxi={1}] (0,-1) to (1,-1); \draw(0,-1) to (1,-1);
        \draw(1,0) to (1,-1);
        \draw(1,-1) to (1.2,-1.2);
        \draw (1,0) to (1.2,0.2);
        \draw(-0.2,-1.2) to (0,-1);
\end{tikzpicture} \cong \ket{G_{a \frac{a}{\sqrt{2}}\frac{a}{\sqrt{2}}0}},\\
    \begin{tikzpicture}[basel={-1},every node/.style={scale=0.6}]
        \draw (-0.2,0.2) to (0,0);
        \draw[markxi={0},markxi={1}] (0,0) to (1,0);
        \draw[marko={0.5}](0,0) to (1,0);
        \draw(0,0) to (0,-1);
        \draw[markxi={0},markxi={1}] (0,-1) to (1,-1); \draw(0,-1) to (1,-1);
        \draw(1,0) to (1,-1);
        \draw(1,-1) to (1.2,-1.2);
        \draw (1,0) to (1.2,0.2);
        \draw(-0.2,-1.2) to (0,-1);
\end{tikzpicture} \cong \ket{L_{a (ia) 0_2}} ,\\
    \begin{tikzpicture}[basel={-2},every node/.style={scale=0.6}]
        \draw (-0.2,0.2) to (0,0);
        \draw[markxi={0},markxi={1}] (0,0) to (1,0);
        \draw(0,0) to (1,0);
        \draw(0,0) to (0,-1);
        \draw[markxi={0},markxi={1}] (0,-1) to (1,-1); \draw(0,-1) to (1,-1);
        \draw(1,0) to (1,-1);
        \draw(1,-1) to (1.2,-1.2);
        \draw (1,0) to (1.2,0.2);
        \draw(-0.2,-1.2) to (0,-1);
        \ball{(0.5,0)};
        \draw (0.5,0.5) node{$ \frac{2\sqrt{ab}}{(a-b)}$};
\end{tikzpicture} \cong \ket{G_{ab\sqrt{ab}\sqrt{ab}}}, 
\end{gathered}
\end{equation*}
where all of these states are labeled according to Table \ref{table:fourqubitsclass}, and the correspondence was found by using the classification scheme shown in \cite{LuqueThibon-tame}, where $\ket{G_{a \frac{a}{\sqrt{2}}\frac{a}{\sqrt{2}}0}}$ is a subfamily of $\ket{G_{abcd}}$ with $b=c=\frac{a}{\sqrt{2}},d=0$, $\ket{L_{a(ia)0_2}}$ is a subfamily of $\ket{L_{abc_2}}$ with $b=ia,c=0$ and $\ket{G_{ab\sqrt{ab}\sqrt{ab}}}$ is another subfamily of $\ket{G_{abcd}}$ with $c=d=\sqrt{ab}$. \\

\noindent We now present another example where an important feature of the construction is made explicit: it is possible to identify clearly from the graph when separable qubits or separable subsets of qubits are obtained in the graph state. Consider the graph represented by:
\begin{equation*}
      \begin{tikzpicture}[baseline={([yshift=-.5ex]current bounding box.center)}]
   \draw[markx={1}] (0,0) -- (0.5,0) ; 
   \draw (0.5,0) -- (1,0.5) ;
   \draw[markx={1}] (0,0) -- (-0.5,0);
   \draw (-1,0.5) -- (-0.5,0);\draw [markx={1}](-0.5,0) -- (0,-1);
   \draw (0.5,0) -- (0,-1);\draw(0,-1.5)--(0,-1); 
\end{tikzpicture},
\end{equation*}
Now, we want to consider the case where one of the stitches is a $\Psi$-stitch, while the other two are $\Phi$-stitches; then, we start with:
\begin{equation*}
    \begin{tikzpicture}[basel=-.5]
   \draw[\lao\rac] (-0.5,0) -- (0.5,0) ; 
   \draw [markx={0}](0.5,0) -- (1,0.5) ;
   \draw[markxi={1}] (0,0) -- (-0.5,0);
   \draw (-1,0.5) -- (-0.5,0);
   \draw[\lao\rac] (0.5,0) -- (0,-1);\draw[markx={1}](0,-1.5)--(0,-1); \draw[\rac] (0,-1) --(-0.5,0); 
   \draw[markxi={0.5}](-0.5,0) to (0,-1);
\end{tikzpicture} \stackrel{iii,iv,vi}{\propto}
  \begin{tikzpicture}[basel=-.5]
   \draw[markx={1}](0,0) -- (0.5,0) ; 
   \draw (0.5,0) -- (1,0.5) ;
   \draw [markx={1}](0,0) -- (-0.5,0);
   \draw (-1,0.5) -- (-0.5,0);
   \draw(0,-1.5)--(0,-1); \draw [markx={0}](0,-1) --(0.5,0); 
   \draw[markx={0.5}](-0.5,0) to (0,-1);
   \ball{(0.25,-0.5)};
\end{tikzpicture} 
\end{equation*}
where we have combined all the steps used before into one step, i.e., we pushed the arrows out (vi), slid and pushed the remaining balls to one line (iii,vi), and then multiplied them to combine them (iv). Now, we use the reordering of $W$ states on the $\bullet$-line (ix):
\begin{equation*}
    \begin{tikzpicture}[basel=-.5]
   \draw[markx={1}](0,0) -- (0.5,0) ; 
   \draw (0.5,0) -- (1,0.5) ;
   \draw [markx={1}](0,0) -- (-0.5,0);
   \draw (-1,0.5) -- (-0.5,0);
   \draw(0,-1.5)--(0,-1); \draw [markx={0}](0,-1) --(0.5,0); 
   \draw[markx={0.5}](-0.5,0) to (0,-1);
   \ball{(0.25,-0.5)};
\end{tikzpicture} \stackrel{ix}{=}\begin{tikzpicture}[basel={-.5}]
   \inangarc{2}{(-0.5,0)}{90};
   \draw[markx={0.5},markx={0},markx={1}] (-0.5,0) --(0,0);
   \draw (0,0) to[bend right=50] (1,0) ;
   \draw (0,0) to[bend left=50] (1,0);
   \ball{(0.5,0.2)};
   \draw [markx={0}](1,0)--(1.5,0);
\end{tikzpicture},
\end{equation*}
to end up with a two-qubit loop on the right side, which can be separated into two cases according to the two-qubit constructions rules. The first one, using rule (xi.1), gives:
\begin{equation*}
  \begin{tikzpicture}[basel={-.5}]
   \inangarc{2}{(-0.5,0)}{90};
   \draw[markx={0.5},markx={0},markx={1}] (-0.5,0) --(0,0);
   \draw (0,0) to[bend right=50] (1,0) ;
   \draw (0,0) to[bend left=50] (1,0);
   \ball{(0.5,0.2)};
   \draw [markx={0}](1,0)--(1.5,0); 
\end{tikzpicture}  \stackrel{xi}{\propto}  \begin{tikzpicture}[basel={-.5}]
   \inangarc{2}{(-0.5,0)}{90};
   \draw[markx={0.3},markx={0},] (-0.5,0) --(0.2,0);
   \ball {(0,0)};  
\end{tikzpicture} \cong \begin{tikzpicture}[basel={-.5}]
   \inangarc{2}{(-0.5,0)}{90};
   \draw[markx={0}] (-0.5,0) --(-0.1,0); 
\end{tikzpicture} ,
\end{equation*}
where in the last step, we dropped the external operations, showing that this state is SLOCC equivalent to $\ket{W_3}$. The second case is when the inner ball is a Z-ball. Using the rule (xi.2), we obtain:
\begin{equation*}
  \begin{tikzpicture}[basel={-.5}]
   \inangarc{2}{(-0.5,0)}{90};
   \draw[markx={0.5},markx={0},markx={1}] (-0.5,0) --(0,0);
   \draw (0,0) to[bend right=50] (1,0) ;
   \draw[marko={0.5}] (0,0) to[bend left=50] (1,0);
   \draw [markx={0}](1,0)--(1.5,0); 
\end{tikzpicture}  \stackrel{xi}{=}  \begin{tikzpicture}[basel={-.5}]
   \inangarc{2}{(-0.5,0)}{90};
   \draw[markx={0},markx={0.8}] (-0.5,0) --(-0.2,0);
   \qubitl{(0,0)};
   \qubitr{(0.5,0)};
\end{tikzpicture}\stackrel{vii}{=}   \begin{tikzpicture}
    \draw[markx={0.5}](0,0) to (0.5,0);
    \qubitr{(0.7,0)};
\end{tikzpicture},
\end{equation*}
where in the last step, we used operations on qubits (vii.2). In this case, the construction shows explicitly that one of the qubits ends up separated from the other two; this is a property of the proposed construction that allows identifying separability of subsets which is in general not obvious to tell from the coefficient expansion of multipartite states.\\

\noindent By using the set of rules, we explored the SLOCC classification problem for the cases of three and four qubits, leading to minimal graphs that can be used to reproduce any of the classes. In the following, we present the graph representatives of the SLOCC classes.

\subsubsection{Two qubits}
For the two qubits case, there are two SLOCC classes: the separable class and the entangled class. Both can be obtained with the same graph but different parameters:
\begin{equation*}
     \begin{tikzpicture}[basel={-1}]
   \draw[markx={1}] (0,0) -- (0.5,0) ; 
   \draw [marko={0.5}](0.5,0)  to [bend left=50] (1.5,0) ;
   \draw  (1.5,0) to [bend left=50] (0.5,0) ;
   \draw [markx={0}] (1.5,0)-- (2,0);
\end{tikzpicture} = \begin{tikzpicture}[basel={-.5}]
    \qubitl{(0,0)}
    \qubitr{(0.8,0)}
    \draw(-0.5,0) to (-0.25,0);
      \draw(1.45,0) to (1.7,0);
\end{tikzpicture}  ,  \quad 
     \begin{tikzpicture}[basel={-1}]
   \draw[markx={1}] (0,0) -- (0.5,0) ; 
   \draw  (1.5,0) to [bend right=50] (0.5,0) ;
   \draw (1.5,0) to [bend left=50] (0.5,0) ;
   \draw[markx={0}]  (1.5,0)-- (2,0);
\end{tikzpicture} \cong \begin{tikzpicture}[basel={-.5}]
\draw (0,0) -- (1,0);   
\end{tikzpicture}
\end{equation*}
\subsubsection{Three qubits}
\label{Threequbits}
Three qubit states are separated into six SLOCC classes as shown in Section \ref{Threequbitsclassification}. Modifications of the same one-loop graph can obtain all of them. The completely separable state can be obtained when using $\Psi$-stitches on all the inner edges:
\begin{equation*}
\begin{tikzpicture}[basel={-.5}]
   \draw[markx={0.5},markx={0},markx={1}]  (-0.5,0) -- (-0.5,-1) ; 
   \draw (-0.5,-1) -- (-1,-1.5) ;
  \draw (-1,0.5)--(-0.5,0);
   \draw [markx={0.5},markx={0},markx={1}]  (0.5,-0.5)--(-0.5,0);
   \draw[markx={0.5}]  (-0.5,-1)--(0.5,-0.5);
   \draw  (0.5,-0.5)--(1,-0.5);
   \end{tikzpicture} =\begin{tikzpicture}[basel={-.5}]
    \qubitl{(0,0.2)}; 
    \qubitl{(0,-0.2)};
    \qubitr{(0.5,0)};
    \draw[markx={1}](-0.5,0.2)to (-0.25,0.2); 
   \draw[markx={1}](-0.5,-0.2)to (-0.25,-0.2); 
     \draw[markx={0}](1.15,0) to (1.4,0); \end{tikzpicture}.
   \end{equation*}
The classes $A-BC,B-AC,C-AB$ can be obtained up to permutations when using one $\Psi$-stitch, two $\Phi$-stitches, and using a inner Z-ball:
\begin{equation*}
\begin{tikzpicture}[basel={-.5}]
   \draw[markx={0.5},markx={0},markx={1}]  (-0.5,0) -- (-0.5,-1) ; 
   \draw (-0.5,-1) -- (-1,-1.5) ;
  \draw (-1,0.5)--(-0.5,0);
   \draw [markx={0},markx={1}]  (0.5,-0.5)--(-0.5,0);
   \draw[marko={0.5}](0.5,-0.5) -- (-0.5,-1);
   \draw  (0.5,-0.5)--(1,-0.5);
   \end{tikzpicture} =\begin{tikzpicture}[basel={-.5}] \draw[markx={0.5}](0,0.25) to [bend left=90](0,-0.25);\qubitr{(0.5,0)};
     \draw(1.15,0) to (1.4,0);
     \end{tikzpicture}.
   \end{equation*}
   The W class is the primitive vertex on this construction; however, it can also be obtained when using two $\Psi$-stitches and one $\Phi$-stitch as:
   \begin{equation*}
\begin{tikzpicture}[basel={-.5}]
   \draw[markx={0.5},markx={0},markx={1}]  (-0.5,0) -- (-0.5,-1) ; 
   \draw (-0.5,-1) -- (-1,-1.5) ;
  \draw (-1,0.5)--(-0.5,0);
   \draw [markx={0},markx={1}]  (0.5,-0.5)--(-0.5,0);
   \draw[markx={0.5}](0.5,-0.5) -- (-0.5,-1);
   \draw  (0.5,-0.5)--(1,-0.5);
   \end{tikzpicture} =\begin{tikzpicture}[basel={-.5}] \draw[markx={1}](-0.5,0.5) to (0,0);
   \draw(-0.5,-0.5) to (0,0);
   \draw(0.5,0) to (0,0);   
     \end{tikzpicture}.
   \end{equation*}
Finally, the $GHZ$ class can be obtained from the one-loop graph when all the stitches are $\Phi$-stitches   
 \begin{equation}
\begin{tikzpicture}[basel={-.5}]
   \draw[markx={0},markx={1}]  (-0.5,0) -- (-0.5,-1) ; 
   \draw (-0.5,-1) -- (-1,-1.5) ;
  \draw (-1,0.5)--(-0.5,0);
   \draw [markx={0},markx={1}]  (0.5,-0.5)--(-0.5,0);
   \draw(0.5,-0.5) -- (-0.5,-1);
   \draw  (0.5,-0.5)--(1,-0.5);
   \end{tikzpicture} \cong \ket{GHZ}.
   \label{eq:triangleGHZ}
     \end{equation}
This last result was obtained from the ZW perspective in \cite{ZW}.
\section{Graph equivalence of four qubit SLOCC families}
\label{FourQubitsVerst}
All SLOCC families for four qubits shown in Table \ref{table:fourqubitsclass} can be obtained from graphs with six $W$ vertices and seven stitches.
The correspondence shown here was obtained using the classification protocol shown in \cite{LuqueThibon-tame}; however, this process is not trivial, and we will describe it here. For this, let us consider again the first example in Section \ref{Examples}. The steps to find the correspondence with SLOCC-families are the following:
\begin{enumerate}[1.]
\item First, we start with the graph that we want to analyze and a choice of stitches to be used. We will choose again the case where all stitches are $\Phi$-stitches:
\begin{equation*}
 \begin{tikzpicture}[basel={-1},every node/.style={scale=0.6}]
        \draw (-0.2,0.2) to (0,0);
        \draw[markxi={0},markxi={1}] (0,0) to (1,0);
        \draw[\lac\rao] (0,0) to (1,0);
        \draw[\lac\rao] (0,0) to (0,-1);
        \draw[markxi={0},markxi={1}] (0,-1) to (1,-1); \draw[\lao\rac] (0,-1) to (1,-1);
        \draw[\lao\rac] (1,0) to (1,-1);
        \draw (1,-1) to (1.2,-1.2);
        \draw (1,0) to (1.2,0.2);
        \draw (-0.2,-1.2) to (0,-1);
\end{tikzpicture}
\end{equation*}
\item Then, we apply the stitching and pushing rules from Section \ref{rules} to clean as many stitching parameters as possible. We have already shown this process, and the resulting graph state is:
\begin{equation*}
\begin{tikzpicture}[basel={-1},every node/.style={scale=0.6}]
        \draw (-0.2,0.2) to (0,0);
        \draw[markxi={0},markxi={1}] (0,0) to (1,0);
        \draw(0,0) to (1,0);
        \draw(0,0) to (0,-1);
        \draw[markxi={0},markxi={1}] (0,-1) to (1,-1); \draw(0,-1) to (1,-1);
        \draw(1,0) to (1,-1);
        \draw(1,-1) to (1.2,-1.2);
        \draw (1,0) to (1.2,0.2);
        \draw(-0.2,-1.2) to (0,-1);

        \ball{(0.5,0)};
        \draw (-0.3,0.3) node {$A$};
        \draw (1.3,0.3) node {$B$};
        \draw (1.3,-1.3) node {$C$};
        \draw (-0.3,-1.3) node {$D$};
\end{tikzpicture}
\end{equation*}

\item The ``cleaned'' state is now calculated in its vector form, which will have as many parameters as the stitching parameters in the graph. For the example, the obtained vector is:
\begin{equation*}
\ket{\psi^{\scalebox{0.3}{\begin{tikzpicture}[basel={-1},every node/.style={scale=0.6}]
        \draw (-0.2,0.2) to (0,0);
        \draw[markxi={0},markxi={1}] (0,0) to (1,0);
        \draw(0,0) to (1,0);
        \draw(0,0) to (0,-1);
        \draw[markxi={0},markxi={1}] (0,-1) to (1,-1); \draw(0,-1) to (1,-1);
        \draw(1,0) to (1,-1);
        \draw(1,-1) to (1.2,-1.2);
        \draw (1,0) to (1.2,0.2);
        \draw(-0.2,-1.2) to (0,-1);
        \ball{(0.5,0)};
\end{tikzpicture}}}_v} = \left(v+\frac{1}{v} \right)\ket{0000}+v \left(\ket{0011}+\ket{0101}+\ket{1100}+\ket{1111} \right) + \frac{1}{v}\ket{1010},
\end{equation*}
where $v$ is a generic parameter for the ball left after cleaning the graph. Here, we introduced a notation that will be recurrent, denoting by $\ket{\psi^{G}_{\vec{\Theta}}}$ to the state obtained when filling the graph state $G$ with the list of parameters $\vec{\Theta}$.
\item For the obtained state, we include a scale factor $s$ that allows us to relate unnormalized states and calculate a complete set of four invariants, $B,L,M, D_{xy}$ \cite{LuqueThibon-tame}(equivalently we can also use the invariants that we propose in Section \ref{Invariants}). For our example, this is:
\begin{equation*}
\begin{gathered}
B\left(s\ket{\psi^{\scalebox{0.1}{\begin{tikzpicture}[basel={-1},every node/.style={scale=0.6}]
        \draw (-0.2,0.2) to (0,0);
        \draw[markxi={0},markxi={1}] (0,0) to (1,0);
        \draw(0,0) to (1,0);
        \draw(0,0) to (0,-1);
        \draw[markxi={0},markxi={1}] (0,-1) to (1,-1); \draw(0,-1) to (1,-1);
        \draw(1,0) to (1,-1);
        \draw(1,-1) to (1.2,-1.2);
        \draw (1,0) to (1.2,0.2);
        \draw(-0.2,-1.2) to (0,-1);
        \ball{(0.5,0)};
\end{tikzpicture}}}_v}\right) = 2(1+v^2)s^2  \\
L\left(s\ket{\psi^{\scalebox{0.1}{\begin{tikzpicture}[basel={-1},every node/.style={scale=0.6}]
        \draw (-0.2,0.2) to (0,0);
        \draw[markxi={0},markxi={1}] (0,0) to (1,0);
        \draw(0,0) to (1,0);
        \draw(0,0) to (0,-1);
        \draw[markxi={0},markxi={1}] (0,-1) to (1,-1); \draw(0,-1) to (1,-1);
        \draw(1,0) to (1,-1);
        \draw(1,-1) to (1.2,-1.2);
        \draw (1,0) to (1.2,0.2);
        \draw(-0.2,-1.2) to (0,-1);
        \ball{(0.5,0)};
\end{tikzpicture}}}_v}\right) = s^4  \\
M\left(s\ket{\psi^{\scalebox{0.1}{\begin{tikzpicture}[basel={-1},every node/.style={scale=0.6}]
        \draw (-0.2,0.2) to (0,0);
        \draw[markxi={0},markxi={1}] (0,0) to (1,0);
        \draw(0,0) to (1,0);
        \draw(0,0) to (0,-1);
        \draw[markxi={0},markxi={1}] (0,-1) to (1,-1); \draw(0,-1) to (1,-1);
        \draw(1,0) to (1,-1);
        \draw(1,-1) to (1.2,-1.2);
        \draw (1,0) to (1.2,0.2);
        \draw(-0.2,-1.2) to (0,-1);
        \ball{(0.5,0)};
\end{tikzpicture}}}_v}\right) =-v^4 s^4 \\
D_{xy} \left(s\ket{\psi^{\scalebox{0.1}{\begin{tikzpicture}[basel={-1},every node/.style={scale=0.6}]
        \draw (-0.2,0.2) to (0,0);
        \draw[markxi={0},markxi={1}] (0,0) to (1,0);
        \draw(0,0) to (1,0);
        \draw(0,0) to (0,-1);
        \draw[markxi={0},markxi={1}] (0,-1) to (1,-1); \draw(0,-1) to (1,-1);
        \draw(1,0) to (1,-1);
        \draw(1,-1) to (1.2,-1.2);
        \draw (1,0) to (1.2,0.2);
        \draw(-0.2,-1.2) to (0,-1);
        \ball{(0.5,0)};
\end{tikzpicture}}}_v}\right)=-2v^4s^6 (1+v^2).
\end{gathered}
\end{equation*}

\item We proceed to apply the classification protocol in section V of \cite{LuqueThibon-tame}; for this, we calculate the roots of the \textit{quartics} $Q_1,Q_2$ and $Q_3$ also introduced in \cite{LuqueThibon-tame}. \\

For the example, we can  focus on $Q_3$ whose general expression is:
\begin{equation*}
Q_3\left(s\ket{\psi^{\scalebox{0.1}{\begin{tikzpicture}[basel={-1},every node/.style={scale=0.6}]
        \draw (-0.2,0.2) to (0,0);
        \draw[markxi={0},markxi={1}] (0,0) to (1,0);
        \draw(0,0) to (1,0);
        \draw(0,0) to (0,-1);
        \draw[markxi={0},markxi={1}] (0,-1) to (1,-1); \draw(0,-1) to (1,-1);
        \draw(1,0) to (1,-1);
        \draw(1,-1) to (1.2,-1.2);
        \draw (1,0) to (1.2,0.2);
        \draw(-0.2,-1.2) to (0,-1);
        \ball{(0.5,0)};
\end{tikzpicture}}}_v} \right) = \left(x-s^2(1+v^2)y \right)^{2} \left(x^2-2s^2(1+v^2)xy+s^4(v^2-1)^2 y^2\right),
\end{equation*}
where three different cases can be identified. 
\begin{enumerate}[i.]
\item The first case, when $v=\pm 1$ ( we will pick $v=1$, for simplicity), where
\begin{equation*}
Q_3\left(s\ket{\psi^{\scalebox{0.1}{\begin{tikzpicture}[basel={-1},every node/.style={scale=0.6}]
        \draw (-0.2,0.2) to (0,0);
        \draw[markxi={0},markxi={1}] (0,0) to (1,0);
        \draw(0,0) to (1,0);
        \draw(0,0) to (0,-1);
        \draw[markxi={0},markxi={1}] (0,-1) to (1,-1); \draw(0,-1) to (1,-1);
        \draw(1,0) to (1,-1);
        \draw(1,-1) to (1.2,-1.2);
        \draw (1,0) to (1.2,0.2);
        \draw(-0.2,-1.2) to (0,-1);
        \ball{(0.5,0)};
\end{tikzpicture}}}_{v=1}} \right) = \left(x-2ys^2  \right)^{2} \left(x^2-4s^2xy \right),
\end{equation*}
whose roots, obtained by fixing $y=1$, $Q_3=0$ and solving for $x$, are:
\begin{equation}
x_1=0,\quad x_2=x_3=2s^2,\quad x_4=4s^2.
\label{eq:roots1}
\end{equation}
Then, when fixing $v=1$, the classification falls in the case (2), where only one of the quartics (in this case, $Q_3$) has a zero root. As there is a double nonzero root ($x_2=x_3=2s^2$), the specific case is (d). Then, we evaluate the covariant $\mathcal{L}$ and get $\mathcal{L}\left(s\ket{\psi^{\scalebox{0.1}{\begin{tikzpicture}[basel={-1},every node/.style={scale=0.6}]
        \draw (-0.2,0.2) to (0,0);
        \draw[markxi={0},markxi={1}] (0,0) to (1,0);
        \draw(0,0) to (1,0);
        \draw(0,0) to (0,-1);
        \draw[markxi={0},markxi={1}] (0,-1) to (1,-1); \draw(0,-1) to (1,-1);
        \draw(1,0) to (1,-1);
        \draw(1,-1) to (1.2,-1.2);
        \draw (1,0) to (1.2,0.2);
        \draw(-0.2,-1.2) to (0,-1);
        \ball{(0.5,0)};
\end{tikzpicture}}}_{v=1}} \right) =0$, meaning that the state belongs to the $G_{abb0}$ subfamily, this  is, the family $G_{abcd}$ with $d=0$ and $c=b$.
\item The second case is when  $v=\pm i$ ( we will pick $v=i$ by simplicity), where
\begin{equation*}
Q_3\left(s\ket{\psi^{\scalebox{0.1}{\begin{tikzpicture}[basel={-1},every node/.style={scale=0.6}]
        \draw (-0.2,0.2) to (0,0);
        \draw[markxi={0},markxi={1}] (0,0) to (1,0);
        \draw(0,0) to (1,0);
        \draw(0,0) to (0,-1);
        \draw[markxi={0},markxi={1}] (0,-1) to (1,-1); \draw(0,-1) to (1,-1);
        \draw(1,0) to (1,-1);
        \draw(1,-1) to (1.2,-1.2);
        \draw (1,0) to (1.2,0.2);
        \draw(-0.2,-1.2) to (0,-1);
        \ball{(0.5,0)};
\end{tikzpicture}}}_{v=i}} \right) = x^2(x^2+4s^4y^2),
\end{equation*}
whose roots are:
\begin{equation}
x_1=x_2=0, \quad x_3=-2i s^2, \quad x_4 =2 i s^2.
\label{eq:roots2}
\end{equation}
Then, when fixing $v=i$, the classification falls in the case (2.b), where one of the quartics ($Q_3$) has a double zero root. Then, by evaluating the two covariants $\mathcal{K}_3$ and $\mathcal{L}$ we get:
\begin{equation*}
\mathcal{K}_3 \left(s\ket{\psi^{\scalebox{0.1}{\begin{tikzpicture}[basel={-1},every node/.style={scale=0.6}]
        \draw (-0.2,0.2) to (0,0);
        \draw[markxi={0},markxi={1}] (0,0) to (1,0);
        \draw(0,0) to (1,0);
        \draw(0,0) to (0,-1);
        \draw[markxi={0},markxi={1}] (0,-1) to (1,-1); \draw(0,-1) to (1,-1);
        \draw(1,0) to (1,-1);
        \draw(1,-1) to (1.2,-1.2);
        \draw (1,0) to (1.2,0.2);
        \draw(-0.2,-1.2) to (0,-1);
        \ball{(0.5,0)};
\end{tikzpicture}}}_{v=i}} \right) \neq 0 , \quad \mathcal{L}\left(s\ket{\psi^{\scalebox{0.1}{\begin{tikzpicture}[basel={-1},every node/.style={scale=0.6}]
        \draw (-0.2,0.2) to (0,0);
        \draw[markxi={0},markxi={1}] (0,0) to (1,0);
        \draw(0,0) to (1,0);
        \draw(0,0) to (0,-1);
        \draw[markxi={0},markxi={1}] (0,-1) to (1,-1); \draw(0,-1) to (1,-1);
        \draw(1,0) to (1,-1);
        \draw(1,-1) to (1.2,-1.2);
        \draw (1,0) to (1.2,0.2);
        \draw(-0.2,-1.2) to (0,-1);
        \ball{(0.5,0)};
\end{tikzpicture}}}_{v=i}} \right)=0,
\end{equation*}
meaning that the state belong to the subfamily $L_{ab0_2}$, which is the family $L_{abc_2}$ with $c=0$. 
\item The last case, is when $v\neq \pm 1$ and $v\neq \pm i$, then the roots of $Q_3$ are:
\begin{equation}
x_1=s^2(v-1)^2, \quad x_2=s^2(1+v)^2 , \quad x_3=x_4=s^2 (1+v^2) .
\label{eq:roots3}
\end{equation}
Then, the classification falls in the case (1.b), that when evaluating the covariant $\mathcal{L}$ we get  $\mathcal{L}\left(s\ket{\psi^{\scalebox{0.1}{\begin{tikzpicture}[basel={-1},every node/.style={scale=0.6}]
        \draw (-0.2,0.2) to (0,0);
        \draw[markxi={0},markxi={1}] (0,0) to (1,0);
        \draw(0,0) to (1,0);
        \draw(0,0) to (0,-1);
        \draw[markxi={0},markxi={1}] (0,-1) to (1,-1); \draw(0,-1) to (1,-1);
        \draw(1,0) to (1,-1);
        \draw(1,-1) to (1.2,-1.2);
        \draw (1,0) to (1.2,0.2);
        \draw(-0.2,-1.2) to (0,-1);
        \ball{(0.5,0)};
\end{tikzpicture}}}_{v}} \right) =0$, meaning that the state belongs to the $G_{abcc}$ subfamily ($G_{abcd}$ with $d=c$).
\end{enumerate}
\item The last step is to find a parametrization of the stitching parameters that makes the map between them and the parameters $a,b,c,d$ explicit. This can be done by equating the roots of the quartics from the graph state with the roots of the quartics of the corresponding family or subfamily. In this example, we want to find $s$ and $v$ as functions of $a,b$, and $c$ (none of the cases have $d$ as a parameter). For each of the cases, we have the following:
\begin{enumerate}[i.]
\item For the first case, the representative is $G_{abb0}$, this state has a zero root in $Q_1$. So, it is enough to equate
\begin{equation*}
r\left[Q_1(\ket{G_{abb0}})\right]=r\left[Q_3\left(s\ket{\psi^{\scalebox{0.1}{\begin{tikzpicture}[basel={-1},every node/.style={scale=0.6}]
        \draw (-0.2,0.2) to (0,0);
        \draw[markxi={0},markxi={1}] (0,0) to (1,0);
        \draw(0,0) to (1,0);
        \draw(0,0) to (0,-1);
        \draw[markxi={0},markxi={1}] (0,-1) to (1,-1); \draw(0,-1) to (1,-1);
        \draw(1,0) to (1,-1);
        \draw(1,-1) to (1.2,-1.2);
        \draw (1,0) to (1.2,0.2);
        \draw(-0.2,-1.2) to (0,-1);
        \ball{(0.5,0)};
\end{tikzpicture}}}_{v=1}} \right) \right],
\end{equation*}
where $r$ refers to the roots of the functions. The roots of $Q_1$ for the representative $G_{abb0}$ are:
\begin{equation*}
x_1=0,\quad x_2=a^2 , \quad x_3=x_4=b^2.
\end{equation*}
The only solution for these roots to be equal to the roots in Equation \eqref{eq:roots1} is:
\begin{equation*}
 b= \frac{a}{\sqrt{2}}, \quad s^2=\frac{a^2}{4}.
\end{equation*}
Note how the resultant subfamily is more restricted than $G_{abb0}$; this is expected because we only have the scale factor to modify our graph state. Then, we can achieve a one-parametric subfamily at most. Finally, we find that:
\begin{equation*}
   \pm \frac{a}{2} \cdot  \begin{tikzpicture}[basel={-1},every node/.style={scale=0.6}]
        \draw (-0.2,0.2) to (0,0);
        \draw[markxi={0},markxi={1}] (0,0) to (1,0);
        \draw(0,0) to (1,0);
        \draw(0,0) to (0,-1);
        \draw[markxi={0},markxi={1}] (0,-1) to (1,-1); \draw(0,-1) to (1,-1);
        \draw(1,0) to (1,-1);
        \draw(1,-1) to (1.2,-1.2);
        \draw (1,0) to (1.2,0.2);
        \draw(-0.2,-1.2) to (0,-1);
\end{tikzpicture} \cong \ket{G_{a \frac{a}{\sqrt{2}}\frac{a}{\sqrt{2}}0}},
\end{equation*}
the SLOCC relation can be achieved strictly with local operations in $SL_2$, and permutations of the parts,  and the scale factor is already known.
\item For the second case, the representative is $L_{ab0_2}$, the zero root is in $Q_1$, whose roots are:
\begin{equation*}
x_1=x_2=0,\quad x_3=a^2 , \quad x_4=b^2.
\end{equation*}
Equating these roots with the ones of  Equation \eqref{eq:roots2} we get:
\begin{equation*}
b^2=- a^2 , \quad s^2=\frac{i a^2}{2} .
\end{equation*}
Again, in this case, the resultant subfamily is more restricted than the one identified using the classification protocol. We obtain finally :
\begin{equation*}
\sqrt{\frac{i}{2}}a \cdot 
  \begin{tikzpicture}[basel={-1},every node/.style={scale=0.6}]
        \draw (-0.2,0.2) to (0,0);
        \draw[markxi={0},markxi={1}] (0,0) to (1,0);
        \draw[marko={0.5}](0,0) to (1,0);
        \draw(0,0) to (0,-1);
        \draw[markxi={0},markxi={1}] (0,-1) to (1,-1); \draw(0,-1) to (1,-1);
        \draw(1,0) to (1,-1);
        \draw(1,-1) to (1.2,-1.2);
        \draw (1,0) to (1.2,0.2);
        \draw(-0.2,-1.2) to (0,-1);
\end{tikzpicture} \cong \ket{L_{a (\pm ia) 0_2}}.
\end{equation*}
\item For the last case, the representative is $G_{abcc}$. As we only have as free parameters $s$ and $v$ from the graph state, we know this representative must be more restricted. We can only achieve, at most, a two-parametric subfamily from the graph state. The roots of the first quartic of the representative are:
\begin{equation*}
x_1=a^2,\quad  x_2=b^2, \quad  x_3=x_4=c^2,
\end{equation*}
equating these roots with the ones in Equation \eqref{eq:roots3} we get as a possible solution:
\begin{equation*}
c=\sqrt{a b}, \quad v= \frac{2 \sqrt{a b}}{a-b}, \quad s= \frac{1}{2}(b-a).
\end{equation*}
So, we have the following equivalence:
\begin{equation*}
\frac{1}{2} (b-a) \cdot 
 \begin{tikzpicture}[basel={-2},every node/.style={scale=0.6}]
        \draw (-0.2,0.2) to (0,0);
        \draw[markxi={0},markxi={1}] (0,0) to (1,0);
        \draw(0,0) to (1,0);
        \draw(0,0) to (0,-1);
        \draw[markxi={0},markxi={1}] (0,-1) to (1,-1); \draw(0,-1) to (1,-1);
        \draw(1,0) to (1,-1);
        \draw(1,-1) to (1.2,-1.2);
        \draw (1,0) to (1.2,0.2);
        \draw(-0.2,-1.2) to (0,-1);
        \ball{(0.5,0)};
        \draw (0.5,0.5) node{$ \frac{2\sqrt{ab}}{(a-b)}$};
\end{tikzpicture} \cong \ket{G_{ab\sqrt{ab}\sqrt{ab}}}. 
\end{equation*}
\end{enumerate}
\end{enumerate}
We did this process for all possible graphs for all possible combinations of stitches, leading to an extensive list of subfamilies that are not worth putting explicitly here. \\

\noindent The first SLOCC family is the $L_{0_{3\oplus \bar{1}}0_{3\oplus \bar{1}}}$ corresponding to the product state obtained from a qubit and a $GHZ$ state. One graph that can achieve this class is:
\begin{equation*}
    \begin{tikzpicture}[basel={-.5}]
    \draw[markx={0.5},markx={0},markx={1}](0,0) to (1,0);
    \draw(1,0) to (1.5,0.5);
    \draw(1,0) to (1.5,-0.5);
    \draw(1.5,0.5) to (1.5,-0.5);
    \draw[markx={0}](1.5,0.5) to (1.75,0.75);
    \draw[markx={0}](1.5,-0.5) to (1.75,-0.75);
    \draw (0,0) to (0,0.25);
    \draw[markx={0.5}](-1,0)  to (0,0);
    \draw[marko={0.5},markx={0},markx={1}](-2,0) to [bend left=50](-1,0);
    \draw(-2,0) to [bend right=50](-1,0);
    \draw (-2,0) to (-2.25,0);
    \end{tikzpicture}= \begin{tikzpicture}[basel={-.5}]
    \draw[markx={0.5},markx={0}](0,0) to (1,0);
    \draw[markx={0},markx={1}](1,0) to (1.5,0.5);
    \draw(1,0) to (1.5,-0.5);
    \draw[markx={1}](1.5,0.5) to (1.5,-0.5);
    \draw(1.5,0.5) to (1.75,0.75);
    \draw(1.5,-0.5) to (1.75,-0.75);
    \draw (0,0) to (0,0.25);
    \draw[markx={1}] (0,0) to (-0.25,0);
    \qubitr{(-0.9,0)};
    \qubitl{(-1.75,0)};
    \draw (-2,0) to (-2.25,0);
    \end{tikzpicture} =
    \begin{tikzpicture}[basel={-.5}]
    \draw (0,0) to (1,0);
    \draw[markx={1}](1,0) to (1.5,0.5);
    \draw[markx={0},markx={1}](1,0) to (1.5,-0.5);
    \draw(1.5,0.5) to (1.5,-0.5);
    \draw(1.5,0.5) to (1.75,0.75);
    \draw(1.5,-0.5) to (1.75,-0.75);
   \qubitl{(-0.75,0)};
    \draw(-1,0) to (-0.75,0);
    \end{tikzpicture} \cong \ket{L_{0_{3\oplus \bar{1}}0_{3\oplus \bar{1}}}}. 
\end{equation*}
This family has no parameters, and it is the only one with separable qubits in the representative state. From the graph structure, it is seen clearly that the resultant state is SLOCC equivalent to $\ket{0}\otimes\ket{GHZ}$, which is the form of the representative state $ \ket{L_{0_{3\oplus \bar{1}}0_{3\oplus \bar{1}}}}$. \\

\noindent Another three families can be obtained from the  graph state 
\begin{equation*}
\ket{\psi^{\scalebox{0.3}{\begin{tikzpicture}[basel={-.5}]
        \draw (0,0) to (-0.25,0);
        \draw[markx={0},markx={1}](0,0) to  (1,1) ;
         \draw(0,0) to  (1,-1) ;
          \draw (1,1) to  (1.5,0.5) ;
          \draw [markx={0},markx={1}](1,-1) to  (1.5,-0.5) ;
          \draw [markx={0},markx={1}](1.5,0.5) to (2,0)  ;
           \draw (1.5,-0.5) to (2,0)  ;
            \draw [\lac \rao](1.5,0.5) to (1.5,-0.5)  ;
            \draw (0.75,1.25) to (1,1);
            \draw (0.75,-1.25) to (1,-1);
            \draw (2.25,0) to (2,0);
    \end{tikzpicture}}}_{v_1,w_1,w_2}} =
    \begin{tikzpicture}[basel={-.5},,every node/.style={scale=0.6}]
        \draw (0,0) to (-0.25,0);
        \draw[markx={0},markx={1}](0,0) to  (1,1) ;
         \draw(0,0) to  (1,-1) ;
          \draw (1,1) to  (1.5,0.5) ;
          \draw [markx={0},markx={1}](1,-1) to  (1.5,-0.5) ;
          \draw [markx={0},markx={1}](1.5,0.5) to (2,0)  ;
           \draw (1.5,-0.5) to (2,0)  ;
            \draw [\lac \rao](1.5,0.5) to (1.5,-0.5)  ;
            \draw (0.75,1.25) to (1,1);
            \draw (0.75,-1.25) to (1,-1);
            \draw (2.25,0) to (2,0);
            \draw (1.5,1) node {$v_1,w_1$};
             \draw (1.5,-0.8) node {$w_2$};            
    \end{tikzpicture}.
\end{equation*}
By changing the parameters of the arrows, two different families can be obtained. Let $\vec{\Theta}=v_1,w_1,w_2$, be the parameters of the white arrow and the black arrow, we have that:
\begin{equation*}
\begin{gathered}    
\ket{\psi^{\scalebox{0.3}{\begin{tikzpicture}[basel={-.5}]
        \draw (0,0) to (-0.25,0);
        \draw[markx={0},markx={1}](0,0) to  (1,1) ;
         \draw(0,0) to  (1,-1) ;
          \draw (1,1) to  (1.5,0.5) ;
          \draw [markx={0},markx={1}](1,-1) to  (1.5,-0.5) ;
          \draw [markx={0},markx={1}](1.5,0.5) to (2,0)  ;
           \draw (1.5,-0.5) to (2,0)  ;
            \draw [\lac \rao](1.5,0.5) to (1.5,-0.5)  ;
            \draw (0.75,1.25) to (1,1);
            \draw (0.75,-1.25) to (1,-1);
            \draw (2.25,0) to (2,0);
    \end{tikzpicture}}}_{1,1,-1}}\cong \ket{L_{0_{5\oplus\bar{3}}}} ,\\
\ket{\psi^{\scalebox{0.3}{\begin{tikzpicture}[basel={-.5}]
        \draw (0,0) to (-0.25,0);
        \draw[markx={0},markx={1}](0,0) to  (1,1) ;
         \draw(0,0) to  (1,-1) ;
          \draw (1,1) to  (1.5,0.5) ;
          \draw [markx={0},markx={1}](1,-1) to  (1.5,-0.5) ;
          \draw [markx={0},markx={1}](1.5,0.5) to (2,0)  ;
           \draw (1.5,-0.5) to (2,0)  ;
            \draw [\lac \rao](1.5,0.5) to (1.5,-0.5)  ;
            \draw (0.75,1.25) to (1,1);
            \draw (0.75,-1.25) to (1,-1);
            \draw (2.25,0) to (2,0);
    \end{tikzpicture}}}_{1,1,1}}\cong  \ket{L_{0_{7\oplus\bar{1}}}} .
\end{gathered}
\end{equation*}
One more family is obtained by using one $\Psi$-stitch:
\begin{equation*}
    \begin{tikzpicture}[basel=-1.5,every node/.style={scale=0.6}]
        \draw (0,0) to (-0.25,0);
        \draw[markx={0},markx={1}](0,0) to  (1,1) ;
         \draw(0,0) to  (1,-1) ;
          \draw [markx={0},markx={1}](1,1) to  (1.5,0.5) ;
          \draw[markx={0.5},markx={0},markx={1}](1,-1) to  (1.5,-0.5) ;
          \draw [markx={0},markx={1}](1.5,0.5) to (2,0)  ;
           \draw (1.5,-0.5) to (2,0)  ;
            \draw [\lao-](1.5,0.5) to (1.5,-0.5)  ;
            \draw (0.75,1.25) to (1,1);
            \draw (0.75,-1.25) to (1,-1);
            \draw (2.25,0) to (2,0);
            \draw (1.5,1) node {$-1$};
    \end{tikzpicture} \cong \ket{L_{a_4}}.
\end{equation*}
This is a one-parametric family, and that parameter is reproduced with a scale factor on the graph state.  
The remaining five SLOCC families can be obtained from the graph:
\begin{equation*}
    \begin{tikzpicture}[basel={-.5}]
        \draw(0,0) to (1,0);
        \draw[markx={0},markx={1}] (0,0) to (0,-1);
        \draw (0,-1) to (1,-1);
        \draw[\lao-] (1,-1) to (2,0);
        \draw[\lac-] (1,0) to (2,-1);
        \draw (-0.25,0.25) to (0,0);
        \draw (0,-1) to (-0.25,-1.25);
        \draw (2,0) to (2.25,0.25);
        \draw[markx={0},markx={1}]  (1,0) to (2,0);
        \draw[markx={0},markx={1}]  (1,-1) to (2,-1);
        \draw (2,-1) to (2.25,-1.25);
        \draw (1,0.3) node {$v_1,w_1$};
        \draw (1,-1.3) node {$w_2$};
    \end{tikzpicture}.
\end{equation*}
For the four non-generic families, we have:
\begin{equation*}
\begin{gathered}    
\ket{\psi^{\scalebox{0.3}{ \begin{tikzpicture}[basel={-.5}]
        \draw(0,0) to (1,0);
        \draw[markx={0},markx={1}] (0,0) to (0,-1);
        \draw (0,-1) to (1,-1);
        \draw[\lao-] (1,-1) to (2,0);
        \draw[\lac-] (1,0) to (2,-1);
        \draw (-0.25,0.25) to (0,0);
        \draw (0,-1) to (-0.25,-1.25);
        \draw (2,0) to (2.25,0.25);
        \draw[markx={0},markx={1}]  (1,0) to (2,0);
        \draw[markx={0},markx={1}]  (1,-1) to (2,-1);
        \draw (2,-1) to (2.25,-1.25);
    \end{tikzpicture}}}_{1,1,0}} \cong \frac{\sqrt{2}}{a} \ket{L_{a_20_{3\oplus\bar{1}}}} ,\\
    \ket{\psi^{\scalebox{0.3}{ \begin{tikzpicture}[basel={-.5}]
        \draw(0,0) to (1,0);
        \draw[markx={0},markx={1}] (0,0) to (0,-1);
        \draw (0,-1) to (1,-1);
        \draw[\lao-] (1,-1) to (2,0);
        \draw[\lac-] (1,0) to (2,-1);
        \draw (-0.25,0.25) to (0,0);
        \draw (0,-1) to (-0.25,-1.25);
        \draw (2,0) to (2.25,0.25);
        \draw[markx={0},markx={1}]  (1,0) to (2,0);
        \draw[markx={0},markx={1}]  (1,-1) to (2,-1);
        \draw (2,-1) to (2.25,-1.25);
    \end{tikzpicture}}}_{\sqrt{\frac{a+b}{a-b}},\sqrt{\frac{a+b}{2a}},0}}\cong \frac{\sqrt{2}}{\sqrt{a(a-b)}} \ket{L_{ab_3}} ,\\
    \ket{\psi^{\scalebox{0.3}{ \begin{tikzpicture}[basel={-.5}]
        \draw(0,0) to (1,0);
        \draw[markx={0},markx={1}] (0,0) to (0,-1);
        \draw (0,-1) to (1,-1);
        \draw[\lao-] (1,-1) to (2,0);
        \draw[\lac-] (1,0) to (2,-1);
        \draw (-0.25,0.25) to (0,0);
        \draw (0,-1) to (-0.25,-1.25);
        \draw (2,0) to (2.25,0.25);
        \draw[markx={0},markx={1}]  (1,0) to (2,0);
        \draw[markx={0},markx={1}]  (1,-1) to (2,-1);
        \draw (2,-1) to (2.25,-1.25);
    \end{tikzpicture}}}_{\sqrt{\frac{a^2+b^2}{a^2 -b^2}},1,0}} \cong \frac{\sqrt{2}}{\sqrt{a^2-b^2}} \ket{L_{a_2b_2}},
 \\
    \ket{\psi^{\scalebox{0.3}{ \begin{tikzpicture}[basel={-.5}]
        \draw(0,0) to (1,0);
        \draw[markx={0},markx={1}] (0,0) to (0,-1);
        \draw (0,-1) to (1,-1);
        \draw[\lao-] (1,-1) to (2,0);
        \draw[\lac-] (1,0) to (2,-1);
        \draw (-0.25,0.25) to (0,0);
        \draw (0,-1) to (-0.25,-1.25);
        \draw (2,0) to (2.25,0.25);
        \draw[markx={0},markx={1}]  (1,0) to (2,0);
        \draw[markx={0},markx={1}]  (1,-1) to (2,-1);
        \draw (2,-1) to (2.25,-1.25);
    \end{tikzpicture}}}_{\sqrt{\frac{c^2-ab}{c^2+ab}},\sqrt{\frac{c^2-a^2-b^2}{2(c^2+ab)}},0}} \cong \frac{\sqrt{2}}{\sqrt{c^2+ab}} \ket{L_{abc_2}}.
\end{gathered}
\end{equation*}
It is clear how the parametrization gets more complicated with the number of parameters of the families. To parametrize the generic state, the same method used for the other eight families led to equation systems that were not solvable analytically. For this state, we simultaneously used equations from invariants, quartics, and roots to achieve a possible parametrization. Even with this, the obtained expressions were too complicated, so we found that a nice parametrization can be obtained by defining the symmetric combinations:
\begin{equation*}
    z_1=(a+b+c+d), \quad z_2=(ab+ac+ad+bc+bd+cd), \quad z_3 = (abc+abd+acd+bcd), \quad z_4=abcd.
\end{equation*}
Now, define $w_1^*,w_2^*$ as the solutions to the system of equations
\begin{equation*}
    w_1^2+w_2^2 = \frac{z_1z_2}{z_1z_2-2z_3} \quad, \quad  w_1^2 w_2^2 =\frac{ 2 (z_1 z_2 z_3-z_1 ^2 z_4 -z_3^2)^2 } {(z_1 z_2 - 2 z_3)^3(z_1^3-2z_1 z_2 +4 z_3)},
\end{equation*}
and define
\begin{equation*}
v_1=\sqrt{\frac{z_1^3-2z_1 z_2 +4 z_3}{2 (z_1 z_2 -2 z_3)}} , \qquad s= \sqrt{\frac{2z_1}{z_1 z_2 -2 z_3}},
\end{equation*}
then, the generic state $G_{abcd}$ can be obtained as
\begin{equation*}
 \ket{\psi^{\scalebox{0.3}{ \begin{tikzpicture}[basel={-.5}]
        \draw(0,0) to (1,0);
        \draw[markx={0},markx={1}] (0,0) to (0,-1);
        \draw (0,-1) to (1,-1);
        \draw[\lao-] (1,-1) to (2,0);
        \draw[\lac-] (1,0) to (2,-1);
        \draw (-0.25,0.25) to (0,0);
        \draw (0,-1) to (-0.25,-1.25);
        \draw (2,0) to (2.25,0.25);
        \draw[markx={0},markx={1}]  (1,0) to (2,0);
        \draw[markx={0},markx={1}]  (1,-1) to (2,-1);
        \draw (2,-1) to (2.25,-1.25);
    \end{tikzpicture}}}_{v_1,w_1^{*},w_2^{*}}} \cong  s \ket{G_{abcd}}.
\end{equation*}
With this, we complete the nine SLOCC families of Verstraette, and therefore, any four qubit state can be obtained from the graphs shown before. \noindent In \cite{SLOCCZW} and \cite{ZW}, a brief exploration of graphs from ZW-calculus are shown for reproducing non-parametric representatives of super-classes \cite{superclasses} of four qubits, which correspond to a different classification of SLOCC classes. What we presented here is a complete scheme to reproduce graphically any four-qubit SLOCC class by showing what graph and what stitching parameters are needed to reproduce the nine SLOCC families \cite{Verstraete} with free parameters $a,b,c,d$. \\

\noindent In the next chapter, we will see how this multi-qubit construction using W states and bipartite states can be exploited to find explicit expressions for general Kronecker states and an orthogonal basis for Kronecker subspaces. One of the requirements for the process shown in the next chapter is that we can only ensure that every Kronecker subspace can be obtained if we can build any multi-qubit state from W states and bipartite states only. Due to the connection with ZW calculus and its completeness, we know that it is always possible; moreover, in this chapter, we have shown how these constructions can be done explicitly for the cases of three and four qubits.
\chapter{Kronecker states from W-Stitching}
\label{Chapter5}
In this chapter, we cover the main objective of this document, developing a systematic procedure to calculate multipartite locally maximally entangled states (LME) that belong to the invariant subspace of the tensor products of irreps of the symmetric group $S_n$, where the Hilbert spaces that are entangled have dimensions that grow exponentially with $n$. We call these states \textit{Kronecker states} and we name the invariant subspace where they belong, \textit{Kronecker subsbace}. Interestingly, Kronecker states form to a vector space of LME states, whose dimension is given by the \textit{Kronecker coefficient}, which is generally greater than one. This makes the Kronecker subspace a promising vector space for applications in quantum information, as quantum secret sharing, or designing codes for quantum error correction. For our construction, we exploit two already discussed facts: first, it is possible to obtain expressions for Kronecker states that appear in the multi-copy decomposition of states in the $W$-class and bipartite states in their Schur basis, and second, any multiqubit state can be obtained by stitching W-states with bipartite states. The main observation in the stitching procedure is that when the multi-particle Schur transform is applied to any graph state, the $GL_2$ and $S_n$ subspaces retain the graph structure. In other words, the Schur transformation of a generic graph state can be identified in the graphical notation with the same graphs in $\{\bm{\lambda}\}$ and in $[\bm{\lambda}]$ representations, making it possible to read explicitly from the graph a way to calculate generic Kronecker states from W-Kronecker states.\\

\noindent We begin this chapter with a recapitulation of the problem's setup and the partial solutions developed for calculating Kronecker states. Then, we explore the limitations of the known approaches and emphasize what can and can not be done with them. Later, we generalize the stitching method to calculate Kronecker states and $\Phi$ states, showing how, with this approach, some of the limitations of the known methods are simple to overcome. We dedicate the rest of this chapter to describing the minimal conditions to calculate any possible Kronecker subspace. We discuss the properties of the construction and how the connection between graph states and Kronecker states gives us information about SLOCC classes from their graph representatives.

\section{State of the art: Kronecker states}
We name Kronecker states $\ket{\mathcal{K}_{\bm{\lambda}}}$ the $N$-partite states that belong to the invariant subspace of tensor products of $N$ irreps of $S_n$:
\begin{equation*}
\ket{\mathcal{K}_{\bm{\lambda}}} \in [\bm{\lambda}]^{S_n},
\end{equation*}
where $\bm{\lambda}=\lambda^1, \lambda^2, \dots ,\lambda^N$ is a set of $N$ partitions of $n$ with at most two parts, $[\bm{\lambda}]=\bigotimes_{i=1}^N [\lambda^i]$, and $V^{S_n}$ represents the invariant subspace of $V$ under the same action of the symmetric group $S_n$ on all the parts. As partitions are restricted to having at most two parts, we usually use a shorthand notation for labeling partitions only with their second index: $\lambda^i= \lambda^i_2$. For the convenience of the reader, here we summarize properties of Kronecker states and Kronecker subspaces that were already discussed in Chapter \ref{Chapter2}:
\begin{itemize}
\item Kronecker states are invariant under the diagonal action of $S_n$:
\begin{equation*}
D^{\bm{\lambda}}(\pi) \ket{\mathcal{K}_{\bm{\lambda}}}=\ket{\mathcal{K}_{\bm{\lambda}}} , \quad \forall \ket{\mathcal{K}_{\bm{\lambda}}} \in [\bm{\lambda}]^{S_n} , \quad D^{\bm{\lambda}}(\pi)=\bigotimes_{i=1}^{N} D^{\lambda^i} (\pi).
\end{equation*}
\item Kronecker states are Locally Maximally Entangled states, due to their individual reduced density matrices being proportional to the identity. Defining $\rho=\ket{\mathcal{K}_{\bm{\lambda}}} \bra{\mathcal{K}_{\bm{\lambda}}} $, then:
\begin{equation*}
\rho_i= \tr_{\bm{\overline{\imath}}} (\rho)=\frac{1}{f^{[\lambda^i]}} \bm{I}_{f^{[\lambda^i]}},
\end{equation*}
where $\bm{\overline{\imath}}$ is the set of parts that are complementary to $i$, and $f^{[\lambda^i]}$ is the dimension of irrep $[\lambda^i]$.
\item The dimension of the Kronecker subspace is given by the Kronecker coefficient, which can be calculated from the $S_n$ characters as:
\begin{equation*}
\dim([\bm{\lambda}]^{S_n})=k_{\bm{\lambda}}=\frac{1}{n!} \sum_{\pi \in S_n} \chi ^{[\bm{\lambda}]}(\pi) , \quad  \chi ^{[\bm{\lambda}]}(\pi) = \prod_{i=1}^{N}\chi ^{[\lambda^i]}(\pi). 
\end{equation*}
\end{itemize}
Kronecker states can be calculated from the Clebsch-Gordan coefficients (CGC) of the symmetric group, as shown in section \ref{CGC}. In particular, when expanding the Kronecker states in its coefficients as:
\begin{equation*}
\ket{\mathcal{K}_{\bm{\lambda}}}= \sum_{\bm{q}} K_{\bm{\lambda} ,\bm{q}} \ket{\bm{\lambda}, \bm{q}},
\end{equation*}
with $\bm{q}=q^{[\lambda^1]}q^{[\lambda^2]}\dots q^{[\lambda^N]}$, a list of basis elements of each irrep, then, it is possible to obtain Kronecker states from CGC as:
\begin{equation}
K_{\bm{\lambda},\bm{q},s}= \frac{1}{\sqrt{f^{[\lambda^i]}}} C^{[\bm{\lambda}^{\bm{\overline{\imath}}}],[\lambda^i],s}_{\bm{q}^{\bm{\lambda}^{\bm{\overline{\imath}}}},q^{\lambda^i}},
\label{eq:KronCGC}
\end{equation}
where $i$ is the chosen part to make the decomposition of CGC, $s$ is a multiplicity label, and $\bm{\overline{\imath}}$ is the set of the complementary parts to $i$. However, as discussed in Section \ref{CGC}, the methods to calculate CGC are inefficient even for small values of $n$ or are not general enough to generate significant Kronecker subspaces. \\

\noindent Alternatively to the mathematical approach of using CGC of $S_n$ to calculate Kronecker states, the quantum information-based approach, motivated by entanglement concentration protocols, also sheds light on the task of calculating Kronecker states when considering copies of qubit systems. The simplest case is the bipartite case, where $n$ copies of a state $\ket{\psi}\in \mathcal{H}^1 \otimes \mathcal{H}^2$ with $\mathcal{H}^i$ the Hilbert space of the $i$-th part, is studied through the Schur-Weyl duality, introduced in section \ref{Schur-Weyl}. Schur-Weyl duality states that the total Hilbert space of $n$ copies can be decomposed simultaneously on irreps of $GL_2$ and $S_n$ in each of the parts as:
\begin{equation*}
(\bm{\mathcal{H}} ^{\otimes n})^{S_n} \cong \bigoplus_{\lambda \vdash n,2} \left(\{\lambda \} \otimes \{\lambda\} \right) \otimes \left([\lambda]\otimes[\lambda] \right) ^{S_n} ,
\end{equation*}
where $\bm{\mathcal{H}}= \mathcal{H}^1\otimes \mathcal{H}^2$, $\{\lambda\}$ are irreps of $GL_2$ and $[\lambda]$ are irreps of $S_n$, both labeled by the same partition $\lambda$, and the Kronecker coefficient is $k_{\lambda\lambda}=1$. For this case, the Schur transform has the general form:
\begin{equation*}
\ket{\psi}^{\otimes n} = \bigoplus_{\lambda} \sqrt{p(\lambda|\psi)} \ket{\Phi_{\lambda\lambda}(\psi)} \ket{\mathcal{K}_{\lambda\lambda}},
\end{equation*}
 and the only Kronecker state for any $\lambda$ can be calculated from the Schur transform to be:
\begin{equation*}
\ket{\mathcal{K}_{\lambda\lambda}}=\frac{1}{\sqrt{f^{[\lambda]}}}\sum_{q} \ket{\lambda ,q}\ket{ \lambda,q }.
\end{equation*} 
When considering the multipartite case, this approach becomes much more complex. In this case, the Schur-Weyl duality applied locally to the $n$ copies of a state $\ket{\psi} \in \mathcal{H}^1 \otimes \dots \otimes \mathcal{H}^N$, states that the total Hilbert space decomposes as:
\begin{equation*}
\left(\bm{\mathcal{H}}^{\otimes n} \right) ^{S_n}  \cong \bigoplus_{\bm{\lambda}\vdash n,2} \{\bm{\lambda}\} \otimes [\bm{\lambda}]^{S_n},
\end{equation*} 
where now $\bm{\mathcal{H}}=\bigotimes_{i=1}^{N} \mathcal{H}^i$, $\{\bm{\lambda}\}=\bigotimes_{i=1}^{N} \{\lambda^i\}$, $[\bm{\lambda}]=\bigotimes_{i=1}^{N} [\lambda^i]$ and $\bm{\lambda}=\lambda^1 \dots \lambda^N$ is a multiple label for local partitions. However, in this case, it is in general not possible to obtain unique Kronecker states by performing the Schur transform; instead, the Schur transform for the general case can be written as:
\begin{equation}
\ket{\psi}^{\otimes n}=\bigoplus_{ \lamtup \vdash n,2} \sqrt{p(\bm{\lambda}|\psi)}\left[\sum_{i=1}^{k_{\lamtup}}\ket{\Phi_{\lamtup,i}(\psi)}\ket{\mathcal{K}_{\lamtup,i}} \right],
\label{eq:GenSchur}
\end{equation}
where $\ket{\Phi_{\lamtup,i}(\psi)}$ is a state in $\{\bm{\lambda}\}$ and $\ket{\mathcal{K}_{\lamtup,i}}$ is a Kronecker state in $[\bm{\lambda}]^{S_n}$. In this case,  the multiplicity of the invariant subspace given by the Kronecker coefficient, $k_{\bm{\lambda}}$, does not permit the separation of the part corresponding to $\{\bm{\lambda}\}$ from the part corresponding to $[\lamtup]^{S_n}$, so, it is not possible to obtain Kronecker states for the multipartite case with the Schur transform and measurement for generic states.\\

\noindent Remarkably, when specializing in states in the multiqubit W-class, separability is achieved again in the multi-copy setup. For this SLOCC class, the state corresponding to $\{\bm{\lambda}\}$ is unique for each set $\bm{\lambda}$; in particular, for the W state of $N$ parts, $\ket{W_N}$, the Schur transform gives the form:
\begin{equation*}
     \ket{W_N}^{\otimes n} =\bigoplus_{\bm{\lambda} \in \Lambda_n^{W}} \eta_{\bm{\lambda}} \ket{\reallywidehat{\Phi}_{\bm{\lambda}} \left(W_N \right)}\ket{\mathcal{K}^W_{\bm{\lambda}}},
\end{equation*}
where $\ket{\reallywidehat{\Phi}_{\bm{\lambda}}\left(W_N\right)}$ is an unnormalized state in $\{\bm{\lambda}\}$ that can be calculated as:
\begin{equation*}
\ket{\reallywidehat{\Phi}_{\bm{\lambda}} (W_N)}=\frac{n!^{-(N-2)/2}}{\sqrt{N^n}} \sum_{\bm{\omega}}  \sqrt{A_{\bm{\lambda},\bm{\omega}}}\ket{\bm{\lambda},\bm{\omega}},
\end{equation*}
with a list of weights $\bm{\omega}=\omega^1\omega^2\dots \omega^N$ restricted by $\sum_{i=0}^N \omega^i =n$, and
 \begin{equation*}
 A_{\bm{\lambda},\bm{\omega}}=\prod_{i=1}^N A_{\lambda^i,\omega^i } \quad , \quad  A_{\lambda,\omega}=\frac{(n-\lambda-\omega)!}{(\omega-\lambda)!}.
 \end{equation*}
The factor $\eta_{\bm{\lambda}}$ relates the unnormalized state $\ket{\reallywidehat{\Phi}_{\bm{\lambda}}(W_N)}$ and the normalized state $\ket{\Phi_{\bm{\lambda}}(W_N)}$ as :
 \begin{equation*}
        \ket{\Phi_{\bm{\lambda} }(W_N)}=\frac{ \eta_{\bm{\lambda }}}{\sqrt{p(\bm{\lambda}|W_N)}} \ket{\reallywidehat{\Phi}_{\bm{\lambda}}(W_N)},
  \end{equation*}
with $\Lambda_n^W$ the polytope that restricts the possible set of partitions $\bm{\lambda}$ to those that belong to the region spanned by the W-class, defined by\cite{Polytopes}:
\begin{equation*}
\Lambda_n ^{W} := \{ \bm{\lambda}: 2\lambda_i \leq \sum_{j=1}^N \lambda_j \leq n \}.
\end{equation*}
\\
\noindent The uniqueness of the state $\ket{\reallywidehat{\Phi}_{\bm{\lambda}}(W_N)}$, permits us to apply the Schur transform on copies of $W_N$ and, through a recurrent construction, to obtain the unique associated Kronecker state. By defining an unnormalized Kronecker state,
\begin{equation*}
\ket{\mathcal{K}^W_{\bm{\lambda}}}=\sum_{\bm{\lambda},\bm{q}} \hat{K}^W_{\bm{\lambda},\bm{q}} \ket{\bm{\lambda},\bm{q}},
\end{equation*}
the unnormalized coefficients can be obtained from the following expression:
\begin{equation}
\hat{K}^W_{\lamtup,\bm{q}}=\prod_{j=2} ^{n} \frac{ j- \sum_{i=1}^N \left[ q_j^{i}(j+1-\lambda^{(i),j})-(1-q_j^{i})\lambda^{i,j}\right]  }{\prod_{i=1}^N \sqrt{n-2\lambda^{i,j}+2q_j^{i}}} ,
\label{eq:WKronCoef}
\end{equation}
where $\lambda^{i,j}=\sum_{k=1}^{j} q_k^{i}$ is the partial partition of the $i$-th irrep at step $j$ in the  construction. With this equation, any Kronecker state appearing in the decomposition of W-class states can be obtained efficiently.\\

\noindent This way of calculating Kronecker states has two problems that make it an incomplete approach to generating Kronecker subspaces. The first problem is that the Kronecker states are obtained only for the W-class, corresponding to the possible set of partitions that belong to $\Lambda_n^W$, which is only a part of all the possible sets $\bm{\lambda}$ where Kronecker states appear. The next problem is that the W-Kronecker state is unique in each set $\bm{\lambda}$, making it impossible to use this construction to build all Kronecker subspaces when their dimensions are	 more than one. Nevertheless, when possible, this algorithm can be used to find a first Kronecker state from a set of $k_{\bm{\lambda}}$ orthogonal Kronecker states, showing one privileged direction on the multiplicity space of the invariant subspace. We will show how the W-state stitching can be understood in the Kronecker subspaces as a W-Kronecker-state stitching, allowing us to overcome the previous limitations.

\section{W-Kronecker-state stitching}
\label{KronStitch}
In the same sense that we build generic qubit states by using W-states and bipartite states as building blocks, we can build generic Kronecker states by using W-Kronecker states and bipartite Kronecker states as building blocks. The first important property of Kronecker states is that they are independent of the parameters of the state. Therefore, we introduce a graphical notation for three parts, W-Kronecker states and bipartite Kronecker states, that are in correspondence with the W-state and bipartite state representations as:
\begin{equation*}
\ket{\mathcal{K}^W_{\lambda ^1 \lambda ^2\lambda ^3}}= \begin{tikzpicture}[basel={-.5}]
  \draw[markx={0}] (0,0) to (0.5,0);
  \draw (-0.5,0.5) to (0,0);
  \draw (-0.5,-0.5) to (0,0);
  \draw (-0.8,0.5) node {$\lambda^1$};  
  \draw (-0.8,-0.5) node {$\lambda^2$};  
  \draw (0.8,0) node {$\lambda^3$};
\end{tikzpicture}, \quad 
\bra{\mathcal{K}_{\mu\mu}}= \begin{tikzpicture}
\draw (0,0) to (1,0);
\draw (0.5,0.3) node{$\mu$};
\end{tikzpicture},
\label{eq:GraphsSn}
\end{equation*}
where now $\lambda^i$ labels the irreps appearing in the $i$-th part and, for the bipartite state, we always have the same partition $\mu$ on both sides, so we put it in the middle. In this context, we do not need to define any operations in the Kronecker subspaces. Now, consider the simple process of stitching together two W-Kronecker states with one bipartite Kronecker state in a graph like the one shown in Figure \ref{fig:Tree4}:
 \begin{equation*}
\begin{tikzpicture}[basel={-.5},every node/.style={scale=0.6}]
  \draw[markx={0}] (0,0) to (0.5,0);
  \draw (-0.5,0.5) to (0,0);
  \draw (-0.5,-0.5) to (0,0);
  \draw (-0.8,0.5) node {$\lambda^1$};  
  \draw (-0.8,-0.5) node {$\lambda^2$};  
  \draw (0.5,0.3) node {$\lambda^3$};
  \draw (1,0) to (1.5,0);\\
    \Genstates{(0.8,0)};
  \Genstates{(1.7,0)};
  \draw[markx={1}] (2,0) to (2.5,0);
  \draw (3,0.5) to (2.5,0);
  \draw (3,-0.5) to (2.5,0);
  \draw (2.5,0.3) node {$\lambda^6$};  
  \draw (3.3,0.5) node {$\lambda^4$};  
  \draw (3.3,-0.5) node {$\lambda^5$};
  \draw (1.25,0.3) node {$\mu$}; 
\end{tikzpicture} = \bra{\mathcal{K}_{\mu\mu}}_{EF} \ket{\mathcal{K}^W_{\lambda ^1 \lambda^2 \lambda ^3}}_{ABE}
\ket{\mathcal{K}^W_{\lambda ^4 \lambda^5 \lambda ^6}}_{CDF},
\end{equation*}
we labeled the parts with letters to clarify what parts we were stitching. By expanding each Kronecker state on the right side in their coefficients, we have:
\begin{equation*}
\begin{gathered}
\begin{tikzpicture}[basel={-.5},every node/.style={scale=0.6}]
  \draw[markx={0}] (0,0) to (0.5,0);
  \draw (-0.5,0.5) to (0,0);
  \draw (-0.5,-0.5) to (0,0);
  \draw (-0.8,0.5) node {$\lambda^1$};  
  \draw (-0.8,-0.5) node {$\lambda^2$};  
  \draw (0.5,0.3) node {$\lambda^3$};
  \draw (1,0) to (1.5,0);
  \draw[markx={1}] (2,0) to (2.5,0);
  \draw (3,0.5) to (2.5,0);
  \draw (3,-0.5) to (2.5,0);
  \draw (2.5,0.3) node {$\lambda^6$};  
  \draw (3.3,0.5) node {$\lambda^4$};  
  \draw (3.3,-0.5) node {$\lambda^5$};
  \draw (1.25,0.3) node {$\mu$}; 
  \Genstates{(0.8,0)};
  \Genstates{(1.7,0)};
\end{tikzpicture} =\left( \frac{1}{f^{[\mu]}} \sum_{q^{[\mu]}} \bra{\mu,q^{[\mu]}}_E\bra{\mu,q^{[\mu]}}_F\right)  \\
\cdot \left(\sum_{q^{[\lambda^1]}q^{[\lambda^2]}q^{[\lambda^3]}}K^{W}_{\lambda^1\lambda^2 \lambda^3,q^{[\lambda^1]}q^{[\lambda^2]}q^{[\lambda^3]}} \ket{\lambda^1,q^{[\lambda^1]}}_A
\ket{\lambda^2,q^{[\lambda^2]}}_B
\ket{\lambda^3,q^{[\lambda^3]}}_E \right)\\\cdot \left(\sum_{q^{[\lambda^4]}q^{[\lambda^5]}q^{[\lambda^6]}}K^{W}_{\lambda^4\lambda^5 \lambda^6,q^{[\lambda^4]}q^{[\lambda^5]}q^{[\lambda^6]}} \ket{\lambda^4,q^{[\lambda^4]}}_C
\ket{\lambda^5,q^{[\lambda^5]}}_D
\ket{\lambda^6,q^{[\lambda^6]}}_F \right)
\end{gathered}
\end{equation*}
By contracting the bras and the kets labeled with the same letters, we get:
\begin{equation*}
\begin{tikzpicture}[basel={-.5},every node/.style={scale=0.6}]
  \draw[markx={0}] (0,0) to (0.5,0);
  \draw (-0.5,0.5) to (0,0);
  \draw (-0.5,-0.5) to (0,0);
  \draw (-0.8,0.5) node {$\lambda^1$};  
  \draw (-0.8,-0.5) node {$\lambda^2$};  
  \draw (0.5,0.3) node {$\lambda^3$};
  \draw (1,0) to (1.5,0);
  \draw[markx={1}] (2,0) to (2.5,0);
  \draw (3,0.5) to (2.5,0);
  \draw (3,-0.5) to (2.5,0);
  \draw (2.5,0.3) node {$\lambda^6$};  
  \draw (3.3,0.5) node {$\lambda^4$};  
  \draw (3.3,-0.5) node {$\lambda^5$};
  \draw (1.25,0.3) node {$\mu$}; 
\end{tikzpicture} =\frac{\delta_{\mu,\lambda^3}\delta_{\mu,\lambda^6}}{f^{[\mu]}} \sum_{q^{[\mu]}\bm{q}^{\bm{\lambda}}} K^{W}_{\lambda^1\lambda^2 \mu,q^{[\lambda^1]}q^{[\lambda^2]}q^{[\mu]}} K^{W}_{\lambda^4\lambda^5 \mu,q^{[\lambda^4]}q^{[\lambda^5]}q^{[\mu]}} \ket{ \bm{\lambda},\bm{q}^{\bm{\lambda}}} .
\end{equation*}
From this calculation, we have the first condition of the construction: Kronecker states can only be contracted when the same partition appears in both the W-Kronecker states and the bipartite Kronecker state used to connect them. In this sense, without loss of generality, we will consider only the graphs satisfying this condition and then the graph state obtained from the stitching can be parametrized by the graph $G$, the list of inner partitions $\mu$, and the list of external partitions $\bm{\lambda}$:
\begin{equation}
\begin{tikzpicture}[basel={-.5},every node/.style={scale=0.6}]
  \draw[markx={0}] (0,0) to (1,0);
  \draw (-0.5,0.5) to (0,0);
  \draw (-0.5,-0.5) to (0,0);
  \draw[markx={1}] (1.5,0.5) to (1,0);
  \draw (1.5,-0.5) to (1,0);  
\draw(-0.8,0.5) node {$\lambda^1$};  
  \draw(-0.8,-0.5) node {$\lambda^2$};
  \draw(1.8,0.5) node {$\lambda^3$};
  \draw(1.8,-0.5) node {$\lambda^4$};
   \draw (0.5,0.3) node {$\mu$}; 
\end{tikzpicture} =\ket{\psi^{\mu,\scalebox{0.3}{\begin{tikzpicture}[basel={-.5},every node/.style={scale=0.6}]
  \draw[markx={0}] (0,0) to (1,0);
  \draw (-0.5,0.5) to (0,0);
  \draw (-0.5,-0.5) to (0,0); 
  \draw[markx={1}] (1.5,0.5) to (1,0);
  \draw (1.5,-0.5) to (1,0);
\end{tikzpicture}}}_{\bm{\lambda}}}.
\label{eq:graphKronEx}
\end{equation}
Now, let us act on this state with the corresponding irreps of $S_n$ in each external edge with a permutation $\pi$:
\begin{equation*}
\begin{gathered}
D^{\lambda^1}(\pi)\otimes D^{\lambda^2}(\pi)\otimes  D^{\lambda^3}(\pi)\otimes D^{\lambda^4}(\pi) \ket{\psi^{\mu,\scalebox{0.3}{\begin{tikzpicture}[basel={-.5},every node/.style={scale=0.6}]
  \draw[markx={0}] (0,0) to (1,0);
  \draw (-0.5,0.5) to (0,0);
  \draw (-0.5,-0.5) to (0,0); 
  \draw[markx={1}] (1.5,0.5) to (1,0);
  \draw (1.5,-0.5) to (1,0);
\end{tikzpicture}}}_{\bm{\lambda}}} = \\
 \bra{\mathcal{K}_{\mu\mu}}_{EF}  \left(D^{\lambda^1}(\pi)\otimes D^{\lambda^2}(\pi)\otimes \bm{I}\ket{\mathcal{K}^W_{\lambda ^1 \lambda^2 \mu}}_{ABE} \right) \left(D^{\lambda^3}(\pi)\otimes D^{\lambda^4}(\pi)\otimes \bm{I}
\ket{\mathcal{K}^W_{\lambda ^3 \lambda^4 \mu}}_{CDF} \right),
\end{gathered}
\end{equation*}
using the invariance of Kronecker states and moving the remaining actions to the bipartite state, we get:
\begin{equation*}
 \left(\bra{\mathcal{K}_{\mu\mu}}_{EF}  D^{[\mu]}(\pi^{-1}) \otimes D^{[\mu]}(\pi^{-1}) \right) \ket{\mathcal{K}^W_{\lambda ^1 \lambda^2 \mu}}_{ABE} 
\ket{\mathcal{K}^W_{\lambda ^3 \lambda^4 \mu}}_{CDF}.
\end{equation*}
Then, due to the invariance of the bipartite Kronecker state, we have:
\begin{equation*}
D^{\lambda^1}(\pi)\otimes D^{\lambda^2}(\pi) \otimes D^{\lambda^3}(\pi)\otimes D^{\lambda^4}(\pi) \ket{\psi^{\mu,\scalebox{0.3}{\begin{tikzpicture}[basel={-.5},every node/.style={scale=0.6}]
  \draw[markx={0}] (0,0) to (1,0);
  \draw (-0.5,0.5) to (0,0);
  \draw (-0.5,-0.5) to (0,0); 
  \draw[markx={1}] (1.5,0.5) to (1,0);
  \draw (1.5,-0.5) to (1,0);
\end{tikzpicture}}}_{\bm{\lambda}}} =\ket{\psi^{\mu, \scalebox{0.3}{\begin{tikzpicture}[basel={-.5},every node/.style={scale=0.6}]
  \draw[markx={0}] (0,0) to (1,0);
  \draw (-0.5,0.5) to (0,0);
  \draw (-0.5,-0.5) to (0,0); 
  \draw[markx={1}] (1.5,0.5) to (1,0);
  \draw (1.5,-0.5) to (1,0);
\end{tikzpicture}}}_{\bm{\lambda}}}.
\end{equation*}
The resultant state is invariant under the diagonal actions of $S_n$, showing that the obtained graph state belongs to the invariant subspace, regardless of the inner partition:
\begin{equation*}
\ket{\psi^{\mu, \scalebox{0.3}{\begin{tikzpicture}[basel={-.5},every node/.style={scale=0.6}]
  \draw[markx={0}] (0,0) to (1,0);
  \draw (-0.5,0.5) to (0,0);
  \draw (-0.5,-0.5) to (0,0); 
  \draw[markx={1}] (1.5,0.5) to (1,0);
  \draw (1.5,-0.5) to (1,0);
\end{tikzpicture}}}_{\bm{\lambda}}} \in [\bm{\lambda}]^{S_n}, \quad \forall \mu \vdash n,2.
\end{equation*}
So, by stitching W-Kronecker states with bipartite Kronecker states, the obtained state is again a Kronecker state. We will call this construction \textit{Graph-Kronecker states}, and we will change the notation as:
\begin{equation*}
\ket{\psi^{\mu,\scalebox{0.3}{\begin{tikzpicture}[basel={-.5},every node/.style={scale=0.6}]
  \draw[markx={0}] (0,0) to (1,0);
  \draw (-0.5,0.5) to (0,0);
  \draw (-0.5,-0.5) to (0,0); 
  \draw[markx={1}] (1.5,0.5) to (1,0);
  \draw (1.5,-0.5) to (1,0);
\end{tikzpicture}}}_{\bm{\lambda}}}\rightarrow 
\ket{\reallywidehat{\mathcal{K}}^{\mu,\scalebox{0.3}{\begin{tikzpicture}[basel={-.5},every node/.style={scale=0.6}]
  \draw[markx={0}] (0,0) to (1,0);
  \draw (-0.5,0.5) to (0,0);
  \draw (-0.5,-0.5) to (0,0); 
  \draw[markx={1}] (1.5,0.5) to (1,0);
  \draw (1.5,-0.5) to (1,0);
\end{tikzpicture}}}_{\bm{\lambda}}},
\end{equation*}
the $\reallywidehat{\cdot}$ symbol is used again to signify that the obtained state is not generally normalized. We will show later that these obtained Kronecker states are not necessarily W-Kronecker states, but first, let us generalize and recapitulate what we did here.\\

\noindent The properties discussed previously for graph Kronecker states are generic for any graph construction. We will state those properties as fundamental in the W-Kronecker-state stitching:
\begin{itemize}
\item Stitching W-Kronecker states with bipartite Kronecker states can only be done when the connecting partitions are the same. Given this condition, the obtained state is a Kronecker state.
\end{itemize}
This construction is very simple once the W-Kronecker states are known. Now, we want to show that with this construction, it is possible to overcome the limitations of W-Kronecker states. In contrast with W-Kronecker states, the external partitions of graph Kronecker states are not restricted to the polytope $\Lambda_n ^W$, even though the sets of partitions appearing in the vertices correspond to W-Kronecker states, which are restricted to the polytope. In Equation \eqref{eq:graphKronEx}, we showed the simplest construction that one can imagine for graph Kronecker states, and we will see how this is already useful to go beyond W-Kronecker states. In this case, we have two vertices, with sets $(\lambda^1 \lambda^2 \mu)$, and $(\lambda^3 \lambda ^4 \mu)$, which corresponding to W-Kronecker states, each belonging to $\Lambda_n^W$, this is, the restriction corresponding to the $W$ class is:
\begin{equation*}
\begin{gathered}
\lambda^1 +\lambda^2 + \mu \leq n, \qquad
\lambda^3+\lambda ^4 +\mu \leq n.
\end{gathered}
\end{equation*}
By using the triangular inequalities of partitions:
\begin{equation}
\lambda^1 - \lambda^2 \leq \mu, \quad \lambda^2 - \lambda^1 \leq \mu , \quad 
\lambda^3 - \lambda^4 \leq \mu , \quad \lambda^4 - \lambda^3 \leq \mu,
\end{equation}
one gets that restrictions of the W-class are transmitted to the external partitions as:
\begin{equation*}
\begin{gathered}
n\geq \lambda^1+\lambda^2+\lambda^3-\lambda^4 ,\\
n\geq \lambda^1+\lambda^2+\lambda^4-\lambda^3,\\
n\geq \lambda^1+\lambda^3+\lambda^4-\lambda^2, \\
n\geq \lambda^2+\lambda^3+\lambda^4-\lambda^1 .\\
\end{gathered}
\end{equation*}
These restrictions are stronger than the polytope $\overline{KRON}$ \cite{Nonzero} defined in \eqref{eq:KRON}, meaning that not all the sets are accessible, but they are also weaker than $\Lambda_n ^W$, meaning that sets of irreps not allowed for the W-class can be obtained with this graph. For example, consider the set $(\lambda^1\lambda^2\lambda^3\lambda^4)=(1111)$ in $n=3$, with $k_{\bm{\lambda}}=3$; this set is not allowed for the W-Kronecker states because:
\begin{equation*}
\lambda^1+\lambda^2 + \lambda^3 + \lambda^4 >n;
\end{equation*}
nevertheless, we can find $\mu$ partitions for $\scalebox{0.4}{\begin{tikzpicture}[basel={-.5},every node/.style={scale=0.6}]
  \draw[markx={0}] (0,0) to (1,0);
  \draw (-0.5,0.5) to (0,0);
  \draw (-0.5,-0.5) to (0,0); 
  \draw[markx={1}] (1.5,0.5) to (1,0);
  \draw (1.5,-0.5) to (1,0);
\end{tikzpicture}}$, such that both sets of partitions on vertices are W-Kronecker states. For example, when taking $\mu=1$, we have two sets $(\lambda^1\lambda^2\mu)=(111)$, and $(\lambda^3\lambda^4\mu)=(111)$ where both belong to $\Lambda^W_n$, so, we can calculate a graph Kronecker state from contracting one part in each of two copies of the W-Kronecker state $\ket{\mathcal{K}^W_{111}}$, that we calculated in Equation \eqref{eq:Kron3ex}, which is the only Kronecker state in that subspace. Then we have for the graph-Kronecker state:
\begin{equation*}
\ket{\reallywidehat{\mathcal{K}}^{1,\scalebox{0.3}{\begin{tikzpicture}[basel={-.5},every node/.style={scale=0.6}]
  \draw[markx={0}] (0,0) to (1,0);
  \draw (-0.5,0.5) to (0,0);
  \draw (-0.5,-0.5) to (0,0); 
  \draw[markx={1}] (1.5,0.5) to (1,0);
  \draw (1.5,-0.5) to (1,0);
\end{tikzpicture}}}_{1111,3}} = \frac{2}{3} \left(\ket{0000} -\ket{0011}  +\ket{0101} +\ket{0110} +\ket{1001}+\ket{1010}-\ket{1100}+\ket{1111} \right),
\end{equation*}
where we omitted the label $\bm{\lambda}=1111$ in the basis elements, and the two basis elements of $[1]$ in $n=3$ were labeled as $\ket{0}=q^{[1]}_1,\ket{1}=q^{[1]}_2$. We can notice that the resultant state is not normalized in general. Then, the normalized graph Kronecker state is:
\begin{equation*}
\ket{\mathcal{K}^{1,\scalebox{0.3}{\begin{tikzpicture}[basel={-.5},every node/.style={scale=0.6}]
  \draw[markx={0}] (0,0) to (1,0);
  \draw (-0.5,0.5) to (0,0);
  \draw (-0.5,-0.5) to (0,0); 
  \draw[markx={1}] (1.5,0.5) to (1,0);
  \draw (1.5,-0.5) to (1,0);
\end{tikzpicture}}}_{1111,3}} = \frac{1}{2\sqrt{2}} \left(\ket{0000} -\ket{0011}  +\ket{0101} +\ket{0110} +\ket{1001}+\ket{1010}-\ket{1100}+\ket{1111} \right).
\end{equation*}
One can check that all the one-part reduced density matrices are proportional to the identity, so, with this construction, we can achieve Kronecker states that are not W-Kronecker states. Moreover, this state corresponds to the generic state of four qubits $G_{abcd}$ shown in table \ref{table:fourqubitsclass}, where the parameters are fixed to $a=c=0,b=d=\frac{1}{\sqrt{2}}$; it has been discussed that representatives of this family are LME states, as expected for Kronecker states.\\

\noindent We can generate many graph Kronecker states for the same external partitions. For this simple example, another possible value of $\mu$ allows the construction: $\mu=0$. When picking the inner partition to be $\mu=0$, the two sets of partitions for the vertices are $(\lambda^1\lambda^2\mu)=(110)$ and $(\lambda^3\lambda^4\mu)=(110)$. In this case, the $[0]$ irrep is the trivial representation, a one-dimensional representation. The Kronecker states for the vertices are, in fact, bipartite Kronecker states:
\begin{equation*}
\ket{\mathcal{K}^W_{110}}= \ket{\mathcal{K}_{11}} \ket{0}_0= \frac{1}{\sqrt{2}} \left(\ket{00}+\ket{11} \right)\ket{0}_0,
\end{equation*}
where we kept the reduced notation for the basis elements of $[1]$, and we labeled the only base element of $[0]$ as $\ket{0}_0$. In this case, the contraction described by the graph Kronecker state is just a product, 
\begin{equation*}
\begin{gathered}
\ket{\mathcal{K}^{0,\scalebox{0.3}{\begin{tikzpicture}[basel={-.5},every node/.style={scale=0.6}]
  \draw[markx={0}] (0,0) to (1,0);
  \draw (-0.5,0.5) to (0,0);
  \draw (-0.5,-0.5) to (0,0); 
  \draw[markx={1}] (1.5,0.5) to (1,0);
  \draw (1.5,-0.5) to (1,0);
\end{tikzpicture}}}_{1111,3}} =\bra{00}_{0} \frac{1}{\sqrt{2}} \left( \ket{00}+\ket{11} \right) \ket{0}_0\cdot \frac{1}{\sqrt{2}} \left( \ket{00}+\ket{11} \right)\ket{0}_0\\
= \frac{1}{2}\left( \ket{0000} + \ket{0011} +\ket{1100} +\ket{1111} \right).
\end{gathered}
\end{equation*}
This graph state is a product state of two entangled states of two parts, but interestingly, it is still an LME state; all the one-part reduced density matrices are proportional to the identity. Then, the W-Kronecker-state stitching procedure not only works for obtaining Kronecker states out of the set of partitions belonging to $\Lambda_n^W$ but also allows us to calculate more than one graph Kronecker state for the same set of partitions; in this sense, this mechanism allows one to build Kronecker subspace up to some point. For the example, we found two different Kronecker states for the same external partitions. In this case, the obtained Kronecker states are orthogonal (the reason for this will be explained later in subsection \ref{one-edge}), but this is not a general construction property. However, with these two states, we already have two basis elements for the Kronecker subspace $\left([1]\otimes[1]\otimes[1]\otimes[1] \right)^{S_3}$, which is a three-dimensional subspace, $k_{\bm{\lambda}}=3$, so we are missing one more. Next, let us highlight the property of all states in the Kronecker subspace of being LME. With the two obtained states, we can find Kronecker states with any complex linear combination. For example, take,
\begin{equation*}
\ket{\reallywidehat{\mathcal{K}}^{\star}_{1111,3}}= \sqrt{2} \cdot \ket{\mathcal{K}^{1,\scalebox{0.3}{\begin{tikzpicture}[basel={-.5},every node/.style={scale=0.6}]
  \draw[markx={0}] (0,0) to (1,0);
  \draw (-0.5,0.5) to (0,0);
  \draw (-0.5,-0.5) to (0,0); 
  \draw[markx={1}] (1.5,0.5) to (1,0);
  \draw (1.5,-0.5) to (1,0);
\end{tikzpicture}}}_{1111,3}}  + \ket{\mathcal{K}^{0,\scalebox{0.3}{\begin{tikzpicture}[basel={-.5},every node/.style={scale=0.6}]
  \draw[markx={0}] (0,0) to (1,0);
  \draw (-0.5,0.5) to (0,0);
  \draw (-0.5,-0.5) to (0,0); 
  \draw[markx={1}] (1.5,0.5) to (1,0);
  \draw (1.5,-0.5) to (1,0);
\end{tikzpicture}}}_{1111,3}}  = \frac{1}{2} \left(\ket{0101}+\ket{0110}+\ket{1001} +\ket{1010}\right) +\ket{0000}+\ket{1111}.
\end{equation*}
By calculating the one-part reduced density matrices of the obtained state, one checks that it is an LME state, and the same will be obtained for any complex linear combination of the obtained graph Kronecker states. However, the invariant subspace in this case is three-dimensional; so, there is one Kronecker state orthogonal to the obtained ones that is still missing to define a basis for the Kronecker subspace. For this, we would need to explore the graph constructions with the hope that at some point, with some graph and some set of inner partitions, another graph Kronecker state, linearly independent to 
$\ket{\mathcal{K}^{0,\scalebox{0.3}{\begin{tikzpicture}[basel={-.5},every node/.style={scale=0.6}]
  \draw[markx={0}] (0,0) to (1,0);
  \draw (-0.5,0.5) to (0,0);
  \draw (-0.5,-0.5) to (0,0); 
  \draw[markx={1}] (1.5,0.5) to (1,0);
  \draw (1.5,-0.5) to (1,0);
\end{tikzpicture}}}_{1111,3}} $ and $\ket{\mathcal{K}^{1,\scalebox{0.3}{\begin{tikzpicture}[basel={-.5},every node/.style={scale=0.6}]
  \draw[markx={0}] (0,0) to (1,0);
  \draw (-0.5,0.5) to (0,0);
  \draw (-0.5,-0.5) to (0,0); 
  \draw[markx={1}] (1.5,0.5) to (1,0);
  \draw (1.5,-0.5) to (1,0);
\end{tikzpicture}}}_{1111,3}} $, can be obtained. However, performing this task without any guiding principle is very inefficient. Later, we will see how such a guiding principle can be obtained from the connection between W-state stitching and W-Kronecker-state stitching. Before delving into this problem, we will summarize and generalize the observations made in this section.

\subsubsection*{Generalities of W-Kronecker-state stitching}
In this section, we have introduced the W-Kronecker-state stitching procedure and obtained some interesting features from a simple example. We will summarize those features here and present how they are understood in a generic graph-Kronecker state construction:

\begin{enumerate}[(i).]
\item W-Kronecker-state stitching can only be done when the partition labels in the contracted parts are the same. Given this, we can separate the labels as inner partitions (or irreps), that are contracting parts, labeled by $\mu^i$, and external partitions, that are the ones that are free after the construction, labeled by $\lambda^i$. We will label $\bm{\lambda}$ to the set of external partitions, $\bm{\mu}$ to the set of inner partitions, and $(\bm{\lambda},\bm{\mu})_v$, to the triplet of partitions appearing in the vertex $v$.

\item The resultant state after stitching W-Kronecker states, it is a Kronecker state in $\bm{\lambda}$, which will be labeled as:
\begin{equation*}
\ket{\reallywidehat{\mathcal{K}}
^{\bm{\mu},G}_{\bm{\lambda}}},
\end{equation*}
where $G$ is the graph used for the construction, and the value of $n$ is made explicit. It should be clear from the notation that the obtained state is generally not normalized.

\item Graph Kronecker states can be obtained for sets $\bm{\lambda}$ outside of $\Lambda_n^W$. However, depending on the graph used, the set $\bm{\lambda}$ may be restricted by the restrictions on the triplets $(\bm{\lambda},\bm{\mu})_v \in \Lambda_n^W$.

\item It is possible to calculate as many Kronecker states as the number of sets of compatible inner partitions such that in all the vertices, $(\bm{\lambda},\bm{\mu})_v \in \Lambda_n^W$. This set is defined for each graph and the external partitions; we will name it $\bm{\mu}^{G}_{\bm{\lambda}}$, and it is defined as:
\begin{equation}
\bm{\mu}^{G}_{\bm{\lambda}}= \{ \bm{\mu}| (\bm{\lambda},\bm{\mu})_v \in \Lambda_n^W , \quad \forall v \in G \}.
\label{eq:mug}
\end{equation}
In general, the obtained Kronecker states from a graph $G$ are not orthogonal, nor even linearly independent. However, from them, it is possible to obtain, using any orthogonalization method, a subset of mutually orthogonal Kronecker states. The size of this subset is the \textit{effective} Kronecker coefficient associated with the graph $k_{\bm{\lambda}}^{G}$ and cannot be greater than $k_{\bm{\lambda}}$, i.e.,
\begin{equation*}
k_{\bm{\lambda}}^{G}\leq k_{\bm{\lambda}}.
\end{equation*}
\end{enumerate} 
Now, we are in a position to ask the next question. Under what conditions can we ensure that for a graph $G$, the obtained graph Kronecker states span completely the Kronecker subspace, i.e.,  $k_{\bm{\lambda}}^{G}=k_{\bm{\lambda}}$ ?. To solve this question, we will exploit again the multi-copy setting, but now on graph states.

\section{Schur transfrom in graph states}
In chapter \ref{Chapter4}, we used the W-state stitching and the pushing rules to identify graphs that allow us to obtain the different SLOCC classes for three and four-qubit systems. In this section, we will show an existing underlying structure on the graph construction for qubit states, and Kronecker states that connect their properties, allowing us to obtain a criterion for a complete construction of Kronecker subspaces. At the same time, it gives us information about the qubit states from their graph structure.\\

\noindent Let us start by recalling once more the Schur transform for W states, in particular for $W_3$ states:
\begin{equation}
\ket{W_3}^{\otimes n} = \bigoplus_{\bm{\lambda}\in \Lambda_n^{W} }\eta_{\bm{\lambda}}\ket{\reallywidehat{\Phi}_{\bm{\lambda}}(W_3)}\ket{\mathcal{K}^W_{\bm{\lambda}}},
\label{eq:WSchur2}
\end{equation}
with $\bm{\lambda}=\lambda^1\lambda^2\lambda^3$ a set of three partitions, $\ket{\mathcal{K}^W_{\bm{\lambda}}}$ the unique W-Kronecker state in the invariant subspace$[\bm{\lambda}]^{S_n}$, and
\begin{equation*}
\ket{\reallywidehat{\Phi}_{\bm{\lambda}} (W_3)}=\frac{1}{\sqrt{3^n n!}} \sum_{\bm{\omega}}  \sqrt{A_{\bm{\lambda},\bm{\omega}}}\ket{\bm{\lambda},\bm{\omega}},
\end{equation*}
is a vector on $\{\bm{\lambda}\}$. On the other hand, for bipartite states, if the state is entangled, it can be obtained by SLOCC actions on the maximally entangled state $\bra{\Phi^{+}}$. As discussed in the previous chapter, this can be expressed with one of two possible parametrizations: the $\Phi$-stitch and the $\Psi$-stitch. We will only represent the $\Phi$-stitch option,
\begin{equation*}
\bra{\varsigma} =\bra{\Phi^+} u(v_1,w_1) \otimes \mathcal{A}(w_2).
\end{equation*}
Nevertheless, it should be understood that the $\Psi$-stitch option is equally valid. The Schur transform of the $n$-fold tensor product of $\bra{\varsigma}$ reads as:
\begin{equation}
\left(\bra{\varsigma} \right)^{\otimes n} =\bigoplus_{\mu\vdash n,2} \sqrt{p(\mu|v_1w_1w_2)} \left(\bra{\Phi_{\mu\mu}(\Phi^{+})} V^{\{\mu\}}(u(v_1,w_1)) \otimes  V^{\{\mu\}}(\mathcal{A}(w_2)) \right) \bra{\mathcal{K}_{\mu\mu}},
\label{eq:Schursigma}
\end{equation}
where $u(v,w)$ is defined in Equation \eqref{eq:udef}, $\mathcal{A}(w)$ is defined in Equation \eqref{eq:Adef}, and $V^{\{\mu\}}(B)$ is the matrix representation of $B$ in the irrep $\{\mu\}$ of $GL_2$. Now, in a similar way as we introduced graphical notations for W-states and for W-Kronecker states that appear in $[\bm{\lambda}]^{S_n}$ in the Schur transform of W-states, we define a graphical notation for the states that appear in $\{\bm{\lambda}\}$ in the Schur transform of W-states, including the $\eta_{\bm{\lambda}}$ factor; namely, $\ket{\reallywidehat{\Phi}_{\bm{\lambda}} (W_3)}$
\begin{equation*}
\eta_{\bm{\lambda}}\ket{\reallywidehat{\Phi}_{\bm{\lambda}} (W_3)}=\begin{tikzpicture}[basel={-.5},every node/.style={scale=0.6}]
  \draw[markx={0}] (0,0) to (0.5,0);
  \draw (-0.5,0.5) to (0,0);
  \draw (-0.5,-0.5) to (0,0);
  \draw (-0.8,0.5) node {$\{\lambda^1\}$};  
  \draw (-0.8,-0.5) node {$\{\lambda^2\}$};  
  \draw (0.8,0) node {$\{\lambda^3\}$};
\end{tikzpicture}.
\label{eq:GraphsGL}
\end{equation*}
Similarly for the state appearing in $\{\mu\}\otimes \{\mu\}$ in the Schur transform of bipartite states, where now the edges have operations on them; in the same way as for the W-state stitching, we use arrows and balls to denote the representation of the respective single-copy operation:
\begin{equation*}
\sqrt{p(\mu|v_1w_1w_2)} \left(\bra{\Phi_{\mu\mu}(\Phi^{+})} V^{\{\mu\}}(u(v_1,w_1)) \otimes  V^{\{\mu\}}(\mathcal{A}(w_2)) \right)  =  \begin{tikzpicture}[every node/.style={scale=0.6}]
\draw[\lac \rao] (0,0) to (2,0);
\draw (1,0.3) node{$\{\mu\}$};
\draw(0,0.3) node {$v_1,w_1$};
\draw(2,0.3) node {$w_2$}; 
\end{tikzpicture}.
\end{equation*}
Having defined those last graphical objects, we can rewrite the Equation \eqref{eq:WSchur2} graphically as:
\begin{equation}
\left(
\begin{tikzpicture}[basel={-.5},every node/.style={scale=0.6}]
  \draw[markx={0}] (0,0) to (0.5,0);
  \draw (-0.5,0.5) to (0,0);
  \draw (-0.5,-0.5) to (0,0);
\end{tikzpicture}  \right)^{\otimes n} = \bigoplus_{\bm{\lambda}\in \Lambda_n^{W}} \begin{tikzpicture}[basel={-.5},every node/.style={scale=0.6}]
  \draw[markx={0}] (0,0) to (0.5,0);
  \draw (-0.5,0.5) to (0,0);
  \draw (-0.5,-0.5) to (0,0);
  \draw (-0.8,0.5) node {$\{\lambda^1\}$};  
  \draw (-0.8,-0.5) node {$\{\lambda^2\}$};  
  \draw (0.8,0) node {$\{\lambda^3\}$};
\end{tikzpicture} \otimes \begin{tikzpicture}[basel={-.5},every node/.style={scale=0.6}]
  \draw[markx={0}] (0,0) to (0.5,0);
  \draw (-0.5,0.5) to (0,0);
  \draw (-0.5,-0.5) to (0,0);
  \draw (-0.8,0.5) node {$\lambda^1$};  
  \draw (-0.8,-0.5) node {$\lambda^2$};  
  \draw (0.8,0) node {$\lambda^3$};
\end{tikzpicture},
\label{eq:WSchurGraph}
\end{equation}
and similarly for Equation \eqref{eq:Schursigma} we have:
\begin{equation}
\left(\begin{tikzpicture}[every node/.style={scale=0.6}]
\draw[\lac \rao] (0,0) to (2,0);
\draw(0,0.3) node {$v_1,w_1$};
\draw(2,0.3) node {$w_2$}; 
\end{tikzpicture} \right) ^{\otimes n} = \bigoplus_{\mu\vdash n,2} \begin{tikzpicture}[every node/.style={scale=0.6}]
\draw[\lac \rao] (0,0) to (2,0);
\draw (1,0.3) node{$\{\mu\}$};
\draw(0,0.3) node {$v_1,w_1$};
\draw(2,0.3) node {$w_2$}; 
\end{tikzpicture} \otimes \begin{tikzpicture}[basel=-1.5,every node/.style={scale=0.6}]
\draw (0,0) to (2,0);
\draw (1,0.3) node{$\mu$};
\end{tikzpicture}.
\label{eq:SigmaSchur}
\end{equation}
Stitching can also be applied to the states of $\{\bm{\lambda}\}$, but we will not explore it deeper than saying that given a graph $G$, we can obtain a state by stitching the states correspondent to $W$ states using the states corresponding to bipartite states. In this sense, the obtained state will be labeled as :
\begin{equation*}
\ket{\reallywidehat{\Phi}^{\bm{\mu},G}_{\bm{\lambda},\vec{\Theta}}},
\end{equation*}
where $\vec{\Theta}$ is the list of parameters used in the stitching, $\bm{\lambda}$ label external partitions of the graph and $\bm{\mu}$ label inner partitions as usual. With this, we have all the tools and definitions to exploit the connections between graph constructions.\\

\noindent First, consider any qubit graph state obtained from W-state stitching. This state is calculated as the contraction of  the $s$ stitches with $w$ W states according to the structure of the graph. Next, we tensorize the state:
\begin{equation}
\ket{\psi^{G}_{\vec{\Theta}}}^{\otimes n}= \left(\bigotimes_{j}^s \bra{\varsigma^{j}} \right) ^{\otimes n}  \left(\bigotimes_{i=1}^{w}\ket{W_3} \right) ^{\otimes n}.
\label{eq:GraphSchurtransfrom}
\end{equation}
The Schur transform on the last expression can thus be understood in two ways. First, we can apply it to the copies of the state $\ket{\psi^G_{\vec{\Theta}}}$ following the generic Schur transform from Equation \eqref{eq:GenSchur}. On the other hand, we can apply the Schur transform on the right hand side of the equation, on the stitches and the W states as in Equations \eqref{eq:WSchurGraph} and \eqref{eq:SigmaSchur}.  The states in the Schur basis can be stitched together to build states in $\{\bm{\lambda}\}$ and $[\bm{\lambda}]^{S_n}$ with the same graph $G$ to obtain:
\begin{equation*}
\bigoplus_{\bm{\lambda}} \sqrt{p\left(\bm{\lambda}|\psi^{G}_{\vec{\Theta}}\right)} \sum_{s=1}^{k_{\bm{\lambda}}}\ket{ \Phi_{\bm{\lambda},s}\left(\psi^{G}_{\vec{\Theta}}\right)} \ket{\mathcal{K}_{\bm{\lambda},s}} = \bigoplus_{\bm{\lambda}} \sum_{\bm{\mu} \in \bm{\mu}^G_{\bm{\lambda}}} \ket{\Phi^{\bm{\mu},G}_{\bm{\lambda},\vec{\Theta}}} \ket{\mathcal{K}^{\bm{\mu},G}_{\bm{\lambda}}} ,
\end{equation*}
where $\bm{\mu}^{G}_{\bm{\lambda}}$ is the set of tuples of inner partitions compatible with the graph $G$ and the external partitions $\bm{\lambda}$, as described in Equation \eqref{eq:mug}. The previous equation states that under the Schur transform, the graph structure is preserved in the spaces $\{\bm{\lambda}\}$ and $[\bm{\lambda}]^{S_n}$. Equating terms with the same $\bm{\lambda}$ on both sides, we obtain:
\begin{equation} \sqrt{p\left(\bm{\lambda}|\psi^{G}_{\vec{\Theta}}\right)} \sum_{s=1}^{k_{\bm{\lambda}}}\ket{ \Phi_{\bm{\lambda},s}\left(\psi^{G}_{\vec{\Theta}}\right)} \ket{\mathcal{K}_{\bm{\lambda},s}} = \sum_{\bm{\mu} \in \bm{\mu}^G_{\bm{\lambda}}} \ket{\Phi^{\bm{\mu},G}_{\bm{\lambda},\vec{\Theta}}} \ket{\mathcal{K}^{\bm{\mu},G}_{\bm{\lambda}}} .
\label{eq:Kronsubsp}
\end{equation}
It is then possible to identify a correspondence between Kronecker graph states and basis states of the Kronecker subspaces as:
\begin{equation}
\ket{\mathcal{K}_{\bm{\lambda},s}} = \sum_{\bm{\mu} \in \bm{\mu}_{\bm{\lambda}}^G} C^{\bm{\mu},G}_{\bm{\lambda},s} \ket{\mathcal{K}^{\bm{\mu},G}_{\bm{\lambda}}}.
\label{eq:SpansKron}
\end{equation}
with $C^{\bm{\mu}, G}_{\bm{\lambda},s}$ a set of coefficients that make explicit the correspondence. The correspondence obtained in the last equation means that any Kronecker state $\ket{\mathcal{K}_{\bm{\lambda},s}}$ appearing in the Schur basis of the graph state $\ket{\psi^{G}_{\vec{\Theta}}}$, can be obtained as a linear combination of the graph Kronecker states, obtained with the same set of external partitions and the same graph $\ket{\mathcal{K}_{\bm{\lambda}}^{\bm{\mu},G}}$. However, as the graph Kronecker states are not linearly independent in general, there are many equivalent ways of expressing the states $\ket{\mathcal{K}_{\bm{\lambda},s}}$ in terms of graph Kronecker states labelled by different sets of inner partitions $\bm{\mu}$. Despite this, the previous equation ensures that if a Kronecker state $\ket{\mathcal{K}_{\bm{\lambda},s}}$ appear in the Schur transform of the multiqubit graph state, this can be obtained as a linear combination of some of the possible graph Kronecker states. In other words, the \textit{span} of the Kronecker subspace in both sets of Kronecker states are the same. Knowing this, if for some graph the effective Kronecker coefficient is not complete, $k_{\bm{\lambda}}^G<k_{\bm{\lambda}}$, then the possible Kronecker states appearing in the Schur basis of the qubit states cannot span the Kronecker subspace completely. Thus, the possible Kronecker states on the left side of Equation \eqref{eq:SpansKron} also depend on the graph structure of the state.\\

\noindent The previous results are very important and give us information on how to generate all Kronecker subspaces and the states related to a graph. It is worth highlighting that this relation is independent of the parameters of the multiqubit graph state. This is the case because  the stitching parameters $\vec{\Theta}$ only appear in the states in $\{\bm{\lambda}\}$. We can conclude that the relation in the Kronecker subspaces includes all the possible graph states that can be obtained from the same graph, regardless of the specific stitches ($\Phi$-stitch or $\Psi$-stitch) and the stitching parameters. We will use this connection to state the first theorem, which gives a guide on how to span completely Kronecker subspaces:

\begin{theorem}
Let $G$ be a graph with $N$ external edges, and $k_{\bm{\lambda}}^G$ the effective Kronecker coefficient in a given set of $N$ external partitions $\bm{\lambda}$, i.e., the dimension of the Kronecker subspace spanned by the graph when considering all the possible sets of inner partitions $\bm{\mu}^{G}_{\bm{\lambda}}$. Then, $G$ generates completely any Kronecker subspace of $N$ parts (i.e., $k_{\bm{\lambda}}^G=k_{\bm{\lambda}}, \, \forall \bm{\lambda} \vdash n,2$), if $G$ can be used to generate all the SLOCC classes corresponding to stable orbits of N qubits.
\label{TheoremI}
\end{theorem}
\begin{proof}
To prove this theorem, we start by recalling the result from \cite{Gour}, where it was shown that the set of states SLOCC equivalent to critical states, or, in other words, the set of SLOCC stable orbits, is dense in the total Hilbert space $\bm{\mathcal{H}}$. Then, the remaining orbits, i.e., strictly semi-stable and unstable orbits, are of measure zero compared with the stable orbits.\\

\noindent Now, note that when integrating $n$ copies of rank-one projectors onto pure qubit states of $N$ parties, with the uniform measure induced by the Haar measure,$\mathcal{D}\psi$, we have:
\begin{equation*}
 \int \mathcal{D}\psi\,\left(\ketbra{\psi}{\psi} \right)^{\otimes n}\cong \bigoplus_{\lamtup\vdash n,2}\mathcal{P}^{sym}_{\{\bm{\lambda}\}}  \otimes \mathcal{P}^{[0]}_{\bm{\lambda}} . \label{eq:HaarSchurtransform}
\end{equation*}
The resultant is a matrix proportional to the identity, due to the Schur lemma. Then, it is clear that the resultant matrix is invariant under the multi-local actions of $GL_2$ and $S_n$. Then, again, by Schur's Lemma, the multi-local Schur transform corresponds to the projector onto the invariant subspace of the tensor product of irreducible representations on both groups. $\mathcal{P}^{sym}_{\{\bm{\lambda}\}}$ is the projector onto the invariant subspace of $\{\bm{\lambda}\}$ under multi-local $GL_2$ actions, and $\mathcal{P}^{[0]}_{\bm{\lambda}}$ is the projector onto the invariant subspace of $[\lamtup]$ under $S_n$ actions, i.e., the Kronecker subspace, which can be obtained from any complete set of Kronecker states, defining the invariant subspace. This projector is given by
 \begin{equation*}
     \mathcal{P}^{[0]}_{\bm{\lambda}}  = \sum_{s=1}^{k_{\lamtup}} \ket{\mathcal{K}_{\lamtup,s}}\bra{\mathcal{K}_{\lamtup,s}}.
     \label{eq:KroneckerProjector}
 \end{equation*}
Thus, the set of pure states of full measure in the total Hilbert space spans completely any Kronecker subspace. Note that the span of the Kronecker subspace from graph-Kronecker states obtained with the graph $G$ is equal to the span of the Kronecker subspace in the Schur basis of all possible qubit states obtained with the same graph $G$. Then, if the obtained graph qubit states from $G$ are dense in the total Hilbert space $\bm{\mathcal{H}}$, the graph $G$ spans completely any Kronecker subspace of $N$ parts.
\end{proof}

\noindent This theorem gives us sufficient conditions on graphs for calculating any Kronecker subspace of $N$ parts: we must refer first to the W-stitching procedure to identify what graphs generate all the SLOCC stable orbits, and then, we can use any of those graph; for example, the graph with less vertices, to build all the graph Kronecker states, $\left\{\ket{\mathcal{K}_{\bm{\lambda}}^{\bm{\mu},G}}, \forall \bm{\mu}\in \bm{\mu}_{\bm{\lambda}}^{G}\right\}$. We can obtain a complete basis for any Kronecker subspace from this set. \\

\noindent We call graphs with the property given in \textit{Theorem} \ref{TheoremI}, i.e., graphs that generate the SLOCC stable orbits of $N$ qubits, \textit{sufficient graphs}. Then, the problem of completely building the Kronecker subspace is solved by finding sufficient graphs. The previous chapter already presented sufficient graphs for $N=3$ and $N=4$. We will show later how the Kronecker subspace construction can be made with those graphs. Moreover, we know from the connection with ZW-calculus that with W-state stitching, it is possible to generate any multiqubit state because ZW-calculus is complete \cite{SLOCCZW}. Then, it will always be possible to find a set of graphs $\{G_i\}$, such that any Kronecker subspace of $N$ parts can be obtained from them.\\

\noindent The dual relation of Kronecker subspaces in Equation \eqref{eq:Kronsubsp} not only allows us to make observations on the graph Kronecker construction from the graph multiqubit construction but also permits us to gain information on the qubit states of a given graph based on the restrictions of the graph Kronecker states. For this, let us define an \textit{incomplete graph} in $\bm{\lambda}$ as a graph $G$ that cannot span completely the Kronecker subspace $[\bm{\lambda}]^{S_n}$, i.e., $k_{\bm{\lambda}}^G<k_{\bm{\lambda}}$; then we have the following theorem:
 \begin{theorem}\label{theoremII}
 Let $G$ be an incomplete graph on a set of partitions $\bm{\lambda}$. Then,  the dimension of the Kronecker subspace spanned by any qubit state that can be obtained from $G$ can be at most $k_{\bm{\lambda}}^G$.
 \end{theorem}
 \begin{proof}
From equation \eqref{eq:SpansKron}, we can see that if some graph-qubit state obtained from a graph $G$ for a given set $\bm{\lambda}$ in its Schur basis has a set of $k_{\bm{\lambda}}^{\psi^G_{\vec{\Theta}}}$ orthogonal Kronecker states, then, each Kronecker state can be obtained as a linear combination of some graph Kronecker states of the same graph $G$, hence,  $k_{\bm{\lambda}}^{\psi^G_{\vec{\Theta}}}\leq k_{\bm{\lambda}}^{G}$.
 \end{proof}

\noindent This theorem was already discussed as a consequence of the relations of the Kronecker subspaces of equation \eqref{eq:Kronsubsp}; however, it is worth pointing it out because it allows to identify SLOCC classes that are different from the $W$ class, but for which nevertheless the entanglement concentration protocol can be applied. This depends on the result of the projection in local irreps as follows. Consider a graph $G$ and the set of tuples of external partitions where the effective Kronecker coefficient is one: $\bm{\lambda}^{G}_{EC}:=\{\bm{\lambda}\mid k^{G}_{\bm{\lambda}}=1\}$. Then, the entanglement concentration protocol can be applied whenever the projection on the local irreps gives, as a result some set $\bm{\lambda}\in\bm{\lambda}^{G}_{EC}$. It will be an interesting task for the future to study the implications of this result.\\

\noindent Theorem \ref{TheoremI} and  \ref{theoremII} show how the stitching structure permits us to infer properties from qubit states to Kronecker states and vice-versa. In the next section, we show how, for a special kind of graph, the graph Kronecker states obtained, and the qubit graph states obtained from such graphs are special, showing that this connection may be deeper what one might guess at a first glance.
 \subsection{One edge reducible graphs}
\label{one-edge}
Some interesting properties can be obtained when focusing on graphs that can be separated in two by ``cutting'' only one edge. We call these  \textit{One-edge reducible} (OER) graphs and denoted as $G=(\straight)$. Such graphs correspond to any construction expressed as two subgraphs $G_1$ and $G_2$ stitched with one edge, called the  \textit{reducible edge}. These graphs can always be represented as :
\begin{equation*}
\begin{tikzpicture}[basel=-.5,every node/.style={scale=0.6}]
\def\array{$1$,$2$,$.$,$.$,$.$,$N_1$}   
    \foreach [count=\n] \x in \array{
        \node at ({90+\n*180/(6+1)}:0.6cm) (n\n) {\x};
        \draw (0,0)--(n\n); 
        };
        \draw (0,0) --(1,0);
\Genstate{(0,0)};
\def\arrax{$N_1+1$,$.$,$.$,$.$,$.$,$N$}   
    \foreach [count=\n] \x in \arrax{
        \node at ($(1,0)+({270+\n*180/(6+1)}:0.6cm)$) (n\n) {\x};
        \draw (1,0)--(n\n); 
        };
\Genstate{(1,0)};
\draw [decorate,decoration={brace,amplitude=5pt,mirror,raise=0ex}]
  (-0.3,-0.8) -- (0.3,-0.8) node[midway,yshift=-2em]{$G_1$};
  \draw [decorate,decoration={brace,amplitude=5pt,mirror,raise=0ex}]
  (0.7,-0.8) -- (1.3,-0.8) node[midway,yshift=-2em]{$G_2$};
\end{tikzpicture},
\end{equation*} 
where $\scalebox{0.5}{\begin{tikzpicture}[basel=-.2]{\inWn{8}{0}{(0,0)};\Genstate{(0,0)};} \end{tikzpicture}}$ represents a generic graph. These kinds of graphs have some interesting properties for Kronecker and qubit states. First, let us consider Kronecker states obtained from an OER, denoted by $\ket{\mathcal{K}_{\bm{\lambda}}^{\bm{\mu},(\epsilon,\straight)}}$, where $\epsilon$ is the inner partition used in the reducible edge. Then, this state can be factored as:
\begin{equation*}
\ket{\mathcal{K}_{\bm{\lambda}}^{\bm{\mu},\epsilon,\straight}} \propto \bra{\mathcal{K}_{\epsilon\epsilon}} \ket{\mathcal{K}_{\bm{\lambda}_1\epsilon}^{\bm{\mu}_1,G_1}}\ket{\mathcal{K}_{\bm{\lambda}_2\epsilon}^{\bm{\mu}_2,G_2}}
\end{equation*}
where $G_{1,2}$ are the two sub-graphs obtained when cutting the reducible edge, and $\bm{\mu}_{i},\bm{\lambda}_{i}$ refer to the sets of inner and outer partitions in the separated graph $G_i$. From the relation between CGC and Kronecker states in Equation \eqref{eq:KronCGC}, this equation can be rewritten as:
\begin{equation*}    \ket{\mathcal{K}_{\bm{\lambda}}^{\bm{\mu},(\epsilon,\straight)}} = \frac{1}{\sqrt{(f^{[\epsilon]})^3}}\sum_{\bm{q}^{\bm{\lambda}},q^{[\epsilon]}} C^{[\bm{\lambda}_1],[\epsilon],({\bm{\mu}_1,G_1})}_{\bm{q}^{\bm{\lambda}_1},q^{[\epsilon]}}C^{[\bm{\lambda}_2],[\epsilon],({\bm{\mu}_2,G_2})}_{\bm{q}^{\bm{\lambda}_2},q^{[\epsilon]}} \ket{\bm{q}^{\bm{\lambda}}}
\end{equation*}
where $({\bm{\mu}_{i},G_i})$ is a label for the CGC associated with the graph from where it is obtained. Next, we pick another Kronecker state for the same set $\bm{\lambda}$, obtained with an OER, but with the partition in the reducible edge being $\epsilon'$ and the same separation $\bm{\lambda_1}\bm{\lambda_2}$. Then, the inner product of these states is:
{\small
\begin{equation*}
   \bra{\mathcal{K}_{\bm{\lambda}}^{\bm{\mu},(\epsilon,\straight)}}  \ket{\mathcal{K}_{\bm{\lambda}}^{{\bm{\mu}'},({\epsilon'},{\straight}')}} \propto 
   \sum_{q^{[\epsilon]},q^{[\epsilon']}} \left(\sum_{\bm{q}^{\bm{\lambda}_1}} C^{[\bm{\lambda}_1],[\epsilon],({\bm{\mu}_1,G_1})}_{\bm{q}^{\bm{\lambda}_1},q^{[\epsilon]}}C^{[\bm{\lambda}_1],[\epsilon],({\bm{\mu}'_1,G'_1})}_{\bm{q}^{\bm{\lambda}_1},q^{[\epsilon']}}\right) \cdot \left( \sum_{\bm{q}^{\bm{\lambda}_2}} C^{[\bm{\lambda}_2],[\epsilon],({\bm{\mu}_2,G_2})}_{\bm{q}^{\bm{\lambda}_2},q^{[\epsilon]}}C^{[\bm{\lambda}_2],[\epsilon'],({\bm{\mu'}_2,G'_2})}_{\bm{q}^{\bm{\lambda}_2},q^{[\epsilon']}}\right).
\end{equation*} }
By using orthogonality relations of CGC, we get that:
\begin{equation}     \bra{\mathcal{K}_{\bm{\lambda}}^{\bm{\mu},(\epsilon,\straight)}}  \ket{\mathcal{K}_{\bm{\lambda}}^{{\bm{\mu}'},({\epsilon'},{\straight}')}} \propto \delta_{\epsilon\epsilon'}.
\label{eq:OrthographKron}
\end{equation}
Then, if the partitions in the reducible edge are different, the obtained graph Kronecker states are orthogonal. With this, a subset of orthogonal Kronecker states can be obtained and lead, for OER Kronecker states, to a meaningful way to label the multiplicities of the Kronecker subspace (at least some of them) with the different $\epsilon$ values. This construction offers a natural meaning for the multiplicities that can be obtained from OER graphs.\\

\noindent There is one more result for the OER graph states. But now for the qubit graph states. Consider an $N$-partite qubit state and the separation between $N_1$ parts from the remaining $N-N_1$ parts, with $N_1\leq N-N_1$. Then, the rank of the reduced density matrix for any of the parties could be at most $2^{N_1}$; however, if the state can be obtained from an OER graph, with the reducible edge separating $N_1$ parts from the remaining $N-N_1$ parts, then, the rank of the reduced density matrix for any of the parties can be at most $2$. This restriction is a result of the Keyl-Werner theorem in Section \ref{Keyl-WernerTheorem}. When taking copies of the state, a projective measurement on the irrep corresponding to the separation, the probability of the irrep measured will be concentrated asymptotically on partitions around the spectrum of the reduced density matrix, $\rho_{\{N_1\}}$, with $\{N_1\}$ the smaller set of separated parts. As the $\epsilon$ irrep can only take values on partitions of at most two parts, i.e., $\epsilon \vdash n,2$, the rank of the reduced density matrix cannot be greater than 2. In this sense, the Kronecker graph structure again provides information on possible qubit graph states.  \\

\noindent Now we dedicate the rest of this chapter to the process of Building Kronecker subspaces in the cases of three and four qubits, for which in the last chapter, we found the graphs that span densely the total Hilbert space, then, those graphs can be used to generate any Kronecker subspace. 
\section{Three-part Kronecker states}

In this section, we will use the W-Kronecker-state stitching to obtain any Kronecker state appearing in the Schur-Weyl decomposition for states of three qubits that, according to theorem \ref{TheoremI}, can be obtained from the graph that generates the full measure SLOCC class of three qubits. As discussed in Section \ref{Threequbitsclassification}, there are six SLOCC classes for systems of three qubits; from them, one is completely separable, three have a separable qubit, and two are genuinely entangled. However, there is only one SLOCC class that corresponds to a stable orbit, which is the GHZ class. To build completely the Kronecker subspace of any set of partitions of three parts, due to theorem \ref{TheoremI}, we need a graph such that GHZ-class can be obtained from it. In the W-stitching procedure for three qubits in Section \ref{Threequbits}, we showed that the graph that allows us to obtain the GHZ class is:
 \begin{equation*}
 \ket{GHZ} \cong
  \begin{tikzpicture}[basel={-.5}]
   \draw (0,0) -- (0.5,0) ; 
   \draw[markx={0}] (0.5,0) -- (1,0.5) ;
   \draw[markx={1}] (0,0) -- (-0.5,0);
   \draw (-1,0.5) -- (-0.5,0);
   \draw (0.5,0) --(0,-1);\draw[markx={1}](0,-1.5)--(0,-1); \draw(0,-1) --(-0.5,0);
\end{tikzpicture}.
 \end{equation*}
Then, when fixing a set of external partitions $\bm{\lambda}=\lambda^1\lambda^2\lambda^3$, we can obtain Kronecker states of three parts for each compatible set of inner partitions $\bm{\mu}=\mu^1\mu^2\mu^3$, by building graph Kronecker states as:
 \begin{equation*}
  \ket{\mathcal{K}_{\bm{\lambda}} ^{\bm{\mu},\Kite{0.2}}} \propto
  \begin{tikzpicture}[basel={-.5}]
   \draw (0,0) -- (0.5,0) ; 
   \draw[markx={0}] (0.5,0) -- (1,0.5) ;
   \draw[markx={1}] (0,0) -- (-0.5,0);
   \draw (-1,0.5) -- (-0.5,0);
   \draw (0.5,0) --(0,-1);\draw[markx={1}](0,-1.5)--(0,-1); \draw(0,-1) --(-0.5,0);
   \draw (-0.5,-0.55) node {$\mu^1$};
    \draw (0.5,-0.5) node {$\mu^2$};
     \draw (0,0.3) node {$\mu^3$};
      \draw (-1.25,0.3) node {$\lambda^1$};
      \draw (1.25,0.3) node {$\lambda^2$};
      \draw (0,-1.7) node {$\lambda^3$};
\end{tikzpicture},
 \end{equation*}
 By using all the set of compatible inner partitions with $\bm{\lambda}$, i.e., $\bm{\mu} \in \bm{\mu}_{\bm{\lambda}}^{\Kite{0.2}}$, the span of the obtained graph Kronecker states will be of dimension $k_{\bm{\lambda}}$, meaning that with those graph Kronecker states it is possible to build an orthogonal basis of $k_{\bm{\lambda}}$ Kronecker states for any Kronecker subspace of three parts with $\bm{\lambda}\vdash n,2$. The explicit equation for obtaining the graph Kronecker states from the W-Kronecker states in this graph is:
\begin{equation*}
\ket{\mathcal{K}_{\bm{\lambda}} ^{\bm{\mu},\Kite{0.2}}} \propto
 \bra{\mathcal{K}_{\mu^1\mu^1}}
 \bra{\mathcal{K}_{\mu^2\mu^2}}
 \bra{\mathcal{K}_{\mu^3\mu^3}}
 \ket{\mathcal{K}^{W}_{\lambda^1\mu^1\mu^3}}
 \ket{\mathcal{K}^{W}_{\lambda^2\mu^2\mu^3}}
 \ket{\mathcal{K}^{W}_{\lambda^3\mu^1\mu^2}},
\end{equation*}
where the repeated partitions give the position for the projections. This equation can be written in a more useful way for doing the computations. By writing the bipartite and W Kronecker states in their coefficient expansions given by Equations \eqref{eq:2partKronecker} and \eqref{eq:Kronbasis}, for this case, we have:
\begin{equation}
\ket{\mathcal{K}_{\bm{\lambda}} ^{\bm{\mu},\Kite{0.2}}} \propto
 \sum_{\bm{q}^{\bm{\lambda}}} \left( \sum_{\bm{q}^{\bm{\mu}}} K^W_{\lambda^1 \mu^1 \mu^3 ,q^{[\lambda^1]}q^{[\mu^1]}q^{[\mu^3]}}K^W_{\lambda^2 \mu^2 \mu^3,q^{[\lambda^2]}q^{[\mu^2]}q^{[\mu^3]}}K^W_{\lambda^3 \mu^1 \mu^2,q^{[\lambda^3]}q^{[\mu^1]}q^{[\mu^2]}} \right) \ket{\bm{\lambda},\bm{q}^{\bm{\lambda}}}.
 \label{eq:kroncoefkite} 
\end{equation}
As the process of W-Kronecker-state stitching leads to an unnormalized state, we can use the unnormalized coefficients of W-Kronecker states shown in Equation \eqref{eq:WKronCoef}, and then, the use of Equation \eqref{eq:kroncoefkite} is straightforward.\\

\noindent We will show the calculation of one of these graph-Kronecker states to give a clearer idea of how the method works. The first case where a Kronecker state out of the restriction of W-states, i.e., with a set of partitions outside of $\Lambda_n^W$, is for $n=4$ with $\bm{\lambda}=222$. Note how $\lambda^1+\lambda^2+\lambda^3 =6>4$, showing that this set is out of the polytope $\Lambda^W_4$. The first step is to find the set of tuples of inner partitions $\bm{\mu}$ such that for the graph $\Kite{0.3}$ with external partitions $\bm{\lambda}$, in each vertex $v$, the triplets of partitions $(\bm{\lambda},\bm{\mu})_v$ belong to $\Lambda_4^W$. In this case, all the triplets in the vertices are:
\begin{equation*}
(\bm{\lambda},\bm{\mu})_1 =2\mu^1\mu^3, \quad (\bm{\lambda},\bm{\mu})_2 =2\mu^2\mu^3, \quad (\bm{\lambda},\bm{\mu})^3 =2\mu^1\mu^2.
\end{equation*}
The only possible set of inner partitions, where each vertex belongs to $\Lambda_4^W$, is $\bm{\mu}=111$. It is possible to build only one graph Kronecker state in this set of external partitions; however, the Kronecker coefficient is $k_{\bm{\lambda}}=1$. Hence, there is only one possible Kronecker state. In fact, due to theorem \ref{TheoremI}, we can state that for any set $\bm{\lambda}$ of three parts, the set of compatible inner partitions in this graph is at least as big as the Kronecker coefficient:
\begin{equation*}
k_{\bm{\lambda}}\leq |\bm{\mu}_{\bm{\lambda}}^{\Kite{0.2}}|.
\end{equation*}
Continuing with the calculation of the graph Kronecker state, we have that the basis elements for each partition are, for the three-dimensional irrep $[1]$:
\begin{equation*}
\{ q_1^{[1]}=\{0,0,0,1\},q_2^{[1]}=\{0,0,1,0\},q_3^{[1]}=\{0,1,0,0\} \} \rightarrow \includegraphics[scale=1.2,valign=c]{Figures/paths31.png},
\end{equation*}
and for the two-dimensional irrep $[2]$: 
\begin{equation*}
\{ q_1^{[2]}=\{0,0,1,1\},q_2^{[2]}=\{0,1,0,1\}\}\rightarrow \includegraphics[scale=1.2,valign=c]{Figures/paths22.png}
\end{equation*}
The W-Kronecker state appearing in all vertices is the same, and we already calculated it in Equation \eqref{eq:WKron112}, but with a different order in the irreps. By reordering the state, we have: 
\begin{equation*}
\begin{gathered}
\ket{\reallywidehat{\mathcal{K}}_{211}^W} = \frac{1}{\sqrt{6}} \left( \ket{q_1^{[2]},q_1^{[1]},q_2^{[1]}}+\ket{q_2^{[2]},q_1^{[1]},q_3^{[1]}}+\ket{q_1^{[2]},q_2^{[1]},q_1^{[1]}}+\ket{q_2^{[2]},q_3^{[1]},q_1^{[1]}}  \right) \\
+\frac{1}{2\sqrt{3}} \left(\ket{q_1^{[2]},q_2^{[1]},q_2^{[1]}}-\ket{q_1^{[2]},q_3^{[1]},q_3^{[1]}}-\ket{q_2^{[2]},q_2^{[1]},q_3^{[1]}}- \ket{q_2^{[2]},q_3^{[1]},q_2^{[1]}} \right).
\end{gathered}
\label{eq:K211W2}
\end{equation*}
where the label $\bm{\lambda}=211$ in the kets was omitted. Now, using Equation \eqref{eq:kroncoefkite}, we have:
\begin{equation*}
\ket{\mathcal{K}^{111,\Kite{0.2}}_{222}} \propto \bra{\mathcal{K}_{11}}\bra{\mathcal{K}_{11}}\bra{\mathcal{K}_{11}}
\ket{\reallywidehat{\mathcal{K}}_{211}^W}\ket{\reallywidehat{\mathcal{K}}_{211}^W}\ket{\reallywidehat{\mathcal{K}}_{211}^W}  .
\end{equation*}
Then, we can obtain an unnormalized state where the first coefficient is:
\begin{equation*}
\reallywidehat{K}_{222,q_1^{[2]}q_1^{[2]}q_1^{[2]}} ^{111,\Kite{0.2}}= \sum_{q^{[\mu^1]}q^{[\mu^2]}q^{[\mu^3]}} \reallywidehat{K}_{211,q_1^{[2]}q^{[\mu^1]}q^{[\mu^3]}}  \reallywidehat{K}^W_{211,q_1^{[2]}q^{[\mu^2]}q^{[\mu^3]}}  \reallywidehat{K}^W_{211,q_1^{[2]}q^{[\mu^1]}q^{[\mu^2]}}= \frac{1}{4\sqrt{3}}.   
\end{equation*}
By doing the same with all the coefficients, we get the unnormalized graph state:
\begin{equation*}
\ket	{\reallywidehat{\mathcal{K}}^{111,\Kite{0.2}}_{222}}=\frac{1}{4\sqrt{3}} \left(\ket{q_1^{[2]}q_1^{[2]}q_1^{[2]}}-\ket{q_1^{[2]}q_2^{[2]}q_2^{[2]}}-\ket{q_2^{[2]}q_1^{[2]}q_2^{[2]}}-\ket{q_2^{[2]}q_2^{[2]}q_1^{[2]}}\right),
\end{equation*}
which, after normalization, corresponds to 
\begin{equation*}
\ket	{\mathcal{K}^{111,\Kite{0.2}}_{222}}=\frac{1}{2} \left(\ket{q_1^{[2]}q_1^{[2]}q_1^{[2]}}-\ket{q_1^{[2]}q_2^{[2]}q_2^{[2]}}-\ket{q_2^{[2]}q_1^{[2]}q_2^{[2]}}-\ket{q_2^{[2]}q_2^{[2]}q_1^{[2]}}\right).
\end{equation*}
This state is the same in \eqref{eq:Kron3ex}, and as discussed, this state is unique when considering three two-dimensional irreps; nevertheless, it is the only example that can be shown in the paper in its coefficient expansion. \\

\begin{figure}
\includegraphics[scale=0.5]{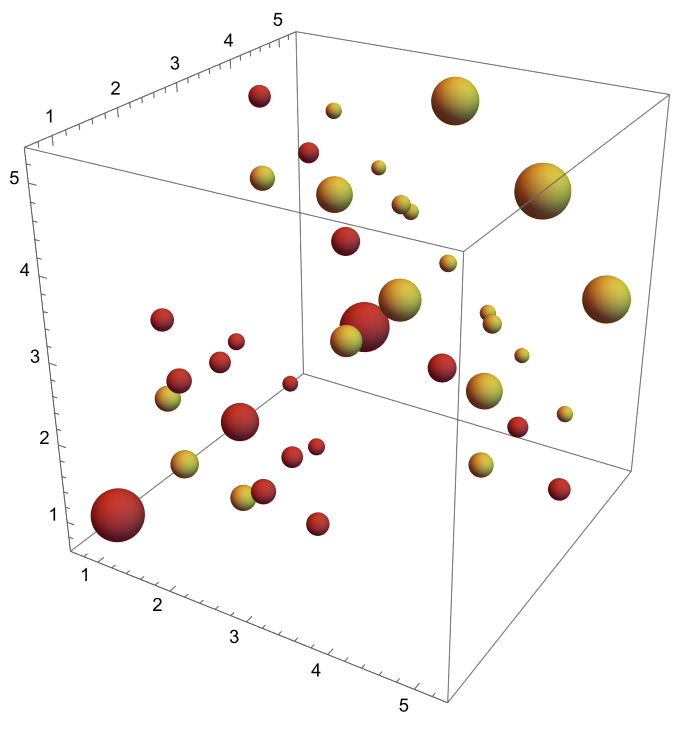}
\caption{Graphical representation of the graph Kronecker state of three parts with $\bm{\lambda}=222$ and $\bm{\mu}=111$. Each axis represents the basis elements of each part, the size of the point represents the magnitude of the coefficients, and the color represents the sign of the coefficient: red for positive and yellow for negative.}
\label{fig:Kron5222}
\end{figure}

\noindent The next set of external partitions that are not achieved with the $W$ class is for $n=5, \bm{\lambda}=222$, and as the dimension of the irrep $[2]$ is $f^{[2]}=5$; states in $[\bm{\lambda}]$ have $125$ coefficients in general. In particular, this Kronecker state only has $39$ no-null coefficients but is still long enough to avoid the coefficient expansion. A useful way to visualize states of three parts used in \cite{Botero} is to represent the coefficients of the state in a three-dimensional graph, where each axis represents the basis elements of the three parts ordered lexicographically with the corresponding Yamanouchi symbols. The size of the point represents the magnitude of the coefficients, and the color represents the sign of the coefficient. By using this representation, we show the state $\ket{\mathcal{K}^{111,\Kite{0.2}}_{222}}$ in Figure \ref{fig:Kron5222}. It is worth highlighting that for this case, the set of compatible inner partitions is not unique, $\bm{\mu}^1=111$ and $\bm{\mu}=112$ are compatible, but as the Kronecker coefficient is $k_{\bm{\lambda}}=1$, both graph-Kronecker states are the same after normalization:
\begin{equation*}
n= 5: \qquad \qquad \ket{\mathcal{K}^{111,\Kite{0.2}}_{222}}=\ket{\mathcal{K}^{112,\Kite{0.2}}_{222}}.
\end{equation*}
These constructions exhibit structural relations between the different ways of obtaining the same Kronecker state that should be studied more deeply in the future, as it . In the next subsection, we will show how to span Kronecker subspaces of three parts for some values of $n$.

\subsubsection{Kronecker subspaces for three parts}
 One of the most interesting applications of this method is to obtain a vector space full of Kronecker states, i.e., a Kronecker subspace. In table \ref{table:threeparts} we show the sets of partitions $\bm{\lambda}$ from $n=6$ up to $n=12$, where the Kronecker coefficient is greater than one, $k_{\bm{\lambda}}>1$, and the column $\Lambda_n^W$ tells whether the set $\bm{\lambda}$ is inside or outside of the polytope of the W class. One useful expression for calculating Kronecker coefficients for triplets of partitions with at most two rows is introduced in \cite{Rosas}. In our notation this expression corresponds to:
\begin{equation}
k_{\lambda^1\lambda^2\lambda^3} =(y-x) (y \geq x), \quad x= \max \left( 0, \ceil*{\frac{\lambda^1+\lambda^2+\lambda^3 -n}{2}} \right ) , \quad y = \ceil*{\frac{\lambda^1 +\lambda ^2 -\lambda ^3 +1}{2}},
\end{equation}
where the partitions are ordered as $\lambda^1\leq \lambda^2 \leq \lambda^3 $.\\

\begin{table}[h]
\begin{tabular}{|c|c|c|c|}
\hline
n & $\bm{\lambda}$ & $k_{\bm{\lambda}}$ & $\Lambda_n ^W$ \\ \hline
6 &  222    & 2     &  \checkmark     \\ \hline
7 &   222     &      2     &  \checkmark     \\ \hline
8 &  222      &       2     &   \checkmark     \\ \hline
8 &  233      &       2     &   \checkmark     \\ \hline
9 &  222      &       2     &   \checkmark     \\ \hline
9 &  233      &       2     &   \checkmark     \\ \hline
9 &  333      &       2     &   \checkmark     \\ \hline
10 &  222      &       2     &   \checkmark     \\ \hline
10 &  233      &       2     &   \checkmark     \\ \hline
10 &  244      &       2     &   \checkmark     \\ \hline
\end{tabular} \, \begin{tabular}{|c|c|c|c|}
\hline
n & $\bm{\lambda}$ & $k_{\bm{\lambda}}$ & $\Lambda_n ^W$ \\ \hline
10 &  333    & 2     &  \checkmark     \\ \hline
10 &  334    & 2     &  \checkmark     \\ \hline
10 &  444    & 2     &  $\times$     \\ \hline
11 &  222    & 2     &  \checkmark     \\ \hline
11 &  233    & 2     &  \checkmark     \\ \hline
11 &  244    & 2     &  \checkmark     \\ \hline
11 &  333    & 2     &  \checkmark     \\ \hline
11 &  334    & 2     &  \checkmark     \\ \hline
11 &  344    & 2     &  \checkmark     \\ \hline
11 &  444    & 2     &  $\times$     \\ \hline
\end{tabular}  \, \begin{tabular}{|c|c|c|c|}
\hline
n & $\bm{\lambda}$ & $k_{\bm{\lambda}}$ & $\Lambda_n ^W$ \\ \hline
12 &  222    & 2     &  \checkmark     \\ \hline
12 &  233    & 2     &  \checkmark     \\ \hline
12 &  244    & 2     &  $\times$     \\ \hline
12 &  255    & 2     &  \checkmark     \\ \hline
12 &  333    & 2     &  \checkmark     \\ \hline
12 &  334    & 2     &  \checkmark     \\ \hline
12 &  344    & 2     &  \checkmark     \\ \hline
12 &  345    & 2     &  \checkmark     \\ \hline
12 &  444    & 3     &  \checkmark     \\ \hline
12 &  455    & 2     &  $\times$     \\ \hline
\end{tabular}
\caption{Triplets of partitions $\bm{\lambda}=\lambda^1\lambda^2\lambda^3$ with Kronecker coefficients greater than 1. The symbol \checkmark  in the last column means that the correspondent triplet is inside the W polytope defined by $\Lambda_n^W$, and the symbol $\times$ means that the triplet is outside of the W polytope.}
\label{table:threeparts}
\end{table}

\begin{figure}
\includegraphics[scale=0.5]{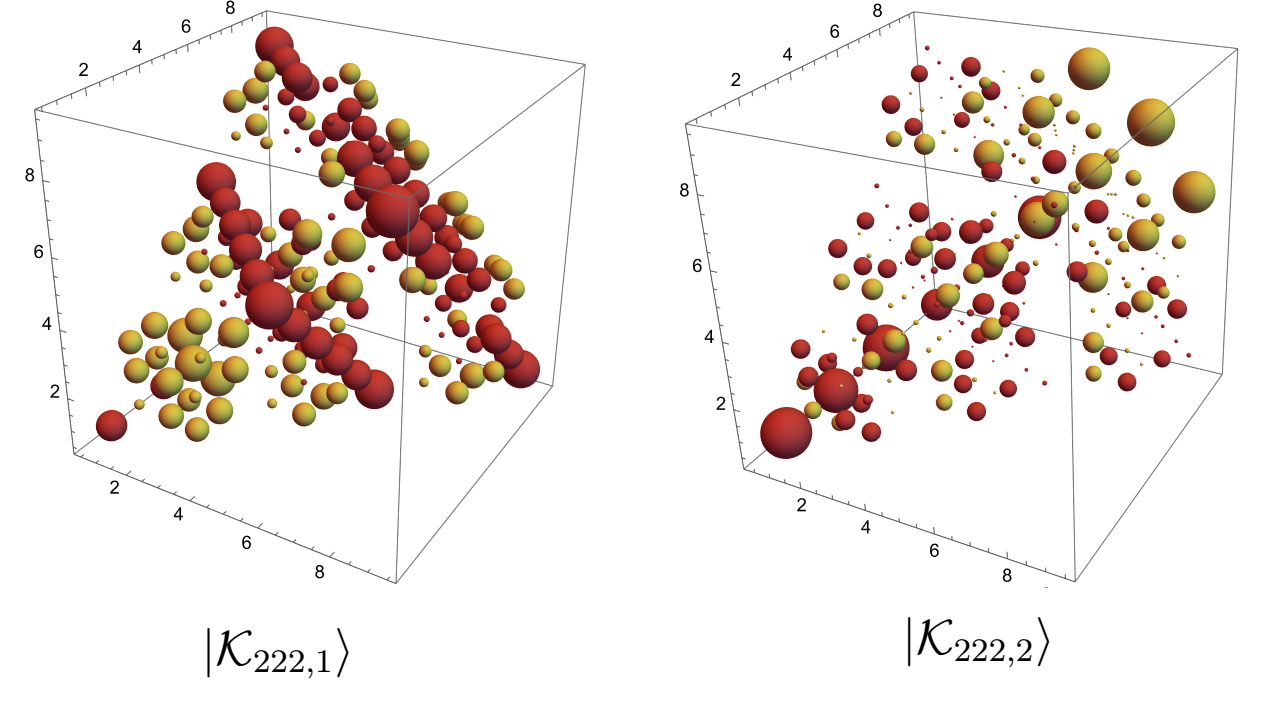}
\caption{Graphical representation of the two orthogonal Kronecker states defining the basis of $([2]\otimes[2]\otimes[2])^{S_6}$}.
\label{fig:Kron6222}
\end{figure}

\noindent The first case where the dimension is greater than 1 $(k_{\bm{\lambda}}>1)$ is for $n=6$ with the partition $\bm{\lambda}=222$ with $k_{\bm{\lambda}}=2$. This partition belongs to $\Lambda_6^W$, then, one of the Kronecker states that we will use to define the basis of the Kronecker subspace is $\ket{\mathcal{K}^{W}_{222}}$. Recalling the notation for the orthogonal basis of Kronecker states $\{\ket{\mathcal{K}_{\bm{\lambda},s}}\}$, this is:
 \begin{equation*}
 n=6:  \qquad \ket{\mathcal{K}_{222,1}} =\ket{\mathcal{K}^{W}_{222}}.
\end{equation*} 
We will use the W-Kronecker construction to obtain one linearly independent Kronecker state. For this set, the compatible inner partitions for the graph $\Kite{0.3}$ are:
 \begin{equation*}
 \bm{\mu}_{222}^{\Kite{0.2}}=\{ 022,111,112,113,122,222\}.
\end{equation*}
The state obtained from the first set is the same W-Kronecker state; with any other inner set, it is possible to build a basis for the Kronecker subspace. In particular, we can choose the state obtained with $\bm{\mu}=111$, then, applying a Gram-Schmidt process we obtain an orthogonal Kronecker state:
\begin{equation*}
n=6: \quad
\ket{\mathcal{K}_{222,2} }= \frac{\sqrt{1961}}{40} \left(\ket{\mathcal{K}_{222}^{111,\Kite{0.2}}}+ \frac{19}{\sqrt{1961}} \ket{\mathcal{K}^W_{222}} \right)
\end{equation*}  
and then $\{\ket{\mathcal{K}_{222,1}},\ket{\mathcal{K}_{222,2}}\}$ define an orthonormal basis for the Kronecker subspace. Unfortunately,  as was discussed before, the dimension of each vector grows exponentially with $n$, which does not allow us to write the vectors explicitly here, for this example, the dimension is given by $f^{[2]}\times f^{[2]}\times f^{[2]} = 9\times 9 \times 9=729$, and the number of no null entries in the Kronecker states used as a basis are $192$ and $231$ respectively. Remarkably, the basis for this Kronecker subspace is obtained on a personal computer in few seconds using this construction. We show in Figure \ref{fig:Kron6222} the graphical representation of the two basis elements for $([2]\otimes[2]\otimes[2])^{S_6}$. \\

\noindent Now we will show how to obtain an orthonormal basis for some Kronecker subspaces with $k_{\bm{\lambda}}=2$ up to $n=9$.

\subsubsection{n=7}
For $n=7$ we have the triplet $\bm{\lambda}=222$, with dimensions $f^{\bm{\lambda}}=(14,14,14)$, by defining:
\begin{equation*}
n=7: \qquad 
 \ket{\mathcal{K}_{222,2}}= \frac{4}{5}\sqrt{\frac{34}{15}} \left( \ket{\mathcal{K}_{222}^{111,\Kite{0.2}}}+ \frac{13}{4\sqrt{34}} \ket{\mathcal{K}^W_{222}} \right) .
\end{equation*}  
Then, an orthonormal basis for the Kronecker subspace is defined by:
 $\{\ket{\mathcal{K}^W_{222}},\ket{\mathcal{K}_{222,2}} \} $.
 
\subsubsection{n=8}
In $n=8$ the first triplet with $k_{\bm{\lambda}}=2$ is $\bm{\lambda}=222$, with dimensions $f^{\bm{\lambda}}=(20,20,20)$, by defining:
\begin{equation*}
n=8: \qquad
 \ket{\mathcal{K}_{222,2}}= \frac{\sqrt{730}}{21} \left( \ket{\mathcal{K}_{222}^{111,\Kite{0.2}}}+ \frac{17}{\sqrt{730}} \ket{\mathcal{K}^W_{222}} \right) .
\end{equation*} 
Then,  an orthonormal basis for the Kronecker subspace $([2]\otimes[2]\otimes[2])^{S_8}$ is defined by: $\{\ket{\mathcal{K}^W_{222}},\ket{\mathcal{K}_{222,2}} \} $.\\

\noindent There is another triplet with $k_{\bm{\lambda}}=2$ given by $\bm{\lambda}=233$, with dimensions $f^{\bm{\lambda}}=(20,28,28)$, defining:
\begin{equation*}
n=8: \qquad
 \ket{\mathcal{K}_{233,2}}= \frac{1}{5}\sqrt{\frac{418}{7}}  \left( \ket{\mathcal{K}_{233}^{122,\Kite{0.2}}}+ 9 \sqrt{\frac{3}{418}}  \ket{\mathcal{K}^W_{233}} \right) .
\end{equation*}
 Then,  an orthonormal basis for the Kronecker subspace $([2]\otimes[3]\otimes[3])^{S_8}$ is defined by: $\{\ket{\mathcal{K}^W_{233}},\ket{\mathcal{K}_{233,2}} \} $.
 \subsubsection{n=9}
 For $n=9$ we have three triplets, the first one is $\bm{\lambda}=222$ with dimensions $f^{\bm{\lambda}}=(27,27,27)$, defining:
 \begin{equation*}
 n=9: \qquad \ket{\mathcal{K}_{222,2}}= \frac{1}{28}\sqrt{\frac{20221}{14}}  \left( \ket{\mathcal{K}_{233}^{122,\Kite{0.2}}}+ 43 \sqrt{\frac{5}{20221}}  \ket{\mathcal{K}^W_{222}} \right) .
\end{equation*}
Then, the orthonormal basis is defined by: $\{\ket{\mathcal{K}^W_{222}},\ket{\mathcal{K}_{222,2}} \} $. The second triplet is $\bm{\lambda}=233$, with dimensions $f^{\bm{\lambda}}=(27,48,48)$, defining:
 \begin{equation*}
 n=9 : \qquad \ket{\mathcal{K}_{233,2}}= \frac{1}{4}\sqrt{\frac{6169}{105}}  \left( \ket{\mathcal{K}_{233}^{122,\Kite{0.2}}}+ \frac{67}{ \sqrt{6169}}  \ket{\mathcal{K}^W_{233}} \right).
\end{equation*}
So, the orthonormal basis is given by: $\{\ket{\mathcal{K}^W_{233}},\ket{\mathcal{K}_{233,2}} \} $. The last triplet is $\bm{\lambda}=333$ with dimensions $f^{\bm{\lambda}}=(48,48,48)$, defining:
\begin{equation*}
n=9:\qquad 
 \ket{\mathcal{K}_{333,2}}= \frac{32}{7\sqrt{15}}  \left( \ket{\mathcal{K}_{333}^{122,\Kite{0.2}}}+ \frac{17}{ 32}  \ket{\mathcal{K}^W_{333}} \right).
 \end{equation*}
Then, the basis is defined by:
$\{\ket{\mathcal{K}^W_{333}},\ket{\mathcal{K}_{333,2}} \} $. \\

\noindent The number of triplets with Kronecker coefficient equal to two, grows fast from this point. Because of this we will jump directly to the first case where the Kronecker coefficient is greater than two.
\subsection*{n=12}
For $n=12$, there are twelve sets of partitions with $k_{\bm{\lambda}}>1$, but the most interesting case is $\bm{\lambda}=444$ that is the first case where the Kronecker coefficient is greater than two, $k_{\bm{\lambda}}=3$. In this set of partitions, the dimensions are: $f^{\bm{\lambda}}=(275,275,275)$. For this case we will show the one possible set for the basis and the graphical representation of the states, that have more than 2.5 million of coefficients. The first basis element is the W-Kronecker state $\ket{\mathcal{K}_{444,1}}=\ket{\mathcal{K}^W_{444}}$, whose graphical representation is:
\begin{equation*}
\ket{\mathcal{K}_{444,1}} = \includegraphics[scale=0.25,valign=c]{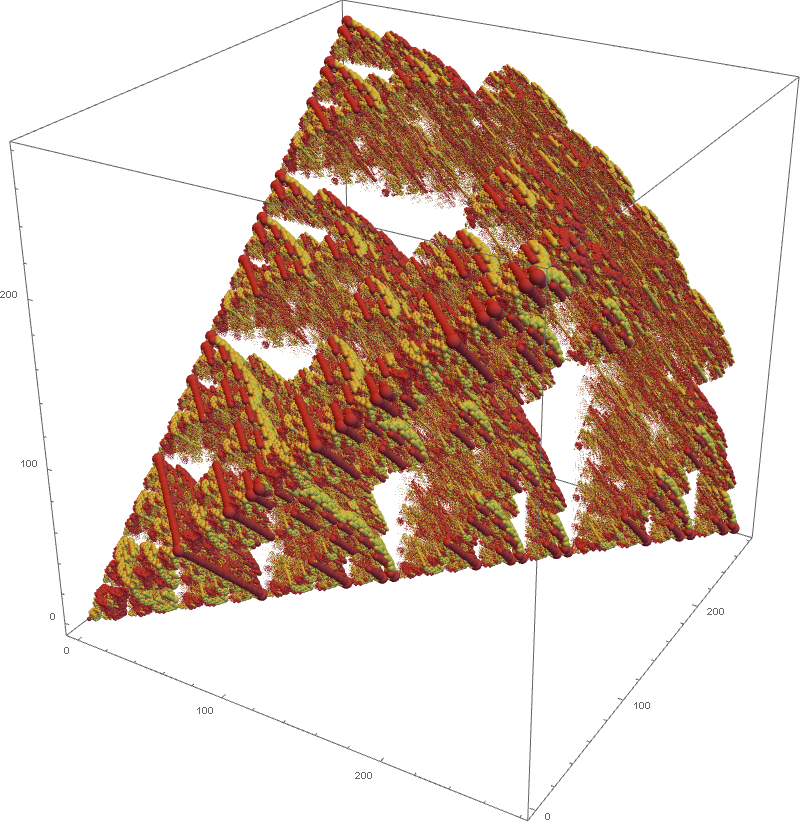}.
\end{equation*}
It would be intriguing to explore the significance of the fractal-like structure in this representation.\\

\noindent The other two basis elements of the Kronecker subspace can be obtained as:
\begin{equation*}
\begin{gathered}
\ket{\mathcal{K}_{444,2}}=\frac{1}{16}\sqrt{\frac{22206361}{85470}} \left(\ket{\mathcal{K}_{444}^{133,\Kite{0.2}}} -\frac{571}{\sqrt{22206361}} \ket{\mathcal{K}^{W}_{444}} \right), \\
\ket{\mathcal{K}_{444,3}}= \frac{1}{252} \sqrt{\frac{339845}{2}} \left(\ket{\mathcal{K}^{222,\Kite{0.2}}_{444}} +13476 \sqrt{\frac{6}{2974663285}}\ket{\mathcal{K}_{444,2}} + 233 \sqrt{\frac{7}{1461751}}\ket{\mathcal{K}^{W}_{444}} \right).
\end{gathered}
\end{equation*}
It is clear that the orthogonalization problem becomes more complicated with higher values of $n$ and $k_{\bm{\lambda}}$. The graphical representation of these states are:
\begin{equation*}
\ket{\mathcal{K}_{444,2}} = \includegraphics[scale=0.2,valign=c]{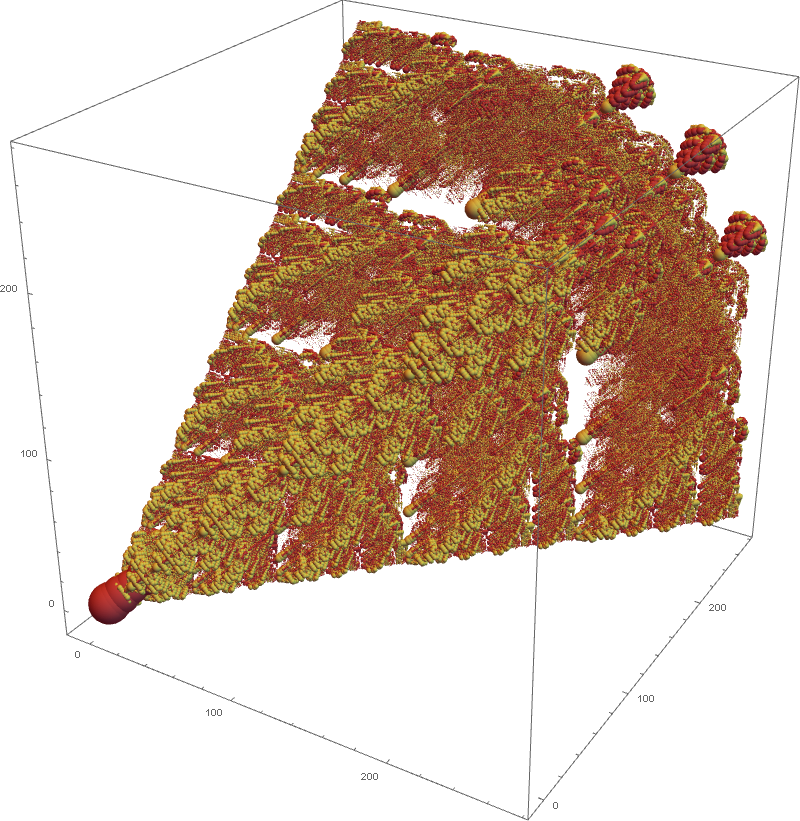}, \quad \ket{\mathcal{K}_{444,3}} = \includegraphics[scale=0.2,valign=c]{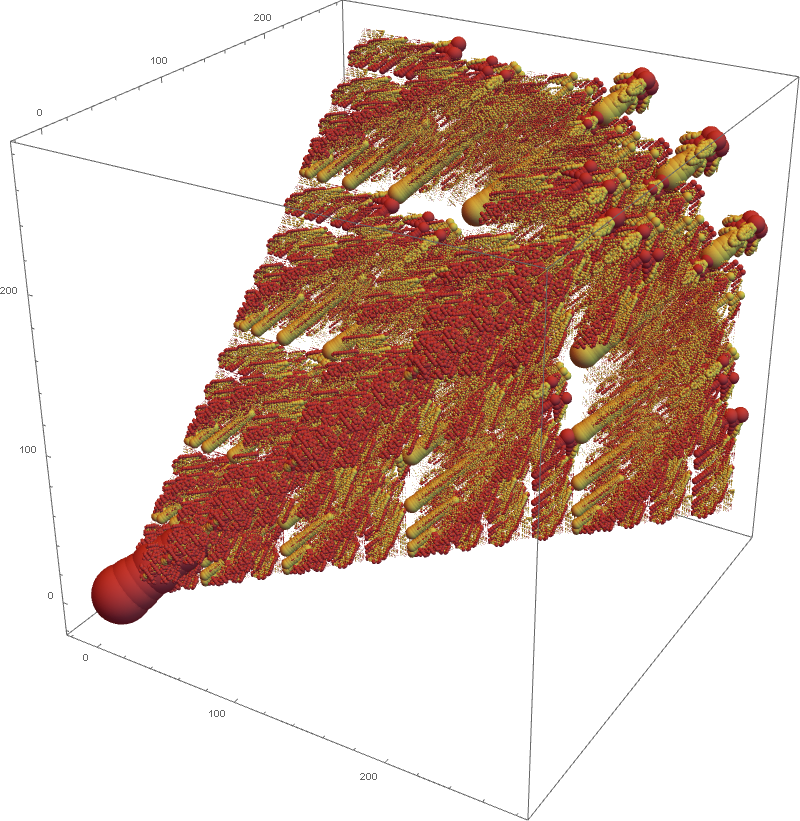}.
\end{equation*}
It can be seen that this graph has a structure that is significantly different from the W-Kronecker state. This structure is less  ``pyramidal'' and some interesting patterns can be noticed on it. It can be noticed that the W-Kronecker states structure is special. When comparing the representations of W-Kronecker states with non-W-Kronecker states this structural difference is always present. It would be interesting to study this characteristic in the future. \\

\noindent With this we finish the exploration in the case of three-parts Kronecker states. All the Kronecker states presented here were calculated in exact form, as square roots of rational numbers. This results highlight the efficiency of the method. Now we show the case of four-part Kronecker states, and explore some interesting results. 
  
\section{Four-part Kronecker states}

When considering the Kronecker states of four parts, the Kronecker coefficient explodes quickly, reaching values greater than $10$ for $n=6$ and more than $100$ for $n=12$. Despite the complexity, the stitching algorithm allows an exhaustive protocol to build completely the Kronecker subspace. For four-qubit states, the full measure SLOCC family is named $G_{abcd}$ class in the classification of \cite{Verstraete}, also discussed in Section \ref{SLOCC}. From the results presented in Section \ref{FourQubitsVerst} and Theorem \ref{TheoremI}, we can conclude that all Kronecker subspaces of four parts can be obtained from the following graph:
\begin{equation*}
    \begin{tikzpicture}[basel={-.5}]
        \draw(0,0) to (1,0);
        \draw[markx={0},markx={1}] (0,0) to (0,-1);
        \draw (0,-1) to (1,-1);
        \draw[\lao-] (1,-1) to (2,0);
        \draw[\lac-] (1,0) to (2,-1);
        \draw (-0.25,0.25) to (0,0);
        \draw (0,-1) to (-0.25,-1.25);
        \draw (2,0) to (2.25,0.25);
        \draw[markx={0},markx={1}]  (1,0) to (2,0);
        \draw[markx={0},markx={1}]  (1,-1) to (2,-1);
        \draw (2,-1) to (2.25,-1.25);
    \end{tikzpicture}.
\end{equation*}
However, there are many possible diagrams for obtaining four-part states, and it is worth exploring their structures and implications in constructing Kronecker states. The simplest diagram for obtaining four qubit states is:
\begin{equation*}
     \begin{tikzpicture}[basel={-0.5}]
   \draw   (1,0) -- (0,0) ;
   \draw[markx={1}](-0.5,0.5) -- (0,0);
   \draw (-0.5,-0.5) -- (0,0);\draw (1.5,0.5) -- (1,0);
   \draw[markx={1}] (1.5,-0.5) -- (1,0);
   \draw (-0.75,0.5) node{$\lambda^1$};
   
   \draw (-0.75,-0.5) node{$\lambda^2$};
   
   \draw (1.8,0.5) node{$\lambda^3$};
   
   \draw (1.8,-0.5) node{$\lambda^4$};
   \draw (0.5, 0.3) node{$\mu$};
\end{tikzpicture}.
\end{equation*}
This graph is simple, but the generated Kronecker subspace is already greater than the $W-$Kronecker subspace, as seen in Section \ref{KronStitch}. Besides the aspects discussed in this graph, we can notice that this is an OER graph. Then, according to the Equation \eqref{eq:OrthographKron}, if we calculate graph Kronecker states for the same set of external partitions $\bm{\lambda}$, with different values of $\mu$, they will be orthogonal between each other. In particular, as there are no more inner partitions than the reducible one, we have that the dimension of the Kronecker subspace spanned by the graph is exactly the number of compatible inner partitions $\mu$ with the set $\bm{\lambda}$,  $k_{\bm{\lambda}}^{\TreeF{0.3}} = |\mu_{\bm{\lambda}}^{\TreeF{0.3}}|$, which is a unique property of this graph. By recalling the second property of OER graphs, we know that any qubit state represented by this graph has reduced density matrices in the partition $12$ - $34$ of rank at most two.   \\

\noindent In Table \ref{table:GraphKronecker} we show the sets $\bm{\lambda}=\lambda^1\lambda^2\lambda^3\lambda^4$ with invariant subspace, $k_{\bm{\lambda}}\neq 0$, up to $n=6$, removing those with some partition $\lambda^i=0$, because they correspond to three parts Kronecker states knowing that $f^{[0]}=1$. The table shows the Kronecker coefficient of each set and the effective Kronecker coefficient of some graphs, obtained by orthogonalizing all the graph Kronecker states from the compatible sets $\bm{\mu}^G_{\bm{\lambda}}$. \\

\begin{table}[]
\begin{tabular}{|c|c|c|c|c|c|c|c|}
\hline
n & $\bm{\lambda}$ & $k_{\bm{\lambda}}$ & $k_{\bm{\lambda}}^{\TreeF{0.2}}$ &$k_{\bm{\lambda}}^{\Kitetree{0.2}}$&$k_{\bm{\lambda}}^{\Cuad{0.2}}=k_{\bm{\lambda}}^{\CuadKite{0.2}}$& $k_{\bm{\lambda}}^{\DobKite{0.2}}$&$k_{\bm{\lambda}}^{\Cam{0.2}}$\\ \hline
3 &   1111  &3 & 2  &2 &3 &2&3\\ \hline
4 &  1111& 4 & 3 &3  &4&3 &4 \\ \hline
4 & 1112  &  2 &1 &1 &2&1 &2\\ \hline
4  & 1122  & 2 & 1&(211111)&(211) &2 &2 \\ \hline
4  & 2222  & 3 & 1&1 &3&2&3 \\ \hline
5 & 1111  & 4 & 3&3 &4&3&4 \\ \hline
5 & 1112 &3 & 2 &2 &3&2&3\\ \hline
5 & 1122& 4 & 2&(322222) &4&(322)&4 \\ \hline
5 & 1222 & 4 &1 &  (222111)&4&2&4 \\ \hline
5 & 2222 & 6 & 2 &2& 6&3& 6\\ \hline
6 & 1111 & 4 & 3 &3& 4 &3& 4 \\ \hline
6 & 1112 &3 & 2 &2& 3&2&3\\ \hline
6 & 1113 &1 & 1 &1& 1&1&1\\ \hline
6 & 1122 &5 & 3 &(433333)&5&(433)&5\\ \hline
6 & 1123 & 2 & 1 &1&2&1&2\\ \hline
6 & 1133 & 2 & 1 & (211111)&2&(211)&2\\ \hline
6 & 1222 & 6 & 2 & (333222)&6&3&6\\ \hline
6 & 1223 & 4 & 1 &(222111)&4&2&4\\ \hline
6 & 1233& 1 &0 & (100000)&1&(100)&1\\ \hline
6 & 1333 & 1 & 0& (111000)&1&1&1 \\ \hline
6 & 2222 & 13 & 3& 4&13&6&13 \\ \hline
6 & 2223 & 4 & 1 &1&4&1&4  \\ \hline
6 & 2233 & 5 & 1 &(211111)&5&(322)&5\\ \hline
6 & 3333 & 4 & 1 &1 &3&2&4\\ \hline
\end{tabular}
\caption{Sets $\bm{\lambda}$, with their respective Kronecker coefficient and the dimension of the Kronecker subspace generated by each graph. The values with parentesis correspond to different orderings of the external partitions in non-symmetric graphs}
\label{table:GraphKronecker}
\end{table}
\noindent It is clear from Table \ref{table:GraphKronecker} that this graph does not completely generate Kronecker subspaces. Some sets $\bm{\lambda}$ cannot be achieved, and the dimension of the generated Kronecker subspace is, in general, lower than $k_{\bm{\lambda}}$ from many of the sets. This characterization also gives information for the qubit states. As with this graph, the subfamilies $L_{a00_2}$ and  $L_{00_3}$ are obtained, then, from the table, we can read what sets of partitions can appear in the Schur transform of any state in these subfamilies. It has to be highlighted that despite this graph is not completely symmetric, the order of the partitions in $\bm{\lambda}$ used for the construction does not affect the results shown in \ref{table:GraphKronecker}.\\

\noindent The next graph to consider is obtained with four W-vertices and four stitches:
\begin{equation*}
 \begin{tikzpicture}[basel={-.5}]
    \draw[markx={0}](0,0) to (1,0);
    \draw(1,0) to (1.5,0.5);
    \draw(1,0) to (1.5,-0.5);
    \draw[markx={0},markx={1}](1.5,0.5) to (1.5,-0.5);
    \draw(1.5,0.5) to (1.75,0.75);
    \draw(1.5,-0.5) to (1.75,-0.75);
    \draw(-0.25,0.25) to (0,0);
    \draw(-0.25,-0.25) to (0,0);
    \draw[markxi={1}](0,0) to (1,0);
     \draw (-0.5,0.25) node {$\lambda^1$};
     \draw (-0.5,-0.25) node {$\lambda^2$};
     \draw (2.05,0.75) node {$\lambda^3$};
     \draw (2.05,-0.75) node {$\lambda^4$};
    \end{tikzpicture}  .
\end{equation*}
Note how this graph is less symmetric than the previous one, and there are six different orderings of external partitions:
\begin{equation*}
\begin{gathered}
 \Kitetree{0.5}_1=\begin{tikzpicture}[basel={-.5}]
    \draw[markx={0}](0,0) to (1,0);
    \draw(1,0) to (1.5,0.5);
    \draw(1,0) to (1.5,-0.5);
    \draw[markx={0},markx={1}](1.5,0.5) to (1.5,-0.5);
    \draw(1.5,0.5) to (1.75,0.75);
    \draw(1.5,-0.5) to (1.75,-0.75);
    \draw(-0.25,0.25) to (0,0);
    \draw(-0.25,-0.25) to (0,0);
    \draw[markxi={1}](0,0) to (1,0);
     \draw (-0.5,0.25) node {$\lambda^1$};
     \draw (-0.5,-0.25) node {$\lambda^2$};
     \draw (2.05,0.75) node {$\lambda^3$};
     \draw (2.05,-0.75) node {$\lambda^4$};
    \end{tikzpicture}  ,  \quad  \Kitetree{0.5}_2=\begin{tikzpicture}[basel={-.5}]
    \draw[markx={0}](0,0) to (1,0);
    \draw(1,0) to (1.5,0.5);
    \draw(1,0) to (1.5,-0.5);
    \draw[markx={0},markx={1}](1.5,0.5) to (1.5,-0.5);
    \draw(1.5,0.5) to (1.75,0.75);
    \draw(1.5,-0.5) to (1.75,-0.75);
    \draw(-0.25,0.25) to (0,0);
    \draw(-0.25,-0.25) to (0,0);
    \draw[markxi={1}](0,0) to (1,0);
     \draw (-0.5,0.25) node {$\lambda^1$};
     \draw (-0.5,-0.25) node {$\lambda^3$};
     \draw (2.05,0.75) node {$\lambda^2$};
     \draw (2.05,-0.75) node {$\lambda^4$};
    \end{tikzpicture} , \quad  \Kitetree{0.5}_3=\begin{tikzpicture}[basel={-.5}]
    \draw[markx={0}](0,0) to (1,0);
    \draw(1,0) to (1.5,0.5);
    \draw(1,0) to (1.5,-0.5);
    \draw[markx={0},markx={1}](1.5,0.5) to (1.5,-0.5);
    \draw(1.5,0.5) to (1.75,0.75);
    \draw(1.5,-0.5) to (1.75,-0.75);
    \draw(-0.25,0.25) to (0,0);
    \draw(-0.25,-0.25) to (0,0);
    \draw[markxi={1}](0,0) to (1,0);
     \draw (-0.5,0.25) node {$\lambda^1$};
     \draw (-0.5,-0.25) node {$\lambda^4$};
     \draw (2.05,0.75) node {$\lambda^2$};
     \draw (2.05,-0.75) node {$\lambda^3$};
    \end{tikzpicture} \\
    \Kitetree{0.5}_4=\begin{tikzpicture}[basel={-.5}]
    \draw[markx={0}](0,0) to (1,0);
    \draw(1,0) to (1.5,0.5);
    \draw(1,0) to (1.5,-0.5);
    \draw[markx={0},markx={1}](1.5,0.5) to (1.5,-0.5);
    \draw(1.5,0.5) to (1.75,0.75);
    \draw(1.5,-0.5) to (1.75,-0.75);
    \draw(-0.25,0.25) to (0,0);
    \draw(-0.25,-0.25) to (0,0);
    \draw[markxi={1}](0,0) to (1,0);
     \draw (-0.5,0.25) node {$\lambda^2$};
     \draw (-0.5,-0.25) node {$\lambda^3$};
     \draw (2.05,0.75) node {$\lambda^1$};
     \draw (2.05,-0.75) node {$\lambda^4$};
    \end{tikzpicture}  ,  \quad  \Kitetree{0.5}_5=\begin{tikzpicture}[basel={-.5}]
    \draw[markx={0}](0,0) to (1,0);
    \draw(1,0) to (1.5,0.5);
    \draw(1,0) to (1.5,-0.5);
    \draw[markx={0},markx={1}](1.5,0.5) to (1.5,-0.5);
    \draw(1.5,0.5) to (1.75,0.75);
    \draw(1.5,-0.5) to (1.75,-0.75);
    \draw(-0.25,0.25) to (0,0);
    \draw(-0.25,-0.25) to (0,0);
    \draw[markxi={1}](0,0) to (1,0);
     \draw (-0.5,0.25) node {$\lambda^2$};
     \draw (-0.5,-0.25) node {$\lambda^4$};
     \draw (2.05,0.75) node {$\lambda^1$};
     \draw (2.05,-0.75) node {$\lambda^3$};
    \end{tikzpicture} , \quad  \Kitetree{0.5}_6=\begin{tikzpicture}[basel={-.5}]
    \draw[markx={0}](0,0) to (1,0);
    \draw(1,0) to (1.5,0.5);
    \draw(1,0) to (1.5,-0.5);
    \draw[markx={0},markx={1}](1.5,0.5) to (1.5,-0.5);
    \draw(1.5,0.5) to (1.75,0.75);
    \draw(1.5,-0.5) to (1.75,-0.75);
    \draw(-0.25,0.25) to (0,0);
    \draw(-0.25,-0.25) to (0,0);
    \draw[markxi={1}](0,0) to (1,0);
     \draw (-0.5,0.25) node {$\lambda^3$};
     \draw (-0.5,-0.25) node {$\lambda^4$};
     \draw (2.05,0.75) node {$\lambda^1$};
     \draw (2.05,-0.75) node {$\lambda^2$};
    \end{tikzpicture} 
    \end{gathered}
\end{equation*}
This graph generally spans a larger Kronecker subspace than the previous one. The restrictions in the external partitions become weaker by the addition of structure, allowing a greater set of compatible inner partitions. As the positions of external partitions become relevant, we show in Table \ref{table:GraphKronecker} a list of effective Kronecker coefficients for each order when they are not all the same. This graph is also an OER and allows us to obtain more orthogonal Kronecker states than the previous graph. Nevertheless, there is still one more OER graph that is optimal in this sense.\\

\noindent The following graph   is another possible construction with four vertices and four stitches:
 \begin{equation*}
        \begin{tikzpicture}[basel={-.5}]
        \draw[markx={1}] (-0.25,0.25) to (0,0);
        \draw (0,0) to (1,0);
        \draw [markx={0}](1,0) to (1.25,0.25);
        \draw (0,0) to (0,-1);
        \draw [markx={1}](-0.25,-1.25) to (0,-1);
        \draw (0,-1) to (1,-1);
        \draw (1,0) to (1,-1);
        \draw [markx={0}](1,-1) to (1.25,-1.25);
        \draw (-0.5,0.25) node {$\lambda_1$};
         \draw (1.55,0.25) node {$\lambda_2$};
          \draw (1.55,-1.25) node {$\lambda_3$};
           \draw (-0.5,-1.25) node {$\lambda_4$};
    \end{tikzpicture} .
\end{equation*}
This graph is the first non-OER graph and leads, as described in Appendix  \ref{Appendix03}, to broader SLOCC families of four qubits. As we show in \ref{table:GraphKronecker}, this fact seems connected with considerable growth in the Kronecker subspaces that can be obtained. Up to $n=6$, all Kronecker subspaces can be obtained completely with this graph, except from the set $\bm{\lambda}=3333$, where only three of four dimensions can be spanned. For larger $n$ values, more sets of external partitions with $k_{\bm{\lambda}}^{\Cuad{0.2}}<k_{\bm{\lambda}}$ appear. \\

\noindent When considering graphs of six vertices and seven stitches, one interesting graph is the following:
\begin{equation*}
 \begin{tikzpicture}[basel={-.5}]
 \draw (-0.5,0.5) to (0,0);
  \draw (-0.5,-0.5) to (0,0);
   \draw (-0.75,0.75) to (-0.5,0.5);
   \draw[markx={0},markx={1}] (-0.5,0.5) to (-0.5,-0.5); 
   \draw (-0.75,-0.75) to (-0.5,-0.5);
    \draw[markx={0}](0,0) to (1,0);
    \draw(1,0) to (1.5,0.5);
    \draw(1,0) to (1.5,-0.5);
    \draw[markx={0},markx={1}](1.5,0.5) to (1.5,-0.5);
    \draw(1.5,0.5) to (1.75,0.75);
    \draw(1.5,-0.5) to (1.75,-0.75);
    \draw[markxi={1}](0,0) to (1,0);
     \draw (-1,0.75) node {$\lambda^1$};
     \draw (-1,-0.75) node {$\lambda^2$};
     \draw (2.05,0.75) node {$\lambda^3$};
     \draw (2.05,-0.75) node {$\lambda^4$};
    \end{tikzpicture},
\end{equation*}
Note that this is an OER graph that is stitching together two $\Kite{0.3}$ graphs, the most general graph of three qubits. This graph is the simpler OER graph, where each subgraph is the most general. Then, with this graph, we can obtain the greatest set of mutually orthogonal graph Kronecker states for any partitions $\bm{\lambda}$. This graph has three inequivalent orderings:
\begin{equation*}
 \begin{tikzpicture}[basel={-.5}]
 \draw (-0.5,0.5) to (0,0);
  \draw (-0.5,-0.5) to (0,0);
   \draw (-0.75,0.75) to (-0.5,0.5);
   \draw[markx={0},markx={1}] (-0.5,0.5) to (-0.5,-0.5); 
   \draw (-0.75,-0.75) to (-0.5,-0.5);
    \draw[markx={0}](0,0) to (1,0);
    \draw(1,0) to (1.5,0.5);
    \draw(1,0) to (1.5,-0.5);
    \draw[markx={0},markx={1}](1.5,0.5) to (1.5,-0.5);
    \draw(1.5,0.5) to (1.75,0.75);
    \draw(1.5,-0.5) to (1.75,-0.75);
    \draw[markxi={1}](0,0) to (1,0);
     \draw (-1,0.75) node {$\lambda^1$};
     \draw (-1,-0.75) node {$\lambda^2$};
     \draw (2.05,0.75) node {$\lambda^3$};
     \draw (2.05,-0.75) node {$\lambda^4$};
    \end{tikzpicture}, \qquad \begin{tikzpicture}[basel={-.5}]
 \draw (-0.5,0.5) to (0,0);
  \draw (-0.5,-0.5) to (0,0);
   \draw (-0.75,0.75) to (-0.5,0.5);
   \draw[markx={0},markx={1}] (-0.5,0.5) to (-0.5,-0.5); 
   \draw (-0.75,-0.75) to (-0.5,-0.5);
    \draw[markx={0}](0,0) to (1,0);
    \draw(1,0) to (1.5,0.5);
    \draw(1,0) to (1.5,-0.5);
    \draw[markx={0},markx={1}](1.5,0.5) to (1.5,-0.5);
    \draw(1.5,0.5) to (1.75,0.75);
    \draw(1.5,-0.5) to (1.75,-0.75);
    \draw[markxi={1}](0,0) to (1,0);
     \draw (-1,0.75) node {$\lambda^1$};
     \draw (-1,-0.75) node {$\lambda^3$};
     \draw (2.05,0.75) node {$\lambda^2$};
     \draw (2.05,-0.75) node {$\lambda^4$};
    \end{tikzpicture}, \qquad \begin{tikzpicture}[basel={-.5}]
 \draw (-0.5,0.5) to (0,0);
  \draw (-0.5,-0.5) to (0,0);
   \draw (-0.75,0.75) to (-0.5,0.5);
   \draw[markx={0},markx={1}] (-0.5,0.5) to (-0.5,-0.5); 
   \draw (-0.75,-0.75) to (-0.5,-0.5);
    \draw[markx={0}](0,0) to (1,0);
    \draw(1,0) to (1.5,0.5);
    \draw(1,0) to (1.5,-0.5);
    \draw[markx={0},markx={1}](1.5,0.5) to (1.5,-0.5);
    \draw(1.5,0.5) to (1.75,0.75);
    \draw(1.5,-0.5) to (1.75,-0.75);
    \draw[markxi={1}](0,0) to (1,0);
     \draw (-1,0.75) node {$\lambda^1$};
     \draw (-1,-0.75) node {$\lambda^4$};
     \draw (2.05,0.75) node {$\lambda^3$};
     \draw (2.05,-0.75) node {$\lambda^2$};
    \end{tikzpicture},
\end{equation*}
in Table \ref{table:GraphKronecker} is shown that this graph generally spans a smaller Kronecker subspace than $\Cuad{0.3}$, but it has the advantage that all graph Kronecker states, with different partitions in the reducible edge, are orthogonal between them. Some other interesting graphs are the following:
\begin{equation*}
\CuadKite{1}, \qquad   ,\DobCuad{1},
\end{equation*}
The first graph has more structure than $\Cuad{0.3}$, but up to $n=6$, the spans of Kronecker subspaces for both are the same. On the other hand, $\DobCuad{0.3}$ is not a sufficient graph, in the sense of Theorem \ref{TheoremI}, but still generates completely any Kronecker subspace up to $n=11$. It would be necessary to reach higher values of $n$ to find at what point $\DobCuad{0.3}$ starts to fail in building  Kronecker subspaces. \\

\noindent Finally, the following graph, with six vertices and seven stitches:
\begin{equation*}
    \begin{tikzpicture}[basel={-.5}]
        \draw(0,0) to (1,0);
        \draw[markx={0},markx={1}] (0,0) to (0,-1);
        \draw (0,-1) to (1,-1);
        \draw (1,-1) to (2,0);
        \draw (1,0) to (2,-1);
        \draw (-0.25,0.25) to (0,0);
        \draw (0,-1) to (-0.25,-1.25);
        \draw (2,0) to (2.25,0.25);
        \draw[markx={0},markx={1}]  (1,0) to (2,0);
        \draw[markx={0},markx={1}]  (1,-1) to (2,-1);
        \draw (2,-1) to (2.25,-1.25);
    \end{tikzpicture},
\end{equation*}
is the graph that generates the full measure family $G_{abcd}$ of four qubits, then, Theorem \ref{TheoremI}, ensures that $k_{\bm{\lambda}}^{\Cam{0.2}}=k_{\bm{\lambda}}$ always. Then, building all the possible graph Kronecker states from the inner partitions $\bm{\mu}^{\Cam{0.2}}_{\bm{\lambda}}$, it is possible to obtain a set of mutually orthogonal Kronecker states to define the base of any invariant subspace. By using this algorithm, we were able to calculate exactly (with roots of rational numbers) a set of bases for any Kronecker subspace of four parts up to $n=9$, where the most remarkable set is $\bm{\lambda}=(3333)$ with $k_{\bm{\lambda}}=39$, each part with a dimension $f^{\lambda}=48$, showing how efficient is the construction. \\

\begin{figure}
\includegraphics[scale=0.45]{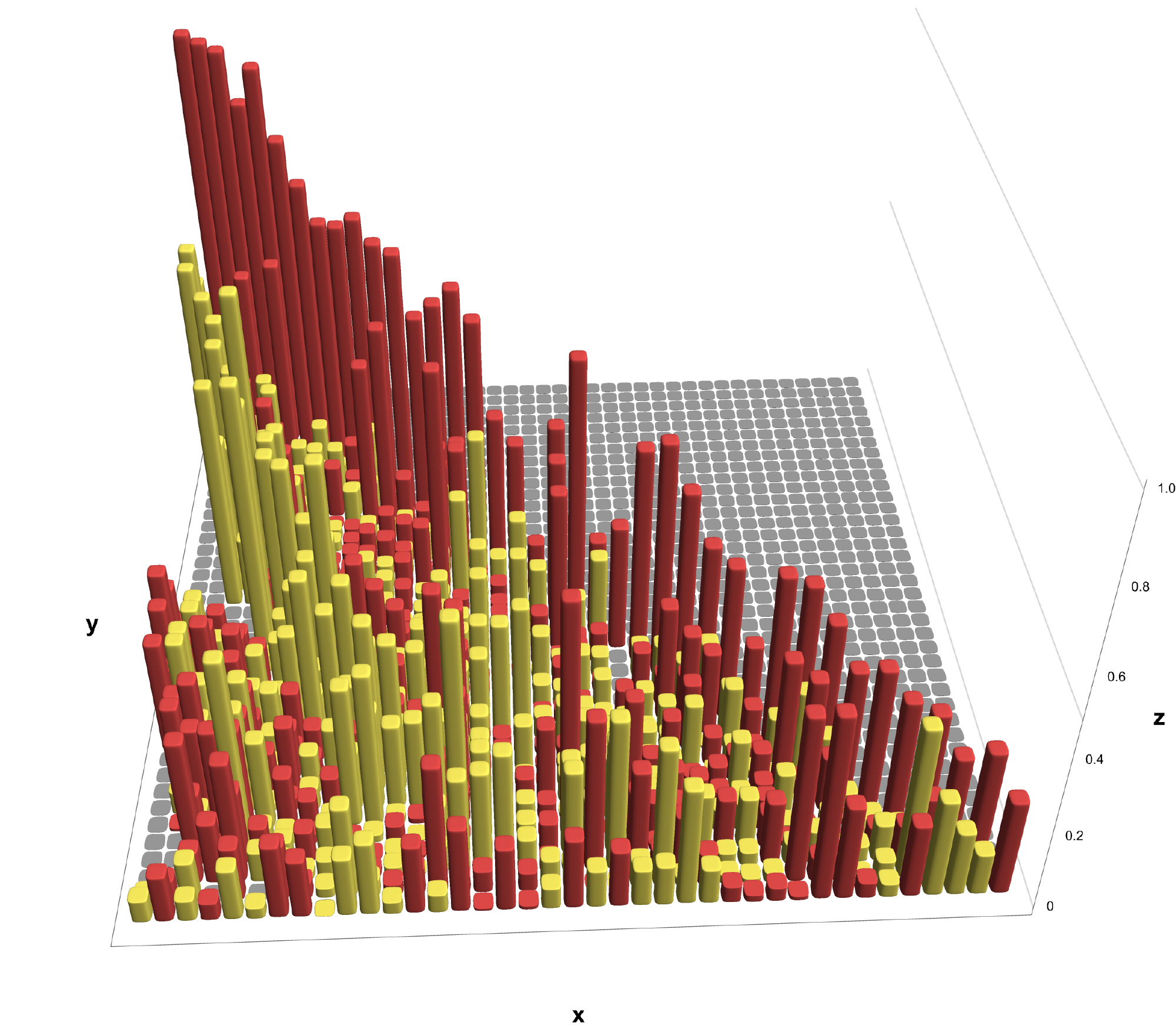}
\caption{Graphical representation of the matrix of inner products between the graph Kronecker states \(\ket{\mathcal{K}^{\bm{\mu}^i,{\protect\Cam{0.2}}}_{\bm{\lambda}}}\) in x axis, and the set of orthogonal Kronecker states \(\ket{\mathcal{K}_{\bm{\lambda},s}}\) in y axis. The states in x axis are ordered according to the sets $\bm{\mu}$ listed in Equation \eqref{eq:mus}. The height of the bar represent the magnitude of the coefficient and the color its sign: red for positive and yellow for negative.}
\label{fig:Grand}
\end{figure}

\noindent Due to the complexity of the obtained states, it is impossible to present their coefficient expansion here, and the graphical representation used for three-part Kronecker states cannot be used here either. Despite this, we will present here the largest case for $n=9$ and state explicitly what sets of inner partitions should be used to generate a complete basis for the Kronecker subspace. For this, we will fix the labels of inner and external partitions as:
 
\begin{equation*}
    \begin{tikzpicture}[basel={-.5},every node/.style={scale=0.6}]
        \draw(0,0) to (1,0);
        \draw[markx={0},markx={1}] (0,0) to (0,-1);
        \draw (0,-1) to (1,-1);
        \draw (1,-1) to (2,0);
        \draw (1,0) to (2,-1);
        \draw (-0.25,0.25) to (0,0);
        \draw (0,-1) to (-0.25,-1.25);
        \draw (2,0) to (2.25,0.25);
        \draw[markx={0},markx={1}]  (1,0) to (2,0);
        \draw[markx={0},markx={1}]  (1,-1) to (2,-1);
        \draw (2,-1) to (2.25,-1.25);
        \draw (-0.5,0.25) node {$\lambda^1$};
        \draw (-0.5,-1.25) node {$\lambda^2$};
        \draw (-0.25,-0.5) node {$\mu^2$};
        \draw (0.5,0.25) node {$\mu^1$};
        \draw (0.5,-1.25) node {$\mu^3$};
        \draw (1.5,0.25) node {$\mu^7$};
        \draw (1,-0.25) node {$\mu^6$};
        \draw (2,-0.25) node {$\mu^5$};
          \draw (1.5,-1.25) node {$\mu^4$};
        \draw (2.5,0.5) node {$\lambda^3$};
          \draw (2.5,-1.25) node {$\lambda^4$};
    \end{tikzpicture},  
\end{equation*}
When using $n=9$ and $\bm{\lambda}=3333$, there are $3074$ possible sets of inner partitions; from those, a minimal set from which can be built the base of the Kronecker subspace is defined by $39$ sets that we will label $\bm{\mu}^i$ with $i \in \{1,\dots,39\}$:
\begin{equation}
\begin{gathered}
\bm{\mu}^1= 0300033 , \quad  \bm{\mu}^2= 0310133, \quad \bm{\mu}^3= 0320233, \quad \bm{\mu}^4= 0321122, \\
\bm{\mu}^5= 0330333 , \quad \bm{\mu}^6= 0331222, \quad \bm{\mu}^7= 1210132, \quad \bm{\mu}^8= 1210133, \\
\bm{\mu}^9= 1210134, \quad \bm{\mu}^{10}= 1220232, \quad \bm{\mu}^{11}= 1220233, \quad \bm{\mu}^{12}= 1220234, \\
\bm{\mu}^{13}= 1221122, \quad \bm{\mu}^{14}=1230332 , \quad \bm{\mu}^{15}= 1230333, \quad
\bm{\mu}^{16}=1231221 , \\
\bm{\mu}^{17}=1231222, \quad \bm{\mu}^{18}= 1231223, \quad 
\bm{\mu}^{19}=1231243, \quad 
\bm{\mu}^{20}=1240432, \\
\bm{\mu}^{21}=1241322, \quad \bm{\mu}^{22}= 1241323, \quad \bm{\mu}^{23}=1241332, \quad 
\bm{\mu}^{24}=2120231, \\
\bm{\mu}^{25}=2120232 , \quad \bm{\mu}^{26}=2120233, \quad 
\bm{\mu}^{27}=2120234,  \quad
\bm{\mu}^{28}= 2121122, \\
\bm{\mu}^{29}= 2121123, \quad \bm{\mu}^{30}= 2130331, \quad 
\bm{\mu}^{31}=2130332, \quad 
\bm{\mu}^{32}=2130333, \\
\bm{\mu}^{33}=2131221, \quad
\bm{\mu}^{34}= 2131222, \quad \bm{\mu}^{35}= 2140431, \quad 
\bm{\mu}^{36}= 2140432, \\\
\bm{\mu}^{37}=2141321 , \quad
\bm{\mu}^{38}=3030330 , \quad \bm{\mu}^{39}=3030331.
 \end{gathered},
 \label{eq:mus}
\end{equation}
We can apply the Gram-Schmidt process to the corresponding $39$ graph Kronecker states to find an orthogonal base of Kronecker states labeled as $\ket{\mathcal{K}_{\bm{\lambda},s}}$ with $s\in\{1,\dots,39\}$. The Figure \ref{fig:Grand} is the graphical representation of the matrix of  inner products between the graph Kronecker states $\ket{\mathcal{K}^{\bm{\mu}^i,\Cam{0.2}}_{\bm{\lambda}}}$ ( in x axis) with the set of orthogonal Kronecker states $\ket{\mathcal{K}_{\bm{\lambda},s}}$ (in y axis).
It can be noticed that in general the graph states are not linearly independent, and the Grand Schmidt process is not trivial. Calculating this basis of $39$ Kronecker states, where most of them have more than 2.5 million coefficients, took a couple of hours, showing the efficiency of the approach. It is also noteworthy from the previous matrix that gray spaces under the diagonal line, represent orthogonality between graph states and the built basis elements of the Kronecker subspace. Then, even in complicated graphs like this, it is possible to obtain  directly orthogonal graph-Kronecker states. Finding all the orthogonality conditions for the graph-Kronecker states is an interesting open problem of the approach. Having such conditions will make this approach even more efficient. For example, for the previous example, we had to look at $2254$ graph-Kronecker states to define completely the basis. \\

\noindent The software developed to calculate the W-Kronecker states, the graph Kronecker states, and the orthogonal Kronecker states can be found in the following link: \\

 \url{https://github.com/waltherlgo/Kronecker-states}  .\\
 
\noindent We expect to soon have a repository with the Clebsh-Gordan coefficients of the symmetric group that can be obtained from this construction.\\

\subsubsection{Five parts case}
We finish this chapter by exploring the case in five parts. By Theorem \ref{TheoremI}, we know that any Kronecker subspace can be obtained from a graph that generates all the SLOCC stable orbits of five qubits. Nevertheless, there is so far no known classification in this case, so there are no criteria to find a sufficient graph firsthand.\\

\noindent A good first approach to this problem is to calculate the polynomial invariants in each graph because it is known that a state in generic families cannot have null invariants \cite{LuqueThibon-tame}. Nevertheless, we will need a way to calculate such invariants and, as far as we know, there are no closed expressions for them; only the structure of the algebra of invariants is described in \cite{5qubits-Thibon}. It is stated that such invariants could be generated by a set of five polynomials of degree $4$, one of degree $6$, five of degree $8$, one of degree $10$, and five of degree $12$, and in the same paper a process is described to find a possible set for those of degree $4$ and $6$, however, due to the complexity of the constructions, the authors stop there.  \\

\noindent In order to ensure that a state has no null invariants, we need to calculate all of the generators. As they define a ring, we must check that they are algebraically independent so that no algebraic function of the invariants is null generically. It turns out that, as a secondary result of this research, we found a practical way to find polynomial invariants from graphical constructions. In Section \ref{Invariants}, we show how it is possible to calculate a set of polynomial invariants according to the structure of generators for the case of five qubits. Moreover, we found that those invariants are algebraically independent and can be used at least to discard non-generic states. \\

\noindent We start by noting that $17$ polynomial invariants define the set of primary generators of the algebra of invariants of five qubits, so if we want all of them to be independent for a given graph, the required parameters of the ``cleaned'' graph must be at least $16$, that along with a scale factor, complete the $17$ parameters. This relation was already seen for the case of three qubits, where four polynomials define the algebra of invariants, and the graph for the generic state after cleaning retains three parameters. Then, when considering a scale factor $s$, any generic state can be reproduced, as shown in the last part of the previous chapter. By requiring that the graph has more than $16$ parameters after cleaning, we found that the generic graph must have at least $15$ vertices. \\

\begin{figure}
\includegraphics[scale=0.4]{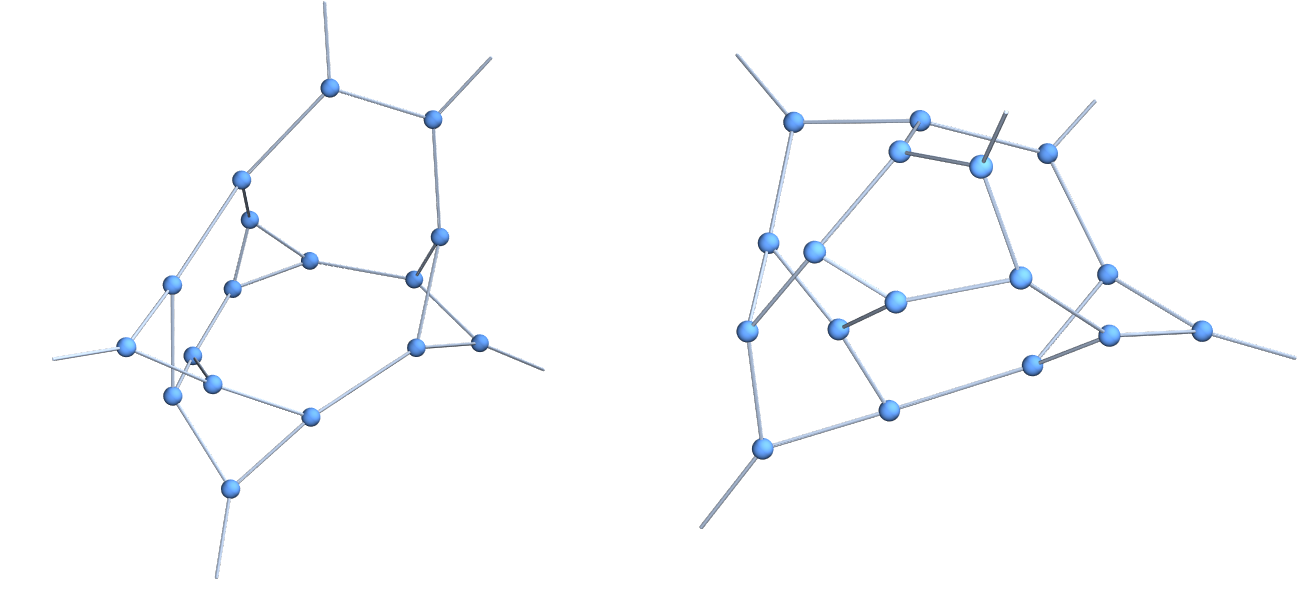}
\caption{Three-dimensional representation of two graphs with $17$ vertices. When building multiqubit states from these graphs, the set of $17$ polynomial invariants presented in Section \ref{Invariants} are algebraically independent, making them good candidates for building Kronecker subspaces of five parts and generic states of five qubits.}
\label{fig:G17}
\end{figure}

\noindent Next, we used the invariants shown in Section \ref{Invariants}, and by calculating the Jacobian matrix, we found that from the more than $60,000$ possible graphs with $15$ vertices, no one has a set of $17$ independent invariants. We attempted to repeat the process with graphs of $17$ vertices, but we rapidly found that the set of $17$ invariants is algebraically independent for many of them. Some of those graphs are presented in Figure \ref{fig:G17} in a three-dimensional representation because the complexity of the graphs makes the two-dimensional representation confusing. It is reasonable to think that Kronecker subspaces of five parts can be built from these graphs as a parametrization for generic five qubit states. \\

\noindent In this chapter, we presented the main results of our research, achieving the goal of a structural method to calculate Multipartite Locally Maximally Entangled states by building the invariant subspace of the tensor product of irreps of the symmetric group $S_n$. For this, we replicate the method of W-state stitching with W-Kronecker states, showing that by doing so, it is possible to overcome the limitations inherent to the W class in the same way that W-states can be used to build generic states. We also show how the connection between both constructions gives a criterion for building any $N$-partite Kronecker subspace by using a graph that generates densely the Hilbert space of $N$ qubits. We expect that these results will allow us to exploit the properties of Kronecker states in quantum information applications and, through the connection with Clebsch Gordan coefficients of the symmetric group $S_n$, also make progress in other research areas such as quantum chemistry and nuclear physics where they are used. The next chapter shows some secondary results that arose in the course of this investigation.
\chapter{Other results}
\label{Chapter6}
Throughout our research, our main motivation was explicitly constructing Kronecker subspaces in multipartite systems. However, during this investigation, we uncovered several other intriguing findings that, while not essential for Kronecker subspace construction, hold significant relevance on their own. In this chapter, we present these noteworthy results.\\

\noindent Firstly, we introduce a novel method for calculating irreducible representations of $S_n$. To the best of our knowledge, this method has not been reported previously. We achieve this by using the well-known Clebsch Gordan Coefficients of $GL_2$ along with Schur-Weyl Duality.\\

\noindent Additionally, we present a graphical framework that helps comprehend SLOCC invariants. This framework offers an intuitive tool for naturally defining invariant polynomials in multipartite scenarios; when combined with the rules outlined in Section \ref{rules}, this graphical approach enables the manual computation of invariants. This construction showcases the efficacy of our approach.\\

\noindent Finally, we show an interesting result on the asymptotic behavior of Kronecker states associated with the multipartite W-SLOCC class, which shows how, for this particular class,   we can achieve a generalization of the  Keyl Werner theorem in subsection \ref{Keyl-WernerTheorem} can be achieved.
\section{Building $S_n$ irreps from Schur-Weyl duality}
\label{myirreps}

We know that Schur-Weyl duality connects the irreducible representations of the general linear group $GL_d$ and the symmetric group $S_n$ with copies of Hilbert spaces in quantum systems. In particular, Schur-Weyl duality says that it is possible, by using the Schur transform, to block-diagonalize the actions of both groups on the Hilbert space into their irreducible representations. In Section  \ref{Schurtransform}, we showed how the Schur transform can be performed on qubit systems only by using the Clebsch Gordan coefficients of $SL_2$, which we showed in Equation \eqref{eq:Gammas}. In the same section, we showed that when applying the Schur transform on elements of $S_n$ acting on the computational basis, such representation can be decomposed in the different irreps $[\lambda]$ of $S_n$, each with a multiplicity given by the dimension of the corresponding irrep $\{\lambda\}$ in $GL_2$. This diagonal decomposition already gives shape to the irreps of $S_n$ without any other algorithm; then, the Schur transform can be used to build any representation of $S_n$ in irreps of at most two rows. \\

\noindent We will show with an example how this construction is achieved. Let us calculate the representation of the permutation $(23)$ in the irrep $[1]$ in $n=3$ (or $[21]$ in the long notation). First, we write the matrix corresponding to this permutation acting on the computational basis of three qubits. This matrix is obtained by applying the permutation to all the basis elements, which gives the map represented by the permutation. In this case, we have the matrix
\begin{equation*}
    (23)= \left( \begin{array}{cccccccc}
     1    & & & & & & & \\
       & & 1& & & & & \\
          &1 & & & & & &\\
              & & & 1& & & &\\
                 & & & &1 & & &\\
                     & & & & & &1 &\\
                         & & & & & 1& &\\
                             & & & & & & &1
    \end{array}\right)
\end{equation*}
which can be seen more easily as 
\begin{equation*}
\begin{gathered}
   (23)=\ket{000}\bra{000}+\ket{001}\bra{010}+\ket{010}\bra{001}+\ket{011}\bra{011} \\
 +  \ket{100}\bra{100}+\ket{101}\bra{110}+\ket{110}\bra{101}+\ket{111}\bra{111} .
\end{gathered}
\end{equation*}
We want to compute the matrix entries that correspond to 
\begin{equation*}
  D^{[1]}(23)_{ij}=  \bra{1,q^{[1]}_i}(23)\ket{1,q^{[1]}_j},
\end{equation*}
where the basis elements of irrep $[1]$ can be represented by the modified Yamanouchi symbols
\begin{equation*}
q^{[1]}_1=\{001 \}, \qquad 
q^{[1]}_2=\{010 \}.
\end{equation*}
Note that from the Schur transform, we know how to calculate the change of basis to the Schur basis $\ket{\lambda,\omega,q}$, that simultaneously decomposes the computational basis into the irreps of $GL_2$ and $S_n$, and what we want is to calculate the change to the basis of irreps of $S_n$, $\ket{\lambda,q}$. We can then fix one value for $\omega$ in the Schur transform, and the obtained representation corresponds to one of the copies of the irreps of $[\lambda]$ in the Schur basis. This is the case because each value of $\omega$ only represents a multiplicity of the same irrep in $S_n$. In this example, by picking $\omega=1$ (or equivalently, any other value for $\omega$ such that $\lambda\leq\omega \leq n-\lambda$), we have the following expression for the matrix entries:
\begin{equation*}
\begin{gathered}
    D^{[1]}(23)_{ij}=  \bra{1,1,q^{[1]}_i}(23)\ket{1,1,q^{[1]}_j}\\
    =\bra{1,1,q^{[1]}_i}\left(\ket{000}\bra{000}+\ket{001}\bra{010}+\ket{010}\bra{001}+\ket{011}\bra{011} \right. \\
 \left. + \ket{100}\bra{100}+\ket{101}\bra{110}+\ket{110}\bra{101}+\ket{111}\bra{111} \right)\ket{1,1,q^{[1]}_j}\\
\end{gathered}.
\end{equation*}
where we just replaced the matrix entries of $(23)$ in the computational basis representation. Note that the value of $\omega=1$ makes all the elements in the computational basis with weight different from one to be canceled. Then, we are left with:
\begin{equation*}
    D^{[1]}(23)_{ij}= \bra{1,1,q^{[1]}_i}\left(\ket{001}\bra{010}+\ket{010}\bra{001}+ \ket{100}\bra{100} \right)\ket{1,1,q^{[1]}_j}.
    \end{equation*}
 Now, let us fix the element of the irrep matrix we want to find, for example, the one corresponding to $q_1^{[1]}q^{[1]}_1$. Then, expanding the previous expression, we get:
\begin{equation*}
    \begin{gathered}
     D^{[1]}(23)_{11}=\bra{1,1,\{0,0,1\}}\ket{001}\bra{010}\ket{1,1,\{0,0,1\}}+\bra{1,1,\{0,0,1\}}\ket{010}\bra{001}\ket{1,1,\{0,0,1\}}\\+ \bra{1,1,\{0,0,1\}}\ket{100}\bra{100} \ket{1,1,\{0,0,1\}}.
     \end{gathered}
\end{equation*}
Now it is clear that each inner product in the last equation is a CGC of the Schur transform as defined in Equations \eqref{eq:Schurtransfsequence}, then we have:
\begin{equation*}    D^{[1]}(23)_{11}=\bm{\Gamma}_{001}^{1,\{001\}}  \bm{\Gamma}_{010}^{1,\{001\}}+\bm{\Gamma}_{010}^{1,\{001\}}\bm{\Gamma}_{001}^{1,\{001\}} +\bm{\Gamma}_{100}^{1,\{001\}}\bm{\Gamma}_{100}^{1,\{001\}}    
\end{equation*}
These values can be obtained from Equation \eqref{eq:Gammas}. Replacing the values from the table and repeating for all the entries of the irrep matrix, we can obtain the following:
\begin{equation*}
\begin{gathered}
D^{[1]}(23)_{11}= \frac{-1}{2}, \qquad  D^{[1]}(23)_{12}= \frac{\sqrt{3}}{2}, \\
D^{[1]}(23)_{21}= \frac{\sqrt{3}}{2}, \qquad D^{[1]}(23)_{22}= \frac{1}{2}.
\end{gathered}
\end{equation*}
With these values, we can build the irrep matrix to be:
\begin{equation*}
    D^{[1]}(23)=\left( \begin{array}{cc}
       - \frac{1}{2}  & \frac{\sqrt{3}}{2} \\
         \frac{\sqrt{3}}{2} & \frac{1}{2}
     \end{array}\right).
\end{equation*}
We used the Young-Yamanouchi algorithm to obtain the same irrep matrix in Equation (\ref{eq:irep2321}). In a general case, this process can be summarized as follows:
\begin{equation*}
    D^{[\lambda]}(\pi)_{i,j}= \sum_{s\in \omega}    \bra{\lambda,\omega,q_i} \ket{s} \bra{\pi s}  \ket{\lambda,\omega,q_j} = \sum_{s\in \omega} \bm{\Gamma}_s^{\lambda,q_i}\bm{\Gamma}_{\pi s}^{\lambda,q_i},  \label{eq:myirreps}
\end{equation*}
where one chooses one possible weight $\lambda\leq \omega \leq n-\lambda$, and the sum is over sequences $s$ in the computational basis with weight equal to $\omega$.\\

\noindent This method allows us to compute directly the matrix for any permutation, not only adjacent transpositions. This difference with the Young-Yamanouchi algorithm is useful when finding specific matrices that do not correspond to adjacent transpositions. In Appendix  \ref{AppendixI}, we show how this construction facilitates the computation of states and matrices in the Schur transform for certain quantum systems.

\section{Invariants}
\label{Invariants}
As discussed in Chapter \ref{Chapter2-5}, the SLOCC invariants are the first filter for classifying the SLOCC classes; however, their construction has been proposed by following a mathematical procedure known as the Omega process \cite{Covariants}\cite{Osterloh}, which requires a basic understanding of covariant theory and are hard to compute in general. In this section, we will show a natural construction for invariants obtained from the graphical point of view, which can be used along with the pushing and stitching rules in \ref{rules} to simplify the calculation of invariants by hand. First, note that the matrix obtained from the $\bullet$ and $Z$-ball corresponds to the two-dimensional \textit{Levi-Civita} tensor:
\begin{equation*}
 \begin{tikzpicture}[basel={-.5},every node/.style={scale=0.6}]
        \draw[markxo={0.5}{0.6}](0,0) to (1,0);
    \end{tikzpicture} = \left(\begin{array}{cc}
 0  & 1 \\
 1 & 0   
\end{array}     \right)\left(\begin{array}{cc}
 1  &0 \\
 0 & -1   
\end{array}     \right) =\left(\begin{array}{cc}
 0  & 1 \\
 -1 & 0   
\end{array}     \right) = \varepsilon_{ij}.
\end{equation*}
Moreover, this object works as an inverter of unit-determinant matrices because $B\varepsilon_{ij}B^T=\varepsilon_{ij}$ with $B\in SL_2$. This process corresponds graphically to moving matrices through the $\varepsilon$ object as:
\begin{equation*}
    \begin{tikzpicture}[basel={-.5},,every node/.style={scale=0.6}]
        \draw[markxo={0.5}{0.6}](0,0) to (1,0);
        \draw[\rac] (1.5,0) to (1,0);
        \draw[\rao] (1.5,0) to (2,0);
        \draw (1,0.3) node {$v_1,w_1$};
        \draw (2,0.3) node {$w_2$};
    \end{tikzpicture} = \begin{tikzpicture}[basel={-.5},,every node/.style={scale=0.6}]
        \draw[markxo={0.5}{0.6}](0,0) to (1,0);
        \draw[\rao] (-0.5,0) to (-1,0);
        \draw[\rac] (-0.5,0) to (0,0);
        \draw (-1,0.3) node {$-w_2$};
        \draw (0,0.3) node {$1/v_1,-w_1$};
    \end{tikzpicture} .
\end{equation*}
Therefore, when the same unit-determinant matrix acts on both sides of the Levi-Civita tensor, we can eliminate both by the previous property:
\begin{equation}
    \begin{tikzpicture}[basel={-1},,every node/.style={scale=0.6}]
    \draw[\rao] (-0.5,0) to (-1,0);
        \draw[\rac] (-0.5,0) to (0,0);
        \draw (-1,0.3) node {$w_2$};
        \draw (0,0.3) node {$(v_1,w_1)$};
        \draw[markxo={0.5}{0.6}](0,0) to (1,0);
        \draw[\rac] (1.5,0) to (1,0);
        \draw[\rao] (1.5,0) to (2,0);
        \draw (1,0.3) node {$(v_1,w_1)$};
        \draw (2,0.3) node {$w_2$};
    \end{tikzpicture} = \begin{tikzpicture}[basel={-.5},,every node/.style={scale=0.6}]
        \draw[markxo={0.5}{0.6}](0,0) to (1,0);
    \end{tikzpicture}.
    \label{eq:XZprop}
\end{equation}
Another useful property is that the Z-ball anti-commutes with $\bullet$:
\begin{equation}
\begin{tikzpicture}[basel={-.5},,every node/.style={scale=0.6}]
        \draw[markxo={0.5}{0.6}](0,0) to (1,0);
    \end{tikzpicture}= -\begin{tikzpicture}[basel={-.5},,every node/.style={scale=0.6}]
        \draw[markxo={0.5}{0.2}](0,0) to (1,0);
    \end{tikzpicture}.
    \label{eq:XZcommute}
\end{equation}
We can exploit these properties to build SLOCC invariants by stitching copies of any qubit state in a way that we always connect the same parts from different copies in each contraction with $\varepsilon$, with no free qubits (external edges) at the end. For example, consider the simplest case of a two-qubit state. We will use the notation for generic graph states, in this case of two qubits:
\begin{equation*}
    \begin{tikzpicture}[basel={-1},every node/.style={scale=0.6}]
    \draw (0,0) to (1,0);
    \Genstate{(0.5,0)};
    \end{tikzpicture}.
\end{equation*}
We can then build an invariant by stitching together two copies of the same state, contracting the first qubit of the first copy, with the first qubit of the second copy, and the same with the second qubit:
\begin{equation}
    \begin{tikzpicture}[basel={-1},,every node/.style={scale=0.6}]
    \draw (0.5,-0.5) to (0.5,0.5);
    \Genstate{(0.5,0)};
    \draw (2,0.5) to (2,-0.5);
    \Genstate{(2,0)};
    \draw[markxo={0.5}{0.6}] (0.5,0.5) to (2,0.5);
    \draw[markxo={0.5}{0.6}] (0.5,-0.5) to (2,-0.5);
    \draw (0.5,0.7) node {$(1)$};
        \draw (0.5,-0.7) node {$(2)$};
    \end{tikzpicture}.
    \label{eq:Inv2}
\end{equation}
where the $(1)$ and $(2)$ label the qubit associated to each stitch. The quantity represented is invariant under SLOCC equivalence, i.e., the local actions of matrices in $SL_2$ on each of the external edges. Any state in the same SLOCC class as the state represented by $\scalebox{0.4}{\tikz{\draw (0,0) to (1,0);
    \Genstate{(0.5,0)};}}$ can be written up to a global factor as some unit-determinant matrices action on qubits one and two:
\begin{equation*}
    \begin{tikzpicture}[basel={-1},,every node/.style={scale=0.6}]
    \draw[\lao \rac] (-0.5,0) to (0.3,0);
    \Genstate{(0.5,0)};
     \draw[\lao \rac] (0.7,0) to (1.5,0);
    \end{tikzpicture}.
\end{equation*}
Then, the construction for this state is:
 \begin{equation*}
    \begin{tikzpicture}[basel={-1},,every node/.style={scale=0.6}]
    \draw[\lao \rac] (0.5,0.2) to (0.5,1);
    \Genstate{(0.5,0)};
     \draw[\lao \rac] (0.5,-0.2) to (0.5,-1);
     \draw[\lao \rac] (2,0.2) to (2,1);
    \Genstate{(2,0)};
     \draw[\lao \rac] (2,-0.2) to (2,-1); 
     \draw[markxo={0.5}{0.6}](0.5,1) to (2,1);
      \draw[markxo={0.5}{0.6}](0.5,-1) to (2,-1); \draw (0.2,0.7) node {$(1)$};
        \draw (0.2,-0.7) node {$(2)$};
    \end{tikzpicture} =
    \begin{tikzpicture}[basel={-1},every node/.style={scale=0.6}]
    \draw (0.5,-0.5) to (0.5,0.5);
    \Genstate{(0.5,0)};
    \draw (2,0.5) to (2,-0.5);
    \Genstate{(2,0)};
    \draw[markxo={0.5}{0.6}] (0.5,0.5) to (2,0.5);
    \draw[markxo={0.5}{0.6}] (0.5,-0.5) to (2,-0.5);
    \draw (0.5,0.7) node {$(1)$};
        \draw (0.5,-0.7) node {$(2)$};
    \end{tikzpicture},
\end{equation*}
where we applied the property in Equation \eqref{eq:XZprop}. As it is the same for any state in the same SLOCC orbit, it is an SLOCC invariant. Furthermore, the resultant quantity is always a polynomial of degree equal to the number of copies of the coefficients of the state. Then, it is, by definition, a SLOCC polynomial invariant.
\subsection*{Invariant for two qubits}
Now we show how the rules in Section \ref{rules} allows to calculate the invariants in some constructions. We also show how the invariant allows us to differentiate between entangled and separable states in the case of two qubits. Let us consider first the separable states, using $\ket{\psi}=\ket{11}$, which is represented by the graph $\scalebox{0.5}{\tikz{\draw(-0.5,0.2) to(0,0.2);\qubitl{(0,0.2)};\draw(-0.5,-0.2) to(0,-0.2);\qubitl{(0,-0.2)}}}$, building the graph representation in \eqref{eq:Inv2}, we have:

\begin{equation*}
    \begin{tikzpicture}[basel={-.5}]
    \qubitr{(-0.4,0)};
     \qubitl{(1,0)};
    \draw[markxo={0.5}{0.6}] (0,0) to (1,0); \qubitr{(-0.4,0.5)};
     \qubitl{(1,0.5)};
    \draw[markxo={0.5}{0.6}] (0,0.5) to (1,0.5);
    \end{tikzpicture} \stackrel{vii.1}{=}0.
\end{equation*}
This invariant is zero because  it corresponds to the operation $(\bra{1} XZ \ket{1})^2=0$. Instead, for an entangled state, for example $\ket	{\phi^+}$,  we have:
\begin{equation*}
    \begin{tikzpicture}[basel={-.5}]
   \draw (0,0) to[bend left=50] (0,0.5); 
    \draw[markxo={0.5}{0.6}] (0,0) to (0.5,0); \draw (0.5,0) to[bend right=50] (0.5,0.5);
    \draw[markxo={0.5}{0.6}] (0,0.5) to (0.5,0.5);
    \end{tikzpicture} \stackrel{iv,iii.3}{=}\begin{tikzpicture}[basel={-.5}]
   \draw (0,0) to[bend left=50] (0,0.5); 
    \draw (0,0) to (0.5,0); \draw (0.5,0) to[bend right=50] (0.5,0.5);
    \draw (0,0.5) to (0.5,0.5);
    \end{tikzpicture} =2.
\end{equation*}
Here, we slid the Z-balls, multiplied them, and repeated the same with $\bullet$s. For the entangled state, the value is not zero as expected and is equal to 2 because it is equivalent to calculating $(\bra{00}+\bra{11}) (\ket{00}+\ket{11})=2$. There is one graph construction where both classes can be obtained depending on one parameter:
\begin{equation*}
    \begin{tikzpicture}[basel={-.5}]
    \draw (0,0) to[bend left=45] (1,0);
          \draw[markx={0},markx={1}] (0,0) to[bend right=45] (1,0);
          \draw (-0.25,0) to (0,0);
           \draw (1.25,0) to (1,0);
           \ball{(0.5,0.2)};
    \end{tikzpicture},
\end{equation*}
which leads to an entangled state when the ball is not $Z$ and a separable state for the $Z-$ball as it is seen in rule (xi.1). When calculating the invariant for a generic parameter of the ball, we have the following:
\begin{equation}   
 \begin{tikzpicture}[basel={-1},every node/.style={scale=0.6}]
    \draw (0.5,0.5) to (0.5,0.5);
   \draw (0.5,-0.5) to [bend left=45] (0.5,0.5);
   \draw[markx={1},markx={0}] (0.5,-0.5) to [bend right=45] (0.5,0.5);
   \draw [markx={1},markx={0}] (2,-0.5) to [bend left=45] (2,0.5);
   \draw (2,-0.5) to [bend right=45] (2,0.5);
    \draw[markxo={0.5}{0.6}] (0.5,0.5) to (2,0.5);
    \draw[markxo={0.5}{0.6}] (0.5,-0.5) to (2,-0.5);
        \ball {(0.3,0)};
         \ball {(2.2,0)};
         \draw (0,0) node {$v$};
    \draw (2.5,0) node {$v$};     
\end{tikzpicture}
\label{eq:twoinv}
 \end{equation}
 To calculate this invariant, some properties from qubit graph construction are very useful. First, note the following equivalence:
 \begin{equation}   
 \begin{tikzpicture}[basel={-1},every node/.style={scale=0.6}]
    \draw (0.5,0.5) to (0.5,0.5);
   \draw[markx={1},markx={0}] (0.5,-0.5) to (0.5,0.5);
   \draw [markx={1},markx={0}] (2,-0.5) to  (2,0.5);
    \draw[markx={0.5}] (0.5,0.5) to (2,0.5);
    \draw[markx={0.5}] (0.5,-0.5) to (2,-0.5);
         \draw (0.25,0.75) to (0.5,0.5);
         \draw (0.25,-0.75) to (0.5,-0.5);
         \draw (2.25,0.75) to (2,0.5);
         \draw (2.25,-0.75) to (2,-0.5);
\end{tikzpicture} =  \begin{tikzpicture}[basel={-1},every node/.style={scale=0.6}]
         \draw (0.25,0) to (0.5,0);
         \draw (0.75,0) to (0.5,0);
         \draw[markx={0},markx={0.5},markx={1}](0.5,0) to (0.5,-0.5);
         \draw (0.5,-0.5) to (0.25,-0.75);         
         \draw (0.5,-0.5) to (0.75,-0.75);
         \draw (0.5,-1) to (0.25,-0.75);         
         \draw (0.5,-1) to (0.75,-0.75);
         \draw[markx={0},markx={0.5},markx={1}](0.5,-1) to (0.5,-1.5);
         \draw (0.25,-1.5) to (0.5,-1.5);
         \draw (0.75,-1.5) to (0.5,-1.5); 
\end{tikzpicture} = \sqrt{2} \begin{tikzpicture}[basel={-1},every node/.style={scale=0.6}]
         \draw (0.25,0) to (0.5,0);
         \draw (0.75,0) to (0.5,0);
         \draw[markx={0},markx={0.5}](0.5,0) to (0.5,-0.5); \draw (0.5,-0.5) to (0.5,-1);
         \ball{(0.5,-0.75)};
         \draw {(0.8,-0.75)} node{$\sqrt{2}$};
         \draw[markx={0.5},markx={1}](0.5,-1) to (0.5,-1.5);
         \draw (0.25,-1.5) to (0.5,-1.5);
         \draw (0.75,-1.5) to (0.5,-1.5); 
\end{tikzpicture} = \sqrt{2} \begin{tikzpicture}[basel={-1},every node/.style={scale=0.6}]
         \draw (0.25,0) to (0.5,0);
         \draw (0.75,0) to (0.5,0);
         \draw[markx={0}](0.5,0) to (0.5,-0.5); \draw (0.5,-0.5) to (0.5,-1);
         \ball{(0.5,-0.75)};
         \draw {(0.8,-0.75)} node{$\frac{1}{\sqrt{2}}$};
         \draw[markx={1}](0.5,-1) to (0.5,-1.5);
         \draw (0.25,-1.5) to (0.5,-1.5);
         \draw (0.75,-1.5) to (0.5,-1.5); 
\end{tikzpicture} .
\label{eq:Squarerule}
 \end{equation}
 In the first equality, we used the rotation of W states in the upper and lower lines, as shown in rule (ix). In the next step, we used the equivalence from the rule (xi.1) with $v=1$. In the last step, we slid the ball through one of the $\bullet$ and then multiplied the $\bullet$s together (rules (iii) and (iv)). Another recurrent object in the invariant calculation is:
    \begin{equation*}
    \begin{tikzpicture}[basel={-1},every node/.style={scale=0.6}]
         \draw[markx={0},markx={1}] (0,0) to (0,-1);
         \draw (0,0) to[bend left=90,distance=1cm] (0,-1);
         \draw (0,0) to[bend right=90,distance=1cm] (0,-1);
         \ball{(0,-0.5)};
         \draw (0.4,-0.5) node {$v$};
    \end{tikzpicture} = \begin{tikzpicture}[basel={-1},every node/.style={scale=0.6}]
         \draw[markx={0},markx={1}] (0,0) to (0,-1);
         \draw (0,0) to[bend left=90,distance=1cm] (0,-1);
         \draw (-0.3,0) to[bend right=90,distance=1cm] (-0.3,-1);
         \ball{(0,-0.5)};
        \draw (-0.1,0) to (0,0);
        \draw (-0.1,-1) to (0,-1); \draw (0.4,-0.5) node {$v$};
    \end{tikzpicture} = \frac{1+v^2}{v} + v =\frac{1+2v^2}{v},
    \end{equation*}
 where the expression obtained in Appendix \ref{Appendix03} in Equation \eqref{eq:twoqubitsball} was used. With these simplifications, we can calculate the invariant of Equation \eqref{eq:twoinv} as:
    \begin{equation*}   
 \begin{tikzpicture}[basel={-1},every node/.style={scale=0.6}]
    \draw (0.5,0.5) to (0.5,0.5);
   \draw (0.5,-0.5) to [bend left=45] (0.5,0.5);
   \draw[markx={1},markx={0}] (0.5,-0.5) to [bend right=45] (0.5,0.5);
   \draw [markx={1},markx={0}] (2,-0.5) to [bend left=45] (2,0.5);
   \draw (2,-0.5) to [bend right=45] (2,0.5);
    \draw[markxo={0.5}{0.6}] (0.5,0.5) to (2,0.5);
    \draw[markxo={0.5}{0.6}] (0.5,-0.5) to (2,-0.5);
        \ball {(0.3,0)};
         \ball {(2.2,0)};
         \draw (0,0) node {$v$};
    \draw (2.5,0) node {$v$};     
\end{tikzpicture}
 =
    \sqrt{2} \cdot \begin{tikzpicture}[basel={-1},every node/.style={scale=0.6}]
         \draw[markx={0},markx={1}] (0,0) to (0,-1);
         \draw (0,0) to[bend left=90,distance=2cm] (0,-1);
         \draw (0,0) to[bend right=90,distance=2cm] (0,-1);
         \ball{(0,-0.5)};
         \ball{(1.5,-0.5)};
         \ball{(-1.5,-0.5)};
         \draw (1.8,-0.5) node {$v$};
         \draw (-1.8,-0.5) node {$v$};
         \draw (0.4,-0.5) node {$\frac{1}{\sqrt{2}}$};
    \end{tikzpicture}= \sqrt{2}v \cdot \begin{tikzpicture}[basel={-1},every node/.style={scale=0.6}]
         \draw[markx={0},markx={1}] (0,0) to (0,-1);
         \draw (0,0) to[bend left=90,distance=1cm] (0,-1);
         \draw (0,0) to[bend right=90,distance=1cm] (0,-1);
         \ball{(0,-0.5)};
         \draw (0.4,-0.5) node {$\frac{1}{\sqrt{2}v}$};
    \end{tikzpicture} = (2+2v^2).
\end{equation*}
Note how the invariant is null for the parameter $v=i$, i.e., the Z-ball, which leads to a separable state, and not null for any other value, which leads to an entangled state as it is shown in rule (xi.1) . This simple example shows how the pushing process leaves only parameters that can be used to modify the SLOCC classes on the graph state. This construction is the only invariant for two qubits, and it can be checked that it is proportional to the concurrence in Equation \eqref{eq:concurrence}. 

\subsection*{Invariant for three qubits}
We can use the same procedure to build an invariant for three qubits. We may be tempted to build an invariant from two copies:
\begin{equation*}
     \begin{tikzpicture}[basel={-.5}] 
    \draw[markxo={0.5}{0.6}] (0,0)  to (2,0);
    \draw[markxo={0.5}{0.6}] (0,0)  to[bend right=50] (2,0);
    \draw[markxo={0.5}{0.6}] (0,0)  to[bend left =50 ] (2,0);
    \Genstate{(0,0)};
   \Genstate{(2,0)};
    \end{tikzpicture};
    \end{equation*}
    however, note how by the anticommutation in Equation \eqref{eq:XZcommute} we have that:
    \begin{equation*}
     \begin{tikzpicture}[basel={-.5}] 
    \draw[markxo={0.5}{0.6}] (0,0)  to (2,0);
    \draw[markxo={0.5}{0.6}] (0,0)  to[bend right=50] (2,0);
    \draw[markxo={0.5}{0.6}] (0,0)  to[bend left =50 ] (2,0);
  \Genstate{(0,0)};
   \Genstate{(2,0)};
    \end{tikzpicture} =-
     \begin{tikzpicture}[basel={-.5}] 
    \draw[markxo={0.5}{0.4}] (0,0)  to (2,0);
    \draw[markxo={0.5}{0.4}] (0,0)  to[bend right=50] (2,0);
    \draw[markxo={0.5}{0.4}] (0,0)  to[bend left =50 ] (2,0);
    \Genstate{(0,0)};
   \Genstate{(2,0)};
    \end{tikzpicture}.
    \end{equation*}
   The graphs on both sides of the previous equation are equal because reordering the copies, or rotating the graph construction relates them. Then, the only way that the previous equation holds is that the invariant is zero:
    \begin{equation*}
     \begin{tikzpicture}[basel={-.5}] 
    \draw[markxo={0.5}{0.6}] (0,0)  to (2,0);
    \draw[markxo={0.5}{0.6}] (0,0)  to[bend right=50] (2,0);
    \draw[markxo={0.5}{0.6}] (0,0)  to[bend left =50 ] (2,0);
   \Genstate{(0,0)};
  \Genstate{(2,0)};
    \end{tikzpicture} =-
     \begin{tikzpicture}[basel={-.5}] 
    \draw[markxo={0.5}{0.4}] (0,0)  to (2,0);
    \draw[markxo={0.5}{0.4}] (0,0)  to[bend right=50] (2,0);
    \draw[markxo={0.5}{0.4}] (0,0)  to[bend left =50 ] (2,0);
   \Genstate{(0,0)};
  \Genstate{(2,0)};
    \end{tikzpicture}= -
     \begin{tikzpicture}[basel={-.5}] 
    \draw[markxo={0.5}{0.6}] (0,0)  to (2,0);
    \draw[markxo={0.5}{0.6}] (0,0)  to[bend right=50] (2,0);
    \draw[markxo={0.5}{0.6}] (0,0)  to[bend left =50 ] (2,0);
    \Genstate{(0,0)};
  \Genstate{(2,0)};
    \end{tikzpicture} =0.
    \end{equation*}
    To build a non-zero invariant, we need to use more copies. With three copies, it is impossible to contract all the parts. Hence, the next option to consider is four copies. We can build the following invariant with four copies:
    \begin{equation*}
        \begin{tikzpicture}[basel={-.5},every node/.style={scale=0.6}]
   \draw[markxo={0.5}{0.6}] (0,0) to  [bend right=50] (2,0);
   \draw[markxo={0.5}{0.6}] (0,0) to  [bend left=50] (2,0);
   \draw[markxo={0.5}{0.6}] (2,0) to (4,0);
     \draw[markxo={0.5}{0.6}] (4,0) to [bend right=50] (6,0);
      \draw[markxo={0.5}{0.6}] (4,0) to [bend left=50] (6,0);
      \draw[markxo={0.5}{0.6}](0,0) to[bend left=50] (6,0); 
    \Genstate{(0,0)};
    \Genstate{(2,0)};
     \Genstate{(4,0)};
      \Genstate{(6,0)};
      \draw (0.5,-0.5) node {$(1)$};
       \draw (5.5,-0.5) node {$(1)$};
       \draw (3,-0.2) node {$(3)$};
       \draw (3,1) node {$(3)$};
       \draw (1.8,0.5) node {$(2)$};
       \draw (4.2,0.5) node {$(2)$};
    \end{tikzpicture}.
   \label{threequbitinv}
    \end{equation*}
    For three-qubit states, it is known that the three-tangle can differentiate between the $W$ state and the $GHZ$ state. We can calculate this proposed invariant in both cases easily using the stitching and pushing rules. For the $W$ state, because it is symmetrical under exchanging the parts, we can remove the labels of the parts. When putting this state in the previous construction, we have:
     \begin{equation*}
        \begin{tikzpicture}[basel={-.5},every node/.style={scale=0.6}]
   \draw[markxo={0.5}{0.6}] (0,0) to  [bend right=50] (2,0);
   \draw[markxo={0.5}{0.6}] (0,0) to  [bend left=50] (2,0);
   \draw[markxo={0.5}{0.6},markx={0},markx={1}] (2,0) to (4,0);
     \draw[markxo={0.5}{0.6}] (4,0) to [bend right=50] (6,0);
      \draw[markxo={0.5}{0.6}] (4,0) to [bend left=50] (6,0);
      \draw[markxo={0.5}{0.6},markx={0}](0,0) to[bend left=50] (6,0);      \draw[ markxi={0}](6,0) to (5,0);
    \end{tikzpicture}= \begin{tikzpicture}[basel={-.5},every node/.style={scale=0.6}]
   \draw[markx={0.5}] (0,0) to  [bend right=50] (1,0);
   \draw[markx={0.5}] (0,0) to  [bend left=50] (1,0);
   \draw[markxo={0.5}{0.6},markx={0},markx={1},marko={0.3}] (1,0) to (2,0);
     \draw[markx={0.5}] (2,0) to [bend right=50] (3,0);
      \draw[markx={0.5}] (2,0) to [bend left=50] (3,0);
      \draw[markxo={0.5}{0.6},markx={0}, marko={0.7}](0,0) to[bend left=50] (3,0);      \draw[ markxi={0}](3,0) to (2,0);
    \end{tikzpicture} = -\begin{tikzpicture}[basel={-.5},every node/.style={scale=0.6}]
   \draw[markx={0.5}] (0,0) to  [bend right=50] (1,0);
   \draw[markx={0.5}] (0,0) to  [bend left=50] (1,0);
   \draw[markx={0.5},markx={0},markx={1}] (1,0) to (2,0);
     \draw[markx={0.5}] (2,0) to [bend right=50] (3,0);
      \draw[markx={0.5}] (2,0) to [bend left=50] (3,0);
      \draw[markx={0.5},markx={0}](0,0) to[bend left=50] (3,0);      \draw[ markxi={0}](3,0) to (2,0);
    \end{tikzpicture}.
    \end{equation*}
   In the previous equation, we slid, pushed, and multiplied the Z-balls, canceling between them. Now we can use the rule (x.3) from Section \ref{rules}, in both loops, with $v=1$ to get:
    \begin{equation*}
         -\begin{tikzpicture}[basel={-.5},every node/.style={scale=0.6}]
   \draw[markx={0.5}] (0,0) to  [bend right=50] (1,0);
   \draw[markx={0.5}] (0,0) to  [bend left=50] (1,0);
   \draw[markx={0.5},markx={0},markx={1}] (1,0) to (2,0);
     \draw[markx={0.5}] (2,0) to [bend right=50] (3,0);
      \draw[markx={0.5}] (2,0) to [bend left=50] (3,0);
      \draw[markx={0.5},markx={0}](0,0) to[bend left=50] (3,0);      \draw[ markxi={0}](3,0) to (2,0);
    \end{tikzpicture}=
         -4 \cdot \begin{tikzpicture}[basel={-.5},every node/.style={scale=0.6}]
         \qubitr{(0,0.3)};
         \draw[markx={0.5},markx={0},markx={1}](0.65,0.3) to (1.75,0.3);
          \draw[markx={0.5},markx={0},markx={1}](0.65,-0.3) to (1.75,-0.3);
         \qubitl{(2,0.3)};
         \qubitr{(0,-0.3)};
         \qubitl{(2,-0.3)};
    \end{tikzpicture} = 0.
    \end{equation*}
    Then, this invariant is null for any state in the $W-$SLOCC class. For the $GHZ-$SLOCC class, we use the triangle graph in Equation\eqref{eq:triangleGHZ} obtaining the following:    
   \begin{equation*}
        \begin{tikzpicture}[basel={-.5},every node/.style={scale=0.6}]
        \draw[markx={0},markx={1}](0,0) to (0.5,0.5);
         \draw[markx={1}](0,0) to (0.5,-0.5);
         \draw(0.5,0.5) to (0.5,-0.5);
         \draw[markxo={0.5}{0.6}](0.5,0.5) to (1,0.5);
         \draw[markxo={0.5}{0.6}](0.5,-0.5) to (1,-0.5);
           \draw[markx={0},markx={1}](1,0.5) to (1,-0.5);
             \draw[markx={1}](1,0.5) to (1.5,0);
             \draw(1,-0.5) to (1.5,0);
             \draw[markxo={0.5}{0.6}](1.5,0) to (2,0);
              \draw[markx={0},markx={1}](2,0) to (2.5,0.5);
               \draw[markx={1}](2,0) to (2.5,-0.5);
               \draw(2.5,0.5) to (2.5,-0.5);\draw[markxo={0.5}{0.6}](2.5,0.5) to (3,0.5);
         \draw[markxo={0.5}{0.6}](2.5,-0.5) to (3,-0.5);
         \draw[markx={0},markx={1}](3,0.5) to (3,-0.5);
          \draw[markx={1}](3,0.5) to (3.5,0);
          \draw(3,-0.5) to (3.5,0);
          \draw[markx={0},markx={1},markxo={0.5}{0.6}](0,0) to[bend left=90] (3.5,0);
    \end{tikzpicture} = - \begin{tikzpicture}[basel={-.5},every node/.style={scale=0.6}]
        \draw[markx={0},markx={1}](0,0) to (0.5,0.5);
         \draw[markx={1}](0,0) to (0.5,-0.5);
         \draw(0.5,0.5) to (0.5,-0.5);
         \draw[markx={0.5}](0.5,0.5) to (1,0.5);
         \draw[markx={0.5}](0.5,-0.5) to (1,-0.5);
           \draw[markx={0},markx={1}](1,0.5) to (1,-0.5);
             \draw[markx={1}](1,0.5) to (1.5,0);
             \draw(1,-0.5) to (1.5,0);
             \draw[markx={0.5}](1.5,0) to (2,0);
              \draw[markx={0},markx={1}](2,0) to (2.5,0.5);
               \draw[markx={1}](2,0) to (2.5,-0.5);
               \draw(2.5,0.5) to (2.5,-0.5);\draw[markx={0.5}](2.5,0.5) to (3,0.5);
         \draw[markx={0.5}](2.5,-0.5) to (3,-0.5);
         \draw[markx={0},markx={1}](3,0.5) to (3,-0.5);
          \draw[markx={1}](3,0.5) to (3.5,0);
          \draw(3,-0.5) to (3.5,0);
          \draw[markx={0},markx={1},markx={0.5}](0,0) to[bend left=90] (3.5,0);
    \end{tikzpicture}.
    \end{equation*}
 Again, we pushed, slid, pushed, and multiplied the Z-balls at first. Then, using the property in Equation \eqref{eq:Squarerule}, pushing the balls that appear and multiplying them together, we obtain:
\begin{equation*}
=-2 \cdot \begin{tikzpicture}[basel={-.5},every node/.style={scale=0.6}] =
        \draw[markx={0},markx={1}](0,0) to (0.5,0.5);
         \draw[markx={1}](0,0) to (0.5,-0.5);
         \draw(0.5,0.5) to (0.5,-0.5);
         \ball{(0.5,0)};
         \draw (0.7,0) node {$\frac{1}{\sqrt{2}}$};
             \draw[markx={1}](0.5,0.5) to (1,0);
             \draw(0.5,-0.5) to (1,0);
             \draw[markx={0.5}](1,0) to (1.5,0);
              \draw[markx={0},markx={1}](1.5,0) to (2,0.5);
               \draw[markx={1}](1.5,0) to (2,-0.5);
               \draw(2,0.5) to (2,-0.5);
          \draw[markx={1}](2,0.5) to (2.5,0);
          \draw(2,-0.5) to (2.5,0);
          \ball{(2,0)};
          \draw (2.2,0) node {$\frac{1}{\sqrt{2}}$};
          \draw[markx={0},markx={1},markx={0.5}](0,0) to[bend left=90] (2.5,0);
    \end{tikzpicture} =-4 \cdot \begin{tikzpicture}[basel={-.5},every node/.style={scale=0.6}] 
        \draw[markx={0},markx={1}](0,0) to (0.5,0.5);
         \draw[markx={1}](0,0) to (0.5,-0.5);
         \draw(0.5,0.5) to (0.5,-0.5);
             \draw[markx={1}](0.5,0.5) to (1,0);
             \draw(0.5,-0.5) to (1,0);
             \draw[markx={0.5}](1,0) to (1.5,0);
              \draw[markx={0},markx={1}](1.5,0) to (2,0.5);
               \draw[markx={1}](1.5,0) to (2,-0.5);
               \draw(2,0.5) to (2,-0.5);
          \draw[markx={1}](2,0.5) to (2.5,0);
          \draw(2,-0.5) to (2.5,0);
          \draw[markx={0},markx={1},markx={0.5}](0,0) to[bend left=90] (2.5,0);
    \end{tikzpicture}.
\end{equation*}
An equivalence that can be obtained by stitching together a triangle with a W state is the following:
\begin{equation*}
\begin{tikzpicture}[basel={-.5},every node/.style={scale=0.6}]
\draw (-0.5,0) to (0,0); \draw[markx={0},markx={1}](0,0) to (0.5,0.5);
         \draw[markx={1}](0,0) to (0.5,-0.5);
         \draw(0.5,0.5) to (0.5,-0.5);
             \draw[markx={1}](0.5,0.5) to (1,0);
             \draw(0.5,-0.5) to (1,0);
             
\draw (1.5,0) to (1,0);
             \end{tikzpicture} = \sqrt{2} \cdot  \begin{tikzpicture}[basel={-2},every node/.style={scale=0.6}]
\draw (0,0) -- (1,0);
\ball{(0.5,0)};
\draw (0.5,0.5) node {$\sqrt{2}$};    
\end{tikzpicture} 
\end{equation*}
Applying this equivalence in the invariant, we get:
\begin{equation*}
-4 \cdot \begin{tikzpicture}[basel={-.5},every node/.style={scale=0.6}] 
        \draw[markx={0},markx={1}](0,0) to (0.5,0.5);
         \draw[markx={1}](0,0) to (0.5,-0.5);
         \draw(0.5,0.5) to (0.5,-0.5);
             \draw[markx={1}](0.5,0.5) to (1,0);
             \draw(0.5,-0.5) to (1,0);
             \draw[markx={0.5}](1,0) to (1.5,0);
              \draw[markx={0},markx={1}](1.5,0) to (2,0.5);
               \draw[markx={1}](1.5,0) to (2,-0.5);
               \draw(2,0.5) to (2,-0.5);
          \draw[markx={1}](2,0.5) to (2.5,0);
          \draw(2,-0.5) to (2.5,0);
          \draw[markx={0},markx={1},markx={0.5}](0,0) to[bend left=90] (2.5,0);
    \end{tikzpicture} = -8 \cdot \begin{tikzpicture}[basel={-.5},every node/.style={scale=0.6}] \draw [markx={0}] (0,0) to[bend left=90] (0,-1);
    \draw [markx={0}](0,-1) to[bend left=90] (0,0);
    \ball {(0.3,-0.5)};
    \ball {(-0.3,-0.5)};
    \end{tikzpicture} = - 8 \cdot \begin{tikzpicture}[basel={-.5}]
   \draw (0,0) to[bend left=50] (0,0.5); 
    \draw (0,0) to (0.5,0); \draw (0.5,0) to[bend right=50] (0.5,0.5);
    \draw (0,0.5) to (0.5,0.5);
    \end{tikzpicture} =-16.
\end{equation*}
We pushed one of the balls through one $\bullet$ and multiplied by the other ball to cancel them. We also slid and multiplied the $\bullet$s to cancel them. It is clear that the invariant is not null for the $GHZ$ class. This process showed how the pushing and stitching rules allow us to calculate invariants easily. We also showed how the graph structures of the states have an intrinsic significance for SLOCC classification.
 The invariant presented in Equation \eqref{threequbitinv} is proportional to the three-tangle $\tau_{123}$ in Equation \eqref{eq:3tangle}. It has to be the case because there is only one invariant for three-qubit systems. 
\subsection*{Invariants for four qubits}
 For the case of four-qubit states, it is known that the invariant ring is defined by one invariant of degree two ($B$), two of degree four ($L,M$), and one of degree six ($D_{xy}$) \cite{LuqueThibon-tame}; however, their deduction is somewhat complicated. With this graphical language, we can define an equivalent set of invariants in a natural way as complete contractions. We will present different constructions for invariants and show the equivalence of the obtained expressions with those of the $B, L, M$ and $D_{xy}$ invariants defined in \cite{LuqueThibon-tame}. The invariant of degree 2 (built from two copies) is:
\begin{equation*}
    B_{0}:= \ \begin{tikzpicture}[basel={-.5}] 
   
    \draw[markxo={0.5}{0.6}] (0,0)  to[bend right=90] (2,0);
    \draw[markxo={0.5}{0.6}] (0,0)  to[bend left =90 ] (2,0);
     \draw[markxo={0.5}{0.6}] (0,0)  to[bend right=20] (2,0);
    \draw[markxo={0.5}{0.6}] (0,0)  to[bend left =20 ] (2,0);
    \Genstate{(0,0)};
    \Genstate{(2,0)};
    \end{tikzpicture} = 2 B.
    \end{equation*}
 To obtain degree-four invariants, we must contract four copies of the state. This can be done in several different ways; for example, we can propose the following two graphs:
 \begin{equation*}
D_1:=\begin{tikzpicture}[basel={-.5},every node/.style={scale=0.6}]  
     \draw[markxo={0.5}{0.6},red] (-1,1) to[bend left=10]  (1,1);
     \draw[markxo={0.5}{0.6},green](-1,1) to[bend left=50]  (1,1);
     \draw[markxo={0.5}{0.6},red] (0.1,0) to (0.1,-1.3); 
     \draw[markxo={0.5}{0.6}] (0,0) to (-1,1);
     \draw[markxo={0.5}{0.6},blue] (0,0) to (1,1); 
     \draw[markxo={0.5}{0.6},green] (-0.1,0) to (-0.1,-1.3);
     \draw[markxo={0.5}{0.6}] (0,-1.3) to[bend right=50]  (1,1);
     \draw[markxo={0.5}{0.6},blue] (0,-1.3) to[bend left=50]  (-1,1);   \Genstate{(0,0)};
            \Genstate{(0,-1.3)};
            \Genstate{(1,1)};
             \Genstate{(-1,1)};
 \end{tikzpicture} = ,\quad   D_2:=
 \begin{tikzpicture}[basel={-.5},every node/.style={scale=0.6}]  
     \draw[markxo={0.5}{0.6},red] (-1,1) to[bend left=10]  (1,1);
     \draw[markxo={0.5}{0.6},blue] (-1,1) to[bend left=50]  (1,1);
     \draw[markxo={0.5}{0.6},red] (0.1,0) to (0.1,-1.3); 
     \draw[markxo={0.5}{0.6}](0,0) to (-1,1);
     \draw[markxo={0.5}{0.6},green] (0,0) to (1,1); 
     \draw[markxo={0.5}{0.6},blue] (-0.1,0) to (-0.1,-1.3);
     \draw[markxo={0.5}{0.6}] (0,-1.3) to[bend right=50]  (1,1);
     \draw[markxo={0.5}{0.6},green] (0,-1.3) to[bend left=50]  (-1,1);  \Genstate{(0,0)};
             \Genstate{(0,-1.3)};
             \Genstate{(1,1)};
            \Genstate{(-1,1)};
 \end{tikzpicture},
 \end{equation*}
 where now we use colors to label each contracted part, according to:
 \begin{equation*}
 (1)=\tikz[baseline=-3.5]{\draw(0,0) to (1,0)}, \quad (2)=\tikz[baseline=-3.5]{\draw[blue](0,0) to (1,0)}, \quad (3)=\tikz[baseline=-3.5]{\draw[green](0,0) to (1,0)}, \quad (4)=\tikz[baseline=-3.5]{\draw[red](0,0) to (1,0)}.
 \end{equation*}
 The invariant $D_1$ is equivalent to a combination of $L$ and $M$ invariants as $D_1=-(8M+4L)$, while $D_2$ corresponds to $D_2=4M+8L$. Some other invariants can be obtained from four copies of the state, but two independent ones are enough to generate the invariant algebra. An invariant of degree six, can be obtained from the graph:
\begin{equation*}
F=
\begin{tikzpicture}[basel={-.5},every node/.style={scale=0.6}]
\draw[markxo={0.5}{0.6},green](0,0) to (6,0);
\draw[markxo={0.5}{0.6},red](0,0) to (2,-2);
\draw[markxo={0.5}{0.6},blue](0,0) to (4,-2);
\draw[markxo={0.5}{0.6}](0,0) to (0,-4);
\draw[markxo={0.5}{0.6}](6,0) to (2,-2);
\draw[markxo={0.5}{0.6},red](6,0) to (4,-2);
\draw[markxo={0.5}{0.6},blue](6,0) to (6,-4);
\draw[markxo={0.5}{0.6},blue](2,-2) to (0,-4);
\draw[markxo={0.5}{0.6},green](2,-2) to (6,-4);
\draw[markxo={0.5}{0.6},green](4,-2) to (0,-4);
\draw[markxo={0.5}{0.6}](4,-2) to (6,-4);
\draw[markxo={0.5}{0.6},red](0,-4) to (6,-4);
 \Genstate{(0,0)};
 \Genstate{(6,0)};
 \Genstate{(6,-4)};
 \Genstate{(0,-4)};
 \Genstate{(2,-2)};
 \Genstate{(4,-2)};
\end{tikzpicture} .
\end{equation*}
This invariant is related to the ones defined in \cite{LuqueThibon-tame} as: $F= 4MB+ 2LB-6D_{xy}$. With this graph invariant and the previous three invariants, we can define the polynomial ring similarly as the invariants $B,L,M$, and $D_{xy}$ do.

\subsection*{Invariants for five qubits}
For five-qubit states, there is no defined set of invariants to the best of our knowledge. In \cite{5qubits-Thibon}, the number of invariants needed to generate the invariant algebra is proposed, and a set for the invariants of lower degrees is proposed. However, due to the complexity of the obtained expressions, the authors only give expressions for some of the invariants. Here, we will present a complete set of invariants that agrees with the degrees proposed in \cite{5qubits-Thibon}. In that paper, it is stated that for five-qubit systems, the set of primary invariants, i.e, a maximal set of algebraically independent invariants, should be composed of five invariants of degree four, one of degree six, five of degree eight, one of degree ten, and five of degree twelve. Here, we present a set of such invariants with the appropriate degrees, conforming to a set of $17$ algebraically independent polynomials. To obtain five invariants of degree four, we can use the following constructions:
\begin{equation*}
    D_{1}:= \ \begin{tikzpicture}[basel={-.5},every node/.style={scale=0.6}] 
   
    \draw[markxo={0.5}{0.6},olive] (0,0)  to[bend right=90] (2,0);
    \draw[,markxo={0.5}{0.6},blue] (0,0)  to[bend left =90 ] (2,0);
     \draw[,markxo={0.5}{0.6},red] (0,0)  to[bend right=20] (2,0);
    \draw[markxo={0.5}{0.6},green] (0,0)  to[bend left =20 ] (2,0);
      
    \draw[markxo={0.5}{0.6},blue] (0,-2)  to[bend left =90 ] (2,-2);
  \draw[markxo={0.5}{0.6},red] (0,-2)  to[bend right=20] (2,-2);
  \draw[markxo={0.5}{0.6},olive] (0,-2)  to[bend right=90] (2,-2);;
    \draw[markxo={0.5}{0.6},green] (0,-2)  to[bend left =20 ] (2,-2);
     \draw[markxo={0.5}{0.6}] (2,0)  to[bend left=90] (2,-2);
     \draw[markxo={0.5}{0.6}] (0,0)  to[bend right=90] (0,-2);
    \Genstate{(0,0)};
   \Genstate{(2,0)};
   \Genstate{(0,-2)};
   \Genstate{(2,-2)};   
    \end{tikzpicture} , \qquad    D_{2}:= \ \begin{tikzpicture}[basel={-.5},every node/.style={scale=0.6}] 
   
    \draw[markxo={0.5}{0.6},olive] (0,0)  to[bend right=90] (2,0);
    \draw[,markxo={0.5}{0.6}] (0,0)  to[bend left =90 ] (2,0);
     \draw[,markxo={0.5}{0.6},red] (0,0)  to[bend right=20] (2,0);
    \draw[markxo={0.5}{0.6},green] (0,0)  to[bend left =20 ] (2,0);
      
    \draw[markxo={0.5}{0.6}] (0,-2)  to[bend left =90 ] (2,-2);
  \draw[markxo={0.5}{0.6},red] (0,-2)  to[bend right=20] (2,-2);
  \draw[markxo={0.5}{0.6},olive] (0,-2)  to[bend right=90] (2,-2);;
    \draw[markxo={0.5}{0.6},green] (0,-2)  to[bend left =20 ] (2,-2);
     \draw[markxo={0.5}{0.6},blue] (2,0)  to[bend left=90] (2,-2);
     \draw[markxo={0.5}{0.6},blue] (0,0)  to[bend right=90] (0,-2);
    \Genstate{(0,0)};
   \Genstate{(2,0)};
   \Genstate{(0,-2)};
   \Genstate{(2,-2)};   
    \end{tikzpicture} , \quad \dots,
    \end{equation*}
where we used olive for the color of the fifth part. We obtain a different invariant for each choice of the part connecting the copies up with the copies down. Then, we can obtain five invariants of degree four that we checked to be algebraically independent. For the invariant of degree six, we propose the following construction:
\begin{equation*}
F_{0}:=
\begin{tikzpicture}[basel={-.5},every node/.style={scale=0.6}]

\draw[markxo={0.5}{0.6},green](0,0) to (6,0);
\draw[markxo={0.5}{0.6},red](0,0) to (2,-2);
\draw[markxo={0.5}{0.6},blue](0,0) to (4,-2);
\draw[markxo={0.5}{0.6}](0,0) to (0,-4);
\draw[markxo={0.5}{0.6}](6,0) to (2,-2);
\draw[markxo={0.5}{0.6},red](6,0) to (4,-2);
\draw[markxo={0.5}{0.6},blue](6,0) to (6,-4);
\draw[markxo={0.5}{0.6},blue](2,-2) to (0,-4);
\draw[markxo={0.5}{0.6},green](2,-2) to (6,-4);
\draw[markxo={0.5}{0.6},green](4,-2) to (0,-4);
\draw[markxo={0.5}{0.6}](4,-2) to (6,-4);
\draw[markxo={0.5}{0.6},red](0,-4) to (6,-4);
\draw[markxo={0.5}{0.6},olive](0,0) to (6,-4);
\draw[markxo={0.5}{0.6},olive](6,0) to (0,-4);
\draw[markxo={0.5}{0.6},olive](2,-2) to (4,-2);
 \Genstate{(0,0)};
 \Genstate{(6,0)};
 \Genstate{(6,-4)};
 \Genstate{(0,-4)};
 \Genstate{(2,-2)};
 \Genstate{(4,-2)};

\end{tikzpicture}.
\end{equation*}
For invariants of degree eight, we propose a  construction based on the invariants of degree four of four qubits:
 \begin{equation*}
H_1:=\begin{tikzpicture}[basel={-.5},every node/.style={scale=0.6}]  
     \draw[markxo={0.5}{0.6},red] (-1,1) to[bend left=10]  (1,1);
     \draw[markxo={0.5}{0.6},green](-1,1) to[bend left=50]  (1,1);
     \draw[markxo={0.5}{0.6},red] (0.1,0) to (0.1,-1.3); 
     \draw[markxo={0.5}{0.6},olive] (0,0) to (-1,1);
     \draw[markxo={0.5}{0.6},blue] (0,0) to (1,1); 
     \draw[markxo={0.5}{0.6},green] (-0.1,0) to (-0.1,-1.3);
     \draw[markxo={0.5}{0.6},olive] (0,-1.3) to[bend right=50]  (1,1);
     \draw[markxo={0.5}{0.6},blue] (0,-1.3) to[bend left=50]  (-1,1); \Genstate{(0,0)};
            \Genstate{(0,-1.3)};
            \Genstate{(1,1)};
             \Genstate{(-1,1)};

             \draw[markxo={0.5}{0.6},red] (3,1) to[bend left=10]  (5,1);
     \draw[markxo={0.5}{0.6},green](3,1) to[bend left=50]  (5,1);
     \draw[markxo={0.5}{0.6},red] (4.1,0) to (4.1,-1.3); 
     \draw[markxo={0.5}{0.6},olive] (4,0) to (3,1);
     \draw[markxo={0.5}{0.6},blue] (4,0) to (5,1); 
     \draw[markxo={0.5}{0.6},green] (3.9,0) to (3.9,-1.3);
     \draw[markxo={0.5}{0.6},olive] (4,-1.3) to[bend right=50]  (5,1);
     \draw[markxo={0.5}{0.6},blue] (4,-1.3) to[bend left=50]  (3,1);  \Genstate{(4,0)};
            \Genstate{(4,-1.3)};
            \Genstate{(5,1)};
             \Genstate{(3,1)};
            \draw[markxo={0.5}{0.6}] (1, 1) to [bend left=40] (5,1);
            \draw[markxo={0.5}{0.6}] (-1, 1) to [bend left=40] (3,1);
            \draw[markxo={0.5}{0.6}] (0, 0) to [bend left=40] (4,0);\draw[markxo={0.5}{0.6}] (0, -1.3) to [bend right=40] (4,-1.3);          
 \end{tikzpicture},
 \end{equation*} 
and similarly for $H_2,H_3,H_4$ and $H_5$, where each choice of the part connecting between sides corresponds to a different invariant. For the invariant of degree ten, we propose the following:
 \begin{equation*}
J_{0}:=\begin{tikzpicture}[basel={-.5},every node/.style={scale=0.6}]
\draw[markxo={0.5}{0.6}] (0,0) to (6,0); 
\draw[markxo={0.5}{0.6},green] (0,0) to (2,-2); 
\draw[markxo={0.5}{0.6},blue] (0,0) to (4,-2); 
\draw[markxo={0.5}{0.6},red] (0,0) to (0,-4); 
\draw[markxo={0.5}{0.6},olive] (0,0) to[bend right=90] (0,-8); 
\draw[markxo={0.5}{0.6},blue] (6,0) to (2,-2);
\draw[markxo={0.5}{0.6},red] (6,0) to (4,-2); 
\draw[markxo={0.5}{0.6},olive] (6,0) to (6,-4); 
\draw[markxo={0.5}{0.6},green] (6,0) to[bend left=90]  (6,-8);
\draw[markxo={0.5}{0.6}] (0,-4) to  (6,-4);  
\draw[markxo={0.5}{0.6},blue] (0,-4) to (2,-6); 
\draw[markxo={0.5}{0.6},green] (0,-4) to  (0,-8); 
\draw[markxo={0.5}{0.6}] (2,-2) to (4,-2); 
\draw[markxo={0.3}{0.6},red] (2,-2) to (2,-6); 
\draw[markxo={0.5}{0.6},olive] (2,-2) to (0,-4);
\draw[markxo={0.5}{0.6},green] (4,-2) to (6,-4);
\draw[markxo={0.3}{0.6},olive] (4,-2) to (4,-6); 
\draw[markxo={0.5}{0.6},blue] (6,-4) to (6,-8);
\draw[markxo={0.5}{0.6},red] (6,-4) to (4,-6); 
\draw[markxo={0.5}{0.6},green] (2,-6) to (4,-6);
\draw[markxo={0.5}{0.6}] (2,-6) to (0,-8); 
\draw[markxo={0.5}{0.6}] (4,-6) to (6,-8); 
\draw[markxo={0.5}{0.6},red] (0,-8) to (6,-8); 
\draw[markxo={0.5}{0.6},olive] (2,-6) to (6,-8); 
\draw[markxo={0.5}{0.6},blue] (4,-6) to (0,-8); 
 \Genstate{(0,0)};
  \Genstate{(6,0)};
   \Genstate{(2,-2)};
    \Genstate{(4,-2)};
     \Genstate{(0,-4)};
      \Genstate{(6,-4)}; \Genstate{(2,-6)};
       \Genstate{(4,-6)};
        \Genstate{(0,-8)}; \Genstate{(6,-8)};
\end{tikzpicture}
 \end{equation*}
The last set of five invariants of degree twelve can be obtained as:
 \begin{equation*}L_1:=
\begin{tikzpicture}[basel={-.5},every node/.style={scale=0.6}]
\draw[markxo={0.5}{0.6},green](0,0) to (6,0);
\draw[markxo={0.5}{0.6},red](0,0) to (2,-2);
\draw[markxo={0.5}{0.6},blue](0,0) to (4,-2);
\draw[markxo={0.5}{0.6},olive](0,0) to (0,-4);
\draw[markxo={0.5}{0.6},olive](6,0) to (2,-2);
\draw[markxo={0.5}{0.6},red](6,0) to (4,-2);
\draw[markxo={0.5}{0.6},blue](6,0) to (6,-4);
\draw[markxo={0.5}{0.6},blue](2,-2) to (0,-4);
\draw[markxo={0.5}{0.6},green](2,-2) to (6,-4);
\draw[markxo={0.5}{0.6},green](4,-2) to (0,-4);
\draw[markxo={0.5}{0.6},olive](4,-2) to (6,-4);
\draw[markxo={0.5}{0.6},red](0,-4) to (6,-4);
\Genstate{(0,0)};
 \Genstate{(6,0)};
 \Genstate{(6,-4)};
 \Genstate{(0,-4)};
 \Genstate{(2,-2)};
 \Genstate{(4,-2)};
\draw[markxo={0.5}{0.6},green](8,0) to (14,0);
\draw[markxo={0.5}{0.6},red](8,0) to (10,-2);
\draw[markxo={0.5}{0.6},blue](8,0) to (12,-2);
\draw[markxo={0.5}{0.6},olive](8,0) to (8,-4);
\draw[markxo={0.5}{0.6},olive](14,0) to (10,-2);
\draw[markxo={0.5}{0.6},red](14,0) to (12,-2);
\draw[markxo={0.5}{0.6},blue](14,0) to (14,-4);
\draw[markxo={0.5}{0.6},blue](10,-2) to (8,-4);
\draw[markxo={0.5}{0.6},green](10,-2) to (14,-4);
\draw[markxo={0.5}{0.6},green](12,-2) to (8,-4);
\draw[markxo={0.5}{0.6},olive](12,-2) to (14,-4);
\draw[markxo={0.5}{0.6},red](8,-4) to (14,-4);
\Genstate{(8,0)};
 \Genstate{(14,0)};
 \Genstate{(14,-4)};
 \Genstate{(8,-4)};
 \Genstate{(10,-2)};
 \Genstate{(12,-2)};

\draw[markxo={0.55}{0.56}](0,0) to[bend left=30] (8,0);
\draw[markxo={0.55}{0.56}](6,0) to[bend left=30] (14,0);
\draw[markxo={0.55}{0.56}](2,-2) to[bend left=30] (10,-2);
\draw[markxo={0.55}{0.56}](4,-2) to[bend left=30] (12,-2);
\draw[markxo={0.55}{0.56}](0,-4) to[bend right=30] (8,-4);
\draw[markxo={0.55}{0.56}](6,-4) to[bend right=30] (14,-4);
\end{tikzpicture}.
\end{equation*}
A different invariant can be obtained by changing the part used to connect the two sides. Using the Jacobian matrix, we checked that the $17$ invariants proposed here are independent and may work as a complete basis for the invariant ring of five qubits. The same approach could be extended for systems of more qubits. Knowing the number of invariants and their respective degrees is very helpful for the construction.

\section{Rate exponents of multi-qubit quantum states in the W SLOCC class}

The last secondary result in this chapter concerns the Kronecker state stitching when specializing to multipartite states in the W class. These states are very interesting because knowing the local spectra of each part uniquely determines the state, which is not a general property of multi-qubit states. Recalling from Section \ref{WKronecker}, $N$-partite states in the W-SLOCC class can be parametrized as:
\begin{equation*}\label{eq:ecstatepsi2}
\ket{\psi^W} = \sqrt{c_{\rtup}^{0}}|\boldsymbol{0} \rangle + \sum_{i=1}^{N}\sqrt{c_{\rtup}^{i}}|\boldsymbol{1}_i \rangle, 
\end{equation*}
where $\boldsymbol{0}$ is a sequence of $N$ zeros, $\boldsymbol{1}_i$  is a sequence with a ``1" at position $i$ and zeros in the remaining $(N-1)$ positions. The coefficients $c_{\rtup}^{i}$ are real with $\sum_{i=0}^{N}c_{\rtup}^{i}=1$, and $\rtup$ labels the local spectra of $\psi^W$. Another property of W-class states is that when separating one of the qubits of the state, it is possible to rewrite the state as:
    \begin{equation*}
        \ket{\psi^W}= \sqrt{c_{\rtup}^0}\ket{0}\ket{\tilde{\bm{0}}}+\sqrt{c_{\rtup}^1}\ket{1}\ket{\tilde{\bm{0}}}+\sqrt{\sum_{i=2}^N c_{\rtup}^i}\ket{0}\ket{\bar{\bm{1}}}, \quad
 \ket{\bar{\bm{1}}}=\frac{1}{\sqrt{\sum_{i=2}^N c_{\rtup}^i}}\left( \sum_{i=2}^N \sqrt{c_{\rtup}^i}\ket{\tilde{\bm{1}}_i} \right),
    \end{equation*}
    where $\ket{\tilde{\bm{1}}_i}$ ($\ket{\tilde{\bm{0}}}$) is the state $\ket{\boldsymbol{1}_i}$($\ket{\bm{0}}$) without the first $\ket{0}$. Note how the state in the previous equation has the same structure as a W-class state of two parts. It is worth noting that $\ket{\tilde{\bm{0}}},\ket{\bar{\bm{1}}}$ are orthonormal, so we can directly find the reduced density matrix in the separated part, by tracing the remaining parts:
    \begin{equation*}
        \rho_1=\bra{\tilde{\bm{0}}}\rho \ket{\tilde{\bm{0}}}+\bra{\bar{\bm{1}}} \rho \ket{\bar{\bm{1}}}=\left(\begin{array}{cc}
    c_{\rtup}^0+\sum_{i=2}^N c_{\rtup}^i &\sqrt{c_{\rtup}^0c_{\rtup}^1}\\\sqrt{c_{\rtup}^0c_{\rtup}^1}         & c_{\rtup}  ^1
    \end{array} \right).
    \end{equation*}
    This structural relation can be generalized; we can group in the same way any subset $S$ of the $N$ parts, $\{S\} \subset \{N\}$. Then by calculating the reduced density matrix for the subset $S$, $\rho_S$, tracing out all the complementary parts to $S$ we obtain:
    \begin{equation}
        \rho_S=\left(\begin{array}{cc}
    c_{\rtup}^0+\sum_{i\notin S} c_{\rtup}^i &\sqrt{c_{\rtup} ^0\sum_{i\in S}c_{\rtup} ^i}\\\sqrt{c_{\rtup}^0\sum_{i\in S}c_{\rtup}^i}         & \sum_{i\in S}c_{\rtup} ^i
    \end{array} \right).
    \label{eq:DensityW}
    \end{equation}
    
\noindent  Then, any reduced density matrix is of rank two. This result can also be obtained from the rule (ix) of the stitching and pushing rules in Section \ref{rules} and the OER properties for qubit states. $N$-partite states in the W class are SLOCC equivalent to $W_N$ states, which can be built by stitching together any two $W$ states with $N_1+1$ and $N-N_1+1$ parts with a $\Psi$ stitch,
\begin{equation*}
  \begin{tikzpicture} [baseline={([yshift=-.5ex]current bounding box.center)},every node/.style={scale=0.6}] \def\array{$1$,$2$,$.$,$.$,$.$,$.$,$.$,$N_1$}    \def\arrayx{$N-N_1$,$.$,$.$,$.$,$.$,$.$,$2$,$1$}
    \foreach [count=\n] \x in \array{
        \node at ({90+\n*180/(8+1)}:1cm) (n\n) {\x};
        \draw (0,0)--(n\n); 
        };
        \draw[markx={0},markx={0.5},markx={1}] (0,0)--(1,0);
         \foreach [count=\n] \x in \arrayx{
        \node at ($(1,0)+({-90+\n*180/(8+1)}:1cm)$) (n\n) {\x};
        \draw (1,0)--(n\n); 
        };
        \draw (0.5,0) node {$\times$};
\end{tikzpicture}  =\begin{tikzpicture} [baseline={([yshift=-.5ex]current bounding box.center)},every node/.style={scale=0.6}] \def\array{$1$,$2$,$.$,$.$,$.$,$.$,$.$,$.$}    \def\arrayx{$.$,$.$,$.$,$.$,$.$,$.$,$.$,$N$}
    \foreach [count=\n] \x in \array{
        \node at ({90+\n*180/(8+1)}:1cm) (n\n) {\x};
        \draw (0,0)--(n\n); 
        };
         \foreach [count=\n] \x in \arrayx{
        \node at ($(0,0)+({-90+\n*180/(8+1)}:1cm)$) (n\n) {\x};
        \draw (0,0)--(n\n); 
        };
        \draw[markxi={0}](0,0) to (0,1);
        \end{tikzpicture}.
\end{equation*}    
Then, $N$-partite states in the W-class have as graph representative an OER graph in each possible separation. Then, the rank of any possible reduced density matrix is of rank at most two. From the density matrices in Equation \eqref{eq:DensityW}, it is possible to obtain the following relation for the eigenvalues:  
    \begin{equation*}
        \lambda_S(1-\lambda_S)=\sum_{i\notin S}c_{\bm{r}}^i\left(\sum_{i\in S}c_{\bm{r}}^i\right)
    \end{equation*}
For the marginal spectra, i.e., the eigenvalues of the one-particle reduced density matrices, we get
\begin{equation}\label{eq:cspec}
r^i_2 (1-r^i _2) =c_{\rtup}^i (1- c_{\rtup}^i -c_{\rtup} ^0).
\end{equation}
Note how the relation between coefficients $c_{\rtup}^i$ and the set of local eigenvalues of $\psi^W$, $\rtup$, is made explicit. This functional relation will allow us to make interesting observations in applying Schur-Weyl duality in this class.\\

\noindent We stated before in subsection \ref{WKronecker} that the Schur transform applied to W-class states corresponds to:
\begin{equation*}
     \ket{\psi^{W}}^{\otimes n} =\bigoplus_{\bm{\lambda} \in \Lambda_n^{W}} \eta_{\bm{\lambda}} \ket{\reallywidehat{\Phi}_{\bm{\lambda}} \left(\psi^{W} \right)}\ket{\mathcal{K}^W_{\bm{\lambda}}},
\end{equation*}
where $\Lambda_n^W$ is the polytope that defines the possible sets of partitions $\bm{\lambda}$ appearing in the Schur basis of W class states, defined in Equation \eqref{eq:WLambda}. 
The factor $\eta_{\bm{\lambda}}$ relates the unnormalized state $\ket{\reallywidehat{\Phi}_{\bm{\lambda}}(\psi^W)}$ and the normalized state $\ket{\Phi_{\bm{\lambda}}(\psi^W)}$ as :
 \begin{equation*}
        \ket{\Phi_{\bm{\lambda} }(\psi^W)}=\frac{ \eta_{\bm{\lambda }}}{\sqrt{p(\bm{\lambda}|\psi^W)}} \ket{\reallywidehat{\Phi}_{\bm{\lambda}}(\psi^W)},
  \end{equation*}
 and $p(\lamtup|\psi)$ is the probability of ending up in a set of partitions $\lamtup$ after a projective measurement in the set of partitions $\lamtup$. The unnormalized state is:
\begin{equation}
\label{eq:phigen23}
\ket{\reallywidehat{\Phi}_{\bm{\lambda}}(\psi^W)}= n!^{-(N-2)/2}
   \sum_{\omega^{0}=0}^{n} \dfrac{(c_{\rtup}^{0})^{\omega^{0}/2} } {\omega^{0}!} \sum_{\omegatup} {\ctup_{\rtup}}^{\omegatup/2} \sqrt{A_{\lamtup,\omegatup}}
 \ket{ \lamtup, \omegatup},
\end{equation}
and we define the following quantities:
 \begin{equation}
 \bm{c}_{\bm{r}}^{\bm{\omega}/2}= \prod_{i=1}^N ({c_{\bm{r}}^{i}}) ^{\omega_i/2}   , \quad    A_{\bm{\lambda},\bm{\omega}}=\prod_{i=1}^N A_{\lambda^i,\omega^i } \quad , \quad  A_{\lambda,\omega}=\frac{(n-\lambda-\omega)!}{(\omega-\lambda)!}.
 \label{Definitions}
 \end{equation}
We want to find the asymptotic behavior of the probability of measuring some set of partitions $\lamtup$ if the state has a set of local spectra $\bm{r}$. In other words, we want to find the exponential rate of the probability:
 \begin{equation}\label{eq:RateP}
    p(\lamtup|\rtup)= e^{-nR(\bar{\lamtup}|\rtup)} \rightarrow R(\overline{\lamtup}|\rtup)=-\frac{1}{n}\log(p(\lamtup|\rtup)),
 \end{equation}
 we will see later how this rate depends on the set of normalized partitions $\bm{\overline{\lambda}}=\lamtup/n$.
Let us now define the norm of the unnormalized state:
 \begin{equation} \label{eq:ZetaDef}
     Z_{\lamtup}(\psi)=\braket{\reallywidehat{\Phi}_{\lamtup}(\psi)}{\reallywidehat{\Phi}_{\lamtup}(\psi)}=n!^{-(N-2)}\sum_{(\omega^0,\bm{\omega})}\frac{{c_{\rtup}^{0}}^{\omega^{0}}}{\omega^{0}!^2}\ctup_{\rtup}^{\omegatup} A_{\lamtup,\bm{\omega}},
 \end{equation}
 that is related with the factor $\eta_{\lamtup}$ and the probability as: \begin{equation} \label{eq:etaDef}
     \eta_{\lamtup}=\sqrt{\frac{p(\lamtup|\psi)}{Z_{\lamtup}(\psi)}}.
 \end{equation}
 \noindent According to  \cite{Botero}, the factor $\eta_{\bm{\lambda}}$ cannot depend on the parameters of the state. Then, the probabilities for two different states in the W class are related by
\begin{equation*}\label{eq:relation2}
    p(\lamtup|\psi) = p(\lamtup|\psi')
  \dfrac{Z_{\lamtup}(\psi)}{Z_{\lamtup}(\psi ')}.
\end{equation*}
We can see that the rate in Equation \eqref{eq:RateP} of two different states are related as:
 \begin{equation} \label{eq:raterel}
     R(\bar{\lamtup}|\rtup)= R(\bar{\lamtup}|\rtup')-  \zeta(\bar{\lamtup}|\rtup')+\zeta(\bar{\lamtup}|\rtup),
 \end{equation}
 where we defined a rate for the norm $Z_{\bm{\lambda}}(\psi)$ as:
 \begin{equation} \label{eq:RateZ}
      \zeta(\bar{\lamtup}|\rtup)=\frac{1}{n}\log(Z_{\lamtup}(\psi)).
 \end{equation}
The rate of $Z_{\bm{\lambda}}(\psi)$ can be calculated from its definition in Equation \eqref{eq:ZetaDef} as:
 \begin{equation*}
      \zeta(\bar{\lamtup}|\rtup)=\frac{1}{n}\log(Z_{\lamtup}(\psi))=\frac{1}{n}\log\left(n!^{-(N-2)}\sum_{(\omega^0,\bm{\omega})}\frac{{c_{\rtup}^{0}}^{\omega^{0}}}{\omega^{0}!^2}\ctup_{\rtup}^{\omegatup} A_{\lamtup,\bm{\omega}} \right).
 \end{equation*}
Using the Stirling formula and the definitions in Equation \eqref{Definitions}, it is possible to show that this rate is given by:
\begin{equation}
\zeta( \overline{\lamtup}|\rtup) = \sup_{\overline{\omegatup} \in \Omega }\left[ H_2(\overline{\omega}^{0}) + \sum_{i=0}^N  \overline{\omega}^{i}\log c^{i}_{\rtup} + \alpha(\overline{\omega}^{i}, \overline{\lambda}^{i})\right]\label{eq:supreme},
\end{equation}
where $H_2(x) = - x \log x - (1-x) \log(1-x)$ is the binary entropy, and
\begin{equation*}
\alpha(\overline{\omega}, \overline{\lambda})  =  (1-\overline{\lambda}-\overline{\omega})\log(1-\overline{\lambda}-\overline{\omega}) - (\overline{\omega}- \overline{\lambda})\log(\overline{\omega}- \overline{\lambda})  .
\end{equation*}
$\overline{\omega}^i=\frac{\omega^i}{n}$ and $\overline{\lambda}^i=\frac{\lambda^i}{n}$ are the normalized weights and partitions, $\bar{\bm{\omega}}=\bar{\omega}^0,\bar{\omega}^1,\dots \bar{\omega}^N$ is the set of normalized weights including the zero-th weight, and we define $\lambda^0=0$.
 $\Omega \subset \mathbb{R}^{N+1}$ is the convex domain 
defined by:
\begin{equation*}
\overline{\lambda}^{i} \leq \overline{\omega}^{i} \leq 1- \overline{\lambda}^{i}, \qquad \sum_{i=0}^{N}\overline{\omega}^{i} = 1.
\end{equation*}
The supremum of equation \eqref{eq:supreme} can be obtained as the solution of the following set of equations:
 \begin{equation*}
     \begin{gathered}
        (\overline{\omega}^0)^2=\kappa c_{\rtup} ^0 ,\\
        \overline{\omega}^i(1-\overline{\omega}^i) =\overline{\lambda}^ i(1-\overline{\lambda} ^i)+ \kappa c^i_{\rtup},
     \end{gathered}
 \end{equation*}
where $\kappa$ is a Lagrange multiplier that is fixed by the condition $\sum_{i=0}^N \overline{\omega}^{i} =1$. Using the solution in the expression for $\zeta(\overline{\lamtup}|\rtup)$ we obtain
\begin{equation*}
     \zeta(\overline{\lamtup}|\rtup)
    = -\log(\kappa)+\sum_i  (1-\overline{\lambda}^{i})\log  (1-\overline{\lambda}^{i} -{\overline{\omega}^{i}})+{\overline{\lambda}^{i}}\log(\overline{\omega}^{i}-\overline{\lambda}^{i}).
 \end{equation*}
Following from Equation \eqref{eq:raterel}, we use the fact that the rate function $R(\overline{\lamtup}| \rtup)$ must vanish where the probability $p(\overline{\lamtup}| \psi)$ is maximized, which, by the Keyl-Werner theorem in subsection \ref{Keyl-WernerTheorem}, must occur when $\overline{\lamtup}= \rtup$. Then, labeling W class states with their sets of eigenvalues, we will pick two states such as $\psi=\psi_{\bm{r}}$, and $\psi'=\psi_{\lamtup}$. Therefore, for a given set of reduced partitions $\overline{\lamtup}$,
\begin{equation*}
\label{wrate}
    R(\overline{\lamtup}| \rtup) =    \zeta(\overline{\lamtup}|\overline{\lamtup}) - \zeta(\overline{\lamtup}|\rtup).
\end{equation*}
When $\rtup = \overline{\lamtup}$ is satisfied, the solution for the extremization problem is $\overline{\omega}^{i} = c^{i}_{\lamtup}$ with $\kappa = c_{{\lamtup}}^{0}$, where the $c^{i}_{{\lamtup}}$ coefficients are solutions to Equation \eqref{eq:cspec} with $\rtup=\overline{\lamtup}$. Hence,
\begin{equation}
  \zeta(\overline{\lamtup}|\overline{\lamtup})
    = -\log(c_{\lamtup}^{0})
    +\sum_i  (1-\overline{\lambda}^{i})\log  (1-\overline{\lambda}^{i} -{{c_{\lamtup}}^{i}})+\overline{\lambda}^{i}\log({c_{\lamtup}}^{i}-\overline{\lambda}^{i}),
    \label{eq:zitall}
\end{equation}
and the rate becomes
\begin{equation*}
    R(\overline{\lamtup}|\rtup)= 
     -\log\left(\frac{\kappa}{c_{\lamtup}^{0}}\right)
     +\sum_i  (1-{\overline{\lambda}^{i}})\log  \left(\frac{1-{\overline{\lambda}^{i}} -{\overline{\omega}^{i}}}{1-{\overline{\lambda}}^{i} -{{c_{\lamtup}}^{i}}}\right)+{\overline{\lambda}^{i}}\log\left(\frac{\overline{\omega}^{i}-\overline{\lambda}^{i}}{{c_{\lamtup}}^{i}-\overline{\lambda}^{i}}\right).
 \end{equation*}
 This solution is a multi-local version of the rate obtained by the Keyl-Werner theorem that corresponds to the relative entropy as shown in Equation \eqref{eq:Relative}.
Achieving this rate is an interesting result on its own; however, we will see how it can be used along with the techniques introduced in this document to understand the construction of the Kronecker subspace spanned by the W-SLOCC class.

\subsection{Asymptotic construction of Kronecker states in the W-SLOCC class}

Now, we want to turn our attention to the representative normalized state of the W-SLOCC class in $N$ qubits:
\begin{equation*}
\ket	{W_N} = \frac{1}{\sqrt{N}} \sum_{i=1}^{N} \ket{\bm{1}_i}.
\end{equation*}
One important property of this state is that it can be obtained by stitching any two W states $\ket{W_{N_1+1}},\ket{W_{N-N_1+1}}$ states using the $\Psi$ stitch, which corresponds to the state $\ket{W_2}$. This property can be seen graphically as:
\begin{equation*}
  \begin{tikzpicture} [baseline={([yshift=-.5ex]current bounding box.center)},every node/.style={scale=0.6}] \def\array{$1$,$2$,$.$,$.$,$.$,$.$,$.$,$N_1$}    \def\arrayx{$N-N_1$,$.$,$.$,$.$,$.$,$.$,$2$,$1$}
    \foreach [count=\n] \x in \array{
        \node at ({90+\n*180/(8+1)}:1cm) (n\n) {\x};
        \draw (0,0)--(n\n); 
        };
        \draw[markx={0},markx={0.5},markx={1}] (0,0)--(1,0);
         \foreach [count=\n] \x in \arrayx{
        \node at ($(1,0)+({-90+\n*180/(8+1)}:1cm)$) (n\n) {\x};
        \draw (1,0)--(n\n); 
        };
        \draw (0.5,0) node {$\times$};
\end{tikzpicture}  =\begin{tikzpicture} [baseline={([yshift=-.5ex]current bounding box.center)},every node/.style={scale=0.6}] \def\array{$1$,$2$,$.$,$.$,$.$,$.$,$.$,$.$}    \def\arrayx{$.$,$.$,$.$,$.$,$.$,$.$,$.$,$N$}
    \foreach [count=\n] \x in \array{
        \node at ({90+\n*180/(8+1)}:1cm) (n\n) {\x};
        \draw (0,0)--(n\n); 
        };
         \foreach [count=\n] \x in \arrayx{
        \node at ($(0,0)+({-90+\n*180/(8+1)}:1cm)$) (n\n) {\x};
        \draw (0,0)--(n\n); 
        };
        \draw[markxi={0}](0,0) to (0,1);
        \end{tikzpicture},
\end{equation*}
where we generalized the notation of $W_3$ to vertices with $N$ edges, as shown in rule (ix) in Section \ref{rules}, to symbolize the $W_N$ state. This decomposition is not unique, and moreover, we could take $W_{N_1+1}$ and separate it as $W_{N_2+1}$ and $W_{N_1-N_2+2}$, and repeat the process of separation as much as we want. We could consider a generic construction from taking $P$ states $W_{N_1+1},W_{N_2+1},\dots,W_{N_P+1}$ with
\begin{equation*}
N=\sum_{i=1}^P N_i,
\end{equation*} 
To be stitched with a $P$ parts state corresponding to 
\begin{equation*}
\bra	{\overline{W}_P}= \bra{{W}_P}\sigma_x ^{\otimes P}.
\end{equation*}
This construction can be seen graphically as:
\begin{equation}
\begin{tikzpicture} [baseline={([yshift=-.5ex]current bounding box.center)},every node/.style={scale=0.6}] \def\array{$1$,$2$,$.$,$.$,$.$,$.$,$.$,$.$}    \def\arrayx{$.$,$.$,$.$,$.$,$.$,$.$,$.$,$N$}
    \foreach [count=\n] \x in \array{
        \node at ({90+\n*180/(8+1)}:1cm) (n\n) {\x};
        \draw (0,0)--(n\n); 
        };
         \foreach [count=\n] \x in \arrayx{
        \node at ($(0,0)+({-90+\n*180/(8+1)}:1cm)$) (n\n) {\x};
        \draw (0,0)--(n\n); 
        };
        \draw[markxi={0}](0,0) to (0,1);
        \end{tikzpicture} = \sqrt{\frac{P \prod_{i=1} ^P (N_i+1)}{N}}\begin{tikzpicture} [baseline={([yshift=-.5ex]current bounding box.center)},every node/.style={scale=0.6}] \def\array{$N_P$,$.$,$.$,$2$,$1$}    \def\arrayx{$N_1$,$.$,$2$,$1$}
        \def\arrayy{$.$,$.$,$.$,$.$}
    \foreach [count=\n] \x in \array{
        \node at ($(0.5,-0.5)+({-150+\n*180/(4+1)}:0.5cm)$) (n\n) {\x};
        \draw (0.5,-0.5)--(n\n); 
        };
         \foreach [count=\n] \x in \arrayx{
        \node at ($(-0.5,-0.5)+({120+\n*180/(3+1)}:0.5cm)$) (n\n) {\x};
        \draw (-0.5,-0.5)--(n\n); 
        };
         \foreach [count=\n] \x in \arrayy{
        \node at ($(0.5,0.5)+({-60+\n*180/(3+1)}:0.5cm)$) (n\n) {\x};
        \draw (0.5,0.5)--(n\n); 
        };
         \foreach [count=\n] \x in \arrayy{
        \node at ($(0,0.7)+({\n*180/(3+1)}:0.5cm)$) (n\n) {\x};
        \draw (0,0.7)--(n\n); 
        };
         \foreach [count=\n] \x in \arrayy{
        \node at ($(-0.7,0)+({10+\n*180/(3+1)}:0.5cm)$) (n\n) {\x};
        \draw (-0.7,0)--(n\n); 
        };
       \draw[markx={0},markx={0.5},markx={1}] (0,0)--(-0.5,-0.5);
        \draw[markx={0.5},markx={1}] (0,0)--(0.5,-0.5);
        \draw[markx={0.5},markx={1}] (0,0)--(-0.7,0);
        \draw[markx={0.5},markx={1}] (0,0)--(0,0.7);
        \draw[markx={0.5},markx={1}] (0,0)--(0.5,0.5);
\end{tikzpicture} .
\label{eq:WNConstr}
\end{equation}
By convenience, we are choosing the $N_i+1$ part as the one to be contracted in each $\ket{W_{N_i+1}}$ state. Now, we take $n$ copies of the state in its two representations:
\begin{equation*}
\left(
\begin{tikzpicture} [baseline={([yshift=-.5ex]current bounding box.center)},every node/.style={scale=0.6}] \def\array{$1$,$2$,$.$,$.$,$.$,$.$,$.$,$.$}    \def\arrayx{$.$,$.$,$.$,$.$,$.$,$.$,$.$,$N$}
    \foreach [count=\n] \x in \array{
        \node at ({90+\n*180/(8+1)}:1cm) (n\n) {\x};
        \draw (0,0)--(n\n); 
        };
         \foreach [count=\n] \x in \arrayx{
        \node at ($(0,0)+({-90+\n*180/(8+1)}:1cm)$) (n\n) {\x};
        \draw (0,0)--(n\n); 
        };
        \draw[markxi={0}](0,0) to (0,1);
        \end{tikzpicture} \right) ^{\otimes n}= \left(\frac{P \prod_{i=1} ^P (N_i+1)}{N}\right) ^{n/2} \left(\begin{tikzpicture} [baseline={([yshift=-.5ex]current bounding box.center)},every node/.style={scale=0.6}] \def\array{$N_P$,$.$,$.$,$2$,$1$}    \def\arrayx{$N_1$,$.$,$2$,$1$}
        \def\arrayy{$.$,$.$,$.$,$.$}
    \foreach [count=\n] \x in \array{
        \node at ($(0.5,-0.5)+({-150+\n*180/(4+1)}:0.5cm)$) (n\n) {\x};
        \draw (0.5,-0.5)--(n\n); 
        };
         \foreach [count=\n] \x in \arrayx{
        \node at ($(-0.5,-0.5)+({120+\n*180/(3+1)}:0.5cm)$) (n\n) {\x};
        \draw (-0.5,-0.5)--(n\n); 
        };
         \foreach [count=\n] \x in \arrayy{
        \node at ($(0.5,0.5)+({-60+\n*180/(3+1)}:0.5cm)$) (n\n) {\x};
        \draw (0.5,0.5)--(n\n); 
        };
         \foreach [count=\n] \x in \arrayy{
        \node at ($(0,0.7)+({\n*180/(3+1)}:0.5cm)$) (n\n) {\x};
        \draw (0,0.7)--(n\n); 
        };
         \foreach [count=\n] \x in \arrayy{
        \node at ($(-0.7,0)+({10+\n*180/(3+1)}:0.5cm)$) (n\n) {\x};
        \draw (-0.7,0)--(n\n); 
        };
       \draw[markx={0},markx={0.5},markx={1}] (0,0)--(-0.5,-0.5);
        \draw[markx={0.5},markx={1}] (0,0)--(0.5,-0.5);
        \draw[markx={0.5},markx={1}] (0,0)--(-0.7,0);
        \draw[markx={0.5},markx={1}] (0,0)--(0,0.7);
        \draw[markx={0.5},markx={1}] (0,0)--(0.5,0.5);
\end{tikzpicture}  \right)^{\otimes n} .
\end{equation*}
Similarly, as in Equation \eqref{eq:GraphSchurtransfrom}, we apply the Schur transform on both sides, obtaining:
\begin{equation}\label{eq:WGraphStitch}
\bigoplus_{\bm{\lambda}} 
\begin{tikzpicture} [baseline={([yshift=-.5ex]current bounding box.center)},every node/.style={scale=0.5}] \def\array{$\{\lambda^1\}$,$\{\lambda^2\}$,$.$,$.$,$.$,$.$,$.$,$.$}    \def\arrayx{$.$,$.$,$.$,$.$,$.$,$.$,$.$,$\{\lambda^N\}$}
    \foreach [count=\n] \x in \array{
        \node at ({90+\n*180/(8+1)}:1cm) (n\n) {\x};
        \draw (0,0)--(n\n); 
        };
         \foreach [count=\n] \x in \arrayx{
        \node at ($(0,0)+({-90+\n*180/(8+1)}:1cm)$) (n\n) {\x};
        \draw (0,0)--(n\n); 
        };
        \draw[markxi={0}](0,0) to (0,1);
        \end{tikzpicture}  \otimes \begin{tikzpicture} [baseline={([yshift=-.5ex]current bounding box.center)},every node/.style={scale=0.5}] \def\array{$\lambda^1$,$\lambda^2$,$.$,$.$,$.$,$.$,$.$,$.$}    \def\arrayx{$.$,$.$,$.$,$.$,$.$,$.$,$.$,$\lambda^N$}
    \foreach [count=\n] \x in \array{
        \node at ({90+\n*180/(8+1)}:1cm) (n\n) {\x};
        \draw (0,0)--(n\n); 
        };
         \foreach [count=\n] \x in \arrayx{
        \node at ($(0,0)+({-90+\n*180/(8+1)}:1cm)$) (n\n) {\x};
        \draw (0,0)--(n\n); 
        };
        \draw[markxi={0}](0,0) to (0,1);
        \end{tikzpicture} =\left(\frac{P \prod_{i=1} ^P (N_i+1)}{N}\right) ^{n/2} \bigoplus_{\bm{\lambda}} \sum_{\bm{\mu}} \begin{tikzpicture} [baseline={([yshift=-.5ex]current bounding box.center)},every node/.style={scale=0.5}] \def\array{$\{\lambda^{N}\}$,$.$,$.$,$.$,$.$}    \def\arrayx{$\{\lambda^{N_1}\}$,$.$,$\{\lambda^2\}$,$\{\lambda^1\}$}
        \def\arrayy{$.$,$.$,$.$,$.$}
    \foreach [count=\n] \x in \array{
        \node at ($(0.5,-0.5)+({-150+\n*180/(4+1)}:0.5cm)$) (n\n) {\x};
        \draw (0.5,-0.5)--(n\n); 
        };
         \foreach [count=\n] \x in \arrayx{
        \node at ($(-0.5,-0.5)+({120+\n*180/(3+1)}:0.5cm)$) (n\n) {\x};
        \draw (-0.5,-0.5)--(n\n); 
        };
         \foreach [count=\n] \x in \arrayy{
        \node at ($(0.5,0.5)+({-60+\n*180/(3+1)}:0.5cm)$) (n\n) {\x};
        \draw (0.5,0.5)--(n\n); 
        };
         \foreach [count=\n] \x in \arrayy{
        \node at ($(0,0.7)+({\n*180/(3+1)}:0.5cm)$) (n\n) {\x};
        \draw (0,0.7)--(n\n); 
        };
         \foreach [count=\n] \x in \arrayy{
        \node at ($(-0.7,0)+({10+\n*180/(3+1)}:0.5cm)$) (n\n) {\x};
        \draw (-0.7,0)--(n\n); 
        };
       \draw[markx={0},markx={0.5},markx={1}] (0,0)--(-0.5,-0.5);
        \draw[markx={0.5},markx={1}] (0,0)--(0.5,-0.5);
        \draw[markx={0.5},markx={1}] (0,0)--(-0.7,0);
        \draw[markx={0.5},markx={1}] (0,0)--(0,0.7);
        \draw[markx={0.5},markx={1}] (0,0)--(0.5,0.5);
       \node at (-0.25,-0.5) {$(\mu_1)$};
        \node at (0.5,-0.3) {$(\mu_P)$};
\end{tikzpicture}  \otimes \begin{tikzpicture} [baseline={([yshift=-.5ex]current bounding box.center)},every node/.style={scale=0.5}] \def\array{$\lambda^{N}$,$.$,$.$,$.$,$.$}    \def\arrayx{$\lambda^{N_1}$,$.$,$\lambda^2$,$\lambda^1$}
        \def\arrayy{$.$,$.$,$.$,$.$}
    \foreach [count=\n] \x in \array{
        \node at ($(0.5,-0.5)+({-150+\n*180/(4+1)}:0.5cm)$) (n\n) {\x};
        \draw (0.5,-0.5)--(n\n); 
        };
         \foreach [count=\n] \x in \arrayx{
        \node at ($(-0.5,-0.5)+({120+\n*180/(3+1)}:0.5cm)$) (n\n) {\x};
        \draw (-0.5,-0.5)--(n\n); 
        };
         \foreach [count=\n] \x in \arrayy{
        \node at ($(0.5,0.5)+({-60+\n*180/(3+1)}:0.5cm)$) (n\n) {\x};
        \draw (0.5,0.5)--(n\n); 
        };
         \foreach [count=\n] \x in \arrayy{
        \node at ($(0,0.7)+({\n*180/(3+1)}:0.5cm)$) (n\n) {\x};
        \draw (0,0.7)--(n\n); 
        };
         \foreach [count=\n] \x in \arrayy{
        \node at ($(-0.7,0)+({10+\n*180/(3+1)}:0.5cm)$) (n\n) {\x};
        \draw (-0.7,0)--(n\n); 
        };
       \draw[markx={0},markx={1}] (0,0)--(-0.5,-0.5);
        \draw[markx={1}] (0,0)--(0.5,-0.5);
        \draw[markx={1}] (0,0)--(-0.7,0);
        \draw[markx={1}] (0,0)--(0,0.7);
        \draw[markx={1}] (0,0)--(0.5,0.5);
          \node at (-0.2,-0.5) {$\mu^1$};
        \node at (0.5,-0.2) {$\mu^P$};
\end{tikzpicture},
\end{equation}
where we are using the graphical notation for $\{\lamtup\}$ and $[\bm{\lambda}]^{S_n}$ parts as introduced in Chapter \ref{Chapter5}. This case is special because we can obtain explicitly the state in $\{\bm{\lambda}\}$ part. As all the objects used in this construction belong to the W-SLOCC class, we can explicitly compute each state using Equation \eqref{eq:phigen23}. The different states are: First, the state in $\{\bm{\lambda}\}$ of the Schur transform of $W_N$: 
\begin{equation*}
\begin{tikzpicture} [baseline={([yshift=-.5ex]current bounding box.center)},every node/.style={scale=0.5}] \def\array{$\{\lambda^1\}$,$\{\lambda^2\}$,$.$,$.$,$.$,$.$,$.$,$.$}    \def\arrayx{$.$,$.$,$.$,$.$,$.$,$.$,$.$,$\{\lambda^N\}$}
    \foreach [count=\n] \x in \array{
        \node at ({90+\n*180/(8+1)}:1cm) (n\n) {\x};
        \draw (0,0)--(n\n); 
        };
         \foreach [count=\n] \x in \arrayx{
        \node at ($(0,0)+({-90+\n*180/(8+1)}:1cm)$) (n\n) {\x};
        \draw (0,0)--(n\n); 
        };
        \draw[markxi={0}](0,0) to (0,1);
        \end{tikzpicture} 
=\eta_{\lamtup}\ket{{\reallywidehat{\Phi}_{\lamtup}(W_N)}} =\eta_{\lamtup} \frac{n!^{-(N-2)/2 }}{\sqrt{N^n}} \sum_{\bm{\omega}} \sqrt{A_{\lamtup,{\bm{\omega}}}}\ket{\lamtup,{\bm{\omega}} } .
\end{equation*}
The second state is the W-Kronecker state of $N$ parts:
\begin{equation*}
\begin{tikzpicture} [baseline={([yshift=-.5ex]current bounding box.center)},every node/.style={scale=0.5}] \def\array{$\lambda^1$,$\lambda^2$,$.$,$.$,$.$,$.$,$.$,$.$}    \def\arrayx{$.$,$.$,$.$,$.$,$.$,$.$,$.$,$\lambda^N$}
    \foreach [count=\n] \x in \array{
        \node at ({90+\n*180/(8+1)}:1cm) (n\n) {\x};
        \draw (0,0)--(n\n); 
        };
         \foreach [count=\n] \x in \arrayx{
        \node at ($(0,0)+({-90+\n*180/(8+1)}:1cm)$) (n\n) {\x};
        \draw (0,0)--(n\n); 
        };
        \draw[markxi={0}](0,0) to (0,1);
        \end{tikzpicture} = \ket{\mathcal{K}^{W}_{\bm{\lambda}}}.
\end{equation*} 
The next is the graph state in $\{\bm{\lambda}\}$, which can be calculated explicitly as:
\begin{equation*}
\begin{gathered}
     \begin{tikzpicture} [baseline={([yshift=-.5ex]current bounding box.center)},every node/.style={scale=0.5}] \def\array{$\{\lambda^{N}\}$,$.$,$.$,$.$,$.$}    \def\arrayx{$\{\lambda^{N_1}\}$,$.$,$\{\lambda^2\}$,$\{\lambda^1\}$}
        \def\arrayy{$.$,$.$,$.$,$.$}
    \foreach [count=\n] \x in \array{
        \node at ($(0.5,-0.5)+({-150+\n*180/(4+1)}:0.5cm)$) (n\n) {\x};
        \draw (0.5,-0.5)--(n\n); 
        };
         \foreach [count=\n] \x in \arrayx{
        \node at ($(-0.5,-0.5)+({120+\n*180/(3+1)}:0.5cm)$) (n\n) {\x};
        \draw (-0.5,-0.5)--(n\n); 
        };
         \foreach [count=\n] \x in \arrayy{
        \node at ($(0.5,0.5)+({-60+\n*180/(3+1)}:0.5cm)$) (n\n) {\x};
        \draw (0.5,0.5)--(n\n); 
        };
         \foreach [count=\n] \x in \arrayy{
        \node at ($(0,0.7)+({\n*180/(3+1)}:0.5cm)$) (n\n) {\x};
        \draw (0,0.7)--(n\n); 
        };
         \foreach [count=\n] \x in \arrayy{
        \node at ($(-0.7,0)+({10+\n*180/(3+1)}:0.5cm)$) (n\n) {\x};
        \draw (-0.7,0)--(n\n); 
        };
       \draw[markx={0},markx={0.5},markx={1}] (0,0)--(-0.5,-0.5);
        \draw[markx={0.5},markx={1}] (0,0)--(0.5,-0.5);
        \draw[markx={0.5},markx={1}] (0,0)--(-0.7,0);
        \draw[markx={0.5},markx={1}] (0,0)--(0,0.7);
        \draw[markx={0.5},markx={1}] (0,0)--(0.5,0.5);
       \node at (-0.25,-0.5) {$(\mu^1)$};
        \node at (0.5,-0.3) {$(\mu^p)$};
\end{tikzpicture} =\eta_{\bm{\mu}}\bra{{\reallywidehat{\Phi}_{\bm{\mu}}(\overline{W_P})}} \bigotimes_{i=1}^{P}\eta_{\bm{\lambda^i},\mu^i} \ket{\reallywidehat{\Phi}_{\lamtup^i,\mu^i}(W_{N_i+1})} \\ = \eta_{\bm{\mu}}\left(\prod_{i=1}^{P} \eta_{\bm{\lambda}^i,\mu^i} \right) \left(\frac{n!^{-(N-2)/2}}{\sqrt{\left(P \prod_{i=1}^P (N_i+1)\right)^n} }\right) \sum_{\bm{\omega}} \sqrt{A_{\lamtup,\bm{\omega}}} \ket{\lamtup,\bm{\omega}}. 
\end{gathered}
\end{equation*} 
The last state is the graph Kronecker state:
\begin{equation*}
\begin{tikzpicture} [baseline={([yshift=-.5ex]current bounding box.center)},every node/.style={scale=0.5}] \def\array{$\lambda^{N}$,$.$,$.$,$.$,$.$}    \def\arrayx{$\lambda^{N_1}$,$.$,$\lambda^2$,$\lambda^1$}
        \def\arrayy{$.$,$.$,$.$,$.$}
    \foreach [count=\n] \x in \array{
        \node at ($(0.5,-0.5)+({-150+\n*180/(4+1)}:0.5cm)$) (n\n) {\x};
        \draw (0.5,-0.5)--(n\n); 
        };
         \foreach [count=\n] \x in \arrayx{
        \node at ($(-0.5,-0.5)+({120+\n*180/(3+1)}:0.5cm)$) (n\n) {\x};
        \draw (-0.5,-0.5)--(n\n); 
        };
         \foreach [count=\n] \x in \arrayy{
        \node at ($(0.5,0.5)+({-60+\n*180/(3+1)}:0.5cm)$) (n\n) {\x};
        \draw (0.5,0.5)--(n\n); 
        };
         \foreach [count=\n] \x in \arrayy{
        \node at ($(0,0.7)+({\n*180/(3+1)}:0.5cm)$) (n\n) {\x};
        \draw (0,0.7)--(n\n); 
        };
         \foreach [count=\n] \x in \arrayy{
        \node at ($(-0.7,0)+({10+\n*180/(3+1)}:0.5cm)$) (n\n) {\x};
        \draw (-0.7,0)--(n\n); 
        };
       \draw[markx={0},markx={1}] (0,0)--(-0.5,-0.5);
        \draw[markx={1}] (0,0)--(0.5,-0.5);
        \draw[,markx={1}] (0,0)--(-0.7,0);
        \draw[markx={1}] (0,0)--(0,0.7);
        \draw[markx={1}] (0,0)--(0.5,0.5);
          \node at (-0.2,-0.5) {$\mu^1$};
        \node at (0.5,-0.2) {$\mu^P$};
\end{tikzpicture} = \bra{\mathcal{K}^W_{\bm{\mu}}} \bigotimes_{i=1}^{P} \ket{\mathcal{K}^W_{\lamtup^i,\mu^i}}.
\end{equation*}
where $\lamtup^i$ is the subset of external partitions in the $i$-th subgroup in the graph construction. Replacing all these states in Equation \eqref{eq:WGraphStitch} and simplifying the proportionality factors we obtain:
\begin{equation*}
\begin{gathered}
\bigoplus_{\lamtup}  \eta_{\lamtup} \sum_{\bm{\omega}} \sqrt{A_{\lamtup,{\bm{\omega}}}}\ket{\lamtup,{\bm{\omega}} }  \ket{\mathcal{K}^W_{\lamtup}}= 
\bigoplus_{\lamtup}  \sum_{\bm{\mu}}\eta_{\bm{\mu}}\left(\prod_{i=1}^{P} \eta_{\bm{\lambda}^i,\mu^i} \right)  \sum_{\bm{\omega}} \sqrt{A_{\lamtup,\bm{\omega}}} \ket{\lamtup,\bm{\omega}} \bra{\mathcal{K}^W_{\bm{\mu}}} \bigotimes_{i=1}^{P} \ket{\mathcal{K}^W_{\lamtup^i,\mu^i}}
\end{gathered}
\end{equation*} 
This expression is simplified because both states in $\{\bm{\lambda}\}$ are proportional. Then, equating terms with the same $\bm{\lambda}$ on both sides, and simplifying we have:
\begin{equation*} \eta_{\lamtup} \ket{\mathcal{K}^W_{\lamtup}}=  \sum_{\bm{\mu}}\eta_{\bm{\mu}}\left(\prod_{i=1}^{P} \eta_{\bm{\lambda}^i,\mu^i} \right) \bra{\mathcal{K}^W_{\bm{\mu}}} \bigotimes_{i=1}^{P} \ket{\mathcal{K}^W_{\lamtup^i,\mu^i}}.
\end{equation*}
For the right side of the equation, we will label the normalized graph Kronecker state as:
\begin{equation*}
\ket{\mathcal{K}^{\bm{\mu}}_{\lamtup^1|\lamtup ^2| \dots|\lamtup^P}}=\sqrt{f^{\bm{\mu}}}\bra{\mathcal{K}^W_{\bm{\mu}}} \bigotimes_{i=1}^{P} \ket{\mathcal{K}^W_{\lamtup^i,\mu^i}},
\end{equation*}
with $f^{\bm{\mu}}=\prod_{i}^P f^{\mu^i} $ and the notation ${\lamtup^1|\lamtup ^2| \dots|\lamtup^P}$ refers that we are interpreting the graph Kronecker state $\ket{\mathcal{K}^W_{\bm{\lambda}^i,\mu^i}}$ as a bipartite Kronecker state of the partition $\mu^i$ and copies of $\mu^i$ in $\bm{\lambda}^i$ for the normalization. Then, we have:
\begin{equation}\ket{\mathcal{K}^W_{\lamtup}}=  \sum_{\bm{\mu}} \frac{\eta_{\bm{\mu}}\left(\prod_{i=1}^{P} \eta_{\bm{\lambda}^i,\mu^i} \right)}{ \sqrt{f^{\bm{\mu}}} \eta_{\lamtup} } \ket{\mathcal{K}^{\bm{\mu}}_{\lamtup^1|\lamtup ^2| \dots|\lamtup^P}}.
\label{eq:KronWparts}
\end{equation}
Note that the W-Kronecker state in the left side of the equation is unique in a set $\bm{\lambda}$; then, all the possible graph Kronecker states obtained with different sets of inner partitions $\bm{\mu}$ must be combined in a special way to keep the structure of W-Kronecker states. So, we define the normalized coefficients
\begin{equation*}
\mathcal{C}^{\bm{\mu}}_{\lamtup^1|\lamtup ^2| \dots|\lamtup^P}= \frac{\eta^2_{\bm{\mu}}\left(\prod_{i=1}^{P} \eta^2_{\bm{\lambda}^i,\mu^i} \right)}{ f^{\bm{\mu}} \eta^2_{\lamtup} } , \quad \sum_{\bm{\mu}}  \mathcal{C}^{\bm{\mu}}_{\lamtup^1|\lamtup ^2| \dots|\lamtup^P} =1.
\end{equation*}
These coefficients tell the weight of the graph state built from the set $\bm{\mu}$ of inner partitions in the construction in the W-Kronecker state. This coefficient can also be interpreted from Equation \eqref{eq:KronWparts} as the square of the probability of having a set of inner irreps $\bm{\mu}$ once that the set of external partitions is determined as $\bm{\lambda}$. Note how the relation in the previous equation must hold for any state in the W-SLOCC class because the Kronecker subspace is independent of the parameters of the state. Our goal now is to find, in the asymptotic limit, the inner partitions that are more relevant in constructing the W-Kronecker state of $N$ parts. For this, let us recall the definition of $\eta_{\lamtup}$ from Equation \eqref{eq:etaDef}. By picking the states to be the ones with the set of marginal spectra according to the set of irreps, the coefficients are given by:
\begin{equation*}
\mathcal{C}^{\bm{\mu}}_{\lamtup^1|\lamtup ^2| \dots|\lamtup^P}= \frac{p(\bm{\mu}|\psi_{\bm{\mu}}) Z_{\lamtup} (\psi_{\lamtup}) \prod_{i=1}^{P} p(\bm{\lambda}^i,\mu^i|\psi_{\bm{\lambda}^i,\mu^i}) }{f^{\bm{\mu}} p(\lamtup|\psi_{\lamtup}) Z_{\bm{\mu}} (\psi_{\bm{\mu}})\prod_{i=1}^P Z_{\bm{\lambda}^i,\mu^i}(\psi_{\bm{\lambda}^i,\mu^i})  }.
\end{equation*}
Defining a rate function for the coefficients as:
\begin{equation}
\mathscr{C}^{\bm{\mu}}_{\lamtup^1|\lamtup ^2| \dots|\lamtup^P} = \lim_{n\rightarrow \infty} -\frac{1}{n} \log( \mathcal{C}^{\bm{\mu}}_{\lamtup^1|\lamtup ^2| \dots|\lamtup^P} ),
\label{eq:coefrate}
\end{equation}
we can use the rate of the dimensions of irreps from Equation \eqref{eq:asymptoticsflambda}, and the rate definitions from \eqref{eq:RateP} and \eqref{eq:RateZ}. By replacing everything we obtain:
\begin{equation*}
\begin{gathered}
\mathscr{C}^{\bm{\mu}}_{\lamtup^1|\lamtup ^2| \dots|\lamtup^P} = R(\overline{\bm{\mu}}|\overline{\bm{\mu}})-\zeta(\overline{\lamtup}|\overline{\lamtup})-R(\overline{\lamtup}|\overline{\lamtup})+\zeta(\overline{\bm{\mu}}|\overline{\bm{\mu}})\\
+ \sum_{i=1}^{P} H(\overline{\mu}^i) + \zeta(\overline{\lambda}^i,\overline{\mu}^i|\overline{	\lambda}^i,\overline{\mu}^i) - R(\overline{\lambda}^i,\overline{\mu}^i| \overline{\lambda}^i,\overline{\mu}^i).
 \end{gathered}
\end{equation*}
By the Keyl-Werner theorem, the rates of the form $R(\overline{\bm{\lambda}}|\overline{\bm{\lambda}})$ vanish, then, we obtain:
\begin{equation*}
\mathscr{C}^{\bm{\mu}}_{\lamtup^1|\lamtup ^2| \dots|\lamtup^P} = -\zeta(\overline{\lamtup}|\overline{\lamtup})+\zeta(\overline{\bm{\mu}}|\overline{\bm{\mu}}) + \sum_{i=1}^{P} H(\overline{\mu}^i) + \zeta(\overline{\lambda}^i,\overline{\mu}^i|\overline{	\lambda}^i,\overline{\mu}^i).
\end{equation*}
We want to calculate the set $\bm{\mu}^*$ that minimizes this rate. In other words, we want to find the set of inner partitions that maximizes asymptotically the coefficient. We will calculate the partial derivative in each inner partition $\mu^i$ to find this set. We can use the Equations \eqref{eq:zitall} and \eqref{eq:cspec} to show that 
\begin{equation*}
\frac{\partial \zeta(\overline{\lamtup}|\overline{\lamtup}) }{\partial \overline{\lambda}^i} = \log \left(\frac{ c^i_{\lamtup} - \overline{\lambda}^i }{1-\overline{\lambda}^i- c^{i}_{\lamtup}}\right).
\end{equation*}
By using this result and the partial derivative in the binary entropy, 
\begin{equation*}
\frac{\partial H(\overline{\mu}^i)}{\partial \overline{\mu} ^i} = \log(\frac{1-\overline{\mu}^i}{\mu^i} ),
\end{equation*}
we get that the derivatives for each inner partition $\mu^i$ of the rate of the coefficients are given by:
\begin{equation*}
\frac{\partial \mathscr{C}^{\bm{\mu}}_{\lamtup^1|\lamtup ^2| \dots|\lamtup^P}}{\partial \overline{\mu}^i}= \log(\frac{(1-\overline{\mu}^i)(c_{\lamtup ^i,\mu^i}^{N_i+1} - \overline{\mu} ^i)(c_{\bm{\mu}}^i - \overline{\mu}^i)}{\overline{\mu}^i(1- \overline{\mu} ^i - c^{N_{i}+1}_{\lamtup ^i,\mu^i})(1- \overline{\mu}^i- c^i_{\bm{\mu}} )}),
\end{equation*}
where we introduced the states $\psi_{\lamtup ^i,\mu^i}$ as the $N_i+1$ parts W class state with marginal spectra $(\lamtup ^i,\mu^i)$ and coefficients $c_{\lamtup ^i,\mu^i}$, and the state $\psi_{\bm{\mu}}$ the $P$ parts W class state with marginal spectra $\bm{\mu}$ and coefficients $c_{\bm{\mu}}$. The expression in the last equation is only zero when
\begin{equation*}
(1-\overline{\mu}^i)(c_{\lamtup ^i,\mu^i}^{N_i+1} - \overline{\mu} ^i)(c_{\bm{\mu}}^i - \overline{\mu}^i)=\overline{\mu}^i(1- \overline{\mu} ^i - c^{N_{i}+1}_{\lamtup ^i,\mu^i})(1- \overline{\mu}^i- c^i_{\bm{\mu}} ).
\end{equation*}
This condition can be simplified to:
\begin{equation*}
(1-2\overline{\mu}^i)(\overline{\mu}^i ( 1- \overline{\mu}^i) -c^{N_{i}+1}_{\lamtup ^i,\mu^i} c^i_{\bm{\mu}} ) =0.
\end{equation*}
Then, each local minimum is achieved for the solution to:
\begin{equation}
\overline{\mu}^i (1- \overline{\mu}^i) =c^{N_{i}+1}_{\lamtup ^i,\mu^i} c^i_{\bm{\mu}} .
\label{eq:coefW1}
\end{equation}
On the other hand, note that $\overline{\mu}^i$ is the local eigenvalue for the $P$ parts state $\psi_{\bm{\mu}}$ in the $i$-th part, then from \eqref{eq:cspec} we have:
\begin{equation}
\overline{\mu}^i (1- \overline{\mu}^i) = c_{\bm{\mu}}^{i} \left( \sum_{j \neq i,1}^{P} c^{j}_{\bm{\mu}} \right).
\label{eq:coefW2}
\end{equation}
Simultaneously, $\overline{\mu}^i$ is the local eigenvalue of the $N_i+1$ part of the state $\psi_{\lamtup ^i,\mu^i}$, then we have:
\begin{equation}
\overline{\mu}^i (1- \overline{\mu}^i) =c^{N_{i}+1}_{\lamtup ^i,\mu^i}\left( \sum_{j =1}^{N_i} c^{j}_{\bm{\lambda}^i,\mu^i} \right).
\label{eq:coefW3}
\end{equation}
Then, by combining the expressions in the last three Equations \eqref{eq:coefW1}, \eqref{eq:coefW2}, and \eqref{eq:coefW3}, we can note that the minimization condition fixes: 
\begin{equation*}
c^{N_{i}+1}_{\lamtup ^i,\mu^i} = \sum_{j\neq i,1}^P c^{j}_{\bm{\mu}} , \quad c^{i}_{\bm{\mu}}= \sum_{j=1}^{N_i} c^{j}_{\lamtup ^i,\mu^i}.
\end{equation*}
With this, we can find that a global minimum is achieved when the condition of the previous equation is satisfied for each part. Then, the global minimum for the rate in Equation \eqref{eq:coefrate} is obtained by the following coefficients:
\begin{equation*}
     c^{N_i+1}_{\lamtup ^i ,\mu^i}=\sum_{j=1|\lambda^j \notin \lamtup ^i}^{N} c^{j}_{\lamtup} \quad,\quad c^{i}_{\bm{\mu}}=\sum_{j=1|\lambda^j \in \lamtup^i}^{N} c^{j}_{\lamtup},
\end{equation*}
where all the minimization conditions in Equation \eqref{eq:coefW1} are satisfied simultaneously for all partitions $\mu^{i}$. We can finally conclude that values $\mathcal{C}^{\bm{\mu}}_{(\lamtup^1 ,\lamtup^2,\dots ,\lamtup ^P)}$ are exponentially concentrated around the set of partitions $\bm{\mu}$ such that for each $\mu^{i}$
\begin{equation}\label{eq:concentration}
\overline{\mu}^{i}(1-\overline{\mu}^{i})=\left(\sum_{j=1|\lambda^j \notin \lamtup^i}^{N} c^{j}_{\lamtup} \right) \left( \sum_{j=1|\lambda^j \in \lamtup^i}^{N} c^{j}_{\lamtup}\right),
\end{equation}
which can be understood as the partitions resulting from the grouping selected to build the full state. This result can be understood better in the next sense: first, recall the construction used to build the $W_N$ state in Equation \eqref{eq:WNConstr}; then, after picking a set $\lamtup$ we have for the Kronecker states:
\begin{equation*}
\begin{tikzpicture} [baseline={([yshift=-.5ex]current bounding box.center)},every node/.style={scale=0.5}] \def\array{$\lambda^1$,$\lambda^2$,$.$,$.$,$.$,$.$,$.$,$.$}    \def\arrayx{$.$,$.$,$.$,$.$,$.$,$.$,$.$,$\lambda^N$}
    \foreach [count=\n] \x in \array{
        \node at ({90+\n*180/(8+1)}:1cm) (n\n) {\x};
        \draw (0,0)--(n\n); 
        };
         \foreach [count=\n] \x in \arrayx{
        \node at ($(0,0)+({-90+\n*180/(8+1)}:1cm)$) (n\n) {\x};
        \draw (0,0)--(n\n); 
        };
        \draw[markxi={0}](0,0) to (0,1);
        \end{tikzpicture} =\sum_{\bm{\mu}} \mathcal{C}^{\bm{\mu}}_{\lamtup^1|\lamtup^2|\dots|\lamtup^P} \sqrt{f^{\bm{\mu}}} \begin{tikzpicture} [baseline={([yshift=-.5ex]current bounding box.center)},every node/.style={scale=0.5}] \def\array{$\lambda^{N}$,$.$,$.$,$.$,$.$}    \def\arrayx{$\lambda^{N_1}$,$.$,$\lambda^2$,$\lambda^1$}
        \def\arrayy{$.$,$.$,$.$,$.$}
    \foreach [count=\n] \x in \array{
        \node at ($(0.5,-0.5)+({-150+\n*180/(4+1)}:0.5cm)$) (n\n) {\x};
        \draw (0.5,-0.5)--(n\n); 
        };
         \foreach [count=\n] \x in \arrayx{
        \node at ($(-0.5,-0.5)+({120+\n*180/(3+1)}:0.5cm)$) (n\n) {\x};
        \draw (-0.5,-0.5)--(n\n); 
        };
         \foreach [count=\n] \x in \arrayy{
        \node at ($(0.5,0.5)+({-60+\n*180/(3+1)}:0.5cm)$) (n\n) {\x};
        \draw (0.5,0.5)--(n\n); 
        };
         \foreach [count=\n] \x in \arrayy{
        \node at ($(0,0.7)+({\n*180/(3+1)}:0.5cm)$) (n\n) {\x};
        \draw (0,0.7)--(n\n); 
        };
         \foreach [count=\n] \x in \arrayy{
        \node at ($(-0.7,0)+({10+\n*180/(3+1)}:0.5cm)$) (n\n) {\x};
        \draw (-0.7,0)--(n\n); 
        };
       \draw[markx={0},markx={1}] (0,0)--(-0.5,-0.5);
        \draw[markx={1}] (0,0)--(0.5,-0.5);
        \draw[markx={1}] (0,0)--(-0.7,0);
        \draw[markx={1}] (0,0)--(0,0.7);
        \draw[markx={1}] (0,0)--(0.5,0.5);
          \node at (-0.2,-0.5) {$\mu^1$};
        \node at (0.5,-0.2) {$\mu^P$};
\end{tikzpicture}.
\end{equation*}
We can find each $\mu^i$  for the concentration by first solving the set of equations:
\begin{equation*}
c_{\lamtup}^i (1- c_{\lamtup}^i -c_{\lamtup} ^0) =\overline{\lambda}^i (1-\overline{\lambda}^i) \quad, \forall i \in \{1,2,\dots ,N\},
\end{equation*}
to get the set $\{ c_{\lamtup}^i\}$. Then, we separate this set according to the separation given by $\mu^i$; for example, for $\mu^1$, we have:
\begin{equation*}
\begin{tikzpicture} [baseline={([yshift=-.5ex]current bounding box.center)},every node/.style={scale=0.5}] \def\array{$\lambda^{N}$,$.$,$.$,$.$,$.$}    \def\arrayx{$\lambda^{N_1}$,$.$,$\lambda^2$,$\lambda^1$}
        \def\arrayy{$.$,$.$,$.$,$.$}
    \foreach [count=\n] \x in \array{
        \node at ($(0.5,-0.5)+({-150+\n*180/(4+1)}:0.5cm)$) (n\n) {\x};
        \draw (0.5,-0.5)--(n\n); 
        };
         \foreach [count=\n] \x in \arrayx{
        \node at ($(-0.5,-0.5)+({120+\n*180/(3+1)}:0.5cm)$) (n\n) {\x};
        \draw (-0.5,-0.5)--(n\n); 
        };
         \foreach [count=\n] \x in \arrayy{
        \node at ($(0.5,0.5)+({-60+\n*180/(3+1)}:0.5cm)$) (n\n) {\x};
        \draw (0.5,0.5)--(n\n); 
        };
         \foreach [count=\n] \x in \arrayy{
        \node at ($(0,0.7)+({\n*180/(3+1)}:0.5cm)$) (n\n) {\x};
        \draw (0,0.7)--(n\n); 
        };
         \foreach [count=\n] \x in \arrayy{
        \node at ($(-0.7,0)+({10+\n*180/(3+1)}:0.5cm)$) (n\n) {\x};
        \draw (-0.7,0)--(n\n); 
        };
       \draw[markx={0},markx={1}] (0,0)--(-0.5,-0.5);
        \draw[markx={1}] (0,0)--(0.5,-0.5);
        \draw[markx={1}] (0,0)--(-0.7,0);
        \draw[markx={1}] (0,0)--(0,0.7);
        \draw[markx={1}] (0,0)--(0.5,0.5);
          \node at (-0.2,-0.5) {$\mu^1$};
        \node at (0.5,-0.2) {$\mu^P$};
\end{tikzpicture} \rightarrow \begin{tikzpicture} [baseline={([yshift=-.5ex]current bounding box.center)},every node/.style={scale=0.5}] \def\array{$\lambda^{N}$,$.$,$.$,$.$,$.$}    \def\arrayx{$\lambda^{N_1}$,$.$,$\lambda^2$,$\lambda^1$}
        \def\arrayy{$.$,$.$,$.$,$.$}
    \foreach [count=\n] \x in \array{
        \node at ($(0.5,-0.5)+({-150+\n*180/(4+1)}:0.5cm)$) (n\n) {\x};
        \draw (0.5,-0.5)--(n\n); 
        };
         \foreach [count=\n] \x in \arrayx{
        \node at ($(-0.5,-0.5)+({120+\n*180/(3+1)}:0.5cm)$) (n\n) {\x};
        \draw (-0.5,-0.5)--(n\n); 
        };
         \foreach [count=\n] \x in \arrayy{
        \node at ($(0.5,0.5)+({-60+\n*180/(3+1)}:0.5cm)$) (n\n) {\x};
        \draw (0.5,0.5)--(n\n); 
        };
         \foreach [count=\n] \x in \arrayy{
        \node at ($(0,0.7)+({\n*180/(3+1)}:0.5cm)$) (n\n) {\x};
        \draw (0,0.7)--(n\n); 
        };
         \foreach [count=\n] \x in \arrayy{
        \node at ($(-0.7,0)+({10+\n*180/(3+1)}:0.5cm)$) (n\n) {\x};
        \draw (-0.7,0)--(n\n); 
        };
       \draw[markx={0},markx={1}] (0,0)--(-0.5,-0.5);
        \draw[markx={1}] (0,0)--(0.5,-0.5);
        \draw[markx={1}] (0,0)--(-0.7,0);
        \draw[markx={1}] (0,0)--(0,0.7);
        \draw[markx={1}] (0,0)--(0.5,0.5);
        \draw[line width=0.5mm,dashed,red] (-1,-0.2) to[bend left=60] (0,-1);
\end{tikzpicture},
\end{equation*}
then, according to \eqref{eq:concentration} we find $\mu^1$ as the solution for:
\begin{equation}
\overline{\mu}^1(1-\overline{\mu}^1) =\left( \sum_{j=1}^{N_1} c_{\lamtup}^{j} \right) \left( \sum_{j=N1+1}^{N} c_{\lamtup} ^{j} \right).
\label{eq:RateMu}
\end{equation}
Note that this is a very strong functional correlation since a similar construction can be performed in any possible separation of sets $\bm{\lambda}^i$, and the relation still holds. This result shows that as a function of $\bm{\mu}$, the coefficient $\mathcal{C}^{\bm{\mu}}_{\lamtup^1|\lamtup ^2| \dots|\lamtup^P}$ concentrates at the reduced partitions $\overline{\mu^i}$ that satisfies a similar relation as the satisfied by the marginal spectra of the W class of  Equation \eqref{eq:cspec}. Remembering that the coefficients $\mathcal{C}^{\bm{\mu}}_{\lamtup^1|\lamtup ^2| \dots|\lamtup^P}$  are the square of the probability of having a set of inner irreps $\bm{\mu}$ when the set of external partitions is fixed as $\bm{\lambda}$. This can be seen easily from Equation \eqref{eq:KronWparts}. The previous result tells that such probability is asymptotically concentrated in the set of inner partitions that maximizes the rate, given by Equation 
\eqref{eq:RateMu}, which exhibits a relation equivalent to the relation of the spectra in W class states.\\

\noindent This result again shows how the W class is a special kind of multipartite entangled states with a nice mathematical structure that allows to perform calculations that are generally harder for states in other classes. In this case, the W class structure allowed us to calculate explicitly the states in $GL_2$, which is much harder for states that do not belong to the W class. Because of this nice structure of the W class, we think that W states must be considered elemental pieces for studying multipartite entanglement, as we did in the previous chapters.\\

\noindent The results presented in this chapter are secondary because they are not needed to calculate Kronecker subspaces. However, they are relevant on their own in the fields of representation theory, SLOCC classification, and the quantum marginal problem, respectively. Because of this, we decided to include them in a separate chapter and not several appendices. This permits us to highlight their relevance independently from the main content of this document.

\chapter{Conclusions and Future work}
\label{Conclusions}
Entanglement is crucial in quantum physics, marking a fundamental difference between classical and quantum behaviors. It is the foundation of various quantum protocols with no classical equivalent. However, understanding the entanglement structure becomes increasingly complex in a multipartite setup. In this thesis, we used the representation theory of the symmetric group, $S_n$, and the General Linear group of dimension two, $GL_2$, for generating vector spaces of multipartite locally maximally entangled states, which we named Kronecker subspaces. Kronecker subspaces appear in the invariant subspace of the decomposition into irreps of the tensor product of irreps of $S_n$. With the method described in this document, particularly in Section \ref{Chapter5}, we calculated maximally entangled states of several systems of three parts. The most impressive case for three-partite states is a three-dimensional Kronecker subspace, where each local Hilbert space is of dimension $275$. For the four-partite case, we made it to calculate a basis for a Kronecker subspace of dimension $39$, where each local Hilbert space is of dimension $48$. All these calculations were made in exact form, obtaining states with coefficients as roots of rational numbers. Numerical approaches can be used to obtain even larger Kronecker subspaces.\\

\noindent In this dissertation, we have presented significant results related to problems in multipartite entanglement. We introduced a graphical construction of multipartite qubit states using $W_3$ states and bipartite states as building blocks. This construction is called the \textit{W-state stitching}, which can be easily understood graphically. We show how the parameters of the bipartite states used in this construction can be decomposed as operations on the $W_3$ states (or vertices) conveniently, and by that permits using the inherent continuous symmetry of $W_3$ states. This symmetry helps to identify the crucial parameters for classifying states under SLOCC. By translating these symmetries into a graphical tool, namely \textit{parameter pushing}, it is possible to associate specific topologies in the stitching procedure to SLOCC classes. This association allows manipulating SLOCC classes using the remaining parameters after the pushing process. This novel approach enhances our comprehension of multipartite entanglement and offers practical tools for analyzing and working with it.\\

\noindent Through stitching and pushing techniques, we have discovered topologies that enable us to obtain any three-qubit pure state and four-qubit pure state. This method has also allowed us to establish explicit connections with the classification of four qubits presented in \cite{Verstraete}. Moreover, our graphical approach provides a natural way to construct Invariants in multi-qubit systems. Typically, Invariants are complex mathematical constructs, but our technique allows us to understand and manually compute many of these objects. This innovative approach not only simplifies the introduction of complex concepts like invariant theory from a physical standpoint but also helps to clarify the role of invariants as initial filters in SLOCC classification. Firstly, we demonstrated how the unique invariant for three qubits arises from this approach. Furthermore, we show how to construct a basis for the invariant ring in four-qubit systems. We propose 17 independent invariants for the five-qubit scenario to conclude our invariant exploration. These invariants, due to their intricate nature, have previously remained unresolved. Importantly, our approach is not limited to specific qubit counts, and the principles established within this construction can be applied to systems with any number of qubits.\\

\noindent In our research, we demonstrate how the stitching process can be used to construct subspaces of multipartite maximally entangled states. This construction involves applying the Schur transform on graph states, which allows for the identification of the $S_n$ and $GL_2$ subspaces linked with the graph. By recognizing that the $S_n$ part corresponds to a subspace of maximally entangled states, which we call the Kronecker subspace, we can build it from the structure of the graph and the easily computable Kronecker states of the $W$ class. This construction enables us to build all Kronecker subspaces present in the decomposition of three and four qubits. We also provide the conditions on the graph to reproduce all the Kronecker subspaces for any number of qubits. \\

\noindent  In this document we presented significant progress in calculating Kronecker states and, hence, in calculating Clebsch-Gordan coefficients of the symmetric group. By endowing the problem with physical significance, we have developed an efficient algorithm that addresses this challenge. This method enables us to calculate three-partite Clebsch-Gordan coefficients exactly up to $n=12$, where a single set of irreps $\bm{\lambda}$ generates three orthonormal vectors with more than 20 million coefficients. Similarly, for four-part systems, we can calculate Clebsch-Gordan coefficients exactly up to $n=9$, resulting in $39$ orthonormal vectors with more than $5\times10^{6}$ coefficients for one set of irreps $\bm{\lambda}$. The primary limitation of this algorithm is the ability to store Kronecker vectors and achieve orthogonalization with exact precision. \\

\noindent Our research has contributed to the mathematical understanding of multipartite entanglement and has provided practical solutions. We have successfully applied our methods to classify entanglement, define and compute invariants, and explicitly construct subspaces of multipartite maximally entangled states. This work sheds light on the complexities of multipartite entanglement and demonstrates the effectiveness of our approaches in addressing the key challenges in this field.

\subsection*{Future work}

The study of multipartite entanglement is given a new perspective through the results and methods presented here, but some interesting questions remain to be answered. In this context, we have compiled a list of some of these questions.\\

\noindent The investigation was motivated by the need to gain a deeper understanding of the mathematical structure of Kronecker states. These states are a particular type of maximally entangled states that establish a vector space structure and have symmetries under the diagonal action of the symmetric group. This intrinsic property makes Kronecker states significant in quantum information protocols, such as Quantum Error Correction, Quantum Secret Sharing, and probing issues like the superadditivity of communication capacity through quantum channels. With a more refined comprehension of the mathematical intricacies of Kronecker states, their applicability and impact in various applications can now be further explored.\\

\noindent The concept of stitching for constructing multi-qubit states leads to intriguing observations. Complex graphs can be simplified to equivalent graphs defined by their SLOCC classes. For instance, any graph describing three-qubit states must correspond to a triangle graph as it encompasses all possible three-qubit states. Identifying such equivalences holds promise for devising classification methods applicable to systems extending beyond four qubits. \\

\noindent When constructing graph Kronecker states, it is common to observe different combinations of inner partitions (denoted by $\bm{\mu}$) leading to orthogonal Kronecker states, even when the corresponding graph is not OER. On the other hand, achieving orthogonality between two graph Kronecker states can be computationally challenging in high-dimensional spaces. Therefore, finding conditions that establish direct orthogonality between Kronecker states is paramount. \\

\noindent In this study, we show that within the realm of qubit systems, $W$ states and bipartite states can serve as fundamental building blocks for constructing any multi-qubit state, spanning the entire spectrum of possible entanglement classes. This perspective provides a pathway to quantify multipartite entanglement by assessing the number of $W$ states and the corresponding graph architecture needed to reproduce a given state accurately. This conceptual framework could lead to formulating a resource theory centered around these ideas, establishing a systematic methodology for understanding the essence of multipartite entanglement.

\appendix
\chapter{{\texorpdfstring{$S_n$}{Sn} representations in Two-type quantum systems}}
\label{AppendixI}
\begin{figure}[ht]
    \centering
    \includegraphics[scale=0.8]{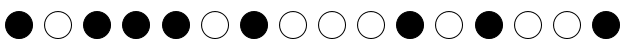}
    \caption{Two-type quantum system}
    \label{fig:Twotypesystem}
\end{figure}

One useful application from the method shown in section \ref{myirreps} for building $S_n$ representations is found when considering a quantum system with two types of quantum states $\rho_0,\rho_1$ whose Schur transformation needs to be computed. Such a system can be considered a linear array of two kinds of particles, as shown in Figure \ref{fig:Twotypesystem}. \\

\begin{figure}[ht]
    \centering
    \includegraphics[scale=0.8]{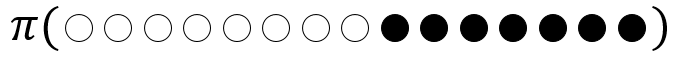}
    \caption{Ordered Two-type quantum system}
    \label{fig:Twotypesystem2}
\end{figure}

\begin{figure}[ht]
    \centering
    \includegraphics{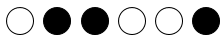}
    \caption{A system with six quantum states of two types}
    \label{fig:sixquantum}
\end{figure}

\noindent The same sequence can always be obtained by applying a permutation $\pi$ on an ordered sequence with the same number of states in each of the two kinds (See Figure \ref{fig:Twotypesystem2}).
Then, the particular distribution is completely defined by the permutation $\pi$ (which is not unique) and the number of particles in each type $n_0,n_1$ with $n=n_0+n_1$. Let us first to consider the ordered system $\rho_{n_0,n_1}$ and its Schur transform
\begin{equation*}
    \rho_{n_0,n_1}=\rho_0^{\otimes n_0} 
 \otimes \rho_1^{\otimes n_1} \sim \bigoplus_{\lambda \vdash n,d} V^{\{\lambda\}}(\rho) \otimes \Omega^{[\lambda]}(n_0,n_1)
\end{equation*}
where $V^{\{\lambda\}}(\rho)$ is the matrix representation of $\rho$ in the irrep $\{\lambda\}$, and $\Omega^{\lambda}(n_0,n_1)$ is a representation in the irrep $[\lambda]$ of $S_n$. As $\rho_0^{\otimes n_0}$ and  $\rho_1^{\otimes n_1}$ are completely symmetric, each of them belongs to the trivial subspace $[n_0]$ and $[n_1]$ in $S_n$ respectively in the correspondent Schur transform. This means that the possible partitions $\lambda$ in the joint Schur transform are restricted to $d=2$, and $\Omega^{[\lambda]}_{(n_0,n_1)}$ must be a rank 1 matrix \cite{Emili}. On the other hand, $\rho$ is an invariant state under all permutations of the first $n_0$ elements and under all permutations of the last $n_1$ elements (or any combination of both). The projectors to each subset of permutations are  
\begin{equation}
    L^{[\lambda]}_{n_0}=\frac{1}{n_0!} \sum_{\pi \in l_{n_0}} D^{[\lambda]}(\pi) \quad, \quad R^{[\lambda]}_{n_1}=\frac{1}{n_1!} \sum_{\pi \in r_{n_1}} D^{[\lambda]}(\pi),
\end{equation}
where $D^{[\lambda]}(\pi)$ is the matrix representation of permutation $\pi$ in irrep $[\lambda]$, $l_{n_0}$ is the set of all permutations in the first $n_0$ elements, and $r_{n_1}$ is the set of all permutations in the last $n_1$ elements. The projector to the invariant subspace associated to $\rho_{n_0,n_1}$ in $[\lambda]$ is $L_{n_0}^{[\lambda]}R_{n_1}^{[\lambda]}$ which is also a rank one projector, due to the invariance of $L$ and $R$. This particular property allows us to identify
\begin{equation}
    \Omega^{\lambda}(n_0,n_1) = L^{\lambda}_{n_0}R^{\lambda}_{n_1}.
    \label{eq:OmegaLR}
\end{equation}
Moreover, this relation tells that the $S_n$ representations of systems with the same permutation symmetry are equal up to normalization, as they belong to the same one-dimensional space. Due to its rank, the matrix $\Omega^{\lambda}(n_0,n_1)$ can be written from its only normalized eigenvector $\ket{\Omega^{[\lambda]}_{n_0,n_1}}$  as:
\begin{equation}
    \Omega^{[\lambda]}(n_0,n_1) =\ket{\Omega^{[\lambda]}_{n_0,n_1}}\bra{\Omega^{[\lambda]}_{n_0,n_1}}
\end{equation}
This vector can be obtained by computing the Schur transform for any pure state with the same symmetry as $\rho_{n_0,n_1}$ and the simplest one is a sequence of qubits with $n_0$ qubits in state $0$ and $n_1$ qubits in state $1$ whose Schur transform can be obtained from Equation \eqref{eq:Schurtransformsequence}. The $S_n$ part is then
\begin{equation}
    \ket{\Omega^{[\lambda]}_{n_0,n_1}} \propto \sum_{q}\prod_{i=1}^{n}\Gamma^{\lambda_{i},\omega_{i},i}_{q_i,s_i} \ket{\lambda,q}
\end{equation}
where $s$ is the ordered qubits sequence  $s=\ket{0^{\otimes n_0}1^{\otimes n_1}}$. All of these arguments for unordered sequences still apply with the difference that projectors cannot be understood as Left and Right projectors. However, the invariant subspace is still rank one, allowing the Schur transform computation using binary sequences. For example, for Figure \ref{fig:Twotypesystem}, the $S_n$ part of the Schur transform can be obtained from the sequence $s=\ket{0100010111010110}$. So, for any two-type system which can be translated to a sequence $s$, its $S_n$ part of Schur transform is obtained with
\begin{equation}
    \ket{\Omega^{[\lambda]}_{s}}=\sqrt{\frac{\lambda!(1+n-\lambda)!}{n_0!(1+n-2\lambda)(n-n_0)!}}\sum_{Y}  \prod_{i=1}^{n}\Gamma^{\lambda_{i},\omega_{i},i}_{y_i,s_i} \ket{\lambda,q}
    \label{eq:VectorOmega}
\end{equation}
where the normalization factor was written explicitly. To understand the implications and simplicity of this approach, we will use it as an example. Consider a system of 6 quantum states of two types as shown in Figure \ref{fig:sixquantum}
for this we can compute the $S_n$ part of Schur transform for partitions $[\lambda]=\{[0],[1],[2],[3]\}$ with the sequence $s=\ket{011001}$. Let us compute explicitly the vector for $\lambda=1$, whose Yamanouchi symbols are
\begin{equation}
\begin{gathered}
    q^{[1]}_1=\{0,0,0,0,0,1\}, \quad q^{[1]}_2=\{0,0,0,0,1,0\}, \quad q^{[1]}_3=\{0,0,0,1,0,0\}, \\ q^{[1]}_4=\{0,0,1,0,0,0\} , \quad q^{[1]}_5=\{0,1,0,0,0,0\}
    \end{gathered}
\end{equation}
Then replacing in \ref{eq:VectorOmega} we obtain
\begin{equation}
    \begin{gathered}
        \ket{\Omega_{011001}^{[1]}}= \sqrt{\frac{1!(1+6-1)!}{3!(1+6-2(1))3!}} \left(\Gamma^{0,0,1}_{0,0}\Gamma^{0,1,2}_{0,1}\Gamma^{0,2,3}_{0,1}\Gamma^{0,2,4}_{0,0}\Gamma^{0,2,5}_{0,0}\Gamma^{1,3,6}_{1,1} \ket{1,\{0,0,0,0,0,1\}} \right. \\
        +\Gamma^{0,0,1}_{0,0}\Gamma^{0,1,2}_{0,1}\Gamma^{0,2,3}_{0,1}\Gamma^{0,2,4}_{0,0}\Gamma^{1,2,5}_{1,0}\Gamma^{1,3,6}_{0,1} \ket{1,\{0,0,0,0,1,0\}} \\
        +\Gamma^{0,0,1}_{0,0}\Gamma^{0,1,2}_{0,1}\Gamma^{0,2,3}_{0,1}\Gamma^{1,2,4}_{1,0}\Gamma^{1,2,5}_{0,0}\Gamma^{1,3,6}_{0,1} \ket{1,\{0,0,0,1,0,0\}} \\
        +\Gamma^{0,0,1}_{0,0}\Gamma^{0,1,2}_{0,1}\Gamma^{1,2,3}_{1,1}\Gamma^{1,2,4}_{0,0}\Gamma^{1,2,5}_{0,0}\Gamma^{1,3,6}_{0,1} \ket{1,\{0,0,1,0,0,0\}} \\
        \left. +\Gamma^{0,0,1}_{0,0}\Gamma^{1,1,2}_{1,1}\Gamma^{1,2,3}_{0,1}\Gamma^{1,2,4}_{0,0}\Gamma^{1,2,5}_{0,0}\Gamma^{1,3,6}_{0,1} \ket{1,\{0,1,0,0,0,0\}} \right)
    \end{gathered}
\end{equation}
By using the Equation \eqref{eq:Gammas} we can calculate the vector to be:
\begin{equation}
    \ket{\Omega^{[1]}_{011001}}= 2 \left(- \frac{1}{2\sqrt{5}} \ket{1,q^{[1]}_1}+\frac{1}{\sqrt{30}} \ket{1,q^{[1]}_2}+\frac{1}{3\sqrt{2}} \ket{1,q^{[1]}_3}-\frac{1}{6} \ket{1,q^{[1]}_4}-\frac{1}{2\sqrt{3}} \ket{1,q^{[1]}_5}\right)
\end{equation}
Being $\ket{\Omega^{[1]}_{011001}}$ proportional to $\lambda$ representation in Schur basis for any six particles pure state with the same symmetry as $011001$. We can easily compute $\Omega$ matrix obtaining
\begin{equation}
    \Omega^{[1]}_{011001}= \ket{\Omega^{[1]}_{011001}} \bra{\Omega^{[1]}_{011001}} = \left(\begin{array}{ccccc}
       \frac{1}{5}  & \frac{-\sqrt{2/3}}{5} &\frac{-\sqrt{2/5}}{3} & \frac{1}{3\sqrt{5}} & \frac{1}{\sqrt{15}} \\
        \frac{-\sqrt{2/3}}{5} & \frac{2}{15} &\frac{2}{3\sqrt{15}} &  \frac{-\sqrt{2/15}}{3} &  \frac{-\sqrt{2/5}}{3} \\
         \frac{-\sqrt{2/3}}{5}  & \frac{2}{3\sqrt{15}} & \frac{2}{9} &-\frac{\sqrt{2}}{9} & \frac{-\sqrt{2/3}}{3} \\
         \frac{1}{3\sqrt{5}} &\frac{-\sqrt{2/15}}{3} &-\frac{\sqrt{2}}{9}  &\frac{1}{9} &\frac{1}{3\sqrt{3}} \\
         \frac{1}{\sqrt{15}} & \frac{-\sqrt{2/5}}{3}  &  \frac{-\sqrt{2/3}}{3}  &\frac{1}{3\sqrt{3}}  &\frac{1}{9} 
    \end{array} \right)
\end{equation}
which is proportional to $\lambda$ representation in Schur basis for any six particles mixed state with the same symmetry as $011001$. Note this system is invariant to permutations $l=\{I,(14),(15),(45),(145),(154)\}$, $r=\{I,(23)(26)(36)(236)(263)\}$ and to any product of them. By using Equation \ref{eq:OmegaLR} we have that
\begin{equation}
    \Omega^{[1]}_{011001}= \left(\frac{1}{3!}  \sum_{\pi\in l} D^{[1]} (\pi) \right) \left(\frac{1}{3!}  \sum_{\pi\in r} D^{[1]} (\pi) \right)
\end{equation}
Doing this same computation by the Young Yamanouchi algorithm in \ref{Young-Yamanouchi} requires first computing the five adjacent transpositions of $S_6$ in irrep $\lambda=1$, then identifying all the subsets of permutations that left the system $011001$ invariant, obtaining each of those permutations as products of adjacent transpositions, compute them,  and finally add them up, a considerably more complicated task.

\chapter{Hands on Stitching and Pushing}
\label{Appendix03}

In this appendix we are going to use the W-state stitching and the parameters pushing to study what states can be obtained from different graphs. Some of the rules and definitions presented in Section \ref{rules} were obtained by this study. We start this with the simplest construction with only one W state.

\subsubsection{One stitch} 
\label{OneStitch}
The most simple case is stitching two parts of the same $W$ state ending up in one free edge, i.e., a one qubit state:
\begin{equation*}
    \begin{tikzpicture}[basel={-0.5}]
        \draw[markx={1}]  (0,0) to (0.5,0);
       \draw[\lao-] (0.5,0)to[bend left=40] (1.5,0);
         \draw[\lac-] (0.5,0)to[bend right=40] (1.5,0);
    \end{tikzpicture} \propto \begin{tikzpicture}[basel={-1.2}]
        \draw[\rac] (0,0) to (0.5,0);
       \draw[markx={0}]  (0.5,0)to[bend left=70] (1.5,0);
         \draw(0.5,0)to[bend right=70] (1.5,0);
         \ball{(1,0.25)};
         \draw (1,0.5) node{$v$};
    \end{tikzpicture}.
\end{equation*}
Here, we pushed both arrows leaving a ball inside. From here we will omit all the outer decorations after pushing , and the equalities will be changed by similarity (under SLOCC) symbols where it corresponds. Note how we can separate the ball in two and making it cross the $\bullet$, then we have:
\begin{equation*}
\begin{tikzpicture}[basel={-1.2}]
        \draw[\rac] (0,0) to (0.5,0);
       \draw[markx={0}]  (0.5,0)to[bend left=70] (1.5,0);
         \draw(0.5,0)to[bend right=70] (1.5,0);
         \ball{(1,0.25)};
         \draw (1,0.5) node{$v$};
    \end{tikzpicture} \cong 
    \begin{tikzpicture}[basel={-1.2}]
        \draw[markx={1}] (0,0) to (0.5,0);
       \draw (0.5,0)to[bend left=70] (1.5,0);
         \draw(0.5,0)to[bend right=70] (1.5,0);
         \ball{(1,0.25)};
         \draw (1,0.5) node{$v$};
    \end{tikzpicture} = \begin{tikzpicture}[basel={-.5}]
        \draw[markx={1}] (0,0) to (0.5,0);
       \draw (0.5,0)to[bend left=70] (1.5,0);
         \draw(0.5,0)to[bend right=70] (1.5,0);
         \ball{(1,0.25)};
         \draw (1,0.5) node{$\sqrt{v}$};
         \ball{(1,-0.25)};
         \draw (1,-0.5) node{$\sqrt{v}$};
    \end{tikzpicture} = \sqrt{v} \cdot \begin{tikzpicture}[basel={-1.1},every node/.style={scale=0.4}]
        \draw[markx={1}] (0,0) to (0.5,0);
        \ball{(0.25,0)};
         \draw (0.25,0.3) node{$\frac{1}{\sqrt{v}}$};
       \draw (0.5,0)to[bend left=70] (1.5,0);
         \draw(0.5,0)to[bend right=70] (1.5,0);
         \ball{(1,0.25)};
         \draw (1,0.5) node{$\sqrt{v}$};
         \ball{(0.7,0.1)};
         \draw (0.7,0.4) node{$\frac{1}{\sqrt{v}}$};
    \end{tikzpicture} \cong \begin{tikzpicture}[basel={-.5},every node/.style={scale=0.4}]
        \draw[markx={1}] (0,0) to (0.5,0);
       \draw (0.5,0)to[bend left=70] (1.5,0);
         \draw(0.5,0)to[bend right=70] (1.5,0);
    \end{tikzpicture} = \ket{1}
\end{equation*}
This is the simplest application of the ball self-destruction rule (rule (viii)). We use the previous symbol for the qubit $\ket{1}$ and from it we can also name $\ket{0}$:
\begin{equation}
     \begin{tikzpicture}
      \qubit{(0,0)}{0} 
      \draw[markx={0.5}](-0.5,0) to 
      (0,0);
   \end{tikzpicture} =  \ket{0} , \quad  \begin{tikzpicture}[basel={-.5}]
      \qubit{(0,0)}{0} ;
       \draw (-0.25,0) to (0,0);
   \end{tikzpicture} = \ket{1} .
   \label{eq:emptyloop}
\end{equation}
We also have to consider the case with the $\Psi$ stitch, this case is:
\begin{equation*}
    \begin{tikzpicture}[basel={-0.5}]
        \draw[markx={1}] (0,0) to (0.5,0);
       \draw[markx={0.9}] (0.5,0)to[bend left=70] (1.5,0);
         \draw[\lac-] (0.5,0)to[bend right=70] (1.5,0);
    \end{tikzpicture}\cong \begin{tikzpicture}[basel={-0.5}]
        \draw[markx={1}] (0,0) to (0.5,0);
       \draw[markx={0.9}] (0.5,0)to[bend left=70] (1.5,0);
         \draw (0.5,0)to[bend right=70] (1.5,0);
          
            \ball{(1,0.25)};
    \end{tikzpicture} = \left(\frac{1+v^2}{v} \right)\ket{0}= \left(\frac{1+v^2}{v} \right) \cdot \begin{tikzpicture}[basel={-.6}]
     \qubit{(0,0)}{0} 
      \draw[markx={0.5}](-0.5,0) to 
      (0,0);
   \end{tikzpicture}.
\end{equation*}
Note how $v$ can be any value, but for the specific values $v=\pm i$ we get a null state. or the specific values $v=\pm i$ the ball corresponds to $\pm i\sigma_z$ with $\sigma_z$ the pauli matrix, we can move out the proportionality factor $\pm i$, then we introduce a special notation for $Z$-balls:
\begin{equation*}
   \begin{tikzpicture}
       \draw (-0.2,0) -- (0.2, 0);
        \ball{(0,0)}
       \draw (0,0.3) node {$\pm i$};
   \end{tikzpicture}= \pm i\begin{tikzpicture}[basel={-1}]
   \draw (-0.2,0) -- (0.2, 0);
       \ball{(0,0.12)}
   \end{tikzpicture} \cong \sigma_z.
\end{equation*}
The Z-ball has the property that it is its own inverse. Then, for the case with the $\Psi$ stitch we have two options:
\begin{equation}
    \begin{tikzpicture}[basel={-0.5}]
        \draw[markx={1}] (0,0) to (0.5,0);
       \draw[markx={0.9}] (0.5,0)to[bend left=70] (1.5,0);
         \draw (0.5,0)to[bend right=70] (1.5,0);
            \ball{(1,0.25)};
    \end{tikzpicture} \propto  \begin{tikzpicture}[basel={-.6}]
      \qubitl{(0,0)} ;
      \draw[markx={0.5}](-0.5,0) to (-0.25,0);
   \end{tikzpicture} , \quad  \begin{tikzpicture}[basel={-0.5}]
        \draw[markx={1}] (0,0) to (0.5,0);
       \draw[markxo={0.9}{0.5}] (0.5,0)to[bend left=70] (1.5,0);
         \draw (0.5,0)to[bend right=70] (1.5,0);
    \end{tikzpicture} =0,
    \label{eq:nonemptyloop}.
\end{equation}
With this we complete the cases of one W state and one stitch. The next graph corresponds to stitching two $W$ states with one stitch, obtaining a four-qubit state, which is
\begin{equation*}
     \begin{tikzpicture}[basel={-0.5}]
   \draw[\lao \rac]    (1,0) -- (0,0) ;
   \draw[markx={1}](-0.5,0.5) -- (0,0);
   \draw (-0.5,-0.5) -- (0,0);\draw (1.5,0.5) -- (1,0);
   \draw[markx={1}] (1.5,-0.5) -- (1,0);
\end{tikzpicture} .
\end{equation*}
We can now push the inner operations to the outer parts
\begin{equation*}
     \begin{tikzpicture}[basel={-0.5}]
   \draw [markx={1},markx={0}]   (0,0) -- (1,0) ; 
   \draw    (1,0) -- (0,0) ;
   \draw (-0.5,0.5) -- (0,0);
   \draw[\rao] (-0.5,-0.5) -- (0,0);\draw [\rac] (1.5,0.5) -- (1,0);
   \draw (1.5,-0.5) -- (1,0); \ball{(1.2,-0.2)} 
\end{tikzpicture} .
\end{equation*}
As the outer actions corresponds to SLOCC operations, we can drop them and get a representative for the obtained class:
\begin{equation*}
     \begin{tikzpicture}[basel={-0.5}]
   \draw [markx={1},markx={0}]   (0,0) -- (1,0) ; 
   \draw    (1,0) -- (0,0) ;
   \draw (-0.5,0.5) -- (0,0);
   \draw (-0.5,-0.5) -- (0,0);\draw  (1.5,0.5) -- (1,0);
   \draw (-0.7,0.5) node{$1$};
    \draw (-0.7,-0.5) node{$2$};
     \draw (1.7,0.5) node{$3$};
      \draw (1.7,-0.5) node{$4$};
   \draw (1.5,-0.5) -- (1,0); 
\end{tikzpicture} = \ket{0000}+ \ket{0101}+\ket{0110}+\ket{1001}+ \ket{1010} =\ket{0000}+ \ket{\Psi^+} \ket{\Psi^+}.
\end{equation*}
Labels in outer edges specify the order of the qubits in the state, however, we omit them when they are not necessary. It can be checked that in the classification of \cite{Verstraete}, this is an state in the $L_{a00_2} (L_{abc_2},b=0,c=0)$ class, in particular one can find the correspondence:
\begin{equation*}
    I \otimes X \otimes \left(\begin{array}{cc}
      0 &a/2  \\
          1 & 0
     \end{array} \right) \otimes \left(\begin{array}{cc}
     2/a &0 \\
          0 & 1
     \end{array} \right)\, \cdot \, \begin{tikzpicture}[basel={-0.5}]
   \draw [markx={1},markx={0}]    (0,0) -- (1,0) ; 
   \draw    (1,0) -- (0,0) ;
   \draw (-0.5,0.5) -- (0,0);
   \draw (-0.5,-0.5) -- (0,0);\draw  (1.5,0.5) -- (1,0);
   \draw (1.5,-0.5) -- (1,0); 
   \draw (-0.7,0.5) node{$1$};
    \draw (-0.7,-0.5) node{$2$};
     \draw (1.7,0.5) node{$3$};
      \draw (1.7,-0.5) node{$4$};
\end{tikzpicture} = \frac{2}{a} \ket{
L_{a00_2}}.
\end{equation*}
 We still have to consider the $\Psi$ stitch. Then, for this case we have:
\begin{equation*}
     \begin{tikzpicture}[basel={-0.5}]
   \draw[\rac]    (0,0) -- (1,0) ; 
   \draw[markxi={1}]  (1,0) -- (0,0) ;
   \draw[markxi={0}](1,0) -- (0,0);
   \draw (-0.5,0.5) -- (0,0);
   \draw (-0.5,-0.5) -- (0,0);\draw (1.5,0.5) -- (1,0);
   \draw (1.5,-0.5) -- (1,0);\draw[markxi={0.5}] (0,0)--(1,0) ;
\end{tikzpicture} \cong  \begin{tikzpicture}[basel={-0.5}]
   \draw [markx={1},markx={0}] (0,0) -- (1,0) ; 
   \draw (-0.5,0.5) -- (0,0);
   \draw (-0.5,-0.5) -- (0,0);\draw (1.5,0.5) -- (1,0);
   \draw (1.5,-0.5) -- (1,0);
   \draw (-0.7,0.5) node{$1$};
    \draw (-0.7,-0.5) node{$2$};
     \draw (1.7,0.5) node{$3$};
      \draw (1.7,-0.5) node{$4$};\draw[markxi={0.5}] (0,0)--(1,0) ;
\end{tikzpicture}  = \ket{0001}+\ket{0010}+\ket{0100}+\ket{1000} \propto \ket{W_4}.
\end{equation*}
This is a very interesting case for us, note that the same $W_4$ state will be obtained if we gather the outer parts in a different way. We now generalize the notation to vertices with $n$ edges to symbolize the $W_n$ state, then we have the following equivalences:
\begin{equation}
    \begin{tikzpicture}[basel={-0.5}]
   \draw[markx={1},markx={0}] (0,0) -- (1,0) ; 
   \draw (1,0) -- (0,0) ;
   \draw (-0.5,0.5) -- (0,0);
   \draw (-0.5,-0.5) -- (0,0);\draw (1.5,0.5) -- (1,0);
   \draw (1.5,-0.5) -- (1,0);\draw[markxi={0.5}] (0,0)--(1,0) ;
   \draw (-0.7,0.5) node{$1$};
    \draw (-0.7,-0.5) node{$2$};
     \draw (1.7,0.5) node{$3$};
      \draw (1.7,-0.5) node{$4$}; 
\end{tikzpicture} =\begin{tikzpicture}[basel={-0.5}]
   \draw (0,0) -- (0.5,0.5) ; 
   \draw (-0.5,0.5) -- (0,0) ;
   \draw[markx={1},markx={0}] (0,0) -- (0,-1);
   \draw (0,-1) -- (0.5,-1.5);\draw (0,-1) -- (-0.5,-1.5);
  \draw[markxi={0.5}] (0,0)--(0,-1) ;
   \draw (-0.7,0.5) node{$1$};
    \draw (-0.7,-1.5) node{$2$};
     \draw (0.7,0.5) node{$3$};
      \draw (0.7,-1.5) node{$4$};
\end{tikzpicture}  = \Wn{4}{45}{ \draw[markxi={0}](0,0) to (0.5,0.5);}.
\end{equation}
This construction and symmetry is the rule (ix). This state in Verstraette classification is $L_{00_3} (L_{ab_3},a=0,b=0)$, the operation that put $W_4$ in the form of the representative in \cite{Verstraete} is given by:
\begin{equation}
   I \otimes I \otimes Y \otimes Y \,\cdot \, \Wn{4}{45}{\draw[markxi={0}](0,0) to (0.5,0.5);} =i \sqrt{2}\ket{ L_{00_3}}
\end{equation}
These are all the possible constructions with  only one stitch. Now we consider the cases with two stitches.

\subsubsection{Two Stitches}
The next diagram consists on stitching two W states, but, in this case, with two stitches, obtaining a two qubits state. The graph for this case is the following:
\begin{equation*}
    \begin{tikzpicture}[basel={-.5}]
   \draw[markx={1}] (0,0) -- (0.5,0) ; 
   \draw [\lao \rac] (1.5,0) to [bend right=50] (0.5,0) ;
   \draw[\lao \rac]  (1.5,0) to [bend left=50] (0.5,0) ;
   \draw[markx={0}]  (1.5,0)-- (2,0);   
\end{tikzpicture} \cong
    \begin{tikzpicture}[basel={-2}]
   \draw[markx={1}] (0,0) -- (0.5,0) ; 
   \draw  (1.5,0) to [bend right=50] (0.5,0) ;
   \draw  (1.5,0) to [bend left=50] (0.5,0) ;
   \draw [markx={0}] (1.5,0)-- (2,0);
   \ball{ (1,0.25) };
    \draw (1.0,0.6) node {$v$}; 
\end{tikzpicture}.
\end{equation*}
In this pushing process, we used the property of pushing black arrows, leaving a ball and an arrow on the other edges, for both black arrows it was chosen to leave balls inside the graph and arrows on the outer edge. Next, both balls can be combined in just one. From now on, white arrows on outer edges will be omitted from the beginning as they can be pushed freely out. In this case, the parameter of the ball works as a switch between two classes. 
Then, the first case with the $Z$- ball is:
\begin{equation*}
     \begin{tikzpicture}[basel={-1}]
   \draw[markx={1}] (0,0) -- (0.5,0) ; 
   \draw [marko={0.5}](0.5,0)  to [bend left=50] (1.5,0) ;
   \draw  (1.5,0) to [bend left=50] (0.5,0) ;
   \draw [markx={0}] (1.5,0)-- (2,0);
\end{tikzpicture} = \ket{11} =  \begin{tikzpicture}[basel={-.5}]
    \qubitl{(0,0)}
    \qubitr{(0.8,0)}
    \draw(-0.5,0) to (-0.25,0);
      \draw(1.45,0) to (1.7,0);
\end{tikzpicture} 
\end{equation*}
For any other value of $v$ we have an entangled state:
\begin{equation}
     \begin{tikzpicture}[basel={-1}]
   \draw[markx={1}] (0,0) -- (0.5,0) ; 
   \draw  (1.5,0) to [bend right=50] (0.5,0) ;
   \draw (1.5,0) to [bend left=50] (0.5,0) ;
   \draw[markx={0}]  (1.5,0)-- (2,0);
   \ball{(1,0.2)};
   \draw {(1,0.5)} node {$v$};
\end{tikzpicture} = \frac{1+v^2}{v} \ket{00}+ v \ket{11} = \sqrt{1+v^2} \begin{tikzpicture}[basel={-3}]
\draw (0,0) -- (1,0);
\ball{(0.5,0)}
\draw (0.5,0.5) node {$\frac{\sqrt{1+v^2}}{v}$};    
\end{tikzpicture}
\label{eq:twoqubitsball}
\end{equation}
One important case is when $v=1$, which means no ball, then we have:
\begin{equation*}
     \begin{tikzpicture}[basel={-1}]
   \draw[markx={1}] (0,0) -- (0.5,0) ; 
   \draw  (1.5,0) to [bend right=50] (0.5,0) ;
   \draw (1.5,0) to [bend left=50] (0.5,0) ;
   \draw[markx={0}]  (1.5,0)-- (2,0);
\end{tikzpicture} = 2\ket{00}+  \ket{11} = \sqrt{2} \cdot   \begin{tikzpicture}[basel={-3}]
\draw (0,0) -- (1,0);
\ball{(0.5,0)}
\draw (0.5,0.5) node {$\sqrt{2}$};    
\end{tikzpicture}.
\end{equation*}
When considering the cases with the $\Psi$ stitch, in this graph we have first with one $\Phi$ stitch and one $\Psi$ stitch:
 \begin{equation*}
        \begin{tikzpicture}[basel={-0},every node/.style={scale=0.6}]
            \draw[markx={1}](0,0) to (0.5,0);
            \draw(0.5,0) to[bend right=90] (1.5,0);
            \draw[markx={0.5}](0.5,0) to[bend left=90]  (1.5,0);
            \draw[markx={0}] (1.5,0) to (2,0); 
            \ball{(1,-0.3)};
            \draw (1,-0.5) node {$v$};
            \end{tikzpicture} \cong \begin{tikzpicture}[basel={-0.6},every node/.style={scale=0.6}]
            \draw[markx={1}](0,0) to (0.5,0);
            \draw(0.5,0) to[bend right=90] (1.5,0);
            \draw[markx={0.5}](0.5,0) to[bend left=90]  (1.5,0);
            \draw [markx={0}](1.5,0) to (2,0); 
            \end{tikzpicture},
    \end{equation*} 
Where we applied the ball self-destruction rule. The unique state (up to SLOCC) obtained here is :
\begin{equation*}
        \begin{tikzpicture}[basel={-0.5}]
            \draw[markx={1}](0,0) to (0.5,0);
            \draw(0.5,0) to[bend right=90] (1.5,0);
            \draw[markx={0.5}](0.5,0) to[bend left=90]  (1.5,0);
            \draw [markx={0}](1.5,0) to (2,0); 
            \end{tikzpicture} =  \begin{tikzpicture}[basel={-0.5}]
            \inangarc{2}{(0,0)}{0};
            \draw[markx={0},markx={1},markx={0.5}](0,0) to (0,-1);
            \draw(0,-1) to[bend right=90] (0,-2);
            \draw(0,-1) to[bend left=90] (0,-2);
            \end{tikzpicture} = \begin{tikzpicture}[basel={-0.5}]
            \draw[markx={0.5}](0,0) to (1,0);
            \end{tikzpicture}.
 \end{equation*}
  In the first equality, we used the rotational property of the $W_4$ state that we obtained before and it corresponds to the rule (ix). In the next step, we used  the operations on qubits from rule (vii). The recurrent use of rules and properties is very important in the construction. Now, we have to consider the case with two $\Psi$ stitches:
 \begin{equation*}
        \begin{tikzpicture}[basel={-0},every node/.style={scale=0.6}]
            \draw[markx={1}](0,0) to (0.5,0);
            \draw[markx={0.5}](0.5,0) to[bend right=90] (1.5,0);
            \draw[markx={0.5}](0.5,0) to[bend left=90]  (1.5,0);
            \draw[markx={0}] (1.5,0) to (2,0); 
            \ball{(0.7,-0.25)};
            \draw (0.7,-0.4) node {$v$};
            \end{tikzpicture}=  \begin{tikzpicture}[basel={-0.5},every node/.style={scale=0.6}]
            \inangarc{2}{(0,0)}{0};
            \draw[markx={0.5},markx={0},markx={1}](0,0) to (0,-1);
            \draw(0,-1) to[bend right=90] (0,-2);
            \draw[markx={0.9}](0,-1) to[bend left=90] (0,-2);
            \ball{(-0.2,-1.8)};
            \draw (-0.2,-2.1) node {$v$};
            \end{tikzpicture} \in \left\{ \begin{tikzpicture}[basel={-0.5},every node/.style={scale=0.6}]
            \inangarc{2}{(0,0)}{0};
            \draw[markx={0.5},markx={0},markx={1}](0,0) to (0,-1);
            \draw(0,-1) to[bend right=90] (0,-2);
            \draw[markxo={0.9}{0.7}](0,-1) to[bend left=90] (0,-2);   \end{tikzpicture}= 0 , \quad  \begin{tikzpicture}[basel={-0.5},every node/.style={scale=0.6}]
            \inangarc{2}{(0,0)}{0};
            \draw[markx={0.5},markx={0},markx={0.8},markx={1}](0,0) to (0,-1);\draw(0,-1) to[bend right=90] (0,-2);
            \draw(0,-1) to[bend left=90] (0,-2);   
            \end{tikzpicture} = \begin{tikzpicture}[basel={-.5}]
    \qubitl{(0,0)}
    \qubitr{(0.8,0)}
    \draw[markx={1}](-0.5,0) to (-0.25,0);
      \draw[markx={0}](1.45,0) to (1.7,0);
\end{tikzpicture}  \right\}.
 \end{equation*}
 The two cases in the previous equation correspond to the two cases from \eqref{eq:nonemptyloop}, or equivalently from rule (x). With this construction we finish the options with two stitches. With two stitches is also possible to obtain five-qubit states; however, there is no known classification for this case. 
 
 \section{Three-qubit states}
 Now, we will use the rules in the section \ref{rules} to analyze the case for stitching three $W$ states with three stitches, obtaining a three-qubit state. The graph we are interested is:
\begin{equation*}
    \begin{tikzpicture}[basel={-.5}]
   \draw[\rac] (0,0) -- (0.5,0) ; 
   \draw[markx={0}] (0.5,0) -- (1,0.5) ;
   \draw[markx={1}] (0,0) -- (-0.5,0);
   \draw (-1,0.5) -- (-0.5,0);
   \draw[\rac] (0.5,0) --(0,-1);\draw[markx={1}](0,-1.5)--(0,-1); \draw[\rac] (0,-1) --(-0.5,0);
\end{tikzpicture}  \cong \begin{tikzpicture}[basel={-.5}]
   \draw [markx={1}](0,0) -- (0.5,0) ; \draw [markx={1}](0,0)--(-0.5,0);
   \draw (0.5,0) -- (1,0.5) ;
   (0,0) -- (-0.5,0);
   \draw (-1,0.5) -- (-0.5,0);
   \draw[\rac] (0.5,0) -- (0,-1);\draw[markx={1}](0,-1.5)--(0,-1); \draw[\rac] (0,-1) --(-0.5,0); \ball{(0.35,-0.3)};
\end{tikzpicture}\cong \begin{tikzpicture}[basel={-.5}]
   \draw [markx={1}](0,0) -- (0.5,0) ; \draw (0,0)--(-0.5,0);
   \draw (0.5,0) -- (1,0.5) ;
   (0,0) -- (-0.5,0);
   \draw[markx={1}] (-1,0.5) -- (-0.5,0);
   \draw[markx={1}](0.5,0) -- (0,-1);\draw(0,-1.5)--(0,-1); \draw[\rac] (0,-1) --(-0.5,0); \ball{(-0.35,-0.3) };
\end{tikzpicture} \cong \begin{tikzpicture}[basel={-.5}]
   \draw[markx={1}] (0,0) -- (0.5,0) ; 
   \draw (0.5,0) -- (1,0.5) ;
   \draw[markx={1}] (0,0) -- (-0.5,0);
   \draw (-1,0.5) -- (-0.5,0);\draw [markx={1}](-0.5,0) -- (0,-1);
   \draw (0.5,0) -- (0,-1);\draw(0,-1.5)--(0,-1); \ball{ (0,0)};
\end{tikzpicture}.
\end{equation*}
 Here, the fact that balls can be moved through the line and absorbed by dark arrows was used (rules iii,iv and vi). For the case with three $\Phi$ stitches, a ball remains after pushing the black arrows. This ball can go through the path with three vertices (an odd number), the ball can be self destroyed. This is a unique state which corresponds to:
\begin{equation*}
    \begin{tikzpicture}[baseline={([yshift=-.5ex]current bounding box.center)}]
   \draw[markx={1}] (0,0) -- (0.5,0) ; 
   \draw (0.5,0) -- (1,0.5) ;
   \draw[markx={1}] (0,0) -- (-0.5,0);
   \draw (-1,0.5) -- (-0.5,0);\draw [markx={1}](-0.5,0) -- (0,-1);
   \draw (0.5,0) -- (0,-1);\draw(0,-1.5)--(0,-1); 
\end{tikzpicture}= \ket{001}+\ket{010}+\ket{100}+\ket{111} ,
\end{equation*}
which is related to the GHZ state as:
\begin{equation}    (\sigma_z H)\otimes(\sigma_z H)\otimes(\sigma_z H) \, \cdot \,  \begin{tikzpicture}[basel={-.5}]
   \draw [markx={1}](0,0) -- (0.5,0) ; 
   \draw (0.5,0) -- (1,0.5) ;
   \draw[markx={1}] (0,0) -- (-0.5,0);
   \draw (-1,0.5) -- (-0.5,0);\draw [markx={1}](-0.5,0) -- (0,-1);
   \draw (0.5,0) -- (0,-1);\draw(0,-1.5)--(0,-1); 
\end{tikzpicture} = \sqrt{2}\ket{GHZ},
\label{eq:GHZstate}
\end{equation}
where $H$ is the Hadamard matrix of order two. Now, we must consider the cases with one $\Psi$ stitch. As the graph is symmetric, we only will consider the $\Psi$ stitch in one position:
\begin{equation*}
    \begin{tikzpicture}[basel=-.5]
   \draw[\rac] (0,0) -- (0.5,0) ; 
   \draw [markx={0}](0.5,0) -- (1,0.5) ;
   \draw[markx={1}] (0,0) -- (-0.5,0);
   \draw (-1,0.5) -- (-0.5,0);
   \draw[\rac] (0.5,0) -- (0,-1);\draw[markx={1}](0,-1.5)--(0,-1); \draw[\rac] (0,-1) --(-0.5,0); 
   \draw[markxi={0.5}](-0.5,0) to (0,-1);
\end{tikzpicture} \cong
  \begin{tikzpicture}[basel=-.5]
   \draw[markx={1}](0,0) -- (0.5,0) ; 
   \draw (0.5,0) -- (1,0.5) ;
   \draw [markx={1}](0,0) -- (-0.5,0);
   \draw (-1,0.5) -- (-0.5,0);
   \draw(0,-1.5)--(0,-1); \draw [markx={0}](0,-1) --(0.5,0); 
   \draw[markx={0.5}](-0.5,0) to (0,-1);
   \ball{(0.25,-0.5)};
\end{tikzpicture} =\begin{tikzpicture}[basel={-.5}]
   \inangarc{2}{(-0.5,0)}{90};
   \draw[markx={0.5},markx={0},markx={1}] (-0.5,0) --(0,0);
   \draw (0,0) to[bend right=50] (1,0) ;
   \draw (0,0) to[bend left=50] (1,0);
   \ball{(0.5,0.2)};
   \draw [markx={0}](1,0)--(1.5,0);
\end{tikzpicture}.
\end{equation*}
In the first part we pushed the black arrows leaving one ball inside. This construction separates in two cases due to rule (xi.1) :
\begin{equation*}
   \begin{tikzpicture}[basel={-.5}]
   \inangarc{2}{(-0.5,0)}{90};
   \draw[markx={0.5},markx={0},markx={1}]  (-0.5,0) --(0,0);
   \draw (0,0) to[bend right=50] (1,0) ;
   \draw (0,0) to[bend left=50] (1,0);
   \ball{(0.5,0.2)};
   \draw[markx={0}] (1,0)--(1.5,0);
\end{tikzpicture} = \begin{tikzpicture}[basel={-.5}]
   \inangarc{2}{(-0.5,0)}{90};
   \draw[markx={0},markx={0.25}] (-0.5,0) --(0.5,0);
   \ball{(0,0)};
\end{tikzpicture} \cong \Wn{3}{0}{\draw[markxi={0}](0,0) to (0.5,0.5);} , 
   \begin{tikzpicture}[basel={-.5}]
   \inangarc{2}{(-0.5,0)}{90};
   \draw[markx={0.5},markx={0},markx={1}]   (-0.5,0) --(0,0);
   \draw (0,0) to[bend right=50] (1,0) ;
   \draw[marko={0.5}] (0,0) to[bend left=50] (1,0);
   \draw[markx={0}](1,0) to (1.25,0);
\end{tikzpicture} =
   \begin{tikzpicture}[basel={-.5}]
   \inangarc{2}{(-0.5,0)}{90};
   \draw[markx={0},markx={0.5}] (-0.5,0) --(0,0);
   \qubitl{(0.25,0)};
   \qubitr{(0.8,0)};
\end{tikzpicture}= \begin{tikzpicture}
    \draw[markx={0.5}](0,0) to (0.5,0);
    \qubitr{(0.7,0)};
\end{tikzpicture}.
\end{equation*}
The first case is an state in the $W$ class, and the second case is an $AB-C$ class state. Note how the position of the $\
Psi$ stitch can change which qubit is separable, allowing the $BC-A$ and the $AC-B$ classes. For the case with two $\Psi$ stitches, the ball is again self-destroyed, getting the unique state corresponding to the W state:
\begin{equation*}
\begin{tikzpicture}[basel={-.5}]
   \draw[markx={1}] (0,0) -- (0.5,0) ; 
   \draw (0.5,0) -- (1,0.5) ;
   \draw[markx={1}] (0,0) -- (-0.5,0);
   \draw (-1,0.5) -- (-0.5,0);
   \draw[markx={1}](0,-1.5)--(0,-1); \draw[markx={0.5}] (0,-1) --(-0.5,0); \draw[markx={0.5}] (0,-1) -- (0.5,0);
   \end{tikzpicture} =
    \begin{tikzpicture}[basel={-.5}]
   \inangarc{2}{(-0.5,0)}{90};
   \draw[markx={0.5},markx={0},markx={1}]  (-0.5,0) --(0,0);
   \draw[markx={0.5}] (0,0) to[bend right=50] (1,0) ;
   \draw (0,0) to[bend left=50] (1,0);
   \draw[markx={0}] (1,0)--(1.5,0);
   \end{tikzpicture} = \Wn{3}{0}{\draw[markxi={0}] (0,0) to (0.5,0.5);}.
\end{equation*}
The last case with three $\Psi$ stitches still contains one ball after pushing the arrows. Then, one has:
\begin{equation*}
\begin{tikzpicture}[basel={-.5}]
   \draw[markx={0.5},markx={0},markx={1}]  (-0.5,0) -- (0.5,0) ; 
   \ball{(0.2,0)};
   \draw (0.5,0) -- (1,0.5) ;
   \draw (-1,0.5) -- (-0.5,0);
   \draw[markx={1}](0,-1.5)--(0,-1); \draw[markx={0.5}] (0,-1) --(-0.5,0); \draw[markx={0.5}] (0,-1) -- (0.5,0);
   \end{tikzpicture} =
    \begin{tikzpicture}[basel={-.5}]
   \inangarc{2}{(-0.5,0)}{90};
   \draw[markx={0.5},markx={0},markx={1}]  (-0.5,0) --(0,0);
   \draw[markx={0.5}] (0,0) to[bend right=50] (1,0) ;
   \draw[markx={0.5}] (0,0) to[bend left=50] (1,0);
   \ball{(0.7,0.2)};
   \draw[markx={0}] (1,0)--(1.5,0);
   \end{tikzpicture} ,
   \end{equation*}
   which separates in two cases according to rule (xi):
   \begin{equation*}
\begin{tikzpicture}[basel={-.5}]
   \inangarc{2}{(-0.5,0)}{90};
   \draw[markx={0.5},markx={0},markx={1}]  (-0.5,0) --(0,0);
   \draw[markx={0.5}] (0,0) to[bend right=50] (1,0) ;
   \draw[markx={0.5}] (0,0) to[bend left=50] (1,0);
   \ball{(0.7,0.2)};
   \draw[markx={0}](1,0)--(1.5,0);
   \end{tikzpicture} = \begin{tikzpicture}[basel={-.5}]
       \inangarc{2}{(-0.5,0)}{90};
   \draw[markx={0.5},markx={0},markx={1}] (-0.5,0) --(0,0);
   \qubitl{(0.25,0)};
    \qubitr{(0.7,0)};
    \draw[markx={0}](1.35,0) to (1.6,0);
   \end{tikzpicture} = \begin{tikzpicture}[basel={-.5}]
    \qubitl{(0,0.2)}; 
    \qubitl{(0,-0.2)};
    \qubitr{(0.5,0)};
    \draw[markx={1}](-0.5,0.2)to (-0.25,0.2); 
   \draw[markx={1}](-0.5,-0.2)to (-0.25,-0.2); 
     \draw[markx={0}](1.15,0) to (1.4,0); \end{tikzpicture},\quad \begin{tikzpicture}[basel={-.5}]
   \inangarc{2}{(-0.5,0)}{90};
   \draw[markx={0.5},markx={0},markx={1}] (-0.5,0) --(0,0);
   \draw[markx={0.5}] (0,0) to[bend right=50] (1,0) ;
   \draw[markxo={0.5}{0.6}] (0,0) to[bend left=50] (1,0);
   \draw[markx={0}] (1,0)--(1.5,0);
   \end{tikzpicture} =0.
\end{equation*}
Obtaining the $A-B-C$ (separable) class and the null state, respectively. With this diagram, all the SLOCC classes for three qubits are obtained. Now we will explore the case of four qubits, which is much more complicated. 

\section{Four-qubit states}
\label{FourQubits}
According to Equation \eqref{eq:Numberofparts}, for obtaining four qubits states we can achieve it with $(s=1,\omega=2),(s=4,\omega=4) , (s=7,\omega=6) \dots$. The cases with one stitch were already analyzed in Section \ref{OneStitch}. When considering four stitches the problem gets more complicated, there are four ways of doing the stitching, i.e., there are four inequivalent graphs with four vertices and four inner edges (Here we discarded the graphs obtained from stitching two parts of the same $W$). Those graphs are:
\begin{equation}
\begin{gathered}
\begin{tikzpicture}[basel={-.5}]
\draw[markx={1}] (0.75,0) to (1,0);
    \draw[\lac \rao](1,0) to[bend right=40] (2,0);
    \draw[\lac \rao](1,0) to[bend left=40] (2,0);
    \draw[\lac \rao](2,0) to (3,0);
    \draw[markxi={0}](2,0) to (3,0); 
    \draw[markxi={0}](4,0) to (3,0);    
    \draw[markxi={1}](4,0) to (3,0);
    \draw(3,0) to (3,-0.25);
    \draw[\lac \rao](3,0) to (4,0);
    \draw (4,0) to (4.25,0.25);
    \draw (4,0) to (4.25,-0.25);
\end{tikzpicture}, \quad 
 \begin{tikzpicture}[basel={-.5}]
 \draw(-0.25,0.25) to (0,0);
    \draw(-0.25,-0.25) to (0,0);
    \draw[\lac \rao](0,0) to (1,0);
    \draw[\lac \rao](1,0) to[bend right=40] (2,0);
    \draw[\lac \rao](1,0) to[bend left=40] (2,0);
    \draw[\lac \rao](2,0) to (3,0);
    \draw(3,0) to (3.25,0.25);
    \draw(3,0) to (3.25,-0.25);
    \draw[markxi={0}](2,0) to (3,0);    
    \draw[markxi={1}](2,0) to (3,0);
    \draw[markxi={0}](0,0) to (3,0);    
    \draw[markxi={1}](0,0) to (1,0);
\end{tikzpicture},\\ 
    \begin{tikzpicture}[basel={-.5}]
    \draw[\lac \rao](0,0) to (1,0);
    \draw[markxi={1}](0,0) to (1,0);
    \draw[markxi={0}](0,0) to (1,0);
    \draw[\lac \rao](1,0) to (1.5,0.5);
    \draw[\lac \rao](1,0) to (1.5,-0.5);
    \draw[\lac \rao](1.5,0.5) to (1.5,-0.5);
    \draw[markx={0}](1.5,0.5) to (1.75,0.75);
    \draw[markx={0}](1.5,-0.5) to (1.75,-0.75);
    \draw(-0.25,0.25) to (0,0);
    \draw(-0.25,-0.25) to (0,0);
    \end{tikzpicture} ,\quad  \begin{tikzpicture}[basel={-.5}]
        \draw[markx={1}] (-0.25,0.25) to (0,0);
        \draw[\lac \rao] (0,0) to (1,0);
        \draw[markx={0}] (1,0) to (1.25,0.25);
        \draw[\lac \rao] (0,0) to (0,-1);
        \draw[markx={1}] (-0.25,-1.25) to (0,-1);
        \draw[\lac \rao] (0,-1) to (1,-1);
        \draw[\lac \rao] (1,0) to (1,-1);
        \draw [markx={0}](1,-1) to (1.25,-1.25);
    \end{tikzpicture}.
    \end{gathered}
    \label{eq:graphs}
\end{equation}
In this part, we will compute the different states that can be obtained from the stitching in each graph. We only emphasize the construction and the map with the classification in \cite{Verstraete} for the first time that a new SLOCC family or subfamily appears. For subfamily, we refer to a state that can only reproduce a family where one or more parameters are not independent. After pushing parameters in the first graph from the previous equation we have the following:
\begin{equation*}
\begin{tikzpicture}[basel={-.5}]
\draw (0.75,0) to (1,0);
    \draw[\lac \rao](1,0) to[bend right=40] (2,0);
    \draw[\lac \rao](1,0) to[bend left=40] (2,0);
    \draw[\lac \rao](2,0) to (3,0);
    \draw(3,0) to (3,-0.25);
    \draw[\lac \rao](3,0) to (4,0);
    \draw (4,0) to (4.25,0.25);
    \draw (4,0) to (4.25,-0.25);
    \draw[markxi={0}](2,0) to (3,0);    
    \draw[markxi={1}](2,0) to (3,0);
    \draw[markxi={0}](1,0) to (3,0);    
    \draw[markxi={1}](0,0) to (4,0);
\end{tikzpicture} \cong \begin{tikzpicture}[basel={-.5}]
\draw (0.75,0) to (1,0);
    \draw(1,0) to[bend right=40] (2,0);
    \draw(1,0) to[bend left=40] (2,0);
    \draw[\lao-](2,0) to (3,0);
    \draw(3,0) to (3,-0.25);
    \draw(3,0) to (4,0);
    \draw (4,0) to (4.25,0.25);
    \draw (4,0) to (4.25,-0.25);
    \ball {(1.5,0.2)};    
    \draw[markxi={0}](2,0) to (3,0);    
    \draw[markxi={1}](2,0) to (3,0);
    \draw[markxi={0}](1,0) to (3,0);    
    \draw[markxi={1}](0,0) to (4,0);
\end{tikzpicture} ,
\end{equation*}
which breaks into two options: the first one for a generic ball:
\begin{equation*}
\begin{tikzpicture}[basel={-.5}]
\draw (0.75,0) to (1,0);
    \draw(1,0) to[bend right=40] (2,0);
    \draw(1,0) to[bend left=40] (2,0);
    \draw[\lao-](2,0) to (3,0);
    \draw(3,0) to (3,-0.25);
    \draw(3,0) to (4,0);
    \draw (4,0) to (4.25,0.25);
    \draw (4,0) to (4.25,-0.25);
    \ball {(1.5,0.2)};
    \draw[markxi={0}](2,0) to (3,0);    
    \draw[markxi={1}](2,0) to (3,0);
    \draw[markxi={0}](1,0) to (3,0);    
    \draw[markxi={1}](0,0) to (4,0);
\end{tikzpicture} =\begin{tikzpicture}[basel={-.5}]
    \draw[\lac-](2,0) to (3,0);
    \draw[markx={0}](3,0) to (3,-0.25);
    \draw[markx={1}](3,0) to (4,0);
    \draw (4,0) to (4.25,0.25);
    \draw (4,0) to (4.25,-0.25);
\end{tikzpicture} \cong \begin{tikzpicture}[basel={-.5}]
 \draw(-0.25,0.25) to (0,0);
    \draw(-0.25,-0.25) to (0,0);
    \draw[markx={0},markx={1}](0,0) to (1,0);
    \draw(1,0) to (1.25,0.25);
    \draw(1,0) to (1.25,-0.25);
\end{tikzpicture}  ,
\end{equation*}
which was already obtained. The second option is when the ball is a $Z$-ball. However, this case separates again in two. One when the arrow is not zero:
\begin{equation*}
\begin{tikzpicture}[basel={-.5}]
\draw [markx={1}](0.75,0) to (1,0);
    \draw[markx={1}](1,0) to[bend right=40] (2,0);
    \draw[marko={0.5}](1,0) to[bend left=40] (2,0);
    \draw[\lao-](2,0) to (3,0);
    \draw(3,0) to (3,-0.25);
    \draw[markx={0}](3,0) to (4,0);
    \draw [markx={0}](4,0) to (4.25,0.25);
    \draw (4,0) to (4.25,-0.25);
\end{tikzpicture} =\begin{tikzpicture}[basel={-.5}]
\draw(0.25,0) to (0.5,0);
   \qubitl{(0.75,0)};
   \qubitr{(1.35,0)};
    \draw[\lao-](2,0) to (3,0);
    \draw(3,0) to (3,-0.25);
    \draw[markx={0},markx={1}](3,0) to (4,0);
    \draw (4,0) to (4.25,0.25);
    \draw[markx={0}] (4,0) to (4.25,-0.25);
\end{tikzpicture} \cong \begin{tikzpicture}[basel={-.5}]
\draw[markx={1}](0.25,0) to (0.5,0);
   \qubitl{(0.75,0)};
    \draw(1.5,0) to (2,0);
    \draw[markx={0}] (2,0) to (2.25,0.25);
    \draw (2,0) to (2.25,-0.25);
\end{tikzpicture} = \ket{L_{0_20_{3\oplus \Bar{1}}}},
\end{equation*}
which corresponds to a separable qubit and a $W_3$ class state. For the classification in \cite{Verstraete}, this state is exactly $L_{0_20_{3\oplus \Bar{1}}}(L_{a_20_{3\oplus \Bar{1}}},a=0)$. The other option is when the arrow is zero (or no arrow):
\begin{equation*}
\begin{tikzpicture}[basel={-.5}]
\draw (0.75,0) to (1,0);
    \draw[markx={0}](1,0) to[bend right=40] (2,0);
    \draw[marko={0.5}](1,0) to[bend left=40] (2,0);
    \draw[markx={0}](2,0) to (3,0);
    \draw[markx={0}](3,0) to (3,-0.25);
    \draw(3,0) to (4,0);
    \draw[markx={0}] (4,0) to (4.25,0.25);
    \draw (4,0) to (4.25,-0.25);
\end{tikzpicture} = \begin{tikzpicture}[basel={-.5}]
\draw(0.25,0) to (0.5,0);
   \qubitl{(0.75,0)};
   \qubitr{(1.35,0)};
    \draw(2,0) to (3,0);
    \draw(3,0) to (3,-0.25);
    \draw[markx={0},markx={1}](3,0) to (4,0);
    \draw (4,0) to (4.25,0.25);
    \draw[markx={0}] (4,0) to (4.25,-0.25);
\end{tikzpicture} = \begin{tikzpicture}[basel={-.5}]
   \draw(0.5,0.2) to (0.75,0.2);   \draw[markx={1}](0.5,-0.2) to (0.75,-0.2);
   \qubitl{(1,0.2)};
   \qubitl{(1,-0.2)};
  \draw[markx={0.5}](2,0.25) to [bend left=90](2,-0.25);
\end{tikzpicture} .
\end{equation*}
This state has two separable qubits and two entangled. It corresponds exactly to the $L_{0_20_2}(L_{a_2b_2},a=0,b=0)$ state. When considering cases with $\Psi$ stitches, only one configuration leads to a different state:
\begin{equation*}
    \begin{tikzpicture}[basel={-.5}]
\draw[markx={1}] (0.75,0) to (1,0);
    \draw[markx={0.5}](1,0) to[bend right=40] (2,0);
    \draw[markx={0.5}](1,0) to[bend left=40] (2,0);
    \draw[markx={0},markx={1},markx={0.5}](2,0) to (3,0);
    \draw(3,0) to (3,-0.25);
    \draw[markx={0.5}](3,0) to (4,0);
    \draw [markx={0}](4,0) to (4.25,0.25);
    \draw (4,0) to (4.25,-0.25);
    \ball {(1.7,0.15)};
\end{tikzpicture}\propto \begin{tikzpicture}[basel={-.5}]
\draw[markx={1}] (0.25,0) to (0.5,0);
    \qubitr{(1.35,0)};
    \qubitl{(0.75,0)};
    \draw[markx={0.5},markx={1},markx={0}](2,0) to (3,0);
    \draw(3,0) to (3,-0.25);
    \draw[markx={0.5}](3,0) to (4,0);
    \draw[markx={0}] (4,0) to (4.25,0.25);
    \draw (4,0) to (4.25,-0.25);
\end{tikzpicture} = \begin{tikzpicture}[basel={-.5}]
    \qubitl{(0,0.2)};
    \draw[markx={1}] (-0.5,0.2) to (-0.25,0.2);
    \draw[markx={1}] (-0.5,-0.2) to (-0.25,-0.2);
    \qubitl{(0,-0.2)};
    \qubitr{(0.8,0.2)};
    \qubitr{(0.8,-0.2)};
    \draw[markx={0}] (1.45,0.2) to (1.7,0.2);
    \draw[markx={0}] (1.45,-0.2) to (1.7,-0.2);
\end{tikzpicture}
\end{equation*}
This is a completely separable state, corresponding to the state $L_{000_2}(L_{abc_2},a=0,b=0,c=0)$ as:
\begin{equation*}
    I \otimes X \otimes X \otimes I \cdot \begin{tikzpicture}[basel={-.5}]
    \qubitl{(0,0.2)};
    \draw[markx={1}] (-0.5,0.2) to (-0.25,0.2);
    \draw[markx={1}] (-0.5,-0.2) to (-0.25,-0.2);
    \qubitl{(0,-0.2)};
    \qubitr{(0.8,0.2)};
    \qubitr{(0.8,-0.2)};
    \draw[markx={0}] (1.45,0.2) to (1.7,0.2);
    \draw[markx={0}] (1.45,-0.2) to (1.7,-0.2);
\end{tikzpicture} = \ket{L_{000_2}}.
\end{equation*}
For the second graph in Equation \eqref{eq:graphs}, the pushing process is:
\begin{equation*}
    \begin{tikzpicture}[basel={-.5}]
 \draw(-0.25,0.25) to (0,0);
    \draw(-0.25,-0.25) to (0,0);
    \draw[\lac \rao](0,0) to (1,0);
    \draw[\lac \rao](1,0) to[bend right=40] (2,0);
    \draw[\lac \rao](1,0) to[bend left=40] (2,0);
    \draw[\lac \rao](2,0) to (3,0);
    \draw(3,0) to (3.25,0.25);
    \draw(3,0) to (3.25,-0.25);
     \draw[markxi={0}](2,0) to (3,0);    
    \draw[markxi={1}](2,0) to (3,0);
    \draw[markxi={0}](1,0) to (3,0);    
    \draw[markxi={0}](0,0) to (4,0);
\end{tikzpicture} \cong
  \begin{tikzpicture}[basel={-.5}]
 \draw(-0.25,0.25) to (0,0);
    \draw(-0.25,-0.25) to (0,0);
    \draw[\rao](0,0) to (1,0);
    \draw(1,0) to[bend right=40] (2,0);
    \draw(1,0) to[bend left=40] (2,0);
    \draw[\lao-](2,0) to (3,0);
    \draw(3,0) to (3.25,0.25);
    \draw(3,0) to (3.25,-0.25);
    \ball{(1.5,0.2)};
    \draw[markxi={0}](2,0) to (3,0);    
    \draw[markxi={1}](2,0) to (3,0);
    \draw[markxi={0}](1,0) to (3,0);    
    \draw[markxi={0}](0,0) to (4,0);
\end{tikzpicture}.
\end{equation*}
From the two qubit rules (xi), we have two options, first, when the ball is not $Z$, we have:
\begin{equation*}
    \begin{tikzpicture}[basel={-.5}]
 \draw(-0.25,0.25) to (0,0);
    \draw(-0.25,-0.25) to (0,0);
    \draw[\rao](0,0) to (1,0);
    \draw(1,0) to[bend right=40] (2,0);
    \draw(1,0) to[bend left=40] (2,0);
    \draw[\lao-](2,0) to (3,0);
    \draw(3,0) to (3.25,0.25);
    \draw(3,0) to (3.25,-0.25);
    \ball{(1.5,0.2)};
    \draw[markxi={0}](2,0) to (3,0);    
    \draw[markxi={1}](2,0) to (3,0);
    \draw[markxi={0}](1,0) to (3,0);    
    \draw[markxi={0}](0,0) to (4,0);
\end{tikzpicture}= \begin{tikzpicture}[basel={-.5}]
 \draw(-0.25,0.25) to (0,0);
    \draw(-0.25,-0.25) to (0,0);
    \draw[\rac](0,0) to (1,0);
    \draw[\lao-](1,0) to (2,0);
    \draw(2,0) to (2.25,0.25);
    \draw(2,0) to (2.25,-0.25);
    \draw[markxi={0}](2,0) to (3,0); 
    \draw[markxi={0}](0,0) to (2,0);
\end{tikzpicture} = \begin{tikzpicture}[basel={-.5}]
 \draw[markx={1}](-0.25,0.25) to (0,0);
    \draw(-0.25,-0.25) to (0,0);
    \draw[\lao \rac](0,0) to (1,0);
    \draw[markx={0}](1,0) to (1.25,0.25);
    \draw(1,0) to (1.25,-0.25);
\end{tikzpicture} ,
\end{equation*}
 the state is equivalent to the one  obtained with one stitch. But for the case with $Z$ we have:
 \begin{equation*}
    \begin{tikzpicture}[basel={-.5}]
 \draw[markx={1}](-0.25,0.25) to (0,0);
    \draw(-0.25,-0.25) to (0,0);
    \draw[\rao](0,0) to (1,0);
    \draw[markx={0},markx={1}](1,0) to[bend right=40] (2,0);
    \draw[marko={0.5}](1,0) to[bend left=40] (2,0);
    \draw[\lao-](2,0) to (3,0);
    \draw[markx={0}](3,0) to (3.25,0.25);
    \draw(3,0) to (3.25,-0.25);
\end{tikzpicture}=\begin{tikzpicture}[basel={-.5}]
 \draw[markx={1}](-0.25,0.25) to (0,0);
    \draw(-0.25,-0.25) to (0,0);
    \draw[\rao](0,0) to (0.5,0);
    \qubitl{(0.75,0)};
    \qubitr{(1.85,0)}{1};    
    \draw[\lao-](2.5,0) to (3,0);
    \draw[markx={0}](3,0) to (3.25,0.25);
    \draw(3,0) to (3.25,-0.25);
\end{tikzpicture} = \left( v_1 \cdot \begin{tikzpicture}[basel={-.5}]
 \draw[markx={0.5}] (0,0.5) to[bend left=90]  (0,-0.5) ; 
\end{tikzpicture}+\begin{tikzpicture}[basel={-.5}]
 \qubitl{(0,0.2)}  ;
 \draw[markx={1}] (-0.5,0.2) to (-0.25,0.2);
  \qubitl{(0,-0.2)}  ;
   \draw[markx={1}] (-0.5,-0.2) to (-0.25,-0.2);
\end{tikzpicture}  \right)\left( v_2 \cdot \begin{tikzpicture}[basel={-.5}]
 \draw[markx={0.5}] (0,0.5) to[bend right=90]  (0,-0.5) ; 
\end{tikzpicture}+\begin{tikzpicture}[basel={-.5}]
 \qubitr{(0,0.2)}  ;
  \draw[markx={1}] (0.9,0.2) to (0.65,0.2);
   \draw[markx={1}] (0.9,-0.2) to (0.65,-0.2);
  \qubitr{(0,-0.2)}  ;
\end{tikzpicture}  \right)
\end{equation*}
This state is new in the construction and can be seen clearly that corresponds to the product of two bipartite entangled states. In the classification of SLOCC families, this state is int the subfamily $G_{a000} (G_{abcd},b=0,c=0,d=0)$ .The correspondence with the representative is given by:
\begin{equation*}
    \left(\begin{array}{cc}
       0  & 1/v_1 \\
       1/v_1 & -1/v_1^2 
    \end{array} \right) \otimes I \otimes \left(\begin{array}{cc}
       0  & 1/v_2 \\
       1/v_2 & -1/v_2^2 
    \end{array} \right)   \otimes I \cdot \left( v_1 \cdot \begin{tikzpicture}[basel={-.5}]
 \draw[markx={0.5}] (0,0.5) to[bend left=90]  (0,-0.5) ; 
\end{tikzpicture}+\begin{tikzpicture}[basel={-.5}]
 \qubitl{(0,0.2)}  ;
 \draw[markx={1}] (-0.5,0.2) to (-0.25,0.2);
  \qubitl{(0,-0.2)}  ;
   \draw[markx={1}] (-0.5,-0.2) to (-0.25,-0.2);
\end{tikzpicture}  \right)\left( v_2 \cdot \begin{tikzpicture}[basel={-.5}]
 \draw[markx={0.5}] (0,0.5) to[bend right=90]  (0,-0.5) ; 
\end{tikzpicture}+\begin{tikzpicture}[basel={-.5}]
 \qubitr{(0,0.2)}  ;
  \draw[markx={1}] (0.9,0.2) to (0.65,0.2);
   \draw[markx={1}] (0.9,-0.2) to (0.65,-0.2);
  \qubitr{(0,-0.2)}  ;
\end{tikzpicture}  \right)= \frac{2}{a} \ket{G_{a000}}.
\end{equation*}
Any other configuration leads to states previously obtained. For the third graph in Equation \eqref{eq:graphs}, the case with all the stitches being $\Phi$ stitches is:
\begin{equation}
 \begin{tikzpicture}[basel={-.5}]
    \draw[\lac \rao](0,0) to (1,0);
    \draw[\lac \rao](1,0) to (1.5,0.5);
    \draw[\lac \rao](1,0) to (1.5,-0.5);
    \draw[\lac \rao](1.5,0.5) to (1.5,-0.5);
    \draw[markx={0}](1.5,0.5) to (1.75,0.75);
    \draw[markx={0}](1.5,-0.5) to (1.75,-0.75);
    \draw[markx={1}](-0.25,0.25) to (0,0);
    \draw(-0.25,-0.25) to (0,0);
    \draw[markxi={1}](0,0) to (1,0);
    \end{tikzpicture}  \cong \begin{tikzpicture}[basel={-.5}]
    \draw[\rao](0,0) to (1,0);
    \draw[markx={0},markx={1}](1,0) to (1.5,0.5);
    \draw[markx={1}](1,0) to (1.5,-0.5);
    \draw(1.5,0.5) to (1.5,-0.5);
    \draw(1.5,0.5) to (1.75,0.75);
    \draw(1.5,-0.5) to (1.75,-0.75);
    \draw[markx={1}](-0.25,0.25) to (0,0);
    \draw(-0.25,-0.25) to (0,0);
    \end{tikzpicture}. 
    \label{eq:KiteTree}
\end{equation}
In this case, the parameter of the arrow works as a switch between subfamilies:
\begin{equation*}
    \begin{tikzpicture}[basel={-.5}]
    \draw[\rao](0,0) to (1,0);
    \draw[markx={0},markx={1}](1,0) to (1.5,0.5);
    \draw[markx={1}](1,0) to (1.5,-0.5);
    \draw(1.5,0.5) to (1.5,-0.5);
    \draw(1.5,0.5) to (1.75,0.75);
    \draw(1.5,-0.5) to (1.75,-0.75);
    \draw[markx={1}](-0.25,0.25) to (0,0);
    \draw(-0.25,-0.25) to (0,0);
    \end{tikzpicture} = \left\{\begin{tikzpicture}[basel={-.5},every node/.style={scale=0.6}]
    \draw[\rao](0,0) to (1,0);
    \draw[markx={0},markx={1}](1,0) to (1.5,0.5);
    \draw[markx={1}](1,0) to (1.5,-0.5);
    \draw(1.5,0.5) to (1.5,-0.5);
    \draw(1.5,0.5) to (1.75,0.75);
    \draw(1.5,-0.5) to (1.75,-0.75);
    \draw[markx={1}](-0.25,0.25) to (0,0);
    \draw(-0.25,-0.25) to (0,0);
    \draw (1,0.5) node {$v=1$};
    \end{tikzpicture}  \in L_{0_4} , \quad \begin{tikzpicture}[basel={-.5},every node/.style={scale=0.6}]
    \draw[\rao](0,0) to (1,0);
    \draw[markx={0},markx={1}](1,0) to (1.5,0.5);
    \draw[markx={1}](1,0) to (1.5,-0.5);
    \draw(1.5,0.5) to (1.5,-0.5);
    \draw(1.5,0.5) to (1.75,0.75);
    \draw(1.5,-0.5) to (1.75,-0.75);
    \draw[markx={1}](-0.25,0.25) to (0,0);
    \draw(-0.25,-0.25) to (0,0);
    \draw (1,0.5) node {$v\neq 1$};
    \end{tikzpicture}  \in L_{a_2a_2}\right\}.
\end{equation*}
The first case in the previous expression belongs to the subfamily $L_{0_4}(L_{a_4},a=0)$, and the second one belongs to $L_{a_2a_2}(L_{a_2,b_2},b=a)$. The mappings for both states with their respective representatives are given by:
\begin{equation*}
    I \otimes  \left( \begin{array}{cc}
      0   & 1 \\
       -1  & 1/2
    \end{array}\right)  \otimes  \left( \begin{array}{cc}
      1   & 1 \\
       -1  & 1
    \end{array}\right) \otimes  \left( \begin{array}{cc}
     2   & -2 \\
       i & i
    \end{array}\right)  \cdot \begin{tikzpicture}[basel={-.5},every node/.style={scale=0.6}]
    \draw[\rao](0,0) to (1,0);
    \draw[markx={0},markx={1}](1,0) to (1.5,0.5);
    \draw[markx={1}](1,0) to (1.5,-0.5);
    \draw(1.5,0.5) to (1.5,-0.5);
    \draw(1.5,0.5) to (1.75,0.75);
    \draw(1.5,-0.5) to (1.75,-0.75);
    \draw[markx={1}](-0.25,0.25) to (0,0);
    \draw(-0.25,-0.25) to (0,0);
    \draw (1,0.5) node {$1$};
    \end{tikzpicture} = U(23) \left( 4\ket{L_{0_4}}\right),
    \end{equation*}
        \begin{equation*}
    \left( \begin{array}{cc}
      0   & \frac{a}{1-v^2} \\
       1  & 0
    \end{array}\right) \otimes  \left( \begin{array}{cc}
      1   & \frac{v}{1-v^2} \\
       0  & \frac{a}{1-v^2}
    \end{array}\right)  \otimes \left( \begin{array}{cc}
     -v  & 1 \\
      1 & -v
    \end{array}\right) \otimes I \cdot \begin{tikzpicture}[basel={-.5},every node/.style={scale=0.6}]
    \draw[\rao](0,0) to (1,0);
    \draw[markx={0},markx={1}](1,0) to (1.5,0.5);
    \draw[markx={1}](1,0) to (1.5,-0.5);
    \draw(1.5,0.5) to (1.5,-0.5);
    \draw(1.5,0.5) to (1.75,0.75);
    \draw(1.5,-0.5) to (1.75,-0.75);
    \draw[markx={1}](-0.25,0.25) to (0,0);
    \draw(-0.25,-0.25) to (0,0);
    \draw (1,0.5) node {$v$};
    \end{tikzpicture} =  U(231) \ket{L_{a_2a_2}}.
    \end{equation*}
    Where $U(\pi)$ refers to a permutation that must be applied over the representative state. We can note how adding more structure to the graph allows for new subfamilies. With this graph, these two are the only new subfamilies. The last graph in Equation \eqref{eq:graphs} can be cleaned as:
    \begin{equation*}
        \begin{tikzpicture}[basel={-.5}]
        \draw[markx={1}] (-0.25,0.25) to (0,0);
        \draw[\lac \rao] (0,0) to (1,0);
        \draw [markx={0}](1,0) to (1.25,0.25);
        \draw[\lac \rao] (0,0) to (0,-1);
        \draw [markx={1}](-0.25,-1.25) to (0,-1);
        \draw[\lac \rao] (0,-1) to (1,-1);
        \draw[\lac \rao] (1,0) to (1,-1);
        \draw [markx={0}](1,-1) to (1.25,-1.25);
    \end{tikzpicture} \cong  \begin{tikzpicture}[basel={-.5}]
        \draw (-0.25,0.25) to (0,0);
        \draw[markx={1},markx={0}] (0,0) to (1,0);
        \draw (1,0) to (1.25,0.25);
        \draw (0,0) to (0,-1);
        \draw (-0.25,-1.25) to (0,-1);
        \draw [markx={1},markx={0}] (0,-1) to (1,-1);
        \draw(1,0) to (1,-1);
        \draw (1,-1) to (1.25,-1.25);
        \ball{ (0.5,0)};
    \end{tikzpicture}.
\end{equation*}
This is the first case where the parameter works not only as a switch between SLOCC families, but also as a continuous parameter. We can achieve three new subfamilies with this graph. The first one when the argument of the ball is 1 (which means no ball):
\begin{equation*}
I \otimes \left( \begin{array}{cc}
      0   & 1 \\
       1/\sqrt{2}  & 0
    \end{array}\right) \otimes \left( \begin{array}{cc}
      1  & 0 \\
       0  & \sqrt{2}
    \end{array}\right) \otimes X 
  \,\cdot \, \begin{tikzpicture}[basel={-.5}]
        \draw (-0.25,0.25) to (0,0);
        \draw[markx={1},markx={0}]  (0,0) to (1,0);
        \draw (1,0) to (1.25,0.25);
        \draw (0,0) to (0,-1);
        \draw (-0.25,-1.25) to (0,-1);
        \draw[markx={1},markx={0}]  (0,-1) to (1,-1);
        \draw (1,0) to (1,-1);
        \draw (1,-1) to (1.25,-1.25);  \end{tikzpicture}= \pi_{(24)}\left( \frac{2}{a} \ket{G_{a(a/\sqrt{2})(a/\sqrt{2})0}}\right).,
\end{equation*}
where $G_{a(a/\sqrt{2})(a/\sqrt{2})0}$ is a subfamily of $G_{abcd}$ where the parameter $d$ is zero, and the parameters $b$ and $c$ depend on $a$. The second subfamily, when the ball is a $Z$-ball:
\begin{equation*}
Y \otimes \left( \begin{array}{cc}
      0   & 1 \\
      \frac{(1+I)a}{2}  & 0
    \end{array}\right) \otimes \left( \begin{array}{cc}
    \frac{(I-1)a}{2} & 0 \\
       0  & 1
    \end{array}\right) \otimes \left(\begin{array}{cc}
    -I & 0 \\
       0  & 1
    \end{array}\right) 
   \,\cdot \, \begin{tikzpicture}[basel={-.5}]
        \draw (-0.25,0.25) to (0,0);
        \draw[markx={1},markx={0},marko={0.5}] (0,0) to (1,0);
        \draw (1,0) to (1.25,0.25);
        \draw (0,0) to (0,-1);
        \draw (-0.25,-1.25) to (0,-1);
        \draw [markx={1},markx={0}] (0,-1) to (1,-1);
        \draw (1,0) to (1,-1);
        \draw (1,-1) to (1.25,-1.25);  \end{tikzpicture} = \pi_{(24)} \ket{L_{a(ia)0_2}},
\end{equation*}
with $L_{a(ia)0_2}$ a subfamily of $L_{abcd}$. And the last case:
{\small
\begin{equation*}
 \left( \begin{array}{cc}      
         1& 1 \\
      1  & -1
    \end{array}\right)  \otimes \left( \begin{array}{cc}
      a   & \frac{a^2-b^2}{4b} \\
       a &  \frac{b^2-a^2}{4b} 
    \end{array}\right) \otimes\left( \begin{array}{cc}
      \frac{1}{a+b}  & \frac{1}{2\sqrt{ab}}\\
        \frac{1}{a+b}   &- \frac{1}{2\sqrt{ab}}
    \end{array}\right) \otimes \left( \begin{array}{cc}
      \frac{1}{2\sqrt{ab}}& \frac{1}{a-b}   \\
         \frac{1}{2\sqrt{ab}}   &\frac{1}{b-a}
    \end{array}\right)   \, \cdot \, \begin{tikzpicture}[basel={-1.5},,every node/.style={scale=0.6}]
        \draw (-0.25,0.25) to (0,0);
        \draw [markx={1},markx={0}] (0,0) to (1,0);
        \draw (1,0) to (1.25,0.25);
        \draw (0,0) to (0,-1);
        \draw (-0.25,-1.25) to (0,-1);
        \draw [markx={1},markx={0}] (0,-1) to (1,-1);
        \draw (1,0) to (1,-1);
        \draw (1,-1) to (1.25,-1.25); \ball{(0.5,0)};
    \draw (0.5,0.5) node {$\frac{2\sqrt{ab}}{(a-b)}$};
    \end{tikzpicture} = \frac{2}{b(a-b)} \ket{G_{ab(\sqrt{ab})(\sqrt{ab})}},
\end{equation*}}
with $G_{ab(\sqrt{ab})(\sqrt{ab})}$ a subfamily of $G_{abcd}$, with $c$ and $d$ as functions of $a$ and $b$. We can see from here how the complexity of the subfamilies and the mapping grows with the complexity of the graphs. Up to this point we have analyzed all the graphs obtained with four stitches, however, no full family of the classification could be obtained with these configurations. When considering graphs with seven stitches and six $W$ states, a considerable big number of subfamilies appear, because of this in Section \ref{FourQubitsVerst} we only show the graphs for each full SLOCC family. It would be interesting to study completely the entanglement properties that share states that can be obtained from the same graph. In terms of the invariants that we propose in Section \ref{Invariants}, it is clear that the topology of the graph is related with the calculation of invariants of the states, which are used to measure the entanglement of the system.
\chapter{From W-state stitching to ZW calculus}
\label{Appendix04}

The graphical process that we introduced for building multiqubit states in \ref{Chapter4} named W-state stitching, is closely related to those known as ZX-calculus \cite{ZX}, ZW-calculus \cite{ZW}, and ZXW calculus \cite{ZXW}. We will focus this appendix on showing how the relation with the ZW-calculus can be made explicit, and hence, we can ensure that any multiqubit state can be obtained with the W-state stitching by using the ZW completeness for qubit systems.\\

\noindent The scheme of the ZW-calculus consists in a set of generators, which will be the most basic diagrams, and a set of rules which indicates the operations that can be performed between the basic diagrams. In this appendix we will show the equivalence between the generators of ZW-calculus and objects in our construction, which are the only necessary pieces to prove the completeness. The set of generators of the ZW-calculus are the following :
\begin{equation}
\begin{gathered}
  \includegraphics[scale=0.1,valign=c]{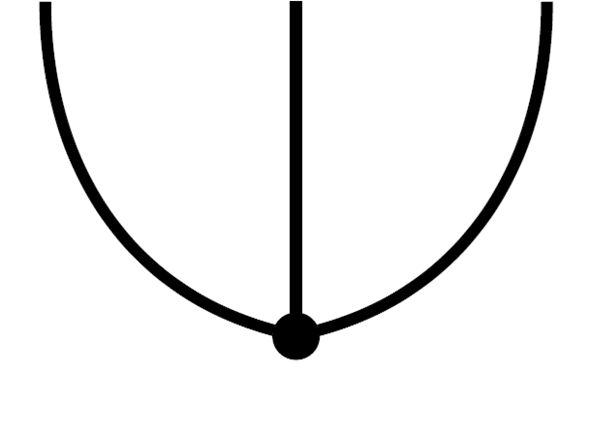} = \ket{001}+\ket{010}+\ket{100}, \\
   \includegraphics[scale=0.1,valign=c]{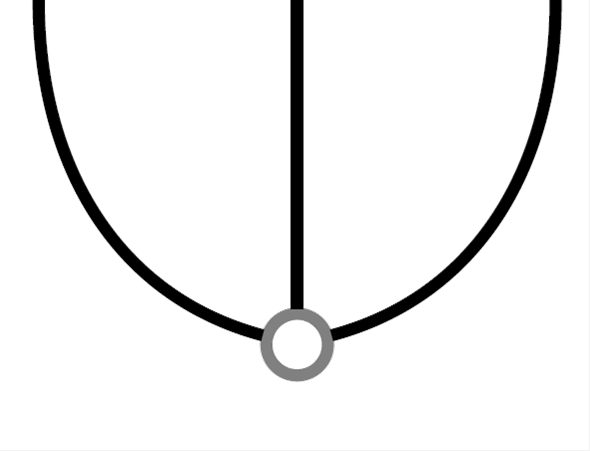} = \ket{000}+\ket{111}, \\
   \includegraphics[scale=0.1,valign=c]{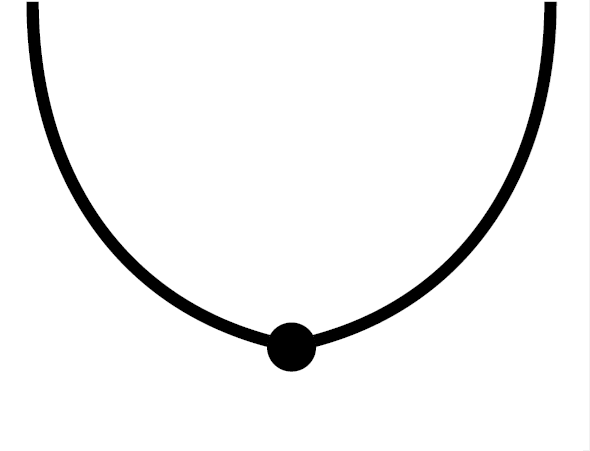} = \ket{01}+\ket{10},\\
   \includegraphics[scale=0.1,valign=c]{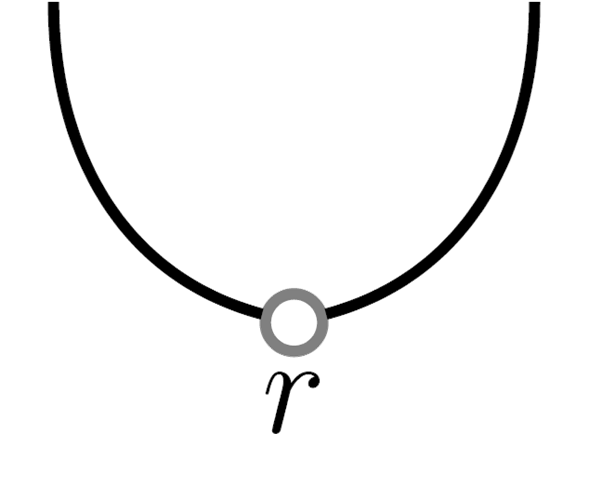} = \ket{00}+r\ket{11}, \\
   \includegraphics[scale=0.1,valign=c]{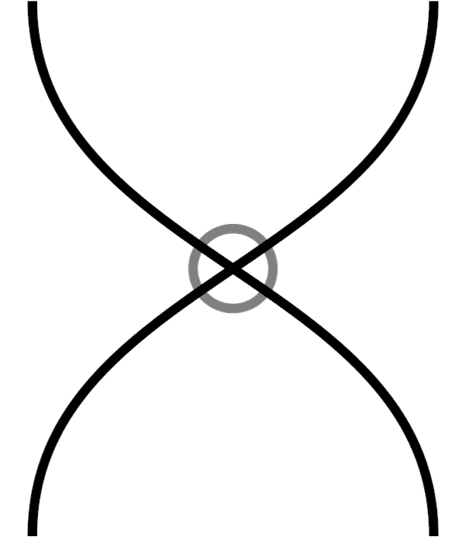} = \ket{00}\bra{00}+\ket{01}\bra{10}+ \ket{10}\bra{01} - \ket{11}\bra{11}.
  \end{gathered}
  \label{eq:Generators}
\end{equation}
In this notation, lines pointing up correspond to indices of a ket, while lines pointing down are indices of a bra. Let us now to express the relation of each of the objects in our notation. The first genertor is simply the W state, that compared with our notation is the same representation besides of the meaning of the direction of the lines:
\begin{equation}
    \includegraphics[scale=0.1,valign=c]{Figures/W.png} \Rightarrow \begin{tikzpicture}[basel={-.5}]
    \draw (-0.5,0.5) to (0,0);
     \draw (-0.5,-0.5) to (0,0);
      \draw[markx={1}] (0.7,0) to (0,0);
    \end{tikzpicture}.
\end{equation}
The second generator in Equation \eqref{eq:Generators} is the GHZ state, that according to Equation \eqref{eq:GHZstate} is obtained from local operations of the graph $\Kite{0.3}$. The local operations are $\sigma_z \cdot H$, which can be parametrized graphically as: 
\begin{equation*}
\sigma_z \cdot H = \frac{1}{\sqrt{2}}\left( \begin{array}{cc}
1 & 1 \\ 
-1 & 1
\end{array} \right) =\begin{tikzpicture}[basel={-1}]
\draw[\lao \rac] (0,0) to (1,0);
\draw (0,0.3) node[scale=0.5] {$-1$};
\draw (1,0.4) node[scale=0.5] {$\frac{1}{\sqrt{2}},\frac{1}{\sqrt{2}} $}; \draw (1.1,0) node[scale=1.5] {$\wr$};
\end{tikzpicture}.
\end{equation*}
Then, the $GHZ$ state can be represented graphically in both notations as:
\begin{equation*}
    \includegraphics[scale=0.1,valign=c]{Figures/GHZ.png} \Rightarrow  \begin{tikzpicture}[baseline={([yshift=-.5ex]current bounding box.center)}]
   \draw[markx={1}] (0,0) -- (0.5,0) ; 
   \draw [\lac \rao](0.5,0) -- (1,0.5) ;
   \draw[markx={1}] (0,0) -- (-0.5,0);
   \draw[\lao \rac] (-1,0.5)to (-0.5,0);\draw [markx={1}](-0.5,0) -- (0,-1);
   \draw (0.5,0) -- (0,-1);\draw[\lao \rac](0,-2)to(0,-1); 
\end{tikzpicture},
\end{equation*}
where all the black arrows have parameters $v=w=\frac{1}{\sqrt{2}}$, while the white arrows have parameters $v=-1$.\\

\noindent The third generator is simply $\ket{W_2}=\ket{\Psi^+}$ that we represent as:
\begin{equation}
 \includegraphics[scale=0.1,valign=c]{Figures/W2.png} \Rightarrow \begin{tikzpicture} \draw[markx={0.5}](0,0) to (1,0);
 \end{tikzpicture}.
\end{equation}
The next generator can be easily built in our notation equally  with a line with a ball as:
\begin{equation}
\includegraphics[scale=0.1,valign=c]{Figures/rball.png} \Rightarrow  \sqrt{r} \cdot  \begin{tikzpicture}
\draw (0,0) to (1,0);
\draw (0.5,0.5) node {$\frac{1}{\sqrt{r}}$};
 \ball{(0.5,0)};
 \end{tikzpicture}.
\end{equation}
The last generator is the fermionic SWAP operation, and it is the hardest one to translate. However, we can achieve this in many different parametrizations. We will only present the simplest one that we found so far:
\begin{equation}
\includegraphics[scale=0.1,valign=c]{Figures/Swap.png} \Rightarrow \frac{1}{\sqrt{2}} \cdot \scalebox{1.4}{
\begin{tikzpicture}[basel={-.5}]
    \draw[\lao \rao](0,0) to (1,0);
    \draw[markx={0},markx={1}](1,0) to (1.5,0.5);
    \draw[markx={1}](1,0) to (1.5,-0.5);
    \draw(1.5,0.5) to (1.5,-0.5);
    \draw[\lao \rao](1.5,0.5) to (2,1);
    \draw[\lac \rao](1.5,-0.5) to (2,-1);
    \draw[markx={1}](-0.25,0.25) to (0,0);
    \draw [markx={1}] (0,0) to (-0.5,-0.5);
    \draw [markx={1}] (0,0) to (-0.5,0.5);
    \draw[\lac \rao] (-0.5,0.5) to (-1,1);
    \draw[\lac \rao] (-0.5,-0.5) to (-1,-1);
    \draw (-0.5,0.5) to (-0.5,-0.5);
    \draw (0,0.2) node [scale=0.5]{$-1$};
     \draw (1,0.3) node[scale=0.5] {$\frac{1}{2}$};
     \draw (-1,1.2) node[scale=0.5] {$\frac{1}{4}$};
     \draw (-0.4,0.8) node[scale=0.5] {$\sqrt{2},\sqrt{2}$};
     
     \draw (-0.4,-0.8) node[scale=0.5] {$\frac{1}{2},\frac{1}{2}$};
      \draw (-1,-1.2) node[scale=0.5] {$-2$};
      
       \draw (2,1.2) node[scale=0.5] {$\frac{1}{2}$};
      \draw (1.5,0.8) node[scale=0.5] {$-1$};
      
       \draw (2,-1.2) node[scale=0.5] {$2$};
      \draw (1.5,-0.8) node[scale=0.5] {$\frac{1}{2},-\frac{1}{2}$};
      \draw (-1,-1) node[scale=1,rotate=45] {$\wr$};
        \draw (2,-1) node[scale=1,rotate=-45] {$\wr$};
    \end{tikzpicture}}.
\end{equation}
With this all the generators of ZW calculus can be translated to our representation. This means that as ZW calculus is complete, and any qubit operation can be implemented, then, with our construction we can build any multiqubit state. We already showed how any state of three and four qubits can be obtained explicitly; but with this connection we can ensure that any state can be reproduced graphically.

\printbibliography

\end{document}